\documentclass[12pt]{article}
\pdfoutput=1

\usepackage[usenames,dvipsnames,svgnames,table]{xcolor} 
\usepackage[obeyspaces,hyphens,spaces]{url}
\usepackage{jcapmod}
\usepackage{verbatim}

\usepackage{booktabs}
\usepackage[english]{babel}
\usepackage{amsmath, amssymb, amsbsy, amstext, amsthm, simplewick}
\usepackage{hyperref}
\usepackage{graphicx}
\usepackage{amsfonts}
\usepackage{upgreek}
\usepackage{framed}
\usepackage{tensor}
\usepackage{pifont}
\usepackage{latexsym, mathrsfs}
\usepackage{array}
\usepackage{hyperref}
\usepackage{xspace}
\usepackage{longtable}
\usepackage{multirow}
\usepackage{cases}
\usepackage{empheq}

\usepackage{bm}
\usepackage{bbm}

\usepackage{afterpage}
\usepackage{setspace}

\usepackage{braket}

\usepackage[load-configurations=astronomy, range-units=brackets, range-phrase=-, per-mode=reciprocal, mode=math]{siunitx}
\usepackage{threeparttable}
\usepackage{subfig}

\usepackage{ragged2e}

\usepackage{hhline}
\usepackage{array}
\newcommand{\hll}[1]{\colorbox{Gainsboro}{$\displaystyle #1$}}
\newcolumntype{P}[1]{>{\centering\arraybackslash}p{#1}}
\usepackage{booktabs}
\usepackage{tikz}
\usetikzlibrary{calc,shadings,patterns,tikzmark,fadings}

\definecolor{Blue}{rgb}{0.25, 0.41, 0.88}
\definecolor{Red}{rgb}{0.92,0.,0.}
\definecolor{darkorange}{rgb}{1.0,0.549,0.}
\definecolor{cobalt}{RGB}{44, 98, 120}
\definecolor{Mathematica1}{rgb}{0.368417, 0.506779, 0.709798}
\definecolor{Mathematica2}{rgb}{0.880722, 0.611041, 0.142051}
\definecolor{Mathematica3}{rgb}{0.560181, 0.691569, 0.194885}
\definecolor{Mathematica4}{rgb}{0.922526, 0.385626, 0.209179}
\definecolor{Mathematica5}{rgb}{0.528488, 0.470624, 0.701351}
\definecolor{Mathematica6}{rgb}{0.772079, 0.431554, 0.102387}
\definecolor{Mathematica7}{rgb}{0.363898, 0.618501, 0.782349}
\definecolor{Mathematica8}{rgb}{1, 0.75, 0}
\definecolor{Mathematica9}{rgb}{0.647624, 0.37816, 0.614037}
\definecolor{plotBlue}{RGB}{94, 130, 181}
\definecolor{plotRed}{RGB}{233, 85, 54}
\definecolor{plotGreen}{RGB}{142, 176, 50}
\definecolor{plotPurple}{RGB}{135, 120, 178}

\newcolumntype{C}[1]{>{\centering\let\newline\\\arraybackslash\hspace{0pt}}m{#1}}

\def\H{{\cal H}}

\def\x{{\bf x}}

\makeatletter
\newlength{\apb@width}
\newcommand{\autoparbox}[2][c]{\settowidth{\apb@width}{#2}\parbox[#1]{\apb@width}{#2}}

\makeatother

\makeatletter
\newsavebox\myboxA
\newsavebox\myboxB
\newlength\mylenA

\newcommand*\xoverline[2][0.75]{
    \sbox{\myboxA}{$\m@th#2$}%
    \setbox\myboxB\null
    \ht\myboxB=\ht\myboxA%
    \dp\myboxB=\dp\myboxA%
    \wd\myboxB=#1\wd\myboxA
    \sbox\myboxB{$\m@th\overline{\copy\myboxB}$}
    \setlength\mylenA{\the\wd\myboxA}
    \addtolength\mylenA{-\the\wd\myboxB}%
    \ifdim\wd\myboxB<\wd\myboxA%
       \rlap{\hskip 0.5\mylenA\usebox\myboxB}{\usebox\myboxA}%
    \else
        \hskip -0.5\mylenA\rlap{\usebox\myboxA}{\hskip 0.5\mylenA\usebox\myboxB}%
    \fi}
\makeatother

\numberwithin{equation}{section}

\def\beq{\begin{equation}}
\def\eeq{\end{equation}}

\def\bea{\begin{eqnarray}}
\def\eea{\end{eqnarray}}

\def\beq{\begin{equation}}
\def\eeq{\end{equation}}
\def\bea{\begin{eqnarray}}
\def\eea{\end{eqnarray}}

\def\H{{\cal H}}

\DeclareRobustCommand{\SkipTocEntry}[4]{}

\usepackage{colortbl}
\definecolor{blue2}{cmyk}{1, 0.1, 0.1, 0.1}

\definecolor{pyBlue}{RGB}{31, 119, 180}
\definecolor{pyRed}{RGB}{214, 39, 40}
\definecolor{pyGreen}{RGB}{44, 160, 44}
\definecolor{pyBlue2}{RGB}{0, 111, 237}
\definecolor{pyRed2}{RGB}{224, 52, 36}

\begin{document}

\pagenumbering{roman}
\begin{titlepage}
\baselineskip=14.5pt \thispagestyle{empty}

\bigskip\

\vspace{0cm}
\begin{center}
{\fontsize{40}{40}\selectfont  \bfseries \textcolor{Sepia}{:THE  COSMOLOGICAL OTOC:}}\\ 
\vspace{0.3cm}
{\fontsize{15.8}{15.9}\selectfont  \bfseries \textcolor{Sepia}{Formulating new cosmological micro-canonical correlation functions for random chaotic fluctuations in \\Out-of-Equilibrium Quantum Statistical Field Theory}}
\end{center}
\vspace{0.01cm}
\begin{center}
{\fontsize{15}{15}\selectfont Sayantan Choudhury
		\footnote{{\it  \textcolor{blue}{ This project is the part of the non-profit virtual international research consortium
“Quantum Structures of the Space-Time \& Matter (QASTM)".}} ${}^{}$}	} 
\end{center}

\begin{center}
\vskip1pt
\textit{Quantum Gravity and Unified Theory and Theoretical Cosmology Group, \\Max Planck Institute for Gravitational Physics (Albert Einstein Institute),\\
	Am M$\ddot{u}$hlenberg 1,
	14476 Potsdam-Golm, Germany.}
	\text{Email:~sayantan.choudhury@aei.mpg.de }


\end{center}

\vspace{0.09cm}
\hrule \vspace{0.09cm}
\begin{center}
\noindent {\bf Abstract}
\end{center} 
The out-of-time-ordered correlation (OTOC) function is an important new probe in quantum field theory which is treated as a significant measure of random quantum correlations. In this paper, with the slogan ``Cosmology meets Condensed Matter Physics" we demonstrate a formalism using which for the first time we compute the Cosmological OTOC during the stochastic particle production during inflation and reheating following canonical quantization technique. In this computation, two dynamical time scales are involved, out of them at one time scale the cosmological perturbation variable and for the other the canonically conjugate momentum is defined, which is the strict requirement to define time scale separated quantum operators for OTOC and perfectly consistent with the general definition of OTOC. Most importantly, using the present formalism not only one can study the quantum correlation during stochastic inflation and reheating, but also study quantum correlation for any random events in Cosmology. Next, using the late time exponential decay of cosmological OTOC with respect to the dynamical time scale of our universe which is associated with the canonically conjugate momentum operator in this formalism we study the phenomena of quantum chaos by computing the expression for {\it Lyapunov spectrum}. Further, using the well known Maldacena Shenker Stanford (MSS) bound, on  Lyapunov exponent, $\lambda\leq 2\pi/\beta$, we propose a lower bound on the equilibrium temperature, $T=1/\beta$, at the very late time scale of the universe. On the other hand, with respect to the other time scale with which the perturbation variable is associated, we find decreasing but not exponentially decaying behaviour, which quantifies the random quantum correlation function at out-of-equilibrium. We have also studied the classical limit of the OTOC check the consistency with the large time limiting behaviour of the correlators. Finally, we prove that the normalized version of OTOC is completely independent of the choice of the preferred definition of the cosmological perturbation variable. 
 
\vskip7pt
\hrule
\vskip7pt

\text{Keywords:~~Cosmology beyond the standard model,  Quantum Dissipative Systems,}\\ \text{Stochastic Processes, Effective Field Theories.}

\end{titlepage}

\thispagestyle{empty}
\setcounter{page}{2}
\begin{spacing}{1.03}
\tableofcontents
\end{spacing}

\clearpage
\pagenumbering{arabic}
\setcounter{page}{1}

\begin{figure}[t!]
    \centering
        \centering
        \includegraphics[width=17.5cm,height=17.5cm]{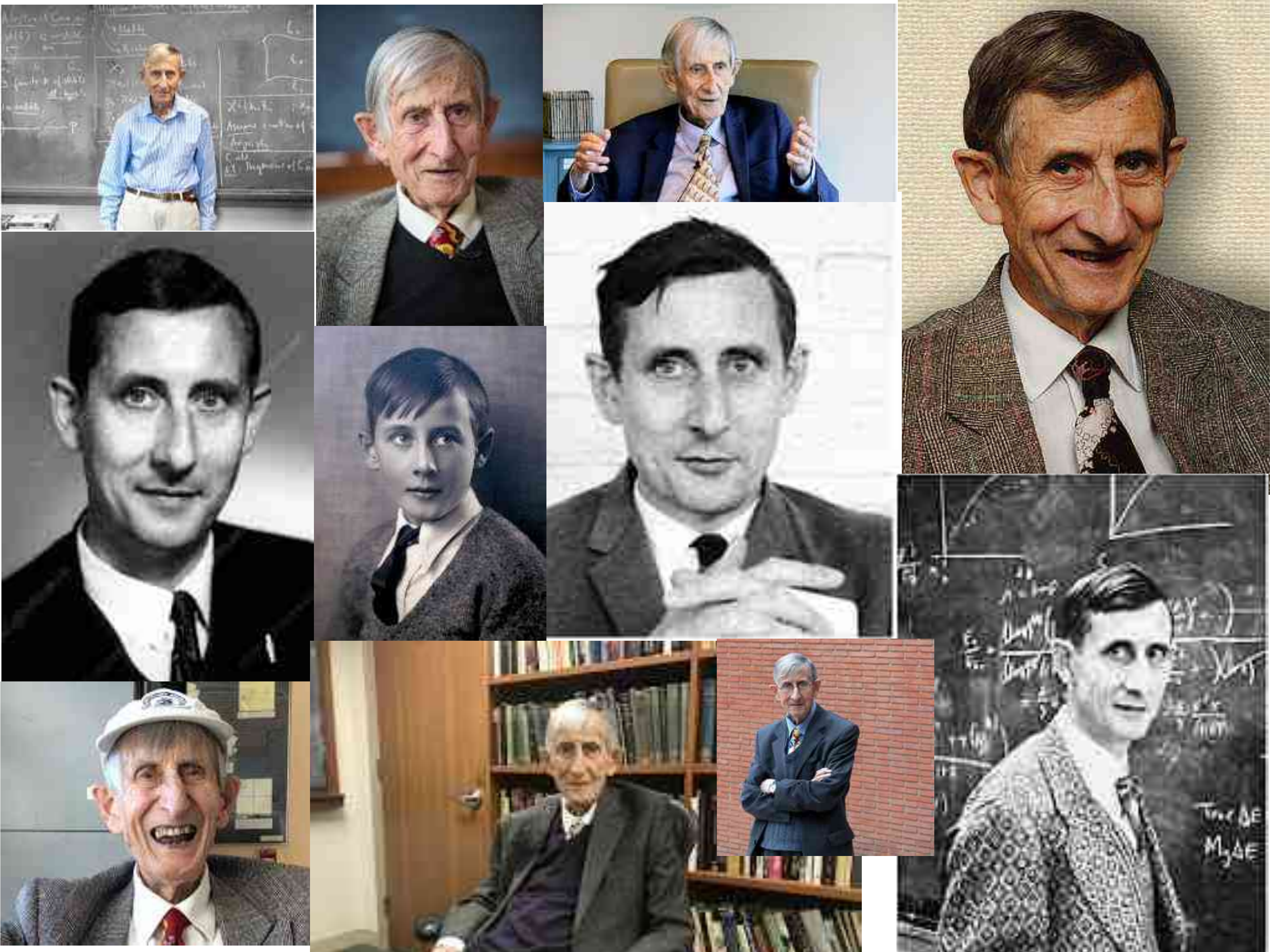}
    This work is written in the memory of the {\it great physicist Professor Freeman J. Dyson} with whom I had the chance to meet personally and discuss my work and understandings during my visit at Institute for Advanced Studies (IAS), Princeton for the purpose of participating at the {\it Workshop on Qubits and Spacetime} which happened on the first week of December, 2019. Here I have to acknowledge the strong support from Professor Juan Martin Maldacena, who had invited me for the mentioned prestigious workshop and also the Max Planck Institute for Gravitational Physics (Albert Einstein Institute), Potsdam, Germany for providing me the financial support to attend this workshop, also for academic visit to the various renowned universities of United States of America (Simons Center for Geometry and Physics-Stony Brook, Princeton, Stanford, Penn State) and to visit at Perimeter Institute for Theoretical Physics (PITP), Ontario, Canada.
      \label{fig:Dyson} 
\end{figure}
\clearpage

\section{Introduction}
\label{sec:introduction}
The out-of-time ordered correlation (OTOC) \cite{Maldacena:2015waa,Hashimoto:2017oit,Chakrabarty:2018dov,Chaudhuri:2018ymp,Chaudhuri:2018ihk,Haehl:2017eob,Akutagawa:2020qbj, Bhagat:2020pcd,Romero-Bermudez:2019vej} functions in the context of quantum field theory at finite temperature is considered as a very strong probe of any kind of stochasticity, randomness and quantum mechanical chaos in the present day research of theoretical physics. Earlier it was only studied in the various condensed matter systems where out-of-equilibrium phenomena plays significant role. The concept of OTOC was first introduced in the computation of superconductivity to describe the vertex correction of current \cite{Larkin}. But for the past few years theoretical physicist are applying this idea to explore various unknown out-of-equilibrium features of various quantum field theories at finite temperature and in the context of bulk gravitational theories. From the very common understanding one can physically interpret OTOC as a quantum mechanical analogue of the  classical version of the sensitiveness against tiny random fluctuations in the initial conditions, and particularly if we get exponential growth in the time dynamics of the OTOC then it is treated as a very strong probe of quantum mechanical chaos~\footnote{In terms of Schr$\ddot{o}$dinger time evolution it is very difficult to describe such quantum phenomena in quantum mechanical systems. }. In terms of the traditional physics one can think of this concept as the theoretical indicator of the energy gaps, and in that connection it is very interesting to investigate that whether or not the OTOC can be treated as a better probe of stochastic randomness and quantum mechanical chaos at out-of-equilibrium. 

To describe this in a more technical way, let us consider two quantum operators $X(t)$ and $Y(0)$, which are separated in time scale and using them OTOC is defined by the following expression \cite{Maldacena:2015waa}:
\begin{equation} \hll{\underline{\textcolor{red}{\bf How~ to~ define~ OTOC?}}~~~C(t):\equiv-\langle \left[X(t),Y(0)\right]^2\rangle_{\beta}=-\frac{1}{Z}~{\rm Tr}\left[\exp\left(-\beta H\right)~\left[X(t),Y(0)\right]^2\right]}~,~~~~~~\end{equation}
where the {\it thermal partition function} is defined as:
\bea \hll{\underline{\textcolor{red}{\bf Partion~function:}}~~~~Z={\rm Tr}\left[\exp\left(-\beta H\right)\right]~~~~{\rm where}~~\beta=\frac{1}{T}~~~{\rm with}~~k_B=1}~.\eea
Specifically in this context commutator of the two time separated operator in quantum mechanics describes the effect of perturbation by the operator $Y$ on the measurement of the operator $X$ on later time scales and the converse statement is also true here. In this construction, we also assume that these two operators have zero one point functions. Now if we fix these operators as, two canonically conjugate operators i.e. in terms of position operator $X(t)=q(t)$ and momentum operator $Y(0)=p(0)$ for a quantum system, then in the semi-classical limit, one can replace the commutator bracket of these two operators, $\left[q(t),p(0)\right]$, will be replaced by the {\it Poisson bracket}, $i\left\{q(t),p(0)\right\}_{\bf PB}=i(\partial q(t)/\partial q(0))$ (in the natural unit system, where we take, $\hbar=1$). For classical system to have chaotic description, $(\partial q(t)/\partial q(0))\sim \exp(\lambda t)$, where $\lambda$ is the {\it Lyapunov exponent}, which is appearing as an outcome of the sensitiveness of the initial conditions. Consequently, if we compute the expression of OTOC out of these two canonically conjugate quantum operators then it will scale with respect to time scale as, $C(t)\sim \exp(2\lambda t)$, to achieve a quantum chaotic description from this set up and here $\lambda$ is treated as {\it quantum Lyapunov exponent}. It is expected from this discussion that, if we can able to quantize the classical chaotic system properly then it can provide a positive numerical value of the {\it quantum Lyapunov exponent} within the framework of OTOC. Following this discussion one can further distinguish between the classical and quantum chaotic system, which comes from the computation of OTOC and the time evolution of this shows that in quantum mechanics OTOC is a quantity which does not grow with the evolutionary time scale of the quantum theory but at the late time scales saturates to a constant value and in this literature this time scale is identified as the {\it Ehrenfest time scale}. one can also describe this characteristic time scale a critical scale after which the quantum mechanical wave function of the theory spreads over the whole system under consideration for this description~\footnote{In a very rough sense, very crudely also this characteristic time scale is identified as the transition time scale which describes the phase transition from a particle description to a wave description of the quantum mechanical wave function.}

 As we have already mentioned in the present day research the concept of OTOC getting more attention due to the fact that, it can be served the purpose of a strong theoretical probe of possible bulk gravitational dual theories, in terms of the framework of AdS/CFT correspondence \cite{Maldacena:1997re,Aharony:1999ti}. One can quote here many many examples to feature the importance of the OTOC's in the context of AdS/CFT. One of the famous examples are the study of the existence of shock waves in the black hole physics \cite{Shenker:2013pqa,Cotler:2016fpe,Klebanov:2016xxf,Roberts:2014isa,Stanford:2014jda,Shenker:2013yza} which can be described by various types of geometries and this specific study finally led to maximum saturation bound of the quantum version of the {\it Lyapunov exponent} ($\lambda$), given by the following expression \cite{Maldacena:2015waa}:
\bea \hll{\underline{\textcolor{red}{\bf M\textcolor{black}{(aldacena)}~S\textcolor{black}{(henker)}~S\textcolor{black}{(tanford)}~bound:}}~~~\lambda \leq \frac{2\pi}{\beta}~~~{\rm with}~~\hbar=1, ~c=1,~k_{B}=1}~.~~~~~\eea
 which is appearing in the expression for OTOCs at finite temperature. This well known bound was established by \textcolor{red}{\it Maldacena, Shenker, Stanford}, which is commonly addressed as \textcolor{red}{\it MSS bound} on quantum chaos these days \cite{Maldacena:2015waa}. In gravitational paradigm, this saturation bound is physically interpreted as the red shift factor near the black hole event horizon having a finite {\it Hawking temperature}. In this description, \textcolor{red}{\it Sachdev-Ye-Kitaev (SYK)} model \cite{Maldacena:2016hyu,Choudhury:2017tax,Fu:2016vas,Witten:2016iux,Li:2017hdt,Turiaci:2017zwd,Rosenhaus:2019mfr,Gross:2017aos,Gross:2017vhb,Gross:2017hcz,Polchinski:2016xgd,Bulycheva:2017ilt,Kim:2019upg,Gurau:2019qag,Gurau:2017qya,Benedetti:2015ara,Gurau:2012vu,Gurau:2012vk,Klebanov:2018fzb} is the most famous example at present which can able to saturate the \textcolor{red}{\it MSS bound} very successfully, which describes the quantum mechanical model of Majorana fermions in presence of infinitely long disorder. In the context of \textcolor{red}{SYK} model the saturation of the \textcolor{red}{\it MSS bound} corresponds to a quantum mechanical description of the black hole paradigm within the framework of AdS/CFT correspondence. Most significantly, it is important to note that the appearance  of {\it quantum Lyapunov exponent} in the computation of OTOC allows to consider the connection with the bulk gravitational physics in the quantum mechanical regime by applying the understanding from AdS/CFT correspondence. From the detailed past study in this literature we already know that any bulk gravitational physics which have their own dual description are described by the strongly quantum mechanical description. So from the present discussion it is evident that, to connect this description with the phenomena of quantum mechanical chaotic picture we explicitly need to have an analogous description of {\it quantum Lyapunov exponent}, which will mimic the role of the well known {\it Lyapunove exponent} as appearing in the description of chaotic dynamical classical systems and to serve this purpose successfully, OTOC is the only strongest probe which can be treated as the physical discriminator of the classical and quantum mechanical description in the present context of discussion.
 
 Now, we will talk about something very unusual from the perspective of applying the framework of the computation of OTOC in a completely different framework in which we are mostly interested in this paper. This is nothing but cosmological application of OTOC from our own universe. Initially anyone can think that this is just a crazy idea and the final result will not give any physically relevant information regarding cosmological OTOC. We are completely ok with this initial thought. But gradually once we proceed in this paper by detailed computation we will try to convince the general readers regarding the strong applicability of OTOC within the framework of Cosmology. Before going to discuss this feature in detail we will first start with some basic understanding from the study of Cosmology, which will surely helps to understand the background physical motivation of this crazy computation of OTOC in cosmological paradigm. In the quantum mechanical description of Cosmology, which is mostly used to describe the early universe physics the most significant quantity that we study are the $N$-point correlation functions of the cosmological scalar fluctuations or the tensor fluctuations or the mixture of them \cite{Maldacena:2002vr, Maldacena:2011nz, Choudhury:2015yna, Choudhury:2014uxa, Choudhury:2017cos, Green:2013rd, Behbahani:2011it, Senatore:2009gt, Senatore:2008wk, Creminelli:2006gc, Lehners:2009qu}. These quantum fluctuations are originated from a geometrical and a very fundamental quantity of the study of Cosmology, which is the well known classical gravitational background metric. To describe the observationally relevant universe in the cosmological scales it is a very general practice to consider the \textcolor{red}{\it Friedmann-Lemaître-Robertson-Walker (FLRW)} background metric of our space-time which is an exact solution of {\it Einstein's field equations} within the framework of {\it General Theory of Relativity} and describes, a {\it homogeneous, isotropic and expanding} universe. However, the most general solution of the metric contains an additional curvature parameter, which one can fix to be zero by considering the present to observational data obtained from various observational probes of Cosmology. So this implies that, \textcolor{red}{\it spatially flat FLRW metric} is sufficient enough to describe most of the observationally consistent aspects within the framework of Cosmology \cite{Baumann:2009ds}. For this reason we will stick to this background metric only for the rest of the discussions of this paper. Using this metric one can study the cosmological perturbation theory from which one can explain the origin of previously mentioned scalar and tensor fluctuations, which are treated to be quantum to describe the physics of early universe and for the late time scale it is considered to be classical in nature. These quantum fluctuations are very fundamental objects in the context of Cosmology from which all $N$-point correlation functions can be computed in Fourier space and these results can be used to probe the physics of early universe as well as to comment on its impact on the present day galaxy and cluster formation in large scale structure non-linear cosmological perturbation theory. Apart from having a great understanding the computation of these correlation functions  of the quantum fluctuations at thermodynamic equilibrium, we have till now have a lot of constraints and limitations from the observational probes. From cosmological observations till now we have information regarding the amplitude of the primordial power spectrum from scalar mode fluctuations~\footnote{This is actually represent the amplitude of the cosmological two-point correlation function of scalar mode quantum fluctuations in Fourier space.} \cite{Aghanim:2018eyx} and about its nearly scale invariance feature with respect to momentum scale~\footnote{This constraint will helps us to determine the feature of the primordial power spectrum for scalar modes in all cosmological momentum scales. Till now from observational probes only the information regarding the spectral index (which is represented by the logarithmic derivative of the logarithm of the power spectrum at the cosmological horizon crossing scale) with high statistical accuracy is available. Any further information, such as running and the running of the running of the scalar spectral index are not available with any significant statistical accuracy.}. Additionally, we have information regarding upper bound on the tensor-to-scalar ratio~\footnote{It represents the ratio of the amplitude of the power spectrum from tensor and scalar modes fluctuation and in terms of quantum description it represents the ratio of the two-point cosmological correlations. This is a very important observable in Cosmology the determination of which with high statistical accuracy will directly confirm the existence of gravitational waves in primordial cosmology and can also further able to fix the origin of primordial fluctuations by exactly determining the scale of inflation.} at the cosmological horizon scale from the observational probes. No other information regarding the higher point (three and four-point etc.) cosmological correlation functions are available till now with significant statistical accuracy which can also be treated as the probe of new physical phenomena through non-Gaussianity in the primordial Cosmology. This implies that, just using the present observational probes one cannot able to distinguish amongst various possible origin of primordial quantum mechanical fluctuations and rule out models which describes inflationary paradigm within the framework of Cosmology. So one can immediately ask about a questions regarding the possible options left to explore the physics of primordial quantum fluctuations:
 \begin{itemize}
 \item \underline{\textcolor{red}{\bf Possibility I:}}\\
 The first possibility pointing towards the future observational aspects which can be probed by different ongoing and upcoming experiments to verify various theoretical features of primordial Cosmology. The most significant quantity using which it is possible to understand the underlying quantum field theory origin of the primordial physics is tensor-to-scalar ratio. Detection of this observable with high statistical accuracy will provide us the information regarding the generation of primordial gravitational waves, which will further confirms the exact origin of the primordial quantum mechanical fluctuations by exactly estimating the scale of inflation. Not only discriminating different  frameworks of inflationary paradigm, but also the existence of  alternatives to inflationary frameworks i.e. bouncing cosmology \cite{Lin:2017fec, Brandenberger:2017pjz, Hipolito-Ricaldi:2016kqq, Ferreira:2016gfg, Brandenberger:2016egn, Li:2014era,Brandenberger:2013zea, Cai:2013vm, Brandenberger:2012zb, Cai:2012va, Cai:2011zx, Karouby:2010wt,Brandenberger:2010dk, Gao:2009wn, Cai:2008qw, Brandenberger:2008zz, Cai:2007zv, Alexander:2007zm, Bars:2011aa, Bars:2012mt, Ijjas:2015hcc, Cook:2020oaj, Lehners:2011ig, Koehn:2013upa, Battarra:2014tga, Fertig:2015dva, Lehners:2015mra, Koehn:2015vvy, Fertig:2016czu, Brandenberger:2009rs,Anabalon:2019equ,Quintin:2019orx, Quintin:2019pbm, Quintin:2015rta, Quintin:2016qro, Li:2016xjb, Bramberger:2019oss,Xue:2011nw,Quintin:2014oea,Cai:2014xxa}, cyclic cosmology \cite{Erickson:2006wc,Boyle:2003km,Steinhardt:2001st,Turok:2004yx,Steinhardt:2001pi,Khoury:2003vb}, ekpyrotic scenario \cite{Bramberger:2017cgf,Lehners:2010ug,Lehners:2015efa,Lehners:2015sia, Battarra:2013cha, Battarra:2014xoa,Fertig:2013kwa,Khoury:2011da, Ijjas:2015zma, Levy:2015awa,Ijjas:2014fja,Khoury:2009my, Gratton:2003pe, Turok:2002ee, Brandenberger:2020eyf,Brandenberger:2020wha} etc. can also be verified by the confirmation of the primordial gravity waves. The next important quantity within the framework of primordial Cosmology is study the existence of non-Gaussian features in the quantum mechanical fluctuations and the probability distribution profile of its related correlation functions. For this purpose the study of bispectrum and trispectrum, which are basically representing the momentum dependent amplitude of the three-point and four-point correlation function are very important. In near future through the upcoming cosmological missions if it is possible to detect these non-Gaussian amplitudes with high statistical accuracy then it is further put more stringent constraint on the primordial physics. This is because of the fact that, the probability distribution of such quantum fluctuations in the primordial universe is almost following Gaussian profile and a small but significant deviation from such Gaussianity will be extremely helpful to discriminate amongst various possible theoretical models of inflation which can able to generate significant amount of non-Gaussian amplitude in the context of three-point and four-point correlation functions. We are very hopeful for the detection of these important observables in near future cosmological experiments. For this purpose one needs to upgrade the present experimental tools and techniques or have to wait for upcoming future advanced experiments, which can able to measure these mentioned observables with high statistical accuracy. Now we all know that for a single field inflationary paradigm the amplitude of the three point function is directly related to scalar spectral index, given by \cite{Maldacena:2002vr}~\footnote{In ref.~\cite{Maldacena:2002vr}, the author have considered an additional contribution which is given by the following expression:
 \bea f_{\rm NL}=\frac{5}{12}(1-n_s-f(k)n_{t}),\eea
 where $n_{t}$ is the spectral index of the tensor power spectrum which is appearing from the perturbation of the primordial gravitational waves and not yet observed through CMB observations. On the other hand the scale dependent factor, $f(k)$, is basically capturing some information regarding the bispectrum and the numerical value of this factor will lie within the window, $0\leq f \leq \frac{5}{6}$. Since $n_t =-2\epsilon$ from inflationary paradigm and the contribution from the factor $fn_{t}$ is very small, then one can easily neglect this terms from the full expression for the non-Gaussian amplitude of the three point scalar fluctuations. in future, if we can able to detect the primordial gravitational waves at CMB B-modes and from that can able to extract the individual contribution of the tensor perturbation then adding the contribution of such terms will be more relevant in the present context of discussion.}:
 \bea f_{\rm NL}=\frac{5}{12}(1-n_s),\eea which in this literature is treated as a very strong theoretical probe, commonly known as {\it Maldacena's consistency relation}. Now just from theoretical understanding of inflationary paradigm from this relation the expectation is that the amplitude of the three point non-Gaussian amplitude has to be very small for single filed slow-roll inflation. To detect this very small value very high precision from observation is required. On the other hand, if the cosmological power spectrum have any additional characteristic features, like the running and running of spectral index then theoretically using the mentioned consistency relationship one can translate the statement in terms of the running or the certain scale dependence of the amplitude of the three point function. This fact can be expressed by the two new derived consistency conditions, which are given by, \bea && \frac{df_{\rm NL}(k)}{d\ln k}=-\frac{5}{12}\frac{dn_{s}(k)}{d\ln k}=-\frac{5}{12}\alpha_{s}(k),\\
 && \frac{d^2 f_{\rm NL}(k)}{d\ln k^2}=-\frac{5}{12}\frac{d^2n_{s}(k)}{d\ln k^2}=-\frac{5}{12}\frac{d\alpha_{s}(k)}{d\ln k}=-\frac{5}{12}\beta_{s}(k),\eea 
 which further allows us to write a generic expansion for a scale dependent three point non-Gaussian amplitude as:
 \bea f_{\rm NL}(k)=f_{\rm NL}(k_{\rm CMB})\Biggl(\frac{k}{k_{\rm CMB}}\Biggr)^{\displaystyle \Delta_{\rm NL}(k)-f_{\rm NL}(k_{\rm CMB})},\eea
 where the scale dependent index $\Delta(k)$ is defined as:
 \bea \Delta(k):&=& -\frac{5}{12}\Biggl\{n_{s}(k_{\rm CMB})-1+\alpha_{s}(k_{\rm CMB})\ln\left(\frac{k}{k_{\rm CMB}}\right)\nonumber\\
 &&~~~~~~~~~~~~~~~~~~~~~~~~~~~~~~~~~~~~~~~~+\frac{\beta_{s}(k_{\rm CMB})}{2}\ln^2\left(\frac{k}{k_{\rm CMB}}\right)+\cdots\Biggr\}.\eea
 Here it is important to note that, $k_{\rm CMB}$ is the momentum scale on which the CMB observation is performed and $f_{\rm NL}(k_{\rm CMB})$ is computed in terms of scalar spectral index using the previously mentioned {\it Maldacena's consistency relation}.
 All these three relations can be further written in terms of the slow-roll parameters, which can be constructed from the field derivative of the inflationary potentials or the time derivative from the Hubble parameter. Now as we have mentioned, if we can verify this relation through cosmological observation with high statistical accuracy, then not only the appearance of non-Gaussian fluctuations in the primordial cosmological perturbation can be immediately tested, but also the appearance of the scale dependence of primordial power spectrum through the running and running of the scalar spectral index can also be verified by considering the above mentioned consistency relations. 
 \item \underline{\textcolor{red}{\bf Possibility II:}}\\
 The second possibility is pointing towards the probing of new physical concepts by incorporating additional but significant features within the framework of early universe Cosmology. Till now all of the features are studied by considering the fact that the quantum fluctuations appearing in the primordial universe is at thermal equilibrium. But if the quantum mechanical system which we use to study to describe the early universe Cosmology are not in thermal equilibrium then quantifying the quantum correlation functions within the framework of finite temperature out-of-equilibrium quantum field theory of Cosmology is extremely difficult to compute and till now in the literature of Cosmology no such framework is available using which one can pursue this specific computation. On the other hand, anybody can ask us here why at all computation and quantification of all such quantum mechanical correlation function for the primordial Cosmology describing a out-of-equilibrium is important at all and if these ideas can be provided theoretically then in which stages of the evolutionary scale of our universe this can be really implemented? The answer is very simple and it is already hidden there in the study of the early universe Cosmology. In the following we now explicitly mention about these phenomena where this methodology can be applicable:
 \begin{enumerate}
 \item \underline{\textcolor{red}{\bf Stochasticity and particle production during inflation:}}\\
 During the epoch of inflation\cite{Deshamukhya:2009wc,Ali:2008ij,Panda:2007ie,Panda:2006mw,Choudhury:2003vr,Mazumdar:2001mm,Choudhury:2015pqa,Choudhury:2014kma,Choudhury:2012ib,Choudhury:2012yh,Choudhury:2011jt,Choudhury:2011rz,Choudhury:2013iaa,Choudhury:2015yna,Baumann:2014nda,Baumann:2014cja,Baumann:2015nta,Baumann:2015xxa,Finelli:2016cyd,Baumann:2017jvh,Assassi:2012et,Assassi:2012zq,Baumann:2010nu,Baumann:2010ys,Baumann:2009ds,Choudhury:2011sq,Lemoine:2008zz,Martin:2007ue,Lorenz:2007ze,Martin:2007bw,Martin:2004yj,Martin:2003bt} we describe the physics of inflationary paradigm with scalar field which have a negligible mass compared to the Hubble scale, which is treated as the reference characteristic scale of early universe Cosmology. However, during inflation many particle produces in a stochastic manner which have masses either of the order of Hubble scale or have mass very heavier than the Hubble scale. These particles are commonly known as partially massless or heavy scalar fields whose origin can be explained from the randomness appearing in the quantum mechanical stochastic fluctuation appearing in the early universe where various unknown out-of-equilibrium features play significant role. But it is extremely difficult to quantify or describing this phenomena correctly at the out-of-equilibrium regime of quantum field theory which consistently describe a quantum mechanical theory of primordial Cosmology. Additionally it is important to note that, during the stochastic particle production during inflation due to the presence of randomness noise plays significant role to describe the quantum mechanical phenomena during this particular epoch in the cosmological evolutionary time scale. However, we have a very small understanding regarding the time dependent random noise within the framework of out-of-equilibrium quantum field theory till now. Only in some specific situations if we provide some additional constraints on the two-point and the one-point correlation of the time dependent noise, which have the Gaussian probability distribution profile one can deal with noise. But such analysis within the framework of quantum field theory is not exactly correct which in turn could not able to gives us the correct predictions from the quantum theory of early universe. One the other hand, we really don't know about the nature of the time dependent noise which we are studying in the present context. This means we don't know that whether the time dependent noise have Gaussian or non-Gaussian probability distribution profile, which information can be directly understandable if we really know the exact computation of $N$-point correlation functions of time dependent noise in cosmology. A lot of attempts have been made within the framework of quantum-field theory to study the quantum mechanical features and the related $N$-point correlation functions to study the exact nature, behaviour and magnitude of the noise. However, apart from having the rigorous attempt it is not completely known about the detailed quantum filed theory structure from which one can reliably compute such correlation functions in quantum mechanics. Here lies one of the prime motivations to write this paper. Our expectation from the presented computation of this paper is that the cosmological version of OTOC defined in a specific quantum mechanical vacuum state, which actually describes the initial condition in early universe Cosmology, describes the randomness and stochastic features of quantum mechanical fluctuations in the out-of-equilibrium regime in a perfectly correct fashion. This is just not an arbitrary claim, but also one can appear at such conclusion by considering the basic understandings of the background physical phenomena which describes the particle production mechanism during the epoch of inflation. Throughout our paper we have explicitly established the framework for the first time in Cosmology literature using which one can explicitly perform the computation of the cosmological correlation functions of the random quantum mechanical fluctuations in the out-of-equilibrium regime of quantum field theory. Instead of studying the quantum correlations with the time ordered or the anti-time ordered physics, in the present context cosmological OTOC functions are playing significant role to describe the underlying hidden features of out-of-equilibrium physics to describe the stochastic randomness during the particle production mechanism during inflation~\footnote{Here it is important to note that, the {\it Random Matrix Theory} is an another alternative framework using which one can compute these quantum mechanical cosmological correlation functions within the framework of out-of-equilibrium quantum field theory \cite{Choudhury:2018rjl,Choudhury:2018bcf}.}.

\begin{figure}[t!]
    \centering
        \centering
        \includegraphics[width=17.5cm,height=18cm]{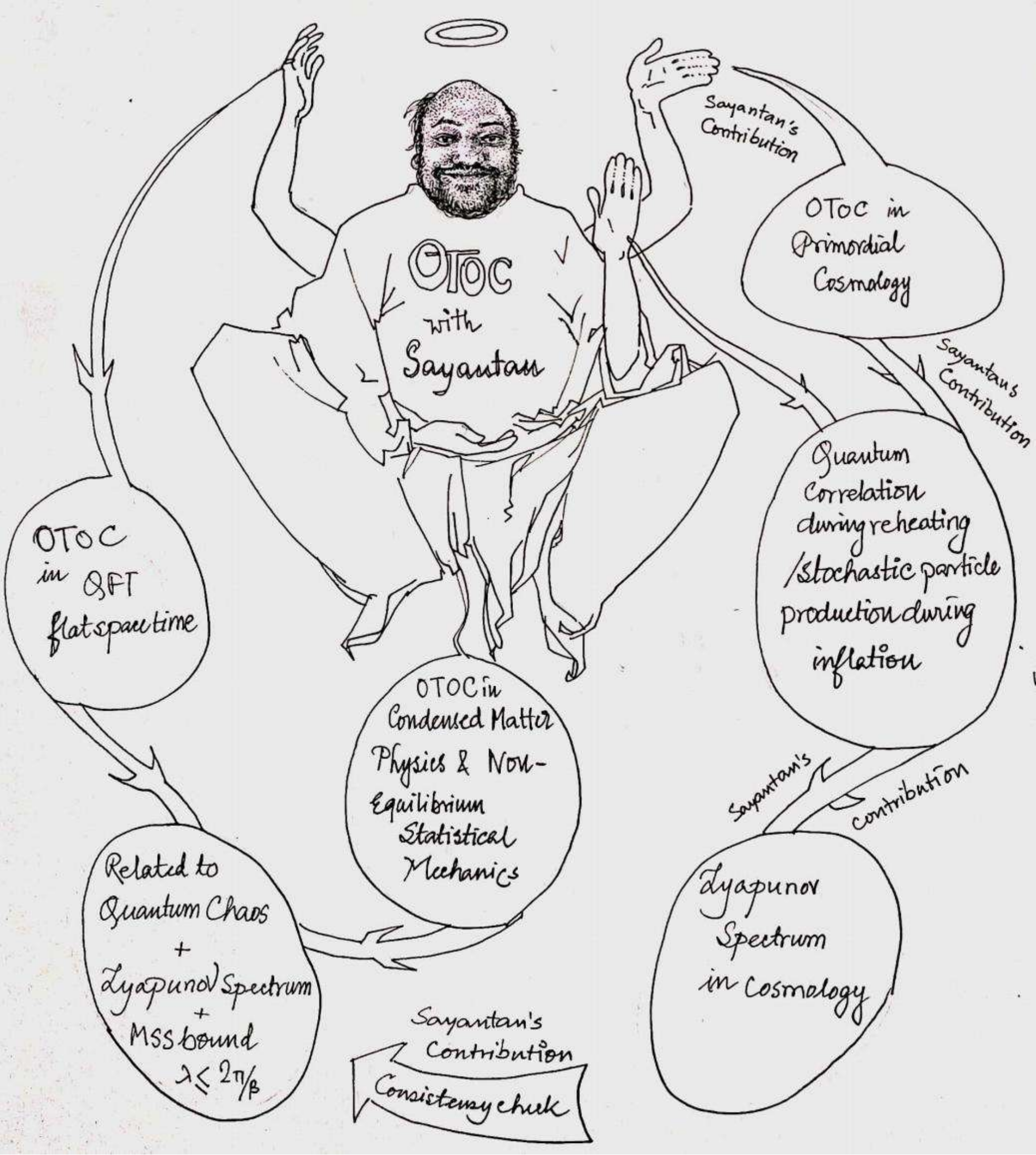}
    I have to thank my student Baibhab Bose from the QASTM group for drawing this interesting diagram for this paper. This picture actually contains the motivation of writing this paper and its connection with various other interesting aspects of quantum field theories. Also this diagram establish the connection among condensed matter physics and quantum statistical mechanics with cosmology. However, Baibhab presented me as a crazy person with four hands, just like lord Ganesha as appearing in Hindu mythology. Only I don't have any elephant trunk.
      \label{fig:A}
\end{figure}

\begin{figure}[t!]
    \centering
        \centering
        \includegraphics[width=17.5cm,height=18.5cm]{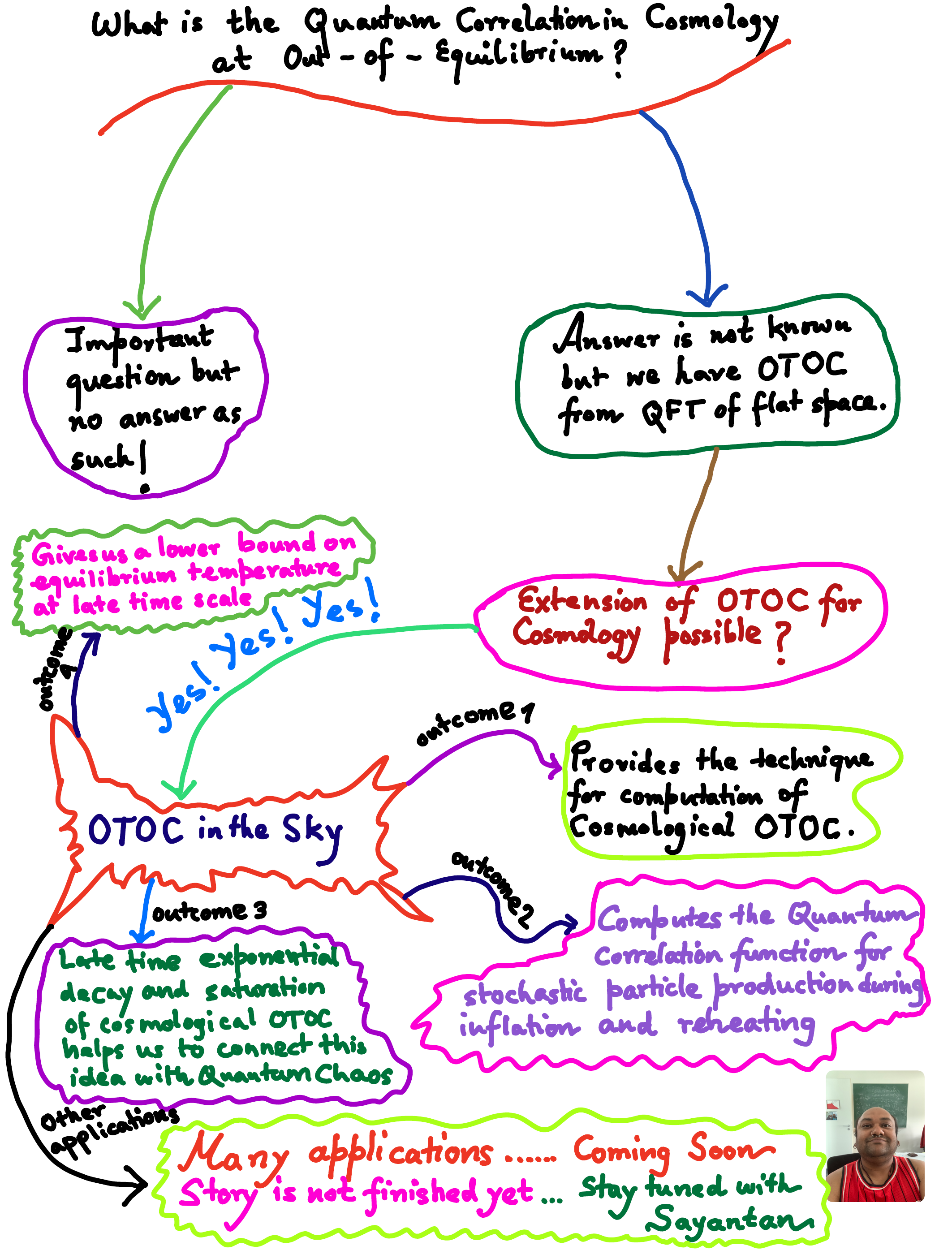}
    The background philosophy of the paper is presented in this diagram. our prime motivation is to find out the quantum correlation functions in the context of Cosmology within the framework of out-of-equilibrium quantum field theory. This diagram also shows our achievements from this paper and how the obtained results can help us to study various unexplored issues in the context of early universe cosmology.
      \label{fig:B}
\end{figure}
 \item \underline{\textcolor{red}{\bf Reheating / Warm~Inflation:}}\\
 Reheating is an epoch in the evolutionary time scale of our universe which appears just after inflation. During inflation the inflaton field started slowly rolling through the valley of the inflationary potential under consideration and it is expected that after a certain time the inflaton field will reach the stable minimum of the potential and the inflationary mechanism just stops there. But is the the end of the story? Obviously it is not the end. Once the inflation ends the inflation field started oscillating around the stable minimum of the potential and started interacting with the valley of the potential as well as some other filed, which we identify as the reheaton, the field who is mostly responsible for reheating~\footnote{It is important to note that, in the Cosmology literature there exists a few models of inflation where reheating is not required and without having any reheating phase in those models one can able to describe the genesis of the dark matter candidate theoretically. However, these models are very small in number and the construction of these models relies on various underlying assumptions, which may not be completely correct as far as the technical rigour of the quantum field theory for Cosmology is concerned. So for our discussion we stick to the general models of inflations which are developed from more reliable version of UV complete theories at very high energy scale and these models has to have the reheating phenomena in the evolutionary cosmological time scale of our universe.}. As a consequence of such interactions enormous amount of heat is generated and the quantum mechanical system that we are studying immediately goes to its out-of-equilibrium phase. From the previous understanding of the subject it was extremely difficult to determine the quantum effects in terms of studying the quantum correlations in this out-of-equilibrium phase. But it is expected from our basic understanding of quantum statistical mechanics that if we wait for long cosmological time scale then the system reaches the equilibrium where it is possible to associate an equilibrium reheating temperature associated with the system. After reaching equilibrium the reheaton field started decaying to some new particle contents which are responsible to describe the genesis of dark matter in the early universe Cosmology. That means without explaining the detailed quantum aspects throughout the total cosmological time scale in detail it is impossible to study the genesis of the dark matter contents of our universe and this actually motivates us to study this hidden unexplored phenomena in this paper.
 Till now we have very less description available regarding the quantum mechanical aspects during the epoch of reheating. Only the quadratic and quartic inflationary models are studied from phenomenological in this context and it is till unknown how to quantify and estimate the quantum mechanical correlations and study the detailed quantum field theory aspects of reheating. On the other hand, as we have already mentioned that it is extremely important to determine the quantum mechanical correlations during this epoch to study the genesis of the dark matter contents where the principles of out-of-equilibrium quantum field theory will play significant role. But till now there is no such significant inputs are available from the theory side which can be able to provide us the relevant tools and techniques to compute such quantum mechanical correlation function and study various other quantum field theory aspects related to reheating at out-of-equilibrium. In this paper, we have established a detailed quantum field theory framework at out-of-equilibrium phase which helps us to quantify the extremely relevant quantum mechanical correlation functions in terms of OTOCs. the methodology presented in this paper not only helps us to quantify the correlation at the out-of-equilibrium phase, but also provides us a detailed understanding about the equilibrium behaviour of the correlation functions at the large time limit in the cosmological time scale. Additionally, this detailed study through the cosmological generalization of OTOCs helps us to determine a lower bound on the reheating temperature in a completely model independent way. From the previously available understanding of the subject it was possible to give an estimate of the reheating temperature in a completely model dependent way after knowing about the relativistic degrees of freedom which are participating during the epoch of reheating. But apart from having a model dependent expression that expression actually relies on the scale of inflation completely, which is not known from the observation side till now. Only an upper bound on the scale of inflation in terms of the tensor-to-scalar ratio is known from the observational probes till now. So the previously described method of determining the reheating temperature is not very reliable and only helps us to know about the upper bound on the reheating temperature in a completely model dependent way. On the other hand, the present study helps us to determine the exact time dependent behaviour of the reheating temperature and additionally provides a lower bound on reheating temperature which we have derived in a completely model independent way. As we proceed through the subject material of the paper one can have a clear understanding about each of our big claim which we have explicitly established with detailed analysis. Apart from this, one can consider another well known framework, which is described by {\it warm inflationary paradigm} \cite{Berera:1995ie,Berera:1998qi}, where the radiation appears due to the stochastic particle production concurrently with inflationary expansion in the early universe evolutionary time scale. Sometimes this framework can be treated as the finite temperature generalization of usual well known zero temperature inflationary paradigm, which is mostly studied in cosmology literature. The presence of effect of radiation during the epoch of inflation implies that there is a possibility exists on which the epoch of inflation could smoothly end directly into a radiation-dominated epoch, particularly without having a separate reheating epoch. This is one of the early universe phenomena apart from reheating where the effect of out-of-equilibrium physics is significantly dominating and the detailed quantum field theoretic phenomena have not been studied yet in detail. Till now from the previous works it is very clear that this phenomena is very model dependent and few a few models only people have been studied this possibility.  This is because of the fact that, handling the out-of-equilibrium dynamics in the quantum regime is extremely complicated to handle. But we are again very hopeful just similarly like reheating because OTOC at finite temperature studied in this paper can able to describe the quantum correlation functions for the {\it warm inflationary scenario} as well, which can able to describe the out-of-equilibrium dynamical phenomena in the quantum regime in a very elegant and model independent way. Though it is true that our computation performed in this paper will be applicable to the {\it warm inflationary scenario}, we will restricted our analysis here to describe only the quantum effects during the epoch of reheating particularly in the very early time scale of evolutionary scale of our universe where the contribution of the out-of-equilibrium physics play pivotal role to control the dynamics.
 
 \item \underline{\textcolor{red}{\bf Stochastic inflation:}}\\
 Stochastic inflationary\cite{Vennin:2015hra,Noorbala:2018zlv,Grain:2017dqa,Vennin:2016wnk} paradigm is a very important aspect of the early universe Cosmology whose technical construction is completely different from the usual inflationary paradigm. To have inflation from a model which is derived from some UV complete high energy quantum field theories we don't need any additional source, time dependent scalar inflaton field slowly rolls down through a inflationary potential and participate in the quantum field theory using which one can explicitly compute the $N$-point correlation functions within the framework of early universe Cosmology. But in the framework of stochastic inflationary 
 paradigm a time dependent stochastic random source function play significant role to study the background construction of the quantum field theory. More precisely, here we have two fields, the inflaton and a stochastic random time dependent noise field, where both of them are participating to carry forward a correct quantum field theory construction of inflationary paradigm within the framework of early universe Cosmology. To construct a correct and consistent version of these type of quantum field theories one needs to start with a theory where the time dependent inflaton and the stochastic random time dependent field coupled to each other. In most of the previous literature, during the construction of these type of quantum field theories the correlation between the time dependent noise using which one can compute the $N$-point correlation function using the time dependent inflation to study the role of quantum mechanical fluctuations in the early universe Cosmology. In this theoretical construction, the correlation between the noise acts as a source of the correlation between the inflaton. But can't say concretely about the exact behaviour of the noise at the starting point of the computation. Most of the computation till now have performed in this literature by assuming the underlying Gaussian behaviour of the probability distribution of the time dependent stochastic noise function. As a consequence of this assumption one can now expect that the one-point function and any odd point correlation function of the time dependent stochastic noise vanish trivially. On the other hand the two-point function has to de proportional to a time translationally invariant Dirac Delta function and the proportionality constant actually determine the amplitude of two-point correlation function, which we identify as the power spectrum of the time dependent stochastic noise in the context of early universe Cosmology. One can also compute any other higher point even correlation functions from this construction, where one can explicitly show that the connected part of these correlations are factorizable in terms of the two-point function or the Green's function of the theory which we are considering to describe the background quantum field theory of early universe Cosmology. If we believe that the initial assumption regarding having a Gaussian probability distribution of time dependent stochastic noise is perfectly correct then everything we have mentioned above are the automatic consequence of that and using these information one can determine the $N$-point correlation function from the inflaton field consistently. Here the type of the noise we have pointed which follow the Gaussian probability distribution is commonly known as the {\it white noise} within the framework of quantum field theory. But unfortunately we really don't have any idea if the assumption that we have taken at the starting point is correct at all or not when one can think of any arbitrary stochastic randomness within the framework of quantum field theory. This allows to think about considering non-Gaussian noise, which commonly identified as the {\it coloured noise} in the present context. However, if we have some time dependent stochastic {\it coloured} non-Gaussian noise then it is extremely complicated to determine the quantum correlations between the noise and hence the $N$-point quantum correlation for inflaton field which is sourced by the {\it coloured noise} time dependent profile. Also it is expected that the random quantum fluctuations in the stochastic noise is the prime source for which the background equilibrium quantum filed theory set up goes to its out-of equilibrium phase, where we have very less information regarding the cosmological correlation functions within the framework of quantum field theory of early universe Cosmology. This actually motivates us to think about some alternative construction of computing the $N$-point quantum correlation functions due to the presence of {\it coloured} time dependent noise profile and in this paper by constructing the cosmological version of OTOC we have tried to address this crucial issue in a alternative way within the framework of out-of-equilibrium quantum field theory.
 
 \item \underline{\textcolor{red}{\bf Quantum quench in Cosmology:}}\\
The study of quantum mechanical quench in presence of time dependent random coupling and its consequences in quantum correlation functions in Cosmology is a very important topic of study in the context of theoretical physics. Till now this is not very well understood and studied in the context of Cosmology. In the earlier literature for various statistical mechanical and condensed matter systems quantum quench have been studied rigorously, but its extension in the framework of early universe Cosmology will provides us the understanding of the thermalization phenomena and the details of the achieving thermal equilibrium from an out-of-equilibrium phase in presence of a random time dependent coupling parameter. One can start with various possibilities here where in each cases theries are minimally coupled with classical FLRW conformally flat cosmological metric in a minimal fashion. The first and the simplest possibility is appearing in free scalar field theory where we consider a time dependent mass, which is a random coupling parameter within the framework of quench. The second possibility appears within the framework of an interacting quantum field theory where one can consider a situation where two scalar fields with constant mass are interacting with each other in presence of a random time dependent coupling parameter. If we treat the interaction term between the two scalar fields are quadratic then constructing the effective theories of each scalar fields becomes simpler after performing the path integration over the other unwanted scalar field for the description. In the quantum description sometimes this is identified to be the partial trace operation when we are describing everything in terms of density matrix and similar approach have been followed in the context of quantum field theories driven by an open quantum system, where the system is non-adiabatically interacting with the environment. One can further generalize this idea for $N$ number of scalar fields which are placed at the thermal bath and interacting with a system which is described by a singe scalar field. In terms of the interaction here one can consider the quadratic or any other non linear interactions. This description one can identified to be the quantum field theory generalization of the well known {\it Feynman-Vernon} model of influence functional theories or the {\it Caldeira Leggett} model which describes the quantum dissipation phenomena in the Quantum Brownian motion picture within the framework of early universe Cosmology. Considering these mentioned frameworks one can explicitly compute the OTOCs using the methodology presented in this paper to study the quantum mechanical $N$-point out-of-time ordered correlation functions in presence of a quantum mechanical quench within the framework of out-of-equilibrium version \cite{weiss,Hu,Rammer,Kamenev} of open quantum field theory of Cosmology \cite{Banerjee,Bhattacherjee:2019eml,Akhtar:2019qdn,Banerjee:2020ljo}.
 \end{enumerate}
 \end{itemize}
 Now, we mention the prime highlights of our obtained results in this paper, which we strongly believe will further help to know about many more unexplored features of cosmological quantum correlations within the framework of out-of-equilibrium version of quantum statistical field theory:
 \begin{itemize}
 \item \underline{\textcolor{red}{\bf Highlight~I:}}\\
 The methodology presented in this paper helps us to quantify the quantum mechanical correlation functions within the framework of Cosmology in presence of random quantum fluctuations. In this connection, we have computed the expressions for the two-point and four-point cosmological OTOC in the quantum regime which will provide the behaviour of the correlation functions in the out-of-equilibrium regime of the quantum field theory of early universe Cosmology.
 
 \item \underline{\textcolor{red}{\bf Highlight~II:}}\\
 We have additionally have studied the classical limiting version of the two-point and four-point cosmological OTOCs which will provide the decaying large time behaviour of the correlation function, which are perfectly matching with the expectations from the chaotic phenomena in the classical regime of the field theory.
 
 \item \underline{\textcolor{red}{\bf Highlight~III:}}\\
 The large time limiting behaviour of the four-point OTOC helps us to comment on the equilibrium behaviour of the quantum correlations and to determine the lower bound of the equilibrium temperature. Using this concept one can further determine the lower bound on the reheating temperature within the framework of early universe Cosmology.
 
 \item \underline{\textcolor{red}{\bf Highlight~IV:}}\\
 We have also studied the quantum {\it Lyapunov spectrum} for Cosmology and computed the associated {\it quantum Lyapunov exponent} to have a consistent chaotic description in the quantum regime from the four-point cosmological OTOC derived in this paper.
 
 \item \underline{\textcolor{red}{\bf Highlight~V:}}\\
 We have explicitly proved from our detailed computation that the results obtained from the normalized version of the four-point cosmological OTOC is completely independent of the choice of the time dependent perturbation variable as appearing in the specific scheme of the cosmological perturbation theory.
 \end{itemize}

By seeing the length of the paper one may feel very scared. Don't worry at all. After reading this paper we strongly believe that the readers can get to know about something very interesting which was not presented earlier in any Cosmology paper. The study material and the obtained results of this paper are organized as follows:
\begin{itemize}
\item In the \underline{\bf section~(\ref{sec:introduction1})}, we discuss how one can formulate the OTOC in the context of Cosmology and what exact quantity we have to compute for this study.

\item In the \underline{\bf section~(\ref{sec:OTO1})}, we discuss about the detailed computation of quantum micro-canonical two-point and four-point OTO amplitudes and the related OTOC in Cosmology.

\item  In the \underline{\bf section~(\ref{sec:OTO2})} and \underline{\bf section~(\ref{sec:OTO3})}, we present the numerical results obtained from the two-point and four-point OTOC and also discuss about its physical significance in Cosmology.

\item In the \underline{\bf section~(\ref{sec:OTO4})} and \underline{\bf section~(\ref{sec:OTO5})}, we present the detailed computation for {\it Lyapunov spectrum}, study the quantum chaotic phenomena, numerically study the obtained results and its physical impact in the context of Cosmology. 

\item In the \underline{\bf section~(\ref{sec:OTO6})}, we discuss about 
classical limit of micro-canonical two-point and four-point OTO amplitudes and the related OTOC in Cosmology.

\end{itemize}

\section{Formulation of OTOC in Cosmology}
\label{sec:introduction1}
\subsection{General remarks on OTOC}
In this section, my prime objective is to study the out-of-equilibrium physics in the inflationary patch of De Sitter space time. In the context of quantum field theory the time dynamics of the out-of-equilibrium physics is described by the out-of-time ordered correlation (OTOC) function, which is typically defined by the following expression:
\begin{equation}\label{e1}
{\textcolor{red}{\bf Thermal~ OTOC:}~~~~~~~C(t):\equiv -\langle \left[X(t),Y(0)\right]^2\rangle_{\beta}}~,~~~~~
\end{equation} 
where $\langle\cdots\rangle_{\beta}$ represents the thermal average which is taken using one parameter family $\alpha$ vacua and Bunch Davies quantum vacuum state in De Sitter space. Here, $X(t)$ and $Y(t)$ are quantum operators are defined at time scale $t$ in the Heisenberg representation. For any quantum operator ${\cal O}(t)$ the thermal average is technically defined as:
\begin{equation}\label{e2}
{\textcolor{red}{\bf Thermal~ average:}~~~~~~~\langle {\cal O}(t)\rangle_{\beta}:\equiv \frac{1}{Z}~{\rm Tr}\left[e^{-\beta H}{\cal O}(t)\right]}~,~~~~~~
\end{equation}
where $Z$ is the thermal Partition Function, which is defined as:
\begin{equation}\label{e3}
{\textcolor{red}{\bf Thermal~ partion ~function:}~~~~~~~Z={\rm Tr}\left[e^{-\beta H}\right]}~.~~~~
\end{equation}
Here, $H$ is the quantum system Hamiltonian under consideration.

Further using Eq~(\ref{e2}) and Eq~(\ref{e3}) in Eq~(\ref{e1}), we get the following simplified expression for the out-of-time ordered correlation (OTOC) function:
\bea
&&{C(t):\equiv -\frac{{\rm Tr}\left[e^{-\beta H} \left[X(t),Y(0)\right]^2\right]}{{\rm Tr}\left[e^{-\beta H}\right]}
=-{\rm Tr}\left[\frac{e^{-\beta H} }{Z}\left[X(t),Y(0)\right]^2\right]
=-{\rm Tr}\left[\rho~\left[X(t),Y(0)\right]^2\right]}~,~~~~~~~
\eea
where we have used the fact that the thermal density matrix is defined by the following expression:
\bea {\textcolor{red}{\bf Thermal~density~matrix:}~~~~~~~\rho=\frac{e^{-\beta H} }{Z}}~.~~~~~\eea
Here it is important to note that the OTOC is defined in terms of the square of the quantum mechanical commutator bracket of two quantum operators separated by a time scale $t$ because its connection to the classical Poisson bracket and the exponentially divergent trajectories expected in the context of classical description of chaos. 

Thermal average of the quantum mechanical commutator bracket of two quantum operators not allowed to describe chaos in the present context. 
To understand the actual physical reason behind this fact let us assume that the commutator bracket is replaced by the Poisson bracket by considering the semi-classical limit. In this situation the Poisson bracket shows an exponential growth with respect to time, $e^{\lambda_{L}t}$, where $\lambda_{L}$ represents the
Lyapunov exponent which quantify chaos. Now if we take the thermal  average of the commutator bracket representing two point OTOC function then both the
contributions are cancelled in the semi-classical limit and will not finally contribute to quantum chaos. On the other hand, from the quantum field theory point of view the two point
thermal averaged function, captures the effect of correlation between the two quantum
Hermitian operators, which decay in the large time limit and cannot
characterise the chaotic behaviour at all. Instead of that if we consider the square of the
commutator bracket, which actually represents the four-point function, after transforming it
to the Poisson bracket in the semi-classical limit it takes positive signature, which implies no
cancellation for both the contributions. Thermal average of this non trivial contribution further quantify quantum chaos. Similarly, in the quantum picture the
four-point thermal averaged function, not decays exponentially with respect to time at the large time
leading order limiting result.

Similarly, to define the quantum chaos the thermal average of the three point as well as any odd point correlation function of the quantum mechanical operators are also not allowed to define OTOC. This can be easily verify using the well known {\it Kubo Martin Schwinger} (KMS) condition, which can be demonstrated by applying Schwinger Keldysh formalism of the closed time path formulation of real time
finite temperature field theory. After applying KMS condition on the any odd point thermal averaged function one can explicitly show that each of the contributions from the odd point function vanish trivially and consequently will not contribute to quantify quantum chaos in terms of odd point OTOC.

A quantum mechanical system is treated as a chaotic system if the quantum mechanical commutator squared exponentially grow with time, which is technically expressed as:
\begin{equation}
{C(t)\sim \frac{1}{N^2}~e^{2\lambda_{L}t}}~,
\end{equation}
where the exponential growth is characterised by the Lyapunov exponent $\lambda_{L}$. Also, $N$ represents the number of degrees of freedom of the system under consideration. In more technical ground this phenomena of the exponential growth is related to the {\it fast scrambling} in the present context. Here the time scale corresponding to {\it fast scrambling} is expected to lie within the interval, $t_{\rm d}\ll t\ll t_{*}$, where 
\begin{equation}
{\textcolor{red}{\bf Dissipation~ time:}~~~~~~~t_{\rm d}\sim \beta=\frac{1}{T}}~~,
\end{equation}
 is the dissipation time scale which is the inverse of the temperature in De Sitter space and the upper bound of the {\it scrambling} time is defined as:
 \begin{equation}
 {\textcolor{red}{\bf Scrambling~ time:}~~~~~~~t_{*}\sim \frac{1}{\lambda_{L}}\log N}~~.
 \end{equation}
 On the physical ground, time scale associated to scrambling represents the associated time scale for a perturbation involving a few degrees of freedom to spread over all the degrees of freedom of the quantum mechanical system under consideration. In this connection it is important to note that, any quantum mechanical operations performed after the time interval for scrambling for a certain number of degrees of freedom can't able to re-track the quantum information associated with the perturbation.
 
 Further, expanding the right hand side of the Eq~(\ref{e1}) we get the following simplified expression for the OTOC, as given by:
 \bea\label{e6}
 {C(t)=\langle X(t)Y(0)Y(0)X(t)\rangle_{\beta}+\langle Y(0)X(t)X(t)Y(0)\rangle_{\beta}-2~ {\rm Re}\left[\langle Y(0)X(t)Y(0)X(t)\rangle_{\beta}\right]}.
 ~~~~~~\eea
It is important to note that the first two terms representing two different thermaql averaged four-point function appearing in the above expression for OTOC can be factorized in terms of the two point thermal averaged function over the dissipation time scale $t_d\sim \beta$ as given by:
\bea &&{\langle X(t)Y(0)Y(0)X(t)\rangle_{\beta} \approx \langle X(t) X(t)\rangle_{\beta}\langle Y(0)Y(0)\rangle_{\beta}+{\cal O}\left(e^{-t/t_{d}}\right)}~,\\
&&{\langle Y(0)X(t)X(t)Y(0)\rangle_{\beta}\approx \langle X(t) X(t)\rangle_{\beta}\langle Y(0)Y(0)\rangle_{\beta}+{\cal O}\left(e^{-t/t_{d}}\right)}~.\eea
Consequently, over the dissipation time scale $t_d\sim \beta$ the full expression for the OTOC can be factorized as:
\bea\label{e6}
 {C(t)=2~\left\{\langle X(t) X(t)\rangle_{\beta}\langle Y(0)Y(0)\rangle_{\beta}-{\rm Re}\left[\langle Y(0)X(t)Y(0)X(t)\rangle_{\beta}\right]\right\}+{\cal O}\left(e^{-t/t_{d}}\right)}.
 ~~~~~~\eea
 Here $\langle X(t) X(t)\rangle_{\beta}$ represents the thermal two point function of the quantum operator $X(t)$ which is actually perturbed by the insertion of the quantum operator $Y(0)$ in terms of the thermal two point function $\langle Y(0)Y(0)\rangle_{\beta}$.  It is important to note that if the insertion energy of the quantum operator $Y(0)$ is small enough, then the quantum state will relax with respect to the time scale and the corresponding expectation value (two point function) of the quantum operator $Y(0)$ will approach to the thermal expectation value multiplied by the norm of the quantum mechanical state~\footnote{It is a very well known fact that the time ordered correlation functions decay over the dissipation time scale $t_d\sim \beta$ to the products of the expectation value of the quantum operator with respect to the thermal quantum mechanical state.}. Consequently, beyond the dissipation time scale $t\gg t_d\sim \beta$, the normalized OTOC can be expressed by the following expression:
 \bea{ {\cal C}(t)=\frac{C(t)}{\langle X(t) X(t)\rangle_{\beta}\langle Y(0)Y(0)\rangle_{\beta}}\approx 2\left\{1-\frac{{\rm Re}\left[\langle Y(0)X(t)Y(0)X(t)\rangle_{\beta}\right]}{\langle X(t) X(t)\rangle_{\beta}\langle Y(0)Y(0)\rangle_{\beta}}\right\}+{\cal O}\left(e^{-t/t_{d}}\right)}.~~~~~~~~~~
 \eea
 Additionally, it is important to note that, late time vanishing behaviour of the OTOC for quantum mechanical systems are equivalent to the saturation of the square of the normalized thermal expectation value of the square of the commutator and this can be expressed by the exponential time dependent growth $e^{\lambda_{L}t}$, where $\lambda_{L}$ is the Lyapunov exponent. Consequently we get:
 \bea {{\cal C}(t)\approx  2\left\{1-\frac{1}{N^2}e^{\lambda_{L}t}+{\cal O}\left(\frac{1}{N^4}\right)\right\}~~~\Longrightarrow~~~\lambda_{L}\approx\frac{1}{t} \ln\left(N^2\frac{{\rm Re}\left[\langle Y(0)X(t)Y(0)X(t)\rangle_{\beta}\right]}{\langle X(t) X(t)\rangle_{\beta}\langle Y(0)Y(0)\rangle_{\beta}}\right)},~~~~~~~~~\eea
 where the number of degrees of freedom $N$ scaled as:
 \bea {N\sim \frac{1}{\sqrt{G_N}}=\sqrt{8\pi}~{\rm in}~M_P=1}~,\eea
 which is a very large number in terms of the energy scale and of the order of the cut-off of the quantum gravity cut-off scale i.e. Planck scale.
 
 Here, the Lyapunov exponent, $\lambda_{L}$, satisfy the following saturation bound for quantum chaos:
 \bea {\textcolor{red}{\bf Bound~on~Lyapunov~exponent:}~~~~\lambda_{L} \leq \frac{2\pi}{\beta}=2\pi T~~~~~{\rm where}~~~\beta=\frac{1}{T}~~{\rm with}~~\hslash=1=c}.~~~~~~~~\eea
 This implies the exponential time dependent growth of the real part of the time dependent thermal correlation function, which is given by the following upper bound:
 \bea {\textcolor{red}{\bf Bound~on~normalised~four~point~function:}~~~~\frac{{\rm Re}\left[\langle Y(0)X(t)Y(0)X(t)\rangle_{\beta}\right]}{\langle X(t) X(t)\rangle_{\beta}\langle Y(0)Y(0)\rangle_{\beta}}\leq \frac{1}{N^2}~e^{\frac{2\pi t}{\beta}}}.~~~~~~~\eea
 \subsection{Eigenstate representation of OTOC in quantum statistical mechanics}
 Now, instead of discussing further about the general definition of OTOC, we now concentrate on the eigenstate representation OTOC, using which many quantum systems can be analysed very easily. In this eigenstate representation we start with two canonically conjugate operators, $q(t)$ and $p(0)$, which are separated in time scale. In this context, at finite temperature the OTOC is defined as:
 \bea {C(t)=-\langle \left[q(t),p(0)\right]^2\rangle_{\beta}}~.\eea
 Here $\beta=1/T$ is the temperature of the quantum system under consideration.  Next considering the energy eigenstates as the required basis of the Hilbert space, we can further rewrite the OTOC as:
 \bea {C(t)=\underbrace{\frac{1}{\underbrace{Z}_{\textcolor{red}{\bf Thermal~partition~function}}}\sum_{n}\underbrace{e^{-\beta E_n}}_{\textcolor{red}{\bf Thermal~Boltzmann~factor}}~\underbrace{g_{n}(t)}_{\textcolor{red}{\bf Microcanonical~OTOC}}}_{\textcolor{red}{\bf Thermal~OTOC}}},\eea
 where the time dependent coefficient $g_{n}(t)$ is defined by the following expressions:
 \bea {\textcolor{red}{\bf Microcanonical~OTOC:}~~~g_{n}(t)\equiv -\langle n| \left[q(t),p(0)\right]^2|n \rangle}~.\eea
 Here $|n\rangle$ is the energy eigenstate of the quantum system under consideration, which satisfy the following eigenvalue equation:
 \bea {\textcolor{red}{\bf Time~independent~Schr\ddot{o}dinger~equation:}~~~~H|n\rangle=E_n|n\rangle}~,\eea
 where $H$ is in general any time independent Hamiltonian of the quantum system under consideration. This is basically the fixed energy eigenstate representation of OTOC for which in this context the time dependent coefficients $g_n(t)$ for a given energy level is identified to be the OTOC computed in the microcanonical statistical ensemble. On the other hand $C(t)$ represents OTOC at finite temperature or thermal OTOC, as in the eigenstate representation an additional Boltzmann factor is involved. This implies that once we compute the expression for the OTOC for a microcanonical type of statistical ensemble then one can easily obtain further the expression for the OTOC at finite temperature after taking the thermal average of $g_n(t)$.
 
 Now to simplify the expression for microcanonical OTOC, $g_n(t)$ it is better to express this in terms of the matrix elements of the previously mentioned canonically conjugate operators, $q(t)$ and $p(0)$, respectively. To implement this strategy we need to use the following completeness relations of the energy eigenstates:
 \bea {\sum_{n}|n\rangle \langle n|= 1}.\eea
 Consequently, the microcannical OTOC can be expressed in terms of the required matrix elements as:
 \bea {g_{n}(t)=\sum_{m}{\cal I}_{nm}(t){\cal I}^{*}_{nm}(t)},\eea
 where the matrix element ${\cal I}_{nm}(t)$ is defined as:
 \bea {{\cal I}_{nm}(t)\equiv -i \langle n|\left[q(t),p(0)\right]|m\rangle}. \eea
 Here one can note a basic property of this matrix ${\cal I}(t)$ is that it is Hermitian, which implies:
 \bea {{\cal I}^{*}_{nm}(t)={\cal I}_{mn}(t)}.\eea
 Consequently, the expression for the microcannical OTOC can be further simplified as:
 \bea {g_{n}(t)=\sum_{m}{\cal I}_{nm}(t){\cal I}_{mn}(t)}.\eea
 Further, considering the operator representation in Heisenberg picture one can write the time dependent operator $q(t)$ as:
 \bea {q(t)=e^{iHt}q(0)e^{-iHt}}.\eea
 Consequently, the matrix element ${\cal I}_{nm}(t)$ can be computed as:
 \bea {{\cal I}_{nm}(t)=-i\sum_{k}\left[e^{i\Delta E_{nk}t}q_{nk}(0)p_{km}(0)-e^{i\Delta E_{km}t}p_{nk}(0)q_{km}(0)\right]},\eea
 where we define $\Delta E_{nm}$, $q_{nm}(0)$ and $p_{nm}(0)$ by the following expressions:
 \bea &&{\Delta E_{nm}=E_{n}-E_{m}},\\
&& {q_{nm}(0)=\langle n|q(0)|m\rangle},\\
&& {p_{nm}(0)=\langle n|p(0)|m\rangle}. \eea
As we know any kind of general any $N$ particle Hamiltonian of a quantum system can be represented by the following expression:
\bea {H=\sum^{N}_{i=1}p^2_i+U(q_1,\cdots,q_N)}.\eea
Here we have assumed that each of the $N$ particle have the same mass, $m_i=1/2~\forall ~i=1,\cdots,N$. Using the above mentioned general form of the Hamiltonian one can further simplify the expression for the matrix element ${\cal I}_{nm}(t)$, which is given by:
 \bea {{\cal I}_{nm}(t)=\frac{1}{2}\sum_{k}q_{nk}(0)q_{km}(0)\left[E_{km}e^{i\Delta E_{nk}t}-E_{nk}e^{i\Delta E_{km}t}\right]},\eea
where I have used the following fact:
\bea {p_{mn}(0)=\langle m| p(0) |n\rangle=\frac{i}{2}\langle m| \left[H(0),q(0)\right] |n\rangle=\frac{i}{2}E_{mn}q_{mn}(0)}.~~~\eea
Consequently, the microcanonical OTOC is computed as:
 \bea &&{g_{n}(t)=\frac{1}{4}\sum_{m}\sum_{k}\sum_{s}q_{nk}(0)q_{km}(0)q_{ms}(0)q_{sn}(0)}\nonumber\\
 &&~~~~~~~~~~~~~~~~~~~~~~~~~{\times\left[E_{km}e^{i\Delta E_{nk}t}-E_{nk}e^{i\Delta E_{km}t}\right]\left[E_{sn}e^{i\Delta E_{ms}t}-E_{ms}e^{i\Delta E_{sn}t}\right]}\nonumber\\
 &&~~~~~~~~{=\frac{1}{4}\sum_{m}\sum_{k}\sum_{s}q_{nk}(0)q_{km}(0)q_{ms}(0)q_{sn}(0)}\nonumber\\
 &&~~~{\times\left[E_{km}E_{sn}e^{i\widetilde{\Delta E_{nkms}}t}+E_{nk}E_{ms}e^{i\widetilde{\Delta E_{kmsn}}t}-E_{nk}E_{sn}e^{i\widetilde{\Delta E_{kmms}}t}-E_{km}E_{ms}e^{i\widetilde{\Delta E_{nksn}}t}\right]}.~~~~~~~~~~\eea
 where we define new energy shifts, $\widetilde{\Delta E_{nkms}}$, $\widetilde{\Delta E_{kmsn}}$, $\widetilde{\Delta E_{kmms}}$ and $\widetilde{\Delta E_{nksn}}$ as:
 \bea &&{\widetilde{\Delta E_{nkms}}=\Delta E_{nk}+\Delta E_{ms}=E_{n}+E_{m}-E_{k}-E_{s}},\\
 &&{\widetilde{\Delta E_{kmsn}}=\Delta E_{km}+\Delta E_{sn}=E_{k}+E_{s}-E_{m}-E_{n}},\\
  &&{\widetilde{\Delta E_{kmms}}=\Delta E_{km}+\Delta E_{ms}=E_{k}-E_{s}}\\
  &&{\widetilde{\Delta E_{nksn}}=\Delta E_{nk}+\Delta E_{sn}=E_{s}-E_{k}}.\eea
   Finally the thermal OTOC can be computed as:
 \bea &&{C(t)=\underbrace{\frac{1}{\underbrace{Z}_{\textcolor{red}{\bf Thermal~partition~function}}}\sum_{n}\underbrace{e^{-\beta E_n}}_{\textcolor{red}{\bf Thermal~Boltzmann~factor}}~\underbrace{g_{n}(t)}_{\textcolor{red}{\bf Microcanonical~OTOC}}}_{\textcolor{red}{\bf Thermal~OTOC}}}\nonumber\\
 &&~~~~~~~{=\left(4\sum_{n}e^{-\beta E_n}\right)^{-1}}\nonumber\\
 &&~~~~~~~~~~~{\times\sum_{n}\sum_{m}\sum_{k}\sum_{s}e^{-\beta E_n}~q_{nk}(0)q_{km}(0)q_{ms}(0)q_{sn}(0)}\nonumber\\
 &&~~~~~~~~~~~~~{\times\left[E_{km}E_{sn}e^{i\widetilde{\Delta E_{nkms}}t}+E_{nk}E_{ms}e^{i\widetilde{\Delta E_{kmsn}}t}-E_{nk}E_{sn}e^{i\widetilde{\Delta E_{kmms}}t}-E_{km}E_{ms}e^{i\widetilde{\Delta E_{nksn}}t}\right]}.~~~~~~~~~~\eea
 Once we determine the shifts in the energy eigen values and the matrix elements of the canonically conjugate variable $q(0)$ in the energy eigen basis, we can explicitly compute the expression for the thermal OTOC in the energy eigen basis itself.
 \subsection{Constructing OTOC in Cosmology}
 One can further map this idea to Quantum Field Theory of Curved Space Time as well. To show this mapping let us start with a theory of $N$ scalar fields in an arbitrary curved gravitational background which is described the following $(d+1)$ dimensional action:
  \bea {S=\int d^{d+1}x \underbrace{\sqrt{-g}\left[-\frac{1}{2}\sum^{N}_{a=1}\sum^{N}_{b=1}g^{\mu\nu}G^{ab}\partial_{\mu}\phi_a\partial_{\nu}\phi_b-U(\phi_a,\phi_b)\right]}_{\textcolor{red}{\bf Lagrangian~density~ in~curved~space }~\equiv {\cal L}(\phi_a,\phi_b,\partial_{\mu}\phi_a,\partial_{\nu}\phi_b,g^{\mu\nu},g)~}~~~~\forall~a,b=1,\cdots,N},~~~~~~\eea
  where in the above action $d$ represents the number of spatial dimension and the gravity is minimally coupled with $N$ scalar fields in an arbitrary background. In general one can consider any arbitrary class of gravitational metric for this calculation. However to avoid mathematical complication for any unwanted reason we restrict ourselves in the class of gravitational metrics which can be expressed in diagonal form. Here $G^{ab}$ takes care of all possible interaction between $N$ scalar fields in the kinetic term. On the other hand, one can consider between all possible interactions in the interaction potential $U(\phi_a,\phi_b)$. In a more generalised physical situation, In a simplest situation where $a=b$ always then in that case, $G^{ab}=\delta^{ab}$, which means that all the off-diagonal components of the matrix is zero. In that specific situation, the above action can be simplified to the following simplified form:
   \bea {S=\int d^{d+1}x \underbrace{\sqrt{-g}\left[-\frac{1}{2}\sum^{N}_{a=1}g^{\mu\nu}\partial_{\mu}\phi_a\partial_{\nu}\phi_a-U(\phi_a)\right]}_{\textcolor{red}{\bf Lagrangian~density~ in~curved~space }~\equiv {\cal L}(\phi_a,\partial_{\mu}\phi_a,g^{\mu\nu},g)~}~~~~\forall~a=1,\cdots,N}~.\eea
   Here first term in the above action represents a very simplest kinetic term for $N$ scalar fields and the second term $U(\phi_a)$ corresponds to the simplest form of the self interacting potential for the $N$ scalar fields.
   
 Now from this most generalised action one can compute the canonically conjugate momenta of the each $N$ scalar fields, which is given by the following expression:
 \bea {\Pi_{\phi_{c}}=\frac{\partial L}{\partial \dot \phi_c}=-\frac{1}{2}\sqrt{-g}g^{00}\left[\sum^{N}_{a=1}\sum^{N}_{b=1}G^{ab}\left(\delta^{c}_{a}\dot{\phi}_b+\delta^{c}_{b}\dot{\phi}_a\right)\right]=-\sqrt{-g}g^{00}\sum^{N}_{a=1}G^{ca}\dot{\phi}_a~~~~~\forall~c=1,\cdots,N}.~~~~~~~\eea
 Also in the simplest situation where $G^{ac}=\delta^{ac}$ we can further simplify the expression for the canonically conjugate field momenta as:
 \bea {\Pi_{\phi_{c}}=-\sqrt{-g}g^{00}\sum^{N}_{a=1}\delta^{ca}\dot{\phi}_a=\sqrt{-g}~\dot{\phi}_c~~\Rightarrow~~\dot{\phi}_c=-\frac{\Pi_{\phi_{c}}}{\sqrt{-g}g^{00}}~~~~~\forall~c=1,\cdots,N}~.~~~~~\eea
 Consequently, for the simplest case non-interacting $N$ scalar fields the Hamiltonian density can be written as:
 \bea &&{{\cal H}=\sum^{N}_{c=1}\Pi_{\phi_c}\dot{\phi}_c-{\cal L}(\phi_a,\partial_{\mu}\phi_a,g^{\mu\nu},g)},\nonumber\\
 &&~~~~{=-\frac{1}{\sqrt{-g}g^{00}}\sum^{N}_{c=1}\Pi^2_{\phi_c}-\sqrt{-g}\left[-\frac{1}{2}\sum^{N}_{a=1}g^{\mu\nu}\partial_{\mu}\phi_a\partial_{\nu}\phi_a-U(\phi_a)\right]}\nonumber\\
 &&~~~~{=-\frac{1}{2}\frac{1}{\sqrt{-g}g^{00}}\sum^{N}_{c=1}\Pi^2_{\phi_c}+\frac{1}{2}\sqrt{-g}g^{ii}\sum^{N}_{c=1}\left(\partial_{i}\phi_c\right)^2+\sqrt{-g}U(\phi_a)}.\eea
 Now we consider a specific situation where space-time is such that the scalar field is independent on space, but only function of time. This type of situation one usually consider in the context of Cosmology. iIn this situation the general structure of the $(d+1)$ dimensional metric can be expressed using the following ansatz:
 \bea {ds^2_{d+1}=g^{00}dt^2+\sum^{d}_{i=1}g^{ii} d{x}_{i}d{x}_{i}}.\eea
 Particularly for $(d+1)$ dimensional De Sitter space the infinitesimal line element in the planar inflationary coordinate is described by:
  \bea {ds^2_{d+1}=-dt^2+a^2(t)\sum^{d}_{i=1} d{x}_{i}d{x}_{i}=-dt^2+a^2(t)d{\bf x}^2_{d}},\eea
 where we have fixed the diagonal component of the metric as:
 \bea {g^{00}=1, ~~~~~g^{ii}=a^2(t)}.\eea
 Also the scale factor$a(t)$ is defined as:
 \bea {a(t)=e^{Ht}},\eea
 where $H$ is the Hubble parameter in the present context.
 
 In this coordinate system, the above mentioned Hamiltonian density of $N$ non-interacting scalar fields can be further simplified as:
  \bea {{\cal H}=\sum^{N}_{c=1}\widetilde{\Pi^2_{\phi_c}}+\widetilde{U(\phi_a)}},\eea
  which is exactly similar like the previously discussed Hamiltonian for $N$ non-interacting system in Quantum Mechanics.
  
  It is important to note that, here we have used the following redefinition:
  \bea &&{\widetilde{\Pi^2_{\phi_c}}=\frac{1}{2a^d(t)}\Pi^2_{\phi_c}},\\
   &&{\widetilde{U(\phi_a)}=a^d(t)U(\phi_a)}.\eea
 In the flat Minkowski space limit one can fix $a(t)=1$.
  
 Now in the present context the Hamiltonian of the $N$ non-interacting scalar fields can be expressed as:
 \bea {H=\int d^{d}x~ {\cal H}=\int d^{d}x \left[\sum^{N}_{c=1}\widetilde{\Pi^2_{\phi_c}}+\widetilde{U(\phi_a)},\right]}.\eea
 However in Quantum Field Theory we don't have any sort of eigenstate representation of the Hamiltonian similar like Quantum Mechanics. So for 
 Quantum Field Theoretic systems representing the thermal OTOC in terms of the microcanonical OTOC is not very straight forward just like Quantum Mechanics. In Quantum Field Theory best possible way is to express the Hamiltonian is in the Fourier space in normal ordered form, as given by:
 \bea {:H:=\int\frac{d^{d}k}{(2\pi)^{d}}E_{k}a^{\dagger}_{\bf k}a_{\bf k}},\eea
 where the energy $E_k$ can be computed from the dispersion relation. Here $a^{\dagger}_{\bf k}$ and $a_{\bf k}$ are the creation and annihilation operators in the present context. In the flat space the corresponding vacuum is known as Minkowski vacuum which is unique,. On the other hand, the corresponding curved space vacuum is not unique in nature. In the context of De Sitter space we use Bunch Davies and $\alpha$ vacua 
 for the computation which are SO(1,4) invariant in nature.
 
 Also in the further computation instead of using the usual time coordinate we use the conformal coordinate in the above mentioned $(d+1)$ dimensional De Sitter metric, which can be expressed using the following ansatz:
  \bea{ ds^2_{d+1}=a^2(\tau)\left(-d\tau^2+\sum^{d}_{i=1} d{x}_{i}d{x}_{i}\right)=a^2(\tau)\left(-d\tau^2+d{\bf x}^2_{d}\right)}.\eea
  Here we have introduced the concept of conformal time which can be expressed in planar De Sitter space as:
 \bea { \underbrace{\tau=\int^{\tau}_{-\infty}\frac{dt}{a(t)}=-\frac{1}{Ha(\tau)}}_{\textcolor{red}{\bf Conformal~time~during~inflation~-\infty<\tau<\tau_{inf}}}~~~~\Longrightarrow~~~~~~\underbrace{a(\tau)=-\frac{1}{H\tau}}_{\textcolor{red}{\bf  Scale~factor~during~inflation}}}.~~~~~~~\eea
 Here $H$ is the Hubble parameter defined in the planar inflationary patch of De Sitter space. This result can be used during stochastic particle production during inflation.
  
 For the reheating case there is no closed form expression exists in literature. When the inflaton oscillates around the potential minimum, the equation of state is that of pressureless matter i.e. $p=0$, so the scale factor would behave accordingly. It is nothing but the non-relativitic matter with equation of state parameter $w=p/\rho=0$. Then, as new light particles are produced, the equation of state will switch to $p = \rho/3 $, and the scale factor will expand according to a radiation dominated phase. In this case the equation of state parameter takes the form, $w=p/\rho=1/3$. The first approximation is then to just match the scale factor (and the conformal time) at these transitions. So it implies that during reheating the scale factor is lying within the window, $0\leq w_{reh} \leq 1/3$.
 Consequently the conformal time on this epoch can be explicitly computed as:
  \begin{equation}
  { \underbrace{\tau=\int^{\tau}_{0}\frac{dt}{a(t)}=\frac{3(1+w_{reh})}{(1+3w_{reh})}\left[a(\tau)\right]^{\frac{(1+3w_{reh})}{2}}}_{\textcolor{red}{\bf Conformal~time~during~reheating~0<\tau<\tau_{reh}}}~\Longrightarrow~\underbrace{a(\tau)=\left[\frac{(1+3w_{reh})}{3(1+w_{reh})}\tau\right]^{\frac{2}{(1+3w_{reh})}}}_{\textcolor{red}{\bf Scale~factor~during~reheating}}}.
\end{equation} 
In the present context, during the computation of OTOC's both of the scale factors computed during inflationary epoch (for stochastic particle production) and reheating epoch are useful. Here it is important to note that, in general the equation of state parameter during the reheating epoch can be expressed as a function of conformal time in general. If we assume that the energy momentum tensor can be expressed using a perfect fluid with pressure $p_{reh}$ and energy density $\rho_{reh}$, then for a time dependent scalar field in De Sitter FLRW background can be written as:
\bea{ \textcolor{red}{\bf Equation ~of~state~parameter~for~reheating:}~~~0\leq w_{reh}(\tau)=\frac{p_{reh}(\tau)}{\rho_{reh}(\tau)}\leq \frac{1}{3}}.~~~~~\eea
For $N$ interacting scalar field the equation of state parameter can be computed as:
\bea {w_{reh}(\tau)=\left[\frac{-\frac{1}{2a^2(\tau)}g^{00}\sum\limits_{a=1}^{N}\sum\limits_{b=1}^{N}G^{ab}\partial_{\tau}\phi_a\partial_{\tau}\phi_b-U(\phi_a)}{-\frac{1}{2a^2(\tau)}g^{00}\sum\limits_{a=1}^{N}\sum\limits_{b=1}^{N}G^{ab}\partial_{\tau}\phi_a\partial_{\tau}\phi_b+U(\phi_a)}\right]},\eea
where the pressure $p_{reh}$ and energy density $\rho_{reh}$ for $N$ interacting scalar field can be written as:
\bea {\textcolor{red}{\bf Pressure:}~~~~~p_{reh}(\tau)=\left[-\frac{1}{2a^2(\tau)}g^{00}\sum\limits_{a=1}^{N}\sum\limits_{b=1}^{N}G^{ab}\partial_{\tau}\phi_a\partial_{\tau}\phi_b-U(\phi_a)\right]}~,\\
{\textcolor{red}{\bf Density:}~~~~~\rho_{reh}(\tau)=\left[-\frac{1}{2a^2(\tau)}g^{00}\sum\limits_{a=1}^{N}\sum\limits_{b=1}^{N}G^{ab}\partial_{\tau}\phi_a\partial_{\tau}\phi_b+U(\phi_a)\right]}~.\eea
Similarly for $N$ non-interacting scalar field the equation of state parameter can be computed as:
\bea {w_{reh}(\tau)=\left[\frac{-\frac{1}{2a^2(\tau)}g^{00}\sum\limits_{a=1}^{N}(\partial_{\tau}\phi_a)^2-U(\phi_a)}{-\frac{1}{2a^2(\tau)}g^{00}\sum\limits_{a=1}^{N}(\partial_{\tau}\phi_a)^2+U(\phi_a)}\right]},\eea
where the pressure $p_{reh}$ and energy density $\rho_{reh}$ for $N$ non-interacting scalar field can be written as:
\bea {\textcolor{red}{\bf Pressure:}~~~~~p_{reh}(\tau)=-\frac{1}{2a^2(\tau)}g^{00}\sum\limits_{a=1}^{N}(\partial_{\tau}\phi_a)^2-U(\phi_a)}~,\\
{\textcolor{red}{\bf Density:}~~~~~\rho_{reh}(\tau)=-\frac{1}{2a^2(\tau)}g^{00}\sum\limits_{a=1}^{N}(\partial_{\tau}\phi_a)^2+U(\phi_a)}~.\eea
For a single scalar field ($N=1$) the equation of state parameter can be further simplified as:
\bea {w_{reh}(\tau)=\left[\frac{-\frac{1}{2a^2(\tau)}g^{00}(\partial_{\tau}\phi)^2-U(\phi)}{-\frac{1}{2a^2(\tau)}g^{00}(\partial_{\tau}\phi)^2+U(\phi)}\right]},\eea
where the pressure $p_{reh}$ and energy density $\rho_{reh}$ for $N$ non-interacting scalar field can be written as:
\bea {\textcolor{red}{\bf Pressure:}~~~~~p_{reh}(\tau)=-\frac{1}{2a^2(\tau)}g^{00}(\partial_{\tau}\phi)^2-U(\phi)}~,\\
{\textcolor{red}{\bf Density:}~~~~~\rho_{reh}(\tau)=-\frac{1}{2a^2(\tau)}g^{00}(\partial_{\tau}\phi)^2+U(\phi)}~.\eea
For FLRW case the time component of the metric $g_{00}=-1=g^{00}$~\footnote{\textcolor{red}{\bf \underline{Statutory~warning:}}~From the above mentioned results it is important to note that, when we are dealing with massless scalar field then we have $U(\phi)=0$ and consequently the equation of state parameter reduces to $w=1\neq w_{reh}$. But since $0\leq w_{reh}\leq \frac{1}{3}$ any values $w_{reh}>\frac{1}{3}$ is not physically allowed for the reheating phenomena. So massless scalar field is not at allowed to describe the reheating phenomena as it predicts $w_{reh}=1$ which is strictly not allowed. So in the context of reheating, partially massless and massive scalar fields are allowed to describe the physical phenomena. This is because of the fact that, in both the cases due to presence of mass term in the effective potential contribution the equation of state parameter computed from this setup expected to be lie within the allowed window, $0\leq w_{reh}\leq \frac{1}{3}$. Because of this reason during the computation of OTOC's during reheating we only restrict on partially massless and massive scalar fields. The details of these issues can be found in the next subsections. But one we introduce the gauge invariant perturbation instead of field variable and translate the whole problem in terms of that new language, then it is explicitly possible to show that, an effective mass $m^2_{eff}=\frac{1}{z(\tau)}\frac{d^2z(\tau)}{d\tau^2}\approx \frac{1}{a(\tau)}\frac{d^2a(\tau)}{d\tau^2}$ and spatial gradient of the gauge invariant perturbation term will be induced. Here $z=\sqrt{2\epsilon}~a$ is Mukhanov Sasaki time dependent variable which controls the mathematical structure of the effective mass term depending on the background physical phenomena in which we are interested in. These phenomena are stochastic particle production during inflation and reheating respectively. There are many more, however we restrict to these two fact only. At the leading order one can show that the effective mass term in the gauge invariant description can be expressed in terms of scale factor of our Universe and its double time derivative as a function of conformal time. Here $a(\tau)$ is the conformal time dependent scale factor and $\epsilon=-\frac{1}{a(\tau)H}\frac{dH}{d\tau}$ captures the rate of Hubble expansion in Cosmology, which also shows that a little bit deviation from constant Hubble parameter as appearing in De Sitter space. This deviation is very small at the particle production during inflation and becomes unity at the end of inflation. On the other hand, at the epoch of reheating this factor becomes large. Due to these all facts, it is expected that the equation of state parameter during the epoch of reheating for massless scalar field in the gauge invariant description will lie within the window, $0\leq w_{reh}\leq \frac{1}{3}$. Details of the computation we will show in the next subsections written for massless, partially massless and heavy scalar field in the gauge invariant perturbed description.}. 

 Now we will comment on some of the important future predictions from our analysis for massless, partially massless and massive scalar field theory in the context of computing OTOC which we will explicitly define in the following subsections. These predictions are appended below:
\begin{enumerate}
\item We are dealing with commutative version of the quantum field theory i.e. all the field momenta and the fields are commutative amongst themselves. Things will change when one deals with the non commutative version of the quantum field theory which means in that in that case all the field momenta and the fields are non-commutative amongst themselves. Usually in both of the versions of quantum field theories we mostly consider the equal time commutation relations at different space points (ETCR) and the unequal time commutation relations (UETCR). In the Fourier transformed versions which can be translated as computing commutation relations at different momenta with same time for ETCR and with different time for UETCR. However, if we look into the specific mathematical structure of OTOC, then we see that the commutators are either ETCR or UETCR with fixed momentum or position.
\item   If we get divergence to compute OTOC, then like the previous cases we deform the contour of integration in Schwinger Keldysh path integral formalism by introducing one or more regulator in the theory. In the technical ground introducing the regulator cut-off means one needs to deform the Schwinger Keldysh path integral contour at finite temperature by introducing a single or more than one parameter.  Consequently, in that context we will get a finite regulated result. On the other hand, in the 
context of non-commutative version of the quantum field theory it is expected that the form OTOC~\footnote{The mathematical definition of OTOC is provided in the next subsections explicitly for massless, partially massless and massive scalar field theories.} we get non vanishing physically relevant answers. During the computation of the OTOC if we get some divergences then like previous case here we also deform the contour of integration by introducing a cut-off to get finite results.
\item The explicit time dependence of the OTOC is very important in the present context, particularly the study the physics of out-of-equilibrium. If the second term of the OTOC grows exponentially with time then one can quantify quantum chaos from the present computation and check whether the final result saturated the chaos bound on the Lyapunov exponent or not. On the other hand, if instead of getting an exponentially growing time dependent behaviour if we get some other non-trivial characteristic behaviour as a function of time that is also very important to study in the present context to study the out-of-equilibrium behaviour of quantum mechanical systems considered in the framework of Cosmology. Particularly in Cosmology, studying the non trivial time dependent behaviour of the OTOC for stochastic particle production during inflation, reheating are very useful to extract physical information from the quantum mechanical systems studied. Since we are trying to compute the OTOC for quantum field theory, it is expected to get some interesting physical outcomes from the correlation functions. However, it is also expected from the studies of the OTOC that at very large time limit the correlation functions reach thermal equilibrium and saturates to a finite value. This may not indicate the saturation of the quantum chaos, but this will also play significant role to study the time dependent behaviour of OTOC. In this context, one can use this information as a limiting boundary condition of OTOC. Apart from that, the early time behaviour of OTOC is also good to know. The early time behaviour is physically important as it gives the idea about the initial condition on the time dependent behaviour of the OTOC. Initial condition OTOC actually fixes from which point in time scale the quantum system goes to out-of-equilibrium. This information in the context of Cosmology is very useful because it will give a physical picture about a particular out-of-equilibrium phenomena in which we are interested in.

\item In this computation we are considering massless ($m<<H$), partially massless ($m=\sqrt{2}H\approx H$) and massive ($m>>H$) scalar field for the computation of OTOC in the context of Cosmology. In the context of Cosmology, the fundamental characteristic scale is described by Hubble scale i.e. $H$. Now compared to this characteristic scale we define massless, partially massless and massive scalar field in the De Sitter gravitational background. In the context of massive scalar field theory one usually consider various time dependent protocols to describe the stochastic particle production phenomena during inflation and reheating process in the context of primordial Cosmology. For a few types of time dependent protocols one can exactly analytically solve the field equations for the scalar field, which are commonly used in the study of quantum quench. In general exact solution of the field equation is very difficult so solve and for this purpose one use the WKB approximation method to find out the analytical solution of the scalar field. These solutions are extremely important to study in De Sitter cosmological background as it provide the key ingredient to compute the previously mentioned OTOC in the present context. In the context of stochastic particle production one consider Dirac Delta type of scatterer as a time dependent protocol, which is nothing but the white noise in the present context. In some cases, one also considers coloured noise time dependent protocol. In the quantum description, the white noise is identified to be non-Markovian and coloured noise is considered as Markovian at the level of correlation function~\footnote{\textcolor{red}{\bf \underline{Quantum analogue:}} ~At the level of quantum correlation function for the white and coloured noise are described by the following correlation functions:
\bea {\underbrace{\textcolor{red}{\bf White ~noise:}~~\langle m(\tau)m(\tau^{'}) \rangle=A(\tau)\delta(\tau-\tau^{'})}_{\textcolor{red}{\bf Non-Markovian}},~~ \underbrace{\textcolor{red}{\bf Coloured ~noise:}~~\langle m(\tau)m(\tau^{'}) \rangle=B(\tau)K(\tau-\tau^{'})}_{\textcolor{red}{\bf Markovian}}}~,~~~~~~\eea
where $A(\tau)$ and $B(\tau)$ are the amplitude for the white and  coloured noise respectively as a function of conformal time. Here $\delta(\tau-\tau^{'})$ represents a localised white noise function at $\tau=\tau^{'}$ time scale. On the other hand, the coloured noise is represented by the general time dependent Kernel $K(\tau-\tau^{'})$ which is not at all localised at $\tau=\tau^{'}$ time scale.}~\footnote{\textcolor{red}{\bf \underline{Classical analogue:}} ~At the classical level Delta type of scatterer mimics the role of white noise appearing in the quantum regime. On the other hand, other non trivial time dependent functions mimics the role of coloured noise at the quantum level. For $N$ number of scatterers at the classical level one can then write:
\bea {\underbrace{\textcolor{red}{\bf White ~noise:}~~ m^2(\tau)=\sum^{N}_{j=1}A(\tau_j)\delta(\tau-\tau_j)}_{\textcolor{red}{\bf Classical~Non-Markovian}},~~ \underbrace{\textcolor{red}{\bf Coloured ~noise:}~~ m^2(\tau)=\sum^{N}_{j=1}B(\tau_j)K(\tau-\tau_j)}_{\textcolor{red}{\bf Classical~Markovian}}}~,~~~~~~\eea
where $A(\tau_j)$ and $B(\tau_j)$ are the amplitudes respectively as a function of conformal time. Here $\delta(\tau-\tau_j)$ represents a localised function at the point of $j$-th scatterer $\tau=\tau_j\forall~j=1,\cdots,N$ time scale. On the other hand, the coloured noise is represented by the general time dependent Kernel $K(\tau-\tau^{'})$ which is not at all localised at $\tau=\tau^{'}$ time scale. However, any reliable physically significant time dependent protocols are allowed in this context to extract the unknown physical information from the OTOC computation.}.

\end{enumerate}
 
 In the next subsections, we will explicitly define the expressions for the thermal OTOC in the context of the massless, partially massless, heavy scalar field and additionally for the stochastic scalar field which is used to study the particle production procedure in Early Universe Cosmology.
\subsubsection{For massless scalar field}
\label{sec:introduction}
In this context we will give the definition of OTOC computed from a massless scalar field. Here the terminology massless is used to consider scalar fields which have mass less than the characteristic Cosmological scale i.e. the Hubble scale ($m<<H$). So here we are not considering exactly massless scalar fields. But due to the approximation $m<<H$ one can easily neglect such contribution in the effective action.
A massless $N$ number of scalar fields can be generalized by the following simplified action:
\bea {S=-\frac{1}{2}\int d^{d+1}x~\sqrt{-g}\sum^{N}_{a=1}\sum^{N}_{b=1}g^{\mu\nu}G^{ab}\partial_{\mu}\phi_a\partial_{\nu}\phi_b}.\eea
In the simplest non-interacting situation one can write, $G^{ab}=\delta^{ab}$ and in that case the action for the $N$ massless scalar field can be simplified as:
\bea {S=-\frac{1}{2}\int d^{d+1}x~\sqrt{-g}\sum^{N}_{a=1}g^{\mu\nu}\partial_{\mu}\phi_a\partial_{\nu}\phi_a}.\eea
From this action one can further compute the Hamiltonian density, which is given by the following expression:
\bea{{\cal H}=\sum^{N}_{c=1}\widetilde{\Pi^2_c}~~~{\rm where}~~~\widetilde{\Pi^2_c}=\frac{1}{2a^d(t)}\Pi^2_c}.\eea
In this case the following normalised OTOC for the non-interacting case are the interesting one, which are given by: 
\bea && {\textcolor{red}{\bf 2-point~OTOC:}~~~~ Y_{cc}(t,\tau)=-\langle \left[\phi_c(t),\widetilde{\Pi_c}(\tau)\right]\rangle_{\beta}},\\
&& {\textcolor{red}{\bf 4-point~OTOC:}~~~~{\cal C}_{cc}(t,\tau)=-\frac{\langle \left[\phi_c(t),\widetilde{\Pi_c}(\tau)\right]^2\rangle_{\beta}}{\sum\limits^{N}_{a=1}\langle \phi_a(t) \phi_a(t)\rangle_{\beta}\sum\limits^{N}_{b=1}\langle \widetilde{\Pi_b}(\tau)\widetilde{\Pi_b}(\tau)\rangle_{\beta}}}.\eea
 For the interacting case this OTOC is further generalised to the following expressions:
 \bea && {\textcolor{red}{\bf 2-point~OTOC:}~~~~ Y_{cm}(t,\tau)=-\langle \left[\phi_c(t),\widetilde{\Pi_m}(\tau)\right]\rangle_{\beta}},\\
&& {\textcolor{red}{\bf 4-point~OTOC:}~~~~
{\cal C}_{cm}(t,\tau)=-\frac{\langle \left[\phi_c(t),\widetilde{\Pi_m}(\tau)\right]^2\rangle_{\beta}}{\sum\limits^{N}_{a=1}\sum\limits^{N}_{n=1}\langle \phi_a(t) \phi_n(t)\rangle_{\beta}\sum\limits^{N}_{b=1}\sum\limits^{N}_{s=1}\langle \widetilde{\Pi_b}(\tau)\widetilde{\Pi_s}(\tau)\rangle_{\beta}}}.\eea
For a single field ($N=1$) case this action can be further simplified as:
\bea {S=\frac{1}{2}\int d^{d+1}x~\sqrt{-g}g^{\mu\nu}\partial_{\mu}\phi\partial_{\nu}\phi=\frac{1}{2}\int d^{d+1}x~\sqrt{-g}~\left(\partial\phi\right)^2},\eea 
where we have expressed the Hamiltonian in $(d+1)$ dimensional De Sitter background, which is given by the following expression:
\bea{{\cal H}=\widetilde{\Pi^2}~~~~{\rm where}~~~\widetilde{\Pi^2}=\frac{1}{2a^d(t)}\Pi^2}.\eea
In this case the following normalised OTOC for the non-interacting case are the interesting one, which are given by:
\bea && {\textcolor{red}{\bf 2-point~OTOC:}~~~~ Y(t,\tau)=-\langle \left[\phi(t),\tilde{\Pi}(\tau)\right]\rangle_{\beta}},\\
&& {\textcolor{red}{\bf 4-point~OTOC:}~~~~{\cal C}(t,\tau)=-\frac{\langle \left[\phi(t),\widetilde{\Pi}(\tau)\right]^2\rangle_{\beta}}{\langle \phi(t) \phi(t)\rangle_{\beta}\langle \widetilde{\Pi}(0)\widetilde{\Pi}(\tau)\rangle_{\beta}}}.\eea
During inflation (when we fix $d=3$) we actually consider massless scalar field i.e. $m<<H$. In this context one needs to consider the following perturbation in the scalar field in the De Sitter background:
\bea {\textcolor{red}{\bf Field~after~perturbation:}~~~\phi({\bf x},t)=\underbrace{\phi(t)}_{\textcolor{red}{\bf Background}}+\underbrace{\delta\phi({\bf x},t)}_{\textcolor{red}{\bf Perturbation}}}\eea
to express the whole dynamics in terms of a gauge invariant description through a variable:
\bea{\footnotesize{\textcolor{red}{\bf Curvature~perturbation:}~~~~\zeta({\bf x},t)=\underbrace{-\underbrace{\frac{H(t)}{\dot{\phi}(t)}}_{\textcolor{red}{\bf Background}}\underbrace{\delta\phi({\bf x},t)}_{\textcolor{red}{\bf perturbation}}}_{\textcolor{red}{\bf First~order~contribution}}+\underbrace{\cdots}_{\textcolor{red}{\bf Higher~order~contribution}}}}~.~~~~~~~~\eea
For the further analysis we restrict ourself only upto the first order contribution and will going to neglect all other higher order contributions as they are sufficiently small enough compared to the first order contribution as appearing in the above equation. In this context, geometrically the curvature perturbation $\zeta({\bf x},t)$ measures the  spatial curvature of constant hypersurface, which is represented by the following equation:
\bea {\textcolor{red}{\bf Spatial~curvature:}~~~R^{(3)}=-\frac{4}{a^{2}(t)}\partial^2\zeta({\bf x},t)=\frac{4}{a^2(t)}\frac{H(t)}{\dot{\phi}(t)}\partial^2\delta\phi({\bf x},t)+\cdots}~.\eea
One can explicitly show that at the level of perturbations of De Sitter metric if we fix the following gauge condition:
\bea && {\delta\phi({\bf x},t)=0},\\
&&{ g_{ij}({\bf x},t)=a^2(t)\left[e^{2\zeta({\bf x},t)}\delta_{ij}+h_{ij}({\bf x},t)\right]\approx a^2(t)\left[\left(1+2\zeta({\bf x},t)\right)\delta_{ij}+h_{ij}({\bf x},t)\right]},\\
&&{\partial_{i}h_{ij}({\bf x},t)=0=h^{i}_{i}({\bf x},t)}.\eea
which actually fix the time and spatial reparameterizations for the dynamical fields. In this gauge  the  spatial curvature of constant hypersurface becomes zero i.e. $R^{(3)}=0$, which implies comoving curvature perturbation $\zeta({\bf x},t)$ is conserved on superhorizon scales. 

It can be explicitly shown that by doing ADM analysis that the second order perturbed action for scalar modes can be expressed by the following action after gauge fixing~\footnote{It is important to note that here if we start with a scalar field theory with non-canonical kinetic term (such as general $P(X,\phi)$ theory, where $X=-1/2~g^{\mu\nu}\partial_{\mu}\phi\partial_{\nu}\phi$ is the kinetic term) then we get an additional factor of square of the sound speed appearing in the action:
\bea {S=\int dt~d^3x~a^3(t)~\frac{\dot{\phi}^2(t)}{2H^2c^2_S}\left[\dot{\zeta}^2({\bf x},t)-\frac{c^2_S}{a^2(t)}\left(\partial_{i}\zeta({\bf x},t)\right)^2\right]}.\eea
If we further fix $P(X,\phi)=X$ for the massless scalar field case then we can get back the sound speed $c_S=1$ as a special case.}:
\bea {S=\int dt~d^3x~a^3(t)~\frac{\dot{\phi}^2(t)}{2H^2}\left[\dot{\zeta}^2({\bf x},t)-\frac{1}{a^2(t)}\left(\partial_{i}\zeta({\bf x},t)\right)^2\right]}.\eea
Now we further define a gauge invariant variable, which is in the cosmology literature sometimes known as {\it Mukhanov} variable, as given by:
\bea {\textcolor{red}{\bf Mukhanov~variable:}~~f({\bf x},t)\equiv z(t)\zeta({\bf x},t),~~{\rm where}~~z(t)=a(t)\frac{\dot{\phi}(t)}{H}=\sqrt{2\epsilon}~a(t)},~~~~~\eea
which serves the purpose of field redefinition in terms of gauge invariant perturbation variable. Here $\epsilon$ is known as the slow-roll parameter, which is defined as, $\epsilon=-\dot{H}/H^2$. In this context, $\dot{H}<0$ and $\epsilon<1$ during inflation and $\epsilon=1$ at the end of inflation.

Translating in terms of the conformal time and applying integration by parts the above mentioned action can be recast in the following form:
\bea {S=\int d\tau ~d^3x~\frac{1}{2} \left[\left(\partial_{\tau}f({\bf x},\tau)\right)^2-\left(\partial_{i}f({\bf x},\tau)\right)^2+\frac{1}{z(\tau)}\frac{d^2z(\tau)}{d\tau^2}\left(f({\bf x},\tau)\right)^2 \right]}~~.\eea
However, dealing this problem in the coordinate space is very difficult as we are further interested in OTOC's in the present context. On the other hand, dealing this problem in Fourier space gives many good underlying physical features for which we do the following Fourier transform:
\bea {\textcolor{red}{\bf Fourier ~transform:}~~~~~f({\bf x},\tau)=\int \frac{d^3{\bf k}}{(2\pi)^3}f_{\bf k}(\tau)~e^{i{\bf k}.{\bf x}}}~,\eea
using which the previously mentioned action can be further simplified as:
\bea {S=\int d\tau ~d^3{\bf k}~\underbrace{\frac{1}{2} \left[\left(\partial_{\tau}f_{{\bf k}}(\tau)\right)^2+\omega^2_{\bf k}(\tau)\left(f_{\bf k}(\tau)\right)^2\right]}_{\textcolor{red}{\bf Lagrangian~density~of~time~dependent~oscillator\equiv {\cal L}(f_{\bf k}(\tau),\partial_{\tau}f_{\bf k}(\tau))}}}~~,\eea
where we define the effective time dependent frequency $\omega_{\bf k}(\tau)$ as:
\bea {\textcolor{red}{\bf Time~dependent~frequency:}~~~~~~\omega^2_{\bf k}(\tau)\equiv \left(k^2-\frac{1}{z(\tau)}\frac{d^2z(\tau)}{d\tau^2}\right)}~.\eea
This means that after gauge fixing in terms of gauge invariant scalar curvature perturbation the problem is transformed to a time dependent parametric oscillator problem. 

From the above mentioned action one can compute the canonically conjugate momentum, which is given by the following expression:
\bea{ \textcolor{red}{\bf Momentum~density:}~~~~\Pi_{\bf k}(\tau)=\frac{\partial {\cal L}(f_{\bf k}(\tau),\partial_{\tau}f_{\bf k}(\tau))}{\partial\left(\partial_{\tau}f_{\bf k}(\tau)\right)}=\partial_{\tau}f_{\bf k}(\tau)}~.\eea
Consequently the Hamiltonian of the system can be expressed in terms of the gauge invariant variables as:
\bea {~~H=\int d^3{\bf k}~\frac{1}{2}\left[\underbrace{{\Pi}^2_{\bf k}(\tau)}_{\textcolor{red}{\bf Kinetic~term}}+~\underbrace{\omega^2_{\bf k}(\tau){f^2_{\bf k}(\tau)}}_{\textcolor{red}{\bf Potential~term}}\right]}~,\eea
Additionally, it is important to mention here that the following constraint have to be satisfied:
\bea {f_{-\bf k}(\tau)=f^{\dagger}_{\bf k}(\tau)=f^{*}_{\bf k}(\tau)}.\eea
Consequently we have used a simplified notations in the above mentioned Hamiltonian in Fourier space, which are given by the following expression:
\bea &&{{\Pi}^2_{\bf k}(\tau)=\Pi_{\bf k}(\tau)\Pi_{-\bf k}(\tau)=(\partial_{\tau}f_{\bf k}(\tau))(\partial_{\tau}f_{-\bf k}(\tau))=f^{'}_{\bf k}(\tau)f^{'}_{-\bf k}(\tau)=|f^{'}_{\bf k}(\tau)|^2=|\Pi_{\bf k}(\tau)|^2},~~~~~~~~~~~~\\
&&{f^2_{\bf k}(\tau)=f_{\bf k}(\tau)f_{-\bf k}(\tau)=|f_{\bf k}(\tau)|^2}.~~~~~~~~\eea
In the present context one can interpret the time dependent effective frequency of the gauge fixed field as the effective time dependent mass of that field: 
\bea {\textcolor{red}{\bf Time~ dependent~mass:}~~~m^2_{\bf E}(\tau)=\omega^2_{\bf k}(\tau)}~.\eea
where we explicitly compute the expression for the second term as:
\bea {\frac{1}{z(\tau)}\frac{d^2z(\tau)}{d\tau^2}= \frac{1}{a(\tau)}\frac{d^2a(\tau)}{d\tau^2}+\underbrace{\frac{1}{2\epsilon}\frac{d^2\epsilon}{d\tau^2}-\frac{1}{2\epsilon^2}\left(\frac{d\epsilon}{d\tau}\right)^2+\frac{1}{a(\tau)\epsilon}\left(\frac{d\epsilon}{d\tau}\right)\left(\frac{da(\tau)}{d\tau}\right)}_{\textcolor{red}{\bf Sub-leading~contributions}}}.~~~~\eea
for the massless scalar field. Now for inflation in the De Sitter background and at the epoch of reheating one can further simplify this above mentioned expression by neglecting the contributions from the sub-leading correction terms as:
\begin{eqnarray}
&&{\frac{1}{z(\tau)}\frac{d^2z(\tau)}{d\tau^2}=\frac{\left(\nu^2-\frac{1}{4}\right)}{\tau^2}= \large \left\{
     \begin{array}{lr}
  \displaystyle  \frac{2}{\tau^2} & \text{\textcolor{red}{\bf De~Sitter}}\\ 
 \displaystyle      \frac{2(1-3w_{reh})}{(1+3w_{reh})^2}\frac{1}{\tau^2},~~~{\rm where}~~0\leq w_{reh}\leq \frac{1}{3} & \text{\textcolor{red}{\bf Reheating}}  \end{array}
   \right.}~~~~~~~~~~~~
\end{eqnarray}
where the parameter $\nu$ is defined as:
\begin{eqnarray}
{\nu= \large \left\{
     \begin{array}{lr}
 \displaystyle   \frac{3}{2} &~~~~~~~~~~~ \text{\textcolor{red}{\bf De~Sitter}}\\ 
  \displaystyle    \sqrt{ \frac{1}{4}+\frac{2(1-3w_{reh})}{(1+3w_{reh})^2}},~~~{\rm where}~~0\leq w_{reh}\leq \frac{1}{3} & \text{\textcolor{red}{\bf Reheating}}  \end{array}
   \right.}~~~~~~~~~
\end{eqnarray}
Now, using this setup in terms of the gauge invariant perturbation variables the following interesting OTOC can be computed explicitly: 
\bea && {\textcolor{red}{\bf 2-point~OTOC:}~~~~ Y^{\zeta}(\tau_1,\tau_2)=-\langle \left[\zeta(\tau_1),{\Pi}_{\zeta}(\tau_2)\right]\rangle_{\beta}},\\
&&{\textcolor{red}{\bf 4-point~OTOC:}~~~~ {\cal C}^{\zeta}(\tau_1,\tau_2)=-\frac{\langle \left[\zeta(\tau_1),{\Pi}_{\zeta}(\tau_2)\right]^2\rangle_{\beta}}{\langle \zeta(\tau_1) \zeta(\tau_1)\rangle_{\beta}\langle {\Pi}_{\zeta}(\tau_2){\Pi}_{\zeta}(\tau_2)\rangle_{\beta}}}.\eea
On the other hand, if we want to express all the OTOC in terms of the gauge invariant perturbation variables and the corresponding rescaled field content and momenta, then one can use the following equivalent definition of OTOC in the present context, which are given by the following expressions:
\bea && {\textcolor{red}{\bf 2-point~OTOC:}~~~~ Y^{f}(\tau_1,\tau_2)=-\langle \left[f(\tau_1),{\Pi}_{f}(\tau_2)\right]\rangle_{\beta}},\\
&&{\textcolor{red}{\bf 4-point~OTOC:}~~~~ {\cal C}^{f}(\tau_1,\tau_2)=-\frac{\langle \left[f(\tau_1),{\Pi}_{f}(\tau_2)\right]^2\rangle_{\beta}}{\langle f(\tau_1) f(\tau_1)\rangle_{\beta}\langle {\Pi}_{f}(\tau_2){\Pi}_{f}(\tau_2)\rangle_{\beta}}}.\eea
For the computation of all of these OTOC we use the SO(1,4) isommetric $\alpha$ vacua and the well known Bunch Davies vacuum state as the quantum vacuum state.

\subsubsection{For partially massless scalar field}
\label{sec:introduction}
Partially massless fields share some intermediate features of massive and massless fields in flat space. On the one hand, it is important to note that, they carry more than two degrees of freedom in the context of quantum field theory. In this context we are interested in the partially massless scalar fields because it is expected that may survive until
the end of inflation and therefore it is expected that during the stochastic particle production mechanism it plays significant role. 
 Here the terminology partially massless is used to consider scalar fields which have mass approximately same with the characteristic Cosmological scale i.e. the Hubble scale ($m\approx H$). To serve this purpose we consider scalar field which is conformally coupled with the gravity and mass of such scalar fields can be written as, $m=\sqrt{2}H$. 
A massless $N$ number of scalar fields can be generalized by the following simplified action:
\bea {S=\frac{1}{2}\int d^{d+1}x~\sqrt{-g}\sum^{N}_{a=1}\sum^{N}_{b=1}\left[g^{\mu\nu}G^{ab}\partial_{\mu}\phi_a\partial_{\nu}\phi_b-(m^2)^{ab}\phi_a\phi_b\right]}.\eea
In the simplest non-interacting situation one can write, $G^{ab}=\delta^{ab}$, $(m^2)^{ab}=c^2H^2\delta^{ab}$ and preferably we consider, $c=\sqrt{2}$. In general $c$ should not be a very large number. In this case the action for the $N$ partially massless scalar field can be simplified as:
\bea {S=\frac{1}{2}\int d^{d+1}x~\sqrt{-g}\sum^{N}_{a=1}\left[g^{\mu\nu}\partial_{\mu}\phi_a\partial_{\nu}\phi_a-c^2H^2\phi^2_a\right]}.\eea
From this action one can further compute the Hamiltonian density, which is given by the following expression:
\bea{{\cal H}=\sum^{N}_{a=1}\widetilde{\Pi^2_a}+\widetilde{U(\phi_a)}~~~{\rm with}~~~\widetilde{\Pi^2_a}=\frac{1}{2a^d(t)}\Pi^2_a,~~~\widetilde{U(\phi_a)}=a^{d}(t)c^2H^2\phi^2_a}.\eea

It is important to note that here all the previously mentioned definition of OTOC's for interacting and non-interacting $N$ number of scalar fields in case of partially massless case look like similar except the thermal Botzmann factor appearing in the thermal average over ensembles . The prime reason is the expression for the Hamiltonian is different for the partially massless case compared to the massless case discussed in the just previous subsection. 

For a single field ($N=1$) case this action can be further simplified as:
\bea {S=\frac{1}{2}\int d^{d+1}x~\sqrt{-g}\left[g^{\mu\nu}\partial_{\mu}\phi\partial_{\nu}\phi-c^2H^2\phi^2\right]=\frac{1}{2}\int d^{d+1}x~\sqrt{-g}~\left[\left(\partial\phi\right)^2-(cH\phi)^2\right]},~~~~~~~~\eea
where we have expressed the Hamiltonian in $(d+1)$ dimensional De Sitter background, which is given by the following expression:
\bea{{\cal H}=\left[\widetilde{\Pi^2}+\widetilde{U(\phi)}\right]~~~{\rm with}~~~\widetilde{\Pi^2}=\frac{1}{2a^d(t)}\Pi^2,~~~\widetilde{U(\phi)}=a^{d}(t)c^2H^2\phi^2}.\eea

To write down everything in terms of gauge invariant perturbations all expressions are similar as appearing in the previous subsection, except the following one:
\bea \displaystyle{\frac{1}{z(\tau)}\frac{d^2z(\tau)}{d\tau^2}= \frac{1}{a(\tau)}\frac{d^2a(\tau)}{d\tau^2}+\underbrace{\frac{1}{2\epsilon}\frac{d^2\epsilon}{d\tau^2}-\frac{1}{2\epsilon^2}\left(\frac{d\epsilon}{d\tau}\right)^2+\frac{1}{a(\tau)\epsilon}\left(\frac{d\epsilon}{d\tau}\right)\left(\frac{da(\tau)}{d\tau}\right)}_{\textcolor{red}{\bf Sub-leading~contributions}}}.~~~~\eea
for the partially massless scalar field. Now for inflation in the De Sitter background and at the epoch of reheating one can further simplify this above mentioned expression by neglecting the contributions from the sub-leading correction terms as:
\begin{eqnarray}
&& {\frac{1}{z(\tau)}\frac{d^2z(\tau)}{d\tau^2}=\frac{\left(\nu^2-\frac{1}{4}\right)}{\tau^2}= \large \left\{
     \begin{array}{lr}
   \displaystyle \left(2-c^2\right)\frac{1}{\tau^2},~~~{\rm where}~~c\geq \sqrt{2} &~ \text{\textcolor{red}{\bf De~Sitter}}\\ 
   \displaystyle    \frac{2(1-3w_{reh})}{(1+3w_{reh})^2}\frac{1}{\tau^2},~~~{\rm where}~~0\leq w_{reh}\leq \frac{1}{3} & \text{\textcolor{red}{\bf Reheating}}  \end{array}
   \right.}~~~~~~~~~~
\end{eqnarray}
where the parameter $\nu$ is defined as:
\begin{eqnarray}
{\nu= \large \left\{
     \begin{array}{lr}
  \displaystyle  \sqrt{\frac{9}{4}-c^2},~~~{\rm where}~~c\geq \sqrt{2} &~~~~~~~~~~~ \text{\textcolor{red}{\bf De~Sitter}}\\ 
   \displaystyle   \sqrt{ \frac{1}{4}+\frac{2(1-3w_{reh})}{(1+3w_{reh})^2}},~~~{\rm where}~~0\leq w_{reh}\leq \frac{1}{3} & \text{\textcolor{red}{\bf Reheating}}  \end{array}
   \right.}~~~~~~~~~~~
\end{eqnarray}
For the conformally coupled partially massless scalar field ($c=2$) it is clearly observed from the above mentioned expressions that, for the De Sitter case we get, $\frac{1}{z(\tau)}\frac{d^2z(\tau)}{d\tau^2}=0$ and $\nu=1/2$. This will be very useful for the further computation of the OTOC's. Apart from the the mathematical definitions of the OTOC's in coordinate space or in momentum space are exactly same apart from the exponential thermal Boltzmann factor where the expression for the Hamiltonian will change in the partially massless scalar field case.
\subsubsection{For massive scalar field}
\label{sec:introduction}
In this section we are interested in to discuss the outcomes from massive scalar field theory. Here the terminology massive is used to consider scalar fields which have mass much heavier the characteristic Cosmological scale i.e. the Hubble scale ($m>>H$). To serve this purpose we consider scalar field which have constant and time dependent profile both. 
A massless $N$ number of scalar fields can be generalized by the following simplified action:
\bea {S=\frac{1}{2}\int d^{d+1}x~\sqrt{-g}\sum^{N}_{a=1}\sum^{N}_{b=1}\left[g^{\mu\nu}G^{ab}\partial_{\mu}\phi_a\partial_{\nu}\phi_b-(m^2(t))^{ab}\phi_a\phi_b\right]}.\eea
In the simplest non-interacting situation one can write, $G^{ab}=\delta^{ab}$, $(m^2)^{ab}=m^2(t)\delta^{ab}$ and in general $m^2(t)$ is a function of time coordinate. But for some special cases one can consider constant time independent profiles as well for the mass function. We will choose this simplest possibility for the further computation. In this case the action for the $N$ massive scalar field can be simplified as:
\bea {S=\frac{1}{2}\int d^{d+1}x~\sqrt{-g}\sum^{N}_{a=1}\left[g^{\mu\nu}\partial_{\mu}\phi_a\partial_{\nu}\phi_a-m^2\phi^2_a\right]}.\eea
From this action one can further compute the Hamiltonian density, which is given by the following expression:
\bea{{\cal H}=\sum^{N}_{a=1}\widetilde{\Pi^2_a}+\widetilde{U(\phi_a)}~~~{\rm with}~~~\widetilde{\Pi^2_a}=\frac{1}{2a^d(t)}\Pi^2_a,~~~\widetilde{U(\phi_a)}=a^{d}(t)m^2\phi^2_a}.\eea

It is important to note that here all the previously mentioned definition of OTOC for interacting and non-interacting $N$ number of scalar fields in case of massive case look like similar except the thermal Boltzmann factor appearing in the thermal average over ensembles. 

For a single field ($N=1$) case this action can be further simplified as:
\bea {S=\frac{1}{2}\int d^{d+1}x~\sqrt{-g}\left[g^{\mu\nu}\partial_{\mu}\phi\partial_{\nu}\phi-m^2(t)\phi^2\right]=\frac{1}{2}\int d^{d+1}x~\sqrt{-g}~\left[\left(\partial\phi\right)^2-(m\phi)^2\right]},~~~~~~~~\eea
where we have expressed the Hamiltonian in $(d+1)$ dimensional De Sitter background, which is given by the following expression:
\bea{{\cal H}=\left[\widetilde{\Pi^2}+\widetilde{U(\phi)}\right]~~~{\rm with}~~~\widetilde{\Pi^2}=\frac{1}{2a^d(t)}\Pi^2,~~~\widetilde{U(\phi)}=a^{d}(t)m^2\phi^2}.\eea

To write down everything in terms of gauge invariant perturbations all expressions are similar as appearing in the previous subsection, except the following one:
\bea {\frac{1}{z(\tau)}\frac{d^2z(\tau)}{d\tau^2}= \frac{1}{a(\tau)}\frac{d^2a(\tau)}{d\tau^2}+\underbrace{\frac{1}{2\epsilon}\frac{d^2\epsilon}{d\tau^2}-\frac{1}{2\epsilon^2}\left(\frac{d\epsilon}{d\tau}\right)^2+\frac{1}{a(\tau)\epsilon}\left(\frac{d\epsilon}{d\tau}\right)\left(\frac{da(\tau)}{d\tau}\right)}_{\textcolor{red}{\bf Sub-leading~contributions}}}.~~~~\eea
for the massive scalar field. Now for inflation in the De Sitter background and at the epoch of reheating one can further simplify this above mentioned expression by neglecting the contributions from the sub-leading correction terms as:
\begin{eqnarray}
&&{\frac{1}{z(\tau)}\frac{d^2z(\tau)}{d\tau^2}=\frac{\left(\nu^2(\tau)-\frac{1}{4}\right)}{\tau^2}= \large \left\{
     \begin{array}{lr}
  \displaystyle  -\left(\frac{m^2(\tau)}{H^2}-2\right)\frac{1}{\tau^2},~~~{\rm where}~~m>>H &~\text{\textcolor{red}{\bf De~Sitter}}\\ 
 \displaystyle      \frac{2(1-3w_{reh})}{(1+3w_{reh})^2}\frac{1}{\tau^2},~~~{\rm where}~~0\leq w_{reh}\leq \frac{1}{3} & \text{\textcolor{red}{\bf Reheating}}  \end{array}
   \right.}~~~~~~~~~
\end{eqnarray}
where the parameter $\nu$ is defined as:
\begin{eqnarray}
{\nu(\tau)= \large \left\{
     \begin{array}{lr}
  \displaystyle  i\sqrt{\frac{m^2}{H^2}-\frac{9}{4}},~~~{\rm where}~~m>> H &~~~~~~~~~~~ \text{\textcolor{red}{\bf De~Sitter}}\\ 
  \displaystyle    \sqrt{ \frac{1}{4}+\frac{2(1-3w_{reh})}{(1+3w_{reh})^2}},~~~{\rm where}~~0\leq w_{reh}\leq \frac{1}{3} & \text{\textcolor{red}{\bf Reheating}}  \end{array}
   \right.}~~~~~~~~~
\end{eqnarray}
 This will be very useful for the further computation of the OTOC's. Apart from the the mathematical definitions of the OTOC's in coordinate space or in momentum space are exactly same apart from the exponential thermal Boltzmann factor where the expression for the Hamiltonian will change in the massive scalar field case.
 
 If the massive scalar field is conformal time dependent then one can use the following conformal time dependent protocols for the computation of OTOC's with the massive scalar fields:
 \begin{eqnarray}
{m^2(\tau)= \large \left\{
     \begin{array}{lr}
 \displaystyle  \delta(\tau-\tau^{'}) &~~~~~~~~~~~ \text{\textcolor{red}{\bf Non-Markovian~White~noise}}\\ 
 \displaystyle   K(\tau-\tau^{'}) & \text{\textcolor{red}{\bf Markovian~Coloured~noise}}  \end{array}
   \right.}
\end{eqnarray}
For the simplicity here we have fixed the amplitudes $A(\tau)=1=B(\tau)$ for both the white and coloured noise conformal time dependent effective mass profile protocols. Now for the Markovian coloured noise case if we fix $\tau^{'}=0$ then then for the phenomenological purpose one can consider the following sets of protocols, which have huge applications in the context of quantum quench, information theory and condensed matter field theory:
  \begin{eqnarray}
{\large \frac{m^2(\tau)}{H^2}=\frac{K(\tau)}{H^2}=  \left\{
     \begin{array}{lr}
   \displaystyle~(a\pm b\tanh(q\tau)) &~~~~~~~~~~~~~\text{\textcolor{red}{\bf Protocol~I}}\\ 
  \displaystyle ~\Theta(\tau\pm\tau_0) & ~~~~~~~~~~~~~~~~~~~~~~\text{\textcolor{red}{\bf Protocol~II}} \\ 
 \displaystyle  ~{\rm sech}^2(q\tau) & ~~~~~~~~~~~~~~~~~\text{\textcolor{red}{\bf Protocol~III}}\\ 
  \displaystyle  ~\left[a\left(\frac{\tau}{\tau_0}-1\right)^2+b\right]& ~~~~~~~~~~~~~~~~~~~~~~\text{\textcolor{red}{\bf Protocol~IV}}\\ 
\displaystyle \tanh^2(q\tau)   & \text{\textcolor{red}{\bf Protocol~V}} \\ 
 \displaystyle(a\pm b~{\rm sech}^2(q\tau))   & \text{\textcolor{red}{\bf Protocol~VI}} 
     \end{array}
   \right.}~~~~~~~~ 
\end{eqnarray}
These Markovian coloured noise profiles are frequently used in the context of quantum mechanical quench in the context of condensed matter field theory and quantum information theory. Since for the above mentioned time dependent protocols the mode function for the scalar perturbations are not exactly solvable the WKB approximated mentioned is very useful to compute the expression for for cosmological OTOC in such cases.Though in this paper we have not computed explicitly OTOC using WKB approximated method, but have explicitly derived the general solution of the WKB approximated solution of general conformal time dependent mass profile in the Appendix, which one can use further to compute the expression for the cosmological OTOC from the present set up. In the next section, we explicitly compute the expressions for the OTOC's that we have defined in this section for massless, partially massless and massive scalar (with time independent mass) field theory.
\section{Quantum micro-canonical OTO amplitudes and OTOC in Cosmology}
\label{sec:OTO1}
In this section we will derive and interpret the results for OTOC introduced and defined in the previous section for massless, partially massless and massive scalar fields in the context of Cosmology, which are useful to study the physical implications of the stochastic process of particle production during the epoch of inflation and also during the reheating epoch as well. We will show that how are derived result will be very useful to provide new information regarding the above mentioned two processes appearing in the context of {\it Cosmology of Early Universe} or {\it Primordial Cosmology}. Instead of computing the expressions for the usual time ordered or the anti-time ordered correlation functions, which is commonly studied in the context of Cosmology, here we are actually interested to study the expressions for the OTOC because in the present context we want to study the physical outcome of the out-of-equilibrium phenomena driven Quantum Field Theory aspects of Cosmology. The stochastic particle production and reheating phenomena are two most important significant facts during which the quantum system studied in this context goes to out-of-equilibrium, and after waiting for a sufficient time scale all such quantum system reaches equilibrium and it is expected in all those cases that the time dependent behaviour of the quantum correlation function saturates for large late time scale in Cosmology. It is expected that when the system goes to the out-of-equilibrium quantum state the behaviour of the quantum correlation function will be solely controlled by the OTOC that we have defined in the earlier sections. Since these type of correlations are not studied in the context of Cosmology, it is expected that many unexplored features of Cosmology at out-of-equilibrium phase can be studied from the outcomes of the OTOC, which we will compute for massless, partially massless and massive heavy scalar field theories. Apart from the application in the context of stochastic particle production during inflation and reheating the present computation can be applicable to the Dark Matter bound-state formation at higher order and many more other places in the evolution of our Universe. Though we have not studied this possibility in this paper, but one can think for doing such calculation in the context of Dark Matter. Additionally it is important to note that, during the computation of OTOC the quantum system studied in the context of the time evolution of our Universe is closed or isolated or adiabatic quantum system. Here the terminology closed is used to describe a quantum system of our Universe is not thermally or any other way interacting with any kind of environment. So it is obvious here that when the system is interacting with the environment thermally or any other way the corresponding quantum system is identified to be a open or non-adiabatic system. So it might be a nice option to carry forward the computation for OTOC for Cosmology with open quantum systems. 

\subsection{Computational strategy}

The steps of computing the OTOC are appended below point-wise:
\begin{enumerate}
\item At first, we need find out the analytic solution of the equation of motion of the scalar field in the corresponding FLRW flat spatial background.
\item From the above mentioned solution we need to find out the canonically conjugate field momentum and using that we need to compute the square of the quantum mechanical commutator bracket, proved during performing this operation we need to quantize the field content in terms of the creation and annihilation operators.
\item From the above mentioned result we need to further compute the thermal average value of the square of the commutator bracket of the field and its canonically conjugate momenta, which physically represents the four-point out-of-time ordered correlation function.
\item Next, we need to compute the partition function the quantum system under consideration by computing the thermal Boltzmann factor and the trace of the corresponding thermal Boltzmann factor. To serve this purpose we need to also compute the expression for the Hamiltonian from the quantum mechanical system under consideration.
\item Further, we need to compute the thermal two point out-of-time ordered correlation functions from the field content and the corresponding canonically conjugate momentum. This will help us to find out the expression for the normalised OTOC which is defined in the earlier sections explicitly.
\item Side wise instead of computing these description in the coordinate space one can do a similar computation of OTOC in Fourier space as well. For this purpose one needs to compute all of the required operators in the Fourier transformed space. Sometimes doing the computation in Fourier transformed space is simpler than the computation in coordinate space. Most importantly, doing the computation in Fourier space is really very helpful in the context of Cosmology. The prime reason is of doing the computation in momentum space (which is the Fourier transformed version of the coordinate space) in the context of Cosmology is more technically interesting. Actually in momentum space most of the quantum correlation functions in the context of Cosmology preserves the conformal symmetry under conformal transformations. In quasi De Sitter space this conformal symmetry is slightly broken in Fourier space by the amount of slow-roll parameter, which we have taken into account in our computations.
\item In the present context it is better to perform the whole computation in a preferred choice of gauge, which is commonly performed in the context of Cosmological Perturbation Theory. Following the same strategy here we work on $\delta\phi=0$ gauge, which makes the computation of OTOC simpler than doing the computation doing without the choice of a preferred gauge.
\item It is important to mention here that, doing the computation of OTOC for $N$ interacting scalar field analytically are extremely complicated. On the other hand, doing the computation with $N$ non-interacting multi scalar field is also not very simple. For this reason we in this paper try to restrict ourself to the non-interacting single field and $N$ field case. For the single field case we also provide the computation of OTOC in the previously mentioned preferred gauge choice.
\item Last but not the least, doing the computation of OTOC in any arbitrary dimensions is not at all possible analytically. So we have restrict our computation in this article in $d=3$ spatial dimension which is compatible with our spatially flat $3+1$ dimensional space time in FLRW cosmological background.
\end{enumerate}

\subsection{Classical mode functions in Cosmology}
After varying the gauged action written in $\delta\phi=0$ gauge with respect to the redefined field we get the following equation of motion in the Fourier transformed space: 
\bea {\left[\frac{d^2}{d\tau^2}+\left(k^2-\frac{\left(\nu^2-\frac{1}{4}\right)}{\tau^2}\right)\right]f_{\bf k}(\tau)=0},\eea
where the parameter $\nu$ is defined for the massless, partially massless and massive heavy scalar field case as:
\begin{eqnarray}
{\nu= \footnotesize \left\{
     \begin{array}{lr}
\displaystyle    \frac{3}{2} &~ \text{\textcolor{red}{\bf DS+massless}}\\ 
    \displaystyle \sqrt{\frac{9}{4}-c^2},~~~{\rm where}~~c\geq \sqrt{2} ~(m\sim H)& \text{\textcolor{red}{\bf DS+partially~massless}}\\ 
  \displaystyle   i\sqrt{\frac{m^2}{H^2}-\frac{9}{4}},~~~{\rm where}~~m>> H &\text{\textcolor{red}{\bf DS+heavy}}\\ 
 \displaystyle     \sqrt{ \frac{1}{4}+\frac{2(1-3w_{reh})}{(1+3w_{reh})^2}},~~~{\rm where}~~0\leq w_{reh}\leq \frac{1}{3} & \text{\textcolor{red}{\bf Reheating}}  \end{array}
   \right.}~~~~~~
\end{eqnarray}
Here for the third case the explicit form of the conformal time dependent mass functions are mentioned in the earlier section. 

One can further generalise the expression for the mass parameter $\nu$ if we introduce a non-minimal coupling parameter $\xi$ to the Ricci scalar curvature term in the action, by replacing the mass term $m^2\phi^2$ with the new rescaled term $(m^2+\xi R)\phi^2$ i.e.
\bea  {m^2\phi^2\longrightarrow m^2_{new}\phi^2=(m^2+\xi R)\phi^2}~.\eea
When we consider the De Sitter space, there the Ricci scalar $R$ is constant and given by $R=12H^2$. In this case the the new mass parameter can be expressed as, \bea{ \nu\longrightarrow \nu_{new}=\sqrt{\frac{9}{4}-\frac{m^2_{new}}{H^2}}=\sqrt{\frac{9}{4}-\frac{m^2+12\xi H^2}{H^2}}=\sqrt{\frac{9}{4}-\left(12\xi+\frac{m^2}{H^2}\right)}}~.\eea Consequently, we will get same expressions everywhere with a newly rescaled mass parameter $\nu$ where non-minimal coupling have been introduced explicitly. So even in the most generalised case where the effective mass term is conformal time dependent, there also if we replace the term $m^2/H^2$ with the $12\xi+m^2/H^2$ i.e.
 \bea {\frac{m^2}{H^2}\longrightarrow \frac{m^2_{new}}{H^2}=\left(12\xi+\frac{m^2}{H^2}\right)}~.\eea
  In the expression for $\nu$ then in the approximated WKB solutions that fact will be propagated. It is important to note that, the same prescription is followed in the context of Higgs inflation where the quadric term in the Higgs field is non-minimally coupled to the gravitational sector through the scalar Ricci scalar curvature term. This is a necessary requirement to have inflation from the Higgs sector within the framework of Einstein gravity. Now one can consider a bit more complicated phenomenological situation where the non minimal coupling in the FLRW spatially flat background itself dependent on conformal time and in principal any arbitrary form of the conformal time dependent functions are allowed for that computation. This will further modify the analytical WKB approximated expression for the mode functions for the scalar perturbation in that case for a given conformal time dependent specific form of the effective mass $m^2(\tau)$ and the non-minimal coupling parameter $\xi(\tau)$ and in that case we have to consider the following rescaling:
   \bea {\frac{m^2(\tau)}{H^2}\longrightarrow \frac{m^2_{new}(\tau)}{H^2}=\left(12\xi(\tau)+\frac{m^2(\tau)}{H^2}\right)}~.\eea
 In this case the the new conformal time dependent mass parameter can be expressed as, \bea{ \nu(\tau)\longrightarrow \nu_{new}(\tau)=\sqrt{\frac{9}{4}-\frac{m^2_{new}(\tau)}{H^2}}=\sqrt{\frac{9}{4}-\frac{m^2(\tau)+12\xi(\tau) H^2}{H^2}}=\sqrt{\frac{9}{4}-\left(12\xi(\tau)+\frac{m^2(\tau)}{H^2}\right)}}~.~~~~~~~~\eea
  
Now the most general solution of the mode equation for any constant mass profile (massless, partially massless and heavy scalar production during inflation and during reheating) is given by the following expression:
\bea {f_{\bf k}(\tau)=\sqrt{-\tau}\left[{\cal C}_1~H^{(1)}_{\nu}(-k\tau)+{\cal C}_2~H^{(2)}_{\nu}(-k\tau)\right]},\eea
where ${\cal C}_1$ and ${\cal C}_2$ are two arbitrary integration constants which are fixed by the choice of the initial quantum vacuum state necessarily needed for this computation. Here $H^{(1)}_{\nu}(-k\tau)$ and $H^{(1)}_{\nu}(-k\tau)$ are the Hankel functions of first and second kind with order $\nu$.

In the general context the mass parameter $\nu$ may be a complex parameter. In this case, the solution for the rescaled scalar perturbation mode can be recast as:
\bea  {f_{\bf k}(\tau)=\sqrt{-\tau}\left[{\cal D}_1~{\cal J}_{\nu}(-k\tau)+{\cal D}_2~{\cal Y}_{\nu}(-k\tau)\right]}, \eea
where the redefined two new arbitrary integration constants, ${\cal D}_1$ and ${\cal D}_2$ are defined in terms of the previously defined two new arbitrary integration constants, ${\cal C}_1$ and ${\cal C}_2$ as, 
${\cal D}_1={\cal C}_1+{\cal C}_2$,and ${\cal D}_2=i\left({\cal C}_1-{\cal C}_2\right)$. For the non-integer value of the mass parameter $\nu$, the solution for the rescaled scalar perturbation mode can be recast as:
\bea  {f_{\bf k}(\tau)=\sqrt{-\tau}\left[{\cal E}_1(\nu)~{\cal J}_{\nu}(-k\tau)+{\cal E}_2(\nu)~{\cal J}_{-\nu}(-k\tau)\right]}, \eea
where the redefined two new arbitrary integration constants, ${\cal E}_1$ and ${\cal E}_2$ are defined in terms of the previously defined two new arbitrary integration constants, ${\cal C}_1$ and ${\cal C}_2$ as:
\bea && {{\cal E}_1(\nu)=\left({\cal C}_1+{\cal C}_2\right)+i\left({\cal C}_1-{\cal C}_2\right)~{\rm cot}~\nu \pi={\cal D}_1+{\cal D}_2~{\rm cot}~\nu \pi}~,\\
&& {{\cal E}_2(\nu)=-i\left({\cal C}_1-{\cal C}_2\right)~{\rm cosec}~\nu\pi=-{\cal D}_2~{\rm cosec}~\nu\pi}~.\eea

In particular, when $\nu$ is an integer or not. In this case, the solution for the rescaled scalar perturbation mode can be recast as:
\bea  {f_{\bf k}(\tau)=\sqrt{-\tau}\left[{\cal G}_1(\nu)~{\cal J}_{\nu}(-k\tau)+{\cal G}_2(\nu)~{\cal J}_{-\nu}(-k\tau)\right]}, \eea
where the redefined two new arbitrary integration constants, ${\cal G}_1$ and ${\cal G}_2$ are defined in terms of the previously defined two new arbitrary integration constants, ${\cal C}_1$ and ${\cal C}_2$ as:
\bea && {{\cal G}_1(\nu)=\left({\cal C}_1+{\cal C}_2\right)+i\left({\cal C}_1-{\cal C}_2\right)~{\rm cot}~\nu\pi={\cal D}_1+{\cal D}_2~{\rm cot}~\nu \pi\neq {\cal E}_1(\nu)}~,\\
&& {{\cal G}_2(\nu)=i~{\rm cosec}~\nu\pi~\left({\cal C}_2-{\cal C}_1\right)=-{\cal D}_2~{\rm cosec}~\nu\pi\neq {\cal E}_{2}(\nu)}~.\eea

The corresponding most general canonically conjugate momentum can be further computed from this derived solution as:
\bea \Pi_{\bf k}(\tau)=\partial_{\tau}f_{\bf k}(\tau)=\frac{1}{2 \sqrt{-\tau }}\left[{\cal C}_1\left( k \tau  H_{\nu -1}^{(1)}(-k \tau )- H_{\nu }^{(1)}(-k \tau )- k \tau  H_{\nu +1}^{(1)}(-k \tau )\right)\nonumber\right.\\ \left.~~~~~~~~~~~~~~~~~~~~~+{\cal C}_2 \left(k \tau  H_{\nu -1}^{(2)}(-k \tau )-H_{\nu }^{(2)}(-k \tau )- k \tau  H_{\nu +1}^{(2)}(-k \tau )\right)\right].\eea

Also one can express the Bessel function of the first kind in terms of the Confluent Hypergeometric limit functions in the present context, using which the most general solution of the above mentioned equation of motion is given by the following expression:
\bea &&{f_{\bf k}(\tau)={\cal E}_1(\nu)\frac{\sqrt{-\tau}}{\Gamma(\nu+1)}\left(-\frac{k\tau}{2}\right)^{\nu}~{}_{0}F_{1}\left(\nu+1;-\frac{(k\tau)^2}{4}\right)}~\nonumber\\
&&~~~~~~~~~~~~~~~~~~~~~~~~~{-{\cal E}_2(\nu)\frac{\sqrt{-\tau}}{\Gamma(1-\nu)}\left(-\frac{k\tau}{2}\right)^{-\nu}~{}_{0}F_{1}\left(1-\nu;-\frac{(k\tau)^2}{4}\right)}~.\eea

Also, the asymptotic solution for the rescaled scalar perturbation can be expressed within the window $0<(-k\tau)<\sqrt{\nu+1}$, as:
 \begin{eqnarray}
&&{ \large f_{\bf k}(\tau)= \left\{
     \begin{array}{lr}
  \displaystyle  \frac{\left({\cal D}_1+{\cal D}_2~{\rm cot}~\nu\pi\right)}{\Gamma(\nu+1)}\left(-\frac{k\tau}{2}\right)^{\nu}-\frac{{\cal D}_2\Gamma(\nu)}{\pi}\left(-\frac{k\tau}{2}\right)^{-\nu} &~ \text{\textcolor{red}{\bf if~$\nu>0$~integer}}\\ 
  \displaystyle \frac{{\cal D}_1(-1)^{\nu}}{(-\nu)!}\left(-\frac{k\tau}{2}\right)^{-\nu}+  \frac{{\cal D}_2(-1)^{\nu+1}\Gamma(-\nu)}{\pi}\left(-\frac{k\tau}{2}\right)^{\nu},& \text{\textcolor{red}{\bf if~$\nu<0$~integer}} \end{array}
   \right.}~~~~~~~~~
\end{eqnarray}
For large real arguments lying within the window, $(-k\tau)>>\left|\nu^2-\frac{1}{4}\right|$, one cannot write an actual asymptotic form for the Bessel functions of the first and second kind (unless in the situation where $\nu$ is a half-integer) because they have zeros all the way out to infinity, which would have to be matched exactly by any asymptotic expansion.
However, for a given value of ${\rm arg} (-k\tau)<\pi$, one can write an equation containing a term of order of $|-k\tau|^{-1}$, given by the following expressions:
\bea && {f_{\bf k}(\tau)=\sqrt{\frac{2}{\pi k}}\left[{\cal C}_1\exp(-ik\tau)\exp\left(-\frac{i\pi}{2}\left(\nu+\frac{1}{2}\right)\right)-{\cal C}_2\exp(ik\tau)\exp\left(\frac{i\pi}{2}\left(\nu+\frac{1}{2}\right)\right)\right]}\nonumber\\
&& ~~~~~~~~~~~~~~~~~~~~~~~~~~~~~~~~~~~~~~~~~~~~~~~~~~~~~{+\sqrt{\frac{2}{\pi k}}\exp\left({\rm Im}(-k\tau)\right){\cal O}\left(\frac{1}{|-k\tau|}\right)\left({\cal D}_1-{\cal D}_2\right)}~.~~~~~~~~~~~~\eea
However, from the general structure of the obtained solution for the rescaled field and for the canonically conjugate momentum it is very difficult to extract the physical information out of that. For this reason the asymptotic solutions are really helpful for physical interpretation in different cosmological scales. These asymptotic limits are $k\tau\rightarrow 0$ and $k\tau\rightarrow- \infty$, where we need to determine the behaviour of the Hankel functions of the first and second kind of order $\nu$. Here $k\tau\rightarrow 0$ and $k\tau\rightarrow -\infty$ asymptotic limiting results are used to describe the superhorizon ($k\tau<<-1$) and subhorizon ($k\tau>>-1$) limiting results in the context of primordial cosmological perturbation scenario. The transition point from the subhorizon to superhorizon regime is identify by $k\tau=-1$ , which in Cosmology known as the horizon exit and play a pivotal role to measure various observables of primordial Universe from different theoretical models.

Combining the behaviour in both the superhorizon and subhorizon  limiting region we get following asymptotic most general solution for the rescaled field and momentum variable computed for the arbitrary quantum initial vacuum can be expressed as:
\bea&&{f_{\bf k}(\tau)=2^{\nu-\frac{3}{2}}\frac{i}{\tau}\frac{1}{\sqrt{2}k^{\frac{3}{2}}}(-k\tau)^{\frac{3}{2}-\nu}\left|\frac{\Gamma(\nu)}{\Gamma\left(\frac{3}{2}\right)}\right|}\nonumber\\
&&~~~{\times\left[{\cal C}_1~(1+ik\tau)~\exp\left(-i\left\{k\tau+\frac{\pi}{2}\left(\nu+\frac{1}{2}\right)\right\}\right)-{\cal C}_2~(1-ik\tau)~\exp\left(i\left\{k\tau+\frac{\pi}{2}\left(\nu+\frac{1}{2}\right)\right\}\right)\right]},\nonumber\\
&&\\ &&\nonumber {\Pi_{\bf k}(\tau)=2^{\nu-\frac{3}{2}}\frac{i(-k\tau)^{\frac{3}{2}-\nu}}{\sqrt{2}k^{\frac{5}{2}}}\left|\frac{\Gamma(\nu)}{\Gamma\left(\frac{3}{2}\right)}\right|}\left[{{\cal C}_1~\left\{\left(\frac{1}{2}-\nu\right)\frac{(1+ik\tau)}{k^2\tau^2}+1\right\}~\exp\left(-i\left\{k\tau+\frac{\pi}{2}\left(\nu+\frac{1}{2}\right)\right\}\right)}\right.\\&& \left.~~~~~~~~~~~~~~~~~~~~~~{-{\cal C}_2~\left\{\left(\frac{1}{2}-\nu\right)\frac{(1-ik\tau)}{k^2\tau^2}+1\right\}~\exp\left(i\left\{k\tau+\frac{\pi}{2}\left(\nu+\frac{1}{2}\right)\right\}\right)}\right],~~~~~~~
\eea
These general asymptotic expressions are extremely important to compute the expressions for the OTOC's in the later subsections. To server this purpose we need to promote both of these classical solutions to the quantum level.

 \subsection{Quantum mode function in Cosmology} 
 
Now, as we have mentioned earlier to explicitly compute OTOC we need to map the classical solution of the equation of motion of the dynamical conformal time dependent rescaled field and the corresponding canonically conjugate momentum to the quantum operator description. Consequently, the classical poisson bracket will be replaced by the quantum mechanical commutator bracket in the present context. The representative map is given by the following expressions:
\bea &&{~~~~~  f_{\bf k}(\tau)~~~~~~ \xrightarrow{\text{\textcolor{red}{\bf Classical~to quantum~map}}}~~~~~~~~~\hat{f}_{\bf k}(\tau)}, \\ \nonumber\\
&&{~~~~~  \Pi_{\bf k}(\tau)~~~~~~ \xrightarrow{\text{\textcolor{red}{\bf Classical~to quantum~map}}}~~~~~~~~~\hat{\Pi}_{\bf k}(\tau)},\\
&&\underbrace{}_{\textcolor{red}{\bf Classical~quantities}}~~~~~~~~~~~~~~~~~~~~~~~~~~~~~~~\underbrace{}_{\textcolor{red}{\bf Quantum~operators}}\nonumber\eea
\bea
&&{ \left\{f_{\bf k}(\tau) , \Pi_{\bf k}(0)\right\}\xrightarrow{\text{\textcolor{red}{\bf Classical~to quantum~map}}}\left[\hat{f}_{\bf k}(\tau),\hat{\Pi}_{\bf k}(0)\right]}, \\ \nonumber\\
&&{ \left\{f_{\bf k}(\tau) ,f_{\bf k}(0)\right\}\xrightarrow{\text{\textcolor{red}{\bf Classical~to quantum~map}}}\left[\hat{f}_{\bf k}(\tau),\hat{f}_{\bf k}(0)\right]}, \\ \nonumber\\
&&{ \left\{\Pi_{\bf k}(\tau) , \Pi_{\bf k}(0)\right\}\xrightarrow{\text{\textcolor{red}{\bf Classical~to quantum~map}}}\left[\hat{f}_{\bf k}(\tau),\hat{\Pi}_{\bf k}(0)\right]},\\
&&\underbrace{}_{\textcolor{red}{\bf Poisson~bracket}}~~~~~~~~~~~~~~~~~~~~~~~~~~~~~~~\underbrace{}_{\textcolor{red}{\bf Commutator~bracket}}\nonumber\eea
Here in this context, the rescaled field operator and the corresponding canonically conjugate momentum in the quantum regime can be expressed as:
\bea &&\hat{f}_{\bf k}(\tau)=f_{\bf k}(\tau)~a_{\bf k}+f^{*}_{\bf -k}(\tau)~a^{\dagger}_{-\bf k},\\
&&\hat{\Pi}_{\bf k}(\tau)=\Pi_{\bf k}(\tau)~a_{\bf k}+\Pi^{*}_{\bf -k}(\tau)~a^{\dagger}_{-\bf k}.\eea
Here $a_{\bf k}$ and $a^{\dagger}_{-\bf k}$ are the annihilation and creation operators of the conformal time dependent parametric oscillators after performing canonical quantization, which satisfy the following commutation relation relations:
\bea {\left[a_{\bf k},a^{\dagger}_{-\bf k'}\right]=(2\pi)^3\delta^{3}({\bf k}+{\bf k}'),~~\left[a_{\bf k},a_{-\bf k'}\right]=0=\left[a^{\dagger}_{\bf k},a^{\dagger}_{-\bf k'}\right]}~.\eea
Consequently the curvature perturbation and the corresponding momentum operator in the quantum regime can be re-expressed as:
\bea && \hat{\zeta}_{\bf k}(\tau)=\frac{\hat{f}_{\bf k}(\tau)}{z}=\frac{f_{\bf k}(\tau)~a_{\bf k}+f^{*}_{\bf -k}(\tau)~a^{\dagger}_{-\bf k}}{z}=\zeta_{\bf k}(\tau)~a_{\bf k}+\zeta^{*}_{\bf -k}(\tau)~a^{\dagger}_{-\bf k},\\
&& \hat{\Pi}_{\zeta,{\bf k}}(\tau)=\partial_{\tau}\left(\frac{\hat{f}_{\bf k}(\tau)}{z}\right)=\frac{ \hat{\Pi}_{\bf k}(\tau)}{z}-\frac{\hat{\zeta}_{\bf k}(\tau)}{z(\tau)}\frac{dz(\tau)}{d\tau}\nonumber\\
&&~~~~~~~~~~~~=\left[\frac{ \hat{\Pi}_{\bf k}(\tau)}{z}-\hat{\zeta}_{\bf k}(\tau)\left(\frac{d\ln a(\tau)}{d\tau}+\underbrace{\frac{1}{2}\frac{d\ln\epsilon(\tau)}{d\tau}}_{\textcolor{red}{\bf Sub-leading}}\right)\right].~~~~~~~~~\eea 
Now neglecting the sub-leading contribution we get the following simplified expression for the canonically conjugate momentum of the quantum curvature perturbation operator:
\bea { \hat{\Pi}_{\zeta,{\bf k}}(\tau)=\left[\left(\Pi_{\zeta,{\bf k}}(\tau)~a_{\bf k}+\Pi^{*}_{\zeta,{\bf -k}}(\tau)~a^{\dagger}_{-\bf k}\right)-\left(\zeta_{\bf k}(\tau)~a_{\bf k}+\zeta^{*}_{\bf -k}(\tau)~a^{\dagger}_{-\bf k}\right)\left(\frac{1}{a(\tau)}\frac{da(\tau)}{d\tau}\right)\right]},~~~~~~~~~\eea
where the scale factor $a(\tau)$ is different for De Sitter inflationary patch and during the epoch of reheating, which is given by the following expression:
\begingroup
\large
\begin{eqnarray}
{a(\tau)= \large \left\{
     \begin{array}{lr}
  \displaystyle  \large-\frac{1}{H\tau} &~~~~~~~~~~~ \text{\textcolor{red}{\bf De~Sitter}}\\ \\
 \displaystyle  \left[\frac{(1+3w_{reh})}{3(1+w_{reh})}\tau\right]^{\frac{2}{(1+3w_{reh})}}~~{\rm with}~~0\leq w_{reh}\leq \frac{1}{3}~ & \text{\textcolor{red}{\bf Reheating}}  \end{array}
   \right.}~~~~~~~~~~~
\end{eqnarray}
\endgroup
Using the above mentioned expressions for the scale factors the corresponding parameter $\epsilon$ for both of the cases can be calculated as:
\begingroup
\large
\begin{eqnarray}
{\epsilon=-\frac{1}{a(\tau)H^2}\frac{dH}{d\tau}= \large \left\{
     \begin{array}{lr}
   \large2 \tau  g'(\tau ) \left(\tau  g'(\tau )-g(\tau )-1\right) &~~~~~~~~~~~ \text{\textcolor{red}{\bf De~Sitter}}\\  \\
  \frac{3}{2}(1+w_{reh}) & \text{\textcolor{red}{\bf Reheating}}  \end{array}
   \right.}~~~~~~~~~~~
\end{eqnarray}
\endgroup
Also the expression for Mukhanov-Sasaki varibale is can be computed for both of the cases as:
\begingroup
\large
\begin{eqnarray}
{z(\tau)=\sqrt{2\epsilon}~a(\tau)= \large \left\{
     \begin{array}{lr}
   \Large-\frac{2}{H}\sqrt{\frac{ g'(\tau ) }{\tau}\left(\tau  g'(\tau )-g(\tau )-1\right)} &~~~~~~~~~~~ \text{\textcolor{red}{\bf De~Sitter}}\\ \\
 \sqrt{3(1+w_{reh})}~  \left[\frac{(1+3w_{reh})}{3(1+w_{reh})}\tau\right]^{\frac{2}{(1+3w_{reh})}} & \text{\textcolor{red}{\bf Reheating}}  \end{array}
   \right.}~~~~~~~~~~~
\end{eqnarray}
\endgroup
Here it is important to note the following crucial facts which are helpful for the further computations of OTOC's:
\begin{enumerate}
\item  In the De Sitter case with FLRW background $g(\tau)$ is a slowly varying conformal time dependent function in general, which physically represents a very slight deviation from the De Sitter solution, where the Hubble parameter $H=\sqrt{\Lambda/3}$ is dominated by Cosmological Constant $\Lambda>0$ in Planckian unit. It is observed that from the computation that, $\epsilon(\tau)$ and $g(\tau)$ both of the parameters captures the effect of such deviation and for this reason sometimes one refer this to be the quasi De Sitter space-time.

\item Another important fact is that since we strictly don't know the explicit conformal tie dependence of the reheating equation of state $w_{reh}(\tau)$ as it is an open issue till date, we have assumed that the time dependence is very slow and for this purpose this crucial parameter exactly behaving like a constant with the restriction that, $0\leq w_{reh}\leq 1/3$ in the present context. This fact is extremely important to mention as it fixes the form of the scale factor $a(\tau)$, parameter $\epsilon(\tau)$~\footnote{The slow-roll parameter $\epsilon=-\dot{H}/H^2=-H^{'}/(aH^2)$,  is commonly used in the context of quasi De Sitter inflationary phase where the scale factor can be expressed as, $a(\tau)=-\frac{1}{H\tau}(1+\epsilon)\sim -\frac{1}{H\tau}(1+g(\tau))$. During inflation $\epsilon<<1$ and at the end of inflationary phase $\epsilon=1$, which is very useful to determine corresponding field value at the end of inflation. So during stochastic particle production during the inflationary phase will respect these constraints on the slow-roll parameter $\epsilon$. On the other hand, since reheating happened after inflation then it is expected that the magnitude of the parameter $\epsilon>1$ during reheating, where the scale factor can be expressed as, $a(\tau)= \left[\frac{(1+3w_{reh})}{3(1+w_{reh})}\tau\right]^{\frac{2}{(1+3w_{reh})}}\sim \left[\left(1-\frac{1}{\epsilon}\right)\tau\right]^{\frac{1}{(\epsilon-1)}}$. This statement can be justified very easily in the context during the epoch of reheating. From the previously derived expression for the parameter $\epsilon$ in the context of reheating we have found that, $\epsilon\sim \frac{3}{2}(1+w_{reh})$, where the equation of state parameter $0\leq w_{reh}\leq 1/3$. This directly implies that during the epoch of reheating the parameter $\epsilon$ will lie within the window, $3/2\leq \epsilon \leq 2$ (alternatively one can say that the Hubble velocity is lying within the window $\frac{3}{2}H^2\leq v_{H}=|\dot{H}|\leq 2H^2$ in the Planckian units), which is obviously a very strong and specific information we get from the  from the present computation.} and the Mukhanov Sasaki variable $z(\tau)$ in terms of conformal time $\tau$ as well as the equation of state parameter $w_{reh}$ in the present context.
\end{enumerate} 
We will strictly follow the above mentioned two facts for the rest of the computations that we have performed to mathematically quantify the expressions for the OTOC's in the context of stocahsic particle production during inflation, which is dominated by quasi De Sitter space-time and during the epoch of reheating. This further implies that the physical outcomes of the out-of-equilibrium studied in terms of OTOC's respect both of the above mentioned assumptions.
\subsection{Canonical quantization of cosmological Hamiltonian: Classical to quantum map}
Now we have to promote the classical Hamiltonian that we have derived earlier to the quantized Hamiltonian. This can be expressed as: 
\bea {H~~ \xrightarrow{\text{\textcolor{red}{\bf Classical~to quantum~map}}}~~ \hat{H}=\int d^3{\bf k}~\frac{1}{2}\left[\underbrace{\hat{\Pi}^2_{\bf k}(\tau)}_{\textcolor{red}{\bf Kinetic~term}}+~\underbrace{\omega^2_{\bf k}(\tau){\hat{f}^2_{\bf k}(\tau)}}_{\textcolor{red}{\bf Potential~term}}\right]}~,\eea
Additionally, it is important to mention here that the following constraint have to be satisfied:
\bea &&\hat{f}^{\dagger}_{\bf k}(\tau)=\left(f_{\bf k}(\tau)~a_{\bf k}+f^{*}_{\bf -k}(\tau)~a^{\dagger}_{-\bf k}\right)^{\dagger}\nonumber\\
&&~~~~~~~~=\left(f_{\bf -k}(\tau)~a_{-\bf k}+f^{*}_{\bf k}(\tau)~a^{\dagger}_{\bf k}\right)=\hat{f}_{-\bf k}(\tau),\eea\bea
&& \hat{\Pi}^{\dagger}_{\bf k}(\tau)=\left(\Pi_{\bf k}(\tau)~a_{\bf k}+\Pi^{*}_{\bf -k}(\tau)~a^{\dagger}_{-\bf k}\right)^{\dagger}\nonumber\\
&&~~~~~~~~=\left(\Pi_{\bf -k}(\tau)~a_{-\bf k}+\Pi^{*}_{\bf k}(\tau)~a^{\dagger}_{\bf k}\right)=\hat{\Pi}_{-\bf k}(\tau).~~~~~\eea 
Consequently we have used a simplified notations in the above mentioned Hamiltonian in Fourier space, which are given by the following expression:
\bea && {\hat{\Pi}^2_{\bf k}(\tau)=\hat{\Pi}_{\bf k}(\tau)\hat{\Pi}_{-\bf k}(\tau)=|\Pi_{\bf k}(\tau)~a_{\bf k}+\Pi^{*}_{\bf -k}(\tau)~a^{\dagger}_{-\bf k}|^2}~~~~~~~~~~~~\nonumber\\
&&~~~~~~~~~{=\left(\Pi_{\bf k}(\tau)~a_{\bf k}+\Pi^{*}_{\bf -k}(\tau)~a^{\dagger}_{-\bf k}\right)^{\dagger}\left(\Pi_{\bf k}(\tau)~a_{\bf k}+\Pi^{*}_{\bf -k}(\tau)~a^{\dagger}_{-\bf k}\right)}\nonumber \\
&&~~~~~~~~~{=\left(\Pi_{\bf -k}(\tau)~a_{-\bf k}+\Pi^{*}_{\bf k}(\tau)~a^{\dagger}_{\bf k}\right)\left(\Pi_{\bf k}(\tau)~a_{\bf k}+\Pi^{*}_{\bf -k}(\tau)~a^{\dagger}_{-\bf k}\right)}\nonumber\\
&&~~~~~~~~~{=2\left(a^{\dagger}_{\bf k}a_{\bf k}+\frac{1}{2}\delta^{3}(0)\right)|\Pi_{\bf k}(\tau)|^2=2\left(a^{\dagger}_{\bf k}a_{\bf k}+\frac{1}{2}\delta^{3}(0)\right)\Pi^2_{\bf k}(\tau)}\\
&&{\hat{f}^2_{\bf k}(\tau)=\hat{f}_{\bf k}(\tau)\hat{f}_{-\bf k}(\tau)=|f_{\bf k}(\tau)~a_{\bf k}+f^{*}_{\bf -k}(\tau)~a^{\dagger}_{-\bf k}|^2}\nonumber\\
&&~~~~~~~~{=\left(f_{\bf k}(\tau)~a_{\bf k}+f^{*}_{\bf -k}(\tau)~a^{\dagger}_{-\bf k}\right)^{\dagger}\left(f_{\bf k}(\tau)~a_{\bf k}+f^{*}_{\bf -k}(\tau)~a^{\dagger}_{-\bf k}\right)}\nonumber\\
&&~~~~~~~~{=\left(f_{\bf -k}(\tau)~a_{-\bf k}+f^{*}_{\bf k}(\tau)~a^{\dagger}_{\bf k}\right)\left(f_{\bf k}(\tau)~a_{\bf k}+f^{*}_{\bf -k}(\tau)~a^{\dagger}_{-\bf k}\right)}\nonumber\\
&&~~~~~~~~{=2\left(a^{\dagger}_{\bf k}a_{\bf k}+\frac{1}{2}\delta^{3}(0)\right)|f_{\bf k}(\tau)|^2=2\left(a^{\dagger}_{\bf k}a_{\bf k}+\frac{1}{2}\delta^{3}(0)\right)f^2_{\bf k}(\tau)}.~~~~~~~~\eea
Consequently, we get the following simplified form of the quantized Hamiltonian:
\bea {\hat{H}=\int d^3{\bf k}~\left(a^{\dagger}_{\bf k}a_{\bf k}+\frac{1}{2}\delta^{3}(0)\right)\left[\hat{\Pi}^2_{\bf k}(\tau)+~\omega^2_{\bf k}(\tau){\hat{f}^2_{\bf k}(\tau)}\right]}~,\eea
After introducing the normal ordering one can neglect the contribution from the zero point energy, which actually gives the divergent contribution. This further simplifies the the expression for the Hamiltonian, which is given by:
\bea {:\hat{H}:=\int d^3{\bf k}~a^{\dagger}_{\bf k}a_{\bf k}~\left[\hat{\Pi}^2_{\bf k}(\tau)+~\omega^2_{\bf k}(\tau){\hat{f}^2_{\bf k}(\tau)}\right]}~.\eea
Here the expression for the time dependent frequency have already mentioned earlier, which can be further expressed in terms of the particle production during inflation in quasi De Sitter phase and during the epoch of reheating.
The above mentioned expressions for the Hamiltonians are extremely useful as it will appear in the expression for the thermal Botzmann factor of all OTOC's. 

\subsection{Cosmological two-point and four-point ``in-in" OTO micro-canonical amplitudes}
In this section our prime objective is to compute the explicit expression for the OTOC.
\subsubsection{OTOC meets Cosmology}
 To compute this explicitly we need the following information in our hand:
\begin{enumerate}
\item \underline{\textcolor{red}{\bf Information~I:}}\\
First of all, one need the quantum operators corresponding to the rescaled field and its canonically conjugate momenta, which is actually written in terms of the classical solution obtained from the mode equation and the creation and annihilation operators. Actually, in the context of Cosmology one can write down the Hamiltonian of the system in terms of an Harmonic oscillator with conformal time dependent frequency in a Fourier space and after constructing this Hamiltonian one needs to integrate over all possible momenta. Using this prescription and using the canonical quantization procedure one can easily quantize the system Hamiltonian in the framework of Cosmology. 

\item \underline{\textcolor{red}{\bf Information~II:}}\\
Next, one needs to compute the expression for the expression for the square of the commutator bracket in the background of FLRW space-time in curved space quantum field theory in coordinate space. This is the prime component which will fix the final expression for OTOC in the framework of Cosmology. We will explicitly show that this contribution can be written in terms of four parts where each of the parts in Fourier transformed space mimics the role of some kind of scattering amplitudes after taking its trace, which are functions of four momenta and two time scale as appearing in the definition of OTOC. Precisely these four momenta are coming due to the fact that in each contributions we are dealing with four quantum operators. Now if we careful about the technicality then we have to say these four-point scattering amplitudes are basically the four-point time dependent correlation functions in the framework of Cosmology as during performing the trace operation we have to use the same quantum vacuum state, which means that the initial and final state both are same, and identified to be ``in" state. So it is basically an in-in amplitude rather than an usual ``in-out" amplitude. The usual ``in-out" amplitude can be usually computed using the idea of $S$ matrix, which is a Schwinger Dyson series in the context of quantum field theory and this is an unitary matrix in the context of a closed system in our Universe when the system under consideration is not exchanging any information with the surroundings. On the other hand, in Cosmological set-up we deal with ``in-in" amplitudes which can be computed using the well known Schwinger Keldysh formalism. Instead of calling this quantity which we want to evaluate as ``in-in" amplitude we call these contributions as ``in-in" quantum correlation functions in the framework of Cosmology.

\item \underline{\textcolor{red}{\bf Information~III:}}\\
Further, we have to fix the definition of trace in the context of quantum field theory in a classical gravitational background, specifically in the context of FLRW Cosmology. To serve this purpose, we need to first use a standard definition of the quantum wave function of our Universe. For these purpose, the most common choice is to use the definition of standard Bunch Davies vacuum state, which is basically a thermal ground state in the framework of Cosmology. Sometimes this vacuum is identified to be the Hartle-Hawking or Cherenkov vacuum state in quantum field theory theory of curved space-time with Cosmological background.  Apart from that the another useful vacua is, $\alpha$ vacua , which is mostly used in the context of interacting quantum field theory in curved space-time with cosmological background. Here in the construction of $\alpha$ vacua it appears as a one real parameter family which can take any continuous value starting from the zero value. The zero value is a special case of $\alpha$ vacua which is basically the well known Bunch Davies quantum vacuum ground state. One the other hand, for the other values of $\alpha$ one can construct infinite number of states which can be considered as some excited states which can be expressed in terms of the known Bunch Davies vacuum state using Bogoliubov transformation.

\item \underline{\textcolor{red}{\bf Information~IV:}}\\
Using the above mentioned $\alpha$ vacua or the Bunch Davies quantum vacuum one can further compute the numerator of OTOC, which is the trace of the square of the commutator bracket of the rescaled field variable and its canonically conjugate momenta along with the thermal Botzmann factor in which the system Hamiltonian is appearing explicitly. Also one can compute the denominator of the OTOC represents the expression for the trace of the only thermal Boltzmann factor, which is physically identified to be the thermal partition function in the Cosmological framework. Then putting a negative sign in front of the computed object one can derive the expression for the OTOC in un normalized form. This overall negative sign is very important as it makes the un normalized OTOC positive and growing with the combined time scale in which both of the above mentioned quantum extended operators are defined. In this computation, the thermal partition function and the trace of the commutator bracket square at finite temperature are evaluated using semi classical approximation, which means that we treat the fluctuation of scalar modes appearing from cosmological perturbation theory in the primordial universe. Precisely the perturbation in the metric in the FLRW background can be written in terms of scalar, vector and tensor modes in Fourier transformed space. This is commonly known as the {\it SVT decomposition} in Cosmology. In this computation, we have restricted our attention to scalar modes which can promoted to be a quantum operator during the computation of OTOC. It is obvious that, the canonically conjugate momenta computed from the scalar modes can also be promoted to be a quantum mechanical operator. But if we look into this problem very carefully, then we observe that the origin of the quantum fluctuations from the scalar modes are coming from metric perturbation in the FLRW background, which we actually treat purely classically. For this reason we will do a semi-classical (not purely quantum or classical) computation for the computation of OTOC in the framework of primordial cosmology.
\item \underline{\textcolor{red}{\bf Information~V:}}\\
Last but not the least, we have to fix the normalization of OTOC. This can be perfectly done using the previously mentioned thermal trace operation in presence of $\alpha$ vacua or Bunch Davies quantum vacuum state in the context of Cosmology. After normalization we can able to construct a dimensionless cosmological four-point ``in-in" amplitude or quantum OTOC. We have computed the normalization factor in the Appendix in detail. Please look into the technical details in the Appendix.
\end{enumerate}
\subsubsection{Fourier space representation of the commutator bracket: Application to two-point OTOC}
Here our job is to compute the following commutator bracket, given by:
\bea && \left[\hat{f}({\bf x},\tau_1),\hat{\Pi}({\bf x},\tau_2)\right]=\underbrace{\hat{f}({\bf x},\tau_1)\hat{\Pi}({\bf x},\tau_2)}_{\textcolor{red}{\bf \equiv \Gamma_{1}({\bf x},\tau_1,\tau_2)}}-\underbrace{\hat{\Pi}({\bf x},\tau_2)\hat{f}({\bf x},\tau_1)}_{\textcolor{red}{\bf \equiv \Gamma_{2}({\bf x},\tau_1,\tau_2)}},\eea
 Now we use the following convention for the Fourier transformation, which is given by:
 \bea &&\hat{f}({\bf x},\tau_1)=\int \frac{d^3{\bf k}}{(2\pi)^3}~\exp(i{\bf k}.{\bf x})~\hat{f}_{{\bf k}}(\tau_1),\eea
 \bea
 &&\hat{\Pi}({\bf x},\tau_1)=\partial_{\tau_1}\hat{f}({\bf x},\tau_1)=\int \frac{d^3{\bf k}}{(2\pi)^3}~\exp(i{\bf k}.{\bf x})~\partial_{\tau_1}\hat{f}_{{\bf k}}(\tau_1)=\int \frac{d^3{\bf k}}{(2\pi)^3}~\exp(i{\bf k}.{\bf x})~\hat{\Pi}_{{\bf k}}(\tau_1),~~~~~~~~~~~~~\eea
 which will be very useful for rest of the computation of this paper. 
 
 Here we can see that the commutator bracket splits the commutator into two parts which we have written in terms of the perturbation field variable and its canonically conjugate momentum, which are appearing in the context of cosmological perturbation theory performed in a specific scheme. In this computation instead of considering the perturbed variable space independent, we start with space-time dependent perturbed cosmological operators as after performing cosmological perturbation all the quantum operators are inhomogeneous and anisotropic in general. Though at the end after performing all the computation we will show that even one start with space-time dependent perturbed cosmological operators, the ultimate result will be independent of space coordinates and in the OTOC only the information regarding the time coordinates will appear explicitly as from the starting point we have started with time scale separated quantities.  The obtained two individual contribution actually represent the two point OTO amplitudes in coordinate space which will finally contribute in the computation of the two-point micro-canonical OTO amplitudes in Fourier space after performing the Fourier transformation following the above mentioned convention. Finally, using this formalism one can explicitly compute the expression for the two point micro-canonical OTOC in the context of Cosmology, which will be only functions of the two conformal time scales in which the quantum operators are associated in the cosmological perturbation theory.
 
 Next, we explicitly compute the expressions for these individual quantum mechanical operators, $\Gamma_{i}({\bf x},\tau_1,\tau_2)$~$\forall~i=1,2$, which can be expressed in Fourier space as:
 \bea \Gamma_{1}({\bf x},\tau_1,\tau_2)&=&\hat{f}({\bf x},\tau_1)\hat{\Pi}({\bf x},\tau_2)\nonumber\\
 &=&\int\frac{d^3{\bf k}_1}{(2\pi)^3}\int\frac{d^3{\bf k}_1}{(2\pi)^3}~\exp(i({\bf k}_1+{\bf k}_2).{\bf x})~\hat{f}_{{\bf k}_1}(\tau_1)\hat{\Pi}_{{\bf k}_2}(\tau_2)\nonumber\\
 &=&\int\frac{d^3{\bf k}_1}{(2\pi)^3}\int\frac{d^3{\bf k}_1}{(2\pi)^3}~\exp(i({\bf k}_1+{\bf k}_2).{\bf x})~\hat{\Delta}_{1}({\bf k}_1,{\bf k}_2;\tau_1,\tau_2),\eea
 where we have introduced a momentum and conformal time dependent quantum mechanical operator $\hat{\Delta}_{1}({\bf k}_1,{\bf k}_2;\tau_1,\tau_2)$, which is defined as:
 \bea \hat{\Delta}_{1}({\bf k}_1,{\bf k}_2;\tau_1,\tau_2)&=&\hat{f}_{{\bf k}_1}(\tau_1)\hat{\Pi}_{{\bf k}_2}(\tau_2)={\cal D}_1 ({\bf k}_1,{\bf k}_2;\tau_1,\tau_2)~a_{{\bf k}_1}a_{{\bf k}_2}+{\cal D}_2 ({\bf k}_1,{\bf k}_2;\tau_1,\tau_2)~a^{\dagger}_{-{\bf k}_1}a_{{\bf k}_2}\nonumber\\
 &&~~~~~~~~~~~~~~~~~~~~~~+{\cal D}_3 ({\bf k}_1,{\bf k}_2;\tau_1,\tau_2)~a_{{\bf k}_1}a^{\dagger}_{-{\bf k}_2}+{\cal D}_4 ({\bf k}_1,{\bf k}_2;\tau_1,\tau_2)~a^{\dagger}_{-{\bf k}_1}a^{\dagger}_{-{\bf k}_2},~~~~~~~~~~~~~~\eea  
 where we have introduced  momentum and time dependent two-point OTO amplitudes, ${\cal D}_i ({\bf k}_1,{\bf k}_2;\tau_1,\tau_2)~~\forall~~i=1,2,3,4$, which are explicitly defined in the Appendix.
  \bea \Gamma_{2}({\bf x},\tau_1,\tau_2)&=&\hat{\Pi}({\bf x},\tau_2)\hat{f}({\bf x},\tau_1)\nonumber\\
 &=&\int\frac{d^3{\bf k}_1}{(2\pi)^3}\int\frac{d^3{\bf k}_1}{(2\pi)^3}~\exp(i({\bf k}_1+{\bf k}_2).{\bf x})~\hat{\Pi}_{{\bf k}_1}(\tau_2)\hat{f}_{{\bf k}_2}(\tau_1)\nonumber\\
 &=&\int\frac{d^3{\bf k}_1}{(2\pi)^3}\int\frac{d^3{\bf k}_1}{(2\pi)^3}~\exp(i({\bf k}_1+{\bf k}_2).{\bf x})~\hat{\Delta}_{2}({\bf k}_1,{\bf k}_2;\tau_1,\tau_2),\eea
 where we have introduced a momentum and conformal time dependent quantum mechanical operator $\hat{\Delta}_{2}({\bf k}_1,{\bf k}_2;\tau_1,\tau_2)$, which is defined as:
 \bea \hat{\Delta}_{2}({\bf k}_1,{\bf k}_2;\tau_1,\tau_2)&=&\hat{\Pi}_{{\bf k}_1}(\tau_2)\hat{f}_{{\bf k}_2}(\tau_1)={\cal L}_1 ({\bf k}_1,{\bf k}_2;\tau_1,\tau_2)~a_{{\bf k}_1}a_{{\bf k}_2}+{\cal L}_2 ({\bf k}_1,{\bf k}_2;\tau_1,\tau_2)~a^{\dagger}_{-{\bf k}_1}a_{{\bf k}_2}\nonumber\\
 &&~~~~~~~~~~~~~~~~~~~~~~+{\cal L}_3 ({\bf k}_1,{\bf k}_2;\tau_1,\tau_2)~a_{{\bf k}_1}a^{\dagger}_{-{\bf k}_2}+{\cal L}_4 ({\bf k}_1,{\bf k}_2;\tau_1,\tau_2)~a^{\dagger}_{-{\bf k}_1}a^{\dagger}_{-{\bf k}_2},~~~~~~~~~~~~~~\eea  
 where we have introduced  momentum and time dependent two-point OTO amplitudes, ${\cal L}_i ({\bf k}_1,{\bf k}_2;\tau_1,\tau_2)~~\forall~~i=1,2,3,4$, which are explicitly defined in the Appendix.
 
  This implies that one can write down the previously mentioned the commutator bracket along with the thermal Boltzmann factor as:
   \bea && e^{-\beta \widehat{H}(\tau_1)}\left[\hat{f}({\bf x},\tau_1),\hat{\Pi}({\bf x},\tau_2)\right]\nonumber\\
   &&=e^{-\beta \widehat{H}(\tau_1)}\left[ {\Gamma}_1({\bf x},\tau_1,\tau_2)- {\Gamma}_2({\bf x},\tau_1,\tau_2)\right]\nonumber\\
 && = e^{-\beta \widehat{H}(\tau_1)}\left\{\int \frac{d^3k_1}{(2\pi)^3}\int \frac{d^3k_2}{(2\pi)^3}\exp\left[i\left({\bf k}_1+{\bf k}_2\right).{\bf x}\right]\left[\hat{\Delta}_1({\bf k}_1,{\bf k}_2;\tau_1,\tau_2)-\hat{\Delta}_2({\bf k}_1,{\bf k}_2;\tau_1,\tau_2)\right]\right\}\nonumber\\
 &&= \int \frac{d^3k_1}{(2\pi)^3}\int \frac{d^3k_2}{(2\pi)^3}\exp\left[i\left({\bf k}_1+{\bf k}_2\right).{\bf x}\right]\left[\hat{\nabla}_1({\bf k}_1,{\bf k}_2;\tau_1,\tau_2;\beta)-\hat{\nabla}_2({\bf k}_1,{\bf k}_2;\tau_1,\tau_2;\beta)\right],~~~~~~~~~~~ \eea
  where we define the new sets of quantum operators, $:\hat{\nabla}_i({\bf k}_1,{\bf k}_2;\tau_1,\tau_2;\beta):~\forall~i=1,2$ as:
  \bea &&{\hat{\nabla}_i({\bf k}_1,{\bf k}_2;\tau_1,\tau_2;\beta)=e^{-\beta \hat{H}(\tau_1)}~\hat{\Delta}_i({\bf k}_1,{\bf k}_2;\tau_1,\tau_2)~~~~\forall ~~~i=1,2}~~~~~~~~~~~\eea
  Here the thermal Boltzmann factor can be expressed in terms of creation and annihilation operator as:
  \bea && {e^{-\beta H(\tau_1)}=\exp\left(-\beta\int d^3{\bf k}~\left(a^{\dagger}_{\bf k}a_{\bf k}+\frac{1}{2}\delta^{3}(0)\right)E_{\bf k}(\tau_1)\right)},~~~\eea
  where we define $E_{\bf k}(\tau_1)$ by the following expressions:
  \bea E_{\bf k}(\tau_1):=\left[|\Pi_{\bf k}(\tau_1)|^2+\omega^2_{\bf k}(\tau_1)|f_{\bf k}(\tau)|^2\right].\eea 
\subsubsection{Fourier space representation of square of the commutator bracket Application to four-point OTOC}
  Now we explicitly compute the following square of the commutator bracket, given by:
  \bea && \left[\hat{f}({\bf x},\tau_1),\hat{\Pi}({\bf x},\tau_2)\right]^2=\underbrace{\hat{f}({\bf x},\tau_1)\hat{\Pi}({\bf x},\tau_2)\hat{f}({\bf x},\tau_1)\hat{\Pi}({\bf x},\tau_2)}_{\textcolor{red}{\equiv~ {\cal K}_1({\bf x},\tau_1,\tau_2)}}\nonumber\\ 
  &&~~~~~~~~~~~~~~~~~~~~~~~~~~~~~~~~~~~~~~~~~~~-\underbrace{\hat{\Pi}({\bf x},\tau_2)\hat{f}({\bf x},\tau_1)\hat{f}({\bf x},\tau_1)\hat{\Pi}({\bf x},\tau_2)}_{\textcolor{red}{\equiv~ {\cal K}_2({\bf x},\tau_1,\tau_2)}}\nonumber\\ &&~~~~~~~~~~~~~~~~~~~~~~~~~~~~~~~~~~~~~~~~~~~-\underbrace{\hat{f}({\bf x},\tau_1)\hat{\Pi}({\bf x},\tau_2)\hat{\Pi}({\bf x},\tau_2)\hat{f}({\bf x},\tau_1)}_{\textcolor{red}{\equiv~ {\cal K}_3({\bf x},\tau_1,\tau_2)}}\nonumber\\ &&~~~~~~~~~~~~~~~~~~~~~~~~~~~~~~~~~~~~~~~~~~~+\underbrace{\hat{\Pi}({\bf x},\tau_2)\hat{f}({\bf x},\tau_1)\hat{\Pi}({\bf x},\tau_2)\hat{f}({\bf x},\tau_1)}_{\textcolor{red}{\equiv~ {\cal K}_4({\bf x},\tau_1,\tau_2)}} \eea
 It is important to note that, in the expression for the square of the commutator bracket defined in terms of the rescaled cosmological perturbation variable for scalar mode fluctuation and its canonically conjugate momenta we actually have introduced few space-time dependent new functions which basically divides the whole expression into four parts. These four quantum operators after taking the thermal average represent the four-point correlation function. However doing the computation in coordinate space is very complicated. For this reason by making use of the mentioned convention of the Fourier transform of each quantum operators we express all of them in momentum space.  Now we mention the explicit structure of these operators, ${\cal K}_i({\bf x},\tau_1,\tau_2)~\forall~i=1,2,3,4$ which are expressed in Fourier space as:
  \bea &&{\cal K}_1({\bf x},\tau_1,\tau_2)\nonumber\\
  &&=\hat{f}({\bf x},\tau_1)\hat{\Pi}({\bf x},\tau_2)\hat{f}({\bf x},\tau_1)\hat{\Pi}({\bf x},\tau_2)\nonumber\\
  &&=\int \frac{d^3k_1}{(2\pi)^3}\int \frac{d^3k_2}{(2\pi)^3}\int \frac{d^3k_3}{(2\pi)^3}\int \frac{d^3k_4}{(2\pi)^3}\exp\left[i\left({\bf k}_1+{\bf k}_2+{\bf k}_3+{\bf k}_4\right).{\bf x}\right]~~~~~~~~\nonumber\\
  &&~~~~~~~~~~~~~~~~~~~~~~~
  ~~~~~~~~~~~~~~~~~~~~~~~~~~~~~~~~~~~~~~~~~~~\hat{f}_{{\bf k}_1}(\tau_1)\hat{\Pi}_{{\bf k}_2}(\tau_2)\hat{f}_{{\bf k}_3}(\tau_1)\hat{\Pi}_{{\bf k}_4}(\tau_2)\nonumber\\
  &&=\int \frac{d^3k_1}{(2\pi)^3}\int \frac{d^3k_2}{(2\pi)^3}\int \frac{d^3k_3}{(2\pi)^3}\int \frac{d^3k_4}{(2\pi)^3}\exp\left[i\left({\bf k}_1+{\bf k}_2+{\bf k}_3+{\bf k}_4\right).{\bf x}\right]~~~~~~~~\nonumber\\
  &&~~~~~~~~~~~~~~~~~~~~~~~
  ~~~~~~~~~~~~~~~~~~~~~~~~~~~~~~~~~~~~~~~~~~~\widehat{\cal T}_1({\bf k}_1,{\bf k}_2,{\bf k}_3,{\bf k}_4;\tau_1,\tau_2),~~~~~~~~~~
 \eea 
  where the function $\widehat{\cal T}_1({\bf k}_1,{\bf k}_2,{\bf k}_3,{\bf k}_4;\tau_1,\tau_2)$ is defined as:
 \bea &&\widehat{\cal T}_1({\bf k}_1,{\bf k}_2,{\bf k}_3,{\bf k}_4;\tau_1,\tau_2)\nonumber\\
 &&=\left[{\cal M}_1({\bf k}_1,{\bf k}_2,{\bf k}_3,{\bf k}_4;\tau_1,\tau_2)~a_{{\bf k}_1}a_{{\bf k}_2}a_{{\bf k}_3}a_{{\bf k}_4}\right.\nonumber\\
  && \left.~~~~~~~+ {\cal M}_2({\bf k}_1,{\bf k}_2,{\bf k}_3,{\bf k}_4;\tau_1,\tau_2)~a^{\dagger}_{-{\bf k}_1}a_{{\bf k}_2}a_{{\bf k}_3}a_{{\bf k}_4}+{\cal M}_3({\bf k}_1,{\bf k}_2,{\bf k}_3,{\bf k}_4;\tau_1,\tau_2)~a_{{\bf k}_1}a^{\dagger}_{-{\bf k}_2}a_{{\bf k}_3}a_{{\bf k}_4}\right.\nonumber\\
  && \left.~~~~~~~+ {\cal M}_4({\bf k}_1,{\bf k}_2,{\bf k}_3,{\bf k}_4;\tau_1,\tau_2)~a^{\dagger}_{-{\bf k}_1}a^{\dagger}_{-{\bf k}_2}a_{{\bf k}_3}a_{{\bf k}_4}+{\cal M}_5({\bf k}_1,{\bf k}_2,{\bf k}_3,{\bf k}_4;\tau_1,\tau_2)~a_{{\bf k}_1}a_{{\bf k}_2}a^{\dagger}_{-{\bf k}_3}a_{{\bf k}_4}\right.\nonumber\\
  && \left.~~~~~~~+ {\cal M}_6({\bf k}_1,{\bf k}_2,{\bf k}_3,{\bf k}_4;\tau_1,\tau_2)~a^{\dagger}_{-{\bf k}_1}a_{{\bf k}_2}a^{\dagger}_{-{\bf k}_3}a_{{\bf k}_4}+{\cal M}_7({\bf k}_1,{\bf k}_2,{\bf k}_3,{\bf k}_4;\tau_1,\tau_2)~a_{{\bf k}_1}a^{\dagger}_{-{\bf k}_2}a^{\dagger}_{-{\bf k}_3}a_{{\bf k}_4}\right.\nonumber\\
  && \left.~~~~~~~+ {\cal M}_8({\bf k}_1,{\bf k}_2,{\bf k}_3,{\bf k}_4;\tau_1,\tau_2)~a^{\dagger}_{-{\bf k}_1}a^{\dagger}_{-{\bf k}_2}a^{\dagger}_{-{\bf k}_3}a_{{\bf k}_4}+{\cal M}_9({\bf k}_1,{\bf k}_2,{\bf k}_3,{\bf k}_4;\tau_1,\tau_2)~a_{{\bf k}_1}a_{{\bf k}_2}a_{{\bf k}_3}a^{\dagger}_{-{\bf k}_4}\right.\nonumber\\
  && \left.~~~~~~~~~~~~~~~~~~~~~~~~~~~~~~~~~~~~+ {\cal M}_{10}({\bf k}_1,{\bf k}_2,{\bf k}_3,{\bf k}_4;\tau_1,\tau_2)~a^{\dagger}_{-{\bf k}_1}a_{{\bf k}_2}a_{{\bf k}_3}a^{\dagger}_{-{\bf k}_4}\right.\nonumber\\
  && \left.+{\cal M}_{11}({\bf k}_1,{\bf k}_2,{\bf k}_3,{\bf k}_4;\tau_1,\tau_2)~a_{{\bf k}_1}a^{\dagger}_{-{\bf k}_2}a_{{\bf k}_3}a^{\dagger}_{-{\bf k}_4}+ {\cal M}_{12}({\bf k}_1,{\bf k}_2,{\bf k}_3,{\bf k}_4;\tau_1,\tau_2)~a^{\dagger}_{-{\bf k}_1}a^{\dagger}_{-{\bf k}_2}a_{{\bf k}_3}a^{\dagger}_{-{\bf k}_4}\right.\nonumber\\
  && \left.+{\cal M}_{13}({\bf k}_1,{\bf k}_2,{\bf k}_3,{\bf k}_4;\tau_1,\tau_2)~a_{{\bf k}_1}a_{{\bf k}_2}a^{\dagger}_{-{\bf k}_3}a^{\dagger}_{-{\bf k}_4}+ {\cal M}_{14}({\bf k}_1,{\bf k}_2,{\bf k}_3,{\bf k}_4;\tau_1,\tau_2)~a^{\dagger}_{-{\bf k}_1}a_{{\bf k}_2}a^{\dagger}_{-{\bf k}_3}a^{\dagger}_{-{\bf k}_4}\right.\nonumber\\
  && \left.+{\cal M}_{15}({\bf k}_1,{\bf k}_2,{\bf k}_3,{\bf k}_4;\tau_1,\tau_2)~a_{{\bf k}_1}a^{\dagger}_{-{\bf k}_2}a^{\dagger}_{-{\bf k}_3}a^{\dagger}_{-{\bf k}_4}+ {\cal M}_{16}({\bf k}_1,{\bf k}_2,{\bf k}_3,{\bf k}_4;\tau_1,\tau_2)~a^{\dagger}_{-{\bf k}_1}a^{\dagger}_{-{\bf k}_2}a^{\dagger}_{-{\bf k}_3}a^{\dagger}_{-{\bf k}_4}\right],~~~~~~~~~~~
  \eea
  where we define, ${\cal M}_{j}({\bf k}_1,{\bf k}_2,{\bf k}_3,{\bf k}_4;\tau_1,\tau_2)~\forall~j=1,\cdots,16$, in the appendix.
 \bea &&{\cal K}_2({\bf x},\tau_1,\tau_2)=\hat{\Pi}({\bf x},\tau_2)\hat{f}({\bf x},\tau_1)\hat{f}({\bf x},\tau_1)\hat{\Pi}({\bf x},\tau_2)\nonumber\\
 &&~~~~~~~~~~~~~~~~=\int \frac{d^3k_1}{(2\pi)^3}\int \frac{d^3k_2}{(2\pi)^3}\int \frac{d^3k_3}{(2\pi)^3}\int \frac{d^3k_4}{(2\pi)^3}\exp\left[i\left({\bf k}_1+{\bf k}_2+{\bf k}_3+{\bf k}_4\right).{\bf x}\right]~~~~~~~~\nonumber\\
  &&~~~~~~~~~~~~~~~~~~~~~~~~~~~~~~~~~~~~~~~~~~~~~~~~~~~~~~\hat{\Pi}_{{\bf k}_1}(\tau_2)\hat{f}_{{\bf k}_2}(\tau_1)\hat{f}_{{\bf k}_3}(\tau_1)\hat{\Pi}_{{\bf k}_4}(\tau_2)\nonumber\\
  &&~~~~~~~~~~~~~~~~=\int \frac{d^3k_1}{(2\pi)^3}\int \frac{d^3k_2}{(2\pi)^3}\int \frac{d^3k_3}{(2\pi)^3}\int \frac{d^3k_4}{(2\pi)^3}\exp\left[i\left({\bf k}_1+{\bf k}_2+{\bf k}_3+{\bf k}_4\right).{\bf x}\right]~~~~~~~~\nonumber\\
  &&~~~~~~~~~~~~~~~~~~~~~~~~~~~~~~~~~~~~~~~~~~~~~~~~~~~~~~~~~~~~~~~~~~~\widehat{\cal T}_2({\bf k}_1,{\bf k}_2,{\bf k}_3,{\bf k}_4;\tau_1,\tau_2),
 \eea 
  where the function $\widehat{\cal T}_2({\bf k}_1,{\bf k}_2,{\bf k}_3,{\bf k}_4;\tau_1,\tau_2)$ is defined as:
 \bea &&\widehat{\cal T}_2({\bf k}_1,{\bf k}_2,{\bf k}_3,{\bf k}_4;\tau_1,\tau_2)\nonumber\\
 &&=\left[{\cal J}_1({\bf k}_1,{\bf k}_2,{\bf k}_3,{\bf k}_4;\tau_1,\tau_2)~a_{{\bf k}_1}a_{{\bf k}_2}a_{{\bf k}_3}a_{{\bf k}_4}+ {\cal J}_2({\bf k}_1,{\bf k}_2,{\bf k}_3,{\bf k}_4;\tau_1,\tau_2)~a^{\dagger}_{-{\bf k}_1}a_{{\bf k}_2}a_{{\bf k}_3}a_{{\bf k}_4}\right.\nonumber\\
  && \left.+{\cal J}_3({\bf k}_1,{\bf k}_2,{\bf k}_3,{\bf k}_4;\tau_1,\tau_2)~a_{{\bf k}_1}a^{\dagger}_{-{\bf k}_2}a_{{\bf k}_3}a_{{\bf k}_4}+ {\cal J}_4({\bf k}_1,{\bf k}_2,{\bf k}_3,{\bf k}_4;\tau_1,\tau_2)~a^{\dagger}_{-{\bf k}_1}a^{\dagger}_{-{\bf k}_2}a_{{\bf k}_3}a_{{\bf k}_4}\right.\nonumber\\
  && \left.+{\cal J}_5({\bf k}_1,{\bf k}_2,{\bf k}_3,{\bf k}_4;\tau_1,\tau_2)~a_{{\bf k}_1}a_{{\bf k}_2}a^{\dagger}_{-{\bf k}_3}a_{{\bf k}_4}+ {\cal J}_6({\bf k}_1,{\bf k}_2,{\bf k}_3,{\bf k}_4;\tau_1,\tau_2)~a^{\dagger}_{-{\bf k}_1}a_{{\bf k}_2}a^{\dagger}_{-{\bf k}_3}a_{{\bf k}_4}\right.\nonumber\\
  && \left.+{\cal J}_7({\bf k}_1,{\bf k}_2,{\bf k}_3,{\bf k}_4;\tau_1,\tau_2)~a_{{\bf k}_1}a^{\dagger}_{-{\bf k}_2}a^{\dagger}_{-{\bf k}_3}a_{{\bf k}_4}+ {\cal J}_8({\bf k}_1,{\bf k}_2,{\bf k}_3,{\bf k}_4;\tau_1,\tau_2)~a^{\dagger}_{-{\bf k}_1}a^{\dagger}_{-{\bf k}_2}a^{\dagger}_{-{\bf k}_3}a_{{\bf k}_4}\right.\nonumber\\
  && \left.+{\cal J}_9({\bf k}_1,{\bf k}_2,{\bf k}_3,{\bf k}_4;\tau_1,\tau_2)~a_{{\bf k}_1}a_{{\bf k}_2}a_{{\bf k}_3}a^{\dagger}_{-{\bf k}_4}+ {\cal J}_{10}({\bf k}_1,{\bf k}_2,{\bf k}_3,{\bf k}_4;\tau_1,\tau_2)~a^{\dagger}_{-{\bf k}_1}a_{{\bf k}_2}a_{{\bf k}_3}a^{\dagger}_{-{\bf k}_4}\right.\nonumber\\
  && \left.+{\cal J}_{11}({\bf k}_1,{\bf k}_2,{\bf k}_3,{\bf k}_4;\tau_1,\tau_2)~a_{{\bf k}_1}a^{\dagger}_{-{\bf k}_2}a_{{\bf k}_3}a^{\dagger}_{-{\bf k}_4}+ {\cal J}_{12}({\bf k}_1,{\bf k}_2,{\bf k}_3,{\bf k}_4;\tau_1,\tau_2)~a^{\dagger}_{-{\bf k}_1}a^{\dagger}_{-{\bf k}_2}a_{{\bf k}_3}a^{\dagger}_{-{\bf k}_4}\right.\nonumber\\
  && \left.+{\cal J}_{13}({\bf k}_1,{\bf k}_2,{\bf k}_3,{\bf k}_4;\tau_1,\tau_2)~a_{{\bf k}_1}a_{{\bf k}_2}a^{\dagger}_{-{\bf k}_3}a^{\dagger}_{-{\bf k}_4}+ {\cal J}_{14}({\bf k}_1,{\bf k}_2,{\bf k}_3,{\bf k}_4;\tau_1,\tau_2)~a^{\dagger}_{-{\bf k}_1}a_{{\bf k}_2}a^{\dagger}_{-{\bf k}_3}a^{\dagger}_{-{\bf k}_4}\right.\nonumber\\
  && \left.+{\cal J}_{15}({\bf k}_1,{\bf k}_2,{\bf k}_3,{\bf k}_4;\tau_1,\tau_2)~a_{{\bf k}_1}a^{\dagger}_{-{\bf k}_2}a^{\dagger}_{-{\bf k}_3}a^{\dagger}_{-{\bf k}_4}+ {\cal J}_{16}({\bf k}_1,{\bf k}_2,{\bf k}_3,{\bf k}_4;\tau_1,\tau_2)~a^{\dagger}_{-{\bf k}_1}a^{\dagger}_{-{\bf k}_2}a^{\dagger}_{-{\bf k}_3}a^{\dagger}_{-{\bf k}_4}\right],~~~~~~~~~~~
  \eea
  where we define, ${\cal J}_{j}({\bf k}_1,{\bf k}_2,{\bf k}_3,{\bf k}_4;\tau_1,\tau_2)~\forall~j=1,\cdots,16$, in the appendix.
 \bea &&{\cal K}_3({\bf x},\tau_1,\tau_2)=\hat{f}({\bf x},\tau_1)\hat{\Pi}({\bf x},\tau_2)\hat{\Pi}({\bf x},\tau_2)\hat{f}({\bf x},\tau_1)\nonumber\\
 &&~~~~~~~~~~~~~~~~=\int \frac{d^3k_1}{(2\pi)^3}\int \frac{d^3k_2}{(2\pi)^3}\int \frac{d^3k_3}{(2\pi)^3}\int \frac{d^3k_4}{(2\pi)^3}\exp\left[i\left({\bf k}_1+{\bf k}_2+{\bf k}_3+{\bf k}_4\right).{\bf x}\right]~~~~~~~~\nonumber\\
  &&~~~~~~~~~~~~~~~~~~~~~~~~~~~~~~~~~~~~~~~~~~~~~~~~~~~~~~~~~~~~~~~~~~\hat{f}_{{\bf k}_1}(\tau_1)\hat{\Pi}_{{\bf k}_2}(\tau_2)\hat{\Pi}_{{\bf k}_3}(\tau_2)\hat{f}_{{\bf k}_4}(\tau_1)\nonumber\\
  &&~~~~~~~~~~~~~~~~=\int \frac{d^3k_1}{(2\pi)^3}\int \frac{d^3k_2}{(2\pi)^3}\int \frac{d^3k_3}{(2\pi)^3}\int \frac{d^3k_4}{(2\pi)^3}\exp\left[i\left({\bf k}_1+{\bf k}_2+{\bf k}_3+{\bf k}_4\right).{\bf x}\right]~~~~~~~~\nonumber\\
  &&~~~~~~~~~~~~~~~~~~~~~~~~~~~~~~~~~~~~~~~~~~~~~~~~~~~~~~~~~~~~~~~~~~~\widehat{\cal T}_3({\bf k}_1,{\bf k}_2,{\bf k}_3,{\bf k}_4;\tau_1,\tau_2),
 \eea 
  where the function $\widehat{\cal T}_3({\bf k}_1,{\bf k}_2,{\bf k}_3,{\bf k}_4;\tau_1,\tau_2)$ is defined as:
 \bea &&\widehat{\cal T}_3({\bf k}_1,{\bf k}_2,{\bf k}_3,{\bf k}_4;\tau_1,\tau_2)=\left[{\cal N}_1({\bf k}_1,{\bf k}_2,{\bf k}_3,{\bf k}_4;\tau_1,\tau_2)~a_{{\bf k}_1}a_{{\bf k}_2}a_{{\bf k}_3}a_{{\bf k}_4}\right.\nonumber\\
  && \left.+ {\cal N}_2({\bf k}_1,{\bf k}_2,{\bf k}_3,{\bf k}_4;\tau_1,\tau_2)~a^{\dagger}_{-{\bf k}_1}a_{{\bf k}_2}a_{{\bf k}_3}a_{{\bf k}_4}+{\cal N}_3({\bf k}_1,{\bf k}_2,{\bf k}_3,{\bf k}_4;\tau_1,\tau_2)~a_{{\bf k}_1}a^{\dagger}_{-{\bf k}_2}a_{{\bf k}_3}a_{{\bf k}_4}\right.\nonumber\\
  && \left.+ {\cal N}_4({\bf k}_1,{\bf k}_2,{\bf k}_3,{\bf k}_4;\tau_1,\tau_2)~a^{\dagger}_{-{\bf k}_1}a^{\dagger}_{-{\bf k}_2}a_{{\bf k}_3}a_{{\bf k}_4}+{\cal N}_5({\bf k}_1,{\bf k}_2,{\bf k}_3,{\bf k}_4;\tau_1,\tau_2)~a_{{\bf k}_1}a_{{\bf k}_2}a^{\dagger}_{-{\bf k}_3}a_{{\bf k}_4}\right.\nonumber\\
  && \left.+ {\cal N}_6({\bf k}_1,{\bf k}_2,{\bf k}_3,{\bf k}_4;\tau_1,\tau_2)~a^{\dagger}_{-{\bf k}_1}a_{{\bf k}_2}a^{\dagger}_{-{\bf k}_3}a_{{\bf k}_4}+{\cal N}_7({\bf k}_1,{\bf k}_2,{\bf k}_3,{\bf k}_4;\tau_1,\tau_2)~a_{{\bf k}_1}a^{\dagger}_{-{\bf k}_2}a^{\dagger}_{-{\bf k}_3}a_{{\bf k}_4}\right.\nonumber\\
  && \left.+ {\cal N}_8({\bf k}_1,{\bf k}_2,{\bf k}_3,{\bf k}_4;\tau_1,\tau_2)~a^{\dagger}_{-{\bf k}_1}a^{\dagger}_{-{\bf k}_2}a^{\dagger}_{-{\bf k}_3}a_{{\bf k}_4}\right.\nonumber\\
  && \left.+{\cal N}_9({\bf k}_1,{\bf k}_2,{\bf k}_3,{\bf k}_4;\tau_1,\tau_2)~a_{{\bf k}_1}a_{{\bf k}_2}a_{{\bf k}_3}a^{\dagger}_{-{\bf k}_4}+ {\cal N}_{10}({\bf k}_1,{\bf k}_2,{\bf k}_3,{\bf k}_4;\tau_1,\tau_2)~a^{\dagger}_{-{\bf k}_1}a_{{\bf k}_2}a_{{\bf k}_3}a^{\dagger}_{-{\bf k}_4}\right.\nonumber\\
  && \left.+{\cal N}_{11}({\bf k}_1,{\bf k}_2,{\bf k}_3,{\bf k}_4;\tau_1,\tau_2)~a_{{\bf k}_1}a^{\dagger}_{-{\bf k}_2}a_{{\bf k}_3}a^{\dagger}_{-{\bf k}_4}\right.\nonumber\\
  && \left.+ {\cal N}_{12}({\bf k}_1,{\bf k}_2,{\bf k}_3,{\bf k}_4;\tau_1,\tau_2)~a^{\dagger}_{-{\bf k}_1}a^{\dagger}_{-{\bf k}_2}a_{{\bf k}_3}a^{\dagger}_{-{\bf k}_4}\right.\nonumber\\
  && \left.+{\cal N}_{13}({\bf k}_1,{\bf k}_2,{\bf k}_3,{\bf k}_4;\tau_1,\tau_2)~a_{{\bf k}_1}a_{{\bf k}_2}a^{\dagger}_{-{\bf k}_3}a^{\dagger}_{-{\bf k}_4}\right.\nonumber\\
  && \left.+ {\cal N}_{14}({\bf k}_1,{\bf k}_2,{\bf k}_3,{\bf k}_4;\tau_1,\tau_2)~a^{\dagger}_{-{\bf k}_1}a_{{\bf k}_2}a^{\dagger}_{-{\bf k}_3}a^{\dagger}_{-{\bf k}_4}\right.\nonumber\\
  && \left.+{\cal N}_{15}({\bf k}_1,{\bf k}_2,{\bf k}_3,{\bf k}_4;\tau_1,\tau_2)~a_{{\bf k}_1}a^{\dagger}_{-{\bf k}_2}a^{\dagger}_{-{\bf k}_3}a^{\dagger}_{-{\bf k}_4}\right.\nonumber\\
  && \left.+ {\cal N}_{16}({\bf k}_1,{\bf k}_2,{\bf k}_3,{\bf k}_4;\tau_1,\tau_2)~a^{\dagger}_{-{\bf k}_1}a^{\dagger}_{-{\bf k}_2}a^{\dagger}_{-{\bf k}_3}a^{\dagger}_{-{\bf k}_4}\right],~~~~~~~~~~~
  \eea
  where we define, ${\cal N}_{j}({\bf k}_1,{\bf k}_2,{\bf k}_3,{\bf k}_4;\tau_1,\tau_2)~\forall~j=1,\cdots,16$, in the appendix.
  \bea &&{\cal K}_4({\bf x},\tau_1,\tau_2)=\int \frac{d^3k_1}{(2\pi)^3}\int \frac{d^3k_2}{(2\pi)^3}\int \frac{d^3k_3}{(2\pi)^3}\int \frac{d^3k_4}{(2\pi)^3}\exp\left[i\left({\bf k}_1+{\bf k}_2+{\bf k}_3+{\bf k}_4\right).{\bf x}\right]~~~~~~~~\nonumber\\
  &&~~~~~~~~~~~~~~~~~~~~~~~~~~~~~~~~~~~~~~~~~~~~~~~~~~~~~~~~~~~~~~~~~~\hat{\Pi}_{{\bf k}_1}(\tau_2)\hat{f}_{{\bf k}_2}(\tau_1)\hat{\Pi}_{{\bf k}_3}(\tau_2)\hat{f}_{{\bf k}_4}(\tau_1)\nonumber\\
  &&~~~~~~~~~~~~~~~~=\int \frac{d^3k_1}{(2\pi)^3}\int \frac{d^3k_2}{(2\pi)^3}\int \frac{d^3k_3}{(2\pi)^3}\int \frac{d^3k_4}{(2\pi)^3}\exp\left[i\left({\bf k}_1+{\bf k}_2+{\bf k}_3+{\bf k}_4\right).{\bf x}\right]~~~~~~~~\nonumber\\
  &&~~~~~~~~~~~~~~~~~~~~~~~~~~~~~~~~~~~~~~~~~~~~~~~~~~~~~~~~~~~~~~~~~~~\widehat{\cal T}_4({\bf k}_1,{\bf k}_2,{\bf k}_3,{\bf k}_4;\tau_1,\tau_2),
 \eea 
  where the function $\widehat{\cal T}_4({\bf k}_1,{\bf k}_2,{\bf k}_3,{\bf k}_4;\tau_1,\tau_2)$ is defined as:
 \bea &&\widehat{\cal T}_4({\bf k}_1,{\bf k}_2,{\bf k}_3,{\bf k}_4;\tau_1,\tau_2)\nonumber\\
 &&=\left[{\cal Q}_1({\bf k}_1,{\bf k}_2,{\bf k}_3,{\bf k}_4;\tau_1,\tau_2)~a_{{\bf k}_1}a_{{\bf k}_2}a_{{\bf k}_3}a_{{\bf k}_4}+ {\cal Q}_2({\bf k}_1,{\bf k}_2,{\bf k}_3,{\bf k}_4;\tau_1,\tau_2)~a^{\dagger}_{-{\bf k}_1}a_{{\bf k}_2}a_{{\bf k}_3}a_{{\bf k}_4}\right.\nonumber\\
  && \left.+{\cal Q}_3({\bf k}_1,{\bf k}_2,{\bf k}_3,{\bf k}_4;\tau_1,\tau_2)~a_{{\bf k}_1}a^{\dagger}_{-{\bf k}_2}a_{{\bf k}_3}a_{{\bf k}_4}+ {\cal Q}_4({\bf k}_1,{\bf k}_2,{\bf k}_3,{\bf k}_4;\tau_1,\tau_2)~a^{\dagger}_{-{\bf k}_1}a^{\dagger}_{-{\bf k}_2}a_{{\bf k}_3}a_{{\bf k}_4}\right.\nonumber\\
  && \left.+{\cal Q}_5({\bf k}_1,{\bf k}_2,{\bf k}_3,{\bf k}_4;\tau_1,\tau_2)~a_{{\bf k}_1}a_{{\bf k}_2}a^{\dagger}_{-{\bf k}_3}a_{{\bf k}_4}+ {\cal Q}_6({\bf k}_1,{\bf k}_2,{\bf k}_3,{\bf k}_4;\tau_1,\tau_2)~a^{\dagger}_{-{\bf k}_1}a_{{\bf k}_2}a^{\dagger}_{-{\bf k}_3}a_{{\bf k}_4}\right.\nonumber\\
  && \left.+{\cal Q}_7({\bf k}_1,{\bf k}_2,{\bf k}_3,{\bf k}_4;\tau_1,\tau_2)~a_{{\bf k}_1}a^{\dagger}_{-{\bf k}_2}a^{\dagger}_{-{\bf k}_3}a_{{\bf k}_4}+ {\cal Q}_8({\bf k}_1,{\bf k}_2,{\bf k}_3,{\bf k}_4;\tau_1,\tau_2)~a^{\dagger}_{-{\bf k}_1}a^{\dagger}_{-{\bf k}_2}a^{\dagger}_{-{\bf k}_3}a_{{\bf k}_4}\right.\nonumber\\
  && \left.+{\cal Q}_9({\bf k}_1,{\bf k}_2,{\bf k}_3,{\bf k}_4;\tau_1,\tau_2)~a_{{\bf k}_1}a_{{\bf k}_2}a_{{\bf k}_3}a^{\dagger}_{-{\bf k}_4}+ {\cal Q}_{10}({\bf k}_1,{\bf k}_2,{\bf k}_3,{\bf k}_4;\tau_1,\tau_2)~a^{\dagger}_{-{\bf k}_1}a_{{\bf k}_2}a_{{\bf k}_3}a^{\dagger}_{-{\bf k}_4}\right.\nonumber\\
  && \left.+{\cal Q}_{11}({\bf k}_1,{\bf k}_2,{\bf k}_3,{\bf k}_4;\tau_1,\tau_2)~a_{{\bf k}_1}a^{\dagger}_{-{\bf k}_2}a_{{\bf k}_3}a^{\dagger}_{-{\bf k}_4}+ {\cal Q}_{12}({\bf k}_1,{\bf k}_2,{\bf k}_3,{\bf k}_4;\tau_1,\tau_2)~a^{\dagger}_{-{\bf k}_1}a^{\dagger}_{-{\bf k}_2}a_{{\bf k}_3}a^{\dagger}_{-{\bf k}_4}\right.\nonumber\\
  && \left.+{\cal Q}_{13}({\bf k}_1,{\bf k}_2,{\bf k}_3,{\bf k}_4;\tau_1,\tau_2)~a_{{\bf k}_1}a_{{\bf k}_2}a^{\dagger}_{-{\bf k}_3}a^{\dagger}_{-{\bf k}_4}+ {\cal Q}_{14}({\bf k}_1,{\bf k}_2,{\bf k}_3,{\bf k}_4;\tau_1,\tau_2)~a^{\dagger}_{-{\bf k}_1}a_{{\bf k}_2}a^{\dagger}_{-{\bf k}_3}a^{\dagger}_{-{\bf k}_4}\right.\nonumber\\
  && \left.+{\cal Q}_{15}({\bf k}_1,{\bf k}_2,{\bf k}_3,{\bf k}_4;\tau_1,\tau_2)~a_{{\bf k}_1}a^{\dagger}_{-{\bf k}_2}a^{\dagger}_{-{\bf k}_3}a^{\dagger}_{-{\bf k}_4}+ {\cal Q}_{16}({\bf k}_1,{\bf k}_2,{\bf k}_3,{\bf k}_4;\tau_1,\tau_2)~a^{\dagger}_{-{\bf k}_1}a^{\dagger}_{-{\bf k}_2}a^{\dagger}_{-{\bf k}_3}a^{\dagger}_{-{\bf k}_4}\right],~~~~~~~~~~~
  \eea
  where we define, ${\cal Q}_{j}({\bf k}_1,{\bf k}_2,{\bf k}_3,{\bf k}_4;\tau_1,\tau_2)~\forall~j=1,\cdots,16$, in the appendix.
 
  This implies that one can write down the previously mentioned square of the commutator bracket along with the thermal Boltzmann factor as:
   \bea && e^{-\beta \widehat{H}(\tau_1)}\left[\hat{f}({\bf x},\tau_1),\hat{\Pi}({\bf x},\tau_2)\right]^2\nonumber\\
   &&=e^{-\beta \widehat{H}(\tau_1)}\left[ {\cal K}_1({\bf x},\tau_1,\tau_2)- {\cal K}_2({\bf x},\tau_1,\tau_2)- {\cal K}_3({\bf x},\tau_1,\tau_2)+ {\cal K}_4({\bf x},\tau_1,\tau_2)\right]\nonumber\\
 && = e^{-\beta \widehat{H}(\tau_1)}\left\{\int \frac{d^3k_1}{(2\pi)^3}\int \frac{d^3k_2}{(2\pi)^3}\int \frac{d^3k_3}{(2\pi)^3}\int \frac{d^3k_4}{(2\pi)^3}\exp\left[i\left({\bf k}_1+{\bf k}_2+{\bf k}_3+{\bf k}_4\right).{\bf x}\right]\right.~~~~~~~~\nonumber\\
  &&\left.~~~~~~~~~~~~~~~~~~~~~~~~~\left[\widehat{\cal T}_1({\bf k}_1,{\bf k}_2,{\bf k}_3,{\bf k}_4;\tau_1,\tau_2)-\widehat{\cal T}_2({\bf k}_1,{\bf k}_2,{\bf k}_3,{\bf k}_4;\tau_1,\tau_2)\right.\right.\nonumber\\ && \left.\left. ~~~~~~~~~~~~~~~~~~~~~~~~~~~~~~ +\widehat{\cal T}_3({\bf k}_1,{\bf k}_2,{\bf k}_3,{\bf k}_4;\tau_1,\tau_2)-\widehat{\cal T}_4({\bf k}_1,{\bf k}_2,{\bf k}_3,{\bf k}_4;\tau_1,\tau_2)\right]\right\}\nonumber\\
 &&= \int \frac{d^3k_1}{(2\pi)^3}\int \frac{d^3k_2}{(2\pi)^3}\int \frac{d^3k_3}{(2\pi)^3}\int \frac{d^3k_4}{(2\pi)^3}\exp\left[i\left({\bf k}_1+{\bf k}_2+{\bf k}_3+{\bf k}_4\right).{\bf x}\right]~~~~~~~~\nonumber\\
  &&~~~~~~~~~~~~~~~~~~~~~~~~~\left[\widehat{\cal V}_1({\bf k}_1,{\bf k}_2,{\bf k}_3,{\bf k}_4;\tau_1,\tau_2;\beta)-\widehat{\cal V}_2({\bf k}_1,{\bf k}_2,{\bf k}_3,{\bf k}_4;\tau_1,\tau_2;\beta)\right.\nonumber\\ && \left.~~~~~~~~~~~~~~~~~~~~~~~~~~~~~~ +\widehat{\cal V}_3({\bf k}_1,{\bf k}_2,{\bf k}_3,{\bf k}_4;\tau_1,\tau_2;\beta)-\widehat{\cal V}_4({\bf k}_1,{\bf k}_2,{\bf k}_3,{\bf k}_4;\tau_1,\tau_2;\beta)\right],~~~~~~~~~~~ \eea
  where we define the new sets of quantum operators, $:\widehat{\cal V}_i({\bf k}_1,{\bf k}_2,{\bf k}_3,{\bf k}_4;\tau_1,\tau_2;\beta):~\forall~i=1,2,3,4$ as:
  \bea &&{\widehat{\cal V}_i({\bf k}_1,{\bf k}_2,{\bf k}_3,{\bf k}_4;\tau_1,\tau_2;\beta)=e^{-\beta \hat{H}(\tau_1)}~\widehat{\cal T}_i({\bf k}_1,{\bf k}_2,{\bf k}_3,{\bf k}_4;\tau_1,\tau_2)~~~~\forall ~~~i=1,2,3,4,}~~~~~~~~~~~\eea
where the thermal Boltzmann factor can be expressed as:
  \bea && {e^{-\beta H(\tau_1)}=\exp\left(-\beta\int d^3{\bf k}~\left(a^{\dagger}_{\bf k}a_{\bf k}+\frac{1}{2}\delta^{3}(0)\right)E_{\bf k}(\tau_1)\right)},~~~\eea
  where we define $E_{\bf k}(\tau_1)$ by the following expressions:
  \bea E_{\bf k}(\tau_1):=\left[|\Pi_{\bf k}(\tau_1)|^2+\omega^2_{\bf k}(\tau_1)|f_{\bf k}(\tau_1)|^2\right].\eea
  Now to find the trace of the square of the commutator bracket along with the thermal Boltzmann factor we need to first define the quantum vacuum state. In the present context, the quantum vacuum is fixed by the initial condition that we choose to define the mode functions obtained from the quantum fluctuations during the particle production during inflation and reheating epoch of the primordial cosmology. To find out the expression for the OTOC here we will concentrate on $SO(1,4)$ isommetric De Sitter vacua. in this category the most famous example is the $\alpha$ -vacua, which is actually described by a one real parameter family $\alpha$ and CPT symmetric. If we fix, $\alpha=0$, then we get the well known Bunch Davies vacuum state. In the context of quantum field theory of curved space time Bunch Davies vacuum actually represents the quantum ground state. On the other hand, the $\alpha$ vacua represent the quantum excited states in the context of quantum field theory of curved space.  For $\alpha$ vacua the parameter $\alpha$ mimics the role of a super-selection quantum number associated
with a different bipartite Hilbert space. But it is still a unresolved issue that whether
the interaction picture of the quantum field theory with any arbitrary value of the parameter $\alpha$ with any arbitrary super-selection rule
are consistent with the underlying physical requirements or not. It might be quite likely that the Hilbert space of $\alpha$ vacua excited states and
the Bunch Davies vacuum coincides with each other in the underlying quantum field theory set up. In such a physical situation it is perfectly consistent
to describe quantum field theory of excited states in terms of Bunch Davies vacuum in the ultraviolet regime. On the other hand,
in the infrared regime of the quantum field theory of curved space, due to the non-removal of physical infinities appearing from various types of interaction, explaining the physics of excited states with the Bunch Davies type of adiabatic vacuum state is not at all a viable good approximation. As a consequence one can write an effective field
theory description from the present set up in the ultraviolet regime of quantum field theory. This further implies that identifying the more appropriate candidate of
quantum $\alpha$ vacua states which are highly fine tuned. However, this only allows us to consider excited quantum states compared to ground state described by the Bunch Davies
vacuum in the context of quantum field theory. Using this prescription apart from describing the inflationary paradigm in (quasi) De Sitter space, one can use it to describe a lot of
unexplored late time physical phenomena, i.e. stochastic particle production phenomena, the process of reheating etc in presence of non-standard quantum vacuum state.
\subsection{Cosmological thermal partition function: Quantum version}
\subsubsection{Quantum vacuum state in Cosmology}
In the context of quantum field theory, one can define the class of all excited $\alpha$ vacua states in terms of the well known adiabatic Bunch Davies vacuum state as:
\bea \hll{|\Psi_{\alpha}\rangle=\frac{1}{\sqrt{|\cosh \alpha|}}~\exp\left(-\frac{i}{2}\tanh \alpha~\int \frac{d^3{\bf k}}{(2\pi)^3}~a^{\dagger}_{\bf k}a^{\dagger}_{\bf k}\right)|\Psi_{\bf BD}\rangle}~,\eea
which satisfy the following constraint condition:
\bea a_{\bf k}|\Psi_{\alpha}\rangle=0~~~~\forall~~{\bf k},~~\alpha~~~\eea
Here one can easily observed that, if we fix $\alpha=0$ then one can easily get back the usual quantum adiabatic Bunch Davies vacuum state. Another important thing we have to mention here that, at the level of solution of the mode function of the quantum fluctuations that we have obtained earlier, one can introduce the concept of these excited $\alpha$ vacua states in the integration constants ${\cal C}_1$ and ${\cal C}_2$ as:
\bea {\cal C}_1=\cosh\alpha,~~~~{\cal C}_2=\sinh\alpha, \eea
which satisfy the following normalization condition:
\bea |{\cal C}_1|^2-|{\cal C}_2|^2=1~~~~\Longrightarrow~~~~\cosh^2\alpha-\sinh^2\alpha=1~~~~\forall~~\alpha~.\eea
If we fix here $\alpha=0$ then we get the following values of the normalization constants for the well known adiabatic quantum Bunch Davies vacuum state:
\bea {\cal C}_1=1,~~~~{\cal C}_2=0.~\eea
Now, using the definition of the above mentioned excited $\alpha$ vacua states, which is explicitly written in terms of the well known adiabatic Bunch Davies vacuum state we compute the expression for previously mentioned OTOC in the present context. For this purpose, we need to first of fix the quantum partition function. 
\subsubsection{Quantum partition function in terms of rescaled field variable} 
In presence of these excited $\alpha$ vacua states the quantum partition function can be expressed as:
\bea &&Z_{\alpha}(\beta;\tau_1)
=\int d\Psi_{\alpha}~\langle \Psi_{\alpha}|e^{-\beta \hat{H}(\tau_1)}|\Psi_{\alpha} \rangle=\frac{1}{|\cosh\alpha|}\int d\Psi_{\bf BD}~\langle \Psi_{\bf BD}|\left\{\exp\left(\frac{i}{2}\tanh \alpha~\int \frac{d^3{\bf k}_1}{(2\pi)^3}~a_{{\bf k}_1}a^{\dagger}_{{\bf k}_1}\right)\right.\nonumber\\
&&\left.~~\exp\left(-\beta\int d^3{\bf k}~\left(a^{\dagger}_{\bf k}a_{\bf k}+\frac{1}{2}\delta^{3}(0)\right)E_{\bf k}(\tau_1)\right)\exp\left(-\frac{i}{2}\tanh \alpha~\int \frac{d^3{\bf k}_2}{(2\pi)^3}~a^{\dagger}_{{\bf k}_2}a_{{\bf k}_2}\right)\right\}|\Psi_{\bf BD}\rangle.~~~~~~~~~~~~\eea
Then the quantum partition function for $\alpha$ vacua can be expressed as: 
\bea Z_{\alpha}(\beta;\tau_1) 
&=&\frac{1}{|\cosh\alpha|}Z_{\bf BD}(\beta;\tau_1),\eea
where $Z_{\bf BD}$ is the quantum partition function computed from adiabatic Bunch Davies vacuum as:
\bea Z_{\bf BD}(\beta;\tau_1)&=&\int d\Psi_{\bf BD}~\langle \Psi_{\bf BD}|\exp\left(-\beta\int d^3{\bf k}~\left(a^{\dagger}_{\bf k}a_{\bf k}+\frac{1}{2}\delta^{3}(0)\right)E_{\bf k}(\tau_1)\right)|\Psi_{\bf BD}\rangle\nonumber\\
&=&\exp\left(-\left(1+\frac{1}{2}\delta^{3}(0)\right)\int d^3{\bf k}~\ln\left(2\sinh \frac{\beta E_{\bf k}(\tau_1)}{2}\right)\right).\eea.
This further implies that the expression for the quantum partition function for $\alpha$ vacua can be simplified as:
\bea \hll{Z_{\alpha}(\beta;\tau_1)
=\frac{1}{|\cosh\alpha|}\exp\left(-\left(1+\frac{1}{2}\delta^{3}(0)\right)\int d^3{\bf k}~\ln\left(2\sinh \frac{\beta E_{\bf k}(\tau_1)}{2}\right)\right)}~.\eea 
Till now we have not derived the expressions for the quantum partition function for the $\alpha$ vacua and the adiabatic Bunch Davies vacuum state by introducing normal ordering which is used to remove unwanted infinities, like here it is coming from the zero point like energy contribution $\delta^{3}(0)/2$, which is a divergent quantity. Another possibility is that, one can also use the Dirac Delta regularization to remove the divergent contribution. 

The normal ordered quantum partition function for Bunch Davies vacuum is given by:
\bea \hll{:Z_{\bf BD}(\beta;\tau_1):=\exp\left(-\int\frac{d^3{\bf k}}{(2\pi)^3}\ln\left(2\sinh \frac{\beta E_{\bf k}(\tau_1)}{2}\right)\right)}~.\eea.
Then the normal ordered quantum partition function for$\alpha$ vacua can be simplified as:
\bea \hll{:Z_{\alpha}(\beta;\tau_1):
=\frac{1}{|\cosh\alpha|}\exp\left(-\int d^3{\bf k}~\ln\left(2\sinh \frac{\beta E_{\bf k}(\tau_1)}{2}\right)\right)}.\eea 
\subsubsection{Quantum partition function in terms of curvature perturbation field variable} 
In this subsection our prime objective is to find out the expression for the partition function in terms of the curvature perturbation field variable. To serve this purpose the time dependent dispersion relation can be expressed in terms of the curvature perturbation variable as:
 \bea E_{{\bf k}}(\tau_1)&=&|\Pi_{\bf k}(\tau_1)|^2+\omega^2_{\bf k}(\tau_1)|f_{\bf k}(\tau_1)|^2\nonumber\\
 &=&z^2(\tau_1)\left\{\left|\Pi^{\zeta}_{\bf k}(\tau_1)+\zeta_{\bf k}(\tau_1)\frac{1}{z(\tau_1)}\frac{dz(\tau_1)}{d\tau_1}\right|^2+\omega^2_{\bf k}(\tau_1)|\zeta_{\bf k}(\tau_1)|^2\right\}\nonumber\\
 &=&z^2(\tau_1)\left\{E_{{\bf k},\zeta}(\tau_1)+\underbrace{\left(\Pi^{\zeta}_{-\bf k}(\tau_1)\zeta_{\bf k}(\tau_1)+\Pi^{\zeta}_{\bf k}(\tau_1)\zeta_{-\bf k}(\tau_1)\right) \left(\frac{1}{z(\tau_1)}\frac{dz(\tau_1)}{d\tau_1}\right)}_{\textcolor{red}{\bf Contribution~from~this~term~is~negligibly~small}}\right\}\nonumber\\
&\approx &z^2(\tau_1)E_{{\bf k},\zeta}(\tau_1),~~~~~~\eea 
where we define the time dependent energy dispersion relation in terms of the curvature perturbation variable as:
\bea E_{{\bf k},\zeta}(\tau_1):&=&\left|\Pi^{\zeta}_{\bf k}(\tau_1)\right|^2+\left(\omega^2_{\bf k}(\tau_1)+\left(\frac{1}{z(\tau_1)}\frac{dz(\tau_1)}{d\tau_1}\right)^2\right)|\zeta_{\bf k}(\tau_1)|^2.\eea
Now, the thermal partition function for cosmology in terms of curvature perturbation computed for $\alpha$ vacua can be expressed as:
 \bea \hll{Z^{\zeta}_{\alpha}(\beta;\tau_1)=\frac{Z^{\zeta}_{\bf BD}(\beta;\tau_1)}{|\cosh\alpha|}}.\eea
where $Z^{\zeta}_{\bf BD}(\beta;\tau_1)$is  thermal partition function for cosmology in terms of curvature perturbation for Bunch Davies vacuum which can be expressed as:
 \bea \hll{Z^{\zeta}_{\bf BD}(\beta;\tau_1)=\exp\left(-\left(1+\frac{1}{2}\delta^{3}(0)\right)\int d^3{\bf k}~\ln\left(2\sinh\frac{\beta z^2(\tau_1)E_{{\bf k},\zeta}(\tau_1)}{2}\right)\right)},~~~~~\eea
 and can be further simplified in the normal ordered form as:
  \bea \hll{:Z^{\zeta}_{\bf BD}(\beta;\tau_1):=\exp\left(-\int d^3{\bf k}~\ln\left(2\sinh\frac{\beta z^2(\tau_1)E_{{\bf k},\zeta}(\tau_1)}{2}\right)\right)}.~~~~~\eea  
    \subsection{Trace of two-point ``in-in" OTO amplitude for Cosmology}
    Now, we will explicitly compute the numerator of the two-point OTOC for quantum $\alpha$ vacua, which is given by:
  \bea && {\rm Tr}\left[e^{-\beta \widehat{H}(\tau_1)}\left[\hat{f}({\bf x},\tau_1),\hat{\Pi}({\bf x},\tau_2)\right]\right]_{(\alpha)}\nonumber\\
 &&= \frac{1}{|\cosh\alpha|}\int d\Psi_{\bf BD}~\prod^{2}_{j=1}\int \frac{d^3k_j}{(2\pi)^3}\exp\left[i{\bf k}_j.{\bf x}\right]\langle\Psi_{\bf BD}|\left[\sum^{2}_{i=1}\hat{\nabla}_i({\bf k}_1,{\bf k}_2;\tau_1,\tau_2;\beta)\right]|\Psi_{\bf BD}\rangle.~~~~~~~~~~~ \eea 
  Further, our aim is to compute the individual contributions which in the normal ordered form is given by the following expression and computed in Appendix:
  \bea &&\int d\Psi_{\bf BD}~\langle\Psi_{\bf BD}|:\hat{\nabla}_i({\bf k}_1,{\bf k}_2;\tau_1,\tau_2;\beta):|\Psi_{\bf BD}\rangle=\int d\Psi_{\bf BD}~ \langle\Psi_{\bf BD}|:e^{-\beta \hat{H}(\tau_1)}~\widehat{\Delta}_i({\bf k}_1,{\bf k}_2;\tau_1,\tau_2):|\Psi_{\bf BD}\rangle\nonumber\\
  &&~~~~~~~~~~~~~~~~~
 ~~~~~~~~~~~~~~~~~ ~~~~~~~~~~~~~~~~~~~~~~~~~~~~~~~~~~~~~~~~~~~~~~~~~~~~~~~~~~~\forall~i=1,2.~~~~~~~~~\eea 
 Further, the trace of sum of these individual two-point ``in-in" OTO micro-canonical amplitudes in normal ordered form can be expressed as:
    \bea &&\int d\Psi_{\bf BD}~\langle\Psi_{\bf BD}|\sum^{2}_{i=1}:\hat{\nabla}_i({\bf k}_1,{\bf k}_2;\tau_1,\tau_2;\beta):|\Psi_{\bf BD}\rangle\nonumber\\
 &&=(2\pi)^3\underbrace{\exp\left(-\int d^3{\bf k}~\ln\left(2\sinh \frac{\beta E_{\bf k}(\tau_1)}{2}\right)\right)}_{\textcolor{red}{\bf Micro-canonical~thermal~partition~function}}~\underbrace{\delta^{3}({\bf k}_1+{\bf k}_2)}_{\textcolor{red}{\bf Momentum~conservation~in~OTO~amplitude}}\nonumber\\
 &&~~~~~~~~\left[ \underbrace{{\cal D}_2({\bf k}_1,{\bf k}_2;\tau_1,\tau_2)}_{\textcolor{red}{\bf Individual~two~point~OTO~amplitude}}~+ \underbrace{{\cal D}_3({\bf k}_1,{\bf k}_2;\tau_1,\tau_2)}_{\textcolor{red}{\bf Individual~two~point~OTO~amplitude}}\right.\nonumber\\
  && \left.~~~~~~~~~~~~-\underbrace{{\cal L}_2({\bf k}_1,{\bf k}_2;\tau_1,\tau_2)}_{\textcolor{red}{\bf Individual~two~point~OTO~amplitude}}~- \underbrace{{\cal L}_3({\bf k}_1,{\bf k}_2;\tau_1,\tau_2)}_{\textcolor{red}{\bf Individual~two~point~OTO~amplitude}}\right]\nonumber\\
  &=&(2\pi)^3\delta^3({\bf k}_1+{\bf k}_2)~{\bf P}({\bf k}_1,{\bf k}_2;\tau_2,\tau_2;\beta).~~~~~~~~~\eea  
  Here we introduce, ${\bf P}({\bf k}_1,{\bf k}_2;\tau_2,\tau_2;\beta)$ is the temperature dependent two-point function, which is defined as:
  \bea {\bf P}({\bf k}_1,{\bf k}_2;\tau_2,\tau_2;\beta):&=&\exp\left(-\int d^3{\bf k}~\ln\left(2\sinh \frac{\beta E_{\bf k}(\tau_1)}{2}\right)\right)\nonumber\\
  &&~~~~~~~~~~~~~\left[ {\cal D}_2({\bf k}_1,{\bf k}_2;\tau_1,\tau_2)+{\cal D}_3({\bf k}_1,{\bf k}_2;\tau_1,\tau_2)\right.\nonumber\\
  && \left.~~~~~~~~~~~~~~~~~~~~~~-{\cal L}_2({\bf k}_1,{\bf k}_2;\tau_1,\tau_2)-{\cal L}_3({\bf k}_1,{\bf k}_2;\tau_1,\tau_2)\right].~~~~~~~~\eea
 \subsection{OTOC from regularised two-point ``in-in" OTO micro-canonical amplitude: rescaled field version}
 
The cosmological OTOC without normalization for $\alpha$ vacua can be expressed as:
    \bea && Y^{f}(\tau_1,\tau_2)=-\frac{1}{Z_{\alpha}(\beta;\tau_1)}{\rm Tr}\left[e^{-\beta \widehat{H}(\tau_1)}\left[\hat{f}({\bf x},\tau_1),\hat{\Pi}({\bf x},\tau_2)\right]\right]_{(\alpha)}\nonumber\\
  &&=-(2\pi)^3\int \frac{d^3{\bf k}_1}{(2\pi)^3}\int \frac{d^3{\bf k}_2}{(2\pi)^3}\exp\left[i\left({\bf k}_1+{\bf k}_2\right).{\bf x}\right]\underbrace{\delta^{3}({\bf k}_1+{\bf k}_2)}_{\textcolor{red}{\bf Momentum~conservation~in~OTO~amplitude}}\nonumber\\
 &&~~~~~~~~\left[ \underbrace{{\cal D}_2({\bf k}_1,{\bf k}_2;\tau_1,\tau_2)}_{\textcolor{red}{\bf Individual~two~point~OTO~amplitude}}~+ \underbrace{{\cal D}_3({\bf k}_1,{\bf k}_2;\tau_1,\tau_2)}_{\textcolor{red}{\bf Individual~two~point~OTO~amplitude}}\right.\nonumber\\
  && \left.~~~~~~~~~~~~-\underbrace{{\cal L}_2({\bf k}_1,{\bf k}_2;\tau_1,\tau_2)}_{\textcolor{red}{\bf Individual~two~point~OTO~amplitude}}~- \underbrace{{\cal L}_3({\bf k}_1,{\bf k}_2;\tau_1,\tau_2)}_{\textcolor{red}{\bf Individual~two~point~OTO~amplitude}}\right]\nonumber\\
  &&=-\int \frac{d^3{\bf k}_1}{(2\pi)^3}{\cal P}({\bf k}_1,-{\bf k}_1;\tau_1,\tau_2),~~~~~~~~~~~ \eea 
  where the two-point OTO micro-canonical amplitude function is  explicitly given by the following expression:
  \bea {\cal P}({\bf k}_1,-{\bf k}_1;\tau_1,\tau_2):&=&\left[ {\cal D}_2({\bf k}_1,-{\bf k}_1;\tau_1,\tau_2)+{\cal D}_3({\bf k}_1,-{\bf k}_1;\tau_1,\tau_2)\right.\nonumber\\
  && \left.~~~~~~~~~~~~~~~~~~~~~~~~~~~~~~~~~~~~~~-{\cal L}_2({\bf k}_1,-{\bf k}_1;\tau_1,\tau_2)- {\cal L}_3({\bf k}_1,-{\bf k}_1;\tau_1,\tau_2)\right]\nonumber\\
  &=&\left[ f^{*}_{{\bf -k}_1}(\tau_1)\Pi_{-{\bf k}_1}(\tau_2)+f_{{\bf k}_1}(\tau_1)\Pi^{*}_{{\bf k}_1}(\tau_2)\right.\nonumber\\
  && \left.~~~~~~~~~~~~~~~~~~~~~~~~~~~~~~~~~~~~~~-\Pi^{*}_{{\bf -k}_1}(\tau_2)f_{-{\bf k}_1}(\tau_1)- \Pi_{{\bf k}_1}(\tau_2)f^{*}_{{\bf k}_1}(\tau_1)\right].~~~~~~~~~~\eea
  Here we define:
 \bea 
 {\cal D}_2 ({\bf k}_1,{\bf k}_2;\tau_1,\tau_2)&=&f^{*}_{{\bf -k}_1}(\tau_1)\Pi_{{\bf k}_2}(\tau_2),\\
 {\cal D}_3 ({\bf k}_1,{\bf k}_2;\tau_1,\tau_2)&=&f_{{\bf k}_1}(\tau_1)\Pi^{*}_{{\bf -k}_2}(\tau_2),\\
 {\cal L}_2 ({\bf k}_1,{\bf k}_2;\tau_1,\tau_2)&=&\Pi^{*}_{{\bf -k}_1}(\tau_2)f_{{\bf k}_2}(\tau_1),\\
 {\cal L}_3 ({\bf k}_1,{\bf k}_2;\tau_1,\tau_2)&=&\Pi_{{\bf k}_1}(\tau_2)f^{*}_{-{\bf k}_2}(\tau_1).\eea
  Now we need to evaluate explicitly by doing the momentum integration over three volume. Now to compute this integral one can express the volume element as:
  \bea \frac{d^3 {\bf k}_1}{(2\pi)^3}=4\pi~k^2_1~dk_1~~~~0<k_1<L.~\eea
  Here we have taken care of the fact that the individual contribution appearing in the two-point OTOC momentum integral is isotropic. Also, we have introduced a momentum finite large cut-off to regulate the contribution of this integral. 
  
  Consequently, one can write the following simplified expressions for the two-point un-normalized OTOC as:
   \bea && \hll{Y^{f}(\tau_1,\tau_2)=-\frac{1}{2\pi^2}{\cal B}(\tau_1,\tau_2)}~,~~~~~~~~~~~ \eea 
   where the conformal time scale dependent regularized integral, ${\cal B}(\tau_1,\tau_2)$ in the above expression, is defined as:
  \bea {\cal B}(\tau_1,\tau_2):&=&\int^{L}_{k_1=0}~k^2_1~dk_1~\left[ f^{*}_{{\bf -k}_1}(\tau_1)\Pi_{-{\bf k}_1}(\tau_2)+f_{{\bf k}_1}(\tau_1)\Pi^{*}_{{\bf k}_1}(\tau_2)\right.\nonumber\\
  && \left.~~~~~~~~~~~~~~~~~~~~~~~~~~~~~~~~~~~~~~-\Pi^{*}_{{\bf -k}_1}(\tau_2)f_{-{\bf k}_1}(\tau_1)- \Pi_{{\bf k}_1}(\tau_2)f^{*}_{{\bf k}_1}(\tau_1)\right] \nonumber\\
  &=&(-\tau_1)^{\frac{1}{2}-\nu}(-\tau_2)^{\frac{3}{2}-\nu}\left[Z_{(1)}(\tau_1,\tau_2)+Z_{(2)}(\tau_1,\tau_2)-Z_{(3)}(\tau_1,\tau_2)-Z_{(4)}(\tau_1,\tau_2)\right],~~~~~~\eea
  where we have introduced the time dependent four individual amplitudes, $Z_{(i)}(\tau_1,\tau_2)~\forall~i=1,2,3,4$:
  \bea && Z_{(1)}(\tau_1,\tau_2):=\int^{L}_{k_1=0}~k^2_1~dk_1~f_{{\bf k}_1}(\tau_1)\Pi^{*}_{{\bf k}_1}(\tau_2),\\
   && Z_{(2)}(\tau_1,\tau_2):=\int^{L}_{k_1=0}~k^2_1~dk_1~ f^{*}_{{\bf -k}_1}(\tau_1)\Pi_{-{\bf k}_1}(\tau_2),\\
   && Z_{(3)}(\tau_1,\tau_2):=\int^{L}_{k_1=0}~k^2_1~dk_1~ \Pi_{{\bf k}_1}(\tau_2)f^{*}_{{\bf k}_1}(\tau_1),\\
   && Z_{(4)}(\tau_1,\tau_2):=\int^{L}_{k_1=0}~k^2_1~dk_1~ \Pi^{*}_{{\bf -k}_1}(\tau_2)f_{-{\bf k}_1}(\tau_1),\eea
  which satisfy the following symmetry properties:
\bea && Z_{(2)}(\tau_1,\tau_2)=(-1)^{-(2\nu+1)}Z_{(1)}(\tau_1,\tau_2),\\
&&Z_{(4)}(\tau_1,\tau_2)=(-1)^{-(2\nu+1)}Z_{(3)}(\tau_1,\tau_2),\eea
using which the simplified form of the momentum integrated time dependent two-point OTOC can be written as:
 \bea \hll{Y^{f}(\tau_1,\tau_2)=-\frac{1}{2\pi^2}{\cal B}(\tau_1,\tau_2)=\frac{(-\tau_1)^{\frac{1}{2}-\nu}(-\tau_2)^{\frac{3}{2}-\nu}}{2\pi^2}\left[1+(-1)^{-(2\nu+1)}\right]\left(Z_{(3)}(\tau_1,\tau_2)-Z_{(2)}(\tau_1,\tau_2)\right)}.~~~~~~~~\eea
 We need to explicitly evaluate this above mentioned integral which will going to fix the final expression for the two-point micro-canonical OTOC for Cosmology. We have presented the detailed computation of this integral in the Appendix.   
 \subsection{OTOC from regularised four-point ``in-in" OTO micro-canonical amplitude: curvature perturbation field version}

Here we need to perform the computation for the two-point OTOC  in terms of the scalar curvature perturbation and the canonically conjugate momentum associated with it, which we have found that is given by the following simplified expression:
\bea && \hll{Y^{\zeta}(\tau_1,\tau_2)=-\frac{1}{Z^{\zeta}_{\alpha}(\beta,\tau_1)}{\rm Tr}\left[e^{-\beta \hat{H}(\tau_1)}\left[\hat{\zeta}({\bf x},\tau_1),\hat{\Pi}({\bf x},\tau_2)\right]\right]_{(\alpha)}=\frac{1}{z(\tau_1)z(\tau_2)}Y^{f}(\tau_1,\tau_2)}.~~~~~~~~~~~~~\eea
Now substituting the explicit for of the two-point function that we have derived in the previous section we get the following expression:
\bea && \hll{Y^{\zeta}(\tau_1,\tau_2)=-\frac{1}{2\pi^2}\frac{{\cal B}(\tau_1,\tau_2)}{z(\tau_1)z(\tau_2)}=\frac{(-\tau_1)^{\frac{1}{2}-\nu}(-\tau_2)^{\frac{3}{2}-\nu}}{2\pi^2~z(\tau_1)z(\tau_2)}\left[1+(-1)^{-(2\nu+1)}\right]\left(Z_{(3)}(\tau_1,\tau_2)-Z_{(2)}(\tau_1,\tau_2)\right)}\nonumber\\
&&\eea
Now additionally, few points we have to mention that from the finally obtained answer for the two-point OTOC's obtained from the two different set-ups:
\begin{enumerate}
\item First of we have explicitly shown that the definition of the two-point OTOC is completely coordinate independent and just only depend on the time scale on which the operators in cosmological perturbation theory are time scale separated. Even we start defining the operators in a specific space point at different times, in the final result all such information are integrated out and we get completely a time dependent two-point OTOC.

\item Next, it is important to note that, for $\nu=0$, which implies, $m/H=3/2$ the two-point OTOC trivially vanishes. But for partially massless and heavy scalar fields, two-point OTOC become non-trivial and carry significant information of the theory.

\item We also have found that, the final obtained result of the two-point OTOC is $\beta$ independent even though we start with a thermal micro-canonical statistical ensemble during definition the two-point OTOC in the trace formula.

\item Also we observe that, if we fix, $\tau_1=\tau_2=\tau$, that means if both the quantum operators representing the cosmological fluctuations are defined at the same time then two-point OTOC explicitly gives diverging contribution in the quantum regime. This is perfectly consistent with the definition of any general OTOC and also support all the physical requirements to construct the foundational set-up to compute OTOC.

\item If we perform an explicit computation, then it is possible to show that, in our set-up the one-point functions in the context of Cosmology are given by the following expressions:
\bea &&\hll{\langle \hat{f}({\bf x},\tau) \rangle_{\beta}=\frac{1}{Z_{\alpha}(\beta;\tau)}{\rm Tr}\left[e^{-\beta \widehat{H}(\tau)}~\hat{f}({\bf x},\tau)\right]_{(\alpha)}=0}~,\nonumber\\
 &&\hll{ \langle \hat{\Pi}({\bf x},\tau) \rangle_{\beta}=\frac{1}{Z_{\alpha}(\beta;\tau)}{\rm Tr}\left[e^{-\beta \widehat{H}(\tau)}~\hat{\Pi}({\bf x},\tau)\right]_{(\alpha)}=0}~.\eea
  The similar statement is also valid if we define the one-point functions for Cosmology in terms of the curvature perturbation field variable and its canonically conjugate momentum. 
  This result implies that, there is no point of performing normalization of the obtained two-point OTOC for Cosmology with respect to the above mentioned one-point functions. So technically, it is sufficient enough to compute the two-point OTOC without normalization in the context of Cosmology.
  
  \item Another important point we have to mention that, not only the one point function but also the three point functions in the present context is trivially zero which can be explicitly proved by making use of the well known {\it Kubo Martin Schwinger} condition in terms of time translational symmetry of the thermal correlation functions computed for Cosmology. These possibilities are explicitly mentioned below:
  \bea  &&\hll{\langle \hat{f}({\bf x},\tau_1) \hat{\Pi}({\bf x},\tau_2)\hat{\Pi}({\bf x},\tau_3)\rangle_{\beta}=\frac{1}{Z_{\alpha}(\beta;\tau)}{\rm Tr}\left[e^{-\beta \widehat{H}(\tau)}~\hat{f}({\bf x},\tau)\hat{\Pi}({\bf x},\tau_2)\hat{\Pi}({\bf x},\tau_3)\right]_{(\alpha)}=0}~,\\
  &&\hll{ \langle \hat{\Pi}({\bf x},\tau_1)\hat{f}({\bf x},\tau_2) \hat{\Pi}({\bf x},\tau_3)\rangle_{\beta}=\frac{1}{Z_{\alpha}(\beta;\tau)}{\rm Tr}\left[e^{-\beta \widehat{H}(\tau)}~\hat{\Pi}({\bf x},\tau_1)\hat{f}({\bf x},\tau_2) \hat{\Pi}({\bf x},\tau_3)\right]_{(\alpha)}=0},\\
  && \hll{ \langle \hat{\Pi}({\bf x},\tau_1)\hat{\Pi}({\bf x},\tau_2)\hat{f}({\bf x},\tau_3) \rangle_{\beta}=\frac{1}{Z_{\alpha}(\beta;\tau)}{\rm Tr}\left[e^{-\beta \widehat{H}(\tau)}~ \hat{\Pi}({\bf x},\tau_1)\hat{\Pi}({\bf x},\tau_2)\hat{f}({\bf x},\tau_3)\right]_{(\alpha)}=0}~,\\
   && \hll{\langle \hat{f}({\bf x},\tau_1) \hat{f}({\bf x},\tau_2)\hat{\Pi}({\bf x},\tau_3)\rangle_{\beta}=\frac{1}{Z_{\alpha}(\beta;\tau)}{\rm Tr}\left[e^{-\beta \widehat{H}(\tau)}~\hat{f}({\bf x},\tau_1) \hat{f}({\bf x},\tau_2)\hat{\Pi}({\bf x},\tau_3)\right]_{(\alpha)}=0}~,\\
   &&\hll{\langle \hat{f}({\bf x},\tau_1) \hat{\Pi}({\bf x},\tau_2)\hat{f}({\bf x},\tau_3)\rangle_{\beta}=\frac{1}{Z_{\alpha}(\beta;\tau)}{\rm Tr}\left[e^{-\beta \widehat{H}(\tau)}~\hat{f}({\bf x},\tau_1) \hat{\Pi}({\bf x},\tau_2)\hat{f}({\bf x},\tau_3)\right]_{(\alpha)}=0}~,\\
   &&\hll{\langle \hat{\Pi}({\bf x},\tau_1) \hat{f}({\bf x},\tau_2)\hat{f}({\bf x},\tau_3)\rangle_{\beta}=\frac{1}{Z_{\alpha}(\beta;\tau)}{\rm Tr}\left[e^{-\beta \widehat{H}(\tau)}~\hat{\Pi}({\bf x},\tau_1) \hat{f}({\bf x},\tau_2)\hat{f}({\bf x},\tau_3)\right]_{(\alpha)}=0}~,\\
    && \hll{\langle \hat{f}({\bf x},\tau_1) \hat{f}({\bf x},\tau_2)\hat{f}({\bf x},\tau_3)\rangle_{\beta}=\frac{1}{Z_{\alpha}(\beta;\tau)}{\rm Tr}\left[e^{-\beta \widehat{H}(\tau)}~\hat{f}({\bf x},\tau_1) \hat{f}({\bf x},\tau_2)\hat{f}({\bf x},\tau_3)\right]_{(\alpha)}=0}~,\\
    &&\hll{ \langle \hat{\Pi}({\bf x},\tau_1) \hat{\Pi}({\bf x},\tau_2)\hat{\Pi}({\bf x},\tau_3)\rangle_{\beta}=\frac{1}{Z_{\alpha}(\beta;\tau)}{\rm Tr}\left[e^{-\beta \widehat{H}(\tau)}~\hat{\Pi}({\bf x},\tau_1) \hat{\Pi}({\bf x},\tau_2)\hat{\Pi}({\bf x},\tau_3)\right]_{(\alpha)}=0}~,~~~~~~~~~~~~~~
   \eea
   which in principle can be further generalized to any odd $N$ point thermal correlation function for micro-canonical ensemble within the framework of Cosmology.
\end{enumerate}
	\newpage
  \subsection{Trace of four-point ``in-in" OTO amplitude for Cosmology}
Now, we will explicitly compute the numerator of the four-point OTOC for quantum $\alpha$ vacua, which is given by:
  \bea && {\rm Tr}\left[e^{-\beta \widehat{H}(\tau_1)}\left[\hat{f}({\bf x},\tau_1),\hat{\Pi}({\bf x},\tau_2)\right]^2\right]_{(\alpha)}\nonumber\\
 &&= \frac{1}{|\cosh\alpha|}\int d\Psi_{\bf BD}~\prod^{4}_{j=1}\int \frac{d^3k_j}{(2\pi)^3}\exp\left[i{\bf k}_j.{\bf x}\right]\langle\Psi_{\bf BD}|\left[\sum^{4}_{i=1}\widehat{\cal V}_i({\bf k}_1,{\bf k}_2,{\bf k}_3,{\bf k}_4;\tau_1,\tau_2;\beta)\right]|\Psi_{\bf BD}\rangle.~~~~~~~~~~~ \eea 
  Further, our aim is to compute the individual contributions which in the normal ordered form are given by the following expression and computed in Appendix:
  \bea &&\int d\Psi_{\bf BD}~\langle\Psi_{\bf BD}|:\widehat{\cal V}_i({\bf k}_1,{\bf k}_2,{\bf k}_3,{\bf k}_4;\tau_1,\tau_2;\beta):|\Psi_{\bf BD}\rangle\nonumber\\
  &&~~~~~~~~~~~~~~~~~= \int d\Psi_{\bf BD}~\langle\Psi_{\bf BD}|:e^{-\beta \hat{H}(\tau_1)}~\widehat{\cal T}_i({\bf k}_1,{\bf k}_2,{\bf k}_3,{\bf k}_4;\tau_1,\tau_2):|\Psi_{\bf BD}\rangle
 ~~~~~\forall~i=1,2,3,4.~~~~~~~~~\eea
 Further, the trace of sum of these individual four-point ``in-in" OTO micro-canonical amplitudes in normal ordered form can be expressed as:
    \bea &&\int d\Psi_{\bf BD}~\langle\Psi_{\bf BD}|\sum^{4}_{i=1}\widehat{\cal V}_i({\bf k}_1,{\bf k}_2,{\bf k}_3,{\bf k}_4;\tau_1,\tau_2;\beta):|\Psi_{\bf BD}\rangle=(2\pi)^6\exp\left(-\int d^3{\bf k}~\ln\left(2\sinh \frac{\beta E_{\bf k}(\tau_1)}{2}\right)\right)\nonumber\\
 &&~~~~~~~~\sum^{4}_{i=1}\left[ \underbrace{{\Theta}^{(i)}_4({\bf k}_1,{\bf k}_2,{\bf k}_3,{\bf k}_4;\tau_1,\tau_2)}_{\textcolor{red}{\bf Individual~four~point~OTO~amplitude}}~\underbrace{\left\{\delta^3\left({\bf k}_1+{\bf k}_4\right)\delta^3\left({\bf k}_2+{\bf k}_3\right)+\delta^3\left({\bf k}_1+{\bf k}_3\right)\delta^3\left({\bf k}_2+{\bf k}_4\right)\right\}}_{\textcolor{red}{\bf Momentum~conservation~in~OTO~amplitude}}\right.\nonumber\\
  && \left.~~~~~~~~~~~~+ \underbrace{{\Theta}^{(i)}_6({\bf k}_1,{\bf k}_2,{\bf k}_3,{\bf k}_4;\tau_1,\tau_2)}_{\textcolor{red}{\bf Individual~four~point~OTO~amplitude}}~\underbrace{\left\{\delta^3\left({\bf k}_1+{\bf k}_2\right)\delta^3\left({\bf k}_3+{\bf k}_4\right)+\delta^3\left({\bf k}_1+{\bf k}_4\right)\delta^3\left({\bf k}_2+{\bf k}_3\right)\right\}}_{\textcolor{red}{\bf Momentum~conservation~in~OTO~amplitude}}\right.\nonumber\\
  && \left.~~~~~~~~~~~~+\underbrace{{\Theta}^{(i)}_7({\bf k}_1,{\bf k}_2,{\bf k}_3,{\bf k}_4;\tau_1,\tau_2)}_{\textcolor{red}{\bf Individual~four~point~OTO~amplitude}}~\underbrace{\left\{\delta^3\left({\bf k}_1+{\bf k}_2\right)\delta^3\left({\bf k}_3+{\bf k}_4\right)+\delta^3\left({\bf k}_1+{\bf k}_3\right)\delta^3\left({\bf k}_2+{\bf k}_4\right)\right\}}_{\textcolor{red}{\bf Momentum~conservation~in~OTO~amplitude}}\right.\nonumber\\
  && \left.~~~~~~~~~~~~+ \underbrace{{\Theta}^{(i)}_{10}({\bf k}_1,{\bf k}_2,{\bf k}_3,{\bf k}_4;\tau_1,\tau_2)}_{\textcolor{red}{\bf Individual~four~point~OTO~amplitude}}~\underbrace{\left\{\delta^3\left({\bf k}_1+{\bf k}_2\right)\delta^3\left({\bf k}_3+{\bf k}_4\right)+\delta^3\left({\bf k}_1+{\bf k}_3\right)\delta^3\left({\bf k}_2+{\bf k}_4\right)\right\}}_{\textcolor{red}{\bf Momentum~conservation~in~OTO~amplitude}}\right.\nonumber\\
  && \left.~~~~~~~~~~~~+\underbrace{{\Theta}^{(i)}_{11}({\bf k}_1,{\bf k}_2,{\bf k}_3,{\bf k}_4;\tau_1,\tau_2)}_{\textcolor{red}{\bf Individual~four~point~OTO~amplitude}}~\underbrace{\left\{\delta^3\left({\bf k}_1+{\bf k}_2\right)\delta^3\left({\bf k}_3+{\bf k}_4\right)+\delta^3\left({\bf k}_1+{\bf k}_4\right)\delta^3\left({\bf k}_2+{\bf k}_3\right)\right\}}_{\textcolor{red}{\bf Momentum~conservation~in~OTO~amplitude}}\right.\nonumber\\
  && \left.~~~~~~~~~~~~+\underbrace{{\Theta}^{(i)}_{13}({\bf k}_1,{\bf k}_2,{\bf k}_3,{\bf k}_4;\tau_1,\tau_2)}_{\textcolor{red}{\bf Individual~four~point~OTO~amplitude}}~\underbrace{\left\{\delta^3\left({\bf k}_1+{\bf k}_3\right)\delta^3\left({\bf k}_2+{\bf k}_4\right)+\delta^3\left({\bf k}_1+{\bf k}_4\right)\delta^3\left({\bf k}_2+{\bf k}_3\right)\right\}}_{\textcolor{red}{\bf Momentum~conservation~in~OTO~amplitude}}\right],\nonumber\\
  &&~~~~~~~~~~~~~~~~~
 ~~~~~~~~~~~~~~~~~ ~~~~~~~~~~~~~~~~~~~~~~~\forall~i= 1 (\equiv{\cal M}), 2 (\equiv -{\cal J}), 3 (\equiv{\cal N}), 4 (\equiv -{\cal Q}).~~~~~~~~~\eea 
 
  \subsection{OTOC from regularised four-point ``in-in" OTO micro-canonical amplitude: rescaled field version}
  \subsubsection{Without normalization}
The cosmological OTOC without normalization for $\alpha$ vacua can be expressed as:
    \bea && C^{f}(\tau_1,\tau_2)=-\frac{1}{Z_{\alpha}(\beta;\tau_1)}{\rm Tr}\left[e^{-\beta \widehat{H}(\tau_1)}\left[\hat{f}({\bf x},\tau_1),\hat{\Pi}({\bf x},\tau_2)\right]^2\right]_{(\alpha)}\nonumber\\
  &&=-(2\pi)^6\int \frac{d^3k_1}{(2\pi)^3}\int \frac{d^3k_2}{(2\pi)^3}\int \frac{d^3k_3}{(2\pi)^3}\int \frac{d^3k_4}{(2\pi)^3}\exp\left[i\left({\bf k}_1+{\bf k}_2+{\bf k}_3+{\bf k}_4\right).{\bf x}\right]~~~~~~~~\nonumber\\
 &&~~~~~~~~~~~~\left[ \underbrace{{\cal E}_4({\bf k}_1,{\bf k}_2,{\bf k}_3,{\bf k}_4;\tau_1,\tau_2)}_{\textcolor{red}{\bf Four-point~OTO~amplitude}}~\underbrace{\left\{\delta^3\left({\bf k}_1+{\bf k}_4\right)\delta^3\left({\bf k}_2+{\bf k}_3\right)+\delta^3\left({\bf k}_1+{\bf k}_3\right)\delta^3\left({\bf k}_2+{\bf k}_4\right)\right\}}_{\textcolor{red}{\bf Momentum~conservation~in~OTO~amplitude}}\right.\nonumber\\
  && \left.~~~~~~~~~~~~+ \underbrace{{\cal E}_6({\bf k}_1,{\bf k}_2,{\bf k}_3,{\bf k}_4;\tau_1,\tau_2)}_{\textcolor{red}{\bf Four-point~OTO~amplitude}}~\underbrace{\left\{\delta^3\left({\bf k}_1+{\bf k}_2\right)\delta^3\left({\bf k}_3+{\bf k}_4\right)+\delta^3\left({\bf k}_1+{\bf k}_4\right)\delta^3\left({\bf k}_2+{\bf k}_3\right)\right\}}_{\textcolor{red}{\bf Momentum~conservation~in~OTO~amplitude}}\right.\nonumber\\
  && \left.~~~~~~~~~~~~+\underbrace{{\cal E}_7({\bf k}_1,{\bf k}_1,{\bf k}_3,{\bf k}_3;\tau_1,\tau_2)}_{\textcolor{red}{\bf Four-point~OTO~amplitude}}~\underbrace{\left\{\delta^3\left({\bf k}_1+{\bf k}_2\right)\delta^3\left({\bf k}_3+{\bf k}_4\right)+\delta^3\left({\bf k}_1+{\bf k}_3\right)\delta^3\left({\bf k}_2+{\bf k}_4\right)\right\}}_{\textcolor{red}{\bf Momentum~conservation~in~OTO~amplitude}}\right.\nonumber\\
  && \left.~~~~~~~~~~~~+ \underbrace{{\cal E}_{10}({\bf k}_1,{\bf k}_2,{\bf k}_3,{\bf k}_4;\tau_1,\tau_2)}_{\textcolor{red}{\bf Four-point~OTO~amplitude}}~\underbrace{\left\{\delta^3\left({\bf k}_1+{\bf k}_2\right)\delta^3\left({\bf k}_3+{\bf k}_4\right)+\delta^3\left({\bf k}_1+{\bf k}_3\right)\delta^3\left({\bf k}_2+{\bf k}_4\right)\right\}}_{\textcolor{red}{\bf Momentum~conservation~in~OTO~amplitude}}\right.\nonumber\\
  && \left.~~~~~~~~~~~~+\underbrace{{\cal E}_{11}({\bf k}_1,{\bf k}_2,{\bf k}_3,{\bf k}_4;\tau_1,\tau_2)}_{\textcolor{red}{\bf Four-point~OTO~amplitude}}~\underbrace{\left\{\delta^3\left({\bf k}_1+{\bf k}_2\right)\delta^3\left({\bf k}_3+{\bf k}_4\right)+\delta^3\left({\bf k}_1+{\bf k}_4\right)\delta^3\left({\bf k}_2+{\bf k}_3\right)\right\}}_{\textcolor{red}{\bf Momentum~conservation~in~OTO~amplitude}}\right.\nonumber\\
  && \left.~~~~~~~~~~~~+\underbrace{{\cal E}_{13}({\bf k}_1,{\bf k}_2,{\bf k}_3,{\bf k}_4;\tau_1,\tau_2)}_{\textcolor{red}{\bf Four-point~OTO~amplitude}}~\underbrace{\left\{\delta^3\left({\bf k}_1+{\bf k}_3\right)\delta^3\left({\bf k}_2+{\bf k}_4\right)+\delta^3\left({\bf k}_1+{\bf k}_4\right)\delta^3\left({\bf k}_2+{\bf k}_3\right)\right\}}_{\textcolor{red}{\bf Momentum~conservation~in~OTO~amplitude}}\right]\nonumber\\
  &&=-\int \frac{d^3k_1}{(2\pi)^3}\int \frac{d^3k_2}{(2\pi)^3}\left\{ {\cal E}_4({\bf k}_1,{\bf k}_2,-{\bf k}_2,-{\bf k}_1;\tau_1,\tau_2)+{\cal E}_4({\bf k}_1,{\bf k}_2,-{\bf k}_1,-{\bf k}_2;\tau_1,\tau_2)\right.\nonumber\\ &&\left.~~~~~~~~~~~~~~~~~~~~~~~~~~~~~+{\cal E}_6({\bf k}_1,{\bf k}_2,-{\bf k}_2,-{\bf k}_1;\tau_1,\tau_2)+{\cal E}_7({\bf k}_1,{\bf k}_2,-{\bf k}_1,-{\bf k}_2;\tau_1,\tau_2)\right.\nonumber\\ &&\left.~~~~~~~~~~~~~~~~~~~~~~~~~~~~~+{\cal E}_{10}({\bf k}_1,{\bf k}_2,-{\bf k}_1,-{\bf k}_2;\tau_1,\tau_2)+{\cal E}_{11}({\bf k}_1,{\bf k}_2,-{\bf k}_2,-{\bf k}_1;\tau_1,\tau_2)\right.\nonumber\\ &&\left.~~~~~~~~~~~~~~~~~~~~~~~~~~~~~+{\cal E}_{13}({\bf k}_1,{\bf k}_2,-{\bf k}_1,-{\bf k}_2;\tau_1,\tau_2)+{\cal E}_{13}({\bf k}_1,{\bf k}_2,-{\bf k}_2,-{\bf k}_1;\tau_1,\tau_2)\right.\nonumber\\
  &&\left.~~ +{\cal E}_7({\bf k}_1,-{\bf k}_1,{\bf k}_2,-{\bf k}_2;\tau_1,\tau_2)+{\cal E}_{10}({\bf k}_1,-{\bf k}_1,{\bf k}_2,-{\bf k}_2;\tau_1,\tau_2)+{\cal E}_{11}({\bf k}_1,-{\bf k}_1,{\bf k}_2,-{\bf k}_2;\tau_1,\tau_2)\right\},~~~~~~~~~~~ \eea 
  where we have introduced new four-point OTO micro-canonical amplitude functions, ${\cal E}_{m}({\bf k}_1,{\bf k}_2,{\bf k}_3,{\bf k}_4;\tau_1,\tau_2)~~\forall~ m=4,6,7,10,11,13$, which are defined as:
  \bea  {\cal E}_{m}({\bf k}_1,{\bf k}_2,{\bf k}_3,{\bf k}_4;\tau_1,\tau_2)&=&\sum^{4}_{i=1}\Theta^{(i)}_{m}({\bf k}_1,{\bf k}_2,{\bf k}_3,{\bf k}_4;\tau_1,\tau_2)\nonumber\\
 &=& {\cal M}_{m}({\bf k}_1,{\bf k}_2,{\bf k}_3,{\bf k}_4;\tau_1,\tau_2)-{\cal J}_{m}({\bf k}_1,{\bf k}_2,{\bf k}_3,{\bf k}_4;\tau_1,\tau_2)\nonumber\\
  &&~~~~~~~~~~~~~~~~~~+{\cal N}_{m}({\bf k}_1,{\bf k}_2,{\bf k}_3,{\bf k}_4;\tau_1,\tau_2)-{\cal Q}_{m}({\bf k}_1,{\bf k}_2,{\bf k}_3,{\bf k}_4;\tau_1,\tau_2).~~~~~~~~~~~~~\eea 
  To understand the structure of these functions more clearly one can further write them in terms of the redefined field and its canonically conjugate momenta as:
  \bea && {\cal E}_{4}({\bf k}_1,{\bf k}_2,-{\bf k}_2,-{\bf k}_1;\tau_1,\tau_2)=f^{*}_{-{\bf k}_1}(\tau_1)\Pi^{*}_{-{\bf k}_2}(\tau_2)f_{-{\bf k}_2}(\tau_1)\Pi_{-{\bf k}_1}(\tau_2)-\Pi^{*}_{-{\bf k}_1}(\tau_2)f^{*}_{-{\bf k}_2}(\tau_1)f_{-{\bf k}_2}(\tau_1)\Pi_{-{\bf k}_1}(\tau_2)\nonumber\\
  &&~~~~~~~~~~~~~~~~~~~~~~~+f^{*}_{-{\bf k}_1}(\tau_1)\Pi^{*}_{-{\bf k}_2}(\tau_2)\Pi_{-{\bf k}_2}(\tau_2)f_{-{\bf k}_1}(\tau_1)-\Pi^{*}_{-{\bf k}_1}(\tau_2)f^{*}_{-{\bf k}_2}(\tau_1)\Pi_{-{\bf k}_2}(\tau_2)f_{-{\bf k}_1}(\tau_1),~~~~~~~~~~~~~\\
  && {\cal E}_{4}({\bf k}_1,{\bf k}_2,-{\bf k}_1,-{\bf k}_2;\tau_1,\tau_2)=f^{*}_{-{\bf k}_1}(\tau_1)\Pi^{*}_{-{\bf k}_2}(\tau_2)f_{-{\bf k}_1}(\tau_1)\Pi_{-{\bf k}_2}(\tau_2)-\Pi^{*}_{-{\bf k}_1}(\tau_2)f^{*}_{-{\bf k}_2}(\tau_1)f_{-{\bf k}_1}(\tau_1)\Pi_{-{\bf k}_2}(\tau_2)\nonumber\\
  &&~~~~~~~~~~~~~~~~~~~~~~~+f^{*}_{-{\bf k}_1}(\tau_1)\Pi^{*}_{-{\bf k}_2}(\tau_2)\Pi_{-{\bf k}_1}(\tau_2)f_{-{\bf k}_2}(\tau_1)-\Pi^{*}_{-{\bf k}_1}(\tau_2)f^{*}_{-{\bf k}_2}(\tau_1)\Pi_{-{\bf k}_1}(\tau_2)f_{-{\bf k}_2}(\tau_1),~~~~~~~~~~~~~\\
&& {\cal E}_{6}({\bf k}_1,{\bf k}_2,-{\bf k}_2,-{\bf k}_1;\tau_1,\tau_2)=f^{*}_{-{\bf k}_1}(\tau_1)\Pi_{{\bf k}_2}(\tau_2)f^{*}_{{\bf k}_2}(\tau_1)\Pi_{-{\bf k}_1}(\tau_2)-\Pi^{*}_{-{\bf k}_1}(\tau_2)f_{{\bf k}_2}(\tau_1)f^{*}_{{\bf k}_2}(\tau_1)\Pi_{-{\bf k}_1}(\tau_2)\nonumber\\
  &&~~~~~~~~~~~~~~~~~~~~~~~+f^{*}_{-{\bf k}_1}(\tau_1)\Pi_{{\bf k}_2}(\tau_2)\Pi^{*}_{{\bf k}_2}(\tau_2)f_{-{\bf k}_1}(\tau_1)-\Pi^{*}_{-{\bf k}_1}(\tau_2)f_{{\bf k}_2}(\tau_1)\Pi^{*}_{{\bf k}_2}(\tau_2)f_{-{\bf k}_1}(\tau_1),~~~~~~~~~~~~~\\
&& {\cal E}_{7}({\bf k}_1,{\bf k}_2,-{\bf k}_1,-{\bf k}_2;\tau_1,\tau_2)=f_{{\bf k}_1}(\tau_1)\Pi^{*}_{-{\bf k}_2}(\tau_2)f^{*}_{{\bf k}_1}(\tau_1)\Pi_{-{\bf k}_2}(\tau_2)-\Pi_{{\bf k}_1}(\tau_2)f^{*}_{-{\bf k}_2}(\tau_1)f^{*}_{{\bf k}_1}(\tau_1)\Pi_{-{\bf k}_2}(\tau_2)\nonumber\\
  &&~~~~~~~~~~~~~~~~~~~~~~~~~~~~~~+f_{{\bf k}_1}(\tau_1)\Pi^{*}_{-{\bf k}_2}(\tau_2)\Pi^{*}_{{\bf k}_1}(\tau_2)f_{-{\bf k}_2}(\tau_1)-\Pi_{{\bf k}_1}(\tau_2)f^{*}_{-{\bf k}_2}(\tau_1)\Pi^{*}_{{\bf k}_1}(\tau_2)f_{-{\bf k}_2}(\tau_1),~~~~~~~~~~~~~\\
&& {\cal E}_{10}({\bf k}_1,{\bf k}_2,-{\bf k}_1,-{\bf k}_2;\tau_1,\tau_2)=f^{*}_{-{\bf k}_1}(\tau_1)\Pi_{{\bf k}_2}(\tau_2)f_{-{\bf k}_1}(\tau_1)\Pi^{*}_{{\bf k}_2}(\tau_2)-\Pi^{*}_{-{\bf k}_1}(\tau_2)f_{{\bf k}_2}(\tau_1)f_{-{\bf k}_1}(\tau_1)\Pi^{*}_{{\bf k}_2}(\tau_2)\nonumber\\
  &&~~~~~~~~~~~~~~~~~~~~~~~~~~~~~~+f^{*}_{-{\bf k}_1}(\tau_1)\Pi_{{\bf k}_2}(\tau_2)\Pi_{-{\bf k}_1}(\tau_2)f^{*}_{{\bf k}_2}(\tau_1)-\Pi^{*}_{-{\bf k}_1}(\tau_2)f_{{\bf k}_2}(\tau_1)\Pi_{{\bf k}_3}(\tau_2)f^{*}_{-{\bf k}_4}(\tau_1),~~~~~~~~~~~~~\\
&& {\cal E}_{11}({\bf k}_1,{\bf k}_2,-{\bf k}_2,-{\bf k}_1;\tau_1,\tau_2)=f_{{\bf k}_1}(\tau_1)\Pi^{*}_{-{\bf k}_2}(\tau_2)f_{-{\bf k}_2}(\tau_1)\Pi^{*}_{{\bf k}_1}(\tau_2)-\Pi_{{\bf k}_1}(\tau_2)f^{*}_{-{\bf k}_2}(\tau_1)f_{-{\bf k}_2}(\tau_1)\Pi^{*}_{{\bf k}_1}(\tau_2)\nonumber\\
  &&~~~~~~~~~~~~~~~~~~~~~~~~~~~~~~+f_{{\bf k}_1}(\tau_1)\Pi^{*}_{-{\bf k}_2}(\tau_2)\Pi_{-{\bf k}_2}(\tau_2)f^{*}_{{\bf k}_1}(\tau_1)-\Pi_{{\bf k}_1}(\tau_2)f^{*}_{-{\bf k}_2}(\tau_1)\Pi_{-{\bf k}_1}(\tau_2)f^{*}_{{\bf k}_2}(\tau_1),~~~~~~~~~~~~~\\
&& {\cal E}_{13}({\bf k}_1,{\bf k}_2,-{\bf k}_1,-{\bf k}_2;\tau_1,\tau_2)=f_{{\bf k}_1}(\tau_1)\Pi_{{\bf k}_2}(\tau_2)f^{*}_{{\bf k}_1}(\tau_1)\Pi^{*}_{{\bf k}_2}(\tau_2)-\Pi_{{\bf k}_1}(\tau_2)f_{{\bf k}_2}(\tau_1)f^{*}_{{\bf k}_1}(\tau_1)\Pi^{*}_{{\bf k}_2}(\tau_2)\nonumber\\
  &&~~~~~~~~~~~~~~~~~~~~~~~~~~~~~~+f_{{\bf k}_1}(\tau_1)\Pi_{{\bf k}_2}(\tau_2)\Pi^{*}_{{\bf k}_1}(\tau_2)f^{*}_{{\bf k}_2}(\tau_1)-\Pi_{{\bf k}_1}(\tau_2)f_{{\bf k}_2}(\tau_1)\Pi^{*}_{{\bf k}_1}(\tau_2)f^{*}_{{\bf k}_2}(\tau_1),~~~~~~~~~~~~~\\
&& {\cal E}_{13}({\bf k}_1,{\bf k}_2,-{\bf k}_2,-{\bf k}_1;\tau_1,\tau_2)=f_{{\bf k}_1}(\tau_1)\Pi_{{\bf k}_2}(\tau_2)f^{*}_{{\bf k}_2}(\tau_1)\Pi^{*}_{{\bf k}_1}(\tau_2)-\Pi_{{\bf k}_1}(\tau_2)f_{{\bf k}_2}(\tau_1)f^{*}_{{\bf k}_2}(\tau_1)\Pi^{*}_{{\bf k}_1}(\tau_2)\nonumber\\
  &&~~~~~~~~~~~~~~~~~~~~~~~~~~~~~~+f_{{\bf k}_1}(\tau_1)\Pi_{{\bf k}_2}(\tau_2)\Pi^{*}_{{\bf k}_2}(\tau_2)f^{*}_{{\bf k}_1}(\tau_1)-\Pi_{{\bf k}_1}(\tau_2)f_{{\bf k}_2}(\tau_1)\Pi^{*}_{{\bf k}_2}(\tau_2)f^{*}_{{\bf k}_1}(\tau_1),~~~~~~~~~~~~~\\
  && {\cal E}_{7}({\bf k}_1,-{\bf k}_1,{\bf k}_2,-{\bf k}_2;\tau_1,\tau_2)=f_{{\bf k}_1}(\tau_1)\Pi^{*}_{{\bf k}_1}(\tau_2)f^{*}_{-{\bf k}_2}(\tau_1)\Pi_{-{\bf k}_2}(\tau_2)-\Pi_{{\bf k}_1}(\tau_2)f^{*}_{{\bf k}_1}(\tau_1)f^{*}_{-{\bf k}_2}(\tau_1)\Pi_{-{\bf k}_2}(\tau_2)\nonumber\\
  &&~~~~~~~~~~~~~~~~~~~~~~~~~~~~~~+f_{{\bf k}_1}(\tau_1)\Pi^{*}_{{\bf k}_1}(\tau_2)\Pi^{*}_{-{\bf k}_2}(\tau_2)f_{-{\bf k}_2}(\tau_1)-\Pi_{{\bf k}_1}(\tau_2)f^{*}_{{\bf k}_1}(\tau_1)\Pi^{*}_{-{\bf k}_2}(\tau_2)f_{-{\bf k}_2}(\tau_1),~~~~~~~~~~~~~\\
&& {\cal E}_{10}({\bf k}_1,-{\bf k}_1,{\bf k}_2,-{\bf k}_2;\tau_1,\tau_2)=f^{*}_{-{\bf k}_1}(\tau_1)\Pi_{-{\bf k}_1}(\tau_2)f_{{\bf k}_2}(\tau_1)\Pi^{*}_{{\bf k}_2}(\tau_2)-\Pi^{*}_{-{\bf k}_1}(\tau_2)f_{-{\bf k}_1}(\tau_1)f_{{\bf k}_2}(\tau_1)\Pi^{*}_{{\bf k}_2}(\tau_2)\nonumber\\
  &&~~~~~~~~~~~~~~~~~~~~~~~~~~~~~~+f^{*}_{-{\bf k}_1}(\tau_1)\Pi_{-{\bf k}_1}(\tau_2)\Pi_{{\bf k}_2}(\tau_2)f^{*}_{{\bf k}_2}(\tau_1)-\Pi^{*}_{-{\bf k}_1}(\tau_2)f_{-{\bf k}_1}(\tau_1)\Pi_{{\bf k}_2}(\tau_2)f^{*}_{{\bf k}_2}(\tau_1),~~~~~~~~~~~~~\\
&& {\cal E}_{11}({\bf k}_1,-{\bf k}_1,{\bf k}_2,-{\bf k}_2;\tau_1,\tau_2)=f_{{\bf k}_1}(\tau_1)\Pi^{*}_{{\bf k}_1}(\tau_2)f_{{\bf k}_2}(\tau_1)\Pi^{*}_{{\bf k}_2}(\tau_2)-\Pi_{{\bf k}_1}(\tau_2)f^{*}_{{\bf k}_1}(\tau_1)f_{{\bf k}_2}(\tau_1)\Pi^{*}_{{\bf k}_2}(\tau_2)\nonumber\\
  &&~~~~~~~~~~~~~~~~~~~~~~~~~~~~~~+f_{{\bf k}_1}(\tau_1)\Pi^{*}_{{\bf k}_1}(\tau_2)\Pi_{{\bf k}_2}(\tau_2)f^{*}_{{\bf k}_2}(\tau_1)-\Pi_{{\bf k}_1}(\tau_2)f^{*}_{{\bf k}_1}(\tau_1)\Pi_{{\bf k}_2}(\tau_2)f^{*}_{{\bf k}_2}(\tau_1),~~~~~~~~~~~~~\eea
  All the above mentioned quantities physically signify the momentum and conformal time dependent amplitude of the OTOC which is written explicitly in terms of the contributions from the four-point correlation function. Here we have to note down few symmetry properties of the above mentioned amplitudes under the exchange of the momenta appearing in the third and fourth position i.e. if we replace $-{\bf k}_2 \rightarrow -{\bf k}_1$ and $-{\bf k}_1 \rightarrow -{\bf k}_2$, which are given by:
  \bea {\cal E}_{4}({\bf k}_1,{\bf k}_2,-{\bf k}_2,-{\bf k}_1;\tau_1,\tau_2)={\cal E}_{4}({\bf k}_1,{\bf k}_2,-{\bf k}_1,-{\bf k}_2;\tau_1,\tau_2),\\
 {\cal E}_{13}({\bf k}_1,{\bf k}_2,-{\bf k}_2,-{\bf k}_1;\tau_1,\tau_2)={\cal E}_{13}({\bf k}_1,{\bf k}_2,-{\bf k}_1,-{\bf k}_2;\tau_1,\tau_2). \eea
  Using these symmetry properties the OTOC can be further simplified as:
   \bea && C(\tau_1,\tau_2)=-\int \frac{d^3{\bf k}_1}{(2\pi)^3}\int \frac{d^3{\bf k}_2}{(2\pi)^3}\left\{ 2\left({\cal E}_4({\bf k}_1,{\bf k}_2,-{\bf k}_2,-{\bf k}_1;\tau_1,\tau_2)+{\cal E}_{13}({\bf k}_1,{\bf k}_2,-{\bf k}_1,-{\bf k}_2;\tau_1,\tau_2)\right)\right.\nonumber\\ &&\left.~~~~~~~~~~~~~~~~~~~~~~~~~~~~~+{\cal E}_6({\bf k}_1,{\bf k}_2,-{\bf k}_2,-{\bf k}_1;\tau_1,\tau_2)+{\cal E}_7({\bf k}_1,{\bf k}_2,-{\bf k}_1,-{\bf k}_2;\tau_1,\tau_2)\right.\nonumber\\ &&\left.~~~~~~~~~~~~~~~~~~~~~~~~~~~~~+{\cal E}_{10}({\bf k}_1,{\bf k}_2,-{\bf k}_1,-{\bf k}_2;\tau_1,\tau_2)+{\cal E}_{11}({\bf k}_1,{\bf k}_2,-{\bf k}_2,-{\bf k}_1;\tau_1,\tau_2)\right.\nonumber\\ 
  &&\left.~~~~ +{\cal E}_7({\bf k}_1,-{\bf k}_1,{\bf k}_2,-{\bf k}_2;\tau_1,\tau_2)+{\cal E}_{10}({\bf k}_1,-{\bf k}_1,{\bf k}_2,-{\bf k}_2;\tau_1,\tau_2)+{\cal E}_{11}({\bf k}_1,-{\bf k}_1,{\bf k}_2,-{\bf k}_2;\tau_1,\tau_2)\right\},~~~~~~~~~~~ \eea 
   In the context of cosmology, we usually integrate over the time scale from $-\infty$ to $0$ to get the momentum dependent trispectrum, which represents the amplitude of the four-point correlation function. Instead of doing the integration over the conformal time here we integrate over the moneta since in the present context we are interested to determine the time dependent behaviour of the OTOC. Now our aim is to compute the explicit form of the momentum integrated four-point amplitudes which will fix the mathematical structure of the time time dependent OTOC in the present context. One of the important point we need to mention here that, the individual parts of the amplitudes before integrating out the momenta is only function of the magnitude of the momenta, not its direction. As a consequence, the volume integrations over these momenta becomes very simpler and one can treat the volume elements of the integrals as given by the following expression:
  \begin{eqnarray}
 &&\displaystyle\prod_{i=1}^{2} \frac{d^3{\bf k}_i}{(2\pi)^6}=\frac{1}{(2\pi)^6}\displaystyle\prod_{i=1}^{2}k^2_i~dk_i~\sin\theta_i~d\theta_i~d\phi_i~,\nonumber\\
 &&~~~~~~~~~~~~~~~~~~~~~~~~~~~~~~~{\rm where}~~0<k_i<\infty,~~~0<\theta_i<\pi,~~~0<\phi_i<2\pi~~~~\forall~i=1,2.~~~~~~~~~~~~
\end{eqnarray}   
Now, it is clearly evident from the the expression for the above mentioned volume element that the angular part $\sin\theta_i~d\theta_i~d\phi_i~\forall~i=1,2$ are the infinitesimal solid angle subtended for $S_2$, which is $d\Omega_{S_2}$ and after performing the integration over the angular coordinates it will give a factor of $4(2\pi)^2$ out of the integral over the volume elements of the momenta appearing in the expression for OTOC. 
Another important point is to note that here it might happen that the integral over the momenta within the range $0<k_i<\infty~~\forall~~i=1,2$ becomes infinite. For this reason we put cut-off $0<k_i<L~~\forall~~i=1,2$ to regulate the integral over the momenta and consequently we can write:
\bea && C(\tau_1,\tau_2)=-\frac{1}{4\pi^4}\int^{L}_{0} k^2_1~dk_1\int^{L}_{0} k^2_2~dk_2~\nonumber\\
&&~~~~~~~~~~~~~~~~~~~~~~~~~~~~~~\left\{ 2\left({\cal E}_4({\bf k}_1,{\bf k}_2,-{\bf k}_2,-{\bf k}_1;\tau_1,\tau_2)+{\cal E}_{13}({\bf k}_1,{\bf k}_2,-{\bf k}_1,-{\bf k}_2;\tau_1,\tau_2)\right)\right.\nonumber\\ &&\left.~~~~~~~~~~~~~~~~~~~~~~~~~~~~~+{\cal E}_6({\bf k}_1,{\bf k}_2,-{\bf k}_2,-{\bf k}_1;\tau_1,\tau_2)+{\cal E}_7({\bf k}_1,{\bf k}_2,-{\bf k}_1,-{\bf k}_2;\tau_1,\tau_2)\right.\nonumber\\ &&\left.~~~~~~~~~~~~~~~~~~~~~~~~~~~~~+{\cal E}_{10}({\bf k}_1,{\bf k}_2,-{\bf k}_1,-{\bf k}_2;\tau_1,\tau_2)+{\cal E}_{11}({\bf k}_1,{\bf k}_2,-{\bf k}_2,-{\bf k}_1;\tau_1,\tau_2)\right.\nonumber\\ 
  &&\left.~~~ +{\cal E}_7({\bf k}_1,-{\bf k}_1,{\bf k}_2,-{\bf k}_2;\tau_1,\tau_2)+{\cal E}_{10}({\bf k}_1,-{\bf k}_1,{\bf k}_2,-{\bf k}_2;\tau_1,\tau_2)+{\cal E}_{11}({\bf k}_1,-{\bf k}_1,{\bf k}_2,-{\bf k}_2;\tau_1,\tau_2)\right\},~~~~~~~~~~~ \eea 
  Further, we define the following momenta integrated time dependent amplitudes:
  \bea && {\cal I}_1(\tau_1,\tau_2):=\int^{L}_{k_1=0} k^2_1~dk_1\int^{L}_{k_2=0} k^2_2~dk_2~2{\cal E}_4({\bf k}_1,{\bf k}_2,-{\bf k}_2,-{\bf k}_1;\tau_1,\tau_2),\\
  && {\cal I}_2(\tau_1,\tau_2):=\int^{L}_{k_1=0} k^2_1~dk_1\int^{L}_{k_2=0} k^2_2~dk_2~2{\cal E}_{13}({\bf k}_1,{\bf k}_2,-{\bf k}_2,-{\bf k}_1;\tau_1,\tau_2),\\
  && {\cal I}_3(\tau_1,\tau_2):=\int^{L}_{k_1=0} k^2_1~dk_1\int^{L}_{k_2=0} k^2_2~dk_2~{\cal E}_6({\bf k}_1,{\bf k}_2,-{\bf k}_2,-{\bf k}_1;\tau_1,\tau_2),\\
  && {\cal I}_4(\tau_1,\tau_2):=\int^{L}_{k_1=0} k^2_1~dk_1\int^{L}_{k_2=0} k^2_2~dk_2~{\cal E}_7({\bf k}_1,{\bf k}_2,-{\bf k}_1,-{\bf k}_2;\tau_1,\tau_2),\\
 &&  {\cal I}_5(\tau_1,\tau_2):=\int^{L}_{k_1=0} k^2_1~dk_1\int^{L}_{k_2=0} k^2_2~dk_2~{\cal E}_{10}({\bf k}_1,{\bf k}_2,-{\bf k}_1,-{\bf k}_2;\tau_1,\tau_2),\\
 && {\cal I}_6(\tau_1,\tau_2):=\int^{L}_{k_1=0} k^2_1~dk_1\int^{L}_{k_2=0} k^2_2~dk_2~{\cal E}_{11}({\bf k}_1,{\bf k}_2,-{\bf k}_2,-{\bf k}_1;\tau_1,\tau_2),\\
 && {\cal I}_7(\tau_1,\tau_2):=\int^{L}_{k_1=0} k^2_1~dk_1\int^{L}_{k_2=0} k^2_2~dk_2~{\cal E}_7({\bf k}_1,-{\bf k}_1,{\bf k}_2,-{\bf k}_2;\tau_1,\tau_2),\\
 && {\cal I}_8(\tau_1,\tau_2):=\int^{L}_{k_1=0} k^2_1~dk_1\int^{L}_{k_2=0} k^2_2~dk_2~{\cal E}_{10}({\bf k}_1,-{\bf k}_1,{\bf k}_2,-{\bf k}_2;\tau_1,\tau_2),\\
 && {\cal I}_9(\tau_1,\tau_2):=\int^{L}_{k_1=0} k^2_1~dk_1\int^{L}_{k_2=0} k^2_2~dk_2~{\cal E}_{10}({\bf k}_1,-{\bf k}_1,{\bf k}_2,-{\bf k}_2;\tau_1,\tau_2).
  \eea 
   Consequently, the OTOC can be expressed in terms of the four-point time dependent amplitudes as:
  \bea \hll{C^{f}(\tau_1,\tau_2)=-\frac{1}{4\pi^4}\sum^{7}_{j=1}w_i{\cal I}_{j}(\tau_1,\tau_2)},\eea
  where our prime motivation is to computed all of these non vanishing time dependent amplitudes. Once we determine all of them then the mathematical structure of the OTOC i.e. the time dependent behaviour in the OTOC will be fixed. Here it is important to note that the weight factors for each individual contributions are given by the following expression:
  \bea w_1=w_2=2,~~~~~w_j=1~~\forall ~~j=3,4,\cdots, 9.\eea
  After computation we have found that:
  \bea  {\cal I}_2(\tau_1,\tau_2)&&=(-\tau_1)^{1-2\nu}(-\tau_2)^{3-2\nu}\sum^{4}_{i=1}X^{(i)}_{1}(\tau_1,\tau_2)=(-1)^{4\nu}{\cal I}_1(\tau_1,\tau_2),~~~\eea\bea
   {\cal I}_3(\tau_1,\tau_2)&&=(-\tau_1)^{1-2\nu}(-\tau_2)^{3-2\nu}\sum^{4}_{i=1}X^{(i)}_{1}(\tau_1,\tau_2)=(-1)^{2\nu}{\cal I}_1(\tau_1,\tau_2),~~~\\
    {\cal I}_4(\tau_1,\tau_2)&&=(-\tau_1)^{1-2\nu}(-\tau_2)^{3-2\nu}\sum^{4}_{i=1}X^{(i)}_{1}(\tau_1,\tau_2)=(-1)^{2\nu}{\cal I}_1(\tau_1,\tau_2),~~~\\
     {\cal I}_5(\tau_1,\tau_2)&&=(-\tau_1)^{1-2\nu}(-\tau_2)^{3-2\nu}\sum^{4}_{i=1}X^{(i)}_{1}(\tau_1,\tau_2)=(-1)^{2\nu}{\cal I}_1(\tau_1,\tau_2),~~~\\
      {\cal I}_6(\tau_1,\tau_2)&&=(-\tau_1)^{1-2\nu}(-\tau_2)^{3-2\nu}\sum^{4}_{i=1}X^{(i)}_{1}(\tau_1,\tau_2)=(-1)^{2\nu}{\cal I}_1(\tau_1,\tau_2),~~~\\
       {\cal I}_7(\tau_1,\tau_2)&&=(-\tau_1)^{1-2\nu}(-\tau_2)^{3-2\nu}\sum^{4}_{i=1}X^{(i)}_{1}(\tau_1,\tau_2)=(-1)^{2\nu}{\cal I}_1(\tau_1,\tau_2),~~~\\
        {\cal I}_8(\tau_1,\tau_2)&&=(-\tau_1)^{1-2\nu}(-\tau_2)^{3-2\nu}\sum^{4}_{i=1}X^{(i)}_{1}(\tau_1,\tau_2)=(-1)^{2\nu}{\cal I}_1(\tau_1,\tau_2),~~~\\
         {\cal I}_9(\tau_1,\tau_2)&&=(-\tau_1)^{1-2\nu}(-\tau_2)^{3-2\nu}\sum^{4}_{i=1}X^{(i)}_{1}(\tau_1,\tau_2)=(-1)^{2\nu}{\cal I}_1(\tau_1,\tau_2).~~~\eea
Consequently, we can further write:
\bea \sum^{7}_{j=1}w_i{\cal I}_{j}(\tau_1,\tau_2)&=&\left[2+2(-1)^{4\nu}+7(-1)^{2\nu}\right]{\cal I}_1(\tau_1,\tau_2)\nonumber\\
&=&2\left[1+(-1)^{4\nu}+\frac{7}{2}(-1)^{2\nu}\right]\frac{(-\tau_1)^{1-2\nu}(-\tau_2)^{3-2\nu}}{(-1)^{4\nu}}\sum^{4}_{i=1}X^{(i)}_{1}(\tau_1,\tau_2),~~~~~~\eea
where the explicit form of the time dependent functions $X^{(i)}_{1}(T,\tau)~~\forall~~i=1,2,3,4$ are defined in the appendix. Further using this result the un-normalised OTOC can be expressed as:
 \begin{equation}
  \hll{C^{f}(\tau_1,\tau_2)=-\frac{1}{4\pi^4}\sum^{7}_{j=1}w_i{\cal I}_{j}(\tau_1,\tau_2)=\left[1+(-1)^{4\nu}+\frac{7}{2}(-1)^{2\nu}\right]\frac{(-\tau_1)^{1-2\nu}(-\tau_2)^{3-2\nu}}{2\pi^4(-1)^{4\nu-1}}\sum^{4}_{i=1}X^{(i)}_{1}(\tau_1,\tau_2)}.
\end{equation} 
\subsubsection{With normalization}
Further, the normalisation factor of OTOC, which is given by the following expression:
\bea \hll{{\cal N}^{f}(\tau_1,\tau_2)=\frac{1}{\langle \hat{f}(\tau_1)\hat{f}(\tau_1)\rangle_{\beta} \langle \hat{\Pi}(\tau_2)\hat{\Pi}(\tau_2)\rangle_{\beta}}=\frac{\pi^4}{{\cal F}_1(\tau_1){\cal F}_2(\tau_2)}}.\eea
where the time dependent functions ${\cal F}_1(\tau_1)$ and ${\cal F}_2(\tau_2)$ are defined in the Appendix. For further details please look into the Appendix for the detailed computation of the normalisation factor of OTOC.

Finally, the normalised OTOC in the present context can be computed as:
\bea &&\hll{{\cal C}^{f}(\tau_1,\tau_2)=\frac{C^{f}(\tau_1,\tau_2)}{\langle \hat{f}(\tau_1)\hat{f}(\tau_1)\rangle_{\beta} \langle \hat{\Pi}(\tau_2)\hat{\Pi}(\tau_2)\rangle_{\beta}}}\nonumber\\
&&~~~~~~~~~~~~~\hll{=-\frac{1}{4{\cal F}_1(\tau_1){\cal F}_2(\tau_2)}\sum^{7}_{j=1}w_j{\cal I}_{j}(\tau_1,\tau_2)}\nonumber\\
&&~~~~~~~~~~~~~\hll{=-\frac{1}{2{\cal F}_1(\tau_1){\cal F}_2(\tau_2)}\left[1+(-1)^{4\nu}+\frac{7}{2}(-1)^{2\nu}\right]{\cal I}_1(\tau_1,\tau_2)}\nonumber\\
&&~~~~~~~~~~~~\hll{=\left[1+(-1)^{4\nu}+\frac{7}{2}(-1)^{2\nu}\right]\frac{(-\tau_1)^{1-2\nu}(-\tau_2)^{3-2\nu}}{2(-1)^{4\nu-1}{\cal F}_1(\tau_1){\cal F}_2(\tau_2)}\sum^{4}_{i=1}X^{(i)}_{1}(\tau_1,\tau_2)},~~~~~~~\eea
which is obviously a new result in the context of primordial cosmology and we are very hopeful that this result will explore various unknown physical phenomena happened in early universe. The detailed explanation of this obtained result will be discussed in the later half of this section.
 \begin{figure}[t!]
    \centering
        \centering
        \includegraphics[width=17.3cm,height=4.8cm]{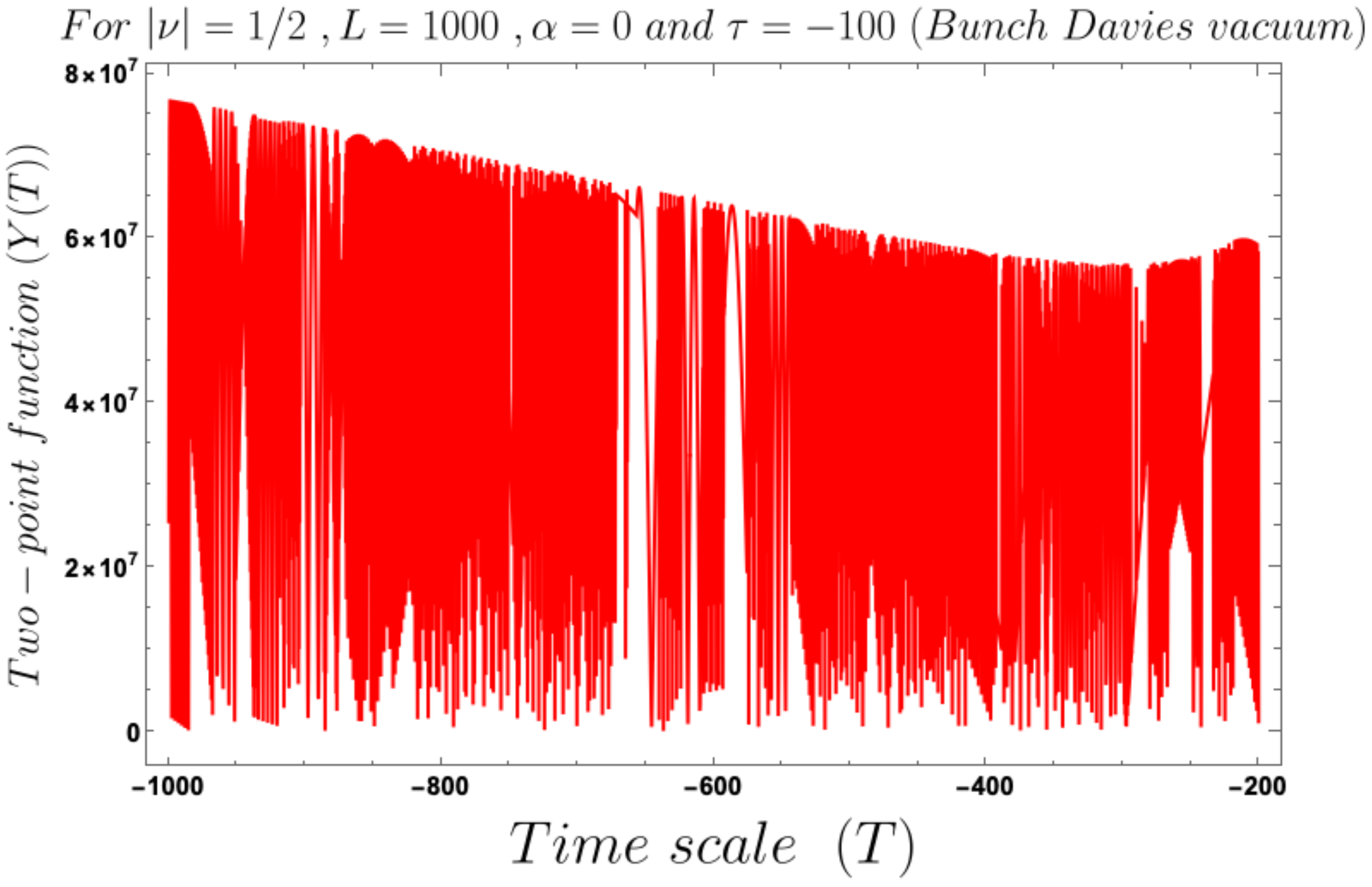}
        \includegraphics[width=16cm,height=4.8cm]{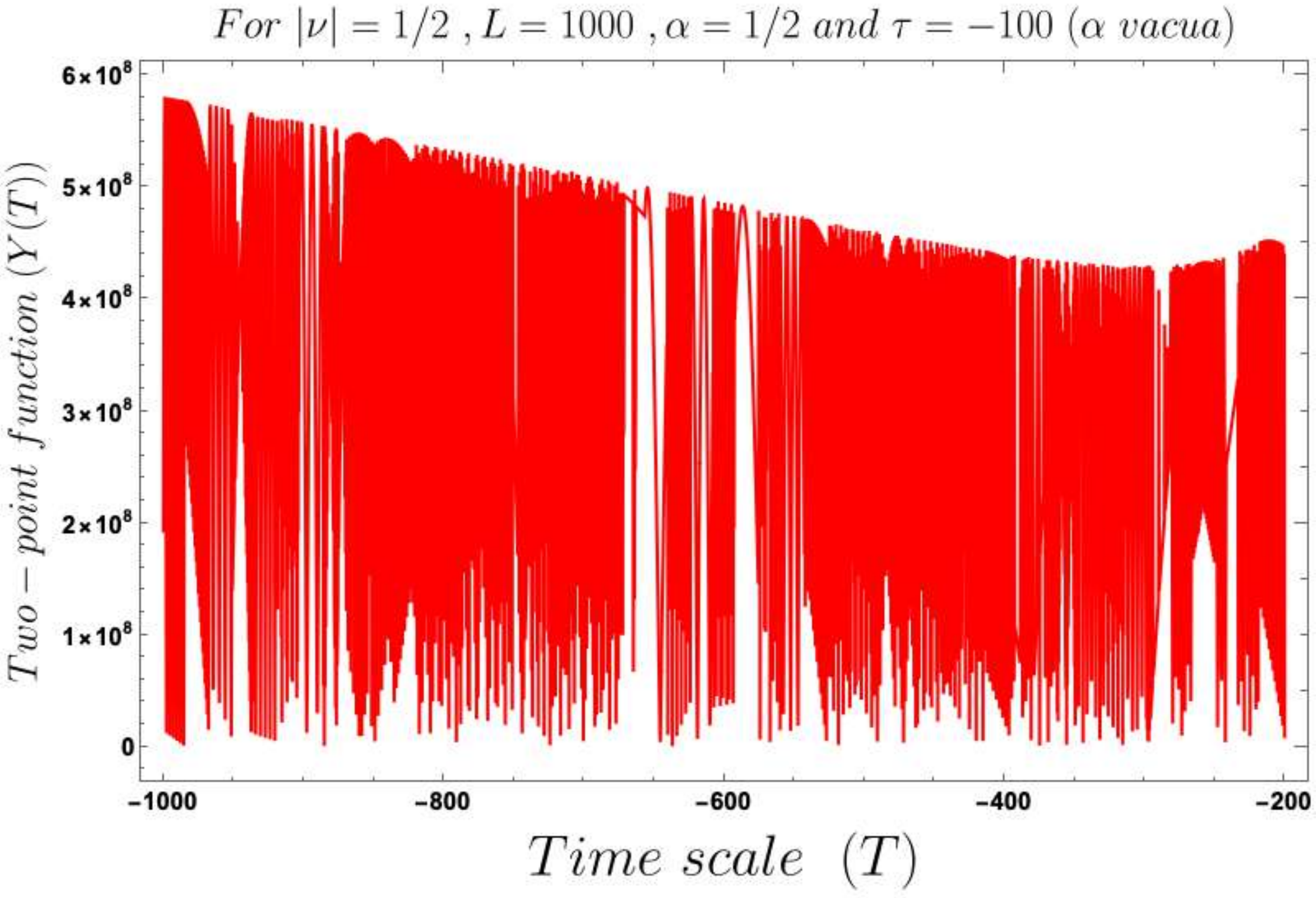}
   \includegraphics[width=17.3cm,height=4.8cm]{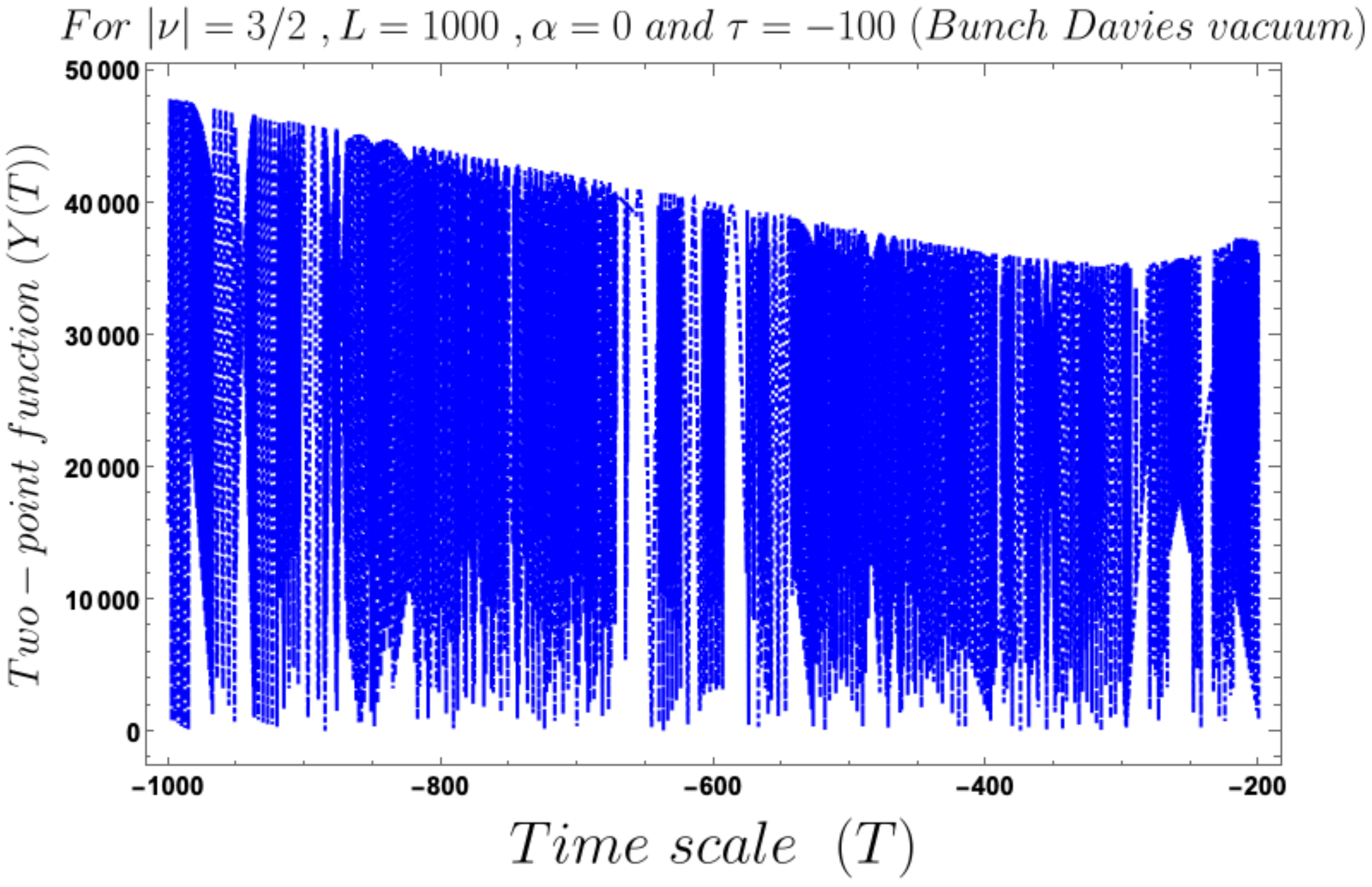}
        \includegraphics[width=16cm,height=4.8cm]{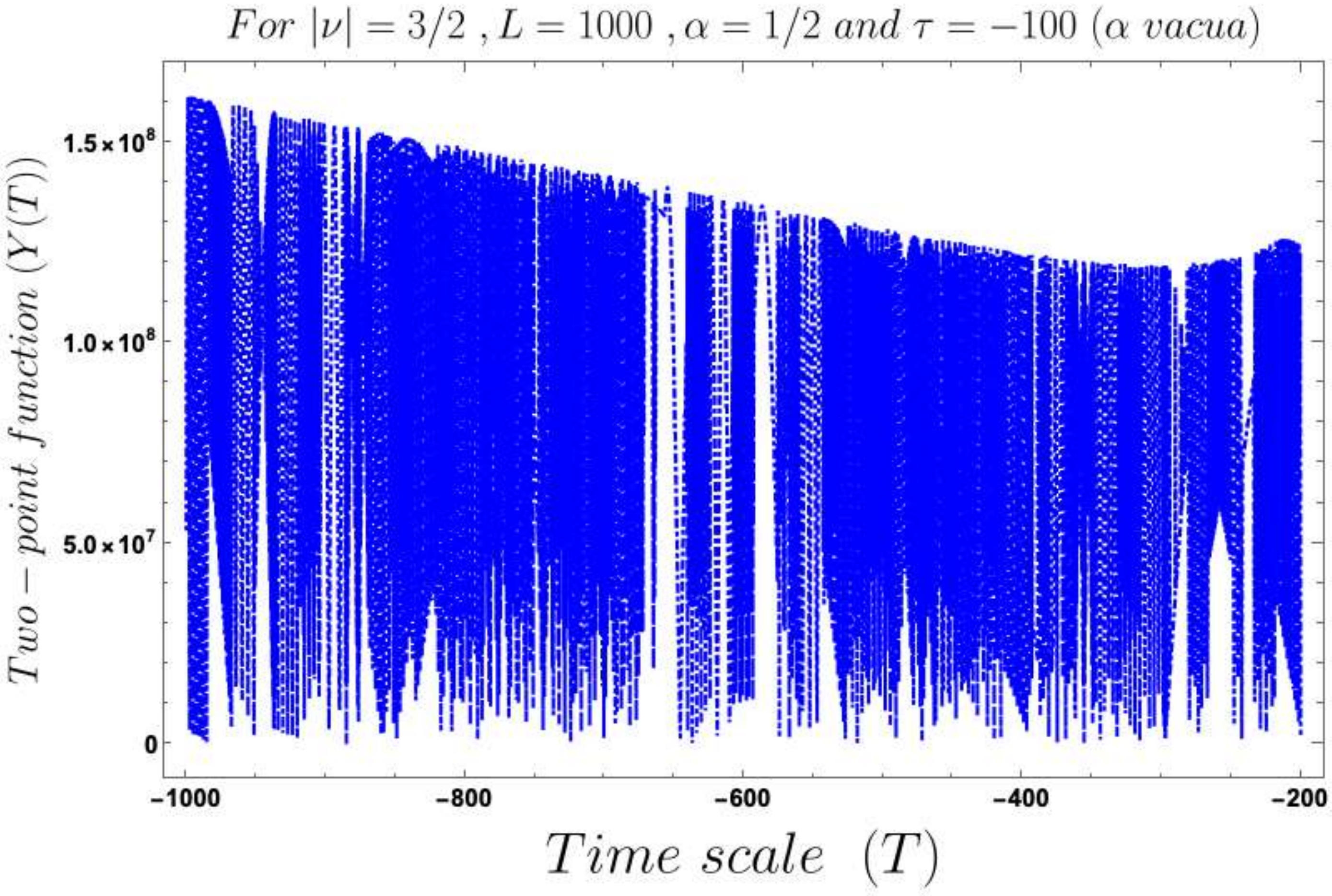} 
    \caption{Behaviour of the two-point function with respect to the $T$ time scale of the theory. Here we fix, mass parameter $\nu=-i/2,-3i/2$, cut-off scale $L=1000$, vacuum parameter $\alpha=0 ~({\rm Bunch~Davies~vacuum)}, 1/2~(\alpha~{\rm vacua})$.}
      \label{fig:12AC}
\end{figure}
\begin{figure}[t!]
    \centering
        \centering
        \includegraphics[width=17.3cm,height=4.8cm]{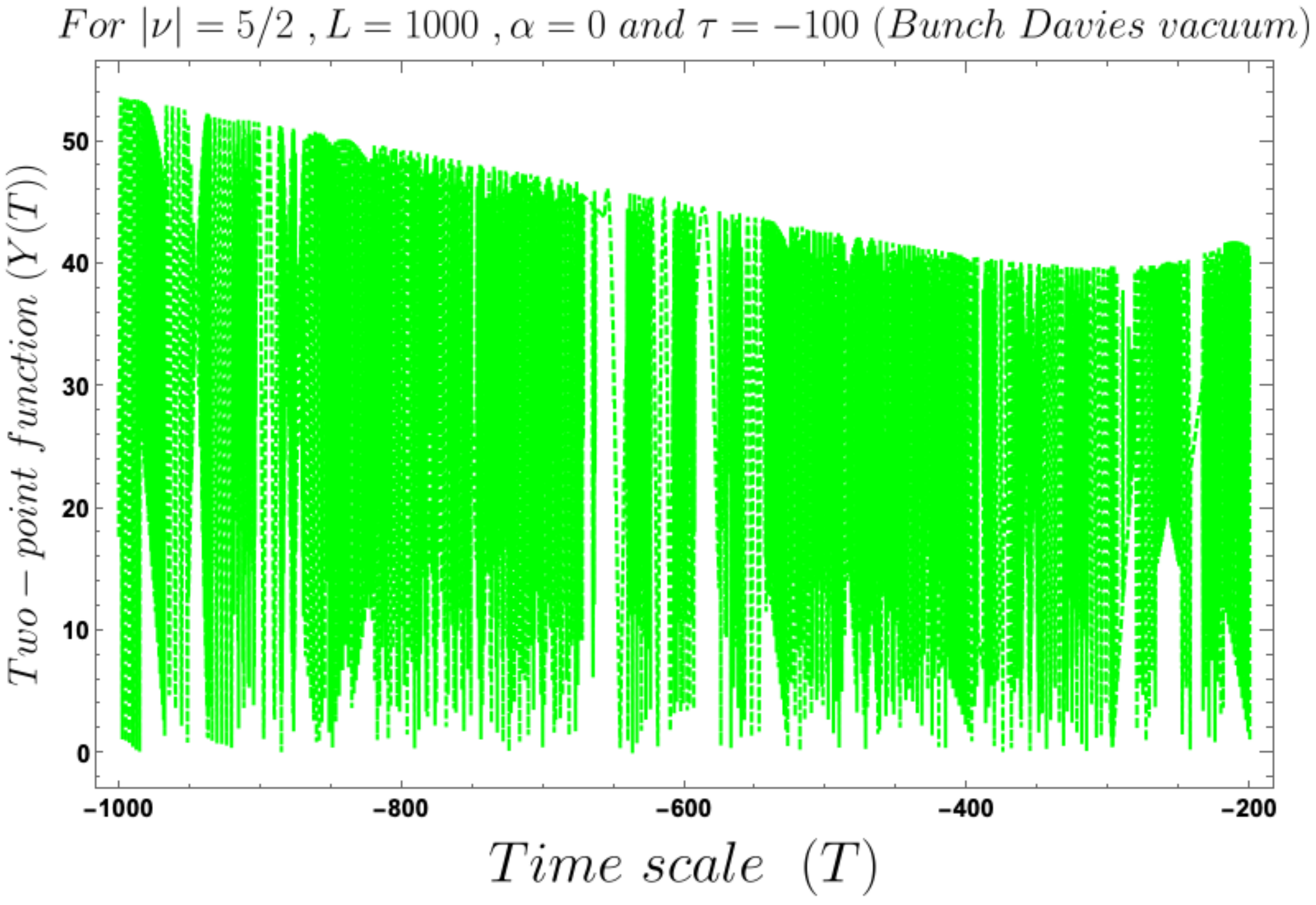}
        \includegraphics[width=16cm,height=4.8cm]{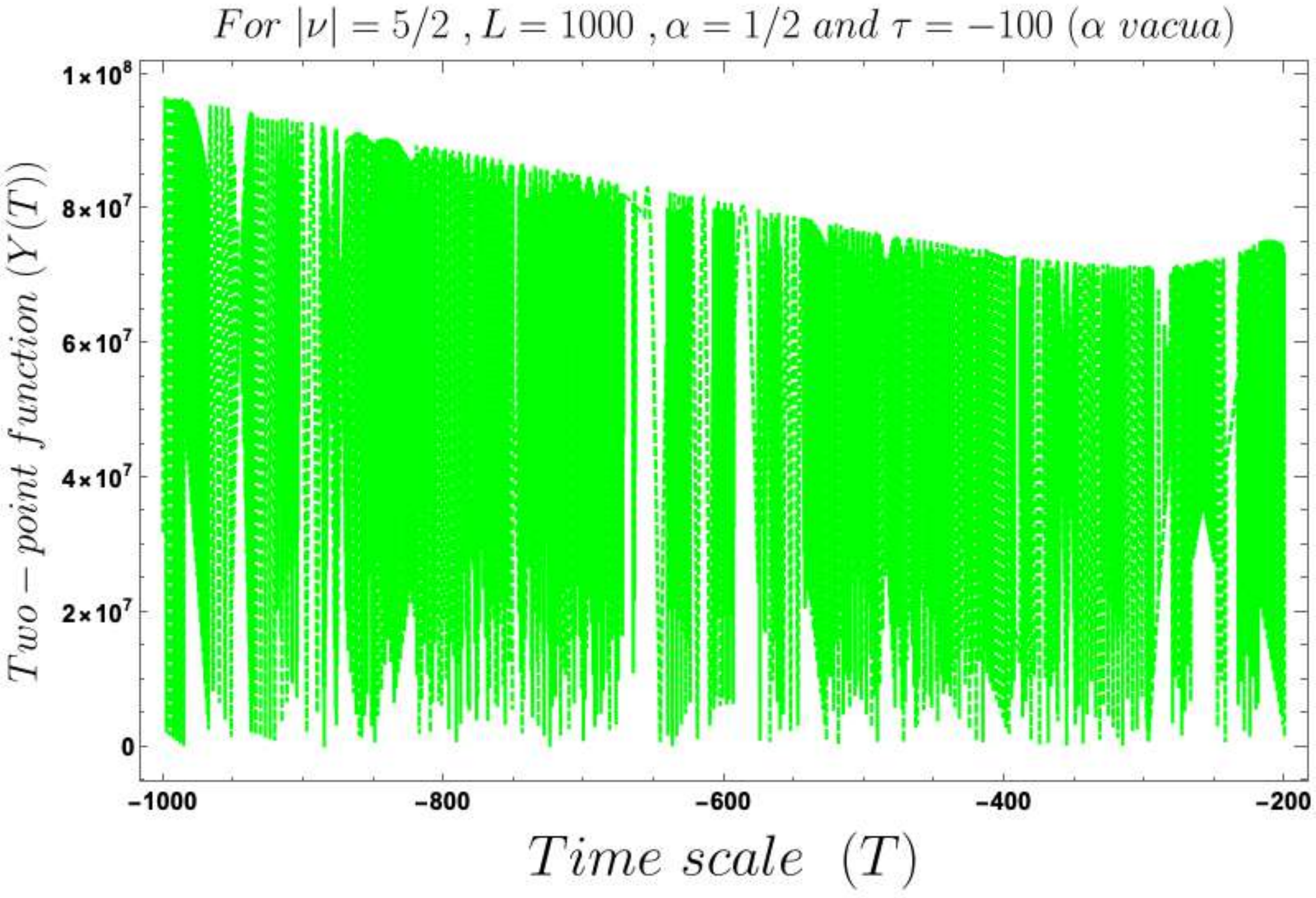}
   \includegraphics[width=17.3cm,height=4.8cm]{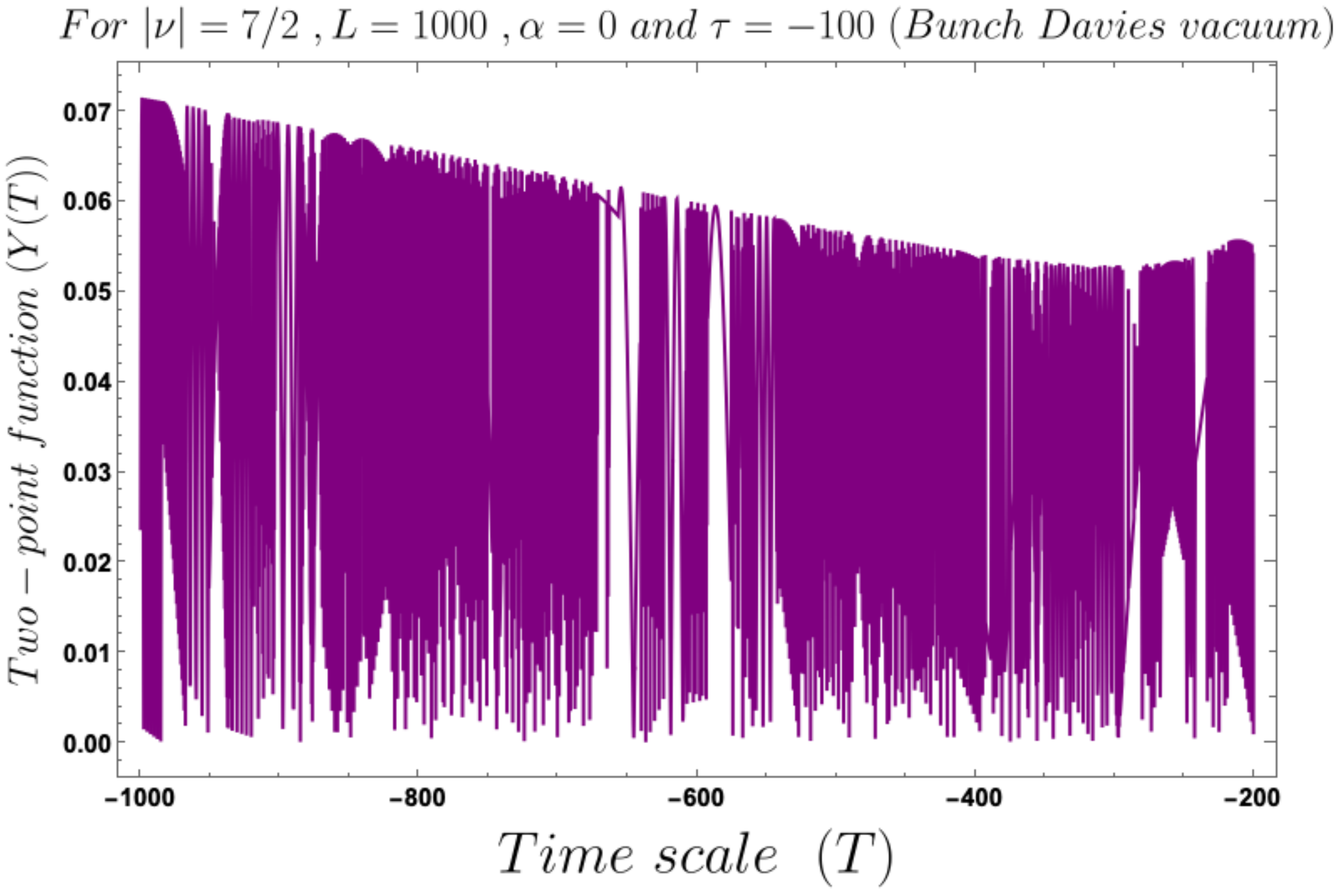}
        \includegraphics[width=16cm,height=4.8cm]{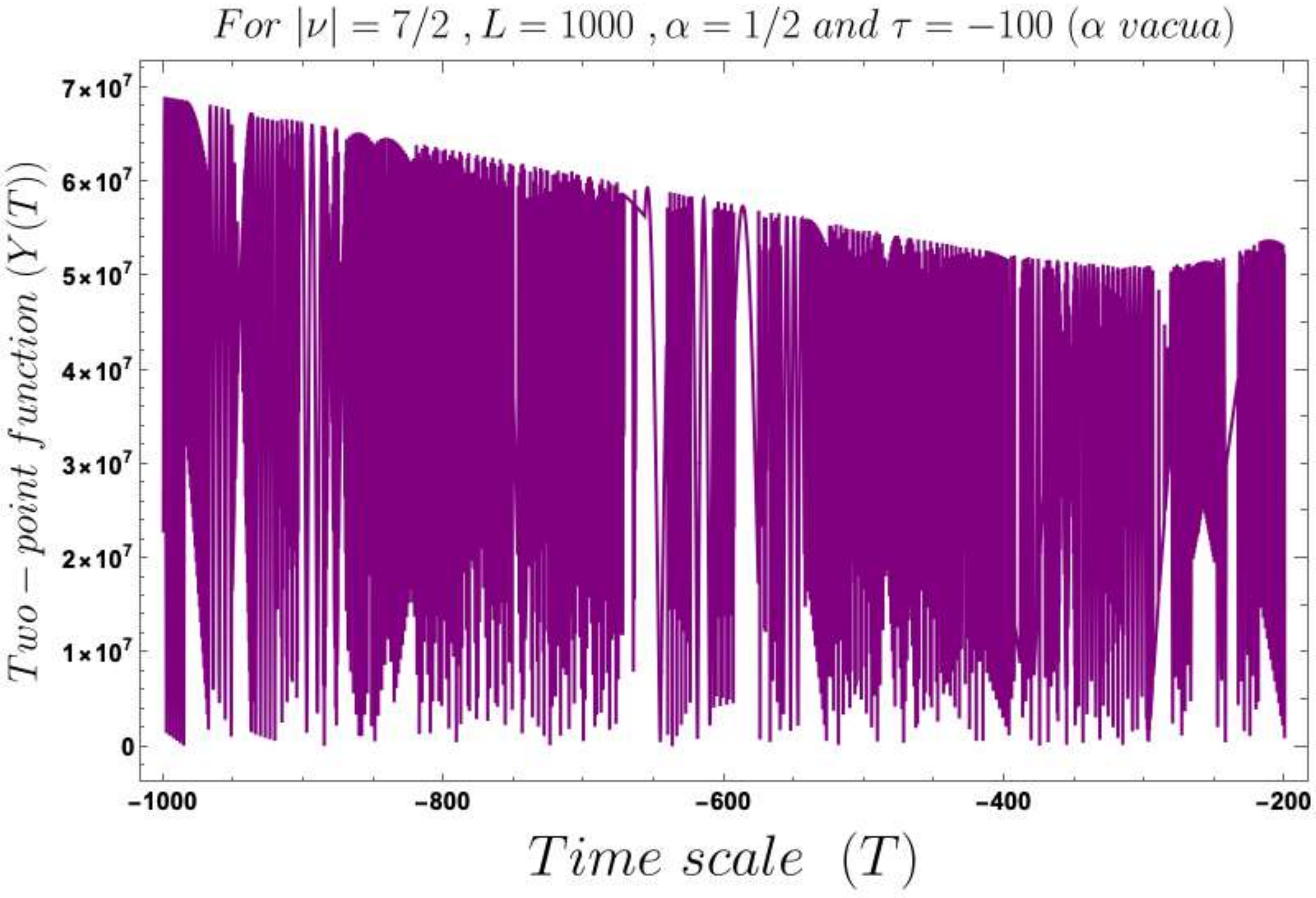}
    \caption{Behaviour of the two-point function with respect to the $T$ time scale of the theory. Here we fix, mass parameter $\nu=-5i/2,-7i/2$, cut-off scale $L=1000$, vacuum parameter $\alpha=0 ~({\rm Bunch~Davies~vacuum)}, 1/2~(\alpha~{\rm vacua})$.}
      \label{fig:13AC}
\end{figure}
\begin{figure}[t!]
    \centering
        \centering
        \includegraphics[width=17.3cm,height=4.8cm]{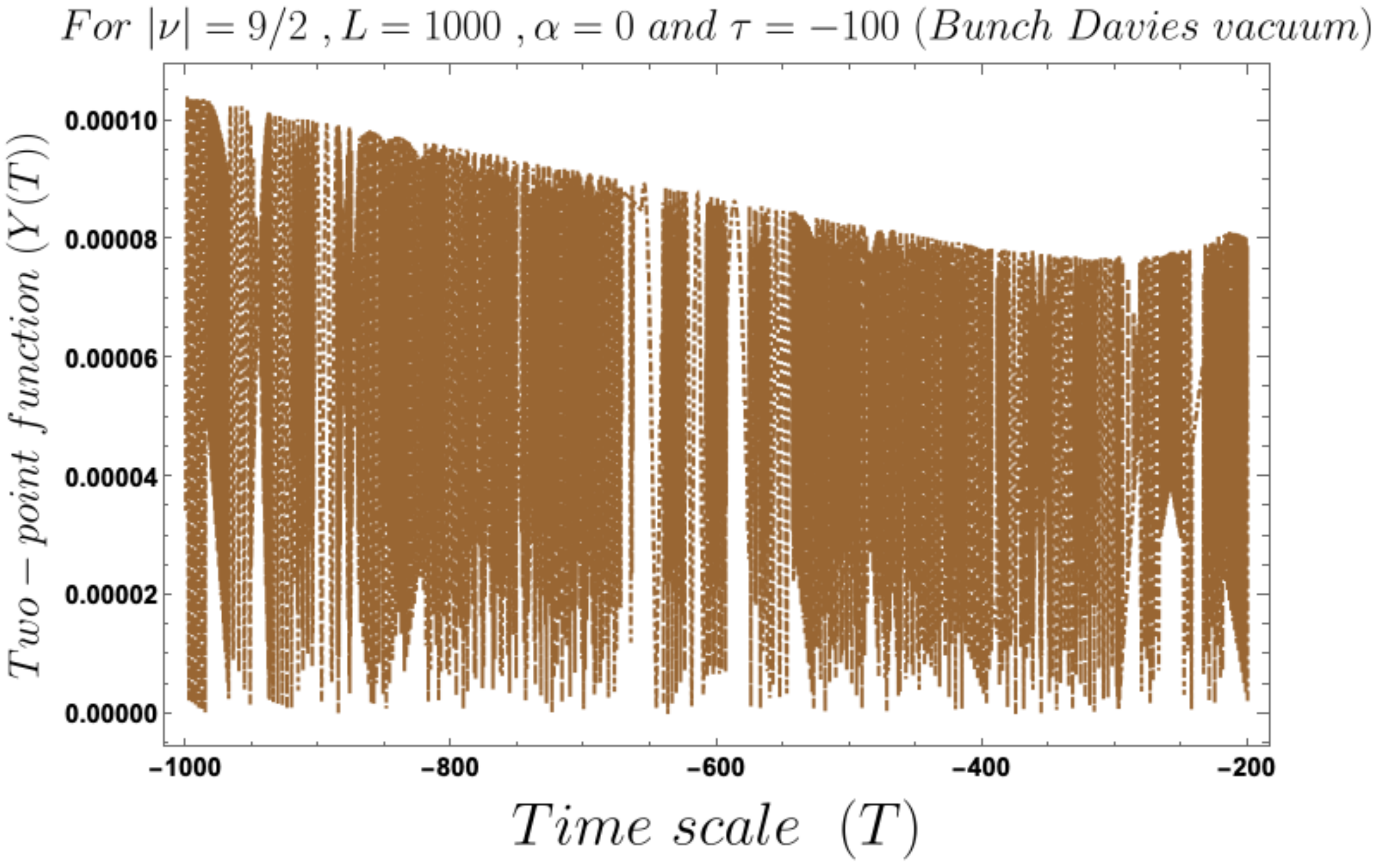}
        \includegraphics[width=16cm,height=4.8cm]{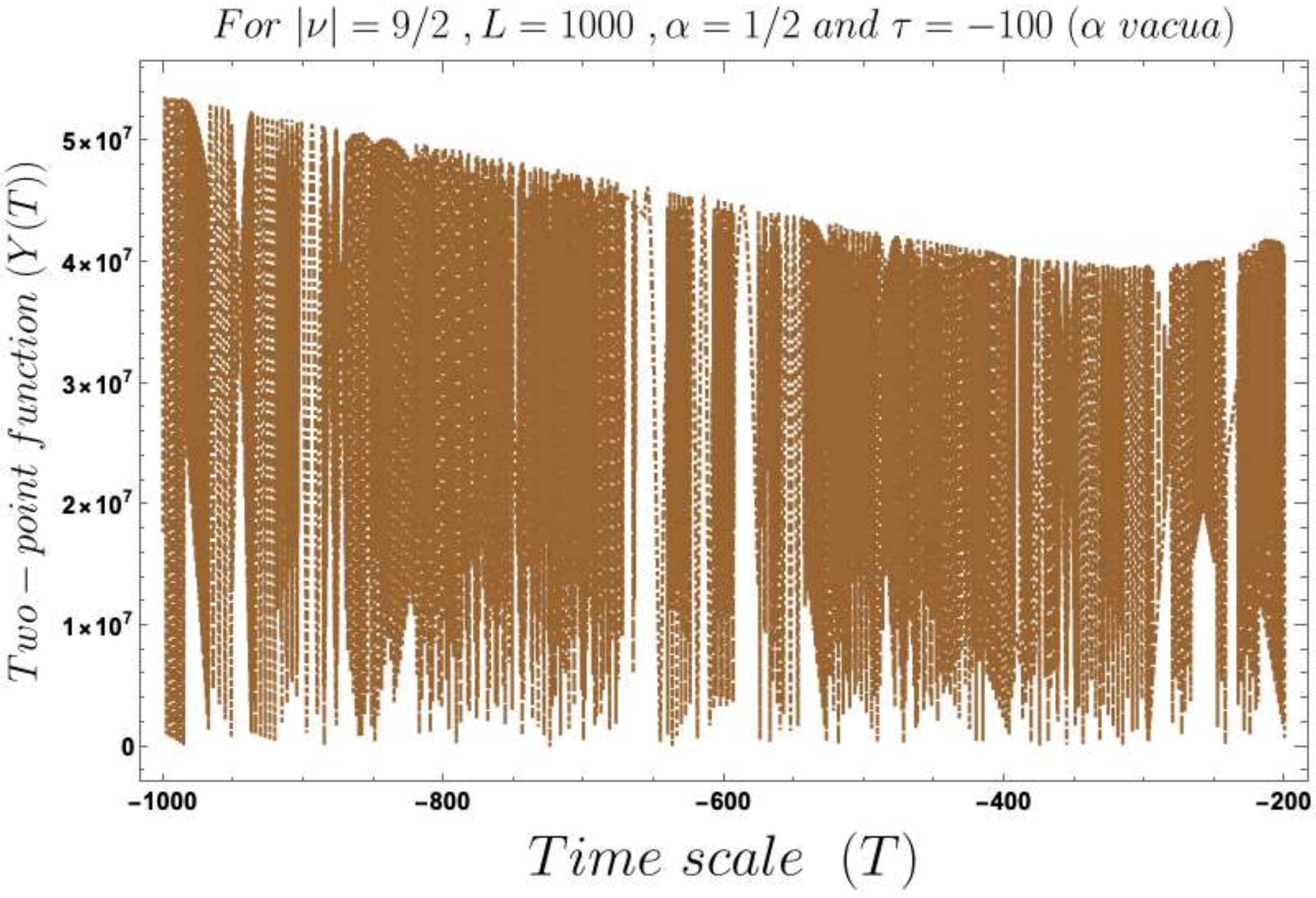}
    \caption{Behaviour of the two-point function with respect to the $T$ time scale of the theory. Here we fix, mass parameter $\nu=-9i/2$, cut-off scale $L=1000$, vacuum parameter $\alpha=0 ~({\rm Bunch~Davies~vacuum)}, 1/2~(\alpha~{\rm vacua})$.}
      \label{fig:14AC}
\end{figure}
\begin{figure}[t!]
    \centering
        \centering
        \includegraphics[width=15.3cm,height=9.6cm]{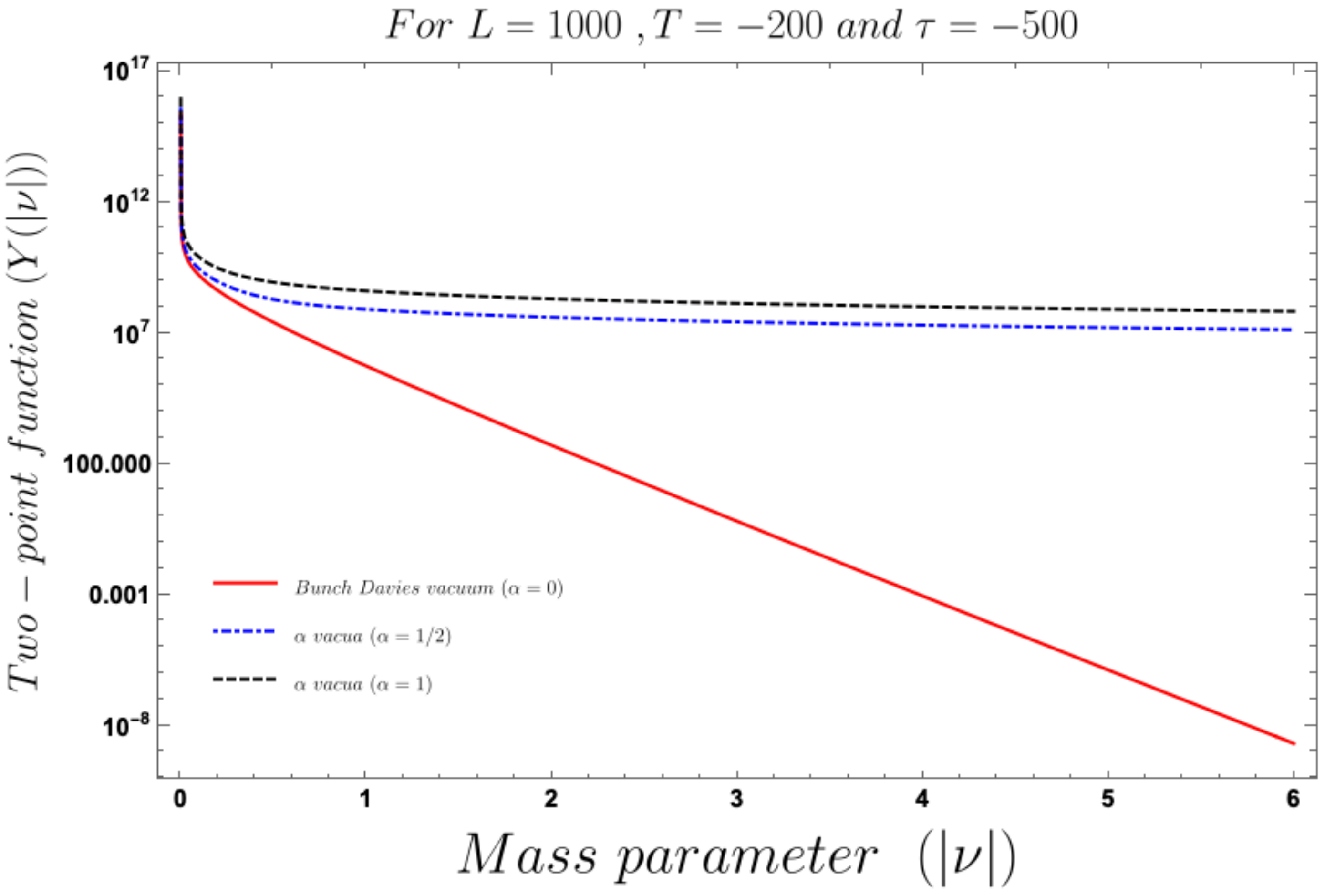}
        \includegraphics[width=16cm,height=9.6cm]{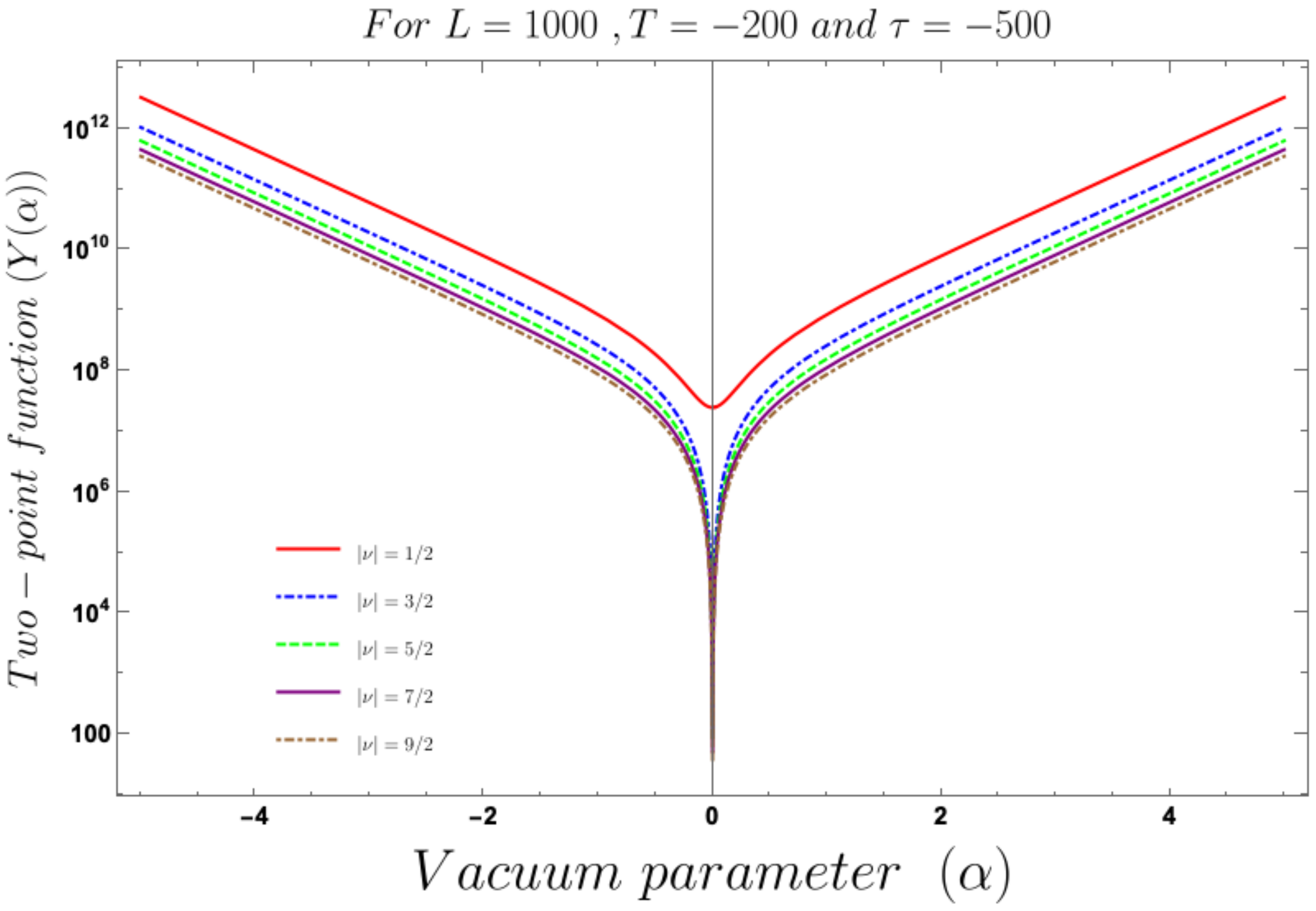}
    \caption{Behaviour of the two-point function with respect to the mass parameter $\nu$ and the vacuum parameter $\alpha$ of the theory. Here we fix, cut-off scale $L=1000$, and the two time parameters, $T=-200$ and $\tau=-500$.}
      \label{fig:14BC}
\end{figure}
\begin{figure}[t!]
    \centering
        \centering
        \includegraphics[width=17.3cm,height=4.8cm]{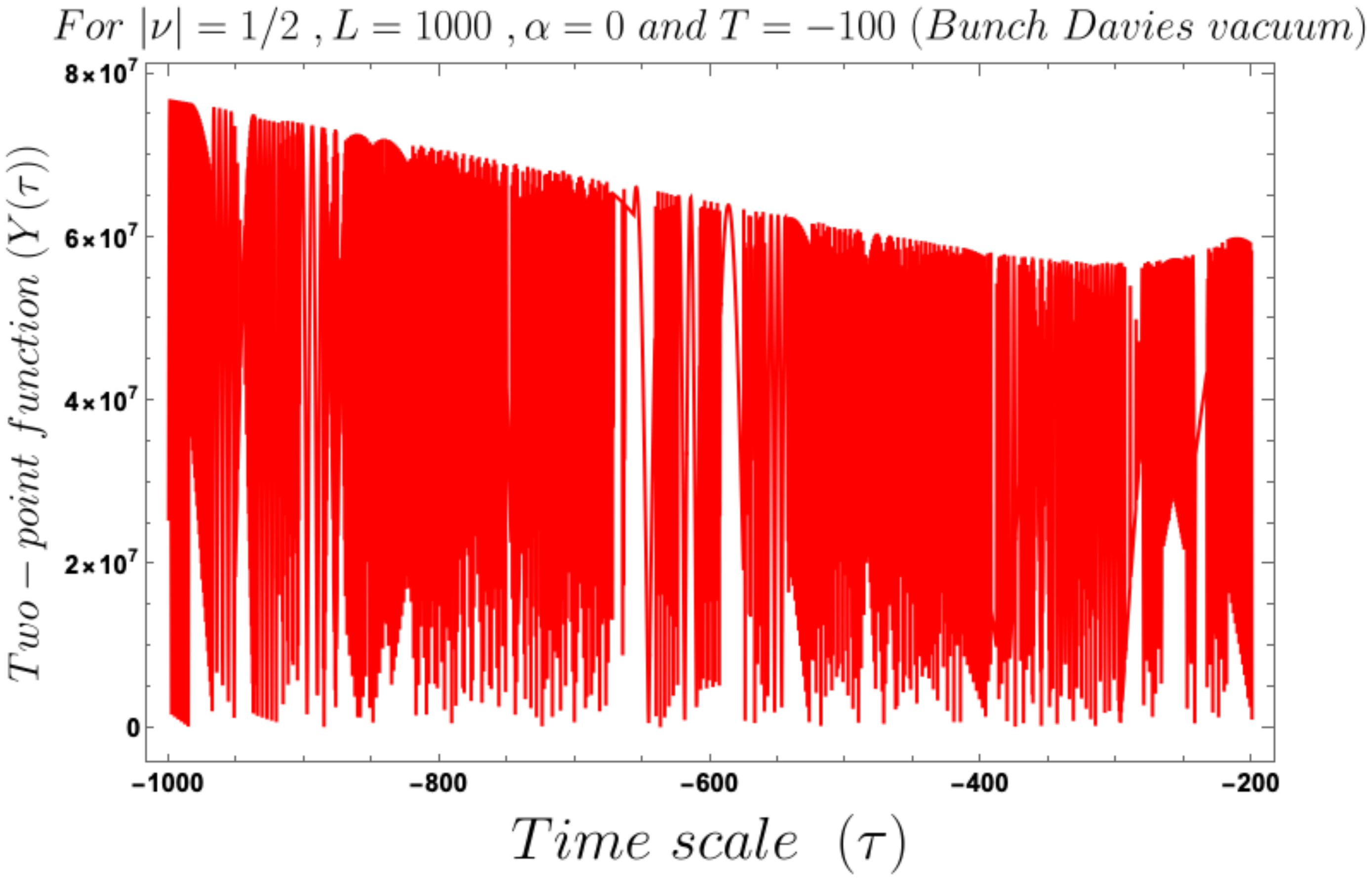}
        \includegraphics[width=16cm,height=4.8cm]{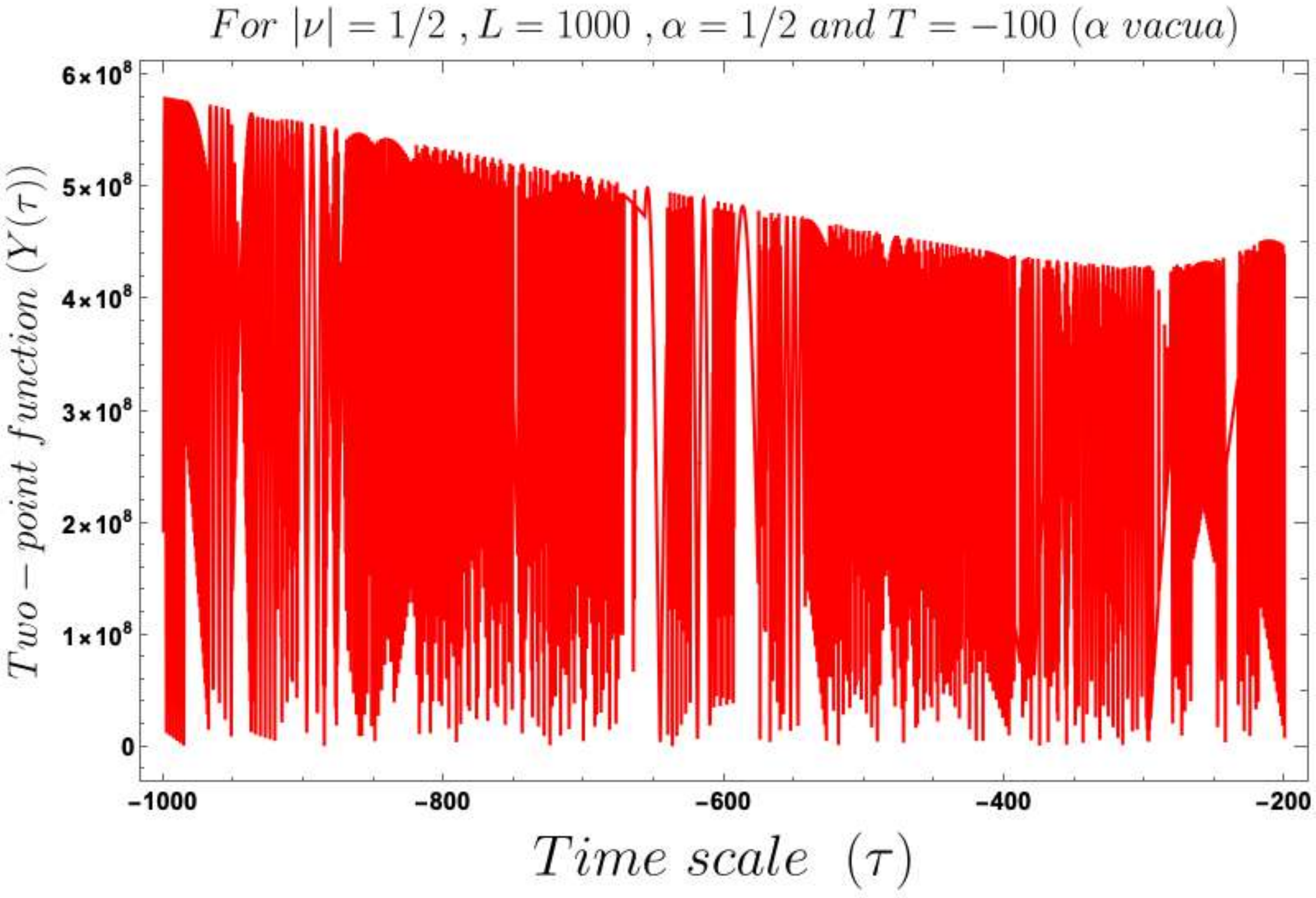}
   \includegraphics[width=17.3cm,height=4.8cm]{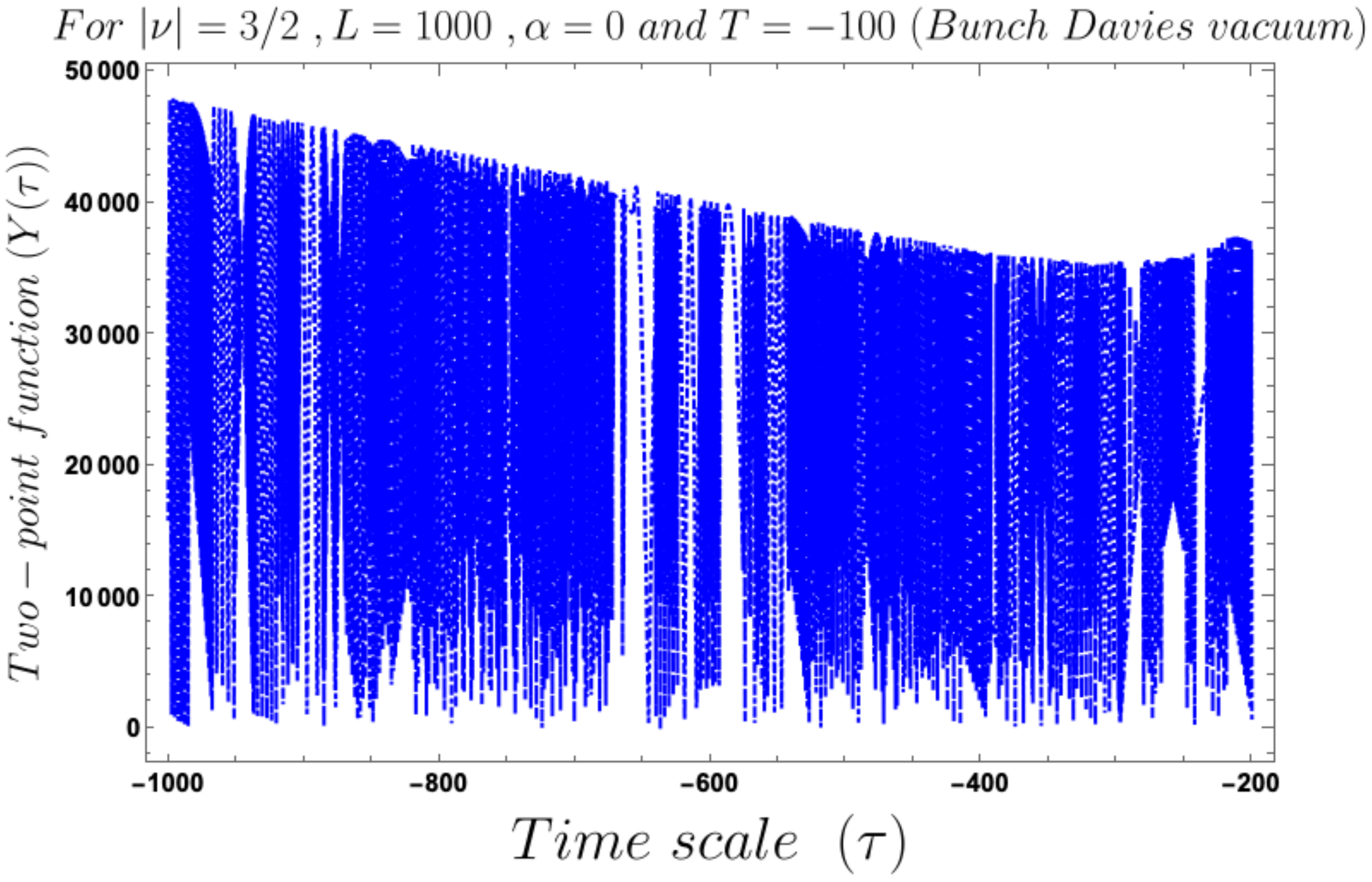}
        \includegraphics[width=16cm,height=4.8cm]{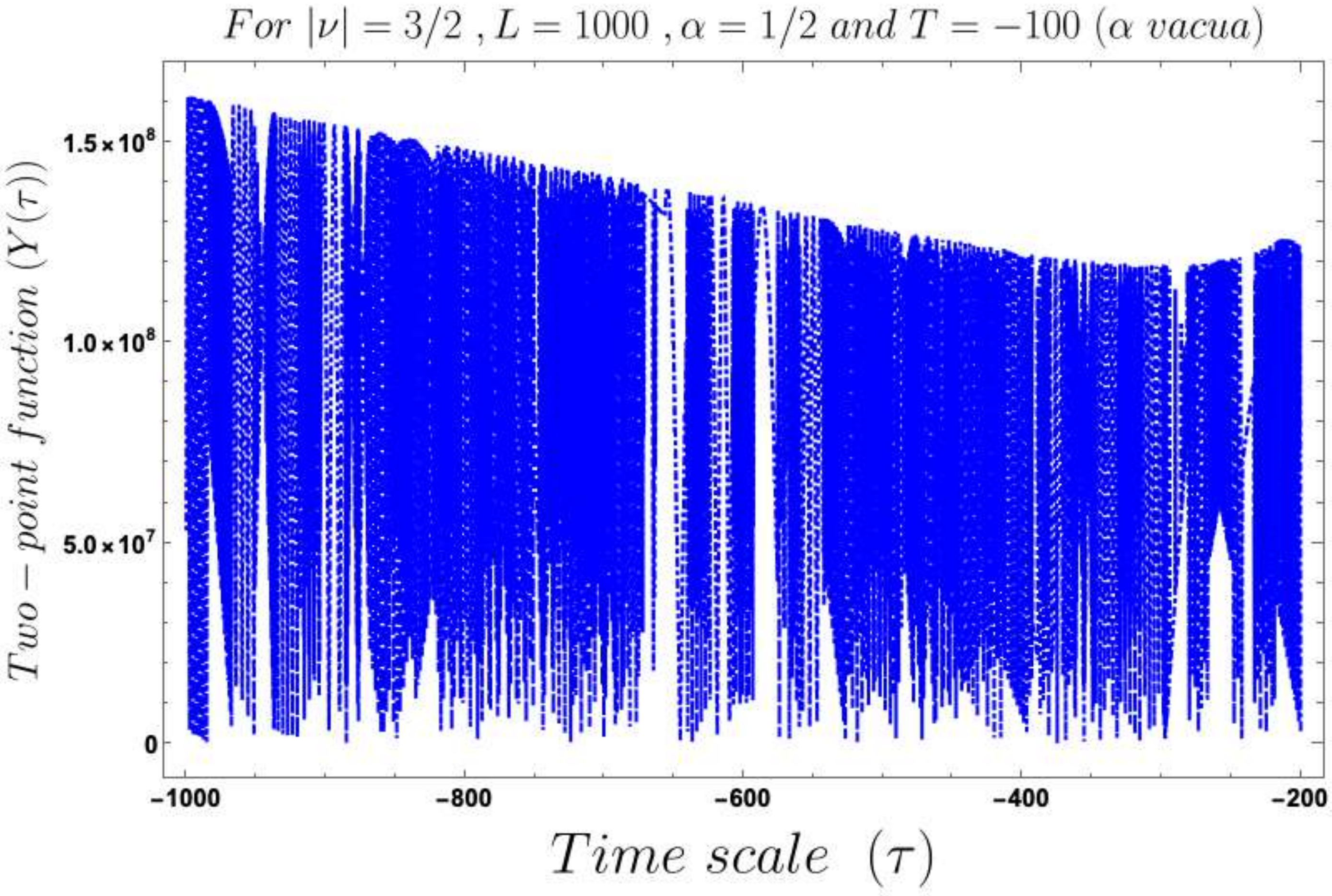}
    \caption{Behaviour of the two-point function with respect to the $\tau$ time scale of the theory. Here we fix, mass parameter $\nu=-i/2,-3i/2$, cut-off scale $L=1000$, vacuum parameter $\alpha=0 ~({\rm Bunch~Davies~vacuum)}, 1/2~(\alpha~{\rm vacua})$.}
      \label{fig:12AB}
\end{figure}
\begin{figure}[t!]
    \centering
        \centering
        \includegraphics[width=17.3cm,height=4.8cm]{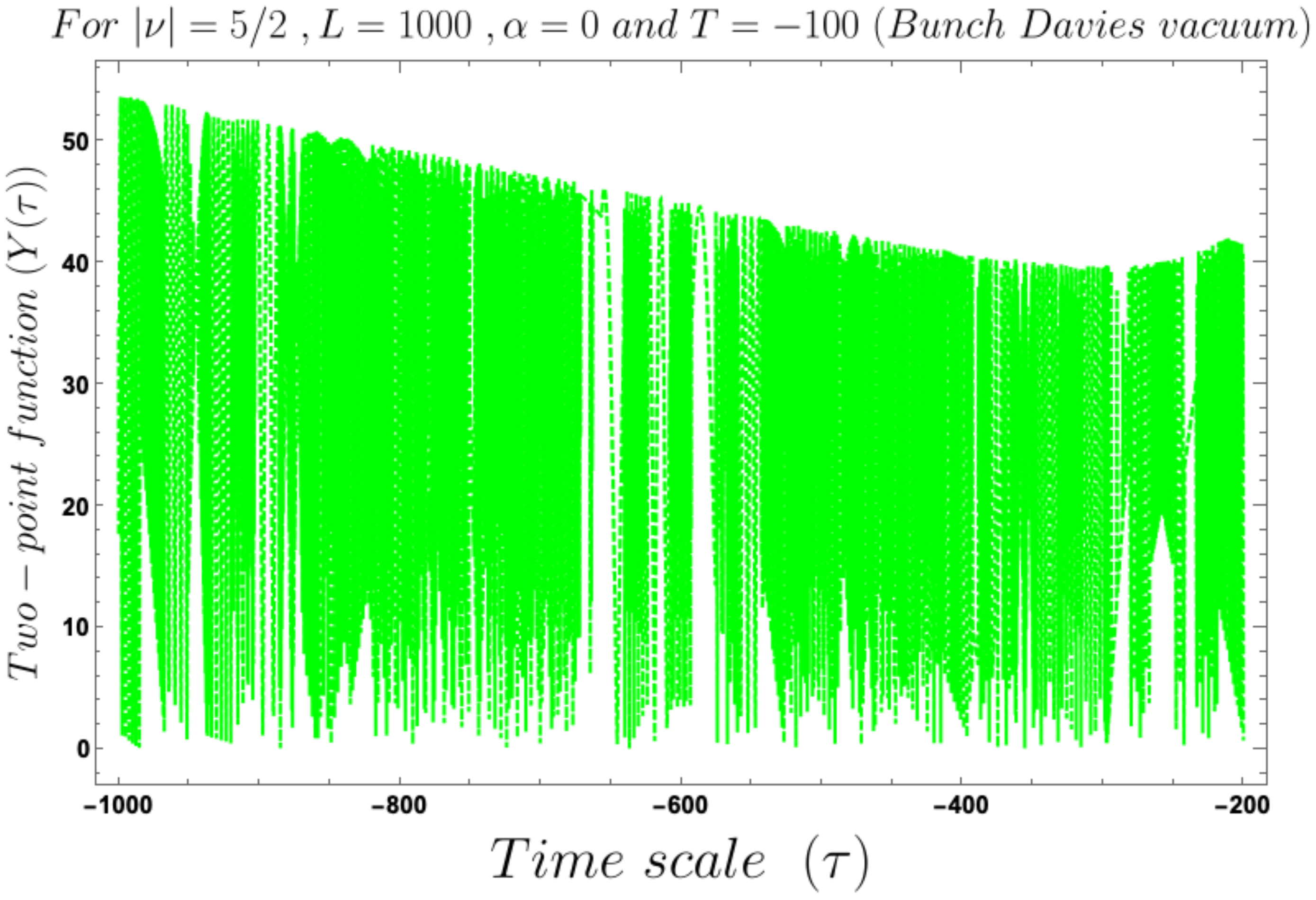}
        \includegraphics[width=16cm,height=4.8cm]{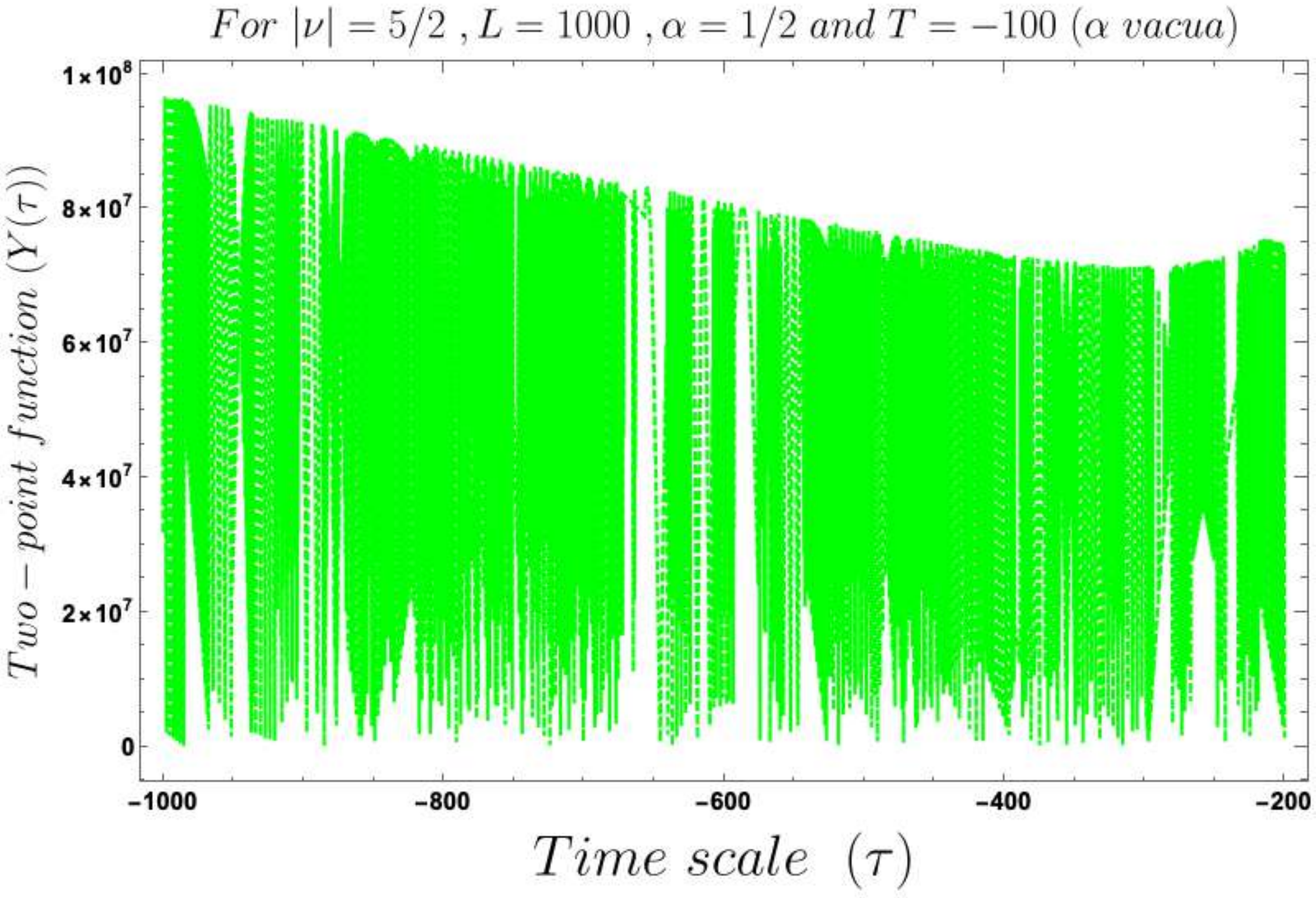}
   \includegraphics[width=17.3cm,height=4.8cm]{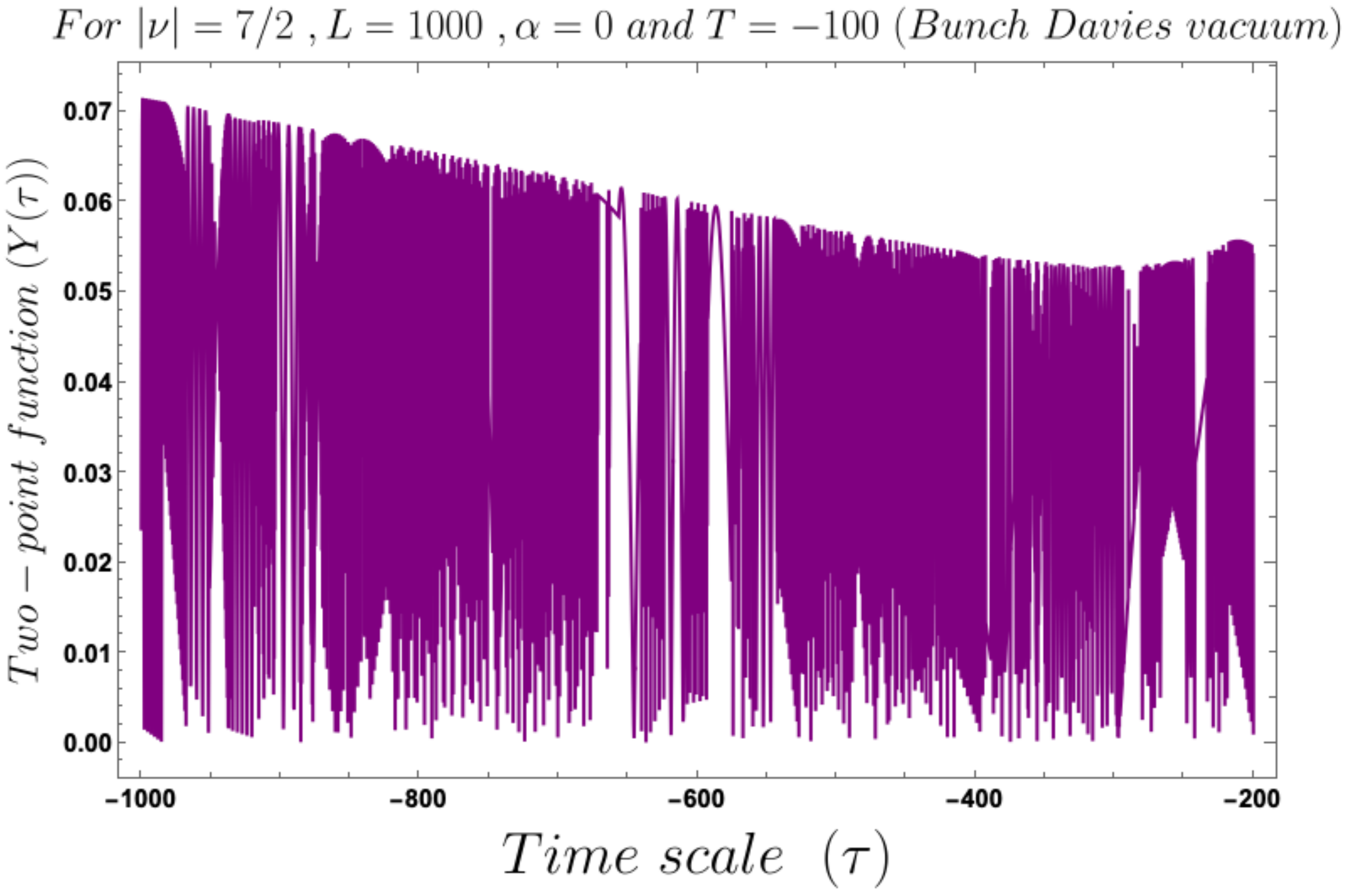}
        \includegraphics[width=16cm,height=4.8cm]{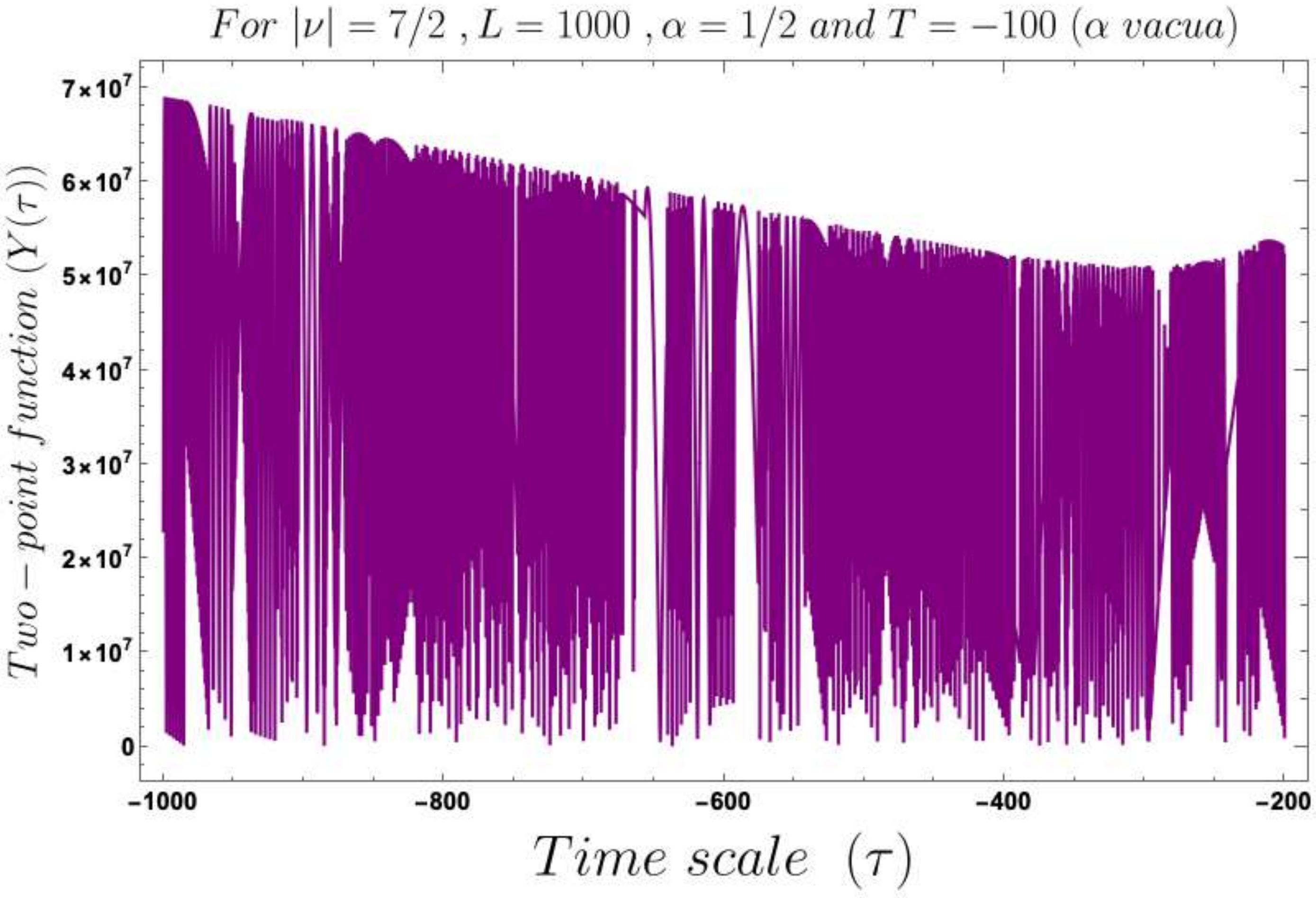}
    \caption{Behaviour of the two-point function with respect to the $\tau$ time scale of the theory. Here we fix, mass parameter $\nu=-5i/2,-7i/2$, cut-off scale $L=1000$, vacuum parameter $\alpha=0 ~({\rm Bunch~Davies~vacuum)}, 1/2~(\alpha~{\rm vacua})$.}
      \label{fig:13AB}
\end{figure}
\begin{figure}[t!]
    \centering
        \centering
        \includegraphics[width=17.3cm,height=4.8cm]{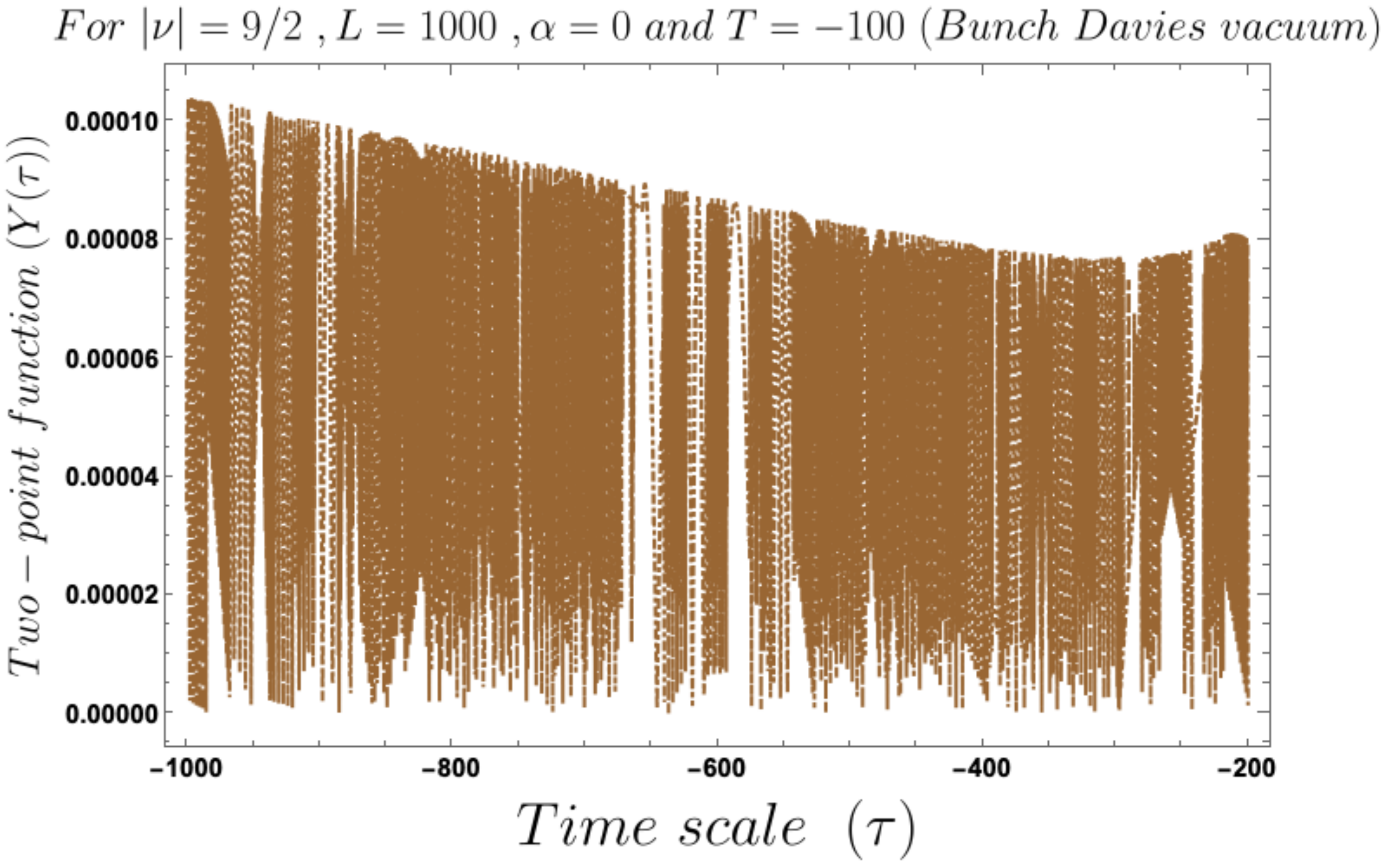}
        \includegraphics[width=16cm,height=4.8cm]{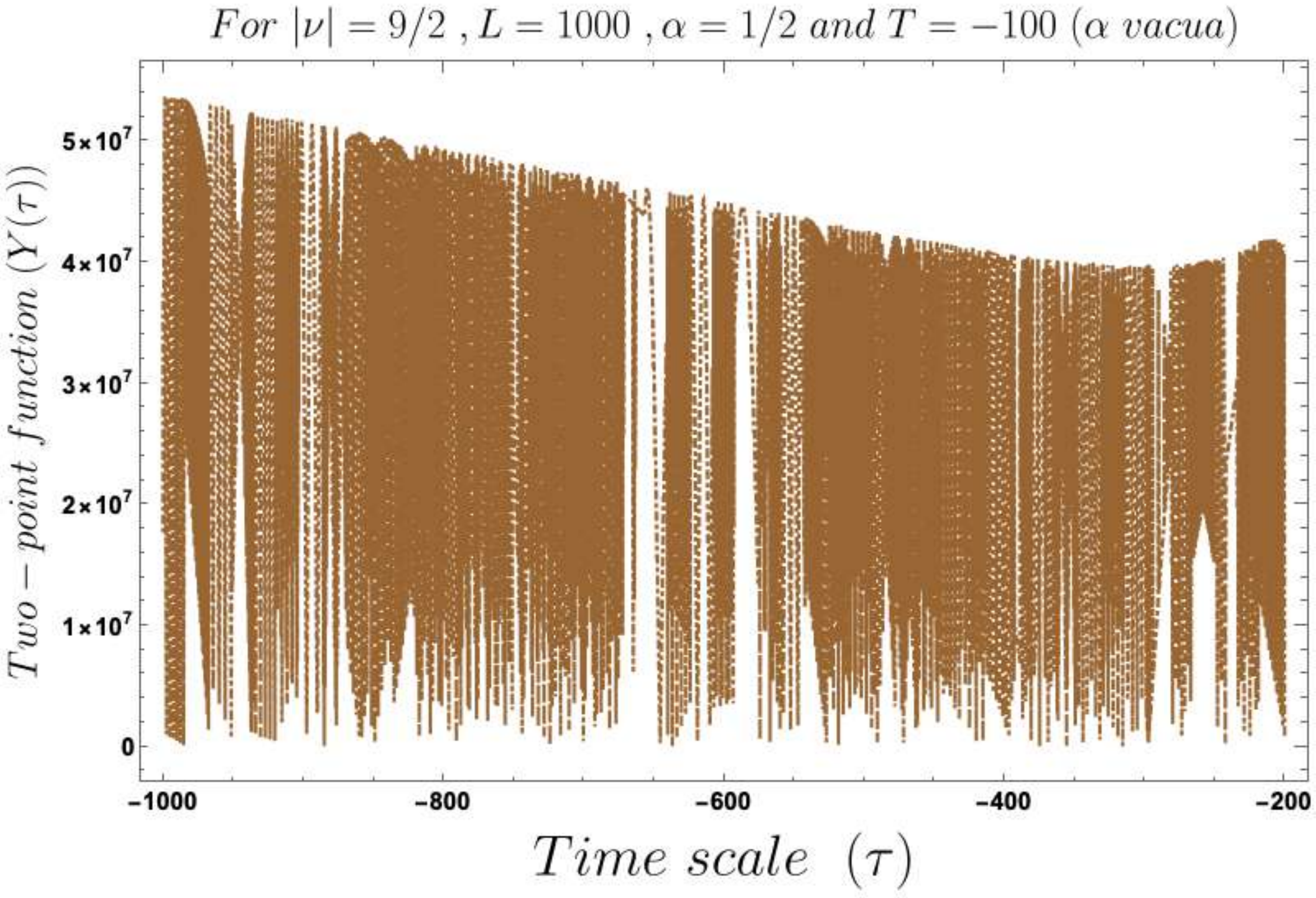}
    \caption{Behaviour of the two-point function with respect to the $\tau$ time scale of the theory. Here we fix, mass parameter $\nu=-9i/2$, cut-off scale $L=1000$, vacuum parameter $\alpha=0 ~({\rm Bunch~Davies~vacuum)}, 1/2~(\alpha~{\rm vacua})$.}
      \label{fig:14ABC}
\end{figure}

\begin{figure}[t!]
    \centering
        \centering
        \includegraphics[width=17.3cm,height=4.8cm]{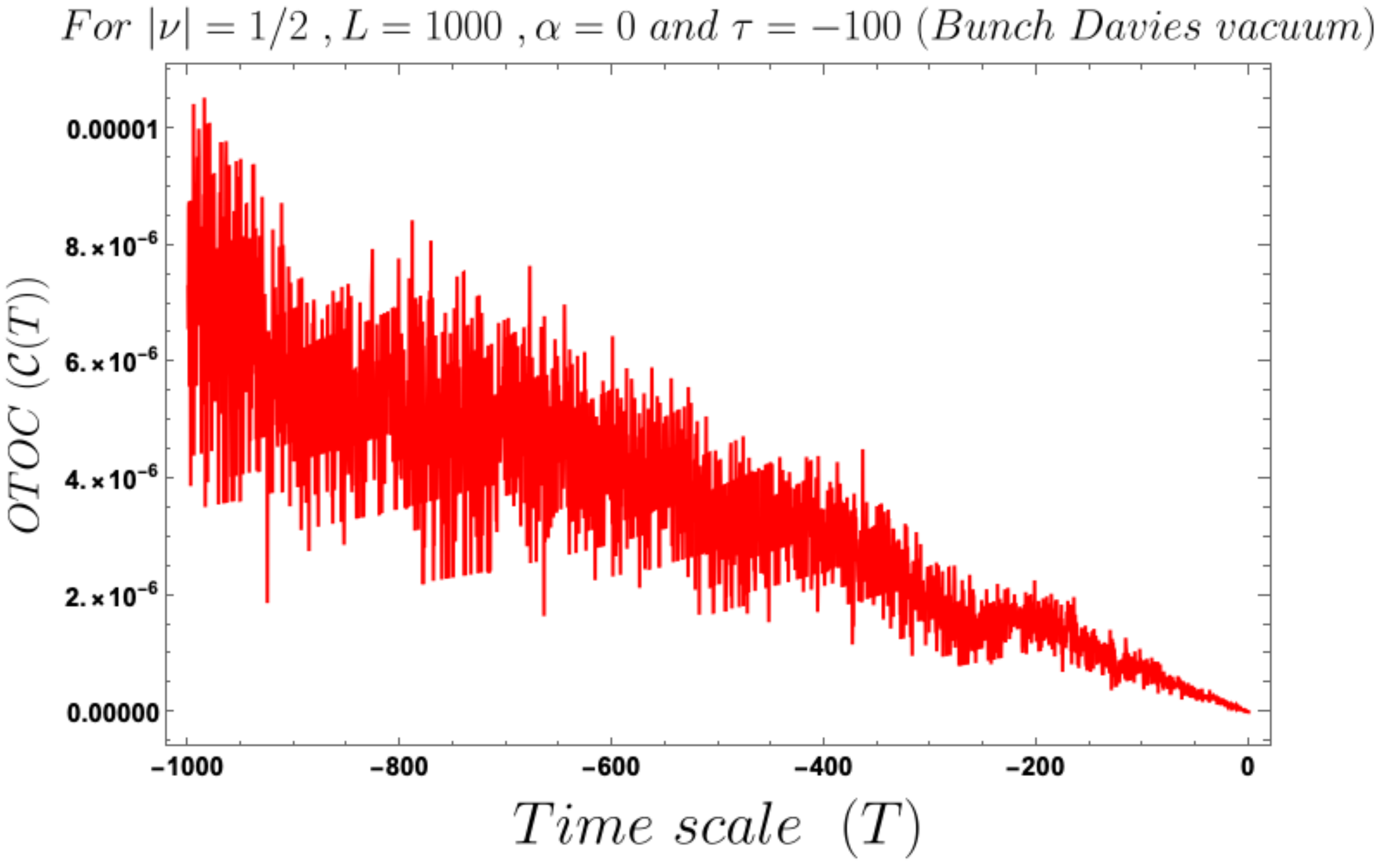}
        \includegraphics[width=16cm,height=4.8cm]{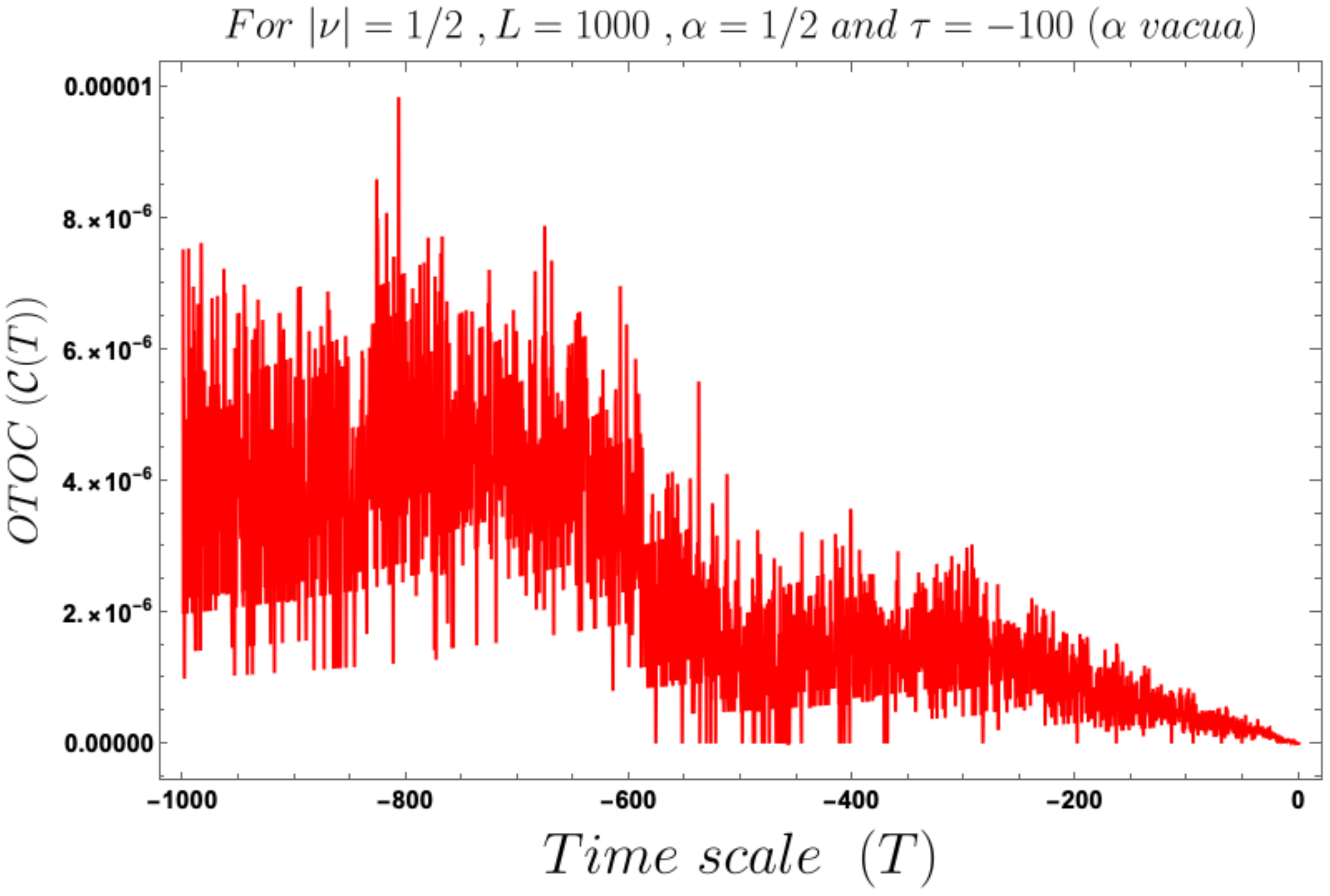}
   \includegraphics[width=17.3cm,height=4.8cm]{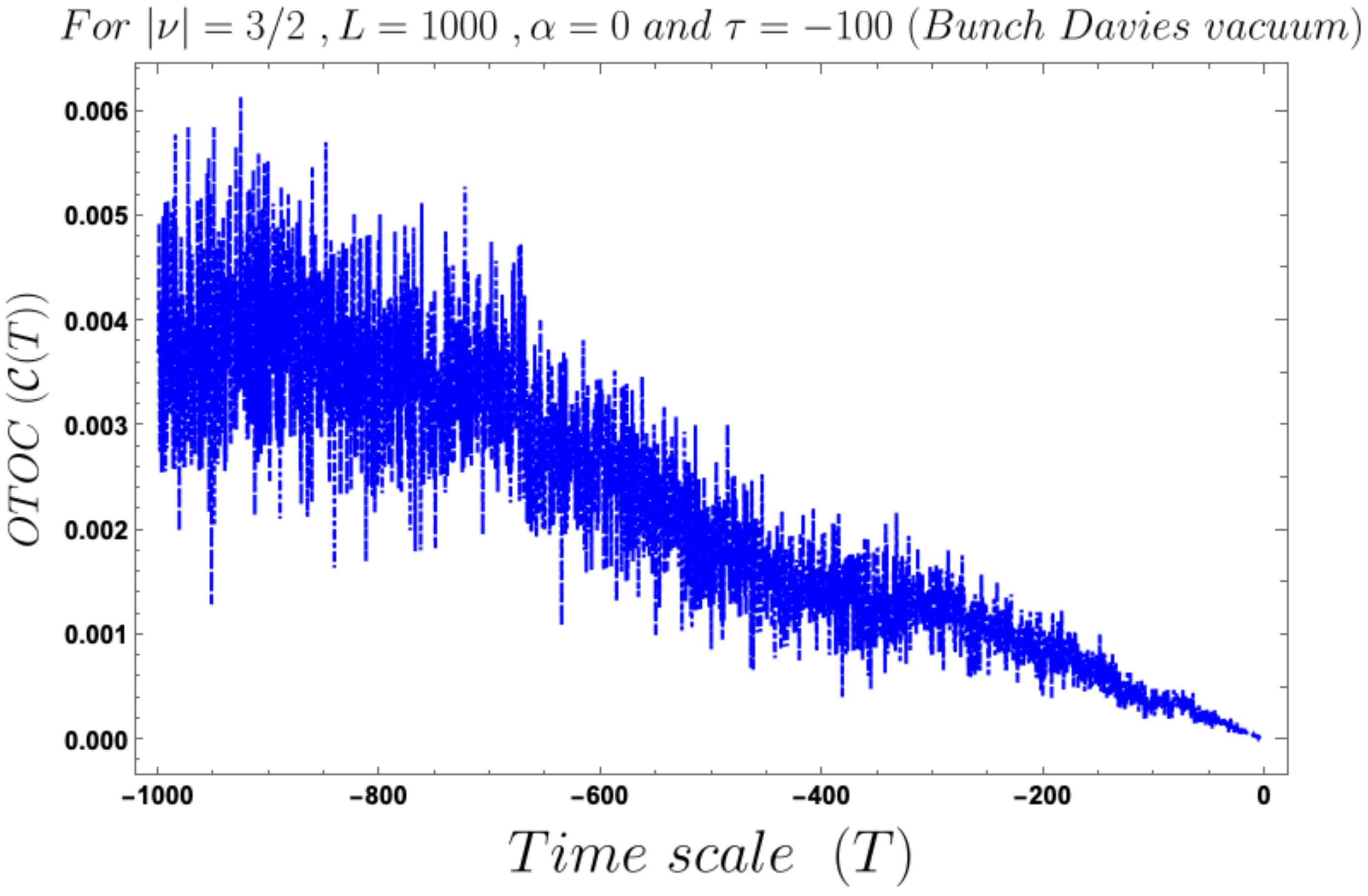}
        \includegraphics[width=16cm,height=4.8cm]{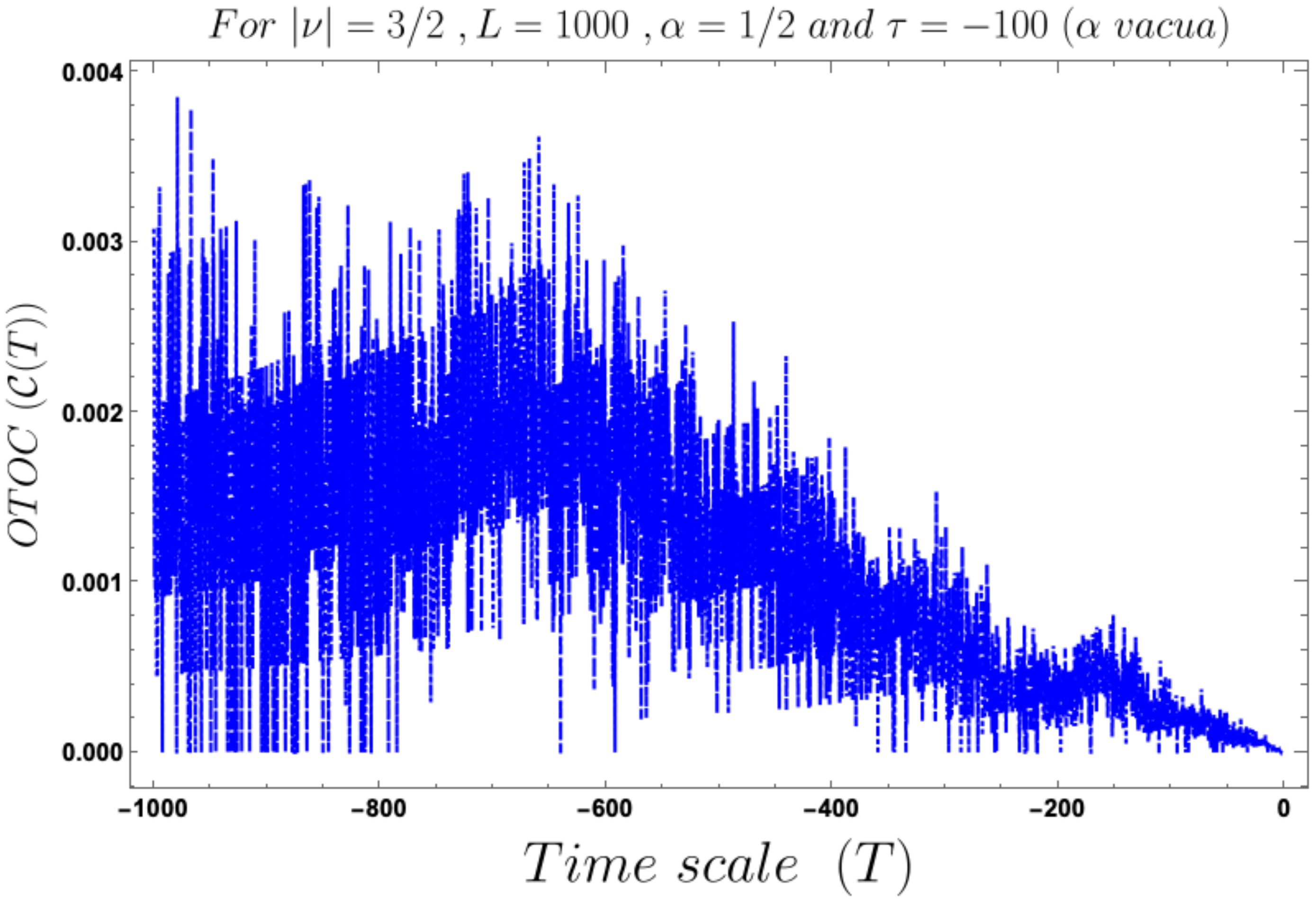}
    \caption{Behaviour of the four-point OTOC with respect to the $T$ time scale of the theory. Here we fix, mass parameter $\nu=-i/2,-3i/2$, cut-off scale $L=1000$, vacuum parameter $\alpha=0 ~({\rm Bunch~Davies~vacuum)}, 1/2~(\alpha~{\rm vacua})$.}
      \label{fig:1}
\end{figure}
 \begin{figure}[t!]
    \centering
        \centering
        \includegraphics[width=17.3cm,height=4.8cm]{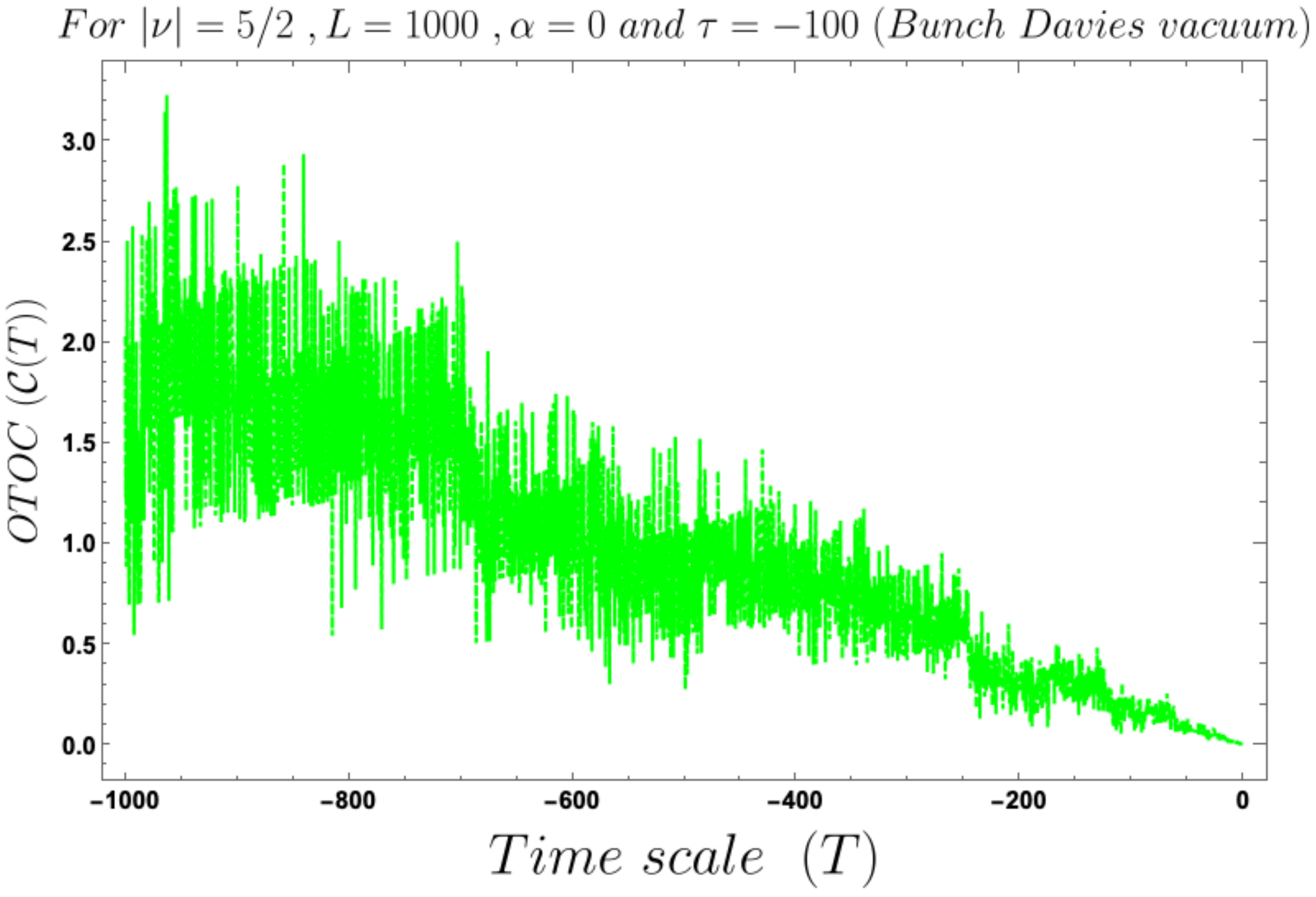}
        \includegraphics[width=16cm,height=4.8cm]{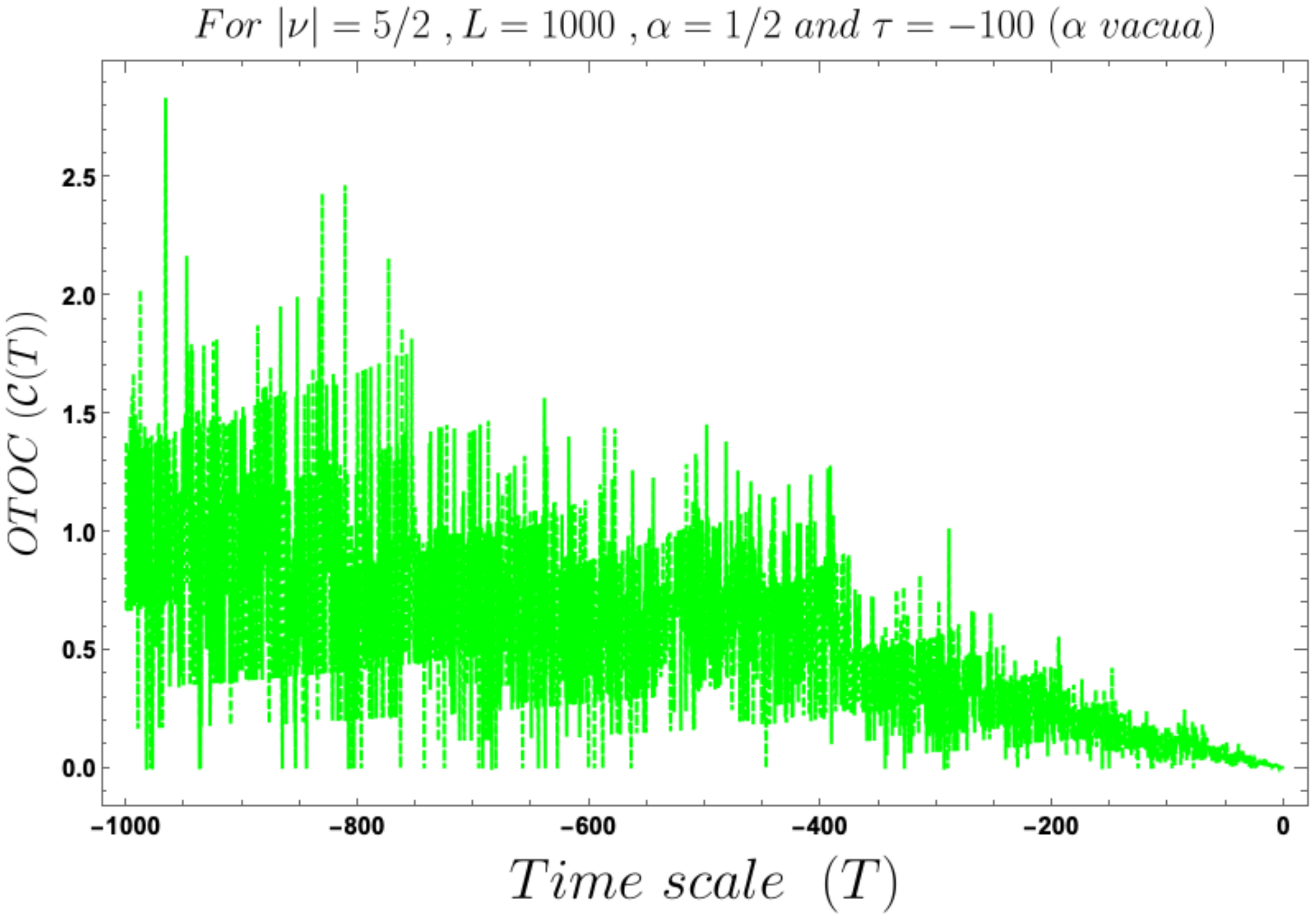}
    \includegraphics[width=17.3cm,height=4.8cm]{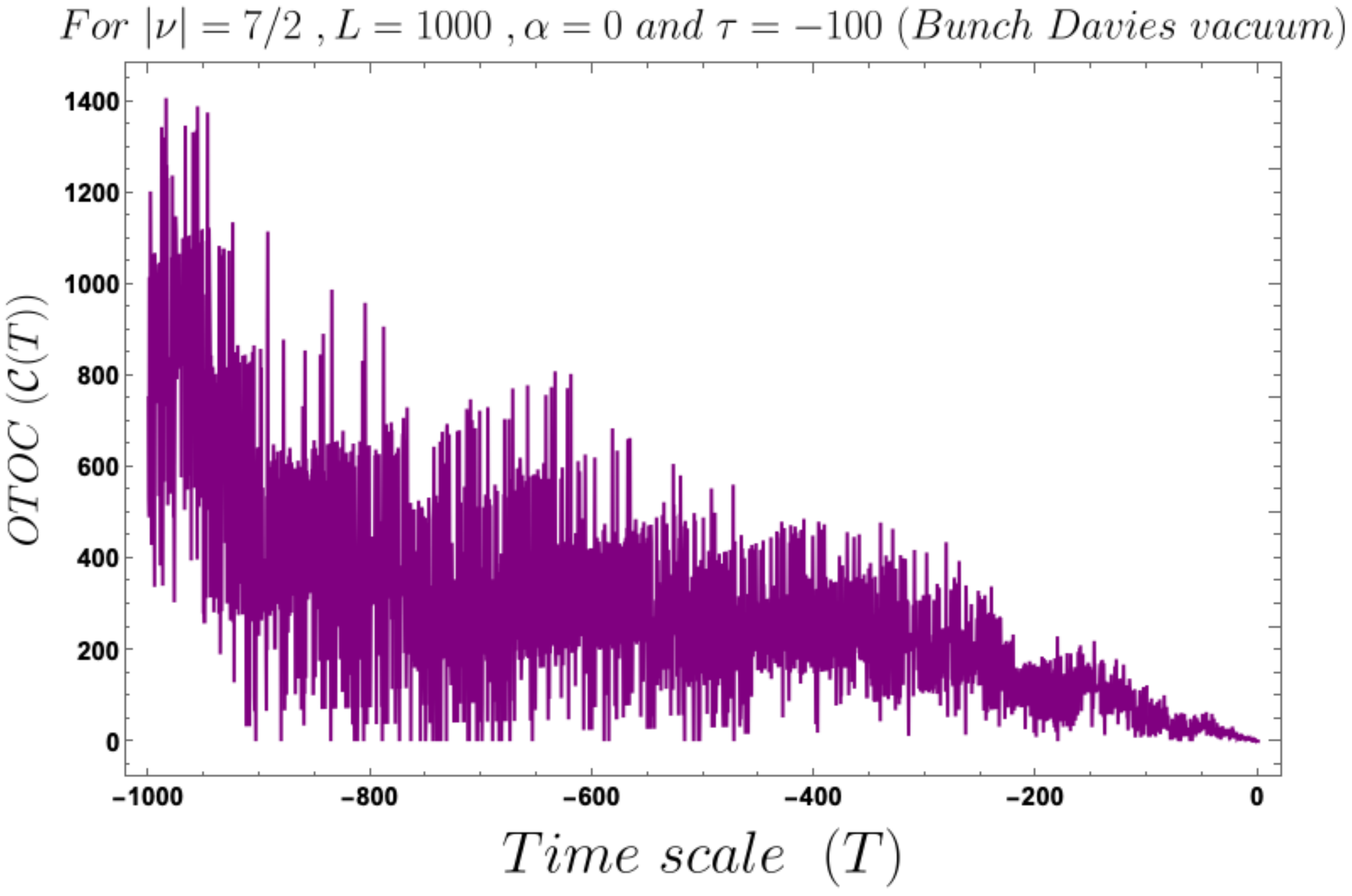}
        \includegraphics[width=16cm,height=4.8cm]{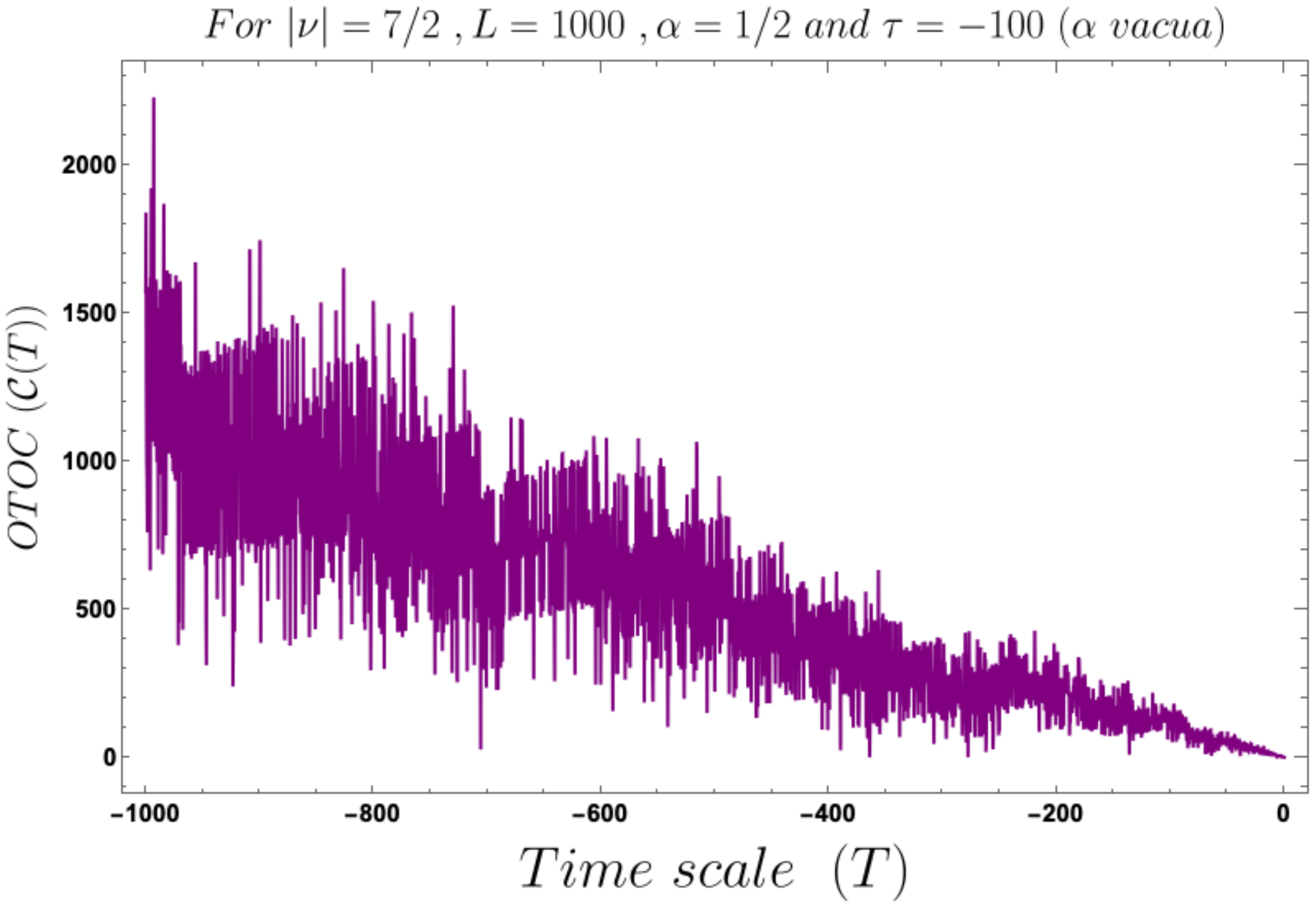}
    \caption{Behaviour of the four-point OTOC with respect to the $T$ time scale of the theory. Here we fix, mass parameter $\nu=-5i/2,-7i/2$, cut-off scale $L=1000$, vacuum parameter $\alpha=0 ~({\rm Bunch~Davies~vacuum)}, 1/2~(\alpha~{\rm vacua})$.}
      \label{fig:2}
\end{figure}
 \begin{figure}[t!]
    \centering
        \centering
        \includegraphics[width=17.3cm,height=5.5cm]{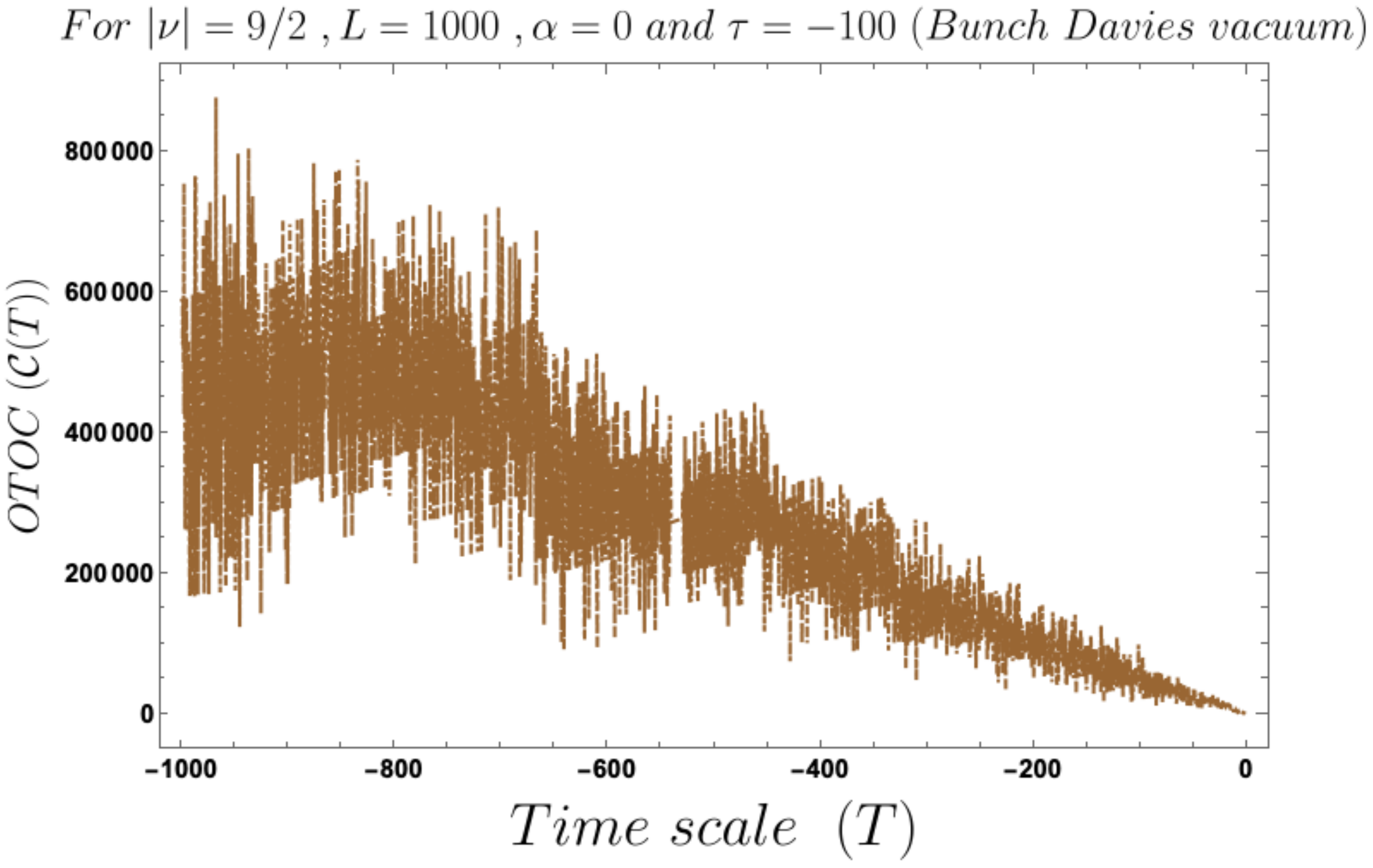}
        \includegraphics[width=16cm,height=5.5cm]{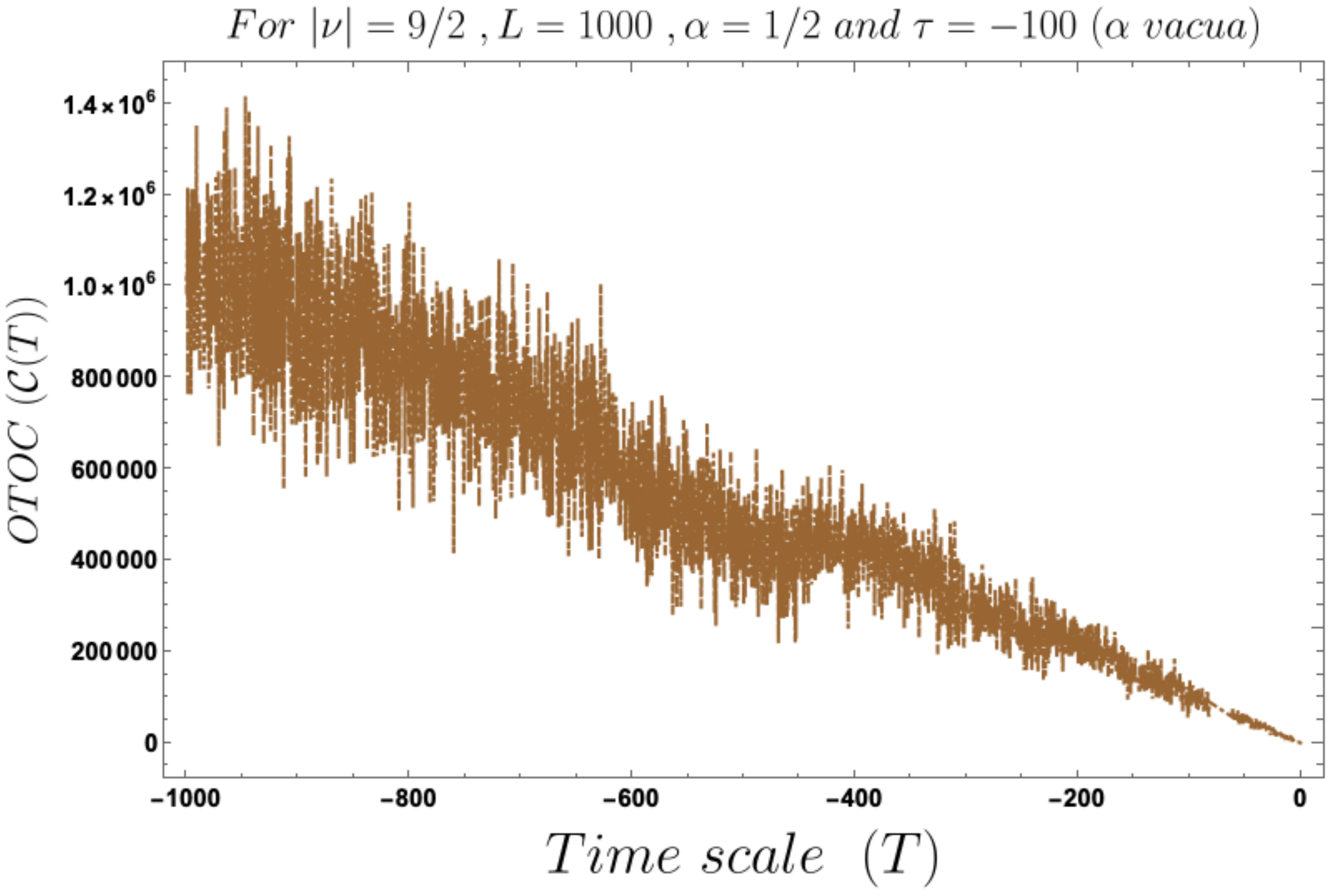}
    \caption{Behaviour of the four-point OTOC with respect to the $T$ time scale of the theory. Here we fix, mass parameter $\nu=-9i/2$, cut-off scale $L=1000$, vacuum parameter $\alpha=0 ~({\rm Bunch~Davies~vacuum)}, 1/2~(\alpha~{\rm vacua})$.}
      \label{fig:3}
\end{figure}
 \begin{figure}[t!]
    \centering
        \centering
        \includegraphics[width=16cm,height=6cm]{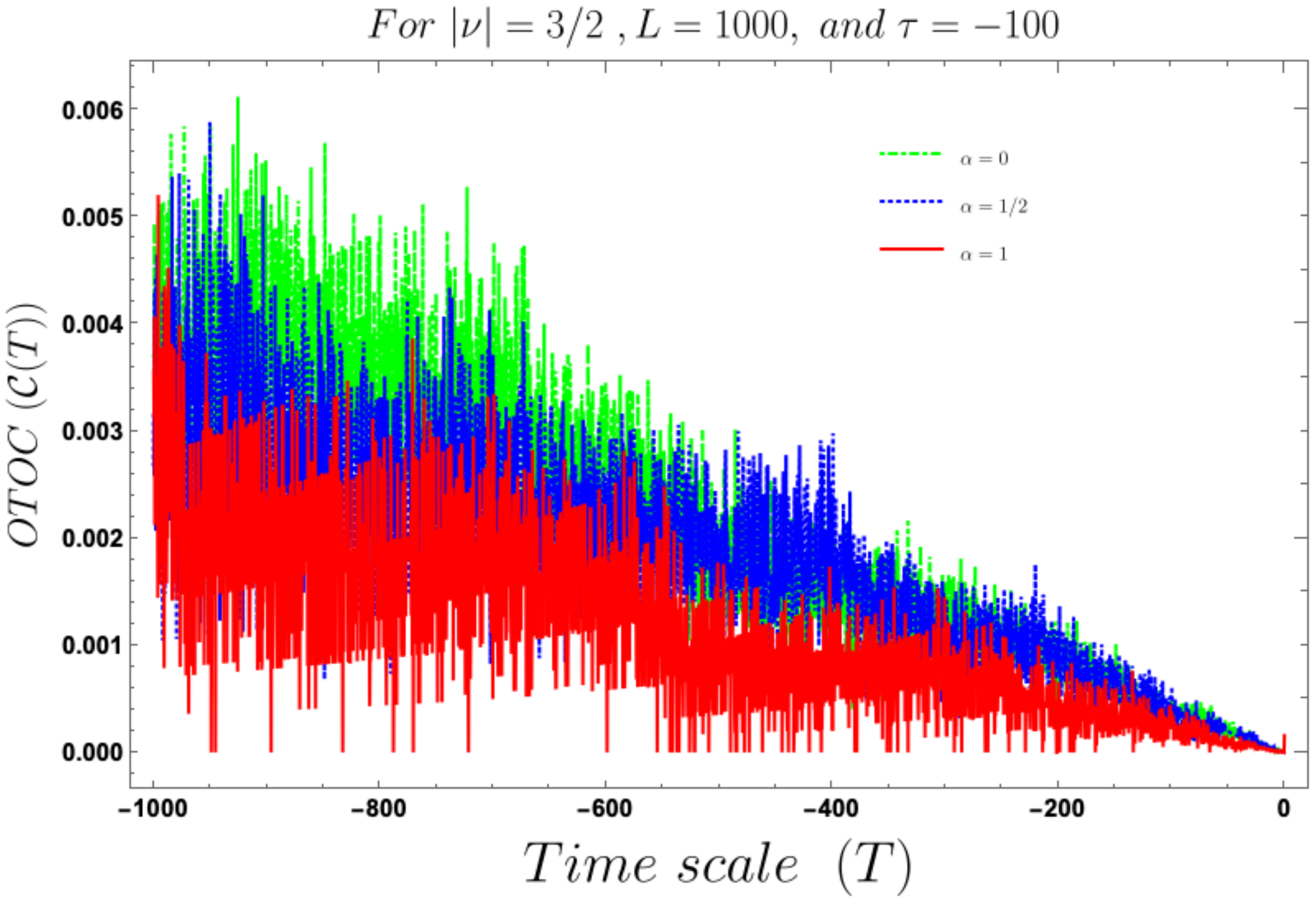}
    \caption{Behaviour of the four-point OTOC with respect to the $T$ time scale of the theory. Here we fix, mass parameter $\nu=-3i/2$, cut-off scale $L=1000$, vacuum parameter $\alpha=0 ~({\rm Bunch~Davies~vacuum)}, 1/2~(\alpha~{\rm vacua}), 1~(\alpha~{\rm vacua})$.  }
      \label{fig:4}
\end{figure}
\begin{figure}[t!]
    \centering
        \centering
        \includegraphics[width=16cm,height=9.7cm]{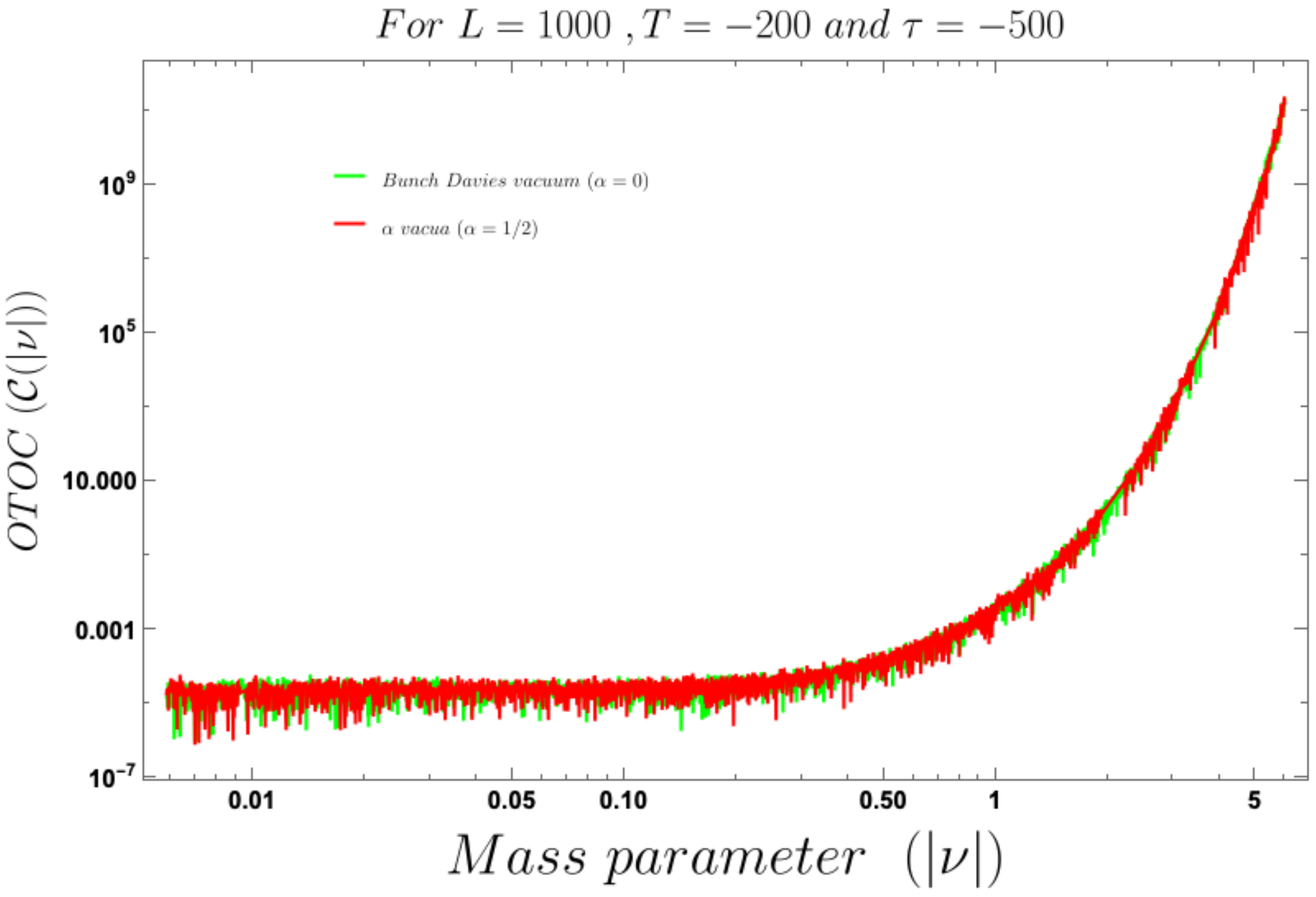}
          \includegraphics[width=16cm,height=9.7cm]{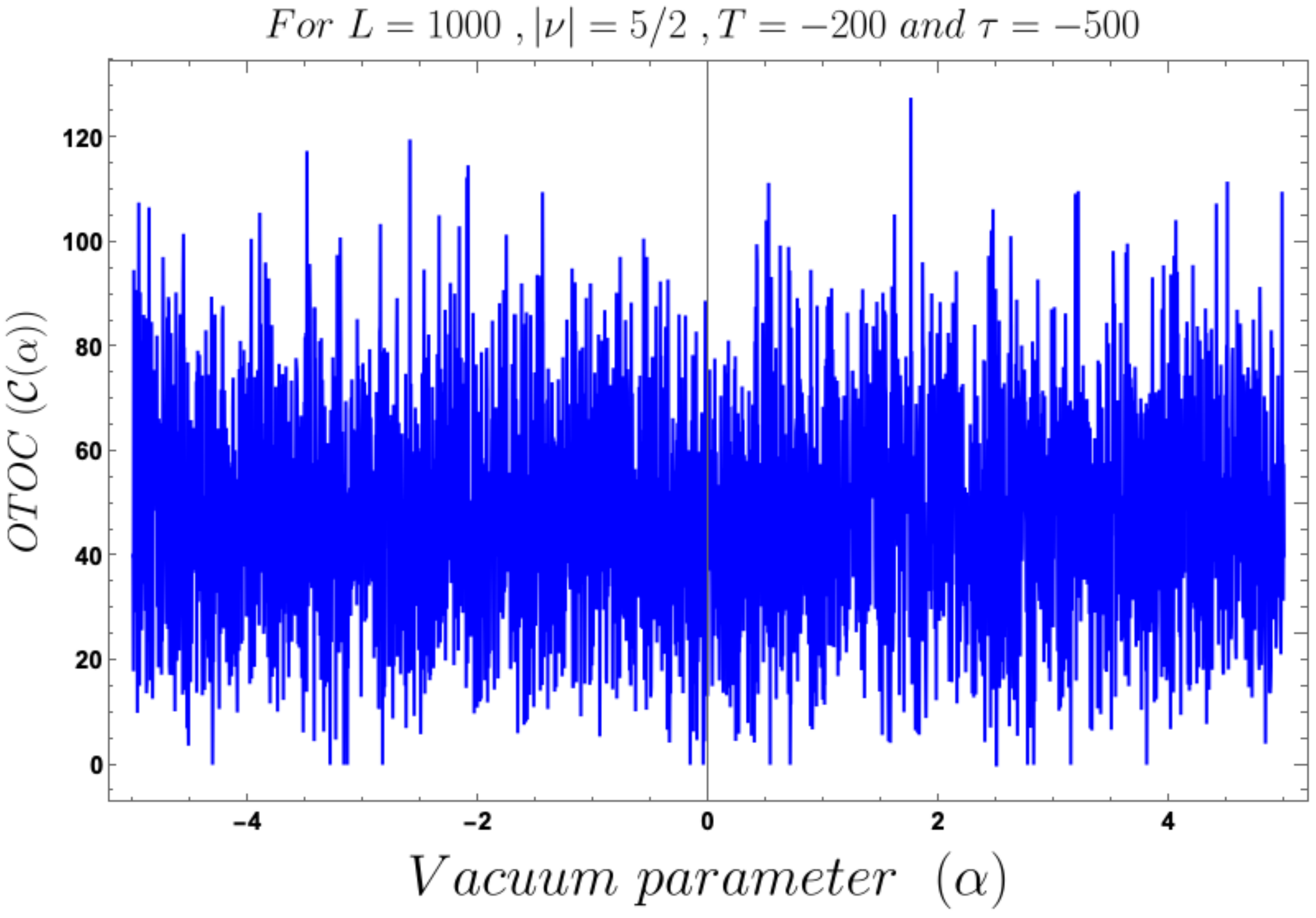}
    \caption{Behaviour of the four-point OTOC with respect to the mass parameter $\nu$ and the vacuum parameter $\alpha$ . Here we fix, time scale $T=-200$, $\tau=-500$, cut-off scale $L=1000$.}
      \label{fig:5}
\end{figure}
\begin{figure}[t!]
    \centering
        \centering
        \includegraphics[width=17.3cm,height=4.8cm]{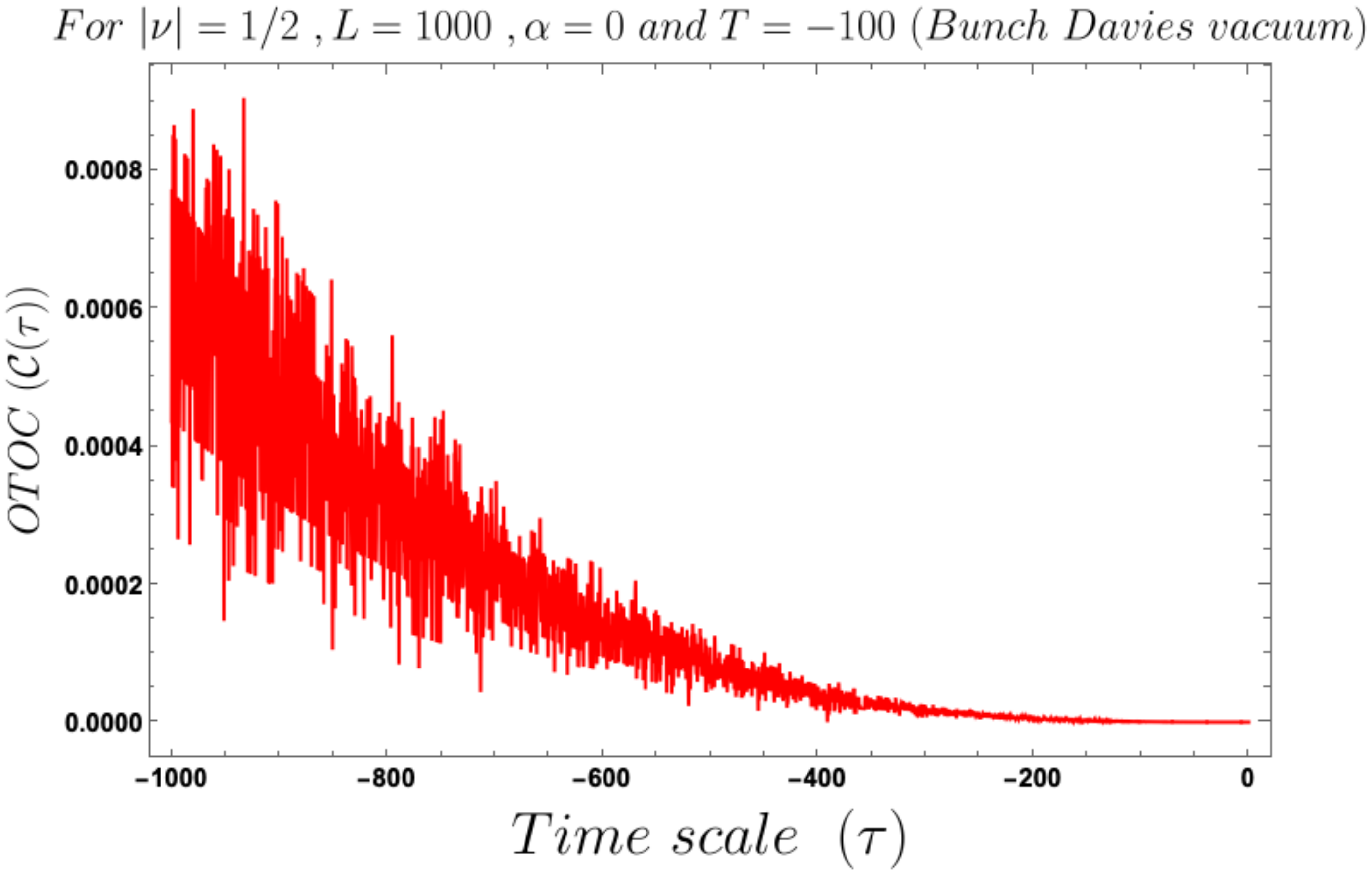}
        \includegraphics[width=16cm,height=4.8cm]{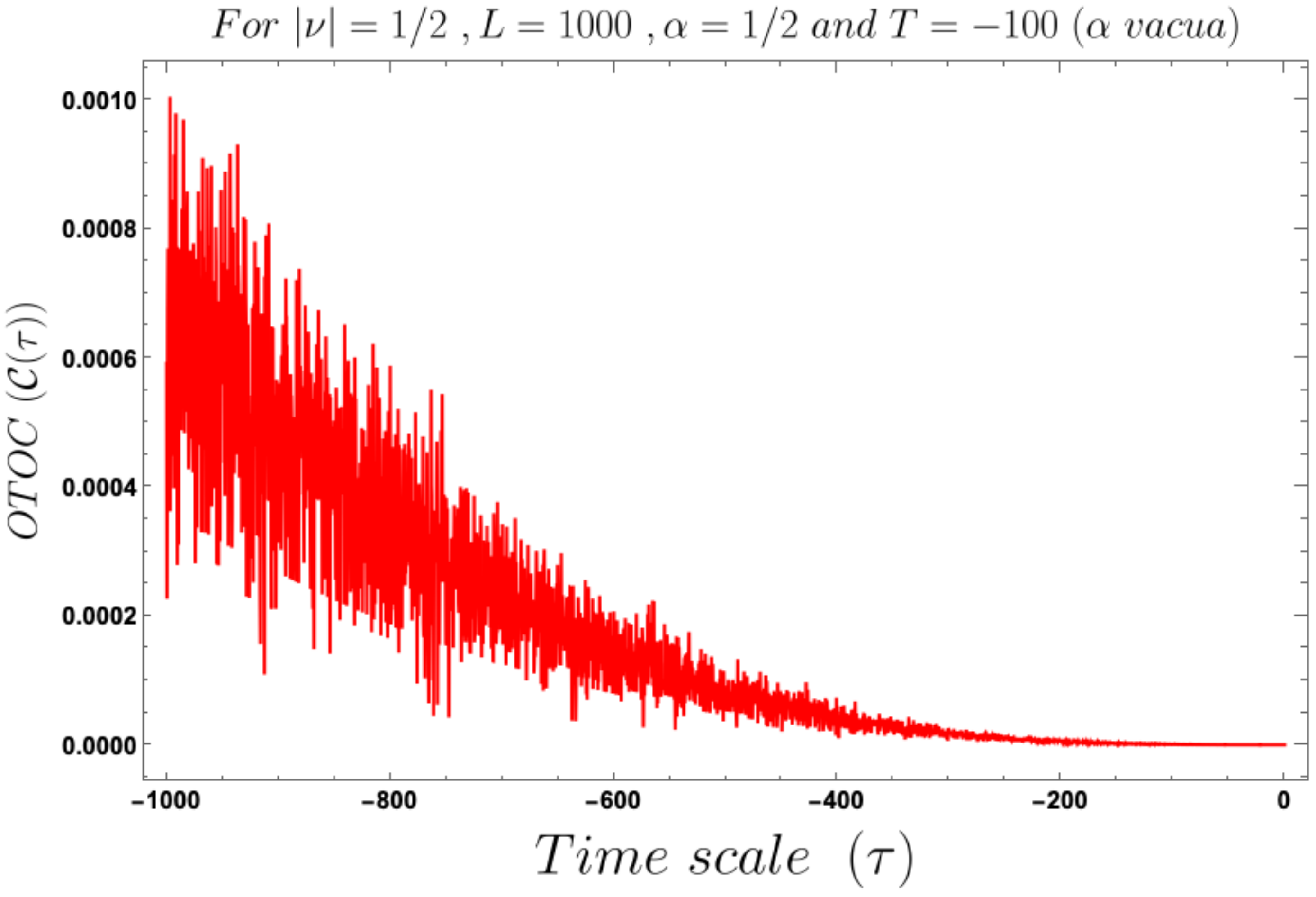}
   \includegraphics[width=17.3cm,height=4.8cm]{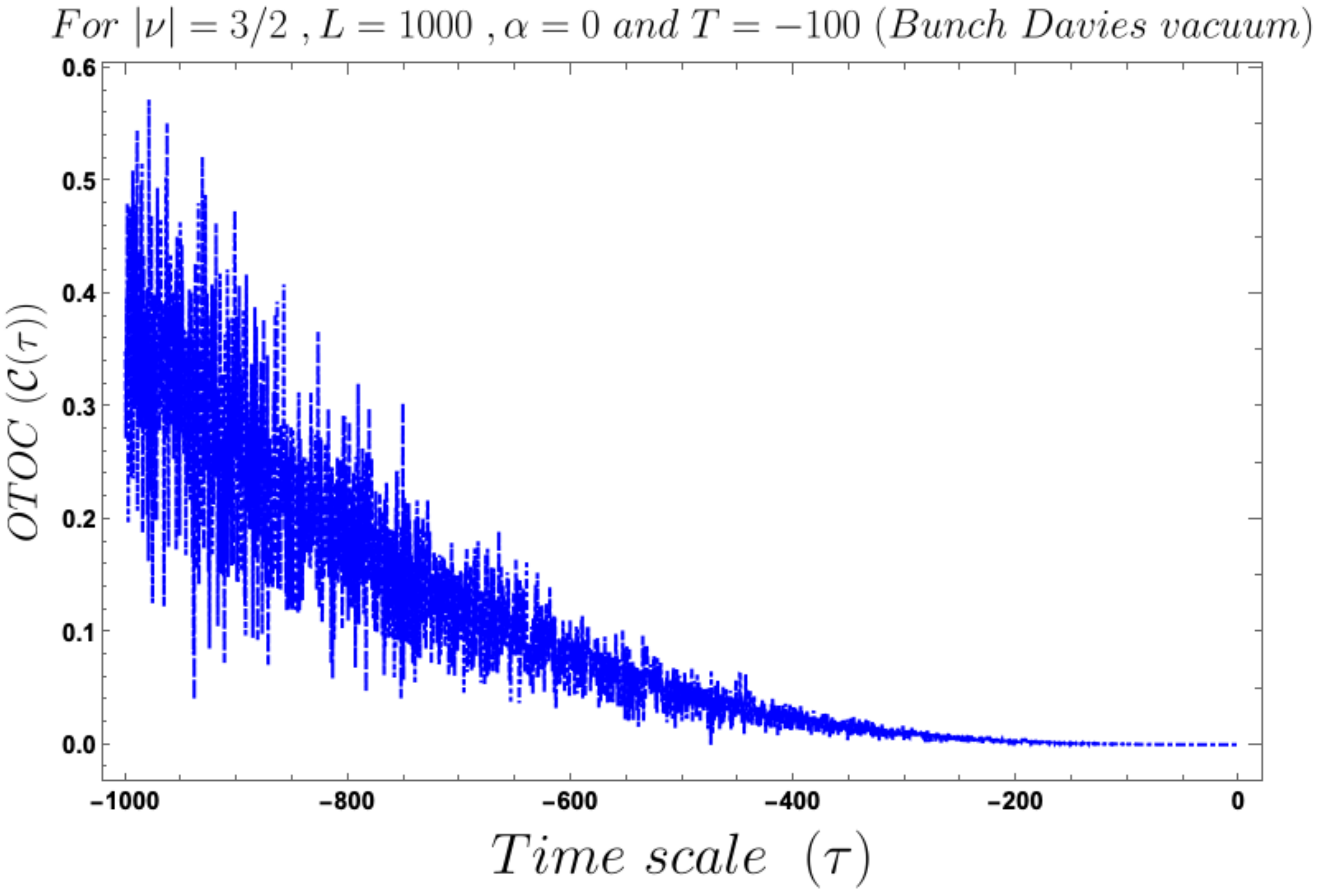}
        \includegraphics[width=16cm,height=4.8cm]{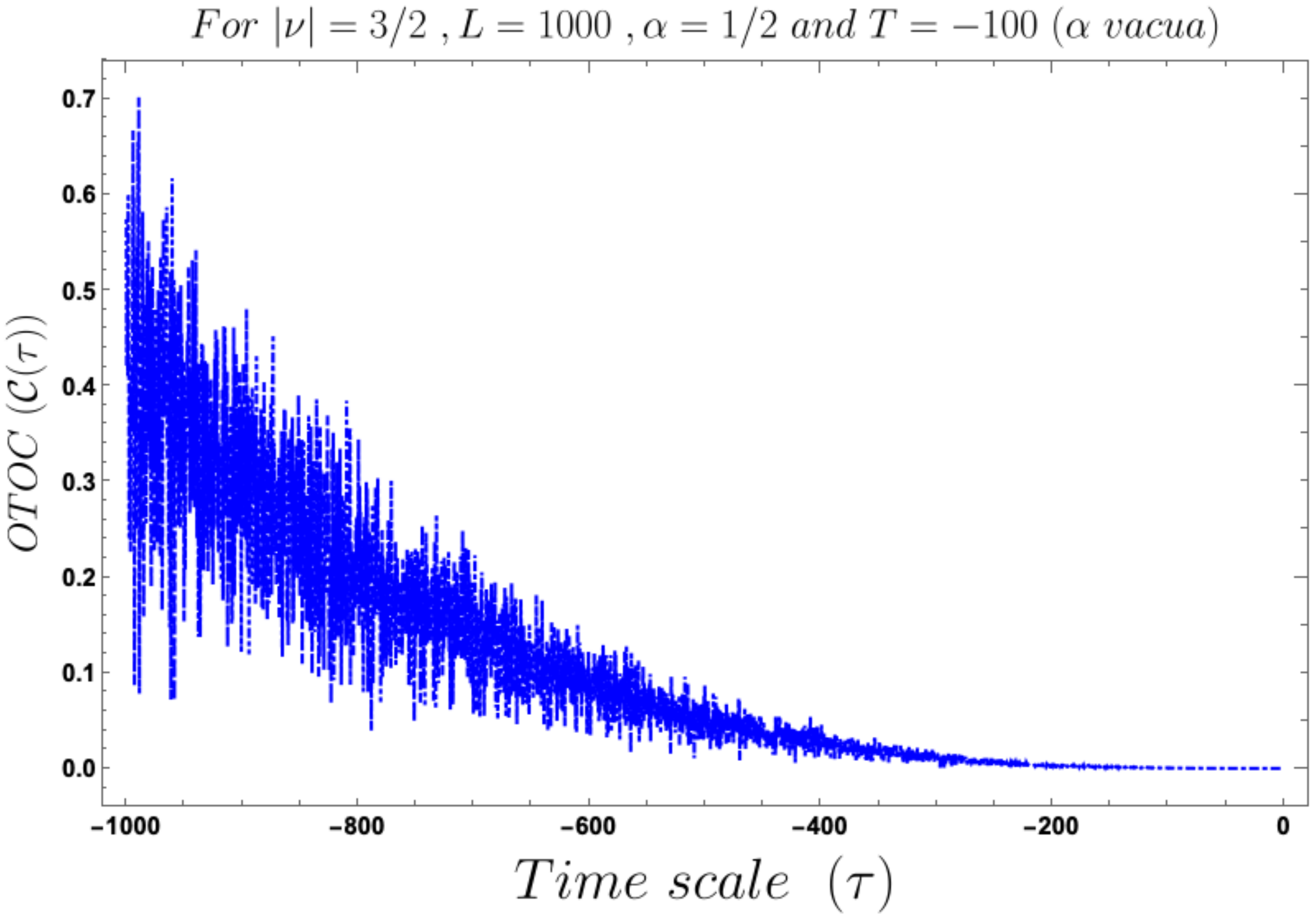}
    \caption{Behaviour of the four-point OTOC with respect to the $\tau$ time scale of the theory. Here we fix, mass parameter $\nu=-i/2,-3i/2$, cut-off scale $L=1000$, vacuum parameter $\alpha=0 ~({\rm Bunch~Davies~vacuum)}, 1/2~(\alpha~{\rm vacua})$.}
      \label{fig:6}
\end{figure}
 \begin{figure}[t!]
    \centering
        \centering
        \includegraphics[width=17.3cm,height=4.8cm]{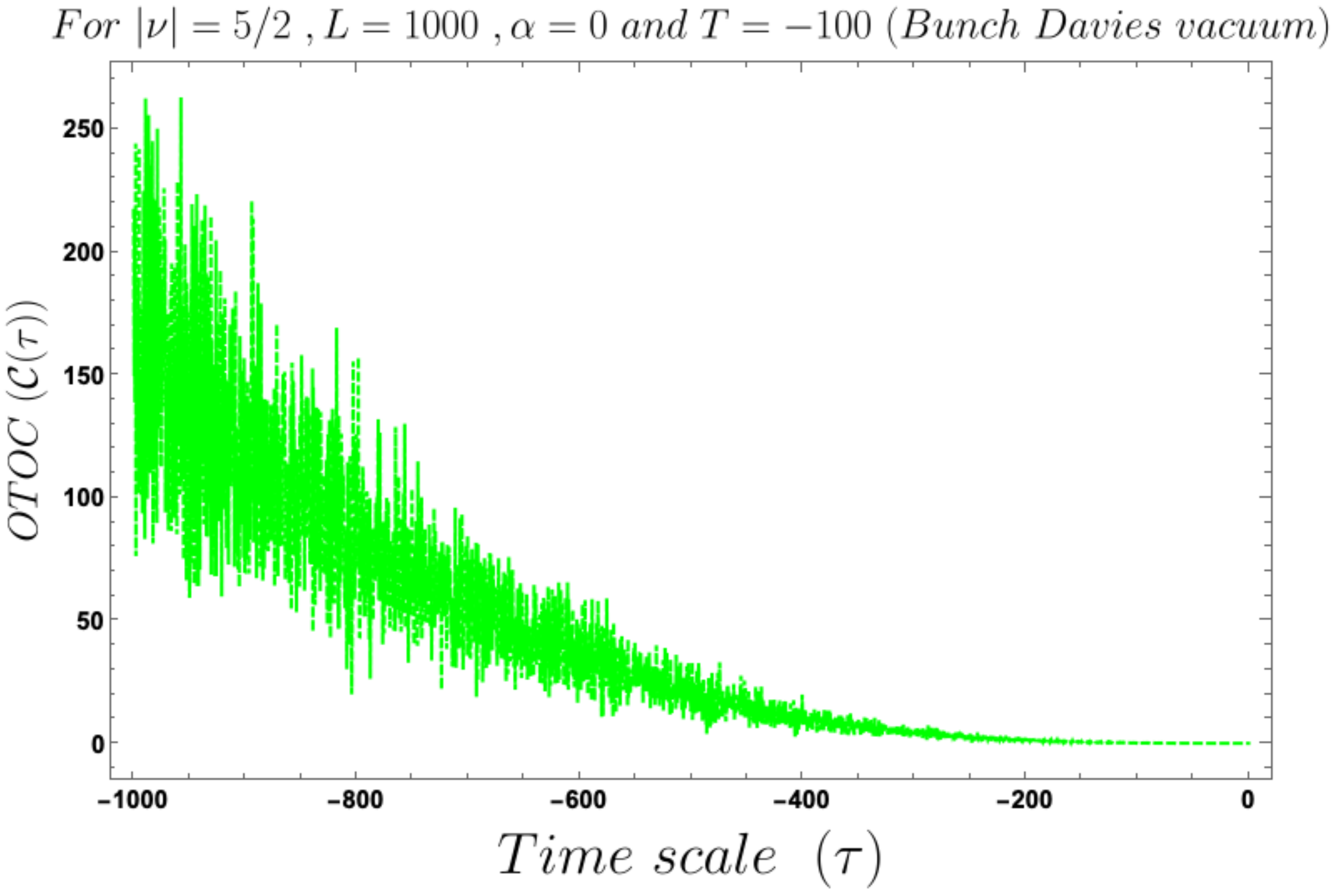}
        \includegraphics[width=16cm,height=4.8cm]{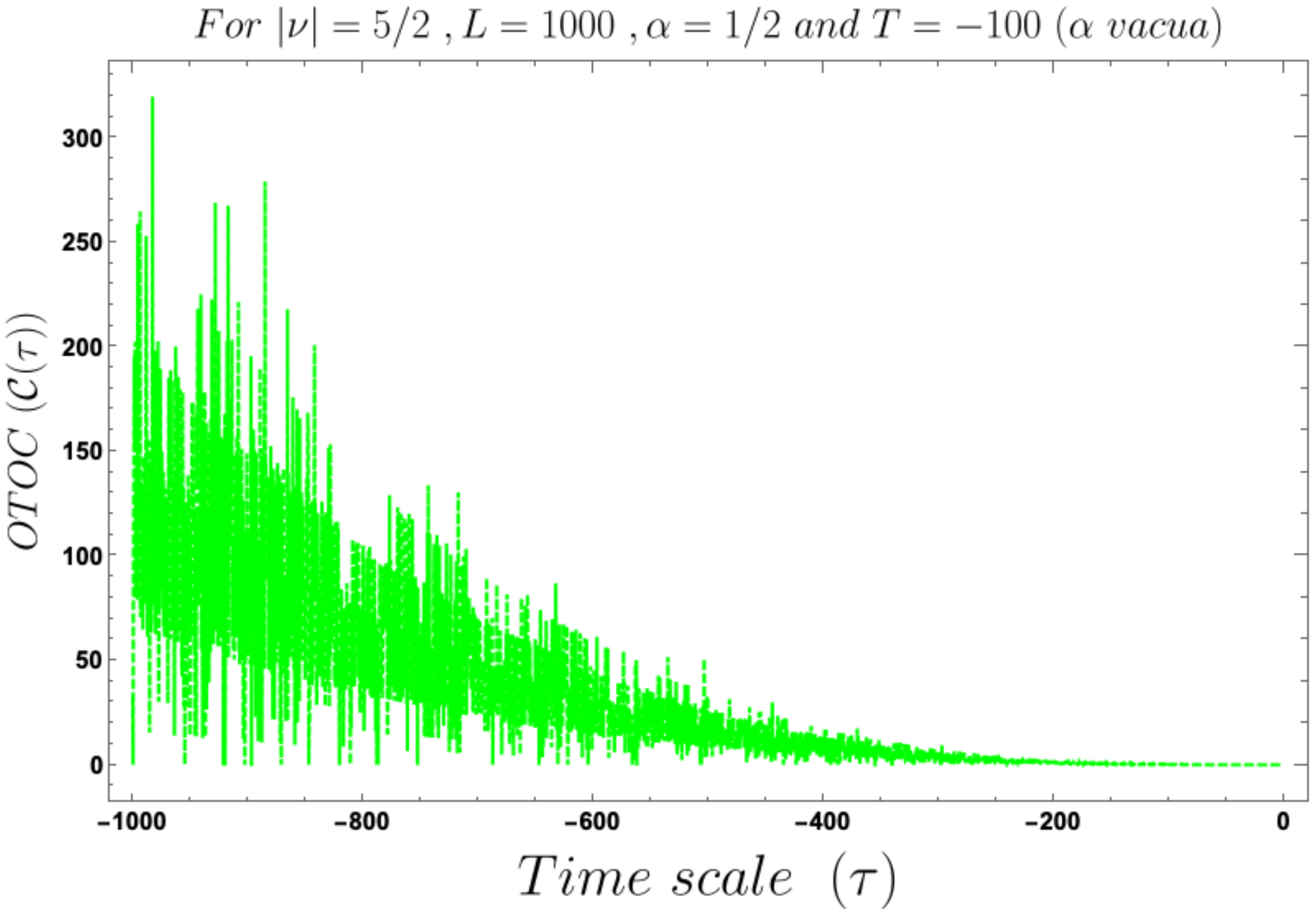}
    \includegraphics[width=17.3cm,height=4.8cm]{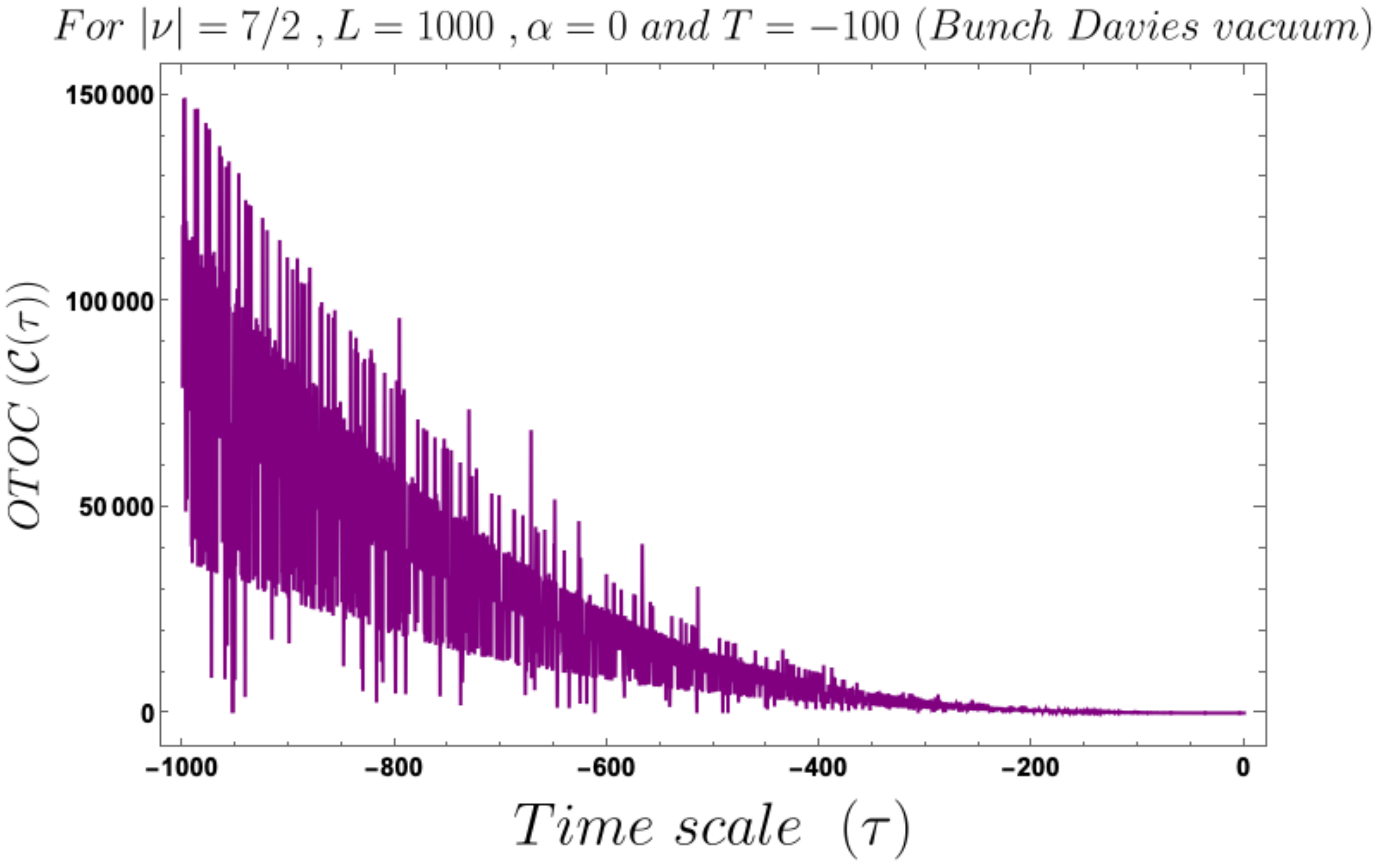}
        \includegraphics[width=16cm,height=4.8cm]{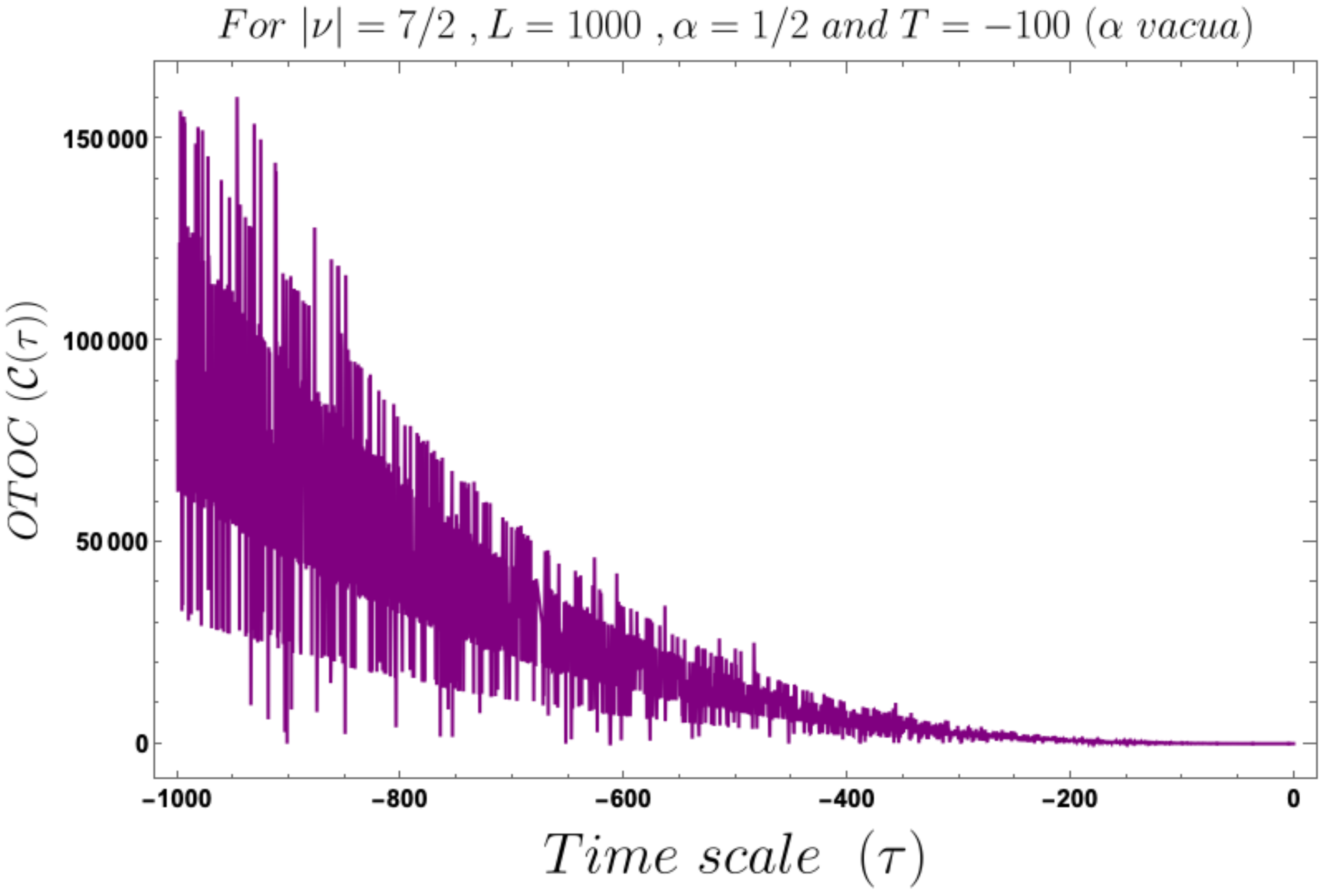}
    \caption{Behaviour of the four-point OTOC with respect to the $\tau$ time scale of the theory. Here we fix, mass parameter $\nu=-5i/2,-7i/2$, cut-off scale $L=1000$, vacuum parameter $\alpha=0 ~({\rm Bunch~Davies~vacuum)}, 1/2~(\alpha~{\rm vacua})$.}
      \label{fig:7}
\end{figure}
 \begin{figure}[t!]
    \centering
        \centering
        \includegraphics[width=17.3cm,height=5.5cm]{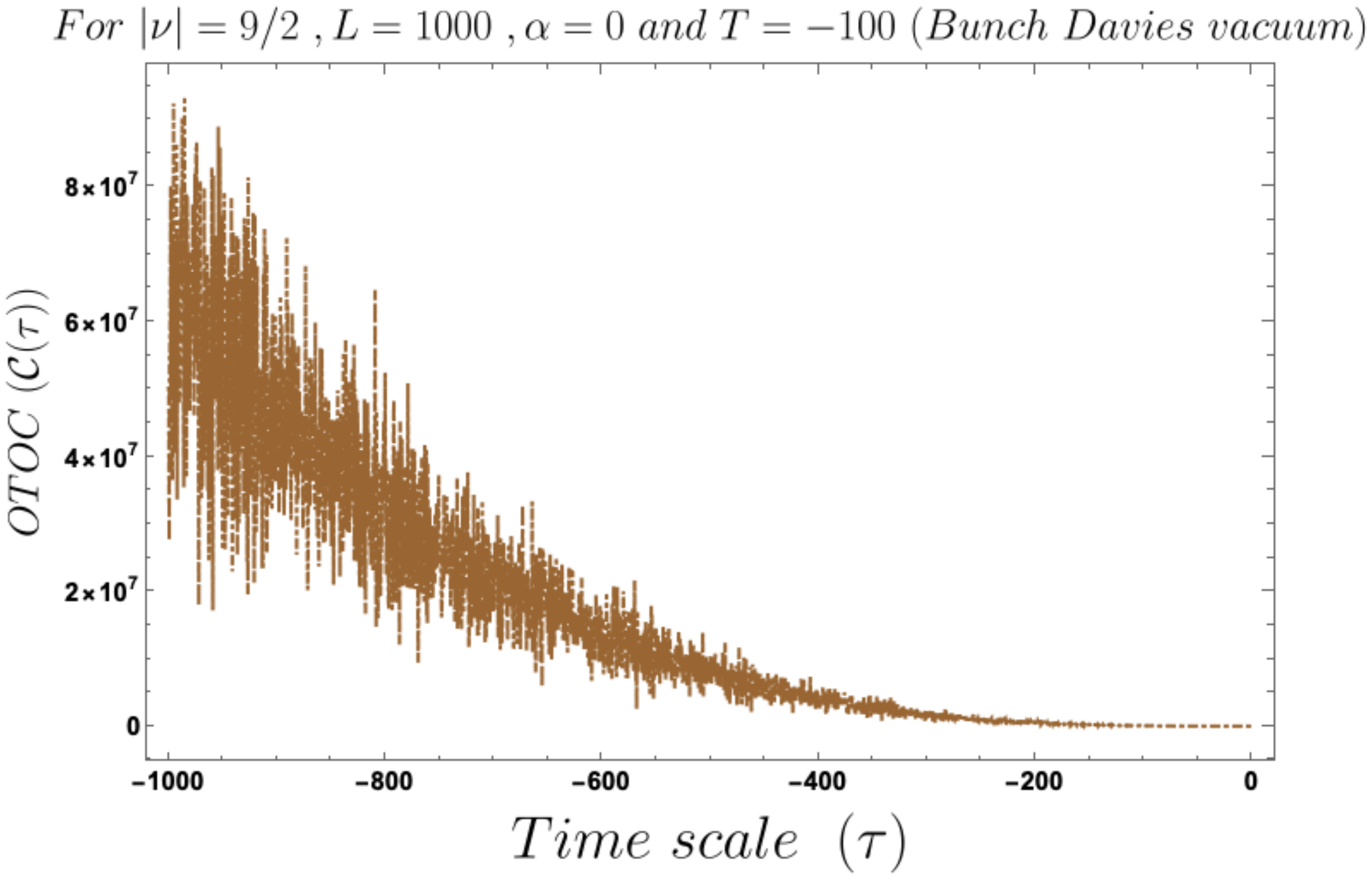}
        \includegraphics[width=17cm,height=5.5cm]{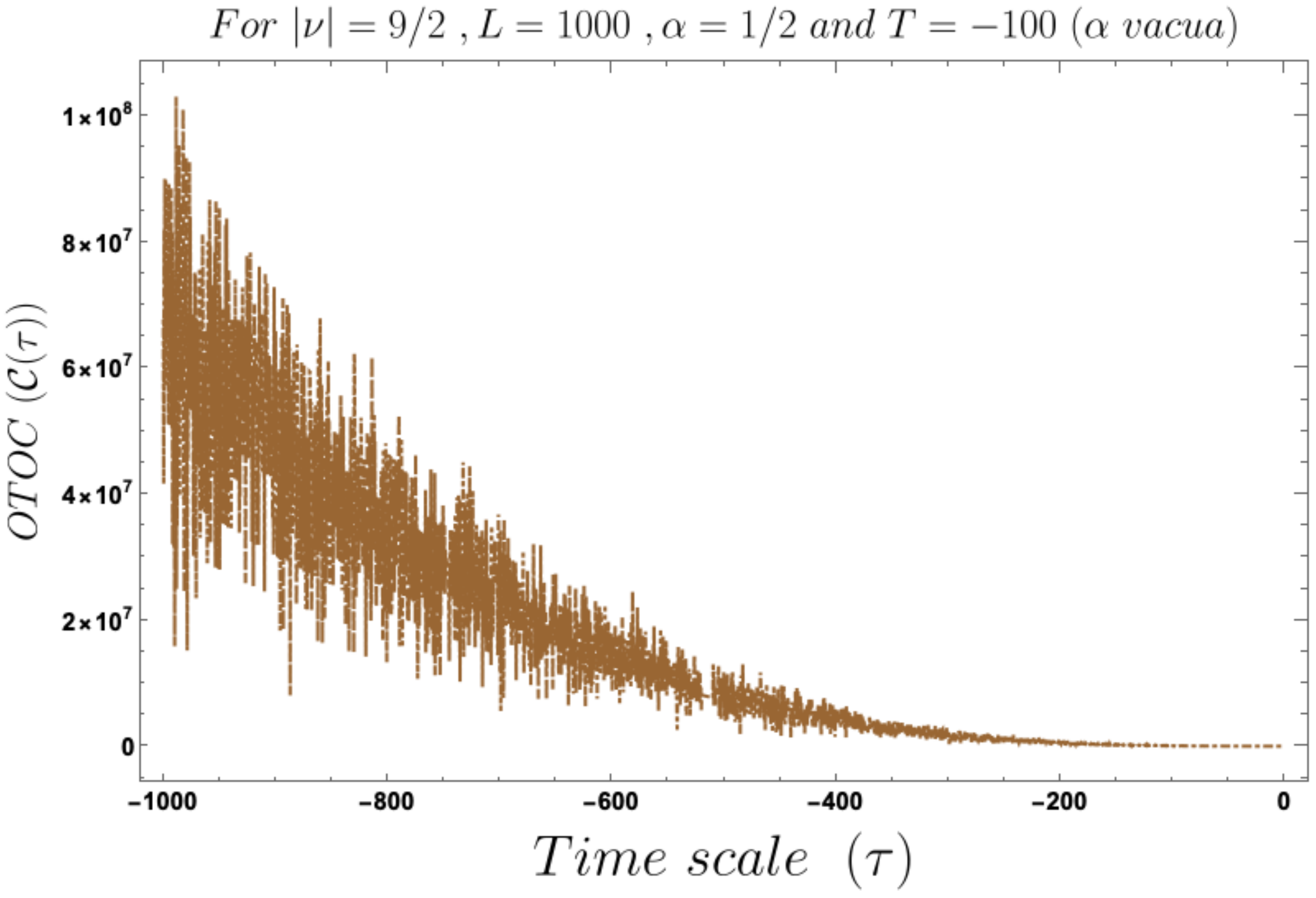}
    \caption{Behaviour of the four-point OTOC with respect to the $\tau$ time scale of the theory. Here we fix, mass parameter $\nu=-9i/2$, cut-off scale $L=1000$, vacuum parameter $\alpha=0 ~({\rm Bunch~Davies~vacuum)}, 1/2~(\alpha~{\rm vacua})$.}
      \label{fig:8}
\end{figure}
\begin{figure}[t!]
    \centering
        \centering
        \includegraphics[width=16cm,height=6cm]{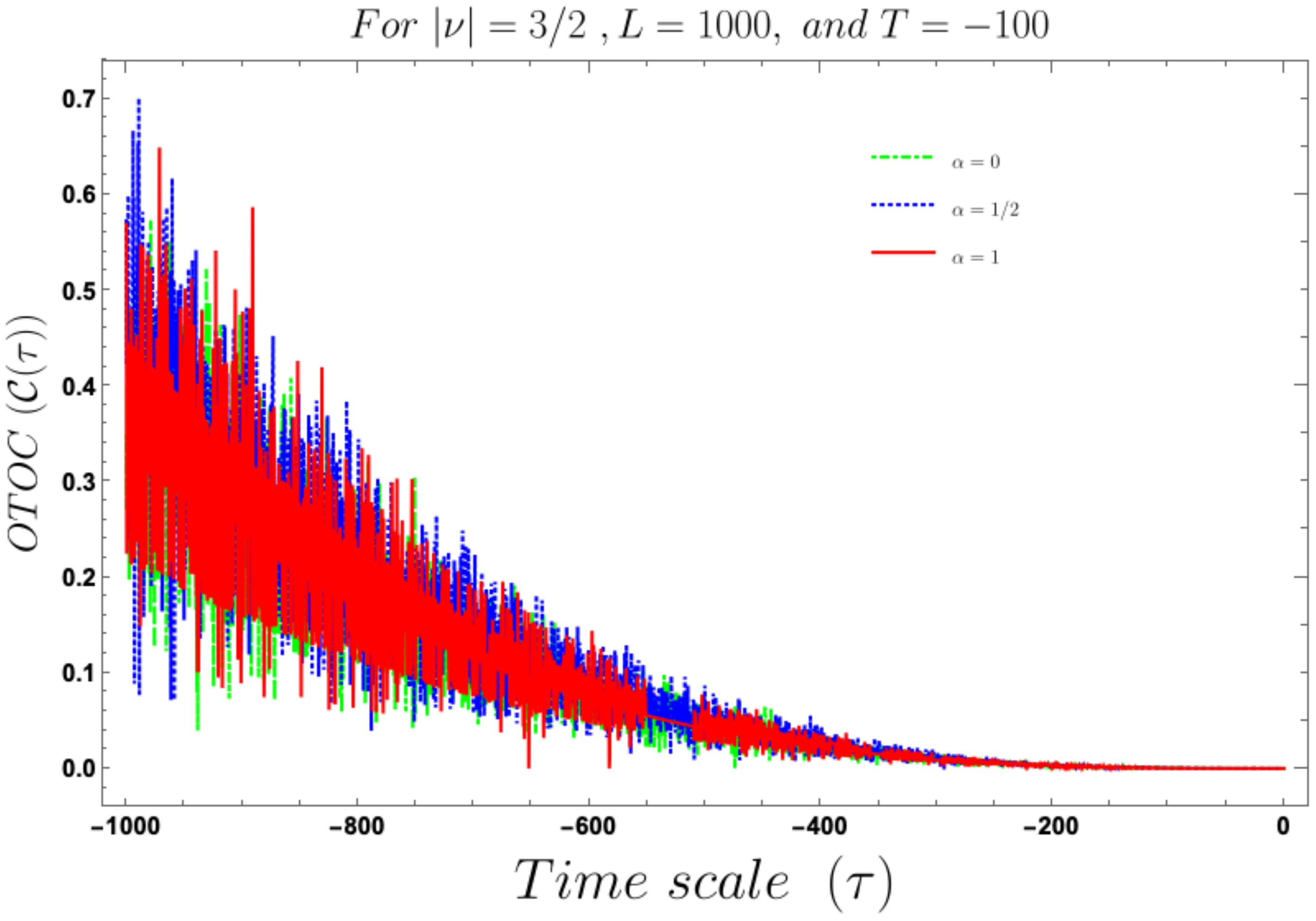}
    \caption{Behaviour of the four-point OTOC with respect to the $\tau$ time scale of the theory. Here we fix, mass parameter $\nu=-3i/2$, cut-off scale $L=1000$, vacuum parameter $\alpha=0 ~({\rm Bunch~Davies~vacuum)}, 1/2~(\alpha~{\rm vacua}), 1~(\alpha~{\rm vacua})$.}
      \label{fig:9}
\end{figure}
\begin{figure}[t!]
    \centering
        \centering
        \includegraphics[width=16cm,height=3.9cm]{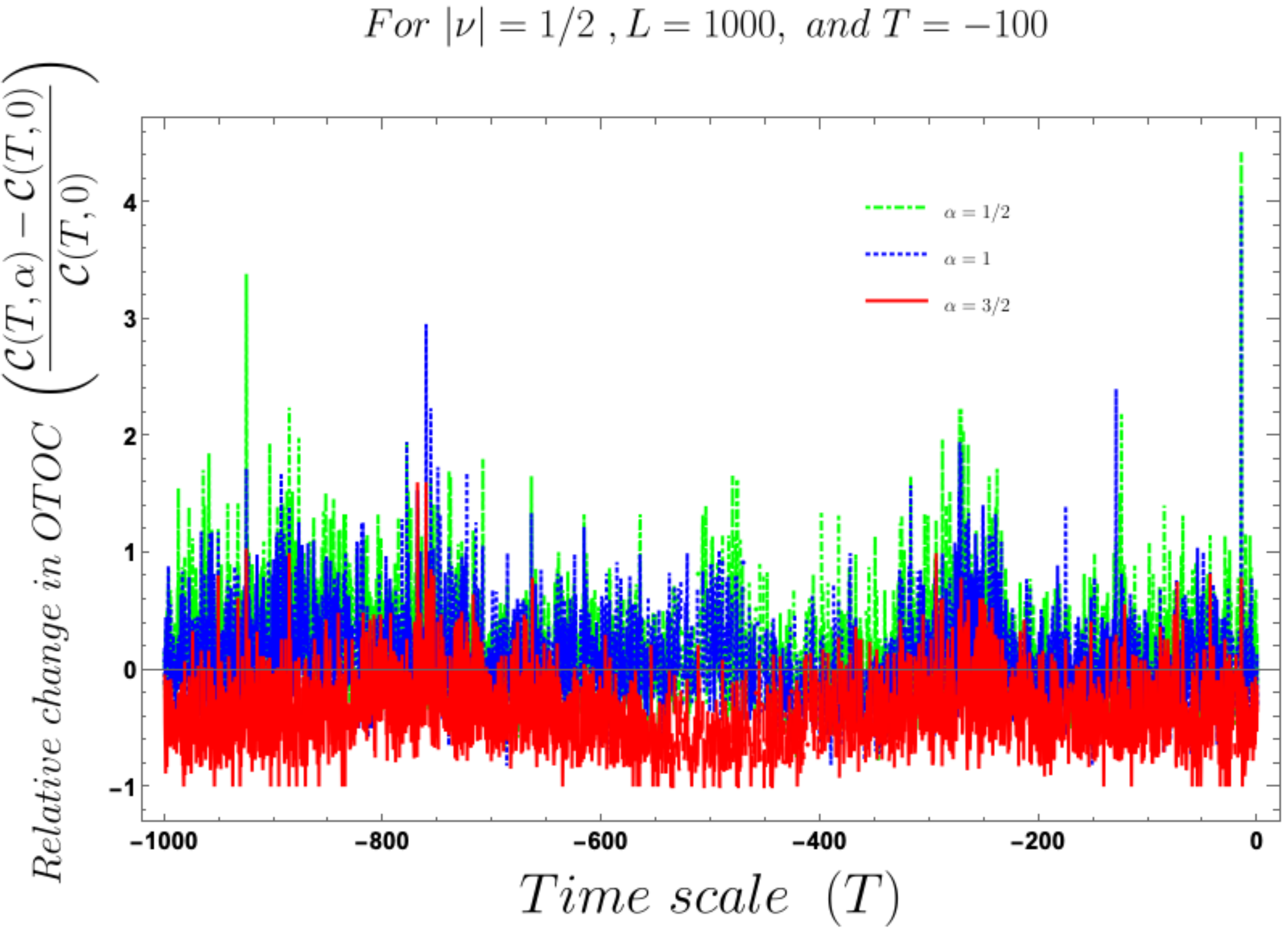}
         \includegraphics[width=16cm,height=3.9cm]{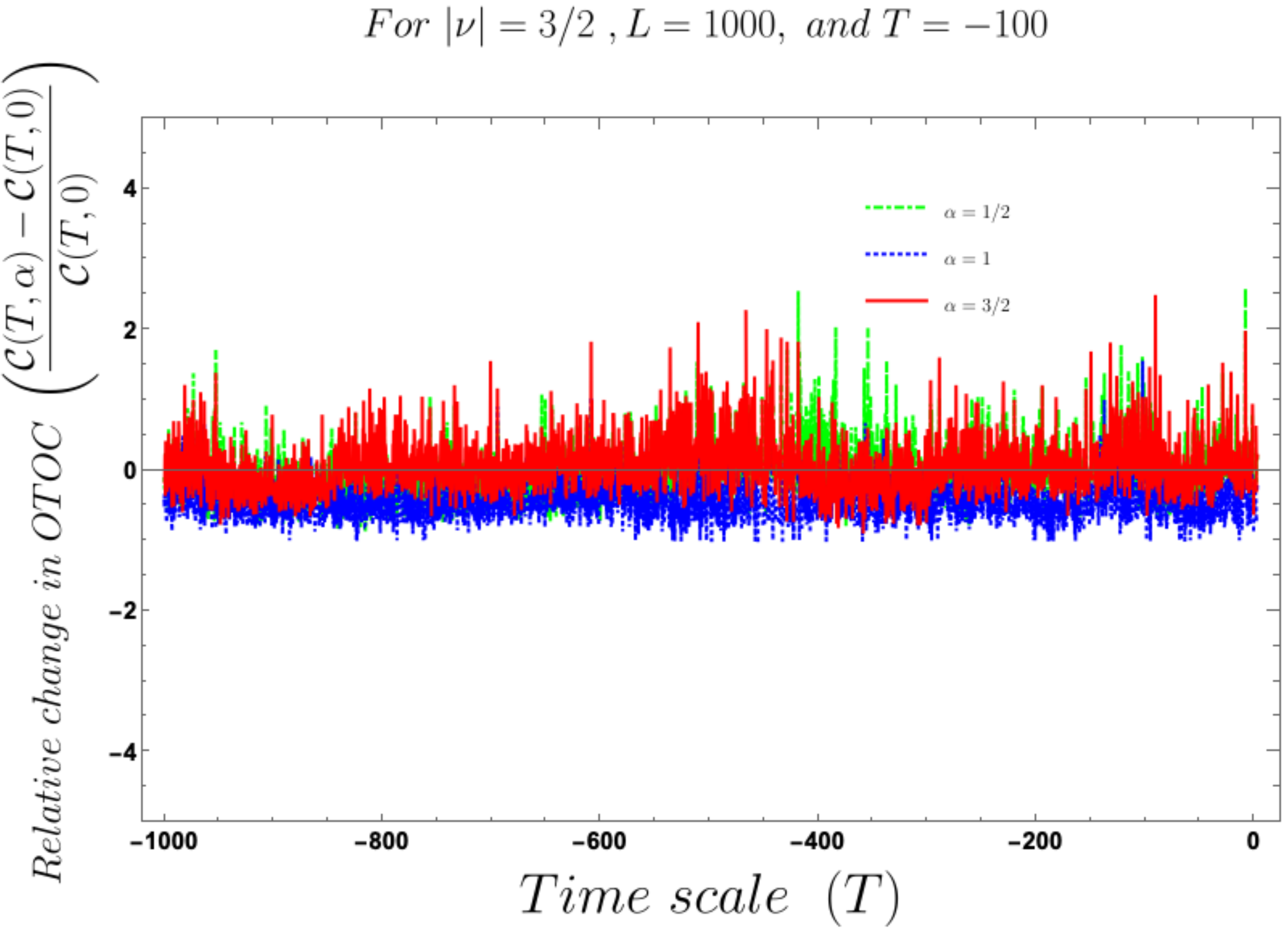}
         \includegraphics[width=16cm,height=3.9cm]{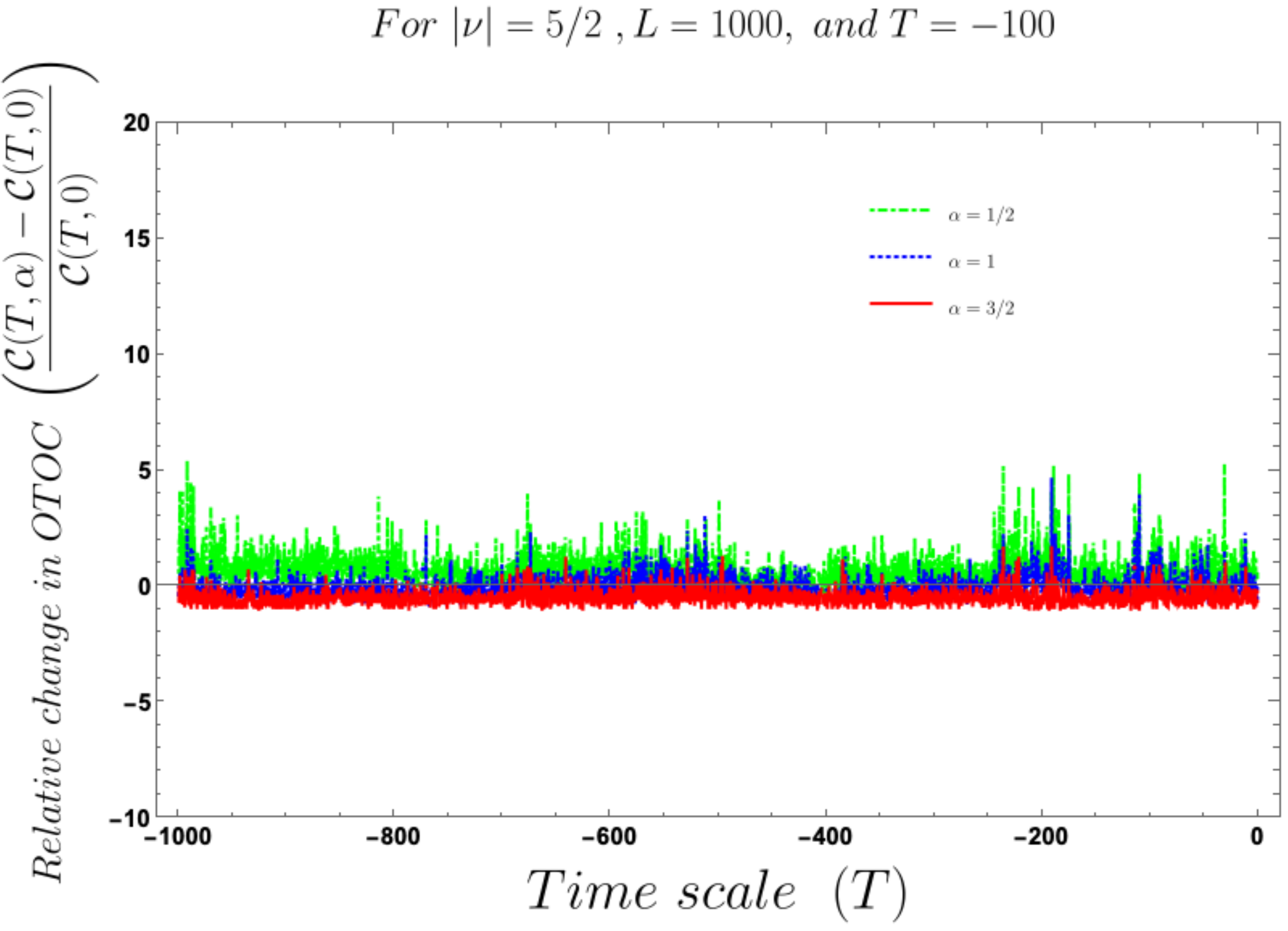}
          \includegraphics[width=16cm,height=3.9cm]{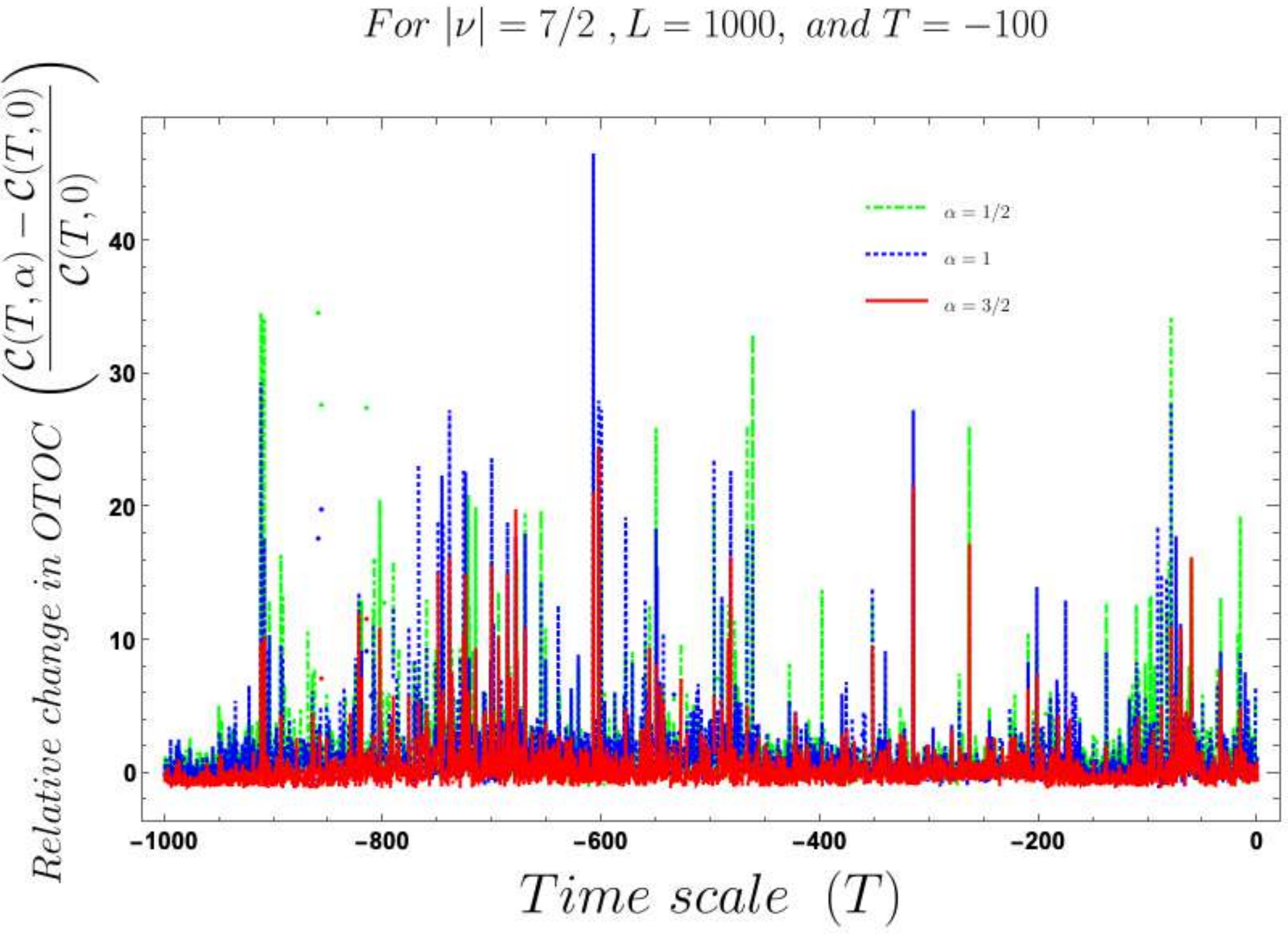}
           \includegraphics[width=16cm,height=3.9cm]{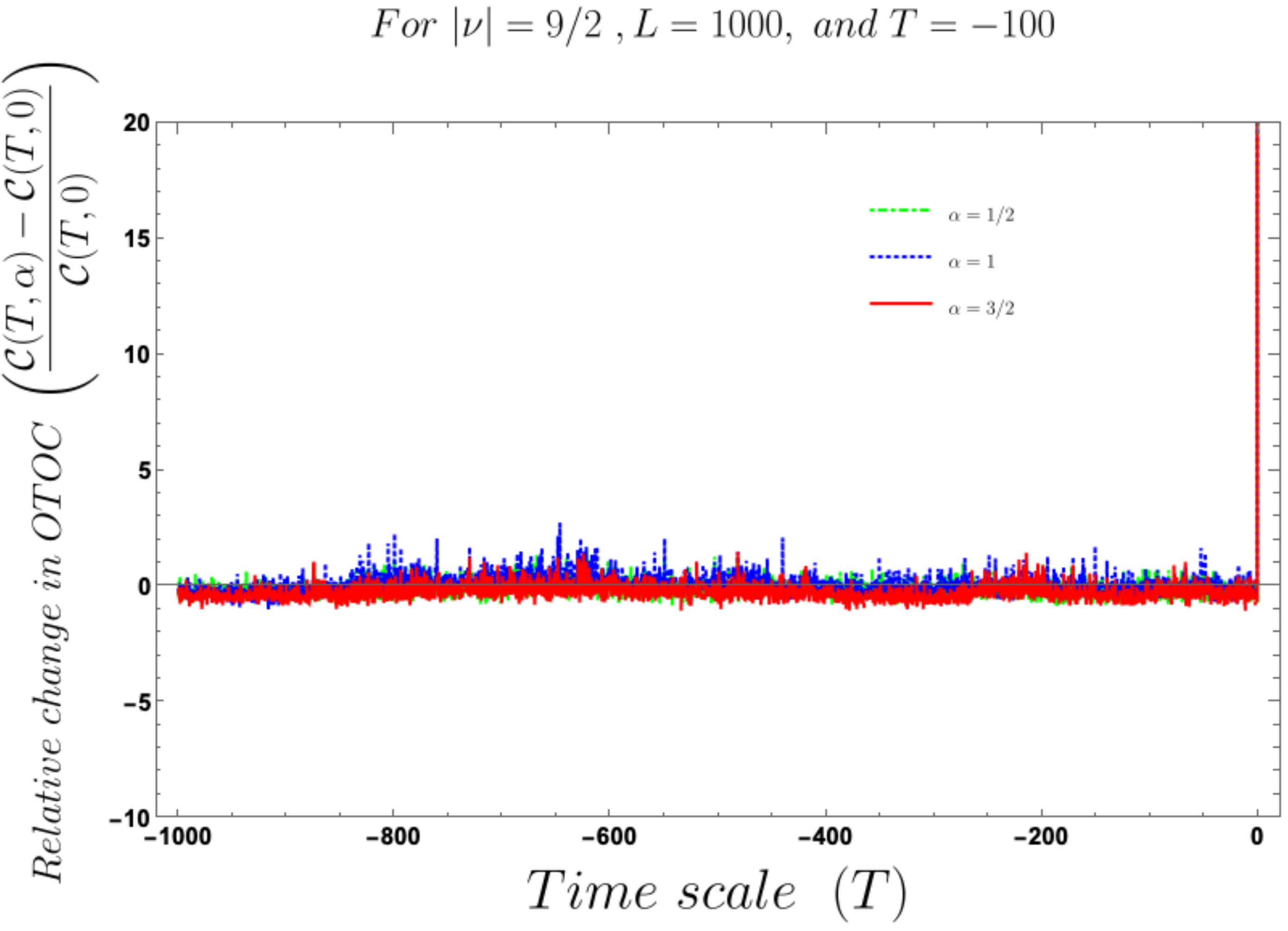}
    \caption{Behaviour of the relative change in four-point OTOC with respect to the two time scale $T$. Here we fix, cut-off scale $L=1000$.}
      \label{fig:10}
\end{figure}
\begin{figure}[t!]
    \centering
        \centering
        \includegraphics[width=16cm,height=3.9cm]{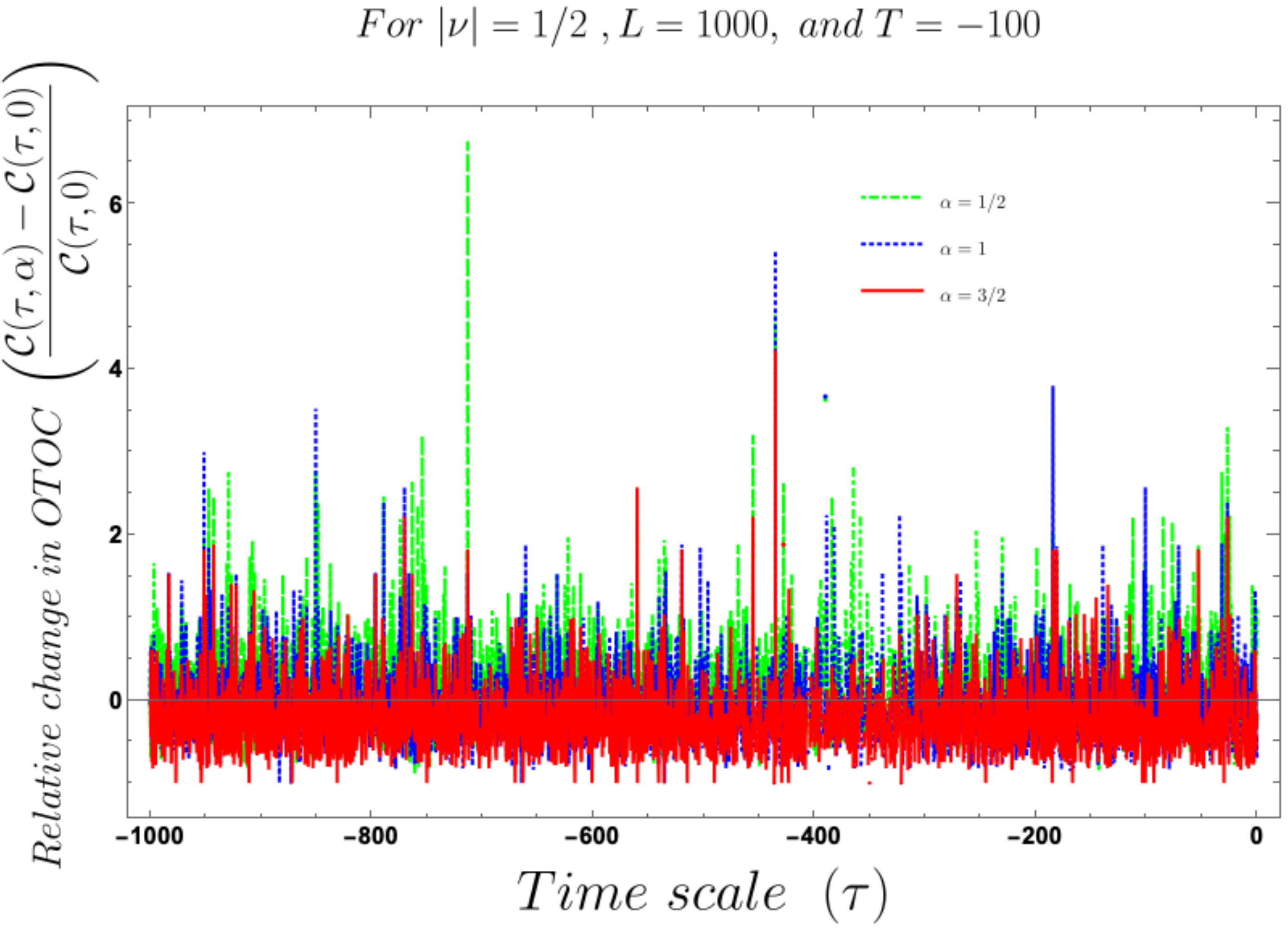}
         \includegraphics[width=16cm,height=3.9cm]{R1.pdf}
         \includegraphics[width=16cm,height=3.9cm]{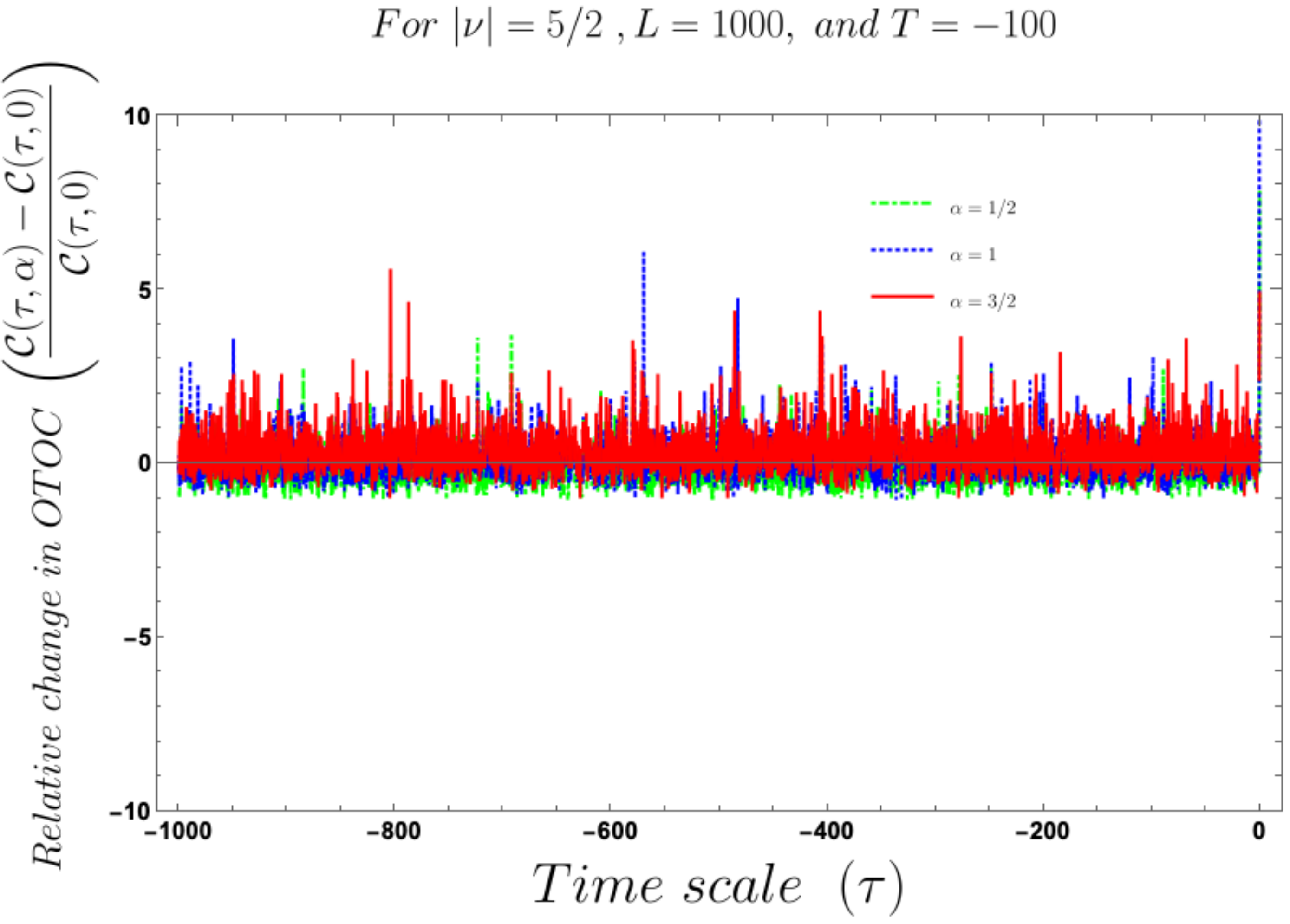}
          \includegraphics[width=16cm,height=3.9cm]{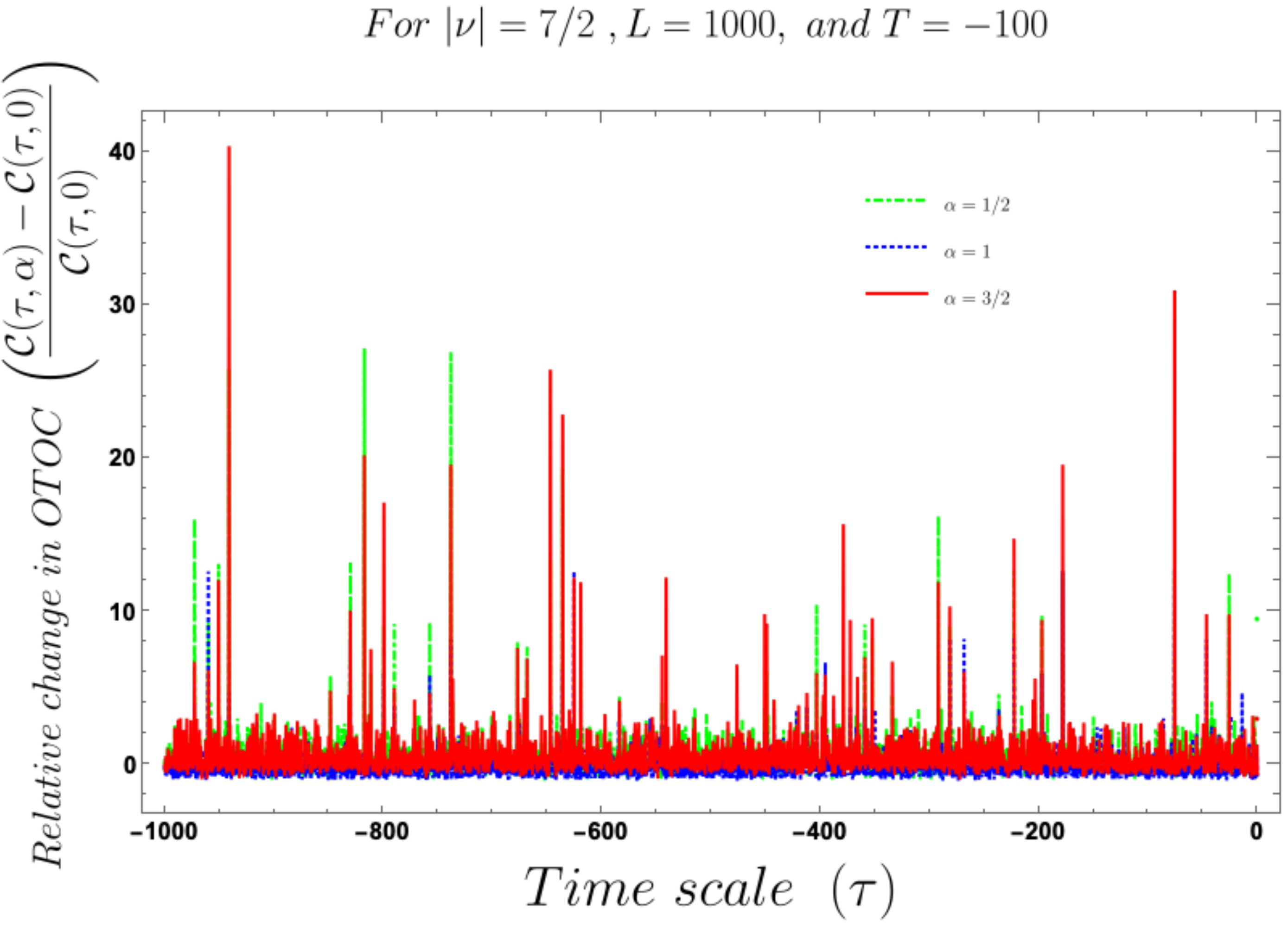}
           \includegraphics[width=16cm,height=3.9cm]{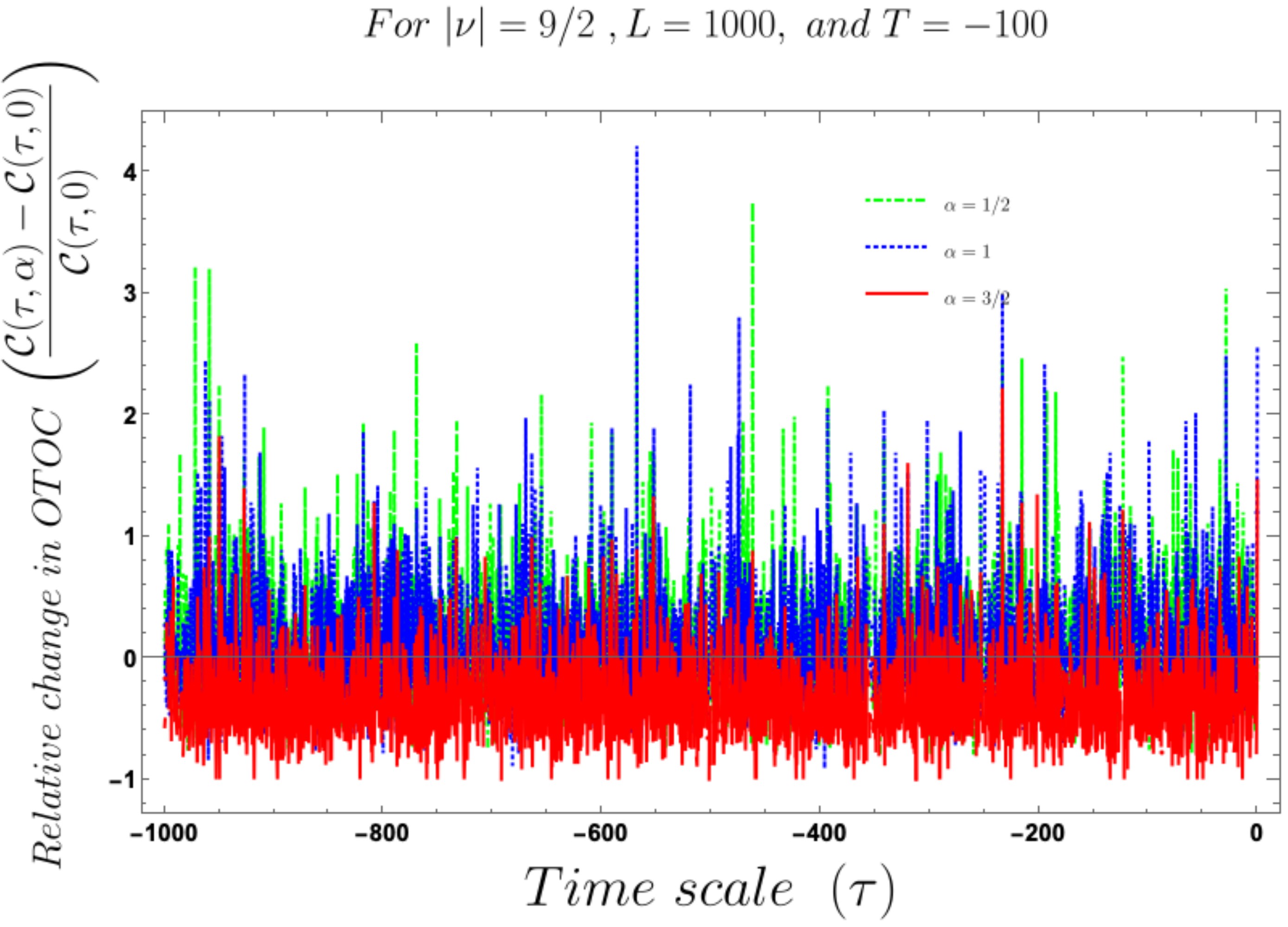}
    \caption{Behaviour of the relative change in four-point OTOC with respect to the two time scale $\tau$. Here we fix, cut-off scale $L=1000$.}
      \label{fig:11}
\end{figure}
  \subsection{OTOC from regularised four-point ``in-in" OTO micro-canonical amplitude: curvature perturbation field version}
  \subsubsection{Without normalization}
Here we need to perform the computation for the un-normalised OTOC  in terms of the scalar curvature perturbation and the canonically conjugate momentum associated with it, which we have found that is given by the following simplified expression:
\bea && \hll{C^{\zeta}(\tau_1,\tau_2)=-\frac{1}{Z^{\zeta}_{\alpha}(\beta,\tau_1)}{\rm Tr}\left[e^{-\beta \hat{H}(\tau_1)}\left[\hat{\zeta}({\bf x},\tau_1),\hat{\Pi}({\bf x},\tau_2)\right]^2\right]_{(\alpha)}=\frac{1}{z^2(\tau_1)z^2(\tau_2)}C^{f}(\tau_1,\tau_2)}.~~~~~~~~~~~~~\eea
\subsubsection{With normalization}
The normalised OTOC in terms of the scalar curvature perturbation and the canonically conjugate momentum associated with it, which is basically the computation of the following normalised OTOC, in the present context:
\bea \hll{{\cal C}^{\zeta}(\tau_1,\tau_2)=\frac{C^{\zeta}(\tau_1,\tau_2)}{\langle \zeta(\tau_1)\zeta(\tau_1)\rangle_{\beta}\langle \Pi_{\zeta}(\tau_1)\Pi_{\zeta}(\tau_1)\rangle_{\beta}}={\cal N}^{\zeta}(\tau_1,\tau_2)~C^{\zeta}(\tau_1,\tau_2)}~, \eea
where the normalisation factor to normalise OTOC is given by:
 \bea \hll{{\cal N}^{\zeta}(\tau_1,\tau_2)=\frac{1}{\langle \zeta(\tau_1)\zeta(\tau_1)\rangle_{\beta}\langle \Pi_{\zeta}(\tau_1)\Pi_{\zeta}(\tau_1)\rangle_{\beta}}=\frac{\pi^4 z^2(\tau_1)z^2(\tau_2)}{{\cal F}_1(\tau_1){\cal F}_2(\tau_2)}=z^2(\tau_1)z^2(\tau_2){\cal N}^{f}(\tau_1,\tau_2)}.~~~~~~~~\eea
 Consequently, the normalised OTOC computed from the curvature perturbation variable is given by the following expression:
 \bea \hll{{\cal C}^{\zeta}(\tau_1,\tau_2)={\cal C}^{f}(\tau_1,\tau_2)=\left[1+(-1)^{4\nu}+\frac{7}{2}(-1)^{2\nu}\right]\frac{(-\tau_1)^{1-2\nu}(-\tau_2)^{3-2\nu}}{2(-1)^{4\nu-1}{\cal F}_1(\tau_1){\cal F}_2(\tau_2)}\sum^{4}_{i=1}X^{(i)}_{1}(\tau_1,\tau_2)}.~~~~~~~~ \eea 
  \section{Numerical results~I: Interpretation of two-point micro-canonical OTOC in Cosmology}
  \label{sec:OTO2}
In this section, our objective is to to numerically study and give physical interpretation of the obtained result of the two-point micro-canonical OTOC obtained for Cosmology in the previous section of this paper.

The detailed interpretation of the two-point micro-canonical OTOC for Cosmology is appended below:
\begin{itemize}
\item  In fig.~(\ref{fig:12AC}), fig.~(\ref{fig:13AC}), fig.~(\ref{fig:14AC}) and fig.~(\ref{fig:12AB}), fig.~(\ref{fig:13AC}) and fig.~(\ref{fig:14ABC}), conformal time dependent behaviour of the two-point function with respect to the two time scale of the Cosmology have explicitly shown. From these plots it is clearly observed that with respect both the time scales the two-point function show random fluctuating behaviour. The amplitude of the fluctuation is small for the result obtained from the Bunch Davies vacuum state and very large for the general $\alpha$ vacua state. 

\item For both the cases it is observed that, we get the significant feature in the time dependent two-point function for partially massless or heavy scalar field those who have the imaginary value of the mass parameter $\nu$. This behaviour of the two-point spectrum is explicitly depicted in fig.~(\ref{fig:14BC}). From this plot one can clearly see that for the result obtained using Bunch Davies initial condition the two-point spectra significantly decay very fast with respect to the magnitude of the mass parameter $\nu$. This implies that, for very heavy or partially massless scalar fields if we increase the mass parameter value then the mentioned two-point correlation between the cosmological perturbation variable and its canonically conjugate momenta decrease very fast in a fixed time scale. On the other hand, if we change the quantum initial condition by changing the initial vacuum state by introducing $\alpha$ vacua then it can be clearly observed from the plot that, if increase the value of the vacuum parameter $\alpha$ gradually then we get significant change compared to the feature observed for the Bunch Davies vacuum state. We have explicitly have shown the behaviour of the two-point spectrum for $\alpha=1/2$ and $\alpha=1$, where it is clearly observed that the correlation decays very slowly with the increase in the magnitude of the mass parameter value and for very large value of $|\nu$ it will saturate to a non-zero large amplitude for a fixed time time scale. There is as such no significant change we have observed from the plots that we have drawn for $\alpha=1/2$ and $\alpha=1$ cases, except form a amplitude shift for the large values of the mass parameter. For very small value of the mass parameter all the different profiles obtained for zero value and non-zero values of the vacuum parameter $\alpha$ start approaching to meet together and at $|\nu|=0$ the amplitude of the two-point correlation of the out-of-time ordered spectrum in Cosmology shoots up with a very large amplitude.
 
\item Also we have observed from the plots that the random fluctuations with respect to the conformal time scale show small but decaying time dependent behaviour upto a very late time scale as far as the amplitude of the spectra are concerned. After crossing that late time region the amplitude start increasing and reaches to a very large value which we have not shown explicitly in these plots. After reaching this maximum value it will again start decaying very fast upto to the present day epoch. This large peak value of the spectrum is obtained at the scale when the two time scales of the theory becomes same and the two operators of cosmological perturbation theory taken to be exactly same. Physically this information is very important for our study in this paper, as it pointing towards the fact that, at this particular time scale we are actually getting zero information from the out-of-time ordered physics in the present context. So at very early time scale of the universe and after crossing the peak we will get the information regarding the out-of-time ordered physics from the present cosmological set up.

\item Last but not the least, in the second plot of fig.~~(\ref{fig:14BC}), we have explicitly depicted the behaviour of the two-point spectrum of the out-of-time ordered aspect of Cosmology with respect the vacuum parameter $\alpha$. Here we get a very interesting symmetric feature for the positive and negative values of the vacuum parameter $\alpha$ around $\alpha=0$. As we increase the value of $\alpha$ along the positive axis, then the two-point function increase very fast with large amplitude. Similar symmetric behaviour is also observed for $\alpha<0$ side as well. As we approach towards $\alpha\rightarrow 0$, we see that the amplitude of the two-point out-of-time ordered spectrum for the fixed values of the time parameters of the theory asymptotically approaches to the zero value from both $\alpha>0$ and $\alpha<0$ side.
\end{itemize}
  \section{Numerical results~II: Interpretation of four-point micro-canonical OTOC in Cosmology}
    \label{sec:OTO3}
 Now comparing the results of the OTOC computed from both rescaled variable and curvature perturbation variable we get the following outcomes:
 \begin{enumerate}
 \item Before normalisation, \underline{OTOC computed from both the 
 sides are not same} and connected via a time dependent {\it Mukhanov Sasaki} variable dependent factor $(z(\tau_1)z(\tau_2))^{-2}$.
 \item After normalisation, \underline{OTOC computed from both the 
 sides are exactly same} and this is really good that after normalisation we don't need to think about the explicit origin or any preferred cosmological perturbation variable.
 \end{enumerate}
 Next, we give the physical interpretation of the obtained result for the normalised OTOC computed in the context of primordial cosmology:
 \begin{itemize}
 \item  In our computed OTOC in the context of Cosmology two time scales are involved which are associated with the two operators, rescaled field variable and its canonically conjugate momenta. During the study of the behaviour of the OTOC we have actually have fixed one time scale and have studied the time dependent dynamical behaviour of OTOC with respect to the other time scale, which have not fixed. We have found that the behaviour with respect to both $\tau_1=T$ and $\tau_2=\tau$, using both Bunch Davies and $\alpha$ vacua as quantum vacuum state. See fig.~(\ref{fig:1}), fig.~(\ref{fig:2}), fig.~(\ref{fig:3}), fig.~(\ref{fig:4}) and fig.~(\ref{fig:6}), fig.~(\ref{fig:7}), fig.~(\ref{fig:8}), fig.~(\ref{fig:9}) to know about the detailed conformal time dependent feature of the cosmological normalised OTOC. One crucial thing we have to mention here that, since we are dealing with conformal time, it varies from, $-\infty$ (Big Bang) to $0$ (present day universe), which will be very useful to understand the physical outcomes of these plots. For this reason we get completely opposite behaviour of the cosmological OTOC compared to the usual quantum mechanical systems. In quantum mechanical random system we actually deal with the usual time scale, which various from $0$ to $\infty$. Here $0$ corresponds to the initial condition on the system when the quantum mechanical system goes to the out-of-equilibrium state and it is expected from our basic understanding of statistical mechanics that if we wait for a very longer time scale then the OTOC will saturates to a specific value where the system actually achieve equilibrium, where one can associate an equilibrium temperature with the quantum system under consideration. Just using the concept of retarded quantum correlators explaining these features are extremely difficult and at the end computation probing those results in experiments are also very complicated job. OTOC plays here a significant role to probe such quantum correlations experimentally in a very simpler fashion. In the cosmological set up things are not same though we study the quantum correlations at out-of-equilibrium case here also. The prime reason for the difference between the study of both types of OTOC is lying in the time scale. In usual quantum mechanical system we are actually interested in the growth of correlation function and if it is an exponential growth then we identify this feature as the signature of quantum chaos. On the other hand, in the cosmological set up we expect the randomness or stochasticity will decrease with respect to the conformal time scale and such decrease is following an exponential decay with respect to the conformal time scale then this is the signature of quantum chaos in the context of primordial universe. In the cosmological set up at far past $-\infty$ the quantum fluctuations appearing during the epoch of reheating or during the stochastic particle production during inflation goes to the out-of-equilibrium state and like usual quantum system it is expected that if we wait for a longer conformal time scale then at late time it will achieve equilibrium, with which similarly one can associate an equilibrium temperature. For this equilibrium case in the context of Cosmology the OTOC if saturates to a constant value then one can identify this phenomena as quantum mechanical chaos. We will discuss about this particular feature in detail in the next section of this paper. In the next point, we will discuss about the features obtained from the numerical studies of the quantum OTOC we have studied from the primordial cosmology set up. We are very positive that this discuss will explore various underlying features of primordial universe when a cosmological system goes to out-of-equilibrium.
 
 \item If we fix the time scale, $\tau_2=\tau$ and study the behaviour of OTOC with respect to the scale $\tau_1=T$  then we found that as we approach from very early universe to the late time scale the normalised OTOC in the context of cosmology shows random decreasing behaviour. This behaviour is quite consistent we the usual expectation. It is observed that in very early universe $\tau_1=T\rightarrow-\infty$, near to Big Bang all such random fluctuations are appearing with large amplitude and very significant to describe the time dependent behaviour of the quantum correlation function during stochastic particle production during inflation and during the epoch of reheating, when the finite temperature out-of-equilibrium physics play significant role. If we wait for longer time, specifically at the late time scale it is observed from the fig.~(\ref{fig:1}), fig.~(\ref{fig:2}), fig.~(\ref{fig:3}) and fig.~(\ref{fig:4}),  that the quantum correlation OTOC decrease very fast and we get very small amplitude for the random or stochastic quantum mechanical fluctuations. Now if we closely look all of these mentioned plots then we see that we decreasing behaviour of OTOC in these cases would not follow any exponential decaying behaviour. So we cannot identify the decreasing behaviour of this OTOC computed from the primordial cosmological set-up with the concept of quantum mechanical chaos. Though it is clear from the plots that the random fluctuations originating from the out-of-equilibrium physics dominates in far past and becomes very very small at the late time universe.
 
 \item  If we fix the time scale, $\tau_1=T$ and study the behaviour of OTOC with respect to the scale $\tau_2=\tau$  then we found that as we approach from very early universe to the late time scale the normalised OTOC in the context of cosmology shows random decreasing on top of it with a exponentially decaying behaviour. This behaviour is quite consistent we the usual expectation of quantum mechanical chaos along with stochastic randomness. It is observed that in very early universe $\tau_2=\tau\rightarrow-\infty$, near to Big Bang all such random fluctuations are appearing with large amplitude and very significant to describe the time dependent behaviour of the OTOC during stochastic particle production during inflation and during the epoch of reheating like the previous case. If we wait for longer time, specifically at the late time scale it is observed from the fig.~(\ref{fig:6}), fig.~(\ref{fig:7}), fig.~(\ref{fig:8}) and fig.~(\ref{fig:9}),  that OTOC decrease very fast with an exponentially decaying amplitude for the random or stochastic fluctuations. Now if we closely look all of these mentioned plots then we see that the behaviour of OTOC in these cases would exactly follow the exponential decaying time dependent behaviour. So we identify the such behaviour of this OTOC computed from the primordial cosmological set-up with the concept of quantum chaos. It is clear from the plots that the quantum chaos, which is a very specific kind of random fluctuations originating in far past and becomes small at the late time scale.
 
\item Additionally we have observed that we can get the expected behaviour from the OTOC with respect to both the time scales when the mass parameter, $\nu$ can be analytically continued to $-i\nu$. So \underline{massless De Sitter} case, which is $\nu=3/2$ is clearly \underline{discarded} here. Consequently, we left with the following possibilities:
\begin{enumerate}
\item \underline{Partially massless De Sitter} case is one of the possibilities, where we can estimate the following parameter from the present understanding:
\bea \hll{-i\nu=\sqrt{\frac{9}{4}-c^2}~~~\Longrightarrow~~~c=\sqrt{\frac{9}{4}+\nu^2}=\frac{3}{2}\sqrt{1+\left(\frac{2\nu}{3}\right)^2}\geq \sqrt{2}},\eea
which immediately implies the following bound on the mass parameter:
\bea \hll{\nu\geq \frac{i}{2} ~~~~\Longrightarrow~~~~|\nu|\geq \frac{1}{2}}~.\eea
This is a very great fact because we have studied the plots for the following values of the mass parameter:
\bea \hll{\nu=\frac{i}{2},~\frac{3i}{2},~\frac{5i}{2},~\frac{7i}{2},~\frac{9i}{2}~~\Longrightarrow~~|\nu|=\frac{1}{2},~\frac{3}{2},~\frac{5}{2},~\frac{7}{2},~\frac{9}{2}}~.\eea

\item Also we have to mention that since a lot of complex gamma function is appearing and mass parameter is $-i\nu$, during the numerical analysis we have taken the absolute value of OTOC during numerical computation. Consequently, the final expression for the cosmological OTOC can be re-expressed for $\alpha$ vacua as:
 \bea \hll{{\cal C}(T,\tau,\alpha)=\left|\left[1+(-1)^{-4i|\nu|}+\frac{7}{2}(-1)^{-2i|\nu|}\right]\frac{(-T)^{1+2i|\nu|}(-\tau)^{3+2i|\nu|}}{2(-1)^{-(4i|\nu|+1)}\widetilde{\cal F}_1(T)\widetilde{\cal F}_2(\tau)}\sum^{4}_{i=1}\widetilde{X}^{(i)}_{1}(T,\tau)\right|}.~~~~\eea 
 Here in the factors appearing with $~~\widetilde{}~~$ we have replaced $\nu$  with $-i|\nu|$.

\item \underline{Massive De Sitter} case is one of the possibilities, where we can estimate the following parameter from the present understanding:
\bea \hll{\nu=i\sqrt{\frac{m^2}{H^2}-\frac{9}{4}}=\frac{im}{H}\sqrt{1-\left(\frac{3H}{2m}\right)^2}\approx\frac{im}{H}\left(1-\frac{1}{2}\left(\frac{3H}{2m}\right)^2+\cdots\right)}~.\eea
Since we have studied the case for $\nu\geq i/2$, we get the following bound on the mass of the heavy field, which is given by the following expression:
\bea \hll{m\geq \sqrt{\frac{5}{2}}~H}~.\eea

\item \underline{Reheating} case is the last possibility, where we can estimate the following parameter from the present understanding:
\bea \hll{\frac{i}{2}\leq \nu\leq \frac{3i}{2}~~~\Longrightarrow~~~\frac{1}{2}\leq |\nu|\leq \frac{3}{2}~~~~~~~~~{\rm for}~~~~0\leq w_{reh}\leq \frac{1}{3}~}~.\eea
\end{enumerate}

\item After fixing both the time scales at far past, the behaviour of OTOC with respect to the mass parameter is depicted in the fig.~(\ref{fig:5}) for Bunch Davies and $\alpha$ vacua as quantum mechanical vacuum state. This plot explicitly shows that if we increase the value of mass parameter then OTOC amplitude increases like an exponential growth. This behaviour is consistent with constraints obtained for the \underline{partially massless scalar}, \underline{massive scalar} and \underline{reheating phenomena} mentioned in the previous point.

\item After fixing both the time scales at far past and fix $\nu=5i/2$, the behaviour of OTOC with respect to the vacuum parameter $\alpha$ is depicted in the fig.~(\ref{fig:5}). This plot explicitly shows that if we increase the value of mass parameter then OTOC amplitude fluctuates like a random stochastic signal. 

\item All the momentum dependent integrals appearing in cosmological OTOC is divergent at $\infty$, for which we have introduced an UV regulator at $L=1000$ a finite but at large value. This makes the final result of OTOC convergent and finite.

\item Further we define a relative change in normalized OTOC where the relativeness is defined with respect to the definition of quantum mechanical vacuum state in the definition of normalized OTOC, given by:
\bea \hll{{\cal R}_{\alpha}(\tau_i)=\left(\frac{{\cal C}(\tau_i,\alpha)-{\cal C}(\tau_i,\alpha=0)}{{\cal C}(\tau_i,\alpha=0)}\right)~~~~~~~~\forall~~ i=1 (T), 2 (\tau)}~.~\eea
In fig.~(\ref{fig:10}) and fig.~(\ref{fig:11}), we have explicitly shown the behaviour of the relative change in normalized OTOC with respect to the two time scales appearing in the computation of OTOC.Here once we see the behaviour of OTOC with respect to one time scale, we keep the other time scale fixed for the computation and numerical estimation of the relative change in the normalized OTOC. Here the relative change we have studies for $\alpha=1/2,1,3/2$ with respect to $\alpha=0$ for all previously allowed values of the mass parameter. this is a very important parameter where one can explicitly see the role of mass parameter $\nu$ and vacuum parameter $\alpha$ together. Also from this relative change one can also study about the fact that in which time scale the stochastic randomness appearing in the quantum fluctuation appearing in cosmological OTOC is large or small. The detailed features obtained  from these plots show that the relative change in cosmological OTOC will not change much starting from Big Bang to the present epoch, apart from the irregular random fluctuations. However, the explicit time dependence of OTOC obtained from Bunch Davies and $\alpha$ vacua is consistent with the expectation from the cosmological set up studied here.

\item Also, we have to mention here that the computed cosmological OTOC is any specific coordinate independent. If we go through the computation then we see that even we have started defining the cosmological OTOC in coordinate space, after taking Fourier transformation and applying the momentum conservation appearing in terms of Dirac Delta functions, the exponential factor which captures both momenta and coordinate will be unity. Finally, we get a momentum integrated cut-off regulated result which only depend on the two dynamical conformal time scale in which we have defined the rescaled perturbation variable and its canonically conjugate momenta. So this implies that the final expression of cosmological OTOC is only conformal time dependent, not depend on any specific choice of coordinate. This is quite impressive outcome and physically consistent with the mathematical framework of cosmological perturbation theory in the early universe. In general, when one define the quantum operators in a specific time and in a specific coordinate system then it always appears in the final result and captures the effect of inhomogeneity in the space-time coordinate. In the context of Cosmology we have found that the final result of OTOC captures the effect of inhomogeneity as we have defined the two quantum cosmological operators in different time scales at the starting point of the definition of OTOC. This have been done in such a way because in cosmology we are actually interested in the time evolution of the quantum cosmological operators which are separated in time scale, which is one of the crucial constraint in the definition of any general OTOC. But such OTOC in cosmology don't capture the effects of inhomogeneity at all. 

\item The final result of cosmological OTOC is not explicitly dependent of the factor $\beta=1/T$ which is appearing in the expression for thermal partition function which we have computed in the context of Cosmology. this implies that the final result is also independent of the thermal partition function computed for Cosmology. This is quite consistent with the basic understanding of out-of-equilibrium aspects of quantum statistical mechanics. Because when a system goes to out-of-equilibrium one cannot associate a temperature with that system as the concept of temperature is always appearing from thermodynamic equilibrium. One can only associate a temperature with the system if we wait for very longer time scale and wait for saturation of OTOC at a finite value. If such feature appears in OTOC then in that specific context this saturation is actually correspond to the thermodynamic equilibrium in which one can define an equilibrium temperature. The similar phenomena we can actually observed in the context of cosmological OTOC as well where we explicitly found that if we wait for very late time scale during the evolution of our universe in conformal time scale then the OTOC exponentially decays with time and saturates to a finite value at late time scales. We can actually find the specific time scale in terms of conformal time coordinate and associate the concept of quantum chaos, where we can give a measure of quantum {\it Lyapunov exponent} from the present framework. In the next section, we will show that even there is no explicit $\beta$ dependence in the calculation of cosmological OTOC, such dependencies will appear once we connect the present concept with quantum chaos through quantum {\it Lyapunov exponent} $\lambda$ and the well known MSS bound on it , i.e. $\lambda\leq 2\pi/\beta$.

 \end{itemize}

\section{Lyapunov spectrum and quantum chaos in Cosmology}
\label{sec:introduction}
  \label{sec:OTO4}
In the previous section, we have found that by keeping the time scale $\tau_1=T$ fixed if we study the time dependent behaviour of cosmological OTOC with respect to the time scale $\tau_2=\tau$, then one can describe the phenomena of quantum chaos using our present prescription. In terms of conformal time, one can express the normalized quantum OTOC in the following generic form:
\bea \hll{{\cal C}(\tau,\alpha):=\epsilon_{\Delta_f ,\Delta_\Pi}\exp\left(-2\lambda_f |\tau|\right)}~,\eea
where we have written the conformal time as, $\tau=-|\tau|$ to avoid all confusion to study the detailed feature of {\it Lyapunov exponent} in the quantum regime as we know very clearly that starting from Big bang to the present day the conformal time varies as, $-\infty<\tau<0$ which means $0<|\tau|<\infty$. This is because of the fact that the negative signature we have already extracted out in the redefinition of conformal time, $\tau=-|\tau|$. Here the proportionality overall factor, $\epsilon_{\Delta_{f}, \Delta_{\Pi}}$, contain the information of the rescaled field variable $f({\bf x},\tau_1=T={\cal T}_{\rm f})$ and its canonically conjugate momenta $\Pi({\bf x},\tau_2=\tau)$. It is expected that this overall proportionality factor exactly matches with the normalization factor of cosmological OTOC that we have computed earlier in this paper. Consequently, we can write the following expression for the proportionality overall factor, $\epsilon_{\Delta_{f}, \Delta_{\Pi}}$, in the context of cosmological OTOC as:
\bea \hll{\epsilon_{\Delta_{f}, \Delta_{\Pi}}={\cal N}^{f}({\cal T}_{\rm f},\tau)=\frac{1}{\langle \hat{f}({\cal T}_{\rm f})\hat{f}({\cal T}_{\rm f})\rangle_{\beta}\langle \hat{\Pi}(\tau)\hat{\Pi}(\tau)\rangle_{\beta}}=\frac{\pi^4}{\widetilde{\cal F}_1({\cal T}_{\rm f}) \widetilde{\cal F}_2(\tau)}}~,\eea
where the explicit mathematical structure of the functions ${\cal F}_1({\cal T}_{\rm f})$ and ${\cal F}_2(\tau)$ are provided in the Appendix of this paper. Now if we replace $\nu$ with $-i|\nu|$ then one can able to get the expressions for $\widetilde{\cal F}_1({\cal T}_{\rm f})$ and  $\widetilde{\cal F}_2(\tau)$, which are basically analytically continued version of the previously mentioned functions for stochastic particle produced in the form of partially massless scalar field and massive scalar field, and also for the prime field component which is participating during the reheating phenomena in the context of primordial cosmology.

Now, here it is important to point that, since the proportionality overall factor contains the explicit information of the quantum operators using which we have defined and computed the cosmological OTOC, in different cosmological perturbation schemes the definition of the quantum operators are different. This fact actually one can explicitly see in the computation of the normalization factor of cosmological OTOC computed in terms of two different cosmological perturbation operators at the quantum, though they are related via conformal time dependent {\it Mukhanov Sasaki} variable. For this specific reason one can write the normalized OTOC as:
\bea \hll{{\cal C}(\tau,\alpha):=\epsilon_{\Delta_{\zeta} ,\Delta_{\Pi_{\zeta}}}\exp\left(-2\lambda_{\zeta}|\tau|\right)}~.\eea
However, the left side of the above expression is exactly same for two different choices of variables in cosmological perturbation scheme. This is a very crucial information which we have proved explicitly in the previous part of this paper.

Here the proportionality overall factor, $\epsilon_{\Delta_{\zeta}, \Delta_{\Pi_{\zeta}}}$, contain the information of the curvature perturbation field variable $\zeta({\bf x},\tau_1=T={\cal T}_{\rm f})$ and its canonically conjugate momenta $\Pi_{\zeta}({\bf x},\tau_2=\tau)$. Exactly just like the previous case here also one can similarly expect that this overall proportionality factor exactly matches with the normalization factor of cosmological OTOC. Consequently, we can write the following expression for the proportionality overall factor, $\epsilon_{\Delta_{\zeta}, \Delta_{\Pi_{\zeta}}}$, in the context of cosmological OTOC as:
\bea \hll{\epsilon_{\Delta_{\zeta}, \Delta_{\Pi_{\zeta}}}={\cal N}^{\zeta}({\cal T}_{\rm f},\tau)=\frac{1}{\langle \hat{\zeta}({\cal T}_{\rm f})\hat{\zeta}({\cal T}_{\rm f})\rangle_{\beta}\langle \hat{\Pi}_{\zeta}(\tau)\hat{\Pi}_{\zeta}(\tau)\rangle_{\beta}}=\frac{\pi^4z^2({\cal T}_{\rm f})z^2 (\tau)}{\widetilde{\cal F}_1({\cal T}_{\rm f}) \widetilde{\cal F}_2(\tau)}}~,~~~~~~~~~~\eea
and consequently the proportionality overall factors in the two formalism are related via the following relation:
\bea\hll{\epsilon_{\Delta_{\zeta}, \Delta_{\Pi_{\zeta}}}=z^2({\cal T}_{\rm f})z^2 (\tau)~\epsilon_{\Delta_{f}, \Delta_{\Pi}}}~. \eea

On the other hand, for fixed $T={\cal T}_{\rm f}$ one can write the cosmological OTOC as:
\bea \hll{{\cal C}(\tau,\alpha)=\left|\left[1+(-1)^{-4i|\nu|}+\frac{7}{2}(-1)^{-2i|\nu|}\right]\frac{(-T_{\rm f})^{1+2i|\nu|}(-\tau)^{3+2i|\nu|}}{2(-1)^{-(4i|\nu|+1)}\widetilde{\cal F}_1(T)\widetilde{\cal F}_2(\tau)}\sum^{4}_{i=1}\widetilde{X}^{(i)}_{1}(T_{\rm f},\tau)\right|}.~~~~\eea 
here $\tau_2=\tau$ is the only dynamical scale left once we fix the another time scale $\tau_1=T={\cal T}_{\rm f}$, which will be useful to study the feature of quantum chaos and the corresponding {\it Lyapunov exponent} in terms of the cosmological OTOC computed in the previous section. The above mentioned result is formalism independent completely.

This implies that, when we are describing the whole computation in terms of the rescaled perturbation variable and its canonically conjugate momenta, the we get the following estimation of the {\it Lyapunov exponent} from cosmological OTOC, given by:
\bea \hll{\lambda_f=\left|\frac{1}{2\tau}\ln\left[\frac{1}{\displaystyle \left[1+(-1)^{-4i|\nu|}+\frac{7}{2}(-1)^{-2i|\nu|}\right]\frac{(-T_{\rm f})^{1+2i|\nu|}(-\tau)^{3+2i|\nu|}}{2\pi^4(-1)^{-(4i|\nu|+1)}}\sum^{4}_{i=1}\widetilde{X}^{(i)}_{1}(T_{\rm f},\tau)}\right]\right|}~.~~~~~~~\eea
On the other hand, when we are describing the whole computation in terms of the curvature perturbation variable and its canonically conjugate momenta, the we get the following estimation of the {\it Lyapunov exponent} from cosmological OTOC, given by:
\bea \hll{\lambda_{\zeta}=\left|\frac{1}{\tau}\ln\left[\frac{z^2({\cal T}_{\rm f})z^2 (\tau)}{\displaystyle \left[1+(-1)^{-4i|\nu|}+\frac{7}{2}(-1)^{-2i|\nu|}\right]\frac{(-T_{\rm f})^{1+2i|\nu|}(-\tau)^{3+2i|\nu|}}{2\pi^4(-1)^{-(4i|\nu|+1)}}\sum^{4}_{i=1}\widetilde{X}^{(i)}_{1}(T_{\rm f},\tau)}\right]\right|}~.\eea
Here $z(\tau)=a(\tau)\sqrt{\epsilon(\tau)}$ is the {\it Mukhanov Sasaki variable}, where $a(\tau)$ is the scale factor and $\epsilon(\tau)$ is very slowly varying function of conformal time scale. Further, substituting the explicit form of the {\it Mukhanov Sasaki variable} in terms of the conformal time scale dependent scale factor one can simplify the above mentioned expression for the {\it Lyapunov exponent} computed from cosmological OTOC in terms of curvature perturbation is given by the following expression:
\bea &&\hll{\lambda_{\zeta}\approx\left|\frac{1}{2\tau}\underbrace{\ln\left[\displaystyle \left[1+(-1)^{-4i|\nu|}+\frac{7}{2}(-1)^{-2i|\nu|}\right]\frac{(-T_{\rm f})^{1+2i|\nu|}(-\tau)^{3+2i|\nu|}}{2\pi^4(-1)^{-(4i|\nu|+1)}}\sum^{4}_{i=1}\widetilde{X}^{(i)}_{1}(T_{\rm f},\tau)\right]^{-1}}_{\textcolor{red}{\bf Dominant~contribution}}\right|}~\nonumber\\
&&~~~~~~~~~~~~~~~~~~~~~~~~~~~\hll{+\left|\frac{1}{2\tau}\left[\underbrace{2\ln(a({\cal T}_{\rm f})a(\tau))}_{\textcolor{red}{\bf Dominant~contribution}}+\underbrace{\ln(\epsilon({\cal T}_{\rm f})\epsilon(\tau))}_{\textcolor{red}{\bf Very~small~contribution}}\right]\right|}.\eea
After neglecting the very small slow-roll contribution we get following simplified expression for the {\it Lyapunov exponent} computed from cosmological OTOC in terms of curvature perturbation:
\bea \hll{\lambda_{\zeta}=\left|\frac{1}{2\tau}\ln\left[\frac{a^2({\cal T}_{\rm f})a^2 (\tau)}{\displaystyle \left[1+(-1)^{-4i|\nu|}+\frac{7}{2}(-1)^{-2i|\nu|}\right]\frac{(-T_{\rm f})^{1+2i|\nu|}(-\tau)^{3+2i|\nu|}}{2\pi^4(-1)^{-(4i|\nu|+1)}}\sum^{4}_{i=1}\widetilde{X}^{(i)}_{1}(T_{\rm f},\tau)}\right]\right|}~.~~~\eea
Now using the explicit form of the scale factor $a(\tau)$ for De Sitter inflationary patch during partially massless and massive scalar field production and also during the epoch of reheating, we get:
\begingroup
\begin{eqnarray}
\hll{\lambda_{\zeta}\approx \left\{
     \begin{array}{lr}
  \displaystyle  \large\left| \frac{1}{2\tau}\ln\left[\frac{1/H^2}{\displaystyle \left[1+(-1)^{-4i|\nu|}+\frac{7}{2}(-1)^{-2i|\nu|}\right]\frac{(-T_{\rm f})^{2i|\nu|-1}(-\tau)^{1+2i|\nu|}}{2\pi^4(-1)^{-(4i|\nu|+1)}}\sum^{4}_{i=1}\widetilde{X}^{(i)}_{1}(T_{\rm f},\tau)}\right]\right| &~~\text{\textcolor{red}{\bf De~Sitter}}\\ 
 \displaystyle   \left|\frac{1}{2\tau}\ln\left[\frac{\left[\frac{(1+3w_{reh})}{3(1+w_{reh})}\right]^{\frac{4}{(1+3w_{reh})}}\left(-{\cal T}_{\rm f}\right)^{\frac{2}{(1+3w_{reh})}}\left(-\tau\right)^{\frac{2}{(1+3w_{reh})}}}{\displaystyle \left[1+(-1)^{-4i|\nu|}+\frac{7}{2}(-1)^{-2i|\nu|}\right]\frac{(-T_{\rm f})^{1+2i|\nu|}(-\tau)^{3+2i|\nu|}}{2\pi^4(-1)^{-(4i|\nu|+1)}}\sum^{4}_{i=1}\widetilde{X}^{(i)}_{1}(T_{\rm f},\tau)}\right]\right|\\
\displaystyle ~~~~~~~~~~~~~~~~~~{\rm with}~~0\leq w_{reh}\leq \frac{1}{3}~ & \text{\textcolor{red}{\bf Reheating}}  \end{array}
   \right.}~~~~~~
\end{eqnarray}
\endgroup 

Next, the relationship between the two {\it Lyapunov exponents} computed from two different, but connecting cosmological perturbation variable is given by the following simplified expression:
\bea \hll{\Delta\lambda:=\lambda_{\zeta}-\lambda_f=\frac{1}{|\tau|}\ln|z({\cal T}_{\rm f})z(\tau)|=\frac{1}{|\tau|}\left[\underbrace{\ln~2+\ln|a({\cal T}_{\rm f})a(\tau)|}_{\textcolor{red}{\bf Dominant~contribution}}+\underbrace{\frac{1}{2}\ln|\epsilon({\cal T}_{\rm f})\epsilon(\tau)|}_{\textcolor{red}{\bf Very~small~contribution}}\right]}~.~~~~~~~\eea
So after neglecting the contribution from small slow-roll contribution we get the following simplified expression for the connecting relationship between the two {\it Lyapunov exponents} computed from two different, but connecting cosmological perturbation variable is given by:
\bea \hll{\Delta\lambda:=\lambda_{\zeta}-\lambda_f\approx\frac{1}{|\tau|}\left[\ln~2+\ln|a({\cal T}_{\rm f})a(\tau)|\right]}~.~~~~~~~\eea
Now using the scale factor $a(\tau)$ for De Sitter inflationary patch during partially massless and massive scalar field production and also during the epoch of reheating, we get:
\begingroup
\large
\begin{eqnarray}
\hll{\Delta\lambda\approx \large\left\{
     \begin{array}{lr}
  \displaystyle  \large \frac{1}{|\tau|}\left[\ln~2-2\ln|H|-\ln|{\cal T}_{\rm f}\tau|\right] &~~\text{\textcolor{red}{\bf De~Sitter}}\\ 
 \displaystyle   \frac{1}{|\tau|}\left[\ln~2+\frac{2}{(1+3w_{reh})}\left\{2\ln\left|\frac{(1+3w_{reh})}{3(1+w_{reh})}\right|+\ln|{\cal T}_{\rm f}\tau|\right\}\right]\\
\displaystyle ~~~~~~~~~~~~~~~~~~{\rm with}~~0\leq w_{reh}\leq \frac{1}{3}~ & \text{\textcolor{red}{\bf Reheating}}  \end{array}
   \right.}~~~~~~~
\end{eqnarray}
\endgroup 
Now, it is important to mention that, if we wait for longer time scale during the evolution of our universe then the conformal time dependent behaviour of {\it Lyapunov spectrum} will reach a saturation value, which is perfectly consistent with our finding in this paper and discussed in detail in the previous section. To interpret our finding in this paper as a signature of quantum mechanical chaos one essentially needs to satisfy the well known {\it MSS bound} on the {\it Lyapunov exponent}.  Consequently, one can write:
\bea \hll{\lambda_{f}(\tau_{\rm Late}) \leq \frac{2\pi}{\beta},~~~~\lambda_{\zeta}(\tau_{\rm Late})\leq \frac{2\pi}{\beta},~~{\rm where}~~\beta=\frac{1}{T}}~\eea
where $T$ represents the equilibrium temperature of our universe when the above condition satisfies at late time scale to achieve quantum mechanical chaos.

Further, one can write down the following simplified bound on the  equilibrium temperature of our universe, given by:
\bea &&\hll{\textcolor{red}{\bf From~f:}~T\geq \left|\frac{1}{4\pi\tau_{\rm Late}}\ln\left[{\cal O}({\cal T}_{\rm f},\tau_{\rm Late},|\nu|)\right]\right|}~.~~~~~~~\\
&&\hll{\textcolor{red}{\bf From~\zeta:}~T\geq \left\{
     \begin{array}{lr}
  \displaystyle \left|\frac{1}{4\pi\tau_{\rm Late}}\ln\left[\frac{1}{H^2}{\cal O}({\cal T}_{\rm f},\tau_{\rm Late},|\nu|)\right]\right|&~~\text{\textcolor{red}{\bf De~Sitter}}\\ \\
 \displaystyle \left|\frac{1}{4\pi\tau_{\rm Late}}\ln\left[{\cal U}({\cal T}_{\rm f},w_{reh}){\cal O}({\cal T}_{\rm f},\tau_{\rm Late},|\nu|)\right]\right|~{\rm with}~~0\leq w_{reh}\leq \frac{1}{3}~~~~ & \text{\textcolor{red}{\bf Reheating}}  \end{array}
   \right.}~~~~~~~~
\eea
where we define a new function, ${\cal O}({\cal T}_{\rm f},\tau_{\rm Late},|\nu|)$, which is given by the following expression:
\bea\displaystyle {\cal O}({\cal T}_{\rm f},\tau_{\rm Late},|\nu|):&=&\frac{1}{\displaystyle\left[1+(-1)^{-4i|\nu|}+\frac{7}{2}(-1)^{-2i|\nu|}\right]\frac{(-{\cal T}_{\rm f})^{1+2i|\nu|}(-\tau_{\rm Late})^{3+2i|\nu|}}{2\pi^4(-1)^{-(4i|\nu|+1)}}\sum^{4}_{i=1}\widetilde{X}^{(i)}_{1}({\cal T}_{\rm f},\tau_{\rm Late})}~,~~~~~~~~~\\
{\cal U}({\cal T}_{\rm f},w_{reh}):&=&\left[\frac{(1+3w_{reh})}{3(1+w_{reh})}\right]^{\frac{4}{(1+3w_{reh})}}\left(-{\cal T}_{\rm f}\right)^{\frac{2}{(1+3w_{reh})}}\left(-\tau_{\rm Late}\right)^{\frac{2}{(1+3w_{reh})}}. \eea
  \section{Numerical results~III: Interpretation of cosmological Lyapunov spectrum}
    \label{sec:OTO5}
 \begin{figure}[t!]
    \centering
        \centering
        \includegraphics[width=17.3cm,height=4.8cm]{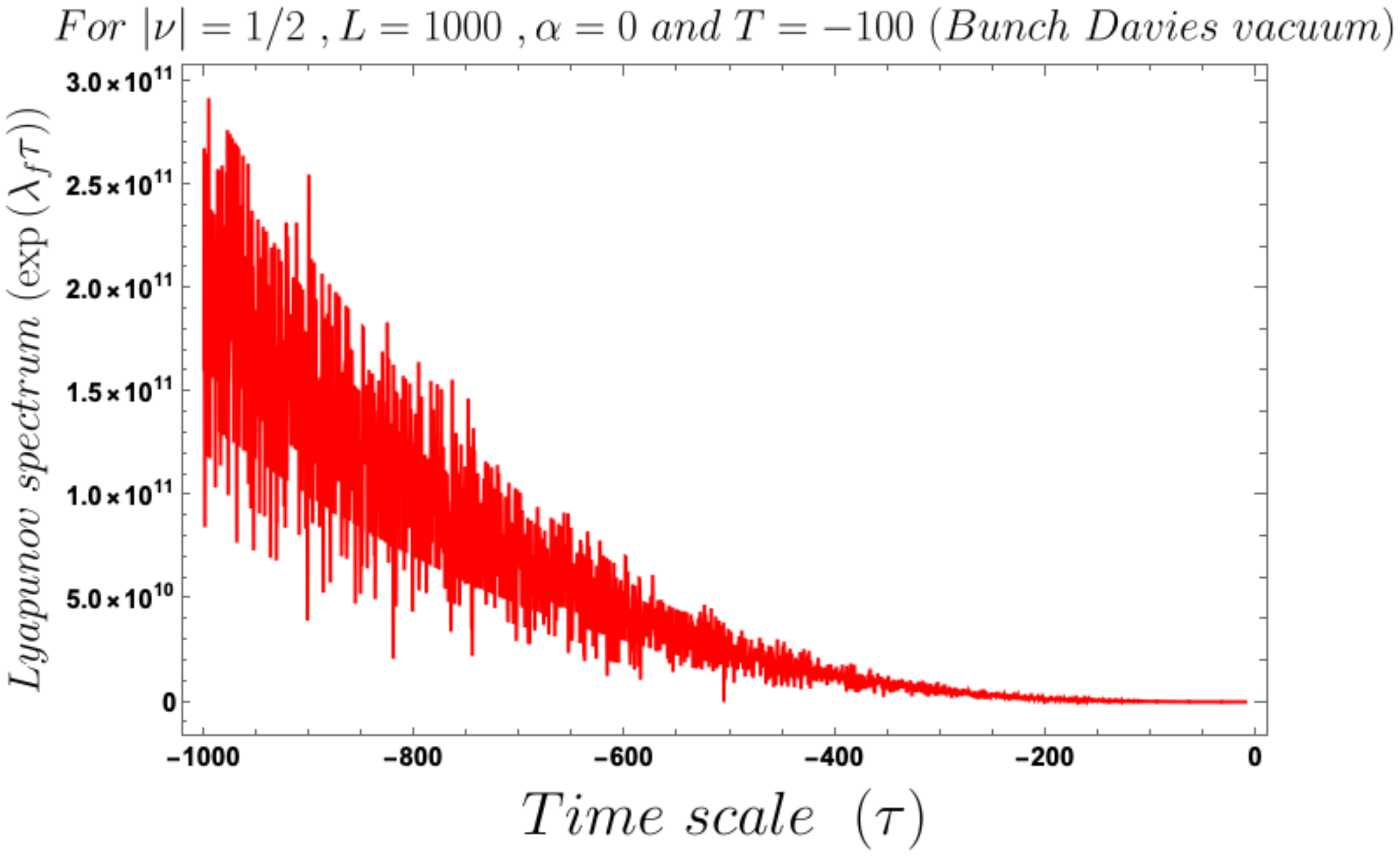}
        \includegraphics[width=16cm,height=4.8cm]{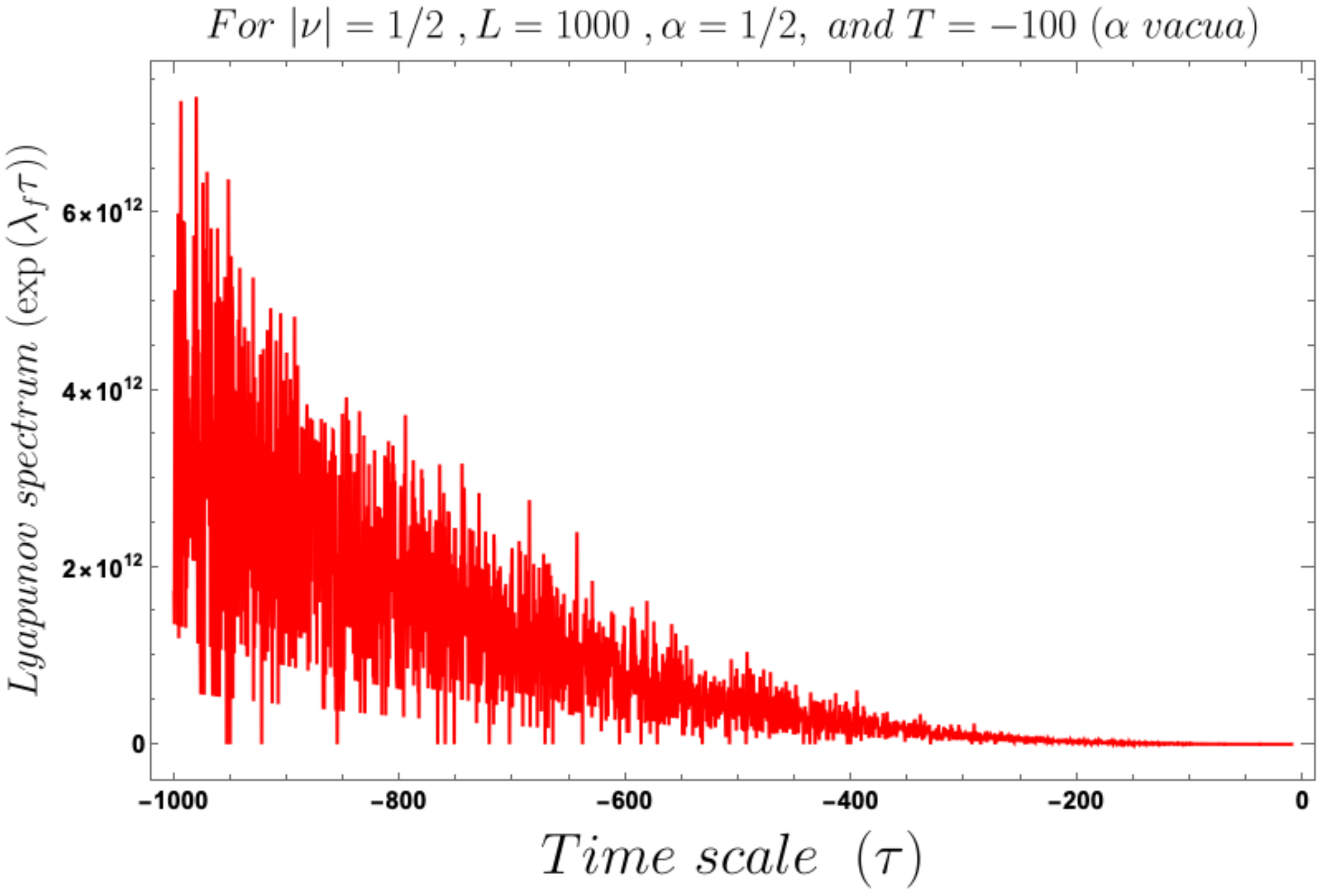}
   \includegraphics[width=17.3cm,height=4.8cm]{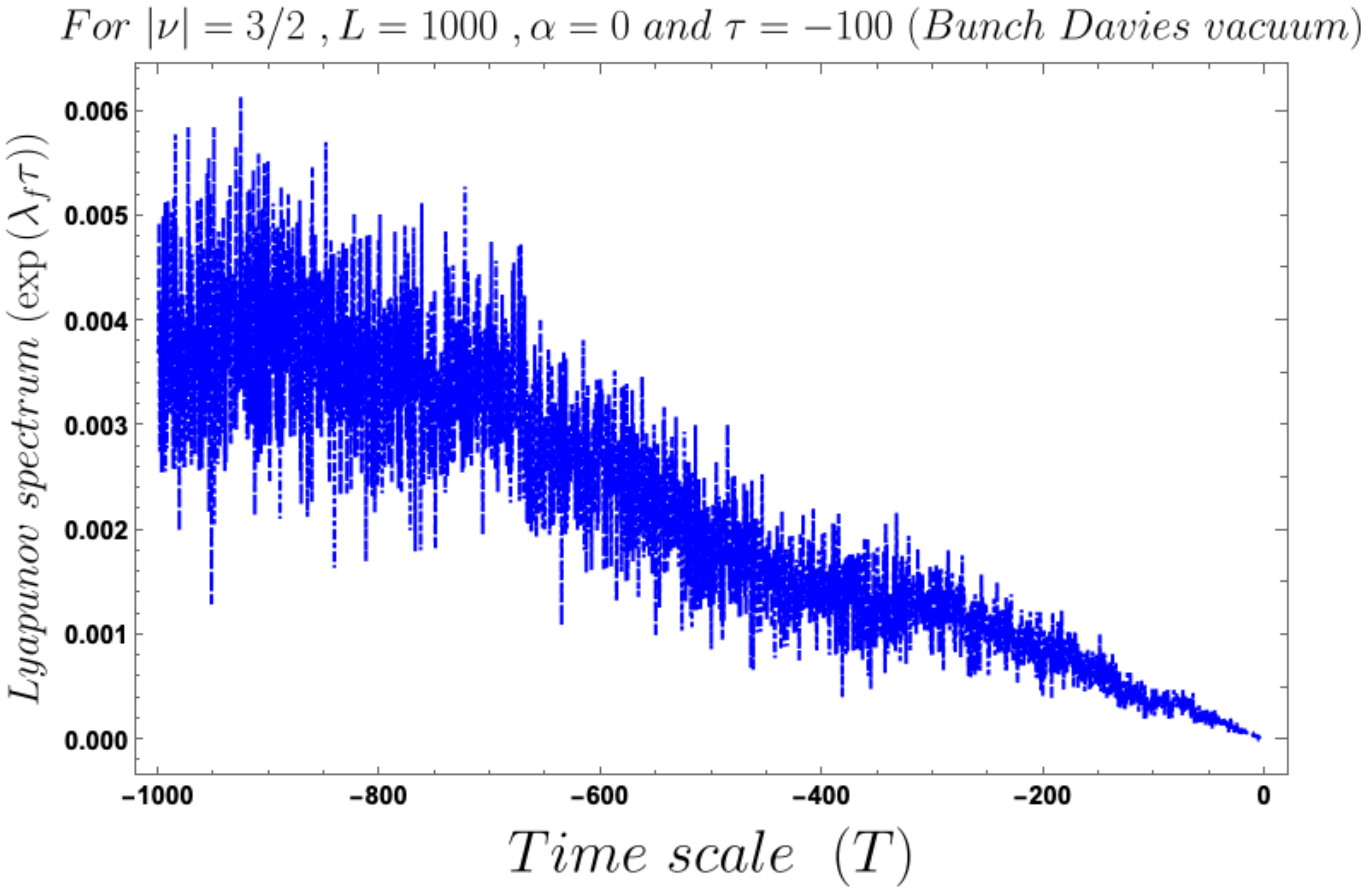}
        \includegraphics[width=16cm,height=4.8cm]{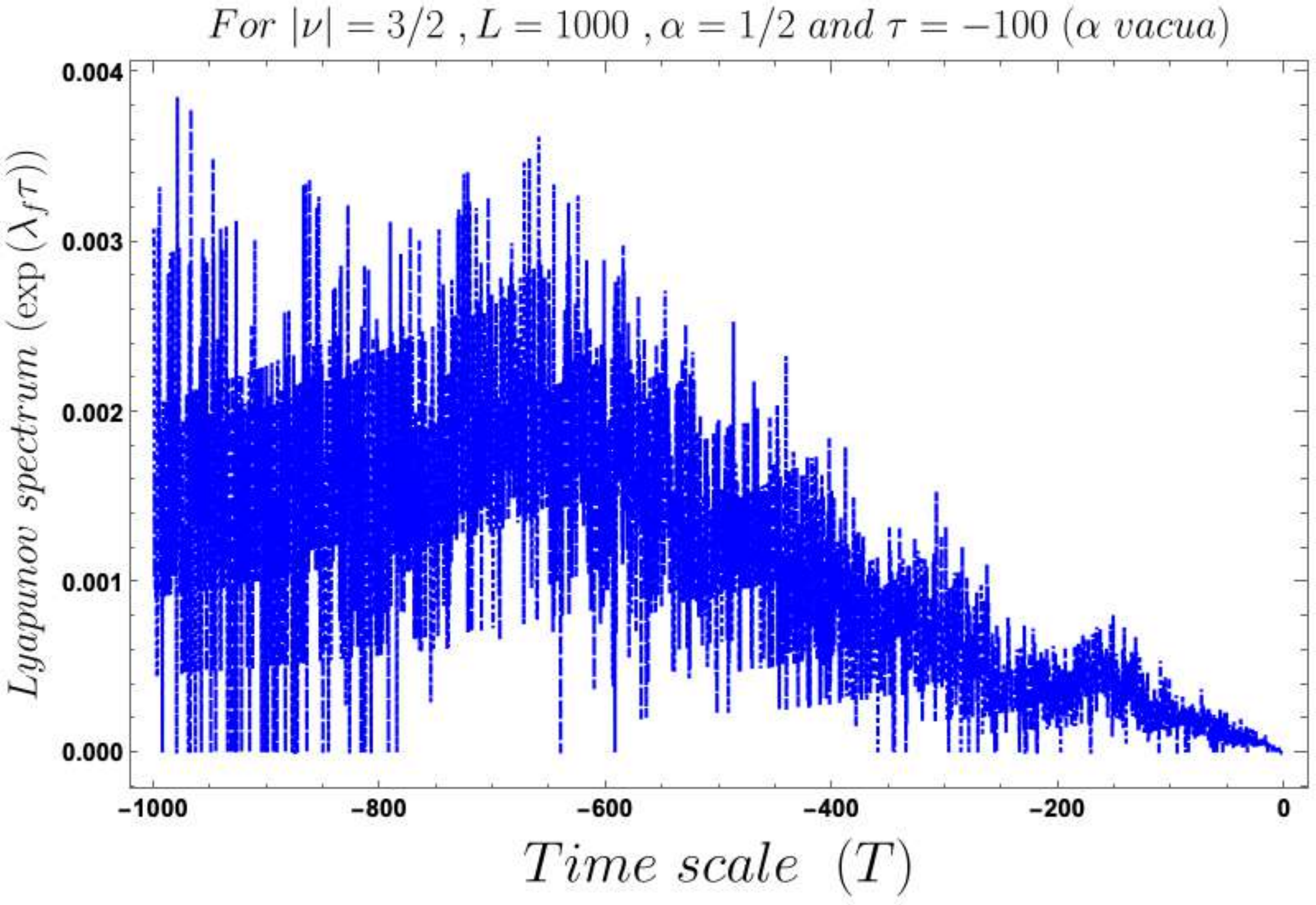}
    \caption{Behaviour of the Lyapunov spectrum with respect to the time scale $T$. Here we fix, mass parameter $\nu=-i/2,-3i/2$, cut-off scale $L=1000$, vacuum parameter $\alpha=0 ~({\rm Bunch~Davies~vacuum)}, 1/2~(\alpha~{\rm vacua})$.} 
      \label{fig:12}
\end{figure}
\begin{figure}[t!]
    \centering
        \centering
        \includegraphics[width=17.3cm,height=4.8cm]{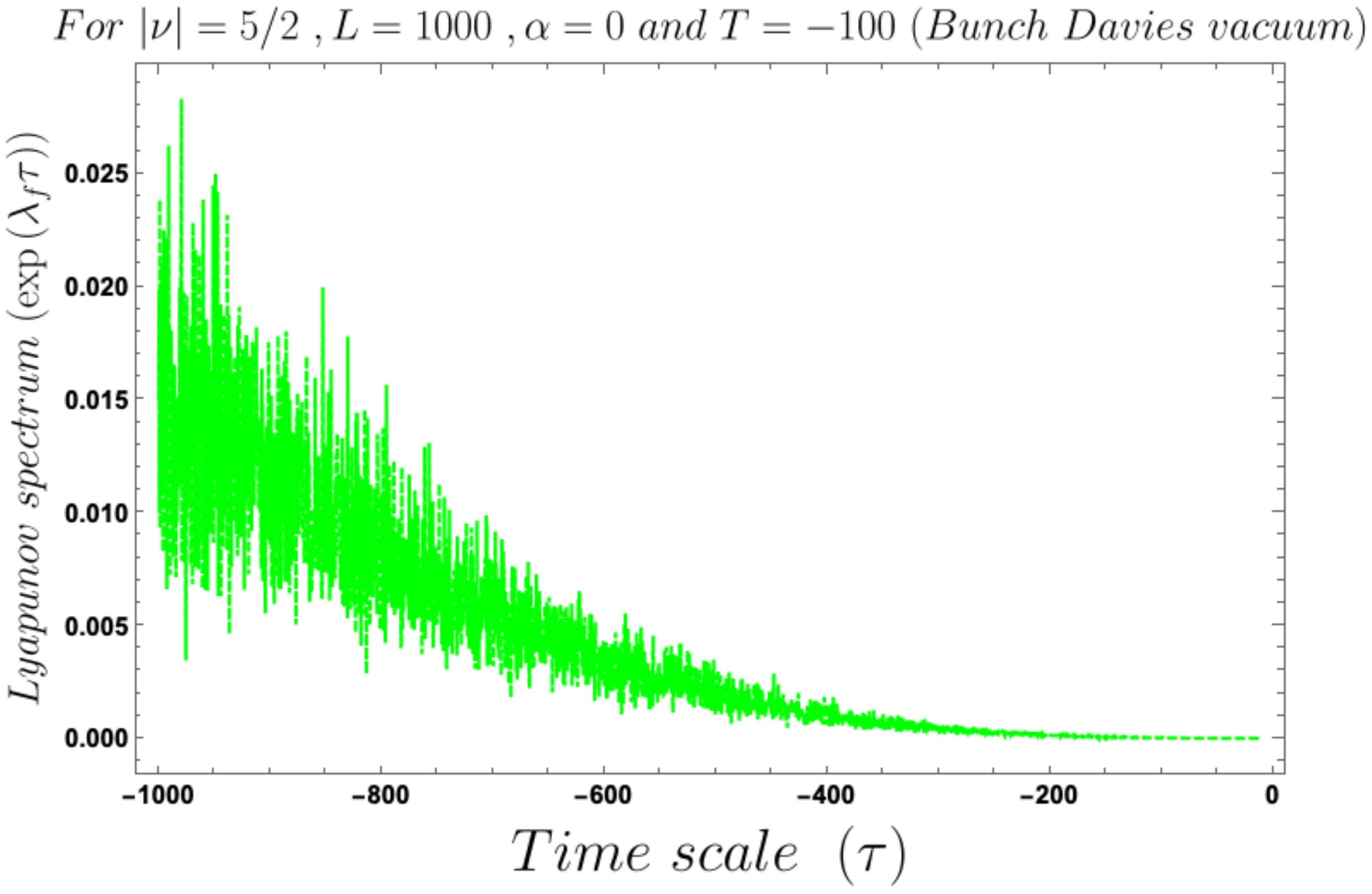}
        \includegraphics[width=16cm,height=4.8cm]{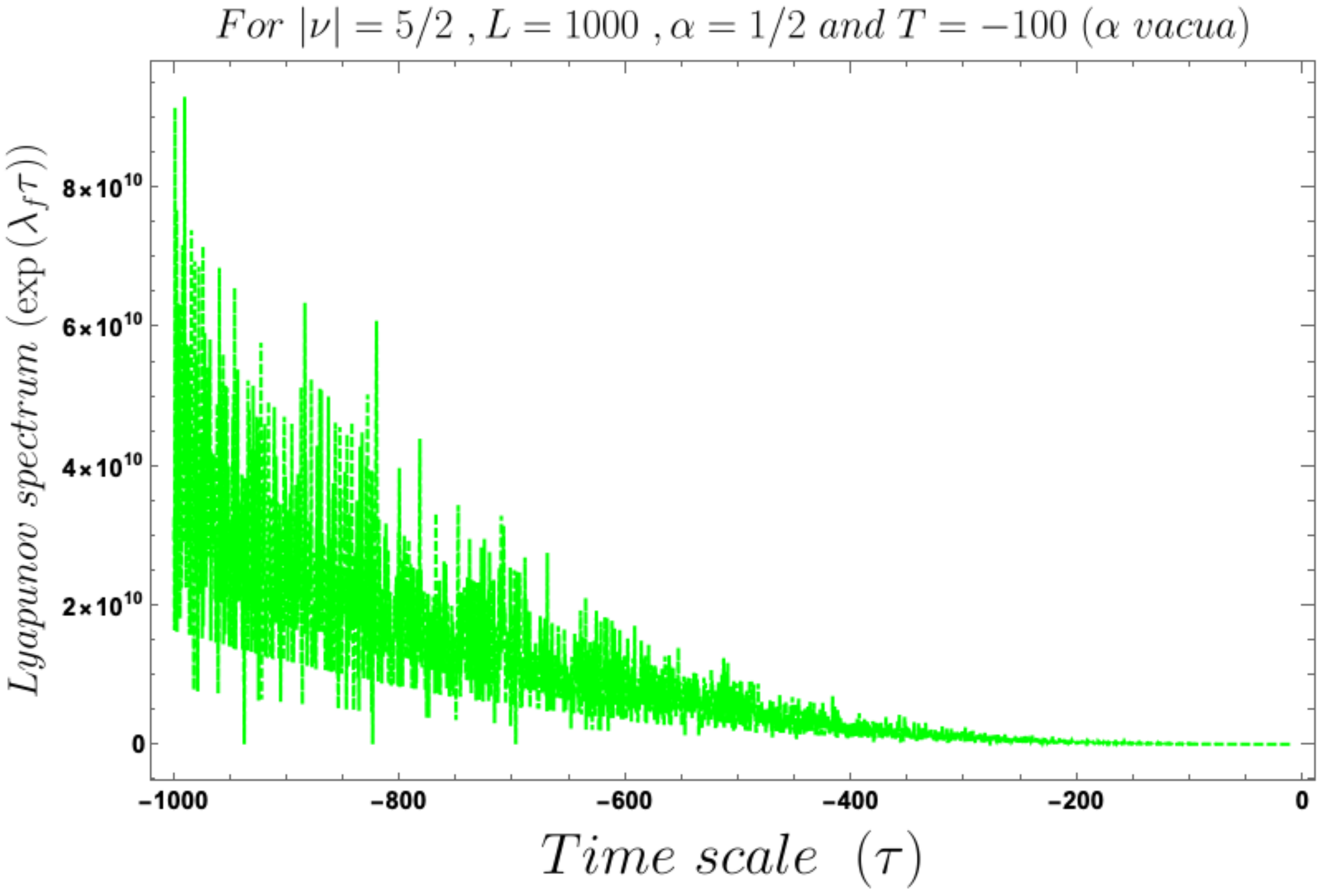}
   \includegraphics[width=16.3cm,height=4.8cm]{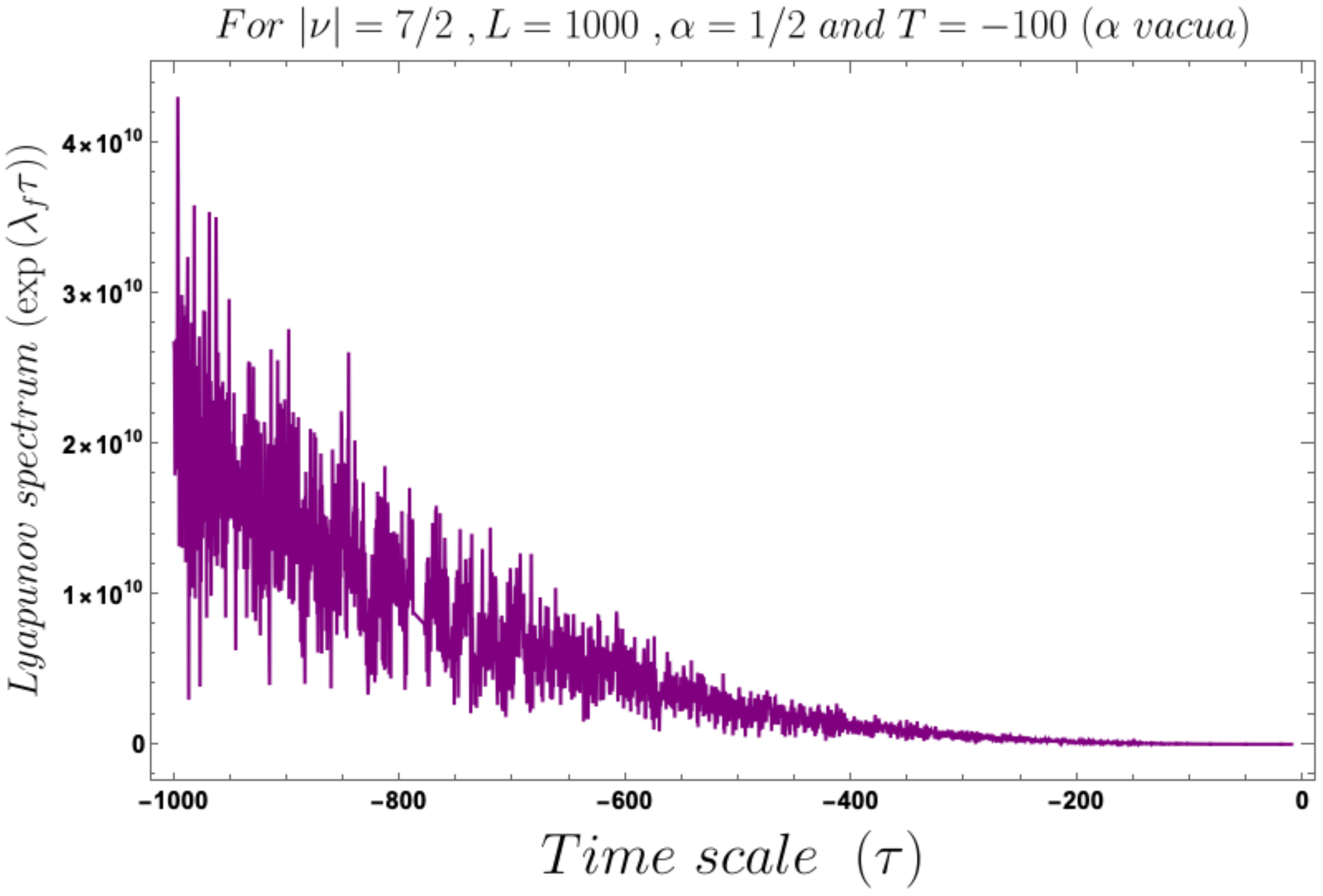}
        \includegraphics[width=17.36cm,height=4.8cm]{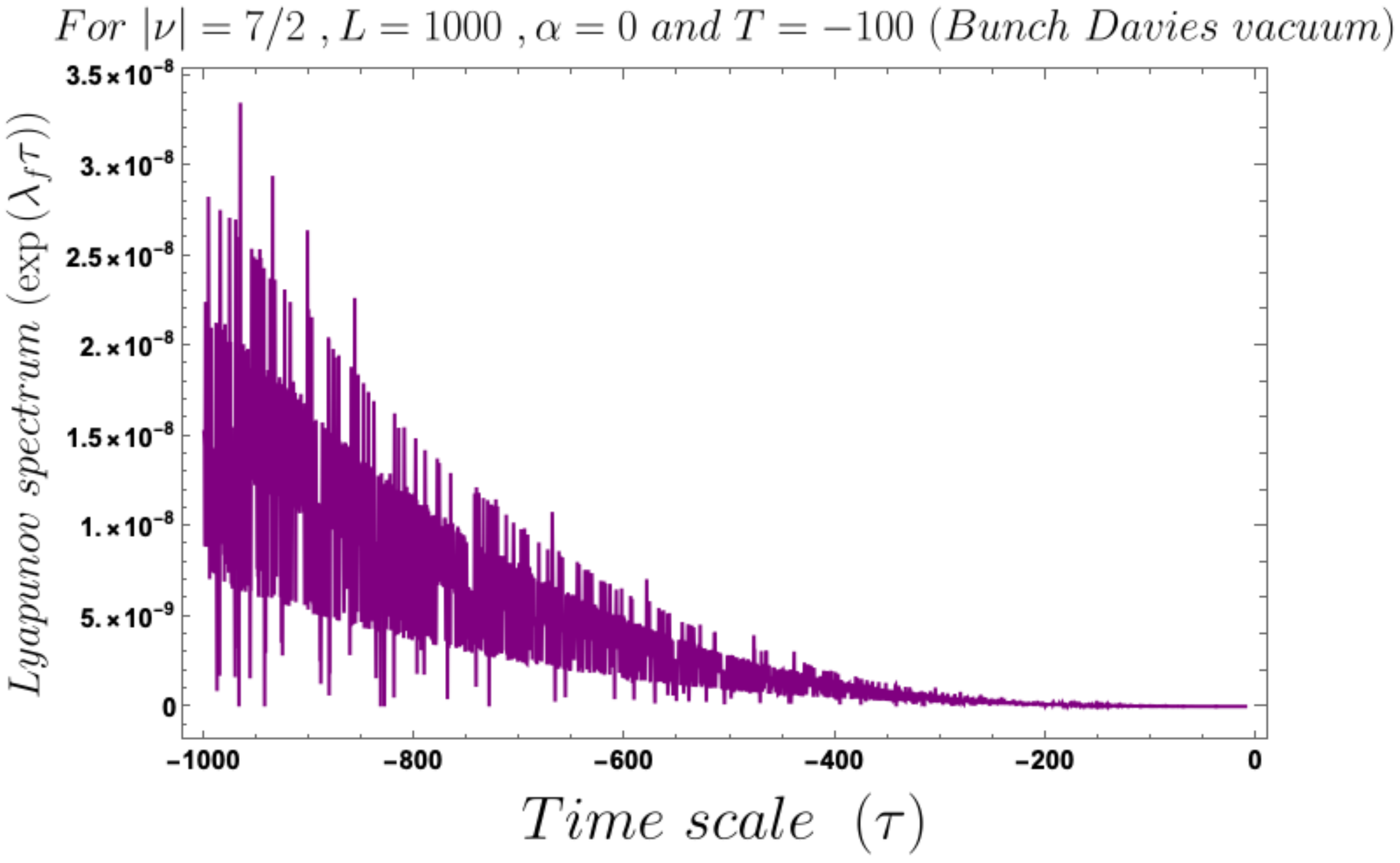}
    \caption{Behaviour of the Lyapunov spectrum with respect to the time scale $T$. Here we fix, mass parameter $\nu=-5i/2,-7i/2$, cut-off scale $L=1000$, vacuum parameter $\alpha=0 ~({\rm Bunch~Davies~vacuum)}, 1/2~(\alpha~{\rm vacua})$.}
      \label{fig:13}
\end{figure}
\begin{figure}[t!]
    \centering
        \centering
        \includegraphics[width=17.3cm,height=4.8cm]{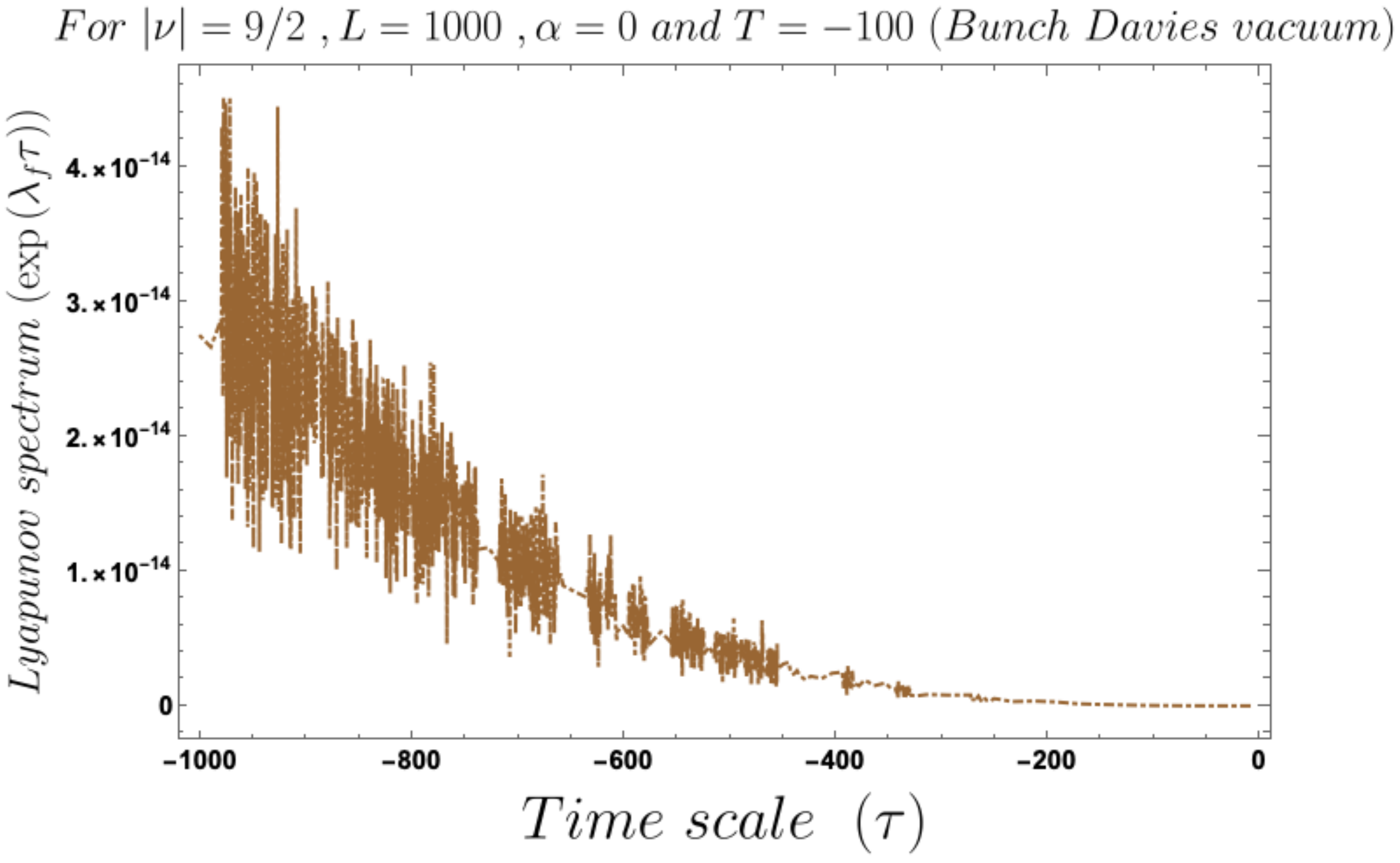}
        \includegraphics[width=16cm,height=4.8cm]{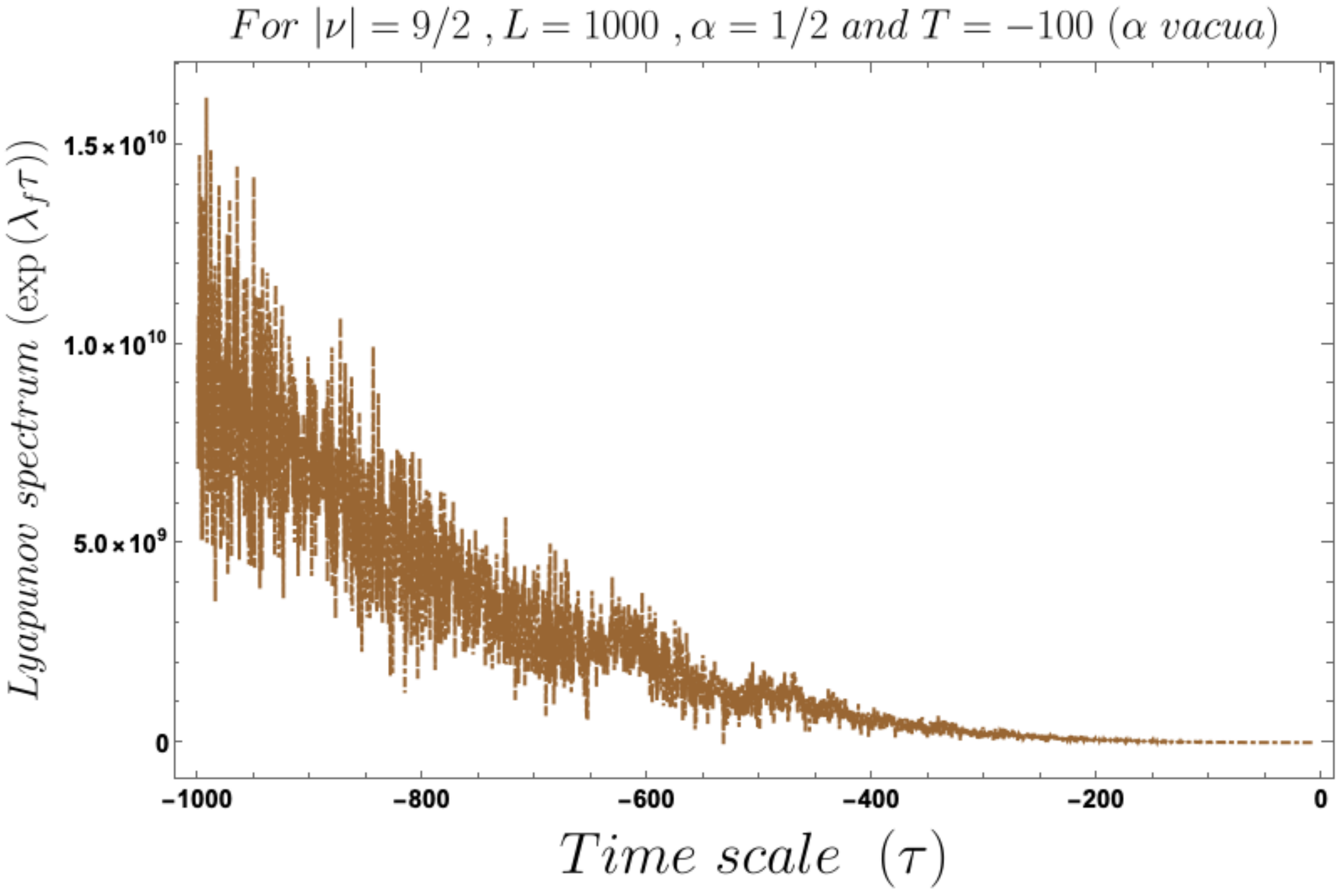}
    \caption{Behaviour of the Lyapunov spectrum with respect to the time scale $T$. Here we fix, mass parameter $\nu=-9i/2$, cut-off scale $L=1000$, vacuum parameter $\alpha=0 ~({\rm Bunch~Davies~vacuum)}, 1/2~(\alpha~{\rm vacua})$.}
      \label{fig:14}
\end{figure}
In this section, we give the physical interpretation of the obtained result for the {\it Lyapunov spectrum}~\footnote{In this context, by the phrase {\it Lyapunov spectrum}, we actually define two exponential factors, $\exp(\lambda_{f}\tau)$ and $\exp(\lambda_{\zeta}\tau))$, for the present computation.} factors computed in the context of primordial cosmology:
 \begin{itemize}
 \item First of all, in the cosmological OTOC when we have considered $\tau$ as the dynamical variable and have fixed the other time scale then for very late time of the cosmological evolution of our universe we have found a exponentially decaying behaviour with respect to the dynamical time scale under consideration. One can identify this phenomena with the quantum chaos, which is actually estimated by the decay coefficient appearing in the exponential factor. This coefficient is known as the {\it Lyapunov exponent} at late time scale. When the cosmological OTOC have started saturating  to a fixed non vanishing value at that point we fix the dynamical time scale and from the OTOC using the exponential decaying behaviour one can explicitly compute the expression for the {\it Lyapunov exponent} at fixed late time scale. But if anyone is interested to find out the whole conformal time dependent {\it Lyapunov spectrum} that can also be found using the cosmological OTOC computed in the previous section of this paper. We have explicitly computed this expression in this section for completeness.

\item Additionally, it is important to note that, since we have used two different cosmological perturbation variable for our present computation, we get two equivalent but not exactly same expression for the {\it Lyapunov spectrum}. If we compute the {\it Lyapunov spectrum} for the rescaled perturbation variable $f$ then we see that in the mass parameter $\nu$ all the information regarding the background cosmological frameworks, like, partially massless or heavy particle production during the epoch of inflation or the phenomena of reheating are encoded. We don't need to think about all of these phenomena separately. On the other hand, if we do the computation in terms of curvature perturbation variable $\zeta$ then we found that the the information regarding the background cosmological frameworks, as mentioned above we have to think separately. This is because in this framework, we have some additional contribution which is actually appearing from the definition of {\it Mukhanov Sasaki variable} which contain the scale factor of the cosmological expansion of De Sitter phase or reheating phase and the very small contribution comping from slow-roll factor, which is basically taking care of the small deviation from exact evolution in De Sitter space or during reheating and interpreted as $SO(1,4)$ conformal symmetry breaking parameter. Now if we neglect such small contribution then one can separately compute the expression for the {\it Lyapunov spectrum} from the variable $\zeta$.

\item Further, we have also have estimated the relative difference between the {\it Lyapunov spectra} computed from two different prescriptions. This difference is different for the stochastic particle production during inflation and reheating epoch respectively.

 \item  In our computed {\it Lyapunov spectrum} in the context of Cosmology two time scales are involved which are associated with the two operators, rescaled field variable and its canonically conjugate momenta. We have found that the behaviour with respect to $\tau_2=\tau$, using both Bunch Davies and $\alpha$ vacua as quantum vacuum state. See fig.~(\ref{fig:12}), fig.~(\ref{fig:13}) and fig.~(\ref{fig:14}) to know about the detailed conformal time dependent feature of the cosmological {\it Lyapunov spectrum}.  In the cosmological set up we expect the {\it Lyapunov spectrum} will decrease with respect to the conformal time scale and such decrease is following an exponential decay with respect to the conformal time scale then this carries the signature of quantum chaos in the context of primordial universe. 
 
 \item  Here we fix the time scale, $\tau_1=T$ and study the behaviour of {\it Lyapunov spectrum} with respect to the scale $\tau_2=\tau$ and we found that as we approach from very early universe to the late time scale the {\it Lyapunov spectrum} in the context of cosmology shows random decreasing on top of it with a exponentially decaying behaviour. This behaviour is quite consistent we the usual expectation of quantum mechanical chaos along with stochastic randomness. It is observed that in very early universe $\tau_2=\tau\rightarrow-\infty$, near to Big Bang all such random fluctuations are appearing with large amplitude and very significant to describe the time dependent behaviour of the {\it Lyapunov spectrum} during stochastic particle production during inflation and during the epoch of reheating like the previous case. If we wait for longer time, specifically at the late time scale it is observed from the fig.~(\ref{fig:12}), fig.~(\ref{fig:13}) and fig.~(\ref{fig:14}),  that {\it Lyapunov spectrum} decrease very fast with an exponentially decaying amplitude for the random or stochastic fluctuations. 
 
\item We have also observed that we can get the expected behaviour from the {\it Lyapunov spectrum} with respect to both the time scales when the mass parameter, $\nu$ can be analytically continued to $-i\nu$. 

\item Another important point we have to mention that, since the {\it Lyapunov spectrum} is computed from the cosmological OTOC in which the information of the initial quantum mechanical vacuum states are encoded, the final expressions for the {\it Lyapunov spectrum} computed for the $\alpha$ vacua and Bunch Davies vacua are different. This non-unique result appears due to the fact that in curved space-time definition of quantum mechanical vacuum state is unique.

\item Last but not the list, using the saturation bound on {\it Lyapunov exponent} obtained from MSS bound we have finally estimated the bound on the equilibrium temperature of our universe at late time scale. Basically this equilibrium temperature correspond to the starting point of saturation of the {\it Lyapunov spectrum} at the late time scale $\tau=\tau_{\rm Late}$.

 \end{itemize} 
 \section{Classical limit of micro-canonical OTO amplitudes and  related OTOC in Cosmology}
   \label{sec:OTO6}
In this section, our prime objective is to study the classical limit of the four-point cosmological OTOC that we have explicitly derived in this paper. This is a very important computation to understand the behaviour of the four-point cosmological OTOC in the classical limit and will gives us information regarding the fact that the result obtained in this section is perfectly consistent with the classical limit of quantum chaos, which is obviously a significant feature which we have clearly observed from the conformal time dependent behaviour of our derived result for cosmological four-point OTOC. In the following subsections we will explicitly compute this result and give a physical interpretation of the derived result in the framework of Cosmology.
\subsection{Computational strategy}
In this subsection we will illustrate our computational strategy to study the classical limit of the cosmological four-point OTOC derived in this paper:
\begin{enumerate}
\item First of all, we have to take the quantum to classical map of all the operators that we have used to compute the expression for cosmological four-point OTOC in this paper. We have already demonstrated clearly in the previous sections that during the computation of cosmological OTOC all the operators that we have considered considered by making use of canonical quantization of these operators. This means that we have written the quantum version of the modes in Fourier space in terms of creation and annihilation operators just like quantum harmonic oscillators and these creation and annihilation operators actually creates or destroy the quantum mechanical vacuum state, which we have taken as the usual well known {\it Bunch Davies vacuum} or the more general $\alpha$ vacua. Now during taking the classical limit of this formulation we all know that there is no concept of vacuum as such exists. As a result there is nothing called creation and annihilation of vacuum state in the context of classical limit of quantum fluctuation of cosmological perturbation theory. This further implies that writing down the quantum to classical map becomes very simple. For this purpose we just have to consider the classical mode function and its canonically conjugate momentum that we have derived already in the earlier section of this paper.

\item Next, instead of using commutator bracket square in the classical limit we have to use the {\it Poisson Bracket} of the classical variables which are appearing as an outcome of cosmological perturbation theory.

\item Further, we have to fix the definition and the representation of the trace in the classical limit. Since we don't have any complete set of basis and any vacuum state in the classical limit we have to be very careful to define the trace operation. Again this can be easily done by making use the basic concepts of classical statistical mechanics. In the classical limit the usual definition of quantum mechanical trace will be replaced by phase space volume, $df_{\bf k}(\tau)d\Pi_{\bf k}(\tau)/2\pi$ which is basically playing the role of path integral measure, ${\cal D}f_{\bf k}(\tau){\cal D}\Pi_{\bf k}(\tau)/2\pi$  in the classical field theory of scalar field in FLRW cosmological background. We set, $\hbar=1=h/2\pi$ in natural units, which implies $h=2\pi$ and this is the origin of the $2\pi$ factor which appearing in the denominator of the phase space volume or in the path integral measure. Also it is important to note that, only one power of $h=2\pi$ is appearing since we are dealing with a one dimensional parametric oscillator problem with conformal time time dependent frequency where the phase space is constructed out of two canonically conjugate variables, the cosmological perturbation variable and iyts conjugate momentum, which are derived from the classical field theory of the scalar fields in conformally and spatially flat FLRW background. As a consequence the phase space trajectory is lying in a two dimensional plane. If we deal with $N$ number of classical scalar fields in that case in the denominator $h^{N}$ factor will appear and in the large $N$ thermodynamic limit we will get physically consistent result in the purely classical limit of quantum version of Cosmology. Additionally, it important to note that, during this computation we have to take care of an additional thermal Boltzmann factor as we are doing the computation for the finite temperature extended version of the classical scalar field theory. 

\item  Next, by following the same procedure mentioned above we compute the expression for the classical thermal partition function for Cosmology, which perfectly matches with the the quantum mechanical thermal partition function for Cosmology.

\item Finally, we compute the expression for the normalisation factor of the classical version of the cosmological four-point OTOC using the above mentioned phase space formalism. This will help us to write down the normalised version of cosmological four-point OTOC in the classical limit. 
\end{enumerate}
\subsection{Classical limit of cosmological two-point ``in-in" OTO micro-canonical amplitudes}
In this subsection, our prime objective is to find out the classical limiting version of the two-point  micro-canonical amplitude computed from the quantum field theory side using the cosmological rescaled field variable and its canonically conjugate momentum as appearing in the context of cosmological perturbation theory.

To compute the classical limit the starting point is the {\it Poisson bracket} of these operators, which is given by the following expression:
\bea \left\{f({\bf x},\tau_1),\Pi({\bf x},\tau_2\right\}_{\bf PB}&=&\int\frac{d^3{\bf k}_1}{(2\pi)^3}\int\frac{d^3{\bf k}_2}{(2\pi)^3}~\exp(i({\bf k}_1+{\bf k}_2).{\bf x})~\nonumber\\
&&\left[\left\{f_{{\bf k}_1}(\tau_1),\Pi_{{\bf k}_2}(\tau_2)\right\}_{\bf PB}+\left\{f_{{\bf k}_2}(\tau_1),\Pi_{{\bf k}_1}(\tau_2)\right\}_{\bf PB}\right].~~~~~~~\eea
In the Fourier transformed space, the {\it Poisson bracket} is given by the following expression:
\bea \left\{f_{{\bf k}_1}(\tau_1),\Pi_{{\bf k}_2}(\tau_2)\right\}_{\bf PB}&=&(2\pi)^3\delta^{3}({\bf k}_1+{\bf k}_2)~{\bf R}(\tau_1,\tau_2),\\
 \left\{f_{{\bf k}_2}(\tau_1),\Pi_{{\bf k}_1}(\tau_2)\right\}_{\bf PB}&=&(2\pi)^3\delta^{3}({\bf k}_2+{\bf k}_1)~{\bf R}(\tau_1,\tau_2), \eea
 which also implies that:
 \bea \left\{f_{{\bf k}_1}(\tau_1),\Pi_{{\bf k}_2}(\tau_2)\right\}_{\bf PB}&=&\left\{f_{{\bf k}_2}(\tau_1),\Pi_{{\bf k}_1}(\tau_2)\right\}_{\bf PB}(2\pi)^3\delta^{3}({\bf k}_1+{\bf k}_2)~{\bf R}(\tau_1,\tau_2),\eea
 since the three dimensional Dirac Delta function is symmetric.
 
Now for  the further purpose we define the two-point random classical correlation function ${\bf R}(\tau_1,\tau_2)$ by the following expression:
\bea {\bf R}(\tau_1,\tau_2):&=&{\bf W}(\tau_1-\tau_2)~\exp\left(-\frac{\lambda_f (|\tau_1|+|\tau_2|)}{2}\right),\eea
where $\lambda_{f}$ is the {\it Lyapunov exponent} in the classical version of the theory, which justifies the decay of the noise correlation function in the classical limit. Here, ${\bf W}(\tau_1-\tau_2)$ is the window function which is defined as:
\bea {\bf W}(\tau_1-\tau_2)&=&\sqrt{\langle \eta_{\bf Noise}(\tau_1)\eta_{\bf Noise}(\tau_2)\rangle},\eea
where the two-point noise correlation at the classical level is given by the following expression:
\bea \langle \eta_{\bf Noise}(\tau_1)\eta_{\bf Noise}(\tau_2):={\bf G}_{\bf kernel}(\tau_1-\tau_2),\eea
where ${\bf G}_{\bf kernel}(\tau_1-\tau_2)$ is known as the noise kernel which is time translational symmetric in nature. The detailed properties of this noise kernel in terms of the Gaussian and non-Gaussian distributions are discussed in the previous subsection and in the Appendix. As a consequence, we get following symmetry property:
\bea {\bf R}(\tau_1,\tau_2)={\bf R}(\tau_2,\tau_1).\eea
\subsection{Classical limit of cosmological four-point ``in-in" OTO micro-canonical amplitudes}
 In this section, our prime objective is to explicitly compute the classical limiting version of the four-point ``in-in" OTO micro-canonical amplitudes appearing in the expression or OTOC. To serve this purpose in the classical limit we explicitly compute the following square of the Poisson brackets instead of computing the commutator brackets which is the key starting point in the quantum regime, given by: 
  \bea  \left\{{f}({\bf x},\tau_1),{\Pi}({\bf x},\tau_2)\right\}^2_{\bf PB}&=&\left\{{f}({\bf x},\tau_1),{\Pi}({\bf x},\tau_2)\right\}_{\bf PB}\left\{{f}({\bf x},\tau_1),{\Pi}({\bf x},\tau_2)\right\}_{\bf PB}.\eea
  Here the subscript {\bf PB} stands 
   for the {\it Poisson bracket} of two classical object of interests which are the rescaled cosmological perturbation variable $f({\bf x},\tau_1)$ and its canonically conjugate momentum $\Pi({\bf x},\tau_2)$ , which are separated in time scale in a specific cosmological perturbation scheme written in, $\delta\phi=0$ gauge. So we are dealing with two space time dependent objects of interest in the framework of classical field theory written in the classically perturbed spatially flat FLRW cosmological background. One important thing we have to mention that, before perturbation the fundamental object (in this context a classical scalar field) is placed in a homogeneous isotropic spatially flat FLRW cosmological background, where as an outcome the field is only time dependent, no space dependence appears. But in the context of cosmological perturbation theory, once we perturb the space-time metric the perturb quantities always become space-time dependent. This is the main reason for which we are considering the two perturbed variables are space-time dependent which is appearing in the {\it Poisson bracket} and consequently in the classical limit of the four-point OTO micro-canonical ampltudes in Cosmology. Now, since we are not studying the theory not in coordinate space, we next perform Fourier transformation on these object of interests to convert them into momentum space.
   
To perform this  Fourier transformation we use the following convention, which is given by:
 \bea &&\hat{f}({\bf x},\tau_1)=\int \frac{d^3{\bf k}}{(2\pi)^3}~\exp(i{\bf k}.{\bf x})~\hat{f}_{{\bf k}}(\tau_1),\\
 &&\hat{\Pi}({\bf x},\tau_1)=\partial_{\tau_1}\hat{f}({\bf x},\tau_1)=\int \frac{d^3{\bf k}}{(2\pi)^3}~\exp(i{\bf k}.{\bf x})~\partial_{\tau_1}\hat{f}_{{\bf k}}(\tau_1)=\int \frac{d^3{\bf k}}{(2\pi)^3}~\exp(i{\bf k}.{\bf x})~\hat{\Pi}_{{\bf k}}(\tau_1),~~~~~~~~~~~~~\eea
 which will be very useful for the computation of the classical limiting result of four-point OTOC in terms of the square of the Poisson bracket.
 
 Consequently, we get the following simplified result:
 \bea && \left\{{f}({\bf x},\tau_1),{\Pi}({\bf x},\tau_2)\right\}^2_{\bf PB}\nonumber\\
 &&=\int \frac{d^3{\bf k}_1}{(2\pi)^3}\int \frac{d^3{\bf k}_2}{(2\pi)^3}\int \frac{d^3{\bf k}_3}{(2\pi)^3}\int \frac{d^3{\bf k}_4}{(2\pi)^3}~\exp\left(i({\bf k}_1+{\bf k}_2+{\bf k}_3+{\bf k}_4).{\bf x}\right)\nonumber\eea
 \bea
 &&~~~~~~\left[\left\{f_{{\bf k}_{1}} (\tau_1),\Pi_{{\bf k}_{2}} (\tau_2)\right\}_{\bf PB}\left\{f_{{\bf k}_{3}}(\tau_1),\Pi_{{\bf k}_4} (\tau_2)\right\}_{\bf PB}+\left\{f_{{\bf k}_{1}} (\tau_1),\Pi_{{\bf k}_{3}} (\tau_2)\right\}_{\bf PB}\left\{f_{{\bf k}_{2}}(\tau_1),\Pi_{{\bf k}_4} (\tau_2)\right\}_{\bf PB}\right.\nonumber\\&& \left.~~~~+\left\{f_{{\bf k}_{1}} (\tau_1),\Pi_{{\bf k}_{4}} (\tau_2)\right\}_{\bf PB}\left\{f_{{\bf k}_{3}}(\tau_1),\Pi_{{\bf k}_2} (\tau_2)\right\}_{\bf PB}+\left\{f_{{\bf k}_{2}} (\tau_1),\Pi_{{\bf k}_{3}} (\tau_2)\right\}_{\bf PB}\left\{f_{{\bf k}_{4}}(\tau_1),\Pi_{{\bf k}_1} (\tau_2)\right\}_{\bf PB}\right.\nonumber\\&& \left.~~~~+\left\{f_{{\bf k}_{2}} (\tau_1),\Pi_{{\bf k}_{1}} (\tau_2)\right\}_{\bf PB}\left\{f_{{\bf k}_{4}}(\tau_1),\Pi_{{\bf k}_3} (\tau_2)\right\}_{\bf PB}+\left\{f_{{\bf k}_{2}} (\tau_1),\Pi_{{\bf k}_{4}} (\tau_2)\right\}_{\bf PB}\left\{f_{{\bf k}_{1}}(\tau_1),\Pi_{{\bf k}_3} (\tau_2)\right\}_{\bf PB}\right.\nonumber\\&& \left.~~~~+\left\{f_{{\bf k}_{3}} (\tau_1),\Pi_{{\bf k}_{1}} (\tau_2)\right\}_{\bf PB}\left\{f_{{\bf k}_{4}}(\tau_1),\Pi_{{\bf k}_2} (\tau_2)\right\}_{\bf PB}+\left\{f_{{\bf k}_{3}} (\tau_1),\Pi_{{\bf k}_{2}} (\tau_2)\right\}_{\bf PB}\left\{f_{{\bf k}_{1}}(\tau_1),\Pi_{{\bf k}_4} (\tau_2)\right\}_{\bf PB}\right.\nonumber\\&& \left.~~~~+\left\{f_{{\bf k}_{3}} (\tau_1),\Pi_{{\bf k}_{4}} (\tau_2)\right\}_{\bf PB}\left\{f_{{\bf k}_{1}}(\tau_1),\Pi_{{\bf k}_2} (\tau_2)\right\}_{\bf PB}+\left\{f_{{\bf k}_{4}} (\tau_1),\Pi_{{\bf k}_{1}} (\tau_2)\right\}_{\bf PB}\left\{f_{{\bf k}_{2}}(\tau_1),\Pi_{{\bf k}_3} (\tau_2)\right\}_{\bf PB}\right.\nonumber\\&& \left.~~~~+\left\{f_{{\bf k}_{4}} (\tau_1),\Pi_{{\bf k}_{2}} (\tau_2)\right\}_{\bf PB}\left\{f_{{\bf k}_{3}}(\tau_1),\Pi_{{\bf k}_1} (\tau_2)\right\}_{\bf PB}+\left\{f_{{\bf k}_{4}} (\tau_1),\Pi_{{\bf k}_{3}} (\tau_2)\right\}_{\bf PB}\left\{f_{{\bf k}_{2}}(\tau_1),\Pi_{{\bf k}_1} (\tau_2)\right\}_{\bf PB}\right]\nonumber\\
 &&=(2\pi)^6 \int \frac{d^3{\bf k}_1}{(2\pi)^3}\int \frac{d^3{\bf k}_2}{(2\pi)^3}\int \frac{d^3{\bf k}_3}{(2\pi)^3}\int \frac{d^3{\bf k}_4}{(2\pi)^3}~\exp\left(i({\bf k}_1+{\bf k}_2+{\bf k}_3+{\bf k}_4).{\bf x}\right)\nonumber\\
 &&~~~~~~~~~~~~~~~~~~~~~~~~~\left[\delta^{3}({\bf k}_1+{\bf k}_2)\delta^3({\bf k}_3+{\bf k}_4)+\delta^{3}({\bf k}_1+{\bf k}_3)\delta^3({\bf k}_2+{\bf k}_4)\right.\nonumber\\&& \left.~~~~~~~~~~~~~~~~~~~~~~~~~+\delta^{3}({\bf k}_1+{\bf k}_4)\delta^3({\bf k}_3+{\bf k}_2)+\delta^{3}({\bf k}_2+{\bf k}_3)\delta^3({\bf k}_4+{\bf k}_1)\right.\nonumber\\&& \left.~~~~~~~~~~~~~~~~~~~~~~~~~+\delta^{3}({\bf k}_2+{\bf k}_1)\delta^3({\bf k}_4+{\bf k}_3)+\delta^{3}({\bf k}_2+{\bf k}_4)\delta^3({\bf k}_1+{\bf k}_3)\right.\nonumber\\&& \left.~~~~~~~~~~~~~~~~~~~~~~~~~+\delta^{3}({\bf k}_3+{\bf k}_1)\delta^3({\bf k}_4+{\bf k}_2)+\delta^{3}({\bf k}_3+{\bf k}_2)\delta^3({\bf k}_1+{\bf k}_4)\right.\nonumber\\&& \left.~~~~~~~~~~+\delta^{3}({\bf k}_3+{\bf k}_4)\delta^3({\bf k}_1+{\bf k}_2)+\delta^{3}({\bf k}_4+{\bf k}_1)\delta^3({\bf k}_2+{\bf k}_3)\right.\nonumber\\&& \left.~~~~~~+\delta^{3}({\bf k}_4+{\bf k}_2)\delta^3({\bf k}_3+{\bf k}_1)+\delta^{3}({\bf k}_4+{\bf k}_3)\delta^3({\bf k}_2+{\bf k}_1)\right]{\bf G}_{\bf Kernel}(\tau_1-\tau_2)\exp\left(-\lambda_{f}[|\tau_1|+|\tau_2|]\right).~~~~~~~~\eea  
 where we have used the following fact:
 \bea &&\left\{f_{{\bf k}_{i}} (\tau_1),\Pi_{{\bf k}_{j}} (\tau_2)\right\}_{\bf PB}=(2\pi)^3\delta^{3}({\bf k}_i+{\bf k}_j){\bf W}(\tau_1-\tau_2)\exp\left(-\frac{\lambda_{f}(|\tau_1|+|\tau_2|)}{2}\right).\nonumber\\
&&~~~~~~~~~~~~~~~~~~~~~~~~~~~~~~~~~~~~~~~~~~~~{\rm where}~~~i\neq j~ \forall~ i,j,=1,2,3,4.~~~~~~~~~~~\eea 
Here $\lambda_{f}$ is the classical version of the {\it Lyapunov exponent} which measure the strength of the chaos in the classical limit of the quantum chaos which we have discussed in the earlier section.

It is important to point that, the individual contributions appearing in the previously mentioned $12$ contributions can be evaluated as:
\bea &&\left\{f_{{\bf k}_{i}} (\tau_1),\Pi_{{\bf k}_{j}} (\tau_2)\right\}_{\bf PB}\left\{f_{{\bf k}_{l}}(\tau_1),\Pi_{{\bf k}_m} (\tau_2)\right\}_{\bf PB}=(2\pi)^6\delta^{3}({\bf k}_i+{\bf k}_j)\delta^3({\bf k}_l+{\bf k}_m)\nonumber\\
&&~~~~~~~~~~~~~~~~~~~~~~~~~~~~~~~~~~~~~~~~~~~~~~~~~~~~~~~~~~~~~~~~~~~~~~~{\bf W}^2(\tau_1-\tau_2)\exp\left(-\lambda_{f}[|\tau_1|+|\tau_2|]\right).\nonumber\\
&&~~~~~~~~~~~~~~~~~~~~~~~~~~~~~~~~~~~~~~~~~~~~~~~~~~~~~~~=(2\pi)^6\delta^{3}({\bf k}_i+{\bf k}_j)\delta^3({\bf k}_l+{\bf k}_m)\nonumber\\
&&~~~~~~~~~~~~~~~~~~~~~~~~~~~~~~~~~~~~~~~~~~~~~~~~
~~~~~~~~~~~~~~~~~~~~~~~{\bf G}_{\bf Kernel}(\tau_1-\tau_2)\exp\left(-\lambda_{f}[|\tau_1|+|\tau_2|]\right).\nonumber\\
&&~~~~~~~~~~~~~~~~~~~~~~~~~~~~~~~~~~~~~~~~~~~~{\rm where}~~~i\neq j \neq l \neq m~ \forall~ i,j,l,m=1,2,3,4.~~~~~~~~~~~\eea
In this computation we introduce a conformal time dependent coloured noise kernel ${\bf G}_{\bf Kernel}(\tau_1-\tau_2)$, which is defined as:
\bea
{\bf G}_{\bf Kernel}(\tau_1-\tau_2)={\bf W}^2(\tau_1-\tau_2).
\eea
Here the coloured noise have the following properties:
\begin{enumerate}
\item For coloured noise two point classical correlation function is time translation invariant.

\item For this random coloured noise, at the classical level we have:
\bea &&\langle \eta_{\bf CN}(\tau_1)\rangle=0,\\
&&\langle \eta_{\bf CN}(\tau_1)\eta_{\bf CN}(\tau_2)\rangle={\bf G}_{\bf Kernel}(\tau_1-\tau_2)=W^2(\tau_1-\tau_2)\rangle\neq 0,\\
&&\langle  \eta_{\bf CN}(\tau_1)\cdots\cdots \eta_{\bf CN}(\tau_N)  \rangle={\bf A}_{\bf kernel}(\tau_1,\cdots,\tau_N)~\delta(\tau_1+\cdots\cdots+\tau_{N})~~~~\forall~N>2.~~~~~~~~\eea
where $\eta_{\bf CN}(\tau)$ is the conformal time dependent coloured noise. 
Here we have assumed that the time dependent kernel is in general have some non-Gaussian probability distribution profile. In the computation, ${\bf G}_{\bf Kernel}(\tau_1-\tau_2)$ is the Green's function which is appearing from the two point classical correlation from the coloured noise defined at two different time scales. In Cosmology, this conformal time dependent function sometimes identified to be the power spectrum as well and for coloured noise it is expected to be non-Gaussian from the above mentioned properties. Also, ${\bf A}_{\bf kernel}(\tau_1,\cdots,\tau_N)$ is the amplitude of the spectrum of any $N>2$ point classical correlations. Here all point amplitudes are non-zero for non-Gaussian coloured noise contributions, which are sourced from random, chaotic and stochastic fluctuations in the classical limit in cosmological paradigm.
\end{enumerate}
On the other hand, if we consider the Gaussian white noise instead of having non-Gaussian coloured noise, in that case we have the following properties:
\begin{enumerate}
\item For white noise two point classical correlation function is time translation invariant, which is exactly same as the coloured noise case.

\item For this random white noise, at the classical level we have:
\bea &&\langle \eta_{\bf WN}(\tau_1)\rangle=0,\\
&&\langle \eta_{\bf WN}(\tau_1)\eta_{\bf WN}(\tau_2)\rangle={\bf G}_{\bf Kernel}(\tau_1-\tau_2)=W^2(\tau_1-\tau_2)\rangle={\bf B}~\delta(\tau_1-\tau_2),\\
&&\langle  \eta_{\bf WN}(\tau_1)\cdots\cdots \eta_{\bf WN}(\tau_N)  \rangle=0~~~~\forall~N=3,5,7,\cdots,\\
&&\langle  \eta_{\bf WN}(\tau_1)\cdots\cdots \eta_{\bf WN}(\tau_N)  \rangle={\bf B}_{\bf kernel}(\tau_1,\cdots,\tau_N)~\delta(\tau_1+\cdots\cdots+\tau_{N})~~~~\forall~N=4,6,8,\cdots.~~~~~~~~\eea
where $\eta_{\bf WN}(\tau)$ is the conformal time dependent white noise. 
Here we have assumed that the time dependent kernel is in general have some Gaussian probability distribution profile. Also, ${\bf B}$ and ${\bf B}_{\bf kernel}(\tau_1,\cdots,\tau_N)$ are the amplitude of the spectrum of any $N=2$ and $N=4,6,8,\cdots$ even point classical correlations. Due to the Gaussian probability distribution profile all the odd point classical correlation of white noise exactly vanish trivially. Consequently, all even point amplitudes are non-zero and all odd point amplitudes become zero for Gaussian white noise contributions, which are again sourced from random, chaotic and stochastic fluctuations in the classical limit in cosmological paradigm.
\end{enumerate}
Here we have actually presented the classical limit of the Fourier transformed version of the four-point cosmological ``in-in" OTO micro-canonical amplitude. From the mentioned result we get the following characteristics:
\begin{enumerate}

\item In the obtained result, we have introduced a conformal time dependent noise kernel phenomenologically to describe the randomness and chaoticity of the fluctuations at the classical limit.

\item Also, we have included an exponentially conformal time dependent decay term as we are interested in to explain the classical limit of the quantum chaotic behaviour that we have obtained in the context of cosmological four-point OTOC at late time scale of our universe. To get a similar behaviour we have included this decaying feature in the classical version of the four-point OTO micro-canonical amplitude. In usual literature this is appearing as an exponential time dependent growth factor. But since we are dealing with conformal time scale in the context of Cosmology, $-\infty<\tau<0$, we write it as, $\tau=-|\tau|$ for the simplicity. So this negative sign in the conformal time scale is responsible for the decaying feature of the classical limit of the four-point cosmological OTOC, instead of having an exponential growth.

\item If we compare the obtained result in the classical limit with the quantum version of the four-point ``in-in" OTO micro-canonical amplitude then we can see that for both the cases we get twelve contributions, but amplitudes in classical limit for all of these contributions are same. On the other hand, in the quantum version of the amplitude we have different individual contributions for these mentioned twelve terms. Additionally, we can observe another similarity in these two results are that, for both the cases momentum conservation is applied through the momentum dependent three dimensional Dirac Delta functions.
\end{enumerate}
\subsection{Cosmological micro-canonical partition function: Classical version}
\subsubsection{Classical micro-canonical partition function in terms of rescaled field variable}
In this section our aim is to derive the expression for the partition function for the cosmology in the classical regime. In terms of the rescaled cosmological perturbation field variable we define the following classical partition function for Cosmology: 
\bea Z_{\bf Classical}(\beta;\tau_1):&=&\int \int \frac{{\cal D}f{\cal D}\Pi}{2\pi}~\exp\left(-\beta H\right)\nonumber\\
&=&\prod_{{\bf k}}\exp\left(-\beta\left[\frac{E_{\bf k}(\tau_1)}{2}+\frac{1}{\beta}\ln\left(1-\exp(-\beta E_{\bf k}(\tau_1)\right)\right]\right)\nonumber\\
&=&\prod_{{\bf k}}\exp\left(-\ln\left(2\sinh\frac{\beta E_{\bf k}(\tau_1)}{2}\right)\right)\nonumber\\
&=&\exp\left(-\int d^3{\bf k} ~\ln\left(2\sinh \frac{\beta E_{\bf k}(\tau_1)}{2}\right)\right).~~~~\eea
This implies that:
\bea \hll{Z_{\bf Classical}(\beta;\tau_1)=:Z_{\bf BD}(\beta;\tau_1):=|\cosh\alpha|~:Z_{\alpha}(\beta;\tau_1):~~~~\forall ~~\alpha}~.\eea
Now from the above expression we found that the expression for the classical partition function and normal ordered partition function for Cosmology is exactly same. 
\subsubsection{Classical micro-canonical partition function in terms of curvature perturbation field variable}
To construct the classical partition function in terms of the curvature perturbation field variable and its conjugate momenta we are going to follow similar procedure and its given by the following expression:
\bea Z^{\zeta}_{\bf Classical}(\beta;\tau_1):&=&\int \int \frac{{\cal D}\zeta{\cal D}\Pi_{\zeta}}{2\pi}~\exp\left(-\beta H\right)\nonumber\\
&=&\prod_{{\bf k}}\exp\left(-\beta\left[\frac{z^2(\tau_1)E^{\zeta}_{\bf k}(\tau_1)}{2}+\frac{1}{\beta}\ln\left(1-\exp(-\beta z^2(\tau_1)E^{\zeta}_{\bf k}(\tau_1)\right)\right]\right)\nonumber\\
&=&\prod_{{\bf k}}\exp\left(-\ln\left(2\sinh\frac{\beta z^2(\tau_1)E^{\zeta}_{\bf k}(\tau_1)}{2}\right)\right)\nonumber\\ 
&=&\exp\left(-\int d^3{\bf k} ~\ln\left(2\sinh \frac{\beta z^2(\tau_1)E^{\zeta}_{\bf k}(\tau_1)}{2}\right)\right).~~~~\eea
This further implies that:
\bea &&\hll{Z^{\zeta}_{\bf Classical}(\beta;\tau_1)=:Z^{\zeta}_{\bf BD}(\beta;\tau_1):=|\cosh\alpha|~:Z_{\alpha}(\beta;\tau_1):~~~~\forall ~~\alpha}~\nonumber\\
&&~~~~~~~~~~~~~~~~~~~~\hll{\neq Z_{\bf Classical}(\beta;\tau_1)}~.\eea 
So the classical partition functions for Cosmology computed in terms of two perturbation variables are not same because of the presence of {\it Mukhanov Sasaki varibale}. On the other hand we have found out the individual results of the classical partition function for Cosmology exactly match with the quantum mechanical thermal partition function derived for Cosmology in the previous section. This observation actually consistent with our expectations as in case of simple harmonic oscillator also we get similar result. In the present context since we are dealing with a scalar field theory in cosmological FLRW background where the effective frequency is conformal time dependent. However, after doing Fourier transformation one can interpret this theory as a parameteric oscillator. For this reason, we get similar type of results except for the Fourier integration over all momentum modes, which are very relevant for Cosmology.
\subsection{Classical limit of cosmological two-point  micro-canonical OTOC: rescaled field version}
In this subsection, our prime objective is to find out the classical limiting version of the two-point  micro-canonical OTOC computed from the quantum field theory side using the cosmological rescaled field variable and its canonically conjugate momentum as appearing in the context of cosmological perturbation theory.

Next, we will compute the two-point OTOC using the above mentioned results. In the classical limit, the two-point function is given by the following expression:
\bea Y^{f}_{\bf Classical}(\tau_1,\tau_2)&=&\frac{1}{Z_{\bf Classical}(\beta;\tau_1)}\int \int \frac{{\cal D}f {\cal D}\Pi}{2\pi}\exp\left(-\beta~H\right)~\left\{{f}({\bf x},\tau_1),{\Pi}({\bf x},\tau_2)\right\}_{\bf PB}\nonumber\\
&=&\frac{1}{Z_{\bf Classical}(\beta;\tau_1)}\underbrace{\int \int \frac{{\cal D}f {\cal D}\Pi}{2\pi}\exp\left(-\beta~H\right)}_{\textcolor{red}{\bf \equiv ~Z_{\bf Classical}(\beta;\tau_1)}}~\nonumber\\
&&~~~~~~~~\times 2(2\pi)^3 \int \frac{d^3{\bf k}_1}{(2\pi)^3}\int \frac{d^3{\bf k}_2}{(2\pi)^3}~\exp\left(i({\bf k}_1+{\bf k}_2).{\bf x}\right)\delta^{3}({\bf k}_1+{\bf k}_2)\nonumber\\
&&~~~~~~~~~~~~~~~~~~~~~~~~{\bf W}(\tau_1-\tau_2)~\exp\left(-\frac{\lambda_f (|\tau_1|+|\tau_2|)}{2}\right)~~~~~~~\nonumber\\
&=&2~{\bf W}(\tau_1-\tau_2)~\exp\left(-\frac{\lambda_f (|\tau_1|+|\tau_2|)}{2}\right)~\int \frac{d^3{\bf k}_1}{(2\pi)^3}.\eea
This result is divergent because of the presence of last terms which represent volume in general. to get the finite contribution out of the above mentioned result derived in the classical limit we regulate the momentum integrals by putting the cut-off scale $L$, for which the momentum range are given by, $0<k_1<L$. By applying this regulator we get:
\bea \int \frac{d^3{\bf k}_1}{(2\pi)^3}=\frac{1}{2\pi^2}\int^{L}_{k_1=0}k^2_1~dk_1=\frac{L^3}{6\pi^2}~.\eea
After substituting the above factor finally we get:
\bea \hll{Y^{f}_{\bf Classical}(\tau_1,\tau_2)=\frac{L^3}{3\pi^2}{\bf W}(\tau_1-\tau_2)~\exp\left(-\frac{\lambda_f (|\tau_1|+|\tau_2|)}{2}\right)}~.\eea
Here $L^3/3\pi^2$, is the overall regulated two-point time independent amplitude of the cosmological two-point OTOC in the classical limit. From the above mentioned result one can consider a situation when we have $\tau_1=\tau_2=\tau$ in the classical limit. In that case we get further simplified answer, which is given by:
\bea \hll{Y^{f}_{\bf Classical}(\tau,\tau)=\frac{L^3}{3\pi^2}~{\bf W}(0)\exp\left(-\lambda_{f}|\tau|\right)}~.~\eea
This result only exists when ${\bf W}(0)$ is finite in the classical limiting case for the coloured non-Gaussian noise and white Gaussian noise respectively.

Now to demonstrate the explicit role of a non-Gaussian coloured noise and Gaussian white noise here we can consider the following mathematical forms of the window functions:
\begin{eqnarray}
&& {{\bf W}(\tau_1-\tau_2)= \large \left\{
     \begin{array}{lr}
   \displaystyle \sqrt{\frac{{\bf A}}{\gamma}}~\exp\left(-\frac{\gamma|\tau_1-\tau_2|}{2}\right), & \text{\textcolor{red}{\bf Coloured~Noise}}\\  \\
   \displaystyle   \lim_{{\bf C}\rightarrow 0}\sqrt{\frac{{\bf B}}{|{\bf C}|\sqrt{\pi}}}~\exp\left(-\frac{(\tau_1-\tau_2)^2}{{{\bf C}^2}}\right) & \text{\textcolor{red}{\bf White~Noise}}  \end{array}
   \right.}~~~~~
\end{eqnarray}
where ${\bf A}$, ${\bf B}$ and ${\bf C}$, represent the conformal time independent amplitudes of the coloured and white random classical noise respectively. Also, $\gamma$ represents the strength of the dissipation in the context of coloured noise.

Now from the general structure of the white noise it is evident that, ${\bf W}(0)\rightarrow \infty$ is giving  diverging contribution for $\tau_1=\tau_2$ case. So appearance of the possibility of the equal time limit is completely discarded as it gives overall diverging contribution in the classical limit of the two-point function. On the other hand, in the equal time limit we have, ${\bf W}(0)=\sqrt{{\bf A}/\gamma}$, for the coloured noise case. This implies for coloured noise equal time limit exists and one can write down the following simplified expression for the classical limit of the two-point function as:
\bea \hll{Y^{f}_{\bf Classical}(\tau,\tau)=\frac{L^3}{3 \pi^2}\sqrt{\frac{{\bf A}}{\gamma}}~\exp\left(-2\lambda_{f}|\tau|\right)}~.~\eea
But for unequal time case both the results exist and we get:
\begin{eqnarray}
&& \hll{Y^{f}_{\bf Classical}(\tau_1,\tau_2)= \left\{
     \begin{array}{lr}
   \displaystyle \frac{L^3}{3\pi^2}\sqrt{\frac{{\bf A}}{\gamma}}~\exp\left(-\frac{\left(\gamma|\tau_1-\tau_2|+\lambda_f (|\tau_1|+|\tau_2|)\right)}{2}\right)~, &~ \text{\textcolor{red}{\bf Coloured~Noise}}\\  \\
   \displaystyle \frac{L^3}{3\pi^2} \lim_{{\bf C}\rightarrow 0}\sqrt{\frac{{\bf B}}{|{\bf C}|\sqrt{\pi}}}~\exp\left(-\left\{\frac{(\tau_1-\tau_2)^2}{{{\bf C}^2}}+\frac{\lambda_f (|\tau_1|+|\tau_2|)}{2}\right\}\right) & \text{\textcolor{red}{\bf White~Noise}}  \end{array}
   \right.}
\end{eqnarray}
\subsection{Classical limit of cosmological two-point  micro-canonical OTOC: curvature perturbation field version}
Here we need to perform the computation for the classical version of the two-point OTOC in terms of the scalar curvature perturbation and the canonically conjugate momentum associated with it, which we have found that is given by the following simplified expression:
\bea && \hll{Y^{\zeta}_{\bf Classical}(\tau_1,\tau_2)=\frac{1}{Z^{\zeta}_{\bf Classical}(\beta,\tau_1)}\int\int\frac{{\cal D}\zeta {\cal D}\Pi_{\zeta}}{2\pi}e^{-\beta H(\tau_1)}\left\{\zeta({\bf x},\tau_1),{\Pi}({\bf x},\tau_2)\right\}_{\bf PB}}\nonumber\\
&&~~~~~~~~~~~~~~~~~~~~~\hll{=\frac{1}{z(\tau_1)z(\tau_2)}Y^{f}_{\bf Classical}(\tau_1,\tau_2)}~.~~~~~~~~~~~~~\eea
In this result, cosmological two-point OTOC depend on the specific choice of the field variable and its conjugate momentum in a given cosmological perturbation scheme due to the presence of an overall factor $(z(\tau_1)z(\tau_2))^{-1}$, which is originated from {\it Mukhanov Sasaki variable} defined in two time scales, $\tau=\tau_1$ and $\tau=\tau_2$ respectively. In terms of the curvature perturbation field variable then one can also justifies the physical consistency of the classical limiting result with the result obtained in the quantum regime in the previous section.
\subsection{Classical limit of cosmological four-point  micro-canonical OTOC: rescaled field version}

\subsubsection{Without normalization}
In this subsection, our aim is to compute the classical limiting version of un-normalized cosmological four-point OTOC, which is given by the following expression:
\bea C^{f}_{\bf Classical}(\tau_1,\tau_2):&=&\frac{1}{Z_{\bf Classical}(\beta;\tau_1)}\int \int \frac{{\cal D}f {\cal D}\Pi}{2\pi}\exp\left(-\beta~H\right)~\left\{{f}({\bf x},\tau_1),{\Pi}({\bf x},\tau_2)\right\}^2_{\bf PB}\nonumber\\
&=&\frac{1}{Z_{\bf Classical}(\beta;\tau_1)}\underbrace{\int \int \frac{{\cal D}f {\cal D}\Pi}{2\pi}\exp\left(-\beta~H\right)}_{\textcolor{red}{\bf \equiv ~Z_{\bf Classical}(\beta;\tau_1)}}~\nonumber\eea\bea
&&~~~~~~~~\times(2\pi)^6 \int \frac{d^3{\bf k}_1}{(2\pi)^3}\int \frac{d^3{\bf k}_2}{(2\pi)^3}\int \frac{d^3{\bf k}_3}{(2\pi)^3}\int \frac{d^3{\bf k}_4}{(2\pi)^3}~\exp\left(i({\bf k}_1+{\bf k}_2+{\bf k}_3+{\bf k}_4).{\bf x}\right)\nonumber\\
 &&~~~~~~~~~~~~~~~~~~~~~~~~~\left[\delta^{3}({\bf k}_1+{\bf k}_2)\delta^3({\bf k}_3+{\bf k}_4)+\delta^{3}({\bf k}_1+{\bf k}_3)\delta^3({\bf k}_2+{\bf k}_4)\right.\nonumber\\&& \left.~~~~~~~~~~~~~~~~~~~~~~~~~+\delta^{3}({\bf k}_1+{\bf k}_4)\delta^3({\bf k}_3+{\bf k}_2)+\delta^{3}({\bf k}_2+{\bf k}_3)\delta^3({\bf k}_4+{\bf k}_1)\right.\nonumber\\&& \left.~~~~~~~~~~~~~~~~~~~~~~~~~+\delta^{3}({\bf k}_2+{\bf k}_1)\delta^3({\bf k}_4+{\bf k}_3)+\delta^{3}({\bf k}_2+{\bf k}_4)\delta^3({\bf k}_1+{\bf k}_3)\right.\nonumber\\&& \left.~~~~~~~~~~~~~~~~~~~~~~~~~+\delta^{3}({\bf k}_3+{\bf k}_1)\delta^3({\bf k}_4+{\bf k}_2)+\delta^{3}({\bf k}_3+{\bf k}_2)\delta^3({\bf k}_1+{\bf k}_4)\right.\nonumber\\&& \left.~~~~~~~~~~~~~~~~~~~~~~~~~+\delta^{3}({\bf k}_3+{\bf k}_4)\delta^3({\bf k}_1+{\bf k}_2)+\delta^{3}({\bf k}_4+{\bf k}_1)\delta^3({\bf k}_2+{\bf k}_3)\right.\nonumber\\&& \left.~~~~~~~~~~~~~~~~~~~~~~~~~+\delta^{3}({\bf k}_4+{\bf k}_2)\delta^3({\bf k}_3+{\bf k}_1)+\delta^{3}({\bf k}_4+{\bf k}_3)\delta^3({\bf k}_2+{\bf k}_1)\right]\nonumber\\
 &&~~~~~~~~~~~~~~~~~~~~~~~~~~~~~~~~~~~~{\bf G}_{\bf Kernel}(\tau_1-\tau_2)\exp\left(-\lambda_{f}[|\tau_1|+|\tau_2|]\right)\eea 
 where the classical version of the thermal partition function in Cosmology is given by the following expression:
 \bea \hll{Z_{\bf Classical}(\beta;\tau_1)=\exp\left(-\int d^3{\bf k} ~\ln\left(2\sinh \frac{\beta E_{\bf k}(\tau_1)}{2}\right)\right)}~.\eea  
Now, after doing a bit of algebraic manipulation we get the following simplified result for the un-normalized version of the classical limit of four-point cosmological OTOC:
\bea \hll{C^{f}_{\bf Classical}(\tau_1,\tau_2)=12~{\bf G}_{\bf Kernel}(\tau_1-\tau_2)\exp\left(-\lambda_{f}[|\tau_1|+|\tau_2|]\right)\int \frac{d^3{\bf k}_1}{(2\pi)^3}\int \frac{d^3{\bf k}_2}{(2\pi)^3}}~.~\eea
This result is divergent because of the presence of last terms which represent volume in general. to get the finite contribution out of the above mentioned result derived in the classical limit we regulate the momentum integrals by putting the cut-off scale $L$, for which the momentum range are given by, $0<k_1<L$ and $0<k_2<L$. By applying this regulator we get:
\bea \int \frac{d^3{\bf k}_1}{(2\pi)^3}\int \frac{d^3{\bf k}_2}{(2\pi)^3}=\frac{1}{4\pi^4}\int^{L}_{k_1=0}k^2_1~dk_1\int^{L}_{k_2=0}k^2_2~dk_2=\frac{L^6}{36\pi^4}~.\eea
After substituting the above factor finally we get:
\bea \hll{C^{f}_{\bf Classical}(\tau_1,\tau_2)=\frac{L^6}{3\pi^4}~{\bf G}_{\bf Kernel}(\tau_1-\tau_2)\exp\left(-\lambda_{f}[|\tau_1|+|\tau_2|]\right)}~.~\eea
Here $L^6/3\pi^4$ the overall regulated four-point time independent amplitude of the cosmological un-normalized version of OTOC in the classical limit. From the above mentioned result one can consider a situation when we have $\tau_1=\tau_2=\tau$ in the classical limit. In that case we get further simplified answer, which is given by:
\bea \hll{C^{f}_{\bf Classical}(\tau,\tau)=\frac{L^6}{3\pi^4}~{\bf G}_{\bf Kernel}(0)\exp\left(-2\lambda_{f}\tau\right)}~.~\eea
This result only exists when ${\bf G}_{\bf Kernel}(0)$ is finite in the classical limiting case for the coloured non-Gaussian noise and white Gaussian noise respectively.

Now to demonstrate the explicit role of a non-Gaussian coloured noise and Gaussian white noise here we can consider the following mathematical forms of the two point classical correlation functions:
\begin{eqnarray}
&& {{\bf G}_{\bf Kernel}(\tau_1-\tau_2)= \large \left\{
     \begin{array}{lr}
   \displaystyle \frac{{\bf A}}{\gamma}~\exp(-\gamma|\tau_1-\tau_2|)~, &~ \text{\textcolor{red}{\bf Coloured~Non-Gaussian~Noise}}\\  \\
   \displaystyle   {\bf B}~\delta(\tau_1-\tau_2) & \text{\textcolor{red}{\bf White~Gaussian~Noise}}  \end{array}
   \right.}~~~~~~~~~~
\end{eqnarray}
where ${\bf A}$ and ${\bf B}$, represent the conformal time independent amplitudes of the coloured non-Gaussian and white Gaussian random classical noise respectively. Here both of them satisfy the previously mentioned crucial properties separately for coloured and white noise. Also, $\gamma$ represents the strength of the dissipation in the context of coloured non-Gaussian noise.

Now from the general structure of the white Gaussian noise it is evident that, ${\bf G}_{\bf Kernel}(0)={\bf B}~\delta(0)\rightarrow \infty$ is giving  diverging contribution for $\tau_1=\tau_2$ case. So appearance of the possibility of the equal time limit is completely discarded as it gives overall diverging contribution in the classical limit of the four-point cosmological OTOC. On the other hand, in the equal time limit we have, ${\bf G}_{\bf Kernel}(0)={\bf A}/\gamma$ for the non-Gaussian coloured noise case. This implies for coloured noise equal time limit exists and one can write down the following simplified expression for the classical limit of the four-point cosmological OTOC as:
\bea \hll{C^{f}_{\bf Classical}(\tau,\tau)=\frac{{\bf A}~L^6}{3\gamma \pi^4}~\exp\left(-2\lambda_{f}\tau\right)}~.~\eea
But for unequal time case both the results exist and we get:
\begin{eqnarray}
&& \hll{C^{f}_{\bf Classical}(\tau_1,\tau_2)= \large \left\{
     \begin{array}{lr}
   \displaystyle \frac{{\bf A}L^6}{3\gamma\pi^4}~\exp\left(-\gamma|\tau_1-\tau_2|-\lambda_{f}[|\tau_1|+|\tau_2|]\right)~, &~ \text{\textcolor{red}{\bf Coloured~Noise}}\\  \\
   \displaystyle  \frac{ {\bf B}L^6}{3\pi^4}~\delta(\tau_1-\tau_2)\exp\left(-\lambda_{f}[|\tau_1|+|\tau_2|]\right) & \text{\textcolor{red}{\bf White~Noise}}  \end{array}
   \right.}~~~~~~~~
\end{eqnarray}
\subsubsection{With normalization}
The normalisation factor of classical limit of OTOC for the rescaled field variable can be computed as:
\bea \hll{{\cal N}^{f}_{\bf Classical}(\tau_1,\tau_2)=\frac{36\pi^4}{L^6{\bf W}^2(0)\exp\left(-\lambda_{f}[|\tau_1|+|\tau_2|]\right)}=\frac{36\pi^4}{L^6{\bf G}_{\bf Kernel}(0)\exp\left(-\lambda_{f}[|\tau_1|+|\tau_2|]\right)}}~.~~~~~~\eea
Now, considering the examples of non-Gaussian coloured noise and Gaussian white noise we get the following answer for the normalization factor:
\begin{eqnarray}
&& \hll{{\cal N}^{f}_{\bf Classical}(\tau_1,\tau_2)= \large \left\{
     \begin{array}{lr}
   \displaystyle\frac{36\gamma \pi^4}{L^6{\bf A}\exp\left(-\lambda_{f}[|\tau_1|+|\tau_2|]\right)}~, &~ \text{\textcolor{red}{\bf Coloured~Noise}}\\  \\
   \displaystyle   0 & \text{\textcolor{red}{\bf White~Noise}}  \end{array}
   \right.}~~~~~~~~~~
\end{eqnarray}
Then the classical limit of normalized four-point OTOC can be expressed as:
\bea \hll{{\cal C}^{f}_{\bf Classical}(\tau_1,\tau_2)={\cal N}^{f}_{\bf Classical}(\tau_1,\tau_2)C^{f}_{\bf Classical}(\tau_1,\tau_2)=12\left(\frac{{\bf G}_{\bf Kernel}(\tau_1-\tau_2)}{{\bf G}_{\bf Kernel}(0)}\right)}~.\eea
Now, considering the examples of non-Gaussian coloured noise and Gaussian white noise we get the following answer for the classical limit of normalized four-point OTOC:
\begin{eqnarray}
&& \hll{{\cal C}^{f}_{\bf Classical}(\tau_1,\tau_2)= \large \left\{
     \begin{array}{lr}
   \displaystyle 12~\exp\left(-\gamma|\tau_1-\tau_2|\right)~, &~ \text{\textcolor{red}{\bf Coloured~Noise}}\\  \\
   \displaystyle  0 & \text{\textcolor{red}{\bf White~Noise}}  \end{array}
   \right.}.~~~~~~~~
\end{eqnarray}
Here for the non-Gaussian coloured noise then the dissipation strength $\gamma$ plays the role of {\it Lyapunov exponent} and in the classical limit the conformal time independent amplitude is $12$. On the other hand, it is clearly seen that Gaussian white noise is not suitable for the study of OTOC as it becomes trivially zero after normalization of OTOC in the classical limit.
\subsection{Classical limit of cosmological four-point micro-canonical OTOC: curvature perturbation field version}
 \subsubsection{Without normalization} 
Here we need to perform the computation for the classical version of the un-normalised OTOC in terms of the scalar curvature perturbation and the canonically conjugate momentum associated with it, which we have found that is given by the following simplified expression:
\bea && \hll{C^{\zeta}_{\bf Classical}(\tau_1,\tau_2)=\frac{1}{Z^{\zeta}_{\bf Classical}(\beta,\tau_1)}\int\int\frac{{\cal D}\zeta {\cal D}\Pi_{\zeta}}{2\pi}e^{-\beta H(\tau_1)}\left\{\zeta({\bf x},\tau_1),{\Pi}({\bf x},\tau_2)\right\}^2_{\bf PB}}\nonumber\\
&&~~~~~~~~~~~~~~~~~~~~~\hll{=\frac{1}{z^2(\tau_1)z^2(\tau_2)}C^{f}_{\bf Classical}(\tau_1,\tau_2)}~.~~~~~~~~~~~~~\eea
Now, if we just compare with the obtained result from the quantum version, we clearly see that that obeys also the similar kind of relation. This connection actually in first hand verifies that the classical limit of the cosmological four-point OTOC obtained from quantum fluctuation in the early universe is physically consistent. It is also important to note that, in the un-normalized version of the classical limiting result the definition of the cosmological four-point OTOC depend on the specific choice of the field variable and its conjugate momentum in a given cosmological perturbation scheme due to the presence of an overall factor $(z(\tau_1)z(\tau_2))^{-2}$, which is originated from {\it Mukhanov Sasaki variable} defined in two time scales, $\tau=\tau_1$ and $\tau=\tau_2$ respectively. This factor actually make a bridge between the two definitions of four-point cosmological OTOC obtained from two different field variables in cosmological perturbation theory both in quantum and classical formalism. 
\subsubsection{With normalization}
The classical version of the normalised four-point cosmological OTOC in terms of the scalar curvature perturbation and the canonically conjugate momentum associated with it, which is basically the computation of the following normalised OTOC, in the present context:
\bea \hll{{\cal C}^{\zeta}_{\bf Classical}(\tau_1,\tau_2)=\frac{C^{\zeta}_{\bf Classical}(\tau_1,\tau_2)}{\langle \zeta(\tau_1)\zeta(\tau_1)\rangle_{\beta}\langle \Pi_{\zeta}(\tau_1)\Pi_{\zeta}(\tau_1)\rangle_{\beta}}={\cal N}^{\zeta}_{\bf Classical}(\tau_1,\tau_2)~C^{\zeta}_{\bf Classical}(\tau_1,\tau_2)}~,~~~~~~~~ \eea 
where the normalisation factor to normalise classical OTOC is given by:
 \bea \hll{{\cal N}^{\zeta}_{\bf Classical}(\tau_1,\tau_2)=\frac{1}{\langle \zeta(\tau_1)\zeta(\tau_1)\rangle_{\beta}\langle \Pi_{\zeta}(\tau_1)\Pi_{\zeta}(\tau_1)\rangle_{\beta}}=z^2(\tau_1)z^2(\tau_2){\cal N}^{f}_{\bf Classical}(\tau_1,\tau_2)}.~~~~~~~~\eea
 Consequently, the classical limit of the normalised four-point cosmological OTOC computed from the curvature perturbation variable is given by the following expression:
 \bea \hll{{\cal C}^{\zeta}_{\bf Classical}(\tau_1,\tau_2)={\cal C}^{f}_{\bf Classical}(\tau_1,\tau_2)=12\left(\frac{{\bf G}_{\bf Kernel}(\tau_1-\tau_2)}{{\bf G}_{\bf Kernel}(0)}\right)}.~~~~~~~~ \eea 
Now, considering the examples of non-Gaussian coloured noise and Gaussian white noise we get the following answer for the classical limit of normalized four-point OTOC:
\begin{eqnarray}
&& \hll{{\cal C}^{\zeta}_{\bf Classical}(\tau_1,\tau_2)={\cal C}^{f}_{\bf Classical}(\tau_1,\tau_2)= \large \left\{
     \begin{array}{lr}
   \displaystyle 12~\exp\left(-\gamma|\tau_1-\tau_2|\right)~, &~ \text{\textcolor{red}{\bf Coloured~Noise}}\\  \\
   \displaystyle  0 & \text{\textcolor{red}{\bf White~Noise}}  \end{array}
   \right.}.~~~~~~~~
\end{eqnarray}

\section{Summary and Outlook}
  \label{sec:OTO7}
To summarize, in this work, we have addressed the following issues to study the OTOC from cosmology set up:
\begin{itemize}
\item First of all in this paper we have provided a detailed formalism using which it is possible now to compute the expression for OTOC in the context of Cosmology. OTOC is a very strong probe to study the quantum correlation in presence of randomness and chaoticity. It was used in different context except for Cosmology. As we all know finding quantum correlation function in the early universe in presence of such randomness or chaoticity was a long standing problem which no body have addressed yet properly. We believe our computation and finding from the cosmological OTOC in this paper will surely be helpful to understand the quantum field theory of various unexplored random cosmological events, i.e. stochastic particle production during inflation, reheating etc. Since we did the computation for the first time from a very simple understandable cosmological perturbation theory, we believe that these results will also be used to study some other random cosmological events appear in the evolutionary time scale of our universe.

\item We have presented the computation of cosmological OTOC by making use of the well know Bunch Davies and $\alpha$ vacua as a choice of initial quantum vacuum state. So it is expected that the final results of cosmological OTOC may show similar feature, but with different amplitude and in different scale. This statement we have verified explicitly by numerically studying the obtained solutions of the cosmological OTOC using both the definition of quantum mechanical vacuum state.

\item It is a very well known fact that in general prescription OTOC's are usually defined with two quantum operators which are separated in time scale. Following the same prescription in Cosmology we have also defined two quantum operators, which are the cosmological perturbation variable describing the quantum fluctuation from scalar modes and its associated canonically conjugate momentum, which are defined in two different time scales in the evolution of our universe. During the computation of cosmological OTOC, the final result for this reason depend on these two time scales. Now, to study the cosmological consequences of the OTOC we fix one of the time scales between the two and study the dynamical feature with respect to the other time variable which we have not fixed. We have found that both of the features of cosmological OTOC with respect to the two different time scales describes the randomness in the quantum correlation function at out-of-equilibrium. On top of that we have additionally found that particularly once we fix the first time scale and study the time evolution behaviour of the cosmological OTOC show exponentially decaying behaviour, which supports the phenomena of quantum chaos in the cosmological paradigm as well. In usual scenario one gets exponential growth with respect to the time scale in OTOC. But, to remind everyone, it is important to note that instead of dealing actual time scale which is varying from $0~({\rm Big~bang})<t<t_0~({\rm Present~ day})$ in the present context we are dealing with the conformal time scale which is varying from $-\infty~({\rm Big~bang})<\tau<0~({\rm Present~day})$, which is the most commonly used time scale in the context of Cosmology. Because of this reason instead of exponentially increasing behaviour we have observed exponentially decreasing behaviour. This behaviour is perfectly consistent with the expectation from the time evolution of the quantum mechanical correlations for the system when it goes to the out-of-equilibrium state.

\item Additionally from the present formalism we have derived the expression for the quantum {\it Lyapunov spectrum} as well the lower bound on the equilibrium temperature for Cosmology. At very early epoch of the evolution of our universe the quantum fluctuations goes to the out-of-equilibrium state for which one cannot associate the concept of temperature, as it is only interpreted when a system under study reaches to equilibrium state. We have derived the expression for cosmological OTOC which is very helpful to understand the quantum mechanical correlation function at out-of-equilibrium state. Now from the exponential decay of this cosmological OTOC one can easily estimate the lower bound on the equilibrium temperature associated with our universe. This can be done in a very simple way. For this purpose we need to identify the exact time scale at which the cosmological OTOC after decaying with respect to the conformal time scale start saturate to a non vanishing small value. From this time-scale , one can further explicitly derive and give an estimate of the temperature of the system when cosmological OTOC saturates to an equilibrium value. After getting an estimate of this lower bound of equilibrium temperature one can give a physical interpretation of this temperature in the cosmological paradigm. One can associate the obtained lower bound on the equilibrium temperature with the temperature of De Sitter space for the stochastic particle production during inflation and with the reheating temperature. Usually in the cosmology literature one actually compute the expression for the reheating temperature by making use of information from quantum statistical mechanics. Using the approximate energy scale for inflation and by making use the information regarding the number of relativistic degrees of freedom one computes a bound on the reheating temperature in the framework of primordial cosmology. Since earlier we don't have any information regarding time dependent behaviour of the cosmological OTOC one don't have any option to compute reheating temperature from the quantum field theory side. So the derived bound on the reheating temperature can be treated as a model independent bound, which is anyway better than  the previously known model dependent bound on reheating temperature.

\item Also, we have found that the cosmological OTOC at finite temperature is dependent on two time scale and independent of any preferred choice of the coordinate system and the parameter $\beta=1/T$. In short, the derived expression for the cosmological OTOC is homogeneous in nature with respect to the space coordinate, or its Fourier transformed momentum coordinate. Also the final result obtained for the cosmological OTOC is independent of the partition function which we have computed for Cosmology. This feature is exactly similar to the OTOC computed from simple harmonic oscillator with time independent frequency. Basically, one can map the stochastic particle production problem during inflation or solving the reheating problem to a parametric oscillator problem whose frequency is function of the magnitude of the momentum in the Fourier transformed space and conformal time scale, where the time is actually associated with the evolution scale of our universe. 

\item By doing the numerical estimations and during the study of detailed time dependent behaviour of the cosmological OTOC we have found that our analysis can give physically consistent result only for partially massless or massive scalar particle production during inflation and during reheating epoch. 

\item We have also studied the classical limit of the two-point and four-point OTOC to check the consistency with the late time behaviour which supports the exponential decay of the quantum correlators in the classical regime.

\item Finally, we have explicitly proved that the definition of cosmological OTOC is completely independent of the choice of preferred perturbation variable in a specific scheme of cosmological perturbation theory. On the other hand, we have also found that, this statement not holds good for the un-normalized cosmological OTOC. We have demonstrated that both the results are related through some factor which is basically function of conformal time dependent {\it Mukhanov Sasaki variable}.

\end{itemize}

The future prospects of this work is as follows.

\begin{itemize}

\item In this we have restricted our analysis only for cosmological spatially flat FLRW space-time, which is more observationally relevant. We are very hopeful that our obtained result in this paper can be probed by various cosmological future missions and if the amplitudes of the cosmological OTOC can able to be measured with significant statistical accuracy then the obtained result for cosmological OTOC can be treated as a standard benchmark using which one can study various unexplored features of early universe cosmology in presence random quantum fluctuations in out-of-equilibrium state. To know about the cosmological consequences of our derived result one can further extend the present methodology for computing OTOC for open and flat FLRW space time as well. This study will give a better understanding to know about the explicit role of spatial curvature in cosmological FLRW space-time to explore the  out-of-equilibrium features. Not only this, the present methodology of computing cosmological OTOC one can further extend any arbitrary spatial dimensions, which will give a clear picture that if one vary the spatial dimensions then at very lower or higher dimensions how the cosmological OTOC behave with the time scale. No one have explored such possibilities yet, so it will be good to study these mentioned aspects in detail.

\item The present methodology of computing the cosmological OTOC is not only restricted to describe stochastic particle production during inflation and during the reheating epoch, but also one can extend this tool to compute the quantum mechanical OTOC in presence of non singular bounce. It may happen that during the non singular bouncing phenomena somehow stochasticity or some random quantum mechanical fluctuations appears and since now it is known how to quantify and compute the quantum correlation functions at out-of-equilibrium state, one can carry forward the similar calculation in this context as well.

\item  Another important aspect one can study, which is the role of quantum entanglement \cite{Choudhury:2018fpj,Choudhury:2017qyl,Choudhury:2017bou,Choudhury:2016cso,Choudhury:2016pfr,Maldacena:2015bha,Maldacena:2012xp,Kanno:2014lma} in cosmological OTOC at out-of-equilibrium. Quantum correlation functions in presence of entanglement was studied at equilibrium in various earlier works. It is good to apply the present methodology to know about cosmological OTOC in presence of quantum entanglement in the early universe. Specifically if one can study the cosmological OTOC in Bell's inequality violating set-up then one can actually study the effect of long range sustainable small amplitude cosmological correlation at out-of-equilibrium.

\end{itemize}

\newpage
	\subsection*{Acknowledgements}
	SC is thankful to Latham Boyle, Andrew R. Liddle, Neil Turok, Nicholas Hunter-Jones, Zohar Komargodski, Alexander Abanov, Douglas Stanford, Juan Martín Maldacena, Paul Joseph Steinhardt, Martin Bojowald, Eugenio Bianchi, Shiraz Minwalla, Subir Sarkar, Sudhakar Panda, Soumitra SenGupta, Justin R. David, Banibrata Mukhopadhyay, Urjit A. Yajnik, S. Umashankar, P. Ramadevi
Zhen Pan, Gabriel Pasquino, Jeremy Peters, Alba Grassi, Mark Mezei, Luigi	Tizzano for helpful discussions and support. SC would like to thank Quantum Gravity and Unified Theory and Theoretical Cosmology
Group, Max Planck Institute for Gravitational Physics, Albert Einstein Institute (AEI) for providing the Post-Doctoral Research Fellowship. SC take this opportunity to thank sincerely to
Jean-Luc Lehners and Axel Kleinschmidt for his constant support and inspiration. SC thank 
 Latham Boyle for inviting at Perimeter Institute for Theoretical Physics (PITP), Zohar Komargodski for inviting at Simons Center for Geometry and Physics (SCGP), Stony Brook University, Leonardo Senatore  for inviting at  Institute for Theoretical Physics, Stanford University, Juan Martín Maldacena for inviting at  Workshop on Qubits and Spacetime, Institute for Advanced Studies (IAS), Princeton, Paul Joseph Steinhardt  for inviting at  Department of Physics, Princeton University, Martin Bojowald  for inviting at 
The Institute for Gravitation and the Cosmos (IGC),  Department of Physics, Eberly College of Science, Pennsylvania State University (University Park campus), Sudhakar Panda  for inviting at School of Physical Sciences, National Insitute of Science Education and Research (NISER), Bhubaneswar, Abhishek Chowdhury  for inviting at Department of Physics, Indian Institute of Technology (IIT), Bhubaneswar, Anjan Sarkar  for inviting at Department of Astrophysics, Raman Research Institute, Bengaluru, Aninda Sinha and Banibrata Mukhopadhyay for inviting at Center for High Energy Physics (CHEP) and Department of Astronomy and Astrophysics, Indian Institute of Science, Bengaluru, Uma Shankar for inviting at Department of Physics, Indian Institute of Technology (IIT), Bombay, Shiraz Minwalla for inviting at Department of Theoretical Physics, Tata Institute of Fundamental Research, Mumbai, Abhishek Mahapatra for inviting at National Institute of Technology (NIT), Rourkela, for official academic visit where the work was done partially. Part of this work was presented as a talk, titled " Cosmology from Condensed Matter Physics: A study of out-of-equilibrium physics" and " Cosmology Meets Condensed Matter Physics" at Perimeter Institute for Theoretical Physics (PITP) (See the link: \textcolor{red}{http://pirsa.org/19110117/}), Simons Center for Geometry and Physics (SCGP), Stony Brook University (See the link: \textcolor{red}{http://scgp.stonybrook.edu/video portal/video.php?id=4358}), Department of Physics, Princeton University, The Institute for Gravitation and the Cosmos (IGC),  Department of Physics, Eberly College of Science, Pennsylvania State University (University Park campus), workshop on "Advances in Astroparticle Physics and Cosmology, AAPCOS-2020" at Saha Institute of Nuclear Physics, Kolkata on the occasion of the 100 years of Saha Ionisation Equation by Prof. Meghnad Saha, Department of Physics, Scottish Church College, Kolkata, School of Physical Sciences, National Insitute of Science Education and Research (NISER), Bhubaneswar, Department of Physics, Indian Institute of Technology (IIT), Bhubaneswar, Department of Astrophysics, Raman Research Institute, Bengaluru, Center for High Energy Physics (CHEP), Indian Institute of Science, Bengaluru, Department of Physics, Indian Institute of Technology (IIT), Bombay, National Institute of Technology (NIT), Rourkela. SC would like to thank Quantum Gravity and Unified Theory and Theoretical Cosmology
Group, Max Planck Institute for Gravitational Physics, Albert Einstein Institute (AEI), Perimeter Institute for Theoretical Physics (PITP), Simons Center for Geometry and Physics (SCGP), Stony Brook University, Institute for Theoretical Physics, Stanford University, 
The Institute for Gravitation and the Cosmos (IGC),  Department of Physics, Eberly College of Science, Pennsylvania State University (University Park campus), School of Physical Sciences, National Insitute of Science Education and Research (NISER), Bhubaneswar, Department of Astrophysics, Raman Research Institute, Bengaluru, Department of Physics, Indian Institute of Technology (IIT), Bombay, Quantum Space-time Group (Earlier known as String Theory and Mathematical Physics Group), Department of Theoretical Physics, Tata Institute of Fundamental Research, Mumbai and National Institute of Technology (NIT), Rourkela for providing financial support for the academic visits at Canada, U.S.A. and India. SC would like to thank the natural beauty of Prague, Dresden, Hamburg, Leipzig, Potsdam, Berlin which inspires to do work very hard during the weekend trips. SC also thank all the members of our newly formed virtual international non-profit consortium ``Quantum Structures of the Space-Time \& Matter" (QASTM) for elaborative discussions. Last but not the least, we would like to acknowledge our debt to 
the people belonging to the various part of the world for their generous and steady support for research in natural sciences.
	\clearpage
	\appendix
\section{Asymptotic behaviour of the cosmological mode functions}
The most general solution of the above mentioned equation of motion is given by the following expression:
\bea {f_{\bf k}(\tau)=\sqrt{-\tau}\left[{\cal C}_1~H^{(1)}_{\nu}(-k\tau)+{\cal C}_2~H^{(2)}_{\nu}(-k\tau)\right]},\eea
where ${\cal C}_1$ and ${\cal C}_2$ are two arbitrary integration constants which are fixed by the choice of the initial quantum vacuum state necessarily needed for this computation. Here $H^{(1)}_{\nu}(-k\tau)$ and $H^{(1)}_{\nu}(-k\tau)$ are the Hankel functions of first and second kind with order $\nu$. In general, both of them can be expressed in terms of the complex linear combination of the Bessel function and the Neumann function of order $\nu$, which are given by the following expressions:
\bea && {H^{(1)}_{\nu}(-k\tau)={\cal J}_{\nu}(-k\tau)+i{\cal Y}_{\nu}(-k\tau)}~,\\
&& {H^{(2)}_{\nu}(-k\tau)={\cal J}_{\nu}(-k\tau)-i{\cal Y}_{\nu}(-k\tau)}~.\eea
In the general context the mass parameter $\nu$ may be a complex parameter. In this case, the solution for the rescaled scalar perturbation mode can be recast as:
\bea  {f_{\bf k}(\tau)=\sqrt{-\tau}\left[{\cal D}_1~{\cal J}_{\nu}(-k\tau)+{\cal D}_2~{\cal Y}_{\nu}(-k\tau)\right]}, \eea
where the redefined two new arbitrary integration constants, ${\cal D}_1$ and ${\cal D}_2$ are defined in terms of the previously defined two new arbitrary integration constants, ${\cal C}_1$ and ${\cal C}_2$ as:
\bea && {{\cal D}_1={\cal C}_1+{\cal C}_2}~,\\
&& {{\cal D}_2=i\left({\cal C}_1-{\cal C}_2\right)}~.\eea
For the non-integer value of the mass parameter $\nu$, the Neumann function or the Bessel function of the second kind  can be written in terms of the usual Bessel function or the Bessel function of the first kind as given by the following expression:
\bea {{\cal Y}_{\nu}(-k\tau)=\frac{1}{\sin \nu\pi}\left[{\cal J}_{\nu}(-k\tau)\cos \nu\pi-{\cal J}_{-\nu}(-k\tau)\right]}~.\eea
In this specific situation, the Hankel functions of first and second kind with order $\nu$ can be expressed in terms of usual Bessel function or the Bessel function of the first kind as:
\bea && {H^{(1)}_{\nu}(-k\tau)={\cal J}_{\nu}(-k\tau)+\frac{i}{\sin \nu\pi}\left[{\cal J}_{\nu}(-k\tau)\cos \nu\pi-{\cal J}_{-\nu}(-k\tau)\right]}\nonumber\\
&&~~~~~~~~~~~~~~{=\left[\left(1+i~{\rm cot}~\nu \pi\right){\cal J}_{\nu}(-k\tau)-i~{\rm cosec}~\nu\pi ~{\cal J}_{-\nu}(-k\tau)\right]}~,\eea\bea
&& {H^{(2)}_{\nu}(-k\tau)={\cal J}_{\nu}(-k\tau)-\frac{i}{\sin \nu\pi}\left[{\cal J}_{\nu}(-k\tau)\cos \nu\pi-{\cal J}_{-\nu}(-k\tau)\right]}\nonumber\\
&&~~~~~~~~~~~~~~{=\left[\left(1-i~{\rm cot}~\nu \pi\right){\cal J}_{\nu}(-k\tau)+i~{\rm cosec}~\nu\pi ~{\cal J}_{-\nu}(-k\tau)\right]}~.\eea 
In this case, the solution for the rescaled scalar perturbation mode can be recast as:
\bea  {f_{\bf k}(\tau)=\sqrt{-\tau}\left[{\cal E}_1(\nu)~{\cal J}_{\nu}(-k\tau)+{\cal E}_2(\nu)~{\cal J}_{-\nu}(-k\tau)\right]}, \eea
where the redefined two new arbitrary integration constants, ${\cal E}_1$ and ${\cal E}_2$ are defined in terms of the previously defined two new arbitrary integration constants, ${\cal C}_1$ and ${\cal C}_2$ as:
\bea && {{\cal E}_1(\nu)=\left({\cal C}_1+{\cal C}_2\right)+i\left({\cal C}_1-{\cal C}_2\right)~{\rm cot}~\nu \pi={\cal D}_1+{\cal D}_2~{\rm cot}~\nu \pi}~,\\
&& {{\cal E}_2(\nu)=-i\left({\cal C}_1-{\cal C}_2\right)~{\rm cosec}~\nu\pi=-{\cal D}_2~{\rm cosec}~\nu\pi}~.\eea

In particular, when, $\nu \notin \mathbb{Z}$, we have the following further simplified expressions for the Hankel functions of first and second kind, can be written as:
\bea && {H^{(1)}_{\nu}(-k\tau)=\frac{1}{i\sin \nu\pi}\left[{\cal J}_{-\nu}(-k\tau)-\exp(-i\nu\pi){\cal J}_{\nu}(-k\tau)\right]}~,\\
&& {H^{(2)}_{\nu}(-k\tau)=\frac{i}{\sin \nu\pi}\left[{\cal J}_{-\nu}(-k\tau)-\exp(i\nu\pi){\cal J}_{\nu}(-k\tau)\right]}~.\eea
This implies the following simplified relations which are very useful for the further computations:
\bea && {H^{(1)}_{-\nu}(-k\tau)=\exp(i\nu\pi)~H^{(1)}_{\nu}(-k\tau)}~,\\
&& {H^{(2)}_{-\nu}(-k\tau)=\exp(-i\nu\pi)~H^{(1)}_{\nu}(-k\tau)}~.\eea 
 The above mentioned relationships are valid, whether $\nu$ is an integer or not. In this case, the solution for the rescaled scalar perturbation mode can be recast as:
\bea  {f_{\bf k}(\tau)=\sqrt{-\tau}\left[{\cal G}_1(\nu)~{\cal J}_{\nu}(-k\tau)+{\cal G}_2(\nu)~{\cal J}_{-\nu}(-k\tau)\right]}, \eea
where the redefined two new arbitrary integration constants, ${\cal G}_1$ and ${\cal G}_2$ are defined in terms of the previously defined two new arbitrary integration constants, ${\cal C}_1$ and ${\cal C}_2$ as:
\bea && {{\cal G}_1(\nu)=\left({\cal C}_1+{\cal C}_2\right)+i\left({\cal C}_1-{\cal C}_2\right)~{\rm cot}~\nu\pi={\cal D}_1+{\cal D}_2~{\rm cot}~\nu \pi\neq {\cal E}_1(\nu)}~,\\
&& {{\cal G}_2(\nu)=i~{\rm cosec}~\nu\pi~\left({\cal C}_2-{\cal C}_1\right)=-{\cal D}_2~{\rm cosec}~\nu\pi\neq {\cal E}_{2}(\nu)}~.\eea

 In a specific situation, where one can express the mass parameter as a half-integer value like, $\nu=\left(n+\frac{1}{2}\right)$ for all non-negative integer $n$, we can write down further the following expressions, which are also very useful when we are dealing with these situations in the present context:
 \bea && {{\cal J}_{-\left(n+\frac{1}{2}\right)}(-k\tau)=(-1)^{n+1}~{\cal Y}_{\left(n+\frac{1}{2}\right)}(-k\tau)}~,\\
 && {{\cal Y}_{-\left(n+\frac{1}{2}\right)}(-k\tau)=(-1)^{n+1}~{\cal J}_{\left(n+\frac{1}{2}\right)}(-k\tau)}~.\eea
The corresponding most general canonically conjugate momentum can be further computed from this derived solution as:
\bea \Pi_{\bf k}(\tau)=\partial_{\tau}f_{\bf k}(\tau)=\frac{1}{2 \sqrt{-\tau }}\left[{\cal C}_1\left( k \tau  H_{\nu -1}^{(1)}(-k \tau )- H_{\nu }^{(1)}(-k \tau )- k \tau  H_{\nu +1}^{(1)}(-k \tau )\right)\nonumber\right.\\ \left.~~~~~~~~~~~~~~~~~~~~~+{\cal C}_2 \left(k \tau  H_{\nu -1}^{(2)}(-k \tau )-H_{\nu }^{(2)}(-k \tau )- k \tau  H_{\nu +1}^{(2)}(-k \tau )\right)\right].\eea
Here we have used the following facts for the derivative of the Hankel function of the first kind and second kind, which are very useful for the computation:
\bea && {\frac{d}{d(-k\tau)}H^{(1)}_{\nu}(-k\tau)=H^{(1)}_{\nu-1}(-k\tau)+\frac{\nu}{k\tau}H^{(1)}_{\nu}(-k\tau)=-H^{(1)}_{\nu+1}(-k\tau)+\frac{\nu}{k\tau}H^{(1)}_{\nu}(-k\tau)}~,~~~~~~~~\\
&& {\frac{d}{d(-k\tau)}H^{(2)}_{\nu}(-k\tau)=H^{(2)}_{\nu-1}(-k\tau)+\frac{\nu}{k\tau}H^{(2)}_{\nu}(-k\tau)=-H^{(2)}_{\nu+1}(-k\tau)+\frac{\nu}{k\tau}H^{(2)}_{\nu}(-k\tau)}~\eea
Also one can express the Bessel function of the first kind in terms of the Confluent Hypergeometric limit functions in the present context, which is given by:
\bea {{\cal J}_{\nu}(-k\tau)=\frac{1}{\Gamma(\nu+1)}\left(-\frac{k\tau}{2}\right)^{\nu}~{}_{0}F_{1}\left(\nu+1;-\frac{(k\tau)^2}{4}\right)}~.\eea
For the non-integer value of the mass parameter $\nu$, the Neumann function or the Bessel function of the second kind can be written in terms of the Confluent Hypergeometric limit functions in the present context, which are given by:
\bea &&{{\cal Y}_{\nu}(-k\tau)=\frac{{\rm cot}~ \nu\pi}{\Gamma(\nu+1)}\left(-\frac{k\tau}{2}\right)^{\nu}~{}_{0}F_{1}\left(\nu+1;-\frac{(k\tau)^2}{4}\right)}~\nonumber\\
&&~~~~~~~~~~~~~~~~~~~~~~{-\frac{{\rm cosec}~ \nu\pi}{\Gamma(1-\nu)}\left(-\frac{k\tau}{2}\right)^{-\nu}~{}_{0}F_{1}\left(1-\nu;-\frac{(k\tau)^2}{4}\right)}~.\eea 
In this specific situation, the Hankel functions of first and second kind with order $\nu$ can be expressed in terms of usual Bessel function or the Bessel function of the first kind as:
\bea && H^{(1)}_{\nu}(-k\tau)=\frac{\left(1+i~{\rm cot}~\nu \pi\right)}{\Gamma(\nu+1)}\left(-\frac{k\tau}{2}\right)^{\nu}~{}_{0}F_{1}\left(\nu+1;-\frac{(k\tau)^2}{4}\right)~\nonumber\\
&&~~~~~~~~~~~~~~~~~~~~~~~~~+i\frac{~~{\rm cosec}~\nu\pi }{\Gamma(1-\nu)}\left(-\frac{k\tau}{2}\right)^{-\nu}~{}_{0}F_{1}\left(1-\nu;-\frac{(k\tau)^2}{4}\right),\eea\bea
&& {H^{(2)}_{\nu}(-k\tau)=\frac{\left(1-i~{\rm cot}~\nu \pi\right)}{\Gamma(\nu+1)}\left(-\frac{k\tau}{2}\right)^{\nu}~{}_{0}F_{1}\left(\nu+1;-\frac{(k\tau)^2}{4}\right)}~\nonumber\\
&&~~~~~~~~~~~~~~~~~~~~~~~~~{-i\frac{~~{\rm cosec}~\nu\pi }{\Gamma(1-\nu)}\left(-\frac{k\tau}{2}\right)^{-\nu}~{}_{0}F_{1}\left(1-\nu;-\frac{(k\tau)^2}{4}\right)}~.\eea 
In this case, the most general solution of the above mentioned equation of motion is given by the following expression:
\bea &&{f_{\bf k}(\tau)={\cal E}_1(\nu)\frac{\sqrt{-\tau}}{\Gamma(\nu+1)}\left(-\frac{k\tau}{2}\right)^{\nu}~{}_{0}F_{1}\left(\nu+1;-\frac{(k\tau)^2}{4}\right)}~\nonumber\\
&&~~~~~~~~~~~~~~~~~~~~~~~~~{-{\cal E}_2(\nu)\frac{\sqrt{-\tau}}{\Gamma(1-\nu)}\left(-\frac{k\tau}{2}\right)^{-\nu}~{}_{0}F_{1}\left(1-\nu;-\frac{(k\tau)^2}{4}\right)}~.\eea
Now, it is important to note that the Bessel function of first and second kind can be expressed in the following asymptotic mathematical form for the small argument lying within the window $0<(-k\tau)<\sqrt{\nu+1}$, given by:
\begin{eqnarray}
&&{ \large {\cal J}_{\nu}(-k\tau)= \left\{
     \begin{array}{lr}
  \displaystyle  \frac{1}{\Gamma(\nu+1)}\left(-\frac{k\tau}{2}\right)^{\nu} &~ \text{\textcolor{red}{\bf if~$\nu>0$~integer}}\\ \\
  \displaystyle  \frac{(-1)^{\nu}}{(-\nu)!}\left(-\frac{k\tau}{2}\right)^{-\nu},~~~&~~~ \text{\textcolor{red}{\bf if~$\nu<0$~~integer}} \end{array}
   \right.}~~~~~~~~~~
\\
&&{ \large {\cal Y}_{\nu}(-k\tau)= \left\{
     \begin{array}{lr}
  \displaystyle  \frac{2}{\pi}\left[\ln\left(-\frac{k\tau}{2}\right)+\gamma\right] &~ \text{\textcolor{red}{\bf if~$\nu=0$}}\\ 
  \displaystyle  -\frac{\Gamma(\nu)}{\pi}\left(-\frac{k\tau}{2}\right)^{-\nu}+\frac{{\rm cot}~\nu\pi}{\Gamma(\nu+1)}\left(-\frac{k\tau}{2}\right)^{\nu},~~~&\text{\textcolor{red}{\bf if~$\nu>0$~integer}} \\ 
  \displaystyle  \frac{(-1)^{\nu+1}\Gamma(-\nu)}{\pi}\left(-\frac{k\tau}{2}\right)^{\nu},~~~& \text{\textcolor{red}{\bf if~$\nu<0$~integer}}\end{array}
   \right.}~~~~~~~~~~
\end{eqnarray}
where $\gamma$ is the {\it Euler–Mascheroni constant}, which is defined as:
\bea &&{ \gamma= \lim_{n\rightarrow\infty}\left(\sum^{n}_{p=1}\frac{1}{p}-\ln n\right)=\lim_{n\rightarrow\infty}\left(\frac{n^{n+\frac{1}{2}}\Gamma\left(\frac{1}{n}\right)\Gamma(n+1)}{\Gamma\left(2+n+\frac{1}{n}\right)}-\frac{n^2}{n+1}\right)}\nonumber\\
&&~~~~~~~~~~~~~~~~~~~~~~~~~~~~~~~~~~{=\sum^{\infty}_{n=2}(-1)^{n}\frac{\zeta(n)}{n}\approx 0.57721}~.\eea
 Consequently, the Hankel function of the first and second kind takes the following asymptotic mathematical form for the small argument lying within the window $0<(-k\tau)<\sqrt{\nu+1}$, as given by:
 \begin{eqnarray}
&& \large H^{(1)}_{\nu}(-k\tau)= \left\{
     \begin{array}{lr}
  \displaystyle  \frac{\left(1+i~{\rm cot}~\nu\pi\right)}{\Gamma(\nu+1)}\left(-\frac{k\tau}{2}\right)^{\nu}-i\frac{\Gamma(\nu)}{\pi}\left(-\frac{k\tau}{2}\right)^{-\nu} &~ \text{\textcolor{red}{\bf if~$\nu>0$~integer}}\\ \\
  \displaystyle  \frac{(-1)^{\nu}}{(-\nu)!}\left(-\frac{k\tau}{2}\right)^{-\nu}+i  \frac{(-1)^{\nu+1}\Gamma(-\nu)}{\pi}\left(-\frac{k\tau}{2}\right)^{\nu},& \text{\textcolor{red}{\bf if~$\nu<0$~~integer}} \end{array}
   \right.~~~~~~~~~
\\
&& \large H^{(2)}_{\nu}(-k\tau)= \left\{
     \begin{array}{lr}
  \displaystyle  \frac{\left(1-i~{\rm cot}~\nu\pi\right)}{\Gamma(\nu+1)}\left(-\frac{k\tau}{2}\right)^{\nu}+i\frac{\Gamma(\nu)}{\pi}\left(-\frac{k\tau}{2}\right)^{-\nu} &~ \text{\textcolor{red}{\bf if~$\nu>0$~integer}}\\ \\
  \displaystyle  \frac{(-1)^{\nu}}{(-\nu)!}\left(-\frac{k\tau}{2}\right)^{-\nu}-i  \frac{(-1)^{\nu+1}\Gamma(-\nu)}{\pi}\left(-\frac{k\tau}{2}\right)^{\nu},& \text{\textcolor{red}{\bf if~$\nu<0$~~integer}} \end{array}
   \right.~~~~~~~~~
\end{eqnarray}
Consequently, the asymptotic solution for the rescaled scalar perturbation can be expressed within the window $0<(-k\tau)<\sqrt{\nu+1}$, as:
 \begin{eqnarray}
&&{ \large f_{\bf k}(\tau)= \left\{
     \begin{array}{lr}
  \displaystyle  \frac{\left({\cal D}_1+{\cal D}_2~{\rm cot}~\nu\pi\right)}{\Gamma(\nu+1)}\left(-\frac{k\tau}{2}\right)^{\nu}-\frac{{\cal D}_2\Gamma(\nu)}{\pi}\left(-\frac{k\tau}{2}\right)^{-\nu} &~ \text{\textcolor{red}{\bf if~$\nu>0$~integer}}\\ \\
  \displaystyle \frac{{\cal D}_1(-1)^{\nu}}{(-\nu)!}\left(-\frac{k\tau}{2}\right)^{-\nu}+  \frac{{\cal D}_2(-1)^{\nu+1}\Gamma(-\nu)}{\pi}\left(-\frac{k\tau}{2}\right)^{\nu},& \text{\textcolor{red}{\bf if~$\nu<0$~integer}} \end{array}
   \right.}~~~~~~~~~
\end{eqnarray}
For large real arguments lying within the window, $(-k\tau)>>\left|\nu^2-\frac{1}{4}\right|$, one cannot write an actual asymptotic form for the Bessel functions of the first and second kind (unless in the situation where $\nu$ is a half-integer) because they have zeros all the way out to infinity, which would have to be matched exactly by any asymptotic expansion.
However, for a given value of ${\rm arg} (-k\tau)<\pi$, one can write an equation containing a term of order of $|-k\tau|^{-1}$, given by the following expressions:
\bea && {{\cal J}_{\nu}(-k\tau)=\sqrt{\frac{2}{\pi}}\frac{1}{\sqrt{-k\tau}}\left[\cos\left(k\tau+\frac{\pi}{2}\left(\nu+\frac{1}{2}\right)\right)+\exp\left({\rm Im}(-k\tau)\right){\cal O}\left(\frac{1}{|-k\tau|}\right)\right]}~,\\ && {{\cal Y}_{\nu}(-k\tau)=-\sqrt{\frac{2}{\pi}}\frac{1}{\sqrt{-k\tau}}\left[\sin\left(k\tau+\frac{\pi}{2}\left(\nu+\frac{1}{2}\right)\right)+\exp\left({\rm Im}(-k\tau)\right){\cal O}\left(\frac{1}{|-k\tau|}\right)\right]}~~~~~~~~~~~~~\eea
Consequently, the Hankel function of the first and second kind can be written in the following asymptotic form:
\bea &&{H^{(1)}_{\nu}(-k\tau)=\sqrt{\frac{2}{\pi}}\frac{1}{\sqrt{-k\tau}}\exp(-ik\tau)\exp\left(-\frac{i\pi}{2}\left(\nu+\frac{1}{2}\right)\right)},\nonumber\\
&&~~~~~~~~~~~~~~~~~~~~~{+\frac{2}{\sqrt{\pi}}\frac{1}{\sqrt{-k\tau}}\exp\left({\rm Im}(-k\tau)-\frac{i\pi}{4}\right){\cal O}\left(\frac{1}{|-k\tau|}\right)},\\
&&{  H^{(2)}_{\nu}(-k\tau)=-\sqrt{\frac{2}{\pi}}\frac{1}{\sqrt{-k\tau}}\exp(ik\tau)\exp\left(\frac{i\pi}{2}\left(\nu+\frac{1}{2}\right)\right)},\nonumber\\
&&~~~~~~~~~~~~~~~~~~~~~{+\frac{2}{\sqrt{\pi}}\frac{1}{\sqrt{-k\tau}}\exp\left({\rm Im}(-k\tau)+\frac{i\pi}{4}\right){\cal O}\left(\frac{1}{|-k\tau|}\right)},\eea
Using these asymptotic solution the general structure of the obtained solution for the rescaled field can be expressed as:
\bea && {f_{\bf k}(\tau)=\sqrt{\frac{2}{\pi k}}\left[{\cal C}_1\exp(-ik\tau)\exp\left(-\frac{i\pi}{2}\left(\nu+\frac{1}{2}\right)\right)-{\cal C}_2\exp(ik\tau)\exp\left(\frac{i\pi}{2}\left(\nu+\frac{1}{2}\right)\right)\right]}\nonumber\\
&& ~~~~~~~~~~~~~~~~~~~~~~~~~~~~~~~~~~~~~~~~~~~~~~~~~~~~~{+\sqrt{\frac{2}{\pi k}}\exp\left({\rm Im}(-k\tau)\right){\cal O}\left(\frac{1}{|-k\tau|}\right)\left({\cal D}_1-{\cal D}_2\right)}~.~~~~~~~~~~~~\eea
However, from the general structure of the obtained solution for the rescaled field and for the canonically conjugate momentum it is very difficult to extract the physical information out of that. For this reason the asymptotic solutions are really helpful for physical interpretation in different cosmological scales. These asymptotic limits are $k\tau\rightarrow 0$ and $k\tau\rightarrow- \infty$, where we need to determine the behaviour of the Hankel functions of the first and second kind of order $\nu$. After taking these asymptotic limits we get the following simplified results:
\bea &&{\lim_{k\tau\rightarrow -\infty} H^{(1)}_{\nu}(-k\tau)=\sqrt{\frac{2}{\pi}}\frac{1}{\sqrt{-k\tau}}\exp(-ik\tau)\exp\left(-\frac{i\pi}{2}\left(\nu+\frac{1}{2}\right)\right)},\\
&&{ \lim_{k\tau\rightarrow -\infty} H^{(2)}_{\nu}(-k\tau)=-\sqrt{\frac{2}{\pi}}\frac{1}{\sqrt{-k\tau}}\exp(ik\tau)\exp\left(\frac{i\pi}{2}\left(\nu+\frac{1}{2}\right)\right)},\\
 &&{ \lim_{k\tau\rightarrow 0} H^{(1)}_{\nu}(-k\tau)=\frac{i}{\pi}\Gamma(\nu)\left(-\frac{k\tau}{2}\right)^{-\nu}},\\
   &&{\lim_{k\tau\rightarrow 0} H^{(2)}_{\nu}(-k\tau)=-\frac{i}{\pi}\Gamma(\nu)\left(-\frac{k\tau}{2}\right)^{-\nu}},\eea
   and
   \bea &&{\lim_{k\tau\rightarrow -\infty} H^{(1)}_{\nu-1}(-k\tau)=\sqrt{\frac{2}{\pi}}\frac{1}{\sqrt{-k\tau}}\exp(-ik\tau)\exp\left(-\frac{i\pi}{2}\left(\nu-\frac{1}{2}\right)\right)},\eea\bea
&&{ \lim_{k\tau\rightarrow -\infty} H^{(2)}_{\nu-1}(-k\tau)=-\sqrt{\frac{2}{\pi}}\frac{1}{\sqrt{-k\tau}}\exp(ik\tau)\exp\left(\frac{i\pi}{2}\left(\nu-\frac{1}{2}\right)\right)},\\
 &&{ \lim_{k\tau\rightarrow 0} H^{(1)}_{\nu-1}(-k\tau)=\frac{i}{\pi}\Gamma(\nu-1)\left(-\frac{k\tau}{2}\right)^{1-\nu}},\\
   &&{\lim_{k\tau\rightarrow 0} H^{(2)}_{\nu-1}(-k\tau)=-\frac{i}{\pi}\Gamma(\nu-1)\left(-\frac{k\tau}{2}\right)^{1-\nu}},\eea
   and
   \bea &&{\lim_{k\tau\rightarrow -\infty} H^{(1)}_{\nu+1}(-k\tau)=\sqrt{\frac{2}{\pi}}\frac{1}{\sqrt{-k\tau}}\exp(-ik\tau)\exp\left(-\frac{i\pi}{2}\left(\nu+\frac{3}{2}\right)\right)},\\
&&{ \lim_{k\tau\rightarrow -\infty} H^{(2)}_{\nu+1}(-k\tau)=-\sqrt{\frac{2}{\pi}}\frac{1}{\sqrt{-k\tau}}\exp(ik\tau)\exp\left(\frac{i\pi}{2}\left(\nu+\frac{3}{2}\right)\right)},\\
 &&{ \lim_{k\tau\rightarrow 0} H^{(1)}_{\nu+1}(-k\tau)=\frac{i}{\pi}\Gamma(\nu+1)\left(-\frac{k\tau}{2}\right)^{-(\nu+1)}},\\
   &&{\lim_{k\tau\rightarrow 0} H^{(2)}_{\nu+1}(-k\tau)=-\frac{i}{\pi}\Gamma(\nu+1)\left(-\frac{k\tau}{2}\right)^{-(\nu+1)}},\eea
which implies that:
\bea && {\lim_{k\tau\rightarrow 0} H^{(1)}_{\nu}(-k\tau)=-\lim_{k\tau\rightarrow 0} H^{(2)}_{\nu}(-k\tau)},\\
&&  {\lim_{k\tau\rightarrow -\infty} H^{(1)}_{\nu}(-k\tau)=- \lim_{k\tau\rightarrow -\infty} H^{(2)}_{\nu}(-k\tau)},
\eea
and
\bea && {\lim_{k\tau\rightarrow 0} H^{(1)}_{\nu-1}(-k\tau)=-\lim_{k\tau\rightarrow 0} H^{(2)}_{\nu-1}(-k\tau)},\\
&&  {\lim_{k\tau\rightarrow -\infty} H^{(1)}_{\nu-1}(-k\tau)=- \lim_{k\tau\rightarrow -\infty} H^{(2)}_{\nu-1}(-k\tau)},
\eea
and
\bea && {\lim_{k\tau\rightarrow 0} H^{(1)}_{\nu+1}(-k\tau)=-\lim_{k\tau\rightarrow 0} H^{(2)}_{\nu+1}(-k\tau)},\\
&&  {\lim_{k\tau\rightarrow -\infty} H^{(1)}_{\nu+1}(-k\tau)=- \lim_{k\tau\rightarrow -\infty} H^{(2)}_{\nu+1}(-k\tau)}. 
\eea
Here $k\tau\rightarrow 0$ and $k\tau\rightarrow -\infty$ asymptotic limiting results are used to describe the superhorizon ($k\tau<<-1$) and subhorizon ($k\tau>>-1$) limiting results in the context of primordial cosmological perturbation scenario. The transition point from the subhorizon to superhorizon regime is identify by $k\tau=-1$ , which in Cosmology known as the horizon exit and play a pivotal role to measure various observables of primordial Universe from different theoretical models.

Now we consider the following {\it Laurent expansion} of the Gamma function as appearing in the superhorizon limiting approximation of the Hankel functions:
\bea &&{\Gamma(\nu)=\frac{2}{\nu}~~\underbrace{-\gamma+\frac{1}{2}\left(\gamma^2+\frac{\pi^2}{6}\right)\nu-\frac{1}{6}\left(\gamma^3+\frac{\gamma\pi^2}{2}+2\zeta(3)\right)\nu^2+{\cal O}(\nu^3)}_{\textcolor{red}{\bf Small~contribution}}},~~~\\ && {\Gamma(\nu-1)=\frac{2}{\nu-1}~~\underbrace{-\gamma+\frac{1}{2}\left(\gamma^2+\frac{\pi^2}{6}\right)(\nu-1)-\frac{1}{6}\left(\gamma^3+\frac{\gamma\pi^2}{2}+2\zeta(3)\right)(\nu-1)^2+{\cal O}((\nu-1)^3)}_{\textcolor{red}{\bf Small~contribution}}},~~~~~~~\\ &&{\Gamma(\nu+1)=\frac{2}{\nu+1}~~\underbrace{-\gamma+\frac{1}{2}\left(\gamma^2+\frac{\pi^2}{6}\right)(\nu+1)-\frac{1}{6}\left(\gamma^3+\frac{\gamma\pi^2}{2}+2\zeta(3)\right)(\nu+1)^2+{\cal O}((\nu+1)^3)}_{\textcolor{red}{\bf Small~contribution}}},~~~~~~\eea
where $\gamma$ is known as the {\it Euler Mascheroni constant} and $\zeta(3)$ is the Riemann zeta function of order 3. Further 
using this result the Hankel functions of the first and second kind of order $\nu$ can be simplified as given by the following expression:
 \bea &&{\lim_{k\tau\rightarrow 0} H^{(1)}_{\nu}(-k\tau)=\frac{i}{\pi}\left(-\frac{k\tau}{2}\right)^{-\nu}\left[\frac{2}{\nu}-\gamma+\frac{1}{2}\left(\gamma^2+\frac{\pi^2}{6}\right)\nu+\cdots\right]},~~~~~~\\
 && {\lim_{k\tau\rightarrow 0} H^{(2)}_{\nu}(-k\tau)=-\frac{i}{\pi}\left(-\frac{k\tau}{2}\right)^{-\nu}\left[\frac{2}{\nu}-\gamma+\frac{1}{2}\left(\gamma^2+\frac{\pi^2}{6}\right)\nu+\cdots\right]},~~~~~~~~\eea
 and
  \bea &&{\lim_{k\tau\rightarrow 0} H^{(1)}_{\nu-1}(-k\tau)=\frac{i}{\pi}\left(-\frac{k\tau}{2}\right)^{1-\nu}\left[\frac{2}{\nu-1}-\gamma+\frac{1}{2}\left(\gamma^2+\frac{\pi^2}{6}\right)(\nu-1)+\cdots\right]},~~~~~~\\
 && {\lim_{k\tau\rightarrow 0} H^{(2)}_{\nu-1}(-k\tau)=-\frac{i}{\pi}\left(-\frac{k\tau}{2}\right)^{1-\nu}\left[\frac{2}{\nu-1}-\gamma+\frac{1}{2}\left(\gamma^2+\frac{\pi^2}{6}\right)(\nu-1)+\cdots\right]},~~~~~~~~\eea
 and
  \bea &&{\lim_{k\tau\rightarrow 0} H^{(1)}_{\nu+1}(-k\tau)=\frac{i}{\pi}\left(-\frac{k\tau}{2}\right)^{-(\nu+1)}\left[\frac{2}{\nu+1}-\gamma+\frac{1}{2}\left(\gamma^2+\frac{\pi^2}{6}\right)(\nu+1)+\cdots\right]},~~~~~~\\
 && {\lim_{k\tau\rightarrow 0} H^{(2)}_{\nu+1}(-k\tau)=-\frac{i}{\pi}\left(-\frac{k\tau}{2}\right)^{-(\nu+1)}\left[\frac{2}{\nu+1}-\gamma+\frac{1}{2}\left(\gamma^2+\frac{\pi^2}{6}\right)(\nu+1)+\cdots\right]}.~~~~~~~~\eea
 Similarly at the horizon exit transition point $k\tau=-1$ the Hankel functions of the first and second kind of order $\nu$ can be simplified as given by the following expression:
 \bea &&{H^{(1)}_{\nu}(-k\tau=1)=\frac{i}{\pi}\left(\frac{1}{2}\right)^{-\nu}\left[\frac{2}{\nu}-\gamma+\frac{1}{2}\left(\gamma^2+\frac{\pi^2}{6}\right)\nu-\frac{1}{6}\left(\gamma^3+\frac{\gamma\pi^2}{2}+2\zeta(3)\right)\nu^2+\cdots\right]},~~~~~~~~~~\eea\bea
 && {H^{(2)}_{\nu}(-k\tau=1)=-\frac{i}{\pi}\left(\frac{1}{2}\right)^{-\nu}\left[\frac{2}{\nu}-\gamma+\frac{1}{2}\left(\gamma^2+\frac{\pi^2}{6}\right)\nu-\frac{1}{6}\left(\gamma^3+\frac{\gamma\pi^2}{2}+2\zeta(3)\right)\nu^2+\cdots\right]},~~~~~~~~~~~~\eea
 and
  \bea &&{H^{(1)}_{\nu-1}(-k\tau=1)=\frac{i}{\pi}\left(\frac{1}{2}\right)^{1-\nu}\left[\frac{2}{\nu-1}-\gamma+\frac{1}{2}\left(\gamma^2+\frac{\pi^2}{6}\right)(\nu-1)+\cdots\right]},~~~~~~\\
 && {H^{(2)}_{\nu}(-k\tau=1)=-\frac{i}{\pi}\left(\frac{1}{2}\right)^{1-\nu}\left[\frac{2}{\nu-1}-\gamma+\frac{1}{2}\left(\gamma^2+\frac{\pi^2}{6}\right)(\nu-1)+\cdots\right]},~~~~~~~~\eea
 and
  \bea &&{H^{(1)}_{\nu+1}(-k\tau=1)=\frac{i}{\pi}\left(-\frac{k\tau}{2}\right)^{-(\nu+1)}\left[\frac{2}{\nu+1}-\gamma+\frac{1}{2}\left(\gamma^2+\frac{\pi^2}{6}\right)(\nu+1)+\cdots\right]},~~~~~~\\
 && {\H^{(1)}_{\nu+1}(-k\tau=1)=-\frac{i}{\pi}\left(-\frac{k\tau}{2}\right)^{-(\nu+1)}\left[\frac{2}{\nu+1}-\gamma+\frac{1}{2}\left(\gamma^2+\frac{\pi^2}{6}\right)(\nu+1)+\cdots\right]}.~~~~~~~~\eea
 
 Now, in the superhorizon limit ($k\tau<<-1$) and subhorizon limit ($k\tau>>-1$) the approximated asymptotic form of the most general solution for the rescaled field variable and the corresponding canonically conjugate momentum computed for the arbitrary quantum initial vacuum can be expressed as:
 \bea &&{\lim_{k\tau\rightarrow 0}f_{\bf k}(\tau)=\sqrt{\frac{2}{k}}\frac{i}{\pi}\Gamma(\nu)\left(-\frac{k\tau}{2}\right)^{\frac{1}{2}-\nu}\left({\cal C}_1-{\cal C}_2\right)},\\
&& {\lim_{k\tau\rightarrow -\infty}f_{\bf k}(\tau)=\sqrt{\frac{2}{\pi k}}\left[{\cal C}_1~\exp\left(-i\left\{k\tau+\frac{\pi}{2}\left(\nu+\frac{1}{2}\right)\right\}\right)-{\cal C}_2~\exp\left(i\left\{k\tau+\frac{\pi}{2}\left(\nu+\frac{1}{2}\right)\right\}\right)\right]}, ~~~~~~~~\eea
and 
 \bea &&{\lim_{k\tau\rightarrow 0}\Pi_{\bf k}(\tau)=\sqrt{\frac{2}{k}}\frac{i}{2\pi k}\left(\nu-\frac{1}{2}\right)\Gamma(\nu)\left(-\frac{k\tau}{2}\right)^{-\left(\nu+\frac{1}{2}\right)}\left({\cal C}_1-{\cal C}_2\right)},\\
&& {\lim_{k\tau\rightarrow -\infty}\Pi_{\bf k}(\tau)=\frac{1}{i}\sqrt{\frac{2k}{\pi }}\left[{\cal C}_1~\exp\left(-i\left\{k\tau+\frac{\pi}{2}\left(\nu+\frac{1}{2}\right)\right\}\right)+{\cal C}_2~\exp\left(i\left\{k\tau+\frac{\pi}{2}\left(\nu+\frac{1}{2}\right)\right\}\right)\right]}. ~~~~~~~~\eea 
Combining the behaviour in both the superhorizon and subhorizon limiting region we get following combined asymptotic most general solution for the rescaled field variable and momenta computed for the arbitrary quantum initial vacuum can be expressed as:
\bea&&\hll{f_{\bf k}(\tau)=2^{\nu-\frac{3}{2}}\frac{1}{i\tau}\frac{1}{\sqrt{2}k^{\frac{3}{2}}}(-k\tau)^{\frac{3}{2}-\nu}\left|\frac{\Gamma(\nu)}{\Gamma\left(\frac{3}{2}\right)}\right|}\nonumber\\
&&~~~\hll{\times\left[{\cal C}_1~(1+ik\tau)~\exp\left(-i\left\{k\tau+\frac{\pi}{2}\left(\nu+\frac{1}{2}\right)\right\}\right)-{\cal C}_2~(1-ik\tau)~\exp\left(i\left\{k\tau+\frac{\pi}{2}\left(\nu+\frac{1}{2}\right)\right\}\right)\right]},\nonumber\\
&&\eea
\bea && \hll{\Pi_{\bf k}(\tau)=2^{\nu-\frac{3}{2}}\frac{1}{\sqrt{2}ik^{\frac{5}{2}}}(-k\tau)^{\frac{3}{2}-\nu}\left|\frac{\Gamma(\nu)}{\Gamma\left(\frac{3}{2}\right)}\right|}\nonumber\\
&&~~~~~~~~~~~~~~\nonumber\times\left[\hll{{\cal C}_1~\left\{\left(\frac{1}{2}-\nu\right)\frac{(1+ik\tau)}{k^2\tau^2}+1\right\}~\exp\left(-i\left\{k\tau+\frac{\pi}{2}\left(\nu+\frac{1}{2}\right)\right\}\right)}\right.\\&& \left.~~~~~~~~~~~~~~~~~~~~~~\hll{-{\cal C}_2~\left\{\left(\frac{1}{2}-\nu\right)\frac{(1-ik\tau)}{k^2\tau^2}+1\right\}~\exp\left(i\left\{k\tau+\frac{\pi}{2}\left(\nu+\frac{1}{2}\right)\right\}\right)}\right],~~~~~~~ 
\eea
These general asymptotic expressions are extremely important to compute the expressions for the OTOC's in the later subsections. To server this purpose we need to promote both of these classical solutions to the quantum level. 
\newpage
\section{WKB solution of the cosmological mode functions for time dependent protocols}

For arbitrary conformal time dependence in the mass parameter $m^2(\tau)$ we use the standard WKB
approximation, using which the general solution can be expressed as:
\bea {f_{\bf k}(\tau)=\left[{\cal C}_1~u_{\bf k}(\tau)+{\cal C}_2~u^{*}_{\bf k}(\tau)\right]},~\eea
where ${\cal C}_1$ and ${\cal C}_2$ are the Bogoliubov coefficients in the two arbitrary integration constants which can be fixed by the choice of the initial quantum vacuum. Here the WKB solutions $u_{\bf k}(\tau)$ and its complex conjugate $u^{*}_{\bf k}(\tau)$ can be written as:
\bea && u_{\bf k}(\tau)=\frac{1}{2\omega_{\bf k}(\tau)}~\exp\left(i\int^{\tau}_{\tau^{'}=-\infty}~d\tau^{'}~\omega_{\bf k}(\tau^{'})\right),~~u^{*}_{\bf k}(\tau)=\frac{1}{2\omega_{\bf k}(\tau)}~\exp\left(-i\int^{\tau}_{\tau^{'}=-\infty}~d\tau^{'}~\omega_{\bf k}(\tau^{'})\right).~~~~~~~\eea
Here the effective time dependent frequency $\omega_{\bf k}(\tau)$ can be expressed as:
\bea {\omega_{k}(\tau)=\sqrt{k^2-\frac{\left(\nu^2(\tau)-\frac{1}{4}\right)}{\tau^2}}}.\eea

Consequently, the canonically conjugate momentum from the above mentioned WKB approximated solution can be expressed as:
 \bea {\Pi_{\bf k}(\tau)=\partial_{\tau}f_{\bf k}(\tau)=\left[{\cal C}_1~v_{\bf k}(\tau)+{\cal C}_2~v^{*}_{\bf k}(\tau)\right]}~,\eea
 where we have introduced two conformal time dependent new functions $v_{\bf k}(\tau)$ and its complex conjugate $v^{*}_{\bf k}(\tau)$ can be expressed as:
 \bea && {v_{\bf k}(\tau)=\partial_{\tau}u_{\bf k}(\tau)=\partial_{\tau}\left(\frac{1}{2\omega_{\bf k}(\tau)}~\exp\left(i\int^{\tau}_{\tau^{'}=-\infty}~d\tau^{'}~\omega_{\bf k}(\tau^{'})\right)\right)}\nonumber \\
 &&~~~~~~~~~~~~~~~~~~~~~~{=\left[-\frac{1}{\omega_{\bf k}(\tau)}\frac{d\omega_{\bf k}(\tau)}{d\tau}+i\frac{d}{d\tau}\left(\int^{\tau}_{\tau^{'}=-\infty}~d\tau^{'}~\omega_{\bf k}(\tau^{'})\right)\right]u_{\bf k}(\tau)}~, \\ && {v^{*}_{\bf k}(\tau)=\partial_{\tau}u^{*}_{\bf k}(\tau)=\partial_{\tau}\left(\frac{1}{2\omega_{\bf k}(\tau)}~\exp\left(-i\int^{\tau}_{\tau^{'}=-\infty}~d\tau^{'}~\omega_{\bf k}(\tau^{'})\right)\right)}\nonumber \\
 &&~~~~~~~~~~~~~~~~~~~~~~{=\left[-\frac{1}{\omega_{\bf k}(\tau)}\frac{d\omega_{\bf k}(\tau)}{d\tau}-i\frac{d}{d\tau}\left(\int^{\tau}_{\tau^{'}=-\infty}~d\tau^{'}~\omega_{\bf k}(\tau^{'})\right)\right]u^{*}_{\bf k}(\tau)}~. \eea
\section{Quantum two-point OTO micro-canonical amplitudes for Cosmology} 
\subsection{Definition of micro-canonical OTO amplitude $\hat{\Delta}_1({\bf k}_1,{\bf k}_2;\tau_1,\tau_2)$}
In this subsection we define a very important momentum and conformal time dependent two-point OTO amplitude, which are given by the following expression:
\bea \hat{\Delta}_{1}({\bf k}_1,{\bf k}_2;\tau_1,\tau_2)&=&\hat{f}_{{\bf k}_1}(\tau_1)\hat{\Pi}_{{\bf k}_2}(\tau_2)\nonumber\\
 &=&f_{{\bf k}_1}(\tau_1)\Pi_{{\bf k}_2}(\tau_2)~a_{{\bf k}_1}a_{{\bf k}_2}+f^{*}_{{\bf -k}_1}(\tau_1)\Pi_{{\bf k}_2}(\tau_2)~a^{\dagger}_{-{\bf k}_1}a_{{\bf k}_2}\nonumber\\
 &&~~~~~~~~+f_{{\bf k}_1}(\tau_1)\Pi^{*}_{{\bf -k}_2}(\tau_2)~a_{{\bf k}_1}a^{\dagger}_{-{\bf k}_2}+f^{*}_{{\bf -k}_1}(\tau_1)\Pi^{*}_{{\bf -k}_2}(\tau_2)~a^{\dagger}_{-{\bf k}_1}a^{\dagger}_{-{\bf k}_2}\nonumber\\
 &=&{\cal D}_1 ({\bf k}_1,{\bf k}_2;\tau_1,\tau_2)~a_{{\bf k}_1}a_{{\bf k}_2}+{\cal D}_2 ({\bf k}_1,{\bf k}_2;\tau_1,\tau_2)~a^{\dagger}_{-{\bf k}_1}a_{{\bf k}_2}\nonumber\\
 &&~~~~~~~~+{\cal D}_3 ({\bf k}_1,{\bf k}_2;\tau_1,\tau_2)~a_{{\bf k}_1}a^{\dagger}_{-{\bf k}_2}+{\cal D}_4 ({\bf k}_1,{\bf k}_2;\tau_1,\tau_2)~a^{\dagger}_{-{\bf k}_1}a^{\dagger}_{-{\bf k}_2},~~~~~~\eea  
 where we have introduced  momentum and time dependent four individual two-point OTO amplitudes, ${\cal D}_i ({\bf k}_1,{\bf k}_2;\tau_1,\tau_2)~~\forall~~i=1,2,3,4$, which are explicitly defined as:
 \bea {\cal D}_1 ({\bf k}_1,{\bf k}_2;\tau_1,\tau_2)&=&f_{{\bf k}_1}(\tau_1)\Pi_{{\bf k}_2}(\tau_2),\\
 {\cal D}_2 ({\bf k}_1,{\bf k}_2;\tau_1,\tau_2)&=&f^{*}_{{\bf -k}_1}(\tau_1)\Pi_{{\bf k}_2}(\tau_2),\\
 {\cal D}_3 ({\bf k}_1,{\bf k}_2;\tau_1,\tau_2)&=&f_{{\bf k}_1}(\tau_1)\Pi^{*}_{{\bf -k}_2}(\tau_2),\\
 {\cal D}_4 ({\bf k}_1,{\bf k}_2;\tau_1,\tau_2)&=&f^{*}_{{\bf -k}_1}(\tau_1)\Pi^{*}_{{\bf -k}_2}(\tau_2).\eea
 These contributions are really helpful to compute the two-point micro-canonical OTO amplitudes and the corresponding momentum integrated OTOC, which we have discussed in this paper.
 \subsection{Definition of micro-canonical OTO amplitude $\hat{\Delta}_2({\bf k}_1,{\bf k}_2;\tau_1,\tau_2)$}
In this subsection we define a very important momentum and conformal time dependent two-point OTO amplitude, which are given by the following expression:
\bea \hat{\Delta}_{2}({\bf k}_1,{\bf k}_2;\tau_1,\tau_2)&=&\hat{f}_{{\bf k}_1}(\tau_1)\hat{\Pi}_{{\bf k}_2}(\tau_2)\nonumber\\
 &=&\Pi_{{\bf k}_1}(\tau_2)f_{{\bf k}_2}(\tau_1)~a_{{\bf k}_1}a_{{\bf k}_2}+\Pi^{*}_{{\bf -k}_1}(\tau_2)f_{{\bf k}_2}(\tau_1)~a^{\dagger}_{-{\bf k}_1}a_{{\bf k}_2}\nonumber\\
 &&~~~~~~~~+\Pi_{{\bf k}_1}(\tau_2)f^{*}_{-{\bf k}_2}(\tau_1)~a_{{\bf k}_1}a^{\dagger}_{-{\bf k}_2}+\Pi^{*}_{{\bf -k}_1}(\tau_2)f^{*}_{{\bf -k}_2}(\tau_1)~a^{\dagger}_{-{\bf k}_1}a^{\dagger}_{-{\bf k}_2}\nonumber\\
 &=&{\cal L}_1 ({\bf k}_1,{\bf k}_2;\tau_1,\tau_2)~a_{{\bf k}_1}a_{{\bf k}_2}+{\cal L}_2 ({\bf k}_1,{\bf k}_2;\tau_1,\tau_2)~a^{\dagger}_{-{\bf k}_1}a_{{\bf k}_2}\nonumber\\
 &&~~~~~~~~+{\cal L}_3 ({\bf k}_1,{\bf k}_2;\tau_1,\tau_2)~a_{{\bf k}_1}a^{\dagger}_{-{\bf k}_2}+{\cal L}_4 ({\bf k}_1,{\bf k}_2;\tau_1,\tau_2)~a^{\dagger}_{-{\bf k}_1}a^{\dagger}_{-{\bf k}_2},~~~~~~\eea  
 where we have introduced  momentum and time dependent four individual two-point OTO amplitudes, ${\cal L}_i ({\bf k}_1,{\bf k}_2;\tau_1,\tau_2)~~\forall~~i=1,2,3,4$, which are explicitly defined as:
 \bea {\cal L}_1 ({\bf k}_1,{\bf k}_2;\tau_1,\tau_2)&=&\Pi_{{\bf k}_1}(\tau_2)f_{{\bf k}_2}(\tau_1),\eea\bea
 {\cal L}_2 ({\bf k}_1,{\bf k}_2;\tau_1,\tau_2)&=&\Pi^{*}_{{\bf -k}_1}(\tau_2)f_{{\bf k}_2}(\tau_1),\\
 {\cal L}_3 ({\bf k}_1,{\bf k}_2;\tau_1,\tau_2)&=&\Pi_{{\bf k}_1}(\tau_2)f^{*}_{-{\bf k}_2}(\tau_1),\\
 {\cal L}_4 ({\bf k}_1,{\bf k}_2;\tau_1,\tau_2)&=&\Pi^{*}_{{\bf -k}_1}(\tau_2)f^{*}_{{\bf -k}_2}(\tau_1).\eea
 These contributions are really helpful to compute the two-point micro-canonical OTO amplitudes and the corresponding momentum integrated OTOC, which we have discussed in this paper.

 \newpage 
\section{Quantum four-point OTO micro-canonical amplitudes for Cosmology} 
\subsection{Definition of micro-canonical OTO amplitude $\widehat{\cal T}_1({\bf k}_1,{\bf k}_2,{\bf k}_3,{\bf k}_4;\tau_1,\tau_2)$}
The function $\widehat{\cal T}_1({\bf k}_1,{\bf k}_2,{\bf k}_3,{\bf k}_4;\tau_1,\tau_2)$ is defined as:
 \bea &&\widehat{\cal T}_1({\bf k}_1,{\bf k}_2,{\bf k}_3,{\bf k}_4;\tau_1,\tau_2)\nonumber\\
 &&=\left[{\cal M}_1({\bf k}_1,{\bf k}_2,{\bf k}_3,{\bf k}_4;\tau_1,\tau_2)~a_{{\bf k}_1}a_{{\bf k}_2}a_{{\bf k}_3}a_{{\bf k}_4}\right.\nonumber\\
  && \left.+ {\cal M}_2({\bf k}_1,{\bf k}_2,{\bf k}_3,{\bf k}_4;\tau_1,\tau_2)~a^{\dagger}_{-{\bf k}_1}a_{{\bf k}_2}a_{{\bf k}_3}a_{{\bf k}_4}+{\cal M}_3({\bf k}_1,{\bf k}_2,{\bf k}_3,{\bf k}_4;\tau_1,\tau_2)~a_{{\bf k}_1}a^{\dagger}_{-{\bf k}_2}a_{{\bf k}_3}a_{{\bf k}_4}\right.\nonumber\\
  && \left.+ {\cal M}_4({\bf k}_1,{\bf k}_2,{\bf k}_3,{\bf k}_4;\tau_1,\tau_2)~a^{\dagger}_{-{\bf k}_1}a^{\dagger}_{-{\bf k}_2}a_{{\bf k}_3}a_{{\bf k}_4}+{\cal M}_5({\bf k}_1,{\bf k}_2,{\bf k}_3,{\bf k}_4;\tau_1,\tau_2)~a_{{\bf k}_1}a_{{\bf k}_2}a^{\dagger}_{-{\bf k}_3}a_{{\bf k}_4}\right.\nonumber\\
  && \left.+ {\cal M}_6({\bf k}_1,{\bf k}_2,{\bf k}_3,{\bf k}_4;\tau_1,\tau_2)~a^{\dagger}_{-{\bf k}_1}a_{{\bf k}_2}a^{\dagger}_{-{\bf k}_3}a_{{\bf k}_4}+{\cal M}_7({\bf k}_1,{\bf k}_2,{\bf k}_3,{\bf k}_4;\tau_1,\tau_2)~a_{{\bf k}_1}a^{\dagger}_{-{\bf k}_2}a^{\dagger}_{-{\bf k}_3}a_{{\bf k}_4}\right.\nonumber\\
  && \left.+ {\cal M}_8({\bf k}_1,{\bf k}_2,{\bf k}_3,{\bf k}_4;\tau_1,\tau_2)~a^{\dagger}_{-{\bf k}_1}a^{\dagger}_{-{\bf k}_2}a^{\dagger}_{-{\bf k}_3}a_{{\bf k}_4}+{\cal M}_9({\bf k}_1,{\bf k}_2,{\bf k}_3,{\bf k}_4;\tau_1,\tau_2)~a_{{\bf k}_1}a_{{\bf k}_2}a_{{\bf k}_3}a^{\dagger}_{-{\bf k}_4}\right.\nonumber\\
  && \left.+ {\cal M}_{10}({\bf k}_1,{\bf k}_2,{\bf k}_3,{\bf k}_4;\tau_1,\tau_2)~a^{\dagger}_{-{\bf k}_1}a_{{\bf k}_2}a_{{\bf k}_3}a^{\dagger}_{-{\bf k}_4}+{\cal M}_{11}({\bf k}_1,{\bf k}_2,{\bf k}_3,{\bf k}_4;\tau_1,\tau_2)~a_{{\bf k}_1}a^{\dagger}_{-{\bf k}_2}a_{{\bf k}_3}a^{\dagger}_{-{\bf k}_4}\right.\nonumber\\
  && \left.+ {\cal M}_{12}({\bf k}_1,{\bf k}_2,{\bf k}_3,{\bf k}_4;\tau_1,\tau_2)~a^{\dagger}_{-{\bf k}_1}a^{\dagger}_{-{\bf k}_2}a_{{\bf k}_3}a^{\dagger}_{-{\bf k}_4}\right.\nonumber\\
  && \left.+{\cal M}_{13}({\bf k}_1,{\bf k}_2,{\bf k}_3,{\bf k}_4;\tau_1,\tau_2)~a_{{\bf k}_1}a_{{\bf k}_2}a^{\dagger}_{-{\bf k}_3}a^{\dagger}_{-{\bf k}_4}+ {\cal M}_{14}({\bf k}_1,{\bf k}_2,{\bf k}_3,{\bf k}_4;\tau_1,\tau_2)~a^{\dagger}_{-{\bf k}_1}a_{{\bf k}_2}a^{\dagger}_{-{\bf k}_3}a^{\dagger}_{-{\bf k}_4}\right.\nonumber\\
  && \left.+{\cal M}_{15}({\bf k}_1,{\bf k}_2,{\bf k}_3,{\bf k}_4;\tau_1,\tau_2)~a_{{\bf k}_1}a^{\dagger}_{-{\bf k}_2}a^{\dagger}_{-{\bf k}_3}a^{\dagger}_{-{\bf k}_4}+ {\cal M}_{16}({\bf k}_1,{\bf k}_2,{\bf k}_3,{\bf k}_4;\tau_1,\tau_2)~a^{\dagger}_{-{\bf k}_1}a^{\dagger}_{-{\bf k}_2}a^{\dagger}_{-{\bf k}_3}a^{\dagger}_{-{\bf k}_4}\right],~~~~~~~~~~~
  \eea
  where we define new sets of functions, ${\cal M}_{j}({\bf k}_1,{\bf k}_2,{\bf k}_3,{\bf k}_4;\tau_1,\tau_2)~\forall~j=1,\cdots,16$, as:
  \bea &&{\cal M}_{1}({\bf k}_1,{\bf k}_2,{\bf k}_3,{\bf k}_4;\tau_1,\tau_2)=f_{{\bf k}_1}(\tau_1)\Pi_{{\bf k}_2}(\tau_2)f_{{\bf k}_3}(\tau_1)\Pi_{{\bf k}_4}(\tau_2),\\
 && {\cal M}_2({\bf k}_1,{\bf k}_2,{\bf k}_3,{\bf k}_4;\tau_1,\tau_2)=f^{*}_{-{\bf k}_1}(\tau_1)\Pi_{{\bf k}_2}(\tau_2)f_{{\bf k}_3}(\tau_1)\Pi_{{\bf k}_4}(\tau_2),\\
 && {\cal M}_{3}({\bf k}_1,{\bf k}_2,{\bf k}_3,{\bf k}_4;\tau_1,\tau_2)=f_{{\bf k}_1}(\tau_1)\Pi^{*}_{-{\bf k}_2}(\tau_2)f_{{\bf k}_1}(\tau_1)\Pi_{{\bf k}_4}(\tau_2),\\
  && {\cal M}_{4}({\bf k}_1,{\bf k}_2,{\bf k}_3,{\bf k}_4;\tau_1,\tau_2)=f^{*}_{-{\bf k}_1}(\tau_1)\Pi^{*}_{-{\bf k}_2}(\tau_2)f_{{\bf k}_3}(\tau_1)\Pi_{{\bf k}_4}(\tau_2),\\
 && {\cal M}_{5}({\bf k}_1,{\bf k}_2,{\bf k}_3,{\bf k}_4;\tau_1,\tau_2)=f_{{\bf k}_1}(\tau_1)\Pi_{{\bf k}_2}(\tau_2)f^{*}_{-{\bf k}_3}(\tau_1)\Pi_{{\bf k}_4}(\tau_2),\\
  && {\cal M}_{6}({\bf k}_1,{\bf k}_2,{\bf k}_3,{\bf k}_4;\tau_1,\tau_2)=f^{*}_{-{\bf k}_1}(\tau_1)\Pi_{{\bf k}_2}(\tau_2)f^{*}_{-{\bf k}_3}(\tau_1)\Pi_{{\bf k}_4}(\tau_2)\\
 && {\cal M}_{7}({\bf k}_1,{\bf k}_2,{\bf k}_3,{\bf k}_4;\tau_1,\tau_2)=f_{{\bf k}_1}(\tau_1)\Pi^{*}_{-{\bf k}_2}(\tau_2)f^{*}_{-{\bf k}_3}(\tau_1)\Pi_{{\bf k}_4}(\tau_2),\\
  && {\cal M}_{8}({\bf k}_1,{\bf k}_2,{\bf k}_3,{\bf k}_4;\tau_1,\tau_2)=f^{*}_{-{\bf k}_1}(\tau_1)\Pi^{*}_{-{\bf k}_2}(\tau_2)f^{*}_{-{\bf k}_3}(\tau_1)\Pi_{{\bf k}_4}(\tau_2)\\
 && {\cal M}_{9}({\bf k}_1,{\bf k}_2,{\bf k}_3,{\bf k}_4;\tau_1,\tau_2)=f_{{\bf k}_1}(\tau_1)\Pi_{{\bf k}_2}(\tau_2)f_{{\bf k}_3}(\tau_1)\Pi^{*}_{-{\bf k}_4}(\tau_2),\\
  && {\cal M}_{10}({\bf k}_1,{\bf k}_2,{\bf k}_3,{\bf k}_4;\tau_1,\tau_2)=f^{*}_{-{\bf k}_1}(\tau_1)\Pi_{{\bf k}_2}(\tau_2)f_{{\bf k}_3}(\tau_1)\Pi^{*}_{-{\bf k}_4}(\tau_2)\\
 && {\cal M}_{11}({\bf k}_1,{\bf k}_2,{\bf k}_3,{\bf k}_4;\tau_1,\tau_2)=f_{{\bf k}_1}(\tau_1)\Pi^{*}_{-{\bf k}_2}(\tau_2)f_{{\bf k}_3}(\tau_1)\Pi^{*}_{-{\bf k}_4}(\tau_2),\\
  && {\cal M}_{12}({\bf k}_1,{\bf k}_2,{\bf k}_3,{\bf k}_4;\tau_1,\tau_2)=f^{*}_{-{\bf k}_1}(\tau_1)\Pi^{*}_{-{\bf k}_2}(\tau_2)f_{{\bf k}_3}(\tau_1)\Pi^{*}_{-{\bf k}_4}(\tau_2)\\
 && {\cal M}_{13}({\bf k}_1,{\bf k}_2,{\bf k}_3,{\bf k}_4;\tau_1,\tau_2)=f_{{\bf k}_1}(\tau_1)\Pi_{{\bf k}_2}(\tau_2)f^{*}_{-{\bf k}_3}(\tau_1)\Pi^{*}_{-{\bf k}_4}(\tau_2),\\
  && {\cal M}_{14}({\bf k}_1,{\bf k}_2,{\bf k}_3,{\bf k}_4;\tau_1,\tau_2)=f^{*}_{-{\bf k}_1}(\tau_1)\Pi_{{\bf k}_2}(\tau_2)f^{*}_{-{\bf k}_3}(\tau_1)\Pi^{*}_{-{\bf k}_4}(\tau_2)\\
 && {\cal M}_{15}({\bf k}_1,{\bf k}_2,{\bf k}_3,{\bf k}_4;\tau_1,\tau_2)=f_{{\bf k}_1}(\tau_1)\Pi^{*}_{-{\bf k}_2}(\tau_2)f^{*}_{-{\bf k}_3}(\tau_1)\Pi^{*}_{-{\bf k}_4}(\tau_2),\\
  && {\cal M}_{16}({\bf k}_1,{\bf k}_2,{\bf k}_3,{\bf k}_4;\tau_1,\tau_2)=f^{*}_{-{\bf k}_1}(\tau_1)\Pi^{*}_{-{\bf k}_2}(\tau_2)f^{*}_{-{\bf k}_3}(\tau_1)\Pi^{*}_{-{\bf k}_4}(\tau_2)\eea
\subsection{Definition of micro-canonical OTO amplitude $\widehat{\cal T}_2({\bf k}_1,{\bf k}_2,{\bf k}_3,{\bf k}_4;\tau_1,\tau_2)$}
The function $\widehat{\cal T}_2({\bf k}_1,{\bf k}_2,{\bf k}_3,{\bf k}_4;\tau_1,\tau_2)$ is defined as:
 \bea &&\widehat{\cal T}_2({\bf k}_1,{\bf k}_2,{\bf k}_3,{\bf k}_4;\tau_1,\tau_2)\nonumber\\
 &&=\left[{\cal J}_1({\bf k}_1,{\bf k}_2,{\bf k}_3,{\bf k}_4;\tau_1,\tau_2)~a_{{\bf k}_1}a_{{\bf k}_2}a_{{\bf k}_3}a_{{\bf k}_4}+ {\cal J}_2({\bf k}_1,{\bf k}_2,{\bf k}_3,{\bf k}_4;\tau_1,\tau_2)~a^{\dagger}_{-{\bf k}_1}a_{{\bf k}_2}a_{{\bf k}_3}a_{{\bf k}_4}\right.\nonumber\\
  && \left.+{\cal J}_3({\bf k}_1,{\bf k}_2,{\bf k}_3,{\bf k}_4;\tau_1,\tau_2)~a_{{\bf k}_1}a^{\dagger}_{-{\bf k}_2}a_{{\bf k}_3}a_{{\bf k}_4}+ {\cal J}_4({\bf k}_1,{\bf k}_2,{\bf k}_3,{\bf k}_4;\tau_1,\tau_2)~a^{\dagger}_{-{\bf k}_1}a^{\dagger}_{-{\bf k}_2}a_{{\bf k}_3}a_{{\bf k}_4}\right.\nonumber\\
  && \left.+{\cal J}_5({\bf k}_1,{\bf k}_2,{\bf k}_3,{\bf k}_4;\tau_1,\tau_2)~a_{{\bf k}_1}a_{{\bf k}_2}a^{\dagger}_{-{\bf k}_3}a_{{\bf k}_4}+ {\cal J}_6({\bf k}_1,{\bf k}_2,{\bf k}_3,{\bf k}_4;\tau_1,\tau_2)~a^{\dagger}_{-{\bf k}_1}a_{{\bf k}_2}a^{\dagger}_{-{\bf k}_3}a_{{\bf k}_4}\right.\nonumber\\
  && \left.+{\cal J}_7({\bf k}_1,{\bf k}_2,{\bf k}_3,{\bf k}_4;\tau_1,\tau_2)~a_{{\bf k}_1}a^{\dagger}_{-{\bf k}_2}a^{\dagger}_{-{\bf k}_3}a_{{\bf k}_4}+ {\cal J}_8({\bf k}_1,{\bf k}_2,{\bf k}_3,{\bf k}_4;\tau_1,\tau_2)~a^{\dagger}_{-{\bf k}_1}a^{\dagger}_{-{\bf k}_2}a^{\dagger}_{-{\bf k}_3}a_{{\bf k}_4}\right.\nonumber\\
  && \left.+{\cal J}_9({\bf k}_1,{\bf k}_2,{\bf k}_3,{\bf k}_4;\tau_1,\tau_2)~a_{{\bf k}_1}a_{{\bf k}_2}a_{{\bf k}_3}a^{\dagger}_{-{\bf k}_4}+ {\cal J}_{10}({\bf k}_1,{\bf k}_2,{\bf k}_3,{\bf k}_4;\tau_1,\tau_2)~a^{\dagger}_{-{\bf k}_1}a_{{\bf k}_2}a_{{\bf k}_3}a^{\dagger}_{-{\bf k}_4}\right.\nonumber\\
  && \left.+{\cal J}_{11}({\bf k}_1,{\bf k}_2,{\bf k}_3,{\bf k}_4;\tau_1,\tau_2)~a_{{\bf k}_1}a^{\dagger}_{-{\bf k}_2}a_{{\bf k}_3}a^{\dagger}_{-{\bf k}_4}+ {\cal J}_{12}({\bf k}_1,{\bf k}_2,{\bf k}_3,{\bf k}_4;\tau_1,\tau_2)~a^{\dagger}_{-{\bf k}_1}a^{\dagger}_{-{\bf k}_2}a_{{\bf k}_3}a^{\dagger}_{-{\bf k}_4}\right.\nonumber\\
  && \left.+{\cal J}_{13}({\bf k}_1,{\bf k}_2,{\bf k}_3,{\bf k}_4;\tau_1,\tau_2)~a_{{\bf k}_1}a_{{\bf k}_2}a^{\dagger}_{-{\bf k}_3}a^{\dagger}_{-{\bf k}_4}+ {\cal J}_{14}({\bf k}_1,{\bf k}_2,{\bf k}_3,{\bf k}_4;\tau_1,\tau_2)~a^{\dagger}_{-{\bf k}_1}a_{{\bf k}_2}a^{\dagger}_{-{\bf k}_3}a^{\dagger}_{-{\bf k}_4}\right.\nonumber\\
  && \left.+{\cal J}_{15}({\bf k}_1,{\bf k}_2,{\bf k}_3,{\bf k}_4;\tau_1,\tau_2)~a_{{\bf k}_1}a^{\dagger}_{-{\bf k}_2}a^{\dagger}_{-{\bf k}_3}a^{\dagger}_{-{\bf k}_4}+ {\cal J}_{16}({\bf k}_1,{\bf k}_2,{\bf k}_3,{\bf k}_4;\tau_1,\tau_2)~a^{\dagger}_{-{\bf k}_1}a^{\dagger}_{-{\bf k}_2}a^{\dagger}_{-{\bf k}_3}a^{\dagger}_{-{\bf k}_4}\right],~~~~~~~~~~~
  \eea
  where we define new sets of functions, ${\cal J}_{j}({\bf k}_1,{\bf k}_2,{\bf k}_3,{\bf k}_4;\tau_1,\tau_2)~\forall~j=1,\cdots,16$, as:
  \bea &&{\cal J}_{1}({\bf k}_1,{\bf k}_2,{\bf k}_3,{\bf k}_4;\tau_1,\tau_2)=\Pi_{{\bf k}_1}(\tau_2)f_{{\bf k}_2}(\tau_1)f_{{\bf k}_3}(\tau_1)\Pi_{{\bf k}_4}(\tau_2),\\
 && {\cal J}_2({\bf k}_1,{\bf k}_2,{\bf k}_3,{\bf k}_4;\tau_1,\tau_2)=\Pi^{*}_{-{\bf k}_1}(\tau_2)f_{{\bf k}_2}(\tau_1)f_{{\bf k}_3}(\tau_1)\Pi_{{\bf k}_4}(\tau_2),\\
 && {\cal J}_{3}({\bf k}_1,{\bf k}_2,{\bf k}_3,{\bf k}_4;\tau_1,\tau_2)=\Pi_{{\bf k}_1}(\tau_2)f^{*}_{-{\bf k}_2}(\tau_1)f_{{\bf k}_3}(\tau_1)\Pi_{{\bf k}_4}(\tau_2),\\
  && {\cal J}_{4}({\bf k}_1,{\bf k}_2,{\bf k}_3,{\bf k}_4;\tau_1,\tau_2)=\Pi^{*}_{-{\bf k}_1}(\tau_2)f^{*}_{-{\bf k}_2}(\tau_1)f_{{\bf k}_3}(\tau_1)\Pi_{{\bf k}_4}(\tau_2),\\
 && {\cal J}_{5}({\bf k}_1,{\bf k}_2,{\bf k}_3,{\bf k}_4;\tau_1,\tau_2)=\Pi_{{\bf k}_1}(\tau_2)f_{{\bf k}_2}(\tau_1)f^{*}_{-{\bf k}_3}(\tau_1)\Pi_{{\bf k}_4}(\tau_2),\\
  && {\cal J}_{6}({\bf k}_1,{\bf k}_2,{\bf k}_3,{\bf k}_4;\tau_1,\tau_2)=\Pi^{*}_{-{\bf k}_1}(\tau_2)f_{{\bf k}_2}(\tau_1)f^{*}_{-{\bf k}_3}(\tau_1)\Pi_{{\bf k}_4}(\tau_2)\\
 && {\cal J}_{7}({\bf k}_1,{\bf k}_2,{\bf k}_3,{\bf k}_4;\tau_1,\tau_2)=\Pi_{{\bf k}_1}(\tau_2)f^{*}_{-{\bf k}_2}(\tau_1)f^{*}_{-{\bf k}_3}(\tau_1)\Pi_{{\bf k}_4}(\tau_2),\\
  && {\cal J}_{8}({\bf k}_1,{\bf k}_2,{\bf k}_3,{\bf k}_4;\tau_1,\tau_2)=\Pi^{*}_{-{\bf k}_1}(\tau_2)f^{*}_{-{\bf k}_2}(\tau_1)f^{*}_{-{\bf k}_3}(\tau_1)\Pi_{{\bf k}_4}(\tau_2)\\
 && {\cal J}_{9}({\bf k}_1,{\bf k}_2,{\bf k}_3,{\bf k}_4;\tau_1,\tau_2)=\Pi_{{\bf k}_1}(\tau_2)f_{{\bf k}_2}(\tau_1)f_{{\bf k}_3}(\tau_1)\Pi^{*}_{-{\bf k}_4}(\tau_2),\\
  && {\cal J}_{10}({\bf k}_1,{\bf k}_2,{\bf k}_3,{\bf k}_4;\tau_1,\tau_2)=\Pi^{*}_{-{\bf k}_1}(\tau_2)f_{{\bf k}_2}(\tau_1)f_{{\bf k}_3}(\tau_1)\Pi^{*}_{-{\bf k}_4}(\tau_2)\\
 && {\cal J}_{11}({\bf k}_1,{\bf k}_2,{\bf k}_3,{\bf k}_4;\tau_1,\tau_2)=\Pi_{{\bf k}_1}(\tau_2)f^{*}_{-{\bf k}_2}(\tau_1)f_{{\bf k}_3}(\tau_1)\Pi^{*}_{-{\bf k}_4}(\tau_2),\\
  && {\cal J}_{12}({\bf k}_1,{\bf k}_2,{\bf k}_3,{\bf k}_4;\tau_1,\tau_2)=\Pi^{*}_{-{\bf k}_1}(\tau_2)f^{*}_{-{\bf k}_2}(\tau_1)f_{{\bf k}_3}(\tau_1)\Pi^{*}_{-{\bf k}_4}(\tau_2)\\
 && {\cal J}_{13}({\bf k}_1,{\bf k}_2,{\bf k}_3,{\bf k}_4;\tau_1,\tau_2)=\Pi_{{\bf k}_1}(\tau_2)f_{{\bf k}_2}(\tau_1)f^{*}_{-{\bf k}_3}(\tau_1)\Pi^{*}_{-{\bf k}_4}(\tau_2),\\
  && {\cal J}_{14}({\bf k}_1,{\bf k}_2,{\bf k}_3,{\bf k}_4;\tau_1,\tau_2)=\Pi^{*}_{-{\bf k}_1}(\tau_2)f_{{\bf k}_2}(\tau_1)f^{*}_{-{\bf k}_3}(\tau_1)\Pi^{*}_{-{\bf k}_4}(\tau_2)\\
 && {\cal J}_{15}({\bf k}_1,{\bf k}_2,{\bf k}_3,{\bf k}_4;\tau_1,\tau_2)=\Pi_{{\bf k}_1}(\tau_2)f^{*}_{-{\bf k}_2}(\tau_1)f^{*}_{-{\bf k}_3}(\tau_1)\Pi^{*}_{-{\bf k}_4}(\tau_2),\\
  && {\cal J}_{16}({\bf k}_1,{\bf k}_2,{\bf k}_3,{\bf k}_4;\tau_1,\tau_2)=\Pi^{*}_{-{\bf k}_1}(\tau_2)f^{*}_{-{\bf k}_2}(\tau_1)f^{*}_{-{\bf k}_3}(\tau_1)\Pi^{*}_{-{\bf k}_4}(\tau_2)\eea
  
\subsection{Definition of micro-canonical OTO amplitude $\widehat{\cal T}_3({\bf k}_1,{\bf k}_2,{\bf k}_3,{\bf k}_4;\tau_1,\tau_2)$}
  The function $\widehat{\cal T}_3({\bf k}_1,{\bf k}_2,{\bf k}_3,{\bf k}_4;\tau_1,\tau_2)$ is defined as:
 \bea &&\widehat{\cal T}_3({\bf k}_1,{\bf k}_2,{\bf k}_3,{\bf k}_4;\tau_1,\tau_2)\nonumber\\
 &&=\left[{\cal N}_1({\bf k}_1,{\bf k}_2,{\bf k}_3,{\bf k}_4;\tau_1,\tau_2)~a_{{\bf k}_1}a_{{\bf k}_2}a_{{\bf k}_3}a_{{\bf k}_4}+ {\cal N}_2({\bf k}_1,{\bf k}_2,{\bf k}_3,{\bf k}_4;\tau_1,\tau_2)~a^{\dagger}_{-{\bf k}_1}a_{{\bf k}_2}a_{{\bf k}_3}a_{{\bf k}_4}\right.\nonumber\\
  && \left.+{\cal N}_3({\bf k}_1,{\bf k}_2,{\bf k}_3,{\bf k}_4;\tau_1,\tau_2)~a_{{\bf k}_1}a^{\dagger}_{-{\bf k}_2}a_{{\bf k}_3}a_{{\bf k}_4}+ {\cal N}_4({\bf k}_1,{\bf k}_2,{\bf k}_3,{\bf k}_4;\tau_1,\tau_2)~a^{\dagger}_{-{\bf k}_1}a^{\dagger}_{-{\bf k}_2}a_{{\bf k}_3}a_{{\bf k}_4}\right.\nonumber\\
  && \left.+{\cal N}_5({\bf k}_1,{\bf k}_2,{\bf k}_3,{\bf k}_4;\tau_1,\tau_2)~a_{{\bf k}_1}a_{{\bf k}_2}a^{\dagger}_{-{\bf k}_3}a_{{\bf k}_4}+ {\cal N}_6({\bf k}_1,{\bf k}_2,{\bf k}_3,{\bf k}_4;\tau_1,\tau_2)~a^{\dagger}_{-{\bf k}_1}a_{{\bf k}_2}a^{\dagger}_{-{\bf k}_3}a_{{\bf k}_4}\right.\nonumber\\
  && \left.+{\cal N}_7({\bf k}_1,{\bf k}_2,{\bf k}_3,{\bf k}_4;\tau_1,\tau_2)~a_{{\bf k}_1}a^{\dagger}_{-{\bf k}_2}a^{\dagger}_{-{\bf k}_3}a_{{\bf k}_4}+ {\cal N}_8({\bf k}_1,{\bf k}_2,{\bf k}_3,{\bf k}_4;\tau_1,\tau_2)~a^{\dagger}_{-{\bf k}_1}a^{\dagger}_{-{\bf k}_2}a^{\dagger}_{-{\bf k}_3}a_{{\bf k}_4}\right.\nonumber\\
  && \left.+{\cal N}_9({\bf k}_1,{\bf k}_2,{\bf k}_3,{\bf k}_4;\tau_1,\tau_2)~a_{{\bf k}_1}a_{{\bf k}_2}a_{{\bf k}_3}a^{\dagger}_{-{\bf k}_4}+ {\cal N}_{10}({\bf k}_1,{\bf k}_2,{\bf k}_3,{\bf k}_4;\tau_1,\tau_2)~a^{\dagger}_{-{\bf k}_1}a_{{\bf k}_2}a_{{\bf k}_3}a^{\dagger}_{-{\bf k}_4}\right.\nonumber\\
  && \left.+{\cal N}_{11}({\bf k}_1,{\bf k}_2,{\bf k}_3,{\bf k}_4;\tau_1,\tau_2)~a_{{\bf k}_1}a^{\dagger}_{-{\bf k}_2}a_{{\bf k}_3}a^{\dagger}_{-{\bf k}_4}+ {\cal N}_{12}({\bf k}_1,{\bf k}_2,{\bf k}_3,{\bf k}_4;\tau_1,\tau_2)~a^{\dagger}_{-{\bf k}_1}a^{\dagger}_{-{\bf k}_2}a_{{\bf k}_3}a^{\dagger}_{-{\bf k}_4}\right.\nonumber\\
  && \left.+{\cal N}_{13}({\bf k}_1,{\bf k}_2,{\bf k}_3,{\bf k}_4;\tau_1,\tau_2)~a_{{\bf k}_1}a_{{\bf k}_2}a^{\dagger}_{-{\bf k}_3}a^{\dagger}_{-{\bf k}_4}+ {\cal N}_{14}({\bf k}_1,{\bf k}_2,{\bf k}_3,{\bf k}_4;\tau_1,\tau_2)~a^{\dagger}_{-{\bf k}_1}a_{{\bf k}_2}a^{\dagger}_{-{\bf k}_3}a^{\dagger}_{-{\bf k}_4}\right.\nonumber\\
  && \left.+{\cal N}_{15}({\bf k}_1,{\bf k}_2,{\bf k}_3,{\bf k}_4;\tau_1,\tau_2)~a_{{\bf k}_1}a^{\dagger}_{-{\bf k}_2}a^{\dagger}_{-{\bf k}_3}a^{\dagger}_{-{\bf k}_4}+ {\cal N}_{16}({\bf k}_1,{\bf k}_2,{\bf k}_3,{\bf k}_4;\tau_1,\tau_2)~a^{\dagger}_{-{\bf k}_1}a^{\dagger}_{-{\bf k}_2}a^{\dagger}_{-{\bf k}_3}a^{\dagger}_{-{\bf k}_4}\right],~~~~~~~~~~~
  \eea
  where we define new sets of functions, ${\cal N}_{j}({\bf k}_1,{\bf k}_2,{\bf k}_3,{\bf k}_4;\tau_1,\tau_2)~\forall~j=1,\cdots,16$, as:
  \bea &&{\cal N}_{1}({\bf k}_1,{\bf k}_2,{\bf k}_3,{\bf k}_4;\tau_1,\tau_2)=f_{{\bf k}_1}(\tau_1)\Pi_{{\bf k}_2}(\tau_2)\Pi_{{\bf k}_3}(\tau_2)f_{{\bf k}_4}(\tau_1),\\
 && {\cal N}_2({\bf k}_1,{\bf k}_2,{\bf k}_3,{\bf k}_4;\tau_1,\tau_2)=\Pi_{{\bf k}_1}(\tau_2)f^{*}_{-{\bf k}_2}(\tau_1)\Pi_{{\bf k}_3}(\tau_2)f_{{\bf k}_4}(\tau_1),\\
 && {\cal N}_{3}({\bf k}_1,{\bf k}_2,{\bf k}_3,{\bf k}_4;\tau_1,\tau_2)=f_{{\bf k}_1}(\tau_1)\Pi^{*}_{-{\bf k}_2}(\tau_2)\Pi_{{\bf k}_1}(\tau_2)f_{{\bf k}_4}(\tau_1),\\
  && {\cal N}_{4}({\bf k}_1,{\bf k}_2,{\bf k}_3,{\bf k}_4;\tau_1,\tau_2)=f^{*}_{-{\bf k}_1}(\tau_1)\Pi^{*}_{-{\bf k}_2}(\tau_2)\Pi_{{\bf k}_3}(\tau_2)f_{{\bf k}_4}(\tau_1),\\
 && {\cal N}_{5}({\bf k}_1,{\bf k}_2,{\bf k}_3,{\bf k}_4;\tau_1,\tau_2)=f_{{\bf k}_1}(\tau_1)\Pi_{{\bf k}_2}(\tau_2)\Pi^{*}_{-{\bf k}_3}(\tau_2)f_{{\bf k}_4}(\tau_1),\\
  && {\cal N}_{6}({\bf k}_1,{\bf k}_2,{\bf k}_3,{\bf k}_4;\tau_1,\tau_2)=f^{*}_{-{\bf k}_1}(\tau_1)\Pi_{{\bf k}_2}(\tau_2)\Pi^{*}_{-{\bf k}_3}(\tau_2)f_{{\bf k}_4}(\tau_1)\\
 && {\cal N}_{7}({\bf k}_1,{\bf k}_2,{\bf k}_3,{\bf k}_4;\tau_1,\tau_2)=f_{{\bf k}_1}(\tau_1)\Pi^{*}_{-{\bf k}_2}(\tau_2)\Pi^{*}_{-{\bf k}_3}(\tau_2)f_{{\bf k}_4}(\tau_1),\\
  && {\cal N}_{8}({\bf k}_1,{\bf k}_2,{\bf k}_3,{\bf k}_4;\tau_1,\tau_2)=f^{*}_{-{\bf k}_1}(\tau_1)\Pi^{*}_{-{\bf k}_2}(\tau_2)\Pi^{*}_{-{\bf k}_3}(\tau_2)f_{{\bf k}_4}(\tau_1)\\
 && {\cal N}_{9}({\bf k}_1,{\bf k}_2,{\bf k}_3,{\bf k}_4;\tau_1,\tau_2)=f_{{\bf k}_1}(\tau_1)\Pi_{{\bf k}_2}(\tau_2)\Pi_{{\bf k}_3}(\tau_2)f^{*}_{-{\bf k}_4}(\tau_1),\\
  && {\cal N}_{10}({\bf k}_1,{\bf k}_2,{\bf k}_3,{\bf k}_4;\tau_1,\tau_2)=f^{*}_{-{\bf k}_1}(\tau_1)\Pi_{{\bf k}_2}(\tau_2)\Pi_{{\bf k}_3}(\tau_2)f^{*}_{-{\bf k}_4}(\tau_1)\\
 && {\cal N}_{11}({\bf k}_1,{\bf k}_2,{\bf k}_3,{\bf k}_4;\tau_1,\tau_2)=f_{{\bf k}_1}(\tau_1)\Pi^{*}_{-{\bf k}_2}(\tau_2)\Pi_{{\bf k}_3}(\tau_2)f^{*}_{-{\bf k}_4}(\tau_1),\\
  && {\cal N}_{12}({\bf k}_1,{\bf k}_2,{\bf k}_3,{\bf k}_4;\tau_1,\tau_2)=f^{*}_{-{\bf k}_1}(\tau_1)\Pi^{*}_{-{\bf k}_2}(\tau_2)\Pi_{{\bf k}_3}(\tau_2)f^{*}_{-{\bf k}_4}(\tau_1)\\
 && {\cal N}_{13}({\bf k}_1,{\bf k}_2,{\bf k}_3,{\bf k}_4;\tau_1,\tau_2)=f_{{\bf k}_1}(\tau_1)\Pi_{{\bf k}_2}(\tau_2)\Pi^{*}_{-{\bf k}_3}(\tau_2)f^{*}_{-{\bf k}_4}(\tau_1),\\
  && {\cal N}_{14}({\bf k}_1,{\bf k}_2,{\bf k}_3,{\bf k}_4;\tau_1,\tau_2)=f^{*}_{-{\bf k}_1}(\tau_1)\Pi_{{\bf k}_2}(\tau_2)\Pi^{*}_{-{\bf k}_3}(\tau_2)f^{*}_{-{\bf k}_4}(\tau_1)\\
 && {\cal N}_{15}({\bf k}_1,{\bf k}_2,{\bf k}_3,{\bf k}_4;\tau_1,\tau_2)=f_{{\bf k}_1}(\tau_1)\Pi^{*}_{-{\bf k}_2}(\tau_2)\Pi^{*}_{-{\bf k}_3}(\tau_2)f^{*}_{-{\bf k}_4}(\tau_1),\\
  && {\cal N}_{16}({\bf k}_1,{\bf k}_2,{\bf k}_3,{\bf k}_4;\tau_1,\tau_2)=f^{*}_{-{\bf k}_1}(\tau_1)\Pi^{*}_{-{\bf k}_2}(\tau_2)\Pi^{*}_{-{\bf k}_3}(\tau_2)f^{*}_{-{\bf k}_4}(\tau_1)\eea  
   
\subsection{Definition of micro-canonical OTO amplitude $\widehat{\cal T}_4({\bf k}_1,{\bf k}_2,{\bf k}_3,{\bf k}_4;\tau_1,\tau_2)$}
  The function $\widehat{\cal T}_4({\bf k}_1,{\bf k}_2,{\bf k}_3,{\bf k}_4;\tau_1,\tau_2)$ is defined as:
 \bea &&\widehat{\cal T}_4({\bf k}_1,{\bf k}_2,{\bf k}_3,{\bf k}_4;\tau_1,\tau_2)\nonumber\\
 &&=\left[{\cal Q}_1({\bf k}_1,{\bf k}_2,{\bf k}_3,{\bf k}_4;\tau_1,\tau_2)~a_{{\bf k}_1}a_{{\bf k}_2}a_{{\bf k}_3}a_{{\bf k}_4}+ {\cal Q}_2({\bf k}_1,{\bf k}_2,{\bf k}_3,{\bf k}_4;\tau_1,\tau_2)~a^{\dagger}_{-{\bf k}_1}a_{{\bf k}_2}a_{{\bf k}_3}a_{{\bf k}_4}\right.\nonumber\\
  && \left.+{\cal Q}_3({\bf k}_1,{\bf k}_2,{\bf k}_3,{\bf k}_4;\tau_1,\tau_2)~a_{{\bf k}_1}a^{\dagger}_{-{\bf k}_2}a_{{\bf k}_3}a_{{\bf k}_4}+ {\cal Q}_4({\bf k}_1,{\bf k}_2,{\bf k}_3,{\bf k}_4;\tau_1,\tau_2)~a^{\dagger}_{-{\bf k}_1}a^{\dagger}_{-{\bf k}_2}a_{{\bf k}_3}a_{{\bf k}_4}\right.\nonumber\\
  && \left.+{\cal Q}_5({\bf k}_1,{\bf k}_2,{\bf k}_3,{\bf k}_4;\tau_1,\tau_2)~a_{{\bf k}_1}a_{{\bf k}_2}a^{\dagger}_{-{\bf k}_3}a_{{\bf k}_4}+ {\cal Q}_6({\bf k}_1,{\bf k}_2,{\bf k}_3,{\bf k}_4;\tau_1,\tau_2)~a^{\dagger}_{-{\bf k}_1}a_{{\bf k}_2}a^{\dagger}_{-{\bf k}_3}a_{{\bf k}_4}\right.\nonumber\\
  && \left.+{\cal Q}_7({\bf k}_1,{\bf k}_2,{\bf k}_3,{\bf k}_4;\tau_1,\tau_2)~a_{{\bf k}_1}a^{\dagger}_{-{\bf k}_2}a^{\dagger}_{-{\bf k}_3}a_{{\bf k}_4}+ {\cal Q}_8({\bf k}_1,{\bf k}_2,{\bf k}_3,{\bf k}_4;\tau_1,\tau_2)~a^{\dagger}_{-{\bf k}_1}a^{\dagger}_{-{\bf k}_2}a^{\dagger}_{-{\bf k}_3}a_{{\bf k}_4}\right.\nonumber\\
  && \left.+{\cal Q}_9({\bf k}_1,{\bf k}_2,{\bf k}_3,{\bf k}_4;\tau_1,\tau_2)~a_{{\bf k}_1}a_{{\bf k}_2}a_{{\bf k}_3}a^{\dagger}_{-{\bf k}_4}+ {\cal Q}_{10}({\bf k}_1,{\bf k}_2,{\bf k}_3,{\bf k}_4;\tau_1,\tau_2)~a^{\dagger}_{-{\bf k}_1}a_{{\bf k}_2}a_{{\bf k}_3}a^{\dagger}_{-{\bf k}_4}\right.\nonumber\\
  && \left.+{\cal Q}_{11}({\bf k}_1,{\bf k}_2,{\bf k}_3,{\bf k}_4;\tau_1,\tau_2)~a_{{\bf k}_1}a^{\dagger}_{-{\bf k}_2}a_{{\bf k}_3}a^{\dagger}_{-{\bf k}_4}+ {\cal Q}_{12}({\bf k}_1,{\bf k}_2,{\bf k}_3,{\bf k}_4;\tau_1,\tau_2)~a^{\dagger}_{-{\bf k}_1}a^{\dagger}_{-{\bf k}_2}a_{{\bf k}_3}a^{\dagger}_{-{\bf k}_4}\right.\nonumber\\
  && \left.+{\cal Q}_{13}({\bf k}_1,{\bf k}_2,{\bf k}_3,{\bf k}_4;\tau_1,\tau_2)~a_{{\bf k}_1}a_{{\bf k}_2}a^{\dagger}_{-{\bf k}_3}a^{\dagger}_{-{\bf k}_4}+ {\cal Q}_{14}({\bf k}_1,{\bf k}_2,{\bf k}_3,{\bf k}_4;\tau_1,\tau_2)~a^{\dagger}_{-{\bf k}_1}a_{{\bf k}_2}a^{\dagger}_{-{\bf k}_3}a^{\dagger}_{-{\bf k}_4}\right.\nonumber\\
  && \left.+{\cal Q}_{15}({\bf k}_1,{\bf k}_2,{\bf k}_3,{\bf k}_4;\tau_1,\tau_2)~a_{{\bf k}_1}a^{\dagger}_{-{\bf k}_2}a^{\dagger}_{-{\bf k}_3}a^{\dagger}_{-{\bf k}_4}+ {\cal Q}_{16}({\bf k}_1,{\bf k}_2,{\bf k}_3,{\bf k}_4;\tau_1,\tau_2)~a^{\dagger}_{-{\bf k}_1}a^{\dagger}_{-{\bf k}_2}a^{\dagger}_{-{\bf k}_3}a^{\dagger}_{-{\bf k}_4}\right],~~~~~~~~~~~
  \eea
  where we define new sets of functions, ${\cal Q}_{j}({\bf k}_1,{\bf k}_2,{\bf k}_3,{\bf k}_4;\tau_1,\tau_2)~\forall~j=1,\cdots,16$, as:
  \bea &&{\cal Q}_{1}({\bf k}_1,{\bf k}_2,{\bf k}_3,{\bf k}_4;\tau_1,\tau_2)=\Pi_{{\bf k}_1}(\tau_2)f_{{\bf k}_2}(\tau_1)\Pi_{{\bf k}_3}(\tau_2)f_{{\bf k}_4}(\tau_1),\\
 && {\cal Q}_2({\bf k}_1,{\bf k}_2,{\bf k}_3,{\bf k}_4;\tau_1,\tau_2)=\Pi^{*}_{-{\bf k}_1}(\tau_2)f_{-{\bf k}_2}(\tau_1)\Pi_{{\bf k}_3}(\tau_2)f_{{\bf k}_4}(\tau_1),\\
 && {\cal Q}_{3}({\bf k}_1,{\bf k}_2,{\bf k}_3,{\bf k}_4;\tau_1,\tau_2)=\Pi_{{\bf k}_1}(\tau_2)f^{*}_{-{\bf k}_2}(\tau_1)\Pi_{{\bf k}_1}(\tau_2)f_{{\bf k}_4}(\tau_1),\\
  && {\cal Q}_{4}({\bf k}_1,{\bf k}_2,{\bf k}_3,{\bf k}_4;\tau_1,\tau_2)=\Pi^{*}_{-{\bf k}_1}(\tau_2)f^{*}_{-{\bf k}_2}(\tau_1)\Pi_{{\bf k}_3}(\tau_2)f_{{\bf k}_4}(\tau_1),\\
 && {\cal Q}_{5}({\bf k}_1,{\bf k}_2,{\bf k}_3,{\bf k}_4;\tau_1,\tau_2)=\Pi_{{\bf k}_1}(\tau_2)f_{{\bf k}_2}(\tau_1)\Pi^{*}_{-{\bf k}_3}(\tau_2)f_{{\bf k}_4}(\tau_1),\\
  && {\cal Q}_{6}({\bf k}_1,{\bf k}_2,{\bf k}_3,{\bf k}_4;\tau_1,\tau_2)=\Pi^{*}_{-{\bf k}_1}(\tau_2)f_{{\bf k}_2}(\tau_1)\Pi^{*}_{-{\bf k}_3}(\tau_2)f_{{\bf k}_4}(\tau_1)\\
 && {\cal Q}_{7}({\bf k}_1,{\bf k}_2,{\bf k}_3,{\bf k}_4;\tau_1,\tau_2)=\Pi_{{\bf k}_1}(\tau_2)f^{*}_{-{\bf k}_2}(\tau_1)f\Pi^{*}_{-{\bf k}_3}(\tau_2)f_{{\bf k}_4}(\tau_1),\\
  && {\cal Q}_{8}({\bf k}_1,{\bf k}_2,{\bf k}_3,{\bf k}_4;\tau_1,\tau_2)=\Pi^{*}_{-{\bf k}_1}(\tau_2)f^{*}_{-{\bf k}_2}(\tau_1)\Pi^{*}_{-{\bf k}_3}(\tau_2)f_{{\bf k}_4}(\tau_1)\\
 && {\cal Q}_{9}({\bf k}_1,{\bf k}_2,{\bf k}_3,{\bf k}_4;\tau_1,\tau_2)=\Pi_{{\bf k}_1}(\tau_2)f_{{\bf k}_2}(\tau_1)\Pi_{{\bf k}_3}(\tau_2)f^{*}_{-{\bf k}_4}(\tau_1),\\
  && {\cal Q}_{10}({\bf k}_1,{\bf k}_2,{\bf k}_3,{\bf k}_4;\tau_1,\tau_2)=\Pi^{*}_{-{\bf k}_1}(\tau_2)f_{{\bf k}_2}(\tau_1)\Pi_{{\bf k}_3}(\tau_2)f^{*}_{-{\bf k}_4}(\tau_1)\\
 && {\cal Q}_{11}({\bf k}_1,{\bf k}_2,{\bf k}_3,{\bf k}_4;\tau_1,\tau_2)=\Pi_{{\bf k}_1}(\tau_2)f^{*}_{-{\bf k}_2}(\tau_1)\Pi_{{\bf k}_3}(\tau_2)f^{*}_{-{\bf k}_4}(\tau_1),\\
  && {\cal Q}_{12}({\bf k}_1,{\bf k}_2,{\bf k}_3,{\bf k}_4;\tau_1,\tau_2)=\Pi^{*}_{-{\bf k}_1}(\tau_2)f^{*}_{-{\bf k}_2}(\tau_1)\Pi_{{\bf k}_3}(\tau_2)f^{*}_{-{\bf k}_4}(\tau_1)\\
 && {\cal Q}_{13}({\bf k}_1,{\bf k}_2,{\bf k}_3,{\bf k}_4;\tau_1,\tau_2)=\Pi_{{\bf k}_1}(\tau_2)f_{{\bf k}_2}(\tau_1)\Pi^{*}_{-{\bf k}_3}(\tau_2)f^{*}_{-{\bf k}_4}(\tau_1),\\
  && {\cal Q}_{14}({\bf k}_1,{\bf k}_2,{\bf k}_3,{\bf k}_4;\tau_1,\tau_2)=\Pi^{*}_{-{\bf k}_1}(\tau_2)f_{{\bf k}_2}(\tau_1)\Pi^{*}_{-{\bf k}_3}(\tau_2)f^{*}_{-{\bf k}_4}(\tau_1)\\
 && {\cal Q}_{15}({\bf k}_1,{\bf k}_2,{\bf k}_3,{\bf k}_4;\tau_1,\tau_2)=\Pi_{{\bf k}_1}(\tau_2)f^{*}_{-{\bf k}_2}(\tau_1)\Pi^{*}_{-{\bf k}_3}(\tau_2)f^{*}_{-{\bf k}_4}(\tau_1),\\
  && {\cal Q}_{16}({\bf k}_1,{\bf k}_2,{\bf k}_3,{\bf k}_4;\tau_1,\tau_2)=\Pi^{*}_{-{\bf k}_1}(\tau_2)f^{*}_{-{\bf k}_2}(\tau_1)\Pi^{*}_{-{\bf k}_3}(\tau_2)f^{*}_{-{\bf k}_4}(\tau_1)\eea 
  
	\newpage
	\section{Computation of classical limit of four-point ``in-in" OTO micro-canonical amplitudes for Cosmology} 
	In this section, our prime objective is to explicitly compute the classical limiting version of the four-point "in-in" OTO micro-canonical amplitudes appearing in the expression or OTOC. To serve this purpose in the classical limit we explicitly compute the following square of the Poisson bracket, given by: 
  \bea  \left\{{f}({\bf x},\tau_1),{\Pi}({\bf x},\tau_2)\right\}^2_{\bf PB}&=&\left\{{f}({\bf x},\tau_1),{\Pi}({\bf x},\tau_2)\right\}_{\bf PB}\left\{{f}({\bf x},\tau_1),{\Pi}({\bf x},\tau_2)\right\}_{\bf PB}.\eea
  
  Now we use the following convention for the Fourier transformation, which is given by:
 \bea &&\hat{f}({\bf x},\tau_1)=\int \frac{d^3{\bf k}}{(2\pi)^3}~\exp(i{\bf k}.{\bf x})~\hat{f}_{{\bf k}}(\tau_1),\\
 &&\hat{\Pi}({\bf x},\tau_1)=\partial_{\tau_1}\hat{f}({\bf x},\tau_1)=\int \frac{d^3{\bf k}}{(2\pi)^3}~\exp(i{\bf k}.{\bf x})~\partial_{\tau_1}\hat{f}_{{\bf k}}(\tau_1)=\int \frac{d^3{\bf k}}{(2\pi)^3}~\exp(i{\bf k}.{\bf x})~\hat{\Pi}_{{\bf k}}(\tau_1),~~~~~~~~~~~~~\eea
 which will be very useful for the computation of the classical limiting result of four-point OTOC in terms of the square of the Poisson bracket.
 
 Consequently, we get the following simplified result:
 \bea && \left\{{f}({\bf x},\tau_1),{\Pi}({\bf x},\tau_2)\right\}^2_{\bf PB}\nonumber\\
 &&=\int \frac{d^3{\bf k}_1}{(2\pi)^3}\int \frac{d^3{\bf k}_2}{(2\pi)^3}\int \frac{d^3{\bf k}_3}{(2\pi)^3}\int \frac{d^3{\bf k}_4}{(2\pi)^3}~\exp\left(i({\bf k}_1+{\bf k}_2+{\bf k}_3+{\bf k}_4).{\bf x}\right)\nonumber\\
 &&~~~~~~\left[\left\{f_{{\bf k}_{1}} (\tau_1),\Pi_{{\bf k}_{2}} (\tau_2)\right\}_{\bf PB}\left\{f_{{\bf k}_{3}}(\tau_1),\Pi_{{\bf k}_4} (\tau_2)\right\}_{\bf PB}+\left\{f_{{\bf k}_{1}} (\tau_1),\Pi_{{\bf k}_{3}} (\tau_2)\right\}_{\bf PB}\left\{f_{{\bf k}_{2}}(\tau_1),\Pi_{{\bf k}_4} (\tau_2)\right\}_{\bf PB}\right.\nonumber\\&& \left.~~~~+\left\{f_{{\bf k}_{1}} (\tau_1),\Pi_{{\bf k}_{4}} (\tau_2)\right\}_{\bf PB}\left\{f_{{\bf k}_{3}}(\tau_1),\Pi_{{\bf k}_2} (\tau_2)\right\}_{\bf PB}+\left\{f_{{\bf k}_{2}} (\tau_1),\Pi_{{\bf k}_{3}} (\tau_2)\right\}_{\bf PB}\left\{f_{{\bf k}_{4}}(\tau_1),\Pi_{{\bf k}_1} (\tau_2)\right\}_{\bf PB}\right.\nonumber\\&& \left.~~~~+\left\{f_{{\bf k}_{2}} (\tau_1),\Pi_{{\bf k}_{1}} (\tau_2)\right\}_{\bf PB}\left\{f_{{\bf k}_{4}}(\tau_1),\Pi_{{\bf k}_3} (\tau_2)\right\}_{\bf PB}+\left\{f_{{\bf k}_{2}} (\tau_1),\Pi_{{\bf k}_{4}} (\tau_2)\right\}_{\bf PB}\left\{f_{{\bf k}_{1}}(\tau_1),\Pi_{{\bf k}_3} (\tau_2)\right\}_{\bf PB}\right.\nonumber\\&& \left.~~~~+\left\{f_{{\bf k}_{3}} (\tau_1),\Pi_{{\bf k}_{1}} (\tau_2)\right\}_{\bf PB}\left\{f_{{\bf k}_{4}}(\tau_1),\Pi_{{\bf k}_2} (\tau_2)\right\}_{\bf PB}+\left\{f_{{\bf k}_{3}} (\tau_1),\Pi_{{\bf k}_{2}} (\tau_2)\right\}_{\bf PB}\left\{f_{{\bf k}_{1}}(\tau_1),\Pi_{{\bf k}_4} (\tau_2)\right\}_{\bf PB}\right.\nonumber\\&& \left.~~~~+\left\{f_{{\bf k}_{3}} (\tau_1),\Pi_{{\bf k}_{4}} (\tau_2)\right\}_{\bf PB}\left\{f_{{\bf k}_{1}}(\tau_1),\Pi_{{\bf k}_2} (\tau_2)\right\}_{\bf PB}+\left\{f_{{\bf k}_{4}} (\tau_1),\Pi_{{\bf k}_{1}} (\tau_2)\right\}_{\bf PB}\left\{f_{{\bf k}_{2}}(\tau_1),\Pi_{{\bf k}_3} (\tau_2)\right\}_{\bf PB}\right.\nonumber\\&& \left.~~~~+\left\{f_{{\bf k}_{4}} (\tau_1),\Pi_{{\bf k}_{2}} (\tau_2)\right\}_{\bf PB}\left\{f_{{\bf k}_{3}}(\tau_1),\Pi_{{\bf k}_1} (\tau_2)\right\}_{\bf PB}+\left\{f_{{\bf k}_{4}} (\tau_1),\Pi_{{\bf k}_{3}} (\tau_2)\right\}_{\bf PB}\left\{f_{{\bf k}_{2}}(\tau_1),\Pi_{{\bf k}_1} (\tau_2)\right\}_{\bf PB}\right].~~~~~~~~~\eea
 Now, here our job is to explicitly compute each of the {\it Poisson brackets}, which are appearing in the above mentioned twelve terms.
 The explicit computation gives the following result:
 \bea \left\{f_{{\bf k}_{i}} (\tau_1),\Pi_{{\bf k}_{j}} (\tau_2)\right\}_{\bf PB}&=&\left(\frac{\partial f_{{\bf k}_i}(\tau_1)}{\partial f_{{\bf k}_j}(\tau_2)}\underbrace{\frac{\partial \Pi_{{\bf k}_j}(\tau_1)}{\partial \Pi_{{\bf k}_j}(\tau_2)}}_{\textcolor{red}{\bf =1}}-\underbrace{\frac{\partial f_{{\bf k}_i}(\tau_1)}{\partial \Pi_{{\bf k}_j}(\tau_2)}\frac{\partial \Pi_{{\bf k}_j}(\tau_2)}{\partial f_{{\bf k}_j}(\tau_2)}}_{\textcolor{red}{\bf =0}}\right)\nonumber\\
 &=&(2\pi)^3\delta^{3}({\bf k}_i+{\bf k}_j)~{\bf R}(\tau_1,\tau_2).~~~~~\forall i\ne j ~~{\rm with}~~ i,j=1,2,3,4.~~~~~~~~\eea
 Here from these computed {\it Poisson bracket} we can extract the following sets of information, which will further helps us to understand more about the classical limit of the four-point OTO micro-canonical amplitude in the present computation for Cosmology:
 \begin{enumerate}
 \item The time dependent part as such very complicated as it contain the information regarding classical statistical version of random chaotic stochastic quantum fluctuations in the primordial universe. In the present context it is hypothesized by a random function ${\bf R}(\tau_1,\tau_2)$ which incorporates the two conformal time scales.
 
 \item Also it is important to not that, ${\bf R}(\tau_1,\tau_2)$ is a homogeneous and isotropic function, which captures the dynamical effect of the classical spatially flat FLRW geometrical background. For this reason the random function, ${\bf R}(\tau_1,\tau_2)$ is completely $i$ and $j$ momentum index independent.
 
 \item Moreover, the interesting to point here that, one can explicitly separately write down the contribution of the inhomogeneity and time dynamics in Fourier space after computing the classical {\it Poisson bracket} in this context.
 
 \item Finally, the appearance of the three dimensional {\it Dirac Delta} function confirms the momentum conservation in the Fourier space in the classical two point OTO micro-canonical amplitude in Cosmology.
 \end{enumerate}
 Further, we compute the square of the {\it Poisson bracket}, which after performing the Fourier transformation can be expressed as:
  \bea \left\{f_{{\bf k}_{i}} (\tau_1),\Pi_{{\bf k}_{j}} (\tau_2)\right\}_{\bf PB}\left\{f_{{\bf k}_{l}} (\tau_1),\Pi_{{\bf k}_{m}} (\tau_2)\right\}_{\bf PB}&=&\left(\frac{\partial f_{{\bf k}_i}(\tau_1)}{\partial f_{{\bf k}_j}(\tau_2)}\underbrace{\frac{\partial \Pi_{{\bf k}_j}(\tau_1)}{\partial \Pi_{{\bf k}_j}(\tau_2)}}_{\textcolor{red}{\bf =1}}-\underbrace{\frac{\partial f_{{\bf k}_i}(\tau_1)}{\partial \Pi_{{\bf k}_j}(\tau_2)}\frac{\partial \Pi_{{\bf k}_j}(\tau_2)}{\partial f_{{\bf k}_j}(\tau_2)}}_{\textcolor{red}{\bf =0}}\right)\nonumber\\
  &&~~~~~~~~~\left(\frac{\partial f_{{\bf k}_l}(\tau_1)}{\partial f_{{\bf k}_m}(\tau_2)}\underbrace{\frac{\partial \Pi_{{\bf k}_l}(\tau_1)}{\partial \Pi_{{\bf k}_l}(\tau_2)}}_{\textcolor{red}{\bf =1}}-\underbrace{\frac{\partial f_{{\bf k}_l}(\tau_1)}{\partial \Pi_{{\bf k}_m}(\tau_2)}\frac{\partial \Pi_{{\bf k}_l}(\tau_2)}{\partial f_{{\bf k}_m}(\tau_2)}}_{\textcolor{red}{\bf =0}}\right)\nonumber\\
 &=&(2\pi)^6\delta^{3}({\bf k}_i+{\bf k}_j)\delta^{3}({\bf k}_l+{\bf k}_m)~{\bf R}^2(\tau_1,\tau_2).~~~~~\nonumber\\
  &&~~~~~~~~~\forall i\neq j \neq l\neq m~~{\rm with}~~ i,j,k,l=1,2,3,4.~~~~~~~~\eea
 Consequently, we get the following simplified result:
  \bea && \left\{{f}({\bf x},\tau_1),{\Pi}({\bf x},\tau_2)\right\}^2_{\bf PB}=(2\pi)^6 \int \frac{d^3{\bf k}_1}{(2\pi)^3}\int \frac{d^3{\bf k}_2}{(2\pi)^3}\int \frac{d^3{\bf k}_3}{(2\pi)^3}\int \frac{d^3{\bf k}_4}{(2\pi)^3}~\exp\left(i({\bf k}_1+{\bf k}_2+{\bf k}_3+{\bf k}_4).{\bf x}\right)\nonumber\\
 &&~~~~~~~~~~~~~~~~~~~~~~~~~~~~~~~~~\underbrace{\sum^{4}_{i,j,l,m=1,i\neq j\neq l \neq m}\delta^{3}({\bf k}_i+{\bf k}_j)\delta^{3}({\bf k}_l+{\bf k}_m)}_{\textcolor{red}{\bf Contribution~from~12~terms}}~{\bf R}^2(\tau_1,\tau_2).\eea
 Here, the explicit computation gives:
 \bea &&\sum^{4}_{i,j,l,m=1,i\neq j\neq l \neq m}\delta^{3}({\bf k}_i+{\bf k}_j)\delta^{3}({\bf k}_l+{\bf k}_m)=\left[\delta^{3}({\bf k}_1+{\bf k}_2)\delta^3({\bf k}_3+{\bf k}_4)+\delta^{3}({\bf k}_1+{\bf k}_3)\delta^3({\bf k}_2+{\bf k}_4)\right.\nonumber\\&& \left.~~~~~~~~~~~~~~~~~~~~~~~~~~~~~~
 ~~~~~~~~~~~~~~~~~~~~~+\delta^{3}({\bf k}_1+{\bf k}_4)\delta^3({\bf k}_3+{\bf k}_2)+\delta^{3}({\bf k}_2+{\bf k}_3)\delta^3({\bf k}_4+{\bf k}_1)\right.\nonumber\\&& \left.~~~~~~~~~~~~~~~~~~~~~~~~~~~~~~
 ~~~~~~~~~~~~~~~~~~~~~+\delta^{3}({\bf k}_2+{\bf k}_1)\delta^3({\bf k}_4+{\bf k}_3)+\delta^{3}({\bf k}_2+{\bf k}_4)\delta^3({\bf k}_1+{\bf k}_3)\right.\nonumber\\&& \left.~~~~~~~~~~~~~~~~~~~~~~~~~~~~~~
 ~~~~~~~~~~~~~~~~~~~~~+\delta^{3}({\bf k}_3+{\bf k}_1)\delta^3({\bf k}_4+{\bf k}_2)+\delta^{3}({\bf k}_3+{\bf k}_2)\delta^3({\bf k}_1+{\bf k}_4)\right.\nonumber\\&& \left.~~~~~~~~~~~~~~~~~~~~~~~~~~~~~~
 ~~~~~~~~~~~~~~~~~~~~~+\delta^{3}({\bf k}_3+{\bf k}_4)\delta^3({\bf k}_1+{\bf k}_2)+\delta^{3}({\bf k}_4+{\bf k}_1)\delta^3({\bf k}_2+{\bf k}_3)\right.\nonumber\\&& \left.~~~~~~~~~~~~~~~~~~~~~~~~~~~~~~
 ~~~~~~~~~~~~~~~~~~~~~+\delta^{3}({\bf k}_4+{\bf k}_2)\delta^3({\bf k}_3+{\bf k}_1)+\delta^{3}({\bf k}_4+{\bf k}_3)\delta^3({\bf k}_2+{\bf k}_1)\right].~~~~~~~~~\eea
 Now, we give the following proposal to quantify the random function ${\bf R}^2(\tau_1,\tau_2)$, which is given by the following expression:
 \bea {\bf R}^2(\tau_1,\tau_2):=\underbrace{\langle \eta_{\bf Noise}(\tau_1)\eta_{\bf Noise}(\tau_2)\rangle}_{\textcolor{red}{\bf Contribution~from~randomness}}~\underbrace{\exp\left(-\lambda_{f}[|\tau_1|+|\tau_2|]\right)}_{\textcolor{red}{\bf Chaos~ and~ decay~of~correlation}},\eea
 where $\lambda_{f}$ is the decay coefficient of the chaotic terms and will saturate after late time scale of universe and $ \eta_{\bf Noise}(\tau_i)~\forall~i=1,2$ represent the conformal time dependent random noise function. Also, the first term represent two point noise kernel, which is non-Gaussian for coloured noise and Gaussian in nature for the white noise. Also it is important to note that the noise kernel is time translation invariant, for which we have written: 
 \bea \langle \eta_{\bf Noise}(\tau_1)\eta_{\bf Noise}(\tau_2)\rangle={\bf G}_{\bf Kernel}(\tau_1,\tau_2):={\bf G}_{\bf Kernel}(\tau_1-\tau_2).\eea
 Additionally, the conformal time dependent noise satisfy the following constraint conditions:
 \bea &&\langle \eta_{\bf Noise}(\tau_i)\rangle=0,~~~~~~~~~~~~~~~~~~~~~{\bf Noise=}~~\textcolor{red}{\bf Gaussian,~~Non-Gaussian},\\
 &&\langle \eta_{\bf Noise}(\tau_1)\eta_{\bf Noise}(\tau_2)\eta_{\bf Noise}(\tau_3)\rangle=0,~~~~~~~~~~~~~~~~~~~{\bf Noise=}~~\textcolor{red}{\bf Gaussian},\\
 &&\langle \eta_{\bf Noise}(\tau_1)\eta_{\bf Noise}(\tau_2).......\eta_{\bf Noise}(\tau_N)\rangle={\bf f}_{\bf Noise}(\tau_1,\tau_2,....,\tau_N)\neq 0~~\forall~N\geq 2,\nonumber\\
 &&~~~~~~~~~~~~~~~~~~~~~~~~~~~~~~~~~~~~~~~~~~~~~~~~~~~~~~~~~{\bf Noise=}~~\textcolor{red}{\bf Non-Gaussian}.~~\eea
 After substituting this result in the previously computed expression for the amplitude we get the following simplified expression:
  \bea && \left\{{f}({\bf x},\tau_1),{\Pi}({\bf x},\tau_2)\right\}^2_{\bf PB}\nonumber\\
 &&=(2\pi)^6\prod^{4}_{p=1} \int \frac{d^3{\bf k}_p}{(2\pi)^3}~\exp\left(i{\bf k}_p.{\bf x}\right)\sum^{4}_{i,j,l,m=1,i\neq j\neq l \neq m}\delta^{3}({\bf k}_i+{\bf k}_j)\delta^{3}({\bf k}_l+{\bf k}_m)\nonumber\\&&~~~~~~~~~~~~~~~~~~~~~~~~~~~~~~~~~~~~~~~~~~~~~~~~~~~~~~~~~{\bf G}_{\bf Kernel}(\tau_1-\tau_2)\exp\left(-\lambda_{f}[|\tau_1|+|\tau_2|]\right).~~~~~~~~~~~~~~\eea  

\newpage 
	
\section{Computation of quantum micro-canonical partition function in Cosmology}  

\subsection{Quantum micro-canonical partition function in terms of rescaled field variable}
In the context of quantum field theory, one can define the class of all excited $\alpha$ vacua states in terms of the well known adiabatic Bunch Davies vacuum state as:
\bea \hll{{|\Psi_{\alpha}\rangle=\frac{1}{\sqrt{|\cosh \alpha|}}~\exp\left(-\frac{i}{2}\tanh \alpha~\int \frac{d^3{\bf k}}{(2\pi)^3}~a^{\dagger}_{\bf k}a_{\bf k}\right)|\Psi_{\bf BD}\rangle}}~,\eea
which satisfy the following constraint condition:
\bea a_{\bf k}|\Psi_{\alpha}\rangle=0~~~~\forall~~{\bf k},~~\alpha~~~\eea
Here one can easily observed that, if we fix $\alpha=0$ then one can easily get back the usual quantum adiabatic Bunch Davies vacuum state. 
In presence of these excited $\alpha$ vacua states the quantum partition function can be expressed as:
\bea &&Z_{\alpha}(\beta;\tau_1)
=\int d\Psi_{\alpha}~\langle \Psi_{\alpha}|e^{-\beta \hat{H}(\tau_1)}|\Psi_{\alpha} \rangle\nonumber\\
&&=\frac{1}{|\cosh\alpha|}\int d\Psi_{\bf BD}~\langle \Psi_{\bf BD}|\left\{\exp\left(\frac{i}{2}\tanh \alpha~\int \frac{d^3{\bf k}_1}{(2\pi)^3}~a_{{\bf k}_1}a_{{\bf k}_1}\right)\right.\nonumber\\
&&\left.~~\exp\left(-\beta\int d^3{\bf k}~\left(a^{\dagger}_{\bf k}a_{\bf k}+\frac{1}{2}\delta^{3}(0)\right)E_{\bf k}(\tau_1)\right)\exp\left(-\frac{i}{2}\tanh \alpha~\int \frac{d^3{\bf k}_2}{(2\pi)^3}~a^{\dagger}_{{\bf k}_2}a^{\dagger}_{{\bf k}_2}\right)\right\}|\Psi_{\bf BD}\rangle.~~~~~~~~~~~~\eea
Now we will explicitly compute the individual contributions, which are  given by:
\bea \exp\left(-\frac{i}{2}\tanh \alpha~\int \frac{d^3{\bf k}_2}{(2\pi)^3}~a^{\dagger}_{{\bf k}_2}a^{\dagger}_{{\bf k}_2}\right)|\Psi_{\bf BD}\rangle &=&\sum^{\infty}_{n=0}\frac{(-1)^n}{n!}\left(\frac{i}{2}\tanh \alpha~\int \frac{d^3{\bf k}_2}{(2\pi)^3}~a^{\dagger}_{{\bf k}_2}a^{\dagger}_{{\bf k}_2}\right)^n|\Psi_{\bf BD}\rangle\nonumber\\
&=&\sum^{\infty}_{n=0}\frac{(-1)^n}{n!}\left(\frac{i}{2}\tanh \alpha~\int \frac{d^3{\bf k}_2}{(2\pi)^3}~\right)^n|\Psi_{\bf BD}\rangle\nonumber\\
&=&\exp\left(-\frac{i}{2}\tanh \alpha~\int \frac{d^3{\bf k}_2}{(2\pi)^3}~\right)|\Psi_{\bf BD}\rangle.,\\
 \langle \Psi_{\bf BD}|\exp\left(\frac{i}{2}\tanh \alpha~\int \frac{d^3{\bf k}_2}{(2\pi)^3}~a_{{\bf k}_2}a_{{\bf k}_2}\right)&=&\left[\exp\left(-\frac{i}{2}\tanh \alpha~\int \frac{d^3{\bf k}_2}{(2\pi)^3}~a^{\dagger}_{{\bf k}_2}a^{\dagger}_{{\bf k}_2}\right)|\Psi_{\bf BD}\rangle\right]^{\dagger}\nonumber\\
&=&\left[\exp\left(-\frac{i}{2}\tanh \alpha~\int \frac{d^3{\bf k}_2}{(2\pi)^3}~\right)|\Psi_{\bf BD}\rangle\right]^{\dagger}\nonumber\\
&=& \langle \Psi_{\bf BD}|\exp\left(\frac{i}{2}\tanh \alpha~\int \frac{d^3{\bf k}_2}{(2\pi)^3}~\right).\eea
Then the quantum partition function for $\alpha$ vacua can be expressed as:
\bea Z_{\alpha}(\beta;\tau_1)
&=&\frac{1}{|\cosh\alpha|}Z_{\bf BD}(\beta;\tau_1),\eea
where $Z_{\bf BD}$ is the quantum partition function computed from adiabatic Bunch Davies vacuum as:
\bea Z_{\bf BD}(\beta;\tau_1)
&=&1+\int d\Psi_{\bf BD}~\sum^{\infty}_{n=1}\sum^{n}_{p=0}{}^{n}C_{p}\left(\frac{1}{2}\delta^{3}(0)\right)^{n-p}\frac{(-1)^n \beta^n}{n!}\left(\int d^3{\bf k}~E_{\bf k}(\tau_1)\right)^n\langle \Psi_{\bf BD}|\left(a^{\dagger}_{\bf k}a^{\dagger}_{\bf k}\right)^{n}|\Psi_{\bf BD}\rangle\nonumber\\
&=&\exp\left(-\left(1+\frac{1}{2}\delta^{3}(0)\right)\int d^3{\bf k}~\ln\left(2\sinh \frac{\beta E_{\bf k}(\tau_1)}{2}\right)\right).\eea.
This further implies that the expression for the quantum partition function for $\alpha$ vacua can be simplified as:
\bea \hll{Z_{\alpha}(\beta;\tau_1)
=\frac{1}{|\cosh\alpha|}\exp\left(-\left(1+\frac{1}{2}\delta^{3}(0)\right)\int d^3{\bf k}~\ln\left(2\sinh \frac{\beta E_{\bf k}(\tau_1)}{2}\right)\right)}.\eea
After introducing the normal ordering the quantum partition function computed from adiabatic Bunch Davies vacuum as:
\bea \hll{:Z_{\bf BD}(\beta;\tau_1):=\exp\left(-\int d^3{\bf k} ~\ln\left(2\sinh \frac{\beta E_{\bf k}(\tau_1)}{2}\right)\right)}.\eea.
Then the normal ordered quantum partition function for the $\alpha$ vacua can be simplified as:
\bea \hll{:Z_{\alpha}(\beta;\tau_1):
=\frac{1}{|\cosh\alpha|}\exp\left(-\int d^3{\bf k}~\ln\left(2\sinh \frac{\beta E_{\bf k}(\tau_1)}{2}\right)\right)}.\eea
\subsection{Quantum micro-canonical partition function in terms of curvature perturbation field variable}In this subsection our prime objective is to find out the expression for the partition function in terms of the curvature perturbation field variable. To serve this purpose the time dependent dispersion relation can be expressed in terms of the curvature perturbation variable as:
 \bea E_{{\bf k}}(\tau_1)&=&|\Pi_{\bf k}(\tau_1)|^2+\omega^2_{\bf k}(\tau_1)|f_{\bf k}(\tau_1)|^2\nonumber\\
 &=&z^2(\tau_1)\left\{\left|\Pi^{\zeta}_{\bf k}(\tau_1)+\zeta_{\bf k}(\tau_1)\frac{1}{z(\tau_1)}\frac{dz(\tau_1)}{d\tau_1}\right|^2+\omega^2_{\bf k}(\tau_1)|\zeta_{\bf k}(\tau_1)|^2\right\}\nonumber\eea\bea
 &=&z^2(\tau_1)\left\{\left|\Pi^{\zeta}_{\bf k}(\tau_1)\right|^2+\left(\omega^2_{\bf k}(\tau_1)+\left(\frac{1}{z(\tau_1)}\frac{dz(\tau_1)}{d\tau_1}\right)^2\right)|\zeta_{\bf k}(\tau_1)|^2\right.\nonumber\\
&&\left.~~~~~~~~~~~~~~~~~~~~+\left(\Pi^{\zeta}_{-\bf k}(\tau_1)\zeta_{\bf k}(\tau_1)+\Pi^{\zeta}_{\bf k}(\tau_1)\zeta_{-\bf k}(\tau_1)\right) \left(\frac{1}{z(\tau_1)}\frac{dz(\tau_1)}{d\tau_1}\right)\right\}\nonumber\\
&=&z^2(\tau_1)\left\{E_{{\bf k},\zeta}(\tau_1)+\underbrace{\left(\Pi^{\zeta}_{-\bf k}(\tau_1)\zeta_{\bf k}(\tau_1)+\Pi^{\zeta}_{\bf k}(\tau_1)\zeta_{-\bf k}(\tau_1)\right) \left(\frac{1}{z(\tau_1)}\frac{dz(\tau_1)}{d\tau_1}\right)}_{\textcolor{red}{\bf Contribution~from~this~term~is~negligibly~small}}\right\}\nonumber\\
&\approx &z^2(\tau_1)E_{{\bf k},\zeta}(\tau_1),~~~~~~\eea 
where we define the time dependent energy dispersion relation in terms of the curvature perturbation variable as:
\bea E_{{\bf k},\zeta}(\tau_1):&=&\left|\Pi^{\zeta}_{\bf k}(\tau_1)\right|^2+\left(\omega^2_{\bf k}(\tau_1)+\left(\frac{1}{z(\tau_1)}\frac{dz(\tau_1)}{d\tau_1}\right)^2\right)|\zeta_{\bf k}(\tau_1)|^2,\nonumber\\
&=&\left|\Pi^{\zeta}_{\bf k}(\tau_1)\right|^2+\left(k^2-\frac{1}{z(\tau_1)}\frac{d^2z(\tau_1)}{d\tau^2_1}+\left(\frac{1}{z(\tau_1)}\frac{dz(\tau_1)}{d\tau_1}\right)^2\right)|\zeta_{\bf k}(\tau_1)|^2.\eea
Now, the thermal partition function for cosmology in terms of curvature perturbation computed for $\alpha$ vacua can be expressed as:
 \bea \hll{Z^{\zeta}_{\alpha}(\beta;\tau_1)=\frac{Z^{\zeta}_{\bf BD}(\beta;\tau_1)}{|\cosh\alpha|}}.\eea
where $Z^{\zeta}_{\bf BD}(\beta;\tau_1)$is  thermal partition function for cosmology in terms of curvature perturbation for Bunch Davies vacuum in terms which we have to compute now. To do this we will now write the partition function for rescaled field variable in terms of the curvature perturbation variable as:
 \bea \hll{Z^{\zeta}_{\bf BD}(\beta;\tau_1)=\exp\left(-\left(1+\frac{1}{2}\delta^{3}(0)\right)\int d^3{\bf k}~\ln\left(2\sinh\frac{\beta z^2(\tau_1)E_{{\bf k},\zeta}(\tau_1)}{2}\right)\right)},~~~~~\eea
 which can be further simplified in the normal ordered form as:
  \bea \hll{:Z^{\zeta}_{\bf BD}(\beta;\tau_1):=\exp\left(-\int d^3{\bf k}~\ln\left(2\sinh\frac{\beta z^2(\tau_1)E_{{\bf k},\zeta}(\tau_1)}{2}\right)\right)}.~~~~~\eea
 \newpage
 \section{Computation of classical micro-canonical partition function in Cosmology}  
\subsection{Classical micro-canonical partition function in terms of rescaled field variable}
In this section our aim is to derive the expression for the partition function for the cosmology in the classical regime which is basically the classical limit of the quantum partition function derived in the previous section. This result will be further helpful to determine the classical limit of the cosmological OTOC that we have derived in this paper.

In terms of the rescaled cosmological perturbation field variable we define the following classical partition function for Cosmology: 
\bea Z_{\bf Classical}(\beta;\tau_1):&=&\int \int \frac{{\cal D}f{\cal D}\Pi}{2\pi}~\exp\left(-\beta H\right)\nonumber\\&=&\prod_{{\bf k}}\int \int \frac{{d}f_{\bf k}(\tau_1) {d}\Pi_{\bf k}(\tau_1)}{2\pi}~\exp\left(- \frac{\beta}{2}\left[|\Pi_{\bf k}(\tau_1)|^2+\omega^2_{\bf k}(\tau_1)|f_{\bf k}(\tau_1)|^2\right]\right)\nonumber\\
&=&\prod_{{\bf k}}\exp\left(-\beta\left[\frac{E_{\bf k}(\tau_1)}{2}+\frac{1}{\beta}\ln\left(1-\exp(-\beta E_{\bf k}(\tau_1)\right)\right]\right)\nonumber\\
&=&\prod_{{\bf k}}\exp\left(\ln\left(\exp\left(-\frac{\beta E_{\bf k}(\tau_1)}{2}\right)\right)-\ln\left(1-\exp(-\beta E_{\bf k}(\tau_1)\right)\right)\nonumber\\
&=&\prod_{{\bf k}}\exp\left(\ln\left(\frac{\exp\left(-\frac{\beta E_{\bf k}(\tau_1)}{2}\right)}{\left(1-\exp\left(-\beta E_{\bf k}(\tau_1)\right)\right)}\right)\right)\nonumber\\
&=&\prod_{{\bf k}}\exp\left(-\ln\left(2\sinh\frac{\beta E_{\bf k}(\tau_1)}{2}\right)\right)\nonumber\\
&=&\exp\left(-\int d^3{\bf k} ~\ln\left(2\sinh \frac{\beta E_{\bf k}(\tau_1)}{2}\right)\right).~~~~\eea
where the conformal time dependent frequency is given by the following expression:
\bea \omega^2_{\bf k}(\tau_1)=\left(k^2-\frac{1}{z(\tau_1)}\frac{d^2z(\tau_1)}{d\tau^2_1}\right).\eea
This implies that:
\bea \hll{Z_{\bf Classical}(\beta;\tau_1)=\exp\left(-\int d^3{\bf k} ~\ln\left(2\sinh \frac{\beta E_{\bf k}(\tau_1)}{2}\right)\right)=:Z_{\bf BD}(\beta;\tau_1):}~.\eea
Now from the above expression we found that the expression for the classical partition function and normal ordered partition function for Cosmology is exactly same. But since we know there in no concept of vacuum exist in classical field theory, for that reason we can write the following expression:
\bea \hll{Z_{\bf Classical}(\beta;\tau_1)=|\cosh\alpha|~:Z_{\alpha}(\beta;\tau_1):~~~~\forall ~~\alpha}. \eea

\subsection{Classical micro-canonical partition function in terms of curvature perturbation field variable}
To construct the classical partition function in terms of the curvature perturbation field variable and its conjugate momenta we are going to follow similar procedure. In this description, the classical partition function for Cosmology can be expressed as:
\bea Z^{\zeta}_{\bf Classical}(\beta;\tau_1):&=&\int \int \frac{{\cal D}\zeta{\cal D}\Pi_{\zeta}}{2\pi}~\exp\left(-\beta H\right)\nonumber\\&=&\prod_{{\bf k}}\int \int \frac{{d}\zeta_{\bf k}(\tau_1) {d}\Pi^{\zeta}_{\bf k}(\tau_1)}{2\pi}~\nonumber\\
&&~~~~~\exp\left(-\frac{\beta z^{2}(\tau_1) }{2}\left[\left|\Pi^{\zeta}_{\bf k}(\tau_1)\right|^2+\left(\omega^2_{\bf k}(\tau_1)+\left(\frac{1}{z(\tau_1)}\frac{dz(\tau_1)}{d\tau_1}\right)^2\right)|\zeta_{\bf k}(\tau_1)|^2\right]\right)\nonumber\\
&=&\prod_{{\bf k}}\exp\left(-\beta\left[\frac{z^2(\tau_1)E^{\zeta}_{\bf k}(\tau_1)}{2}+\frac{1}{\beta}\ln\left(1-\exp(-\beta z^2(\tau_1)E^{\zeta}_{\bf k}(\tau_1)\right)\right]\right)\nonumber\\
&=&\prod_{{\bf k}}\exp\left(\ln\left(\exp\left(-\frac{\beta z^2(\tau_1)E^{\zeta}_{\bf k}(\tau_1)}{2}\right)\right)-\ln\left(1-\exp(-\beta z^2(\tau_1)E^{\zeta}_{\bf k}(\tau_1)\right)\right)\nonumber\\
&=&\prod_{{\bf k}}\exp\left(\ln\left(\frac{\exp\left(-\frac{\beta z^2(\tau_1)E^{\zeta}_{\bf k}(\tau_1)}{2}\right)}{\left(1-\exp\left(-\beta z^2(\tau_1)E^{\zeta}_{\bf k}(\tau_1)\right)\right)}\right)\right)\nonumber\\
&=&\prod_{{\bf k}}\exp\left(-\ln\left(2\sinh\frac{\beta z^2(\tau_1)E^{\zeta}_{\bf k}(\tau_1)}{2}\right)\right)\nonumber\\
&=&\exp\left(-\int d^3{\bf k} ~\ln\left(2\sinh \frac{\beta z^2(\tau_1)E^{\zeta}_{\bf k}(\tau_1)}{2}\right)\right).~~~~\eea
This further implies that:
\bea &&\hll{Z^{\zeta}_{\bf Classical}(\beta;\tau_1)=\exp\left(-\int d^3{\bf k} ~\ln\left(2\sinh \frac{\beta z^2(\tau_1)E^{\zeta}_{\bf k}(\tau_1)}{2}\right)\right)}\nonumber\\
&&~~~~~~~~~~~~~~~~~~~~\hll{=:Z^{\zeta}_{\bf BD}(\beta;\tau_1):=|\cosh\alpha|~:Z_{\alpha}(\beta;\tau_1):~~~~\forall ~~\alpha\neq Z_{\bf Classical}(\beta;\tau_1)}~.~~~~~~~~~~~~~~\eea
The derived classical partition functions for Cosmology computed in terms of two perturbation variables are not same because of the presence of {\it Mukhanov Sasaki varibale}. 
\newpage
\section{Computation of the trace of the two-point amplitude in micro-canonical OTOC}
Now, we will explicitly compute the numerator of the OTOC for quantum $\alpha$ vacua,which is given by the following expression:
  \bea && {\rm Tr}\left[e^{-\beta \widehat{H}(\tau_1)}\left[\hat{f}({\bf x},\tau_1),\hat{\Pi}({\bf x},\tau_2)\right]\right]_{(\alpha)}\nonumber\\
 &&= \frac{1}{|\cosh\alpha|}\int d\Psi_{\bf BD}~\int \frac{d^3{\bf k}_1}{(2\pi)^3}\int \frac{d^3{\bf k}_2}{(2\pi)^3}\exp\left[i\left({\bf k}_1+{\bf k}_2\right).{\bf x}\right]~~~~~~~~\nonumber\\
  &&~~~~~~~~~~~~~~~~~~~~~~~~~~ \langle\Psi_{\bf BD}|\left[\hat{\nabla}_1({\bf k}_1,{\bf k}_2;\tau_1,\tau_2;\beta) -\hat{\nabla}_2({\bf k}_1,{\bf k}_2;\tau_1,\tau_2;\beta)\right]|\Psi_{\bf BD}\rangle.~~~~~~~~~~~ \eea 
  Further, our aim is to compute the individual contributions which are given by:
  \bea &&\int d\Psi_{\bf BD}~\langle\Psi_{\bf BD}|\hat{\nabla}_1({\bf k}_1,{\bf k}_2;\tau_1,\tau_2;\beta)|\Psi_{\bf BD}\rangle=\int d\Psi_{\bf BD}~ \langle\Psi_{\bf BD}|e^{-\beta \hat{H}(\tau_1)}~\hat{\Delta}_1({\bf k}_1,{\bf k}_2;\tau_1,\tau_2)|\Psi_{\bf BD}\rangle,~~~~~~~~~~~~\\
 &&\int d\Psi_{\bf BD}~\langle\Psi_{\bf BD}|\hat{\nabla}_2({\bf k}_1,{\bf k}_2;\tau_1,\tau_2;\beta)|\Psi_{\bf BD}\rangle= \int d\Psi_{\bf BD}~\langle\Psi_{\bf BD}|e^{-\beta \hat{H}(\tau_1)}~\hat{\Delta}_2({\bf k}_1,{\bf k}_2;\tau_1,\tau_2)|\Psi_{\bf BD}\rangle.\eea
  Let us evaluate one by one each of the contributions, which are given by:
  \bea &&\int d\Psi_{\bf BD}~\langle\Psi_{\bf BD}|e^{-\beta \hat{H}(\tau_1)}~a_{{\bf k}_1}a_{{\bf k}_2}|\Psi_{\bf BD}\rangle=0,\\
 &&\int d\Psi_{\bf BD}~\langle\Psi_{\bf BD}|e^{-\beta \hat{H}(\tau_1)}~a_{{\bf k}_1}a^{\dagger}_{-{\bf k}_2}|\Psi_{\bf BD}\rangle\nonumber\\
   &&~~~~~~~~~~~~~~~~~=(2\pi)^3\exp\left(-\left(1+\frac{1}{2}\delta^{3}(0)\right)\int d^3{\bf k}~\ln\left(2\sinh \frac{\beta E_{\bf k}(\tau_1)}{2}\right)\right)\delta^3\left({\bf k}_1+{\bf k}_2\right),~~~~~~~~~~~\\
 &&\int d\Psi_{\bf BD}~\langle\Psi_{\bf BD}|e^{-\beta \hat{H}(\tau_1)}~a^{\dagger}_{-{\bf k}_1}a_{{\bf k}_2}|\Psi_{\bf BD}\rangle\nonumber\\
   &&~~~~~~~~~~~~~~~~~=(2\pi)^3\exp\left(-\left(1+\frac{1}{2}\delta^{3}(0)\right)\int d^3{\bf k}~\ln\left(2\sinh \frac{\beta E_{\bf k}(\tau_1)}{2}\right)\right)\delta^3\left({\bf k}_1+{\bf k}_2\right),~~~~~~~~~~~\\
   &&\int d\Psi_{\bf BD}~\langle\Psi_{\bf BD}|e^{-\beta \hat{H}(\tau_1)}~a^{\dagger}_{-{\bf k}_1}a^{\dagger}_{-{\bf k}_2}|\Psi_{\bf BD}\rangle=0.\eea 
Further, introducing the normal ordering we get:
 \bea &&\int d\Psi_{\bf BD}~\langle\Psi_{\bf BD}|:e^{-\beta \hat{H}(\tau_1)}~a_{{\bf k}_1}a_{{\bf k}_2}:|\Psi_{\bf BD}\rangle=0,\\
 &&\int d\Psi_{\bf BD}~\langle\Psi_{\bf BD}|:e^{-\beta \hat{H}(\tau_1)}~a_{{\bf k}_1}a^{\dagger}_{-{\bf k}_2}:|\Psi_{\bf BD}\rangle\nonumber\\
   &&~~~~~~~~~~~~~~~~~~~~~~~~~~~~~~=(2\pi)^3\exp\left(-\int d^3{\bf k}~\ln\left(2\sinh \frac{\beta E_{\bf k}(\tau_1)}{2}\right)\right)\delta^3\left({\bf k}_1+{\bf k}_2\right),~~~~~~~~~~~\eea\bea
 &&\int d\Psi_{\bf BD}~\langle\Psi_{\bf BD}|:e^{-\beta \hat{H}(\tau_1)}~a^{\dagger}_{-{\bf k}_1}a_{{\bf k}_2}:|\Psi_{\bf BD}\rangle\nonumber\\
   &&~~~~~~~~~~~~~~~~~~~~~~~~~~~~~~=(2\pi)^3\exp\left(-\int d^3{\bf k}~\ln\left(2\sinh \frac{\beta E_{\bf k}(\tau_1)}{2}\right)\right)\delta^3\left({\bf k}_1+{\bf k}_2\right),~~~~~~~~~~~\\
   &&\int d\Psi_{\bf BD}~\langle\Psi_{\bf BD}|:e^{-\beta \hat{H}(\tau_1)}~a^{\dagger}_{-{\bf k}_1}a^{\dagger}_{-{\bf k}_2}:|\Psi_{\bf BD}\rangle=0.\eea
   Consequently, the individual contributions can be computed as: 
    \bea &&\int d\Psi_{\bf BD}~\langle\Psi_{\bf BD}|\hat{\nabla}_1({\bf k}_1,{\bf k}_2;\tau_1,\tau_2;\beta)|\Psi_{\bf BD}\rangle\nonumber\\
 &&=(2\pi)^3\exp\left(-\left(1+\frac{1}{2}\delta^{3}(0)\right)\int d^3{\bf k}~\ln\left(2\sinh \frac{\beta E_{\bf k}(\tau_1)}{2}\right)\right)\delta^{3}({\bf k}_1+{\bf k}_2)\nonumber\\
 &&~~~~~~~~~~~~~~~~~~~~~~~~~~\left[ {\cal D}_2({\bf k}_1,{\bf k}_2;\tau_1,\tau_2)+ {\cal D}_3({\bf k}_1,{\bf k}_2;\tau_1,\tau_2)\right],~~~~~~~~~~~\\
  &&\int d\Psi_{\bf BD}~\langle\Psi_{\bf BD}|\hat{\nabla}_2({\bf k}_1,{\bf k}_2;\tau_1,\tau_2;\beta)|\Psi_{\bf BD}\rangle\nonumber\\
 &&=(2\pi)^3\exp\left(-\left(1+\frac{1}{2}\delta^{3}(0)\right)\int d^3{\bf k}~\ln\left(2\sinh \frac{\beta E_{\bf k}(\tau_1)}{2}\right)\right)\delta^{3}({\bf k}_1+{\bf k}_2)\nonumber\\
 &&~~~~~~~~~~~~~~~~~~~~~~~~~~\left[ {\cal L}_2({\bf k}_1,{\bf k}_2;\tau_1,\tau_2)+ {\cal L}_3({\bf k}_1,{\bf k}_2;\tau_1,\tau_2)\right],~~~~~~~~~~~\eea
 Consequently, the individual contributions can be computed in the normal ordered form as: 
 \bea &&\int d\Psi_{\bf BD}~\langle\Psi_{\bf BD}|\hat{\nabla}_1({\bf k}_1,{\bf k}_2;\tau_1,\tau_2;\beta)|\Psi_{\bf BD}\rangle\nonumber\\
 &&=(2\pi)^3\exp\left(-\int d^3{\bf k}~\ln\left(2\sinh \frac{\beta E_{\bf k}(\tau_1)}{2}\right)\right)\delta^{3}({\bf k}_1+{\bf k}_2)\nonumber\\
 &&~~~~~~~~~~~~~~~~~~~~~~~~~~\left[ {\cal D}_2({\bf k}_1,{\bf k}_2;\tau_1,\tau_2)+ {\cal D}_3({\bf k}_1,{\bf k}_2;\tau_1,\tau_2)\right],~~~~~~~~~~~\\
  &&\int d\Psi_{\bf BD}~\langle\Psi_{\bf BD}|\hat{\nabla}_2({\bf k}_1,{\bf k}_2;\tau_1,\tau_2;\beta)|\Psi_{\bf BD}\rangle\nonumber\\
 &&=(2\pi)^3\exp\left(-\int d^3{\bf k}~\ln\left(2\sinh \frac{\beta E_{\bf k}(\tau_1)}{2}\right)\right)\delta^{3}({\bf k}_1+{\bf k}_2)\nonumber\\
 &&~~~~~~~~~~~~~~~~~~~~~~~~~~\left[ {\cal L}_2({\bf k}_1,{\bf k}_2;\tau_1,\tau_2)+ {\cal L}_3({\bf k}_1,{\bf k}_2;\tau_1,\tau_2)\right],~~~~~~~~~~~\eea
	\newpage
\section{Computation of the trace of the four-point amplitude in micro-canonical OTOC}
Now, we will explicitly compute the numerator of the OTOC for quantum $\alpha$ vacua,which is given by the following expression:
  \bea && {\rm Tr}\left[e^{-\beta \widehat{H}(\tau_1)}\left[\hat{f}({\bf x},\tau_1),\hat{\Pi}({\bf x},\tau_2)\right]^2\right]_{(\alpha)}\nonumber\\
 &&= \frac{1}{|\cosh\alpha|}\int d\Psi_{\bf BD}~\int \frac{d^3{\bf k}_1}{(2\pi)^3}\int \frac{d^3{\bf k}_2}{(2\pi)^3}\int \frac{d^3{\bf k}_3}{(2\pi)^3}\int \frac{d^3{\bf k}_4}{(2\pi)^3}\nonumber\\
 &&~~~~~~~~~~~~~~~~~~~~~~~~~\exp\left[i\left({\bf k}_1+{\bf k}_2+{\bf k}_3+{\bf k}_4\right).{\bf x}\right]~~~~~~~~\nonumber\\
  &&~~~~~~~~~~~~~~~~~~~~~~~~~~ \langle\Psi_{\bf BD}|\left[\widehat{\cal V}_1({\bf k}_1,{\bf k}_2,{\bf k}_3,{\bf k}_4;\tau_1,\tau_2;\beta)\right.\nonumber\\ && \left.~~~~~~~~~~~~~~~~~~~~~~~~~~~~~~~~~~~ -\widehat{\cal V}_2({\bf k}_1,{\bf k}_2,{\bf k}_3,{\bf k}_4;\tau_1,\tau_2;\beta)\right.\nonumber\\ && \left.~~~~~~~~~~~~~~~~~~~~~~~~~~~~~~~~~~ +\widehat{\cal V}_3({\bf k}_1,{\bf k}_2,{\bf k}_3,{\bf k}_4;\tau_1,\tau_2;\beta)\right.\nonumber\\ && \left.~~~~~~~~~~~~~~~~~~~~~~~~~~~~~~~~~~~~~ -\widehat{\cal V}_4({\bf k}_1,{\bf k}_2,{\bf k}_3,{\bf k}_4;\tau_1,\tau_2;\beta)\right]|\Psi_{\bf BD}\rangle.~~~~~~~~~~~ \eea 
  Further, our aim is to compute the individual contributions which are given by:
  \bea &&\int d\Psi_{\bf BD}~\langle\Psi_{\bf BD}|\widehat{\cal V}_1({\bf k}_1,{\bf k}_2,{\bf k}_3,{\bf k}_4;\tau_1,\tau_2;\beta)|\Psi_{\bf BD}\rangle\nonumber\\
  &&~~~~~~~~~~~~~~~~~~= \int d\Psi_{\bf BD}~\langle\Psi_{\bf BD}|e^{-\beta \hat{H}(\tau_1)}~\widehat{\cal T}_1({\bf k}_1,{\bf k}_2,{\bf k}_3,{\bf k}_4;\tau_1,\tau_2)|\Psi_{\bf BD}\rangle,~~~~~~~~~~~~~~~~\\
 &&\int d\Psi_{\bf BD}~\langle\Psi_{\bf BD}|\widehat{\cal V}_2({\bf k}_1,{\bf k}_2,{\bf k}_3,{\bf k}_4;\tau_1,\tau_2;\beta)|\Psi_{\bf BD}\rangle\nonumber\\
  &&~~~~~~~~~~~~~~~~~~=\int d\Psi_{\bf BD}~\langle\Psi_{\bf BD}|e^{-\beta \hat{H}(\tau_1)}~\widehat{\cal T}_2({\bf k}_1,{\bf k}_2,{\bf k}_3,{\bf k}_4;\tau_1,\tau_2)|\Psi_{\bf BD}\rangle,\\ 
&&\int d\Psi_{\bf BD}~\langle\Psi_{\bf BD}|\widehat{\cal V}_3({\bf k}_1,{\bf k}_2,{\bf k}_3,{\bf k}_4;\tau_1,\tau_2;\beta)|\Psi_{\bf BD}\rangle\nonumber\\
  &&~~~~~~~~~~~~~~~~~~= \int d\Psi_{\bf BD}~\langle\Psi_{\bf BD}|e^{-\beta \hat{H}(\tau_1)}~\widehat{\cal T}_3({\bf k}_1,{\bf k}_2,{\bf k}_3,{\bf k}_4;\tau_1,\tau_2)|\Psi_{\bf BD}\rangle,\\
 &&\int d\Psi_{\bf BD}~\langle\Psi_{\bf BD}|\widehat{\cal V}_4({\bf k}_1,{\bf k}_2,{\bf k}_3,{\bf k}_4;\tau_1,\tau_2;\beta)|\Psi_{\bf BD}\rangle\nonumber\\
  &&~~~~~~~~~~~~~~~~~~= \int d\Psi_{\bf BD}~\langle\Psi_{\bf BD}|e^{-\beta \hat{H}(\tau_1)}~\widehat{\cal T}_4({\bf k}_1,{\bf k}_2,{\bf k}_3,{\bf k}_4;\tau_1,\tau_2)|\Psi_{\bf BD}\rangle.\eea
  Let us evaluate one by one each of the contributions, which are given by:
  \bea &&\int d\Psi_{\bf BD}~\langle\Psi_{\bf BD}|e^{-\beta \hat{H}(\tau_1)}~a_{{\bf k}_1}a_{{\bf k}_2}a_{{\bf k}_3}a_{{\bf k}_4}|\Psi_{\bf BD}\rangle=0,\\
 &&\int d\Psi_{\bf BD}~\langle\Psi_{\bf BD}|e^{-\beta \hat{H}(\tau_1)}~a^{\dagger}_{-{\bf k}_1}a_{{\bf k}_2}a_{{\bf k}_3}a_{{\bf k}_4}|\Psi_{\bf BD}\rangle=0,
 \eea\bea
 &&\int d\Psi_{\bf BD}~\langle\Psi_{\bf BD}|e^{-\beta \hat{H}(\tau_1)}~a_{{\bf k}_1}a^{\dagger}_{-{\bf k}_2}a_{{\bf k}_3}a_{{\bf k}_4}|\Psi_{\bf BD}\rangle=0,\\
 &&\int d\Psi_{\bf BD}~\langle\Psi_{\bf BD}|e^{-\beta \hat{H}(\tau_1)}~a^{\dagger}_{-{\bf k}_1}a^{\dagger}_{-{\bf k}_2}a_{{\bf k}_3}a_{{\bf k}_4}|\Psi_{\bf BD}\rangle\nonumber\\
   &&~~~~~~~~~~~~~~~~~~~~~~~~~~~~~~=(2\pi)^6\exp\left(-\left(1+\frac{1}{2}\delta^{3}(0)\right)\int d^3{\bf k}~\ln\left(2\sinh \frac{\beta E_{\bf k}(\tau_1)}{2}\right)\right)\nonumber\\
 &&~~~~~~~~~~~~~~~~~~~~~~~~~~~~~~~~~~~~~~~~~~~~~~~~\left[\delta^3\left({\bf k}_1+{\bf k}_4\right)\delta^3\left({\bf k}_2+{\bf k}_3\right)+\delta^3\left({\bf k}_1+{\bf k}_3\right)\delta^3\left({\bf k}_2+{\bf k}_4\right)\right],~~~~~~~~~~~\\ &&\int d\Psi_{\bf BD}~\langle\Psi_{\bf BD}|e^{-\beta \hat{H}(\tau_1)}~a_{{\bf k}_1}a_{{\bf k}_2}a^{\dagger}_{-{\bf k}_3}a_{{\bf k}_4}|\Psi_{\bf BD}\rangle=0,\\
 &&\int d\Psi_{\bf BD}~\langle\Psi_{\bf BD}|e^{-\beta \hat{H}(\tau_1)}~a^{\dagger}_{-{\bf k}_1}a_{{\bf k}_2}a^{\dagger}_{-{\bf k}_3}a_{{\bf k}_4}|\Psi_{\bf BD}\rangle\nonumber\\
   &&~~~~~~~~~~~~~~~~~~~~~~~~~~~~~~=(2\pi)^6\exp\left(-\left(1+\frac{1}{2}\delta^{3}(0)\right)\int d^3{\bf k}~\ln\left(2\sinh \frac{\beta E_{\bf k}(\tau_1)}{2}\right)\right)\nonumber\\
 &&~~~~~~~~~~~~~~~~~~~~~~~~~~~~~~~~~~~~~~~~~~~~~~~~\left[\delta^3\left({\bf k}_1+{\bf k}_2\right)\delta^3\left({\bf k}_3+{\bf k}_4\right)+\delta^3\left({\bf k}_1+{\bf k}_4\right)\delta^3\left({\bf k}_2+{\bf k}_3\right)\right],~~~~~~~~~~~,\eea\bea
  &&\int d\Psi_{\bf BD}~\langle\Psi_{\bf BD}|e^{-\beta \hat{H}(\tau_1)}~a_{{\bf k}_1}a^{\dagger}_{-{\bf k}_2}a^{\dagger}_{-{\bf k}_3}a_{{\bf k}_4}|\Psi_{\bf BD}\rangle\nonumber\\
   &&~~~~~~~~~~~~~~~~~~~~~~~~~~~~~~=(2\pi)^6\exp\left(-\left(1+\frac{1}{2}\delta^{3}(0)\right)\int d^3{\bf k}~\ln\left(2\sinh \frac{\beta E_{\bf k}(\tau_1)}{2}\right)\right)\nonumber\\
 &&~~~~~~~~~~~~~~~~~~~~~~~~~~~~~~~~~~~~~~~~~~~~~~~~\left[\delta^3\left({\bf k}_1+{\bf k}_2\right)\delta^3\left({\bf k}_3+{\bf k}_4\right)+\delta^3\left({\bf k}_1+{\bf k}_3\right)\delta^3\left({\bf k}_2+{\bf k}_4\right)\right],~~~~~~~~~~~,\\ &&\int d\Psi_{\bf BD}~\langle\Psi_{\bf BD}|e^{-\beta \hat{H}(\tau_1)}~a^{\dagger}_{-{\bf k}_1}a^{\dagger}_{-{\bf k}_2}a^{\dagger}_{-{\bf k}_3}a_{{\bf k}_4}|\Psi_{\bf BD}\rangle=0,\\
  &&\int d\Psi_{\bf BD}~\langle\Psi_{\bf BD}|e^{-\beta \hat{H}(\tau_1)}~a_{{\bf k}_1}a_{{\bf k}_2}a_{{\bf k}_3}a^{\dagger}_{-{\bf k}_4}|\Psi_{\bf BD}\rangle=0,\\
   &&\int d\Psi_{\bf BD}~\langle\Psi_{\bf BD}|e^{-\beta \hat{H}(\tau_1)}~a^{\dagger}_{-{\bf k}_1}a_{{\bf k}_2}a_{{\bf k}_3}a^{\dagger}_{-{\bf k}_4}|\Psi_{\bf BD}\rangle\nonumber\\
   &&~~~~~~~~~~~~~~~~~~~~~~~~~~~~~~=(2\pi)^6\exp\left(-\left(1+\frac{1}{2}\delta^{3}(0)\right)\int d^3{\bf k}~\ln\left(2\sinh \frac{\beta E_{\bf k}(\tau_1)}{2}\right)\right)\nonumber\\
 &&~~~~~~~~~~~~~~~~~~~~~~~~~~~~~~~~~~~~~~~~~~~~~~~~\left[\delta^3\left({\bf k}_1+{\bf k}_2\right)\delta^3\left({\bf k}_3+{\bf k}_4\right)+\delta^3\left({\bf k}_1+{\bf k}_3\right)\delta^3\left({\bf k}_2+{\bf k}_4\right)\right],~~~~~~~~~~~,\\
   &&\int d\Psi_{\bf BD}~\langle\Psi_{\bf BD}|e^{-\beta \hat{H}(\tau_1)}~a_{{\bf k}_1}a^{\dagger}_{-{\bf k}_2}a_{{\bf k}_3}a^{\dagger}_{-{\bf k}_4}|\Psi_{\bf BD}\rangle\nonumber\\
   &&~~~~~~~~~~~~~~~~~~~~~~~~~~~~~~=(2\pi)^6\exp\left(-\left(1+\frac{1}{2}\delta^{3}(0)\right)\int d^3{\bf k}~\ln\left(2\sinh \frac{\beta E_{\bf k}(\tau_1)}{2}\right)\right)\nonumber\\
 &&~~~~~~~~~~~~~~~~~~~~~~~~~~~~~~~~~~~~~~~~~~~~~~~~\left[\delta^3\left({\bf k}_1+{\bf k}_2\right)\delta^3\left({\bf k}_3+{\bf k}_4\right)+\delta^3\left({\bf k}_1+{\bf k}_4\right)\delta^3\left({\bf k}_2+{\bf k}_3\right)\right],~~~~~~~~~~~,\\
   &&\int d\Psi_{\bf BD}~\langle\Psi_{\bf BD}|e^{-\beta \hat{H}(\tau_1)}~a^{\dagger}_{-{\bf k}_1}a^{\dagger}_{-{\bf k}_2}a_{{\bf k}_3}a^{\dagger}_{-{\bf k}_4}|\Psi_{\bf BD}\rangle=0,\eea\bea 
   &&\int d\Psi_{\bf BD}~\langle\Psi_{\bf BD}|e^{-\beta \hat{H}(\tau_1)}~a_{{\bf k}_1}a_{{\bf k}_2}a^{\dagger}_{-{\bf k}_3}a^{\dagger}_{-{\bf k}_4}|\Psi_{\bf BD}\rangle\nonumber\\
   &&~~~~~~~~~~~~~~~~~~~~~~~~~~~~~~=(2\pi)^6\exp\left(-\left(1+\frac{1}{2}\delta^{3}(0)\right)\int d^3{\bf k}~\ln\left(2\sinh \frac{\beta E_{\bf k}(\tau_1)}{2}\right)\right)\nonumber\\
 &&~~~~~~~~~~~~~~~~~~~~~~~~~~~~~~~~~~~~~~~~~~~~~~~~\left[\delta^3\left({\bf k}_1+{\bf k}_3\right)\delta^3\left({\bf k}_2+{\bf k}_4\right)+\delta^3\left({\bf k}_1+{\bf k}_4\right)\delta^3\left({\bf k}_2+{\bf k}_3\right)\right],~~~~~~~~~~~,\\
  &&\int d\Psi_{\bf BD}~\langle\Psi_{\bf BD}|e^{-\beta \hat{H}(\tau_1)}~a^{\dagger}_{-{\bf k}_1}a_{{\bf k}_2}a^{\dagger}_{-{\bf k}_3}a^{\dagger}_{-{\bf k}_4}|\Psi_{\bf BD}\rangle=0,\\ 
  &&\int d\Psi_{\bf BD}~\langle\Psi_{\bf BD}|e^{-\beta \hat{H}(\tau_1)}~a_{{\bf k}_1}a^{\dagger}_{-{\bf k}_2}a^{\dagger}_{-{\bf k}_3}a^{\dagger}_{-{\bf k}_4}|\Psi_{\bf BD}\rangle=0,\\  
   &&\int d\Psi_{\bf BD}~\langle\Psi_{\bf BD}|e^{-\beta \hat{H}(\tau_1)}~a^{\dagger}_{-{\bf k}_1}a^{\dagger}_{-{\bf k}_2}a^{\dagger}_{-{\bf k}_3}a^{\dagger}_{-{\bf k}_4}|\Psi_{\bf BD}\rangle=0.\eea
Further, introducing the normal ordering we get:
 \bea &&\int d\Psi_{\bf BD}~\langle\Psi_{\bf BD}|:e^{-\beta \hat{H}(\tau_1)}~a_{{\bf k}_1}a_{{\bf k}_2}a_{{\bf k}_3}a_{{\bf k}_4}:|\Psi_{\bf BD}\rangle=0,\\
 &&\int d\Psi_{\bf BD}~\langle\Psi_{\bf BD}|:e^{-\beta \hat{H}(\tau_1)}~a^{\dagger}_{-{\bf k}_1}a_{{\bf k}_2}a_{{\bf k}_3}a_{{\bf k}_4}:|\Psi_{\bf BD}\rangle=0,
 \\ 
 &&\int d\Psi_{\bf BD}~\langle\Psi_{\bf BD}|:e^{-\beta \hat{H}(\tau_1)}~a_{{\bf k}_1}a^{\dagger}_{-{\bf k}_2}a_{{\bf k}_3}a_{{\bf k}_4}:|\Psi_{\bf BD}\rangle=0,\\
 &&\langle\Psi_{\bf BD}|:e^{-\beta \hat{H}(\tau_1)}~a^{\dagger}_{-{\bf k}_1}a^{\dagger}_{-{\bf k}_2}a_{{\bf k}_3}a_{{\bf k}_4}:|\Psi_{\bf BD}\rangle\nonumber\\
   &&~~~~~~~~~~~~~~~~~~~~~~~~~~~~~~=(2\pi)^6\exp\left(-\int d^3{\bf k}~\ln\left(2\sinh \frac{\beta E_{\bf k}(\tau_1)}{2}\right)\right)\nonumber\\
 &&~~~~~~~~~~~~~~~~~~~~~~~~~~~~~~~~~~~~~~~~~~~~~~~~\left[\delta^3\left({\bf k}_1+{\bf k}_4\right)\delta^3\left({\bf k}_2+{\bf k}_3\right)+\delta^3\left({\bf k}_1+{\bf k}_3\right)\delta^3\left({\bf k}_2+{\bf k}_4\right)\right],~~~~~~~~~~~\\ &&\int d\Psi_{\bf BD}~\langle\Psi_{\bf BD}|:e^{-\beta \hat{H}(\tau_1)}~a_{{\bf k}_1}a_{{\bf k}_2}a^{\dagger}_{-{\bf k}_3}a_{{\bf k}_4}:|\Psi_{\bf BD}\rangle=0,\eea\bea
 &&\int d\Psi_{\bf BD}~\langle\Psi_{\bf BD}|:e^{-\beta \hat{H}(\tau_1)}~a^{\dagger}_{-{\bf k}_1}a_{{\bf k}_2}a^{\dagger}_{-{\bf k}_3}a_{{\bf k}_4}:|\Psi_{\bf BD}\rangle\nonumber\\
   &&~~~~~~~~~~~~~~~~~~~~~~~~~~~~~~=(2\pi)^6\exp\left(-\int d^3{\bf k}~\ln\left(2\sinh \frac{\beta E_{\bf k}(\tau_1)}{2}\right)\right)\nonumber\\
 &&~~~~~~~~~~~~~~~~~~~~~~~~~~~~~~~~~~~~~~~~~~~~~~~~\left[\delta^3\left({\bf k}_1+{\bf k}_2\right)\delta^3\left({\bf k}_3+{\bf k}_4\right)+\delta^3\left({\bf k}_1+{\bf k}_4\right)\delta^3\left({\bf k}_2+{\bf k}_3\right)\right],~~~~~~~~~~~,\\
  &&\int d\Psi_{\bf BD}~\langle\Psi_{\bf BD}|:e^{-\beta \hat{H}(\tau_1)}~a_{{\bf k}_1}a^{\dagger}_{-{\bf k}_2}a^{\dagger}_{-{\bf k}_3}a_{{\bf k}_4}:|\Psi_{\bf BD}\rangle\nonumber\\
   &&~~~~~~~~~~~~~~~~~~~~~~~~~~~~~~=(2\pi)^6\exp\left(-\int d^3{\bf k}~\ln\left(2\sinh \frac{\beta E_{\bf k}(\tau_1)}{2}\right)\right)\nonumber\\
 &&~~~~~~~~~~~~~~~~~~~~~~~~~~~~~~~~~~~~~~~~~~~~~~~~\left[\delta^3\left({\bf k}_1+{\bf k}_2\right)\delta^3\left({\bf k}_3+{\bf k}_4\right)+\delta^3\left({\bf k}_1+{\bf k}_3\right)\delta^3\left({\bf k}_2+{\bf k}_4\right)\right],~~~~~~~~~~~,\\ &&\int d\Psi_{\bf BD}~\langle\Psi_{\bf BD}|:e^{-\beta \hat{H}(\tau_1)}~a^{\dagger}_{-{\bf k}_1}a^{\dagger}_{-{\bf k}_2}a^{\dagger}_{-{\bf k}_3}a_{{\bf k}_4}:|\Psi_{\bf BD}\rangle=0,\eea\bea
  &&\int d\Psi_{\bf BD}~\langle\Psi_{\bf BD}|:e^{-\beta \hat{H}(\tau_1)}~a_{{\bf k}_1}a_{{\bf k}_2}a_{{\bf k}_3}a^{\dagger}_{-{\bf k}_4}:|\Psi_{\bf BD}\rangle=0,\\
   &&\int d\Psi_{\bf BD}~\langle\Psi_{\bf BD}|:e^{-\beta \hat{H}(\tau_1)}~a^{\dagger}_{-{\bf k}_1}a_{{\bf k}_2}a_{{\bf k}_3}a^{\dagger}_{-{\bf k}_4}:|\Psi_{\bf BD}\rangle\nonumber\\
   &&~~~~~~~~~~~~~~~~~~~~~~~~~~~~~~=(2\pi)^6\exp\left(-\int d^3{\bf k}~\ln\left(2\sinh \frac{\beta E_{\bf k}(\tau_1)}{2}\right)\right)\nonumber\\
 &&~~~~~~~~~~~~~~~~~~~~~~~~~~~~~~~~~~~~~~~~~~~~~~~~\left[\delta^3\left({\bf k}_1+{\bf k}_2\right)\delta^3\left({\bf k}_3+{\bf k}_4\right)+\delta^3\left({\bf k}_1+{\bf k}_3\right)\delta^3\left({\bf k}_2+{\bf k}_4\right)\right],~~~~~~~~~~~,\\
   &&\int d\Psi_{\bf BD}~\langle\Psi_{\bf BD}|:e^{-\beta \hat{H}(\tau_1)}~a_{{\bf k}_1}a^{\dagger}_{-{\bf k}_2}a_{{\bf k}_3}a^{\dagger}_{-{\bf k}_4}:|\Psi_{\bf BD}\rangle\nonumber\\
   &&~~~~~~~~~~~~~~~~~~~~~~~~~~~~~~=(2\pi)^6\exp\left(-\int d^3{\bf k}~\ln\left(2\sinh \frac{\beta E_{\bf k}(\tau_1)}{2}\right)\right)\nonumber\\
 &&~~~~~~~~~~~~~~~~~~~~~~~~~~~~~~~~~~~~~~~~~~~~~~~~\left[\delta^3\left({\bf k}_1+{\bf k}_2\right)\delta^3\left({\bf k}_3+{\bf k}_4\right)+\delta^3\left({\bf k}_1+{\bf k}_4\right)\delta^3\left({\bf k}_2+{\bf k}_3\right)\right],~~~~~~~~~~~,\\
   &&\int d\Psi_{\bf BD}~\langle\Psi_{\bf BD}|:e^{-\beta \hat{H}(\tau_1)}~a^{\dagger}_{-{\bf k}_1}a^{\dagger}_{-{\bf k}_2}a_{{\bf k}_3}a^{\dagger}_{-{\bf k}_4}:|\Psi_{\bf BD}\rangle=0,\\
   &&\int d\Psi_{\bf BD}~\langle\Psi_{\bf BD}|:e^{-\beta \hat{H}(\tau_1)}~a_{{\bf k}_1}a_{{\bf k}_2}a^{\dagger}_{-{\bf k}_3}a^{\dagger}_{-{\bf k}_4}:|\Psi_{\bf BD}\rangle\nonumber\\
   &&~~~~~~~~~~~~~~~~~~~~~~~~~~~~~~=(2\pi)^6\exp\left(-\int d^3{\bf k}~\ln\left(2\sinh \frac{\beta E_{\bf k}(\tau_1)}{2}\right)\right)\nonumber\\
 &&~~~~~~~~~~~~~~~~~~~~~~~~~~~~~~~~~~~~~~~~~~~~~~~~\left[\delta^3\left({\bf k}_1+{\bf k}_3\right)\delta^3\left({\bf k}_2+{\bf k}_4\right)+\delta^3\left({\bf k}_1+{\bf k}_4\right)\delta^3\left({\bf k}_2+{\bf k}_3\right)\right],~~~~~~~~~~~,\\
  &&\int d\Psi_{\bf BD}~\langle\Psi_{\bf BD}|:e^{-\beta \hat{H}(\tau_1)}~a^{\dagger}_{-{\bf k}_1}a_{{\bf k}_2}a^{\dagger}_{-{\bf k}_3}a^{\dagger}_{-{\bf k}_4}:|\Psi_{\bf BD}\rangle=0,\\ 
  &&\int d\Psi_{\bf BD}~\langle\Psi_{\bf BD}|:e^{-\beta \hat{H}(\tau_1)}~a_{{\bf k}_1}a^{\dagger}_{-{\bf k}_2}a^{\dagger}_{-{\bf k}_3}a^{\dagger}_{-{\bf k}_4}:|\Psi_{\bf BD}\rangle=0,\\  
   &&\int d\Psi_{\bf BD}~\langle\Psi_{\bf BD}|:e^{-\beta \hat{H}(\tau_1)}~a^{\dagger}_{-{\bf k}_1}a^{\dagger}_{-{\bf k}_2}a^{\dagger}_{-{\bf k}_3}a^{\dagger}_{-{\bf k}_4}:|\Psi_{\bf BD}\rangle=0.\eea
   Consequently, the individual contributions can be computed as: 
    \bea &&\int d\Psi_{\bf BD}~\langle\Psi_{\bf BD}|\widehat{\cal V}_1({\bf k}_1,{\bf k}_2,{\bf k}_3,{\bf k}_4;\tau_1,\tau_2;\beta)|\Psi_{\bf BD}\rangle\nonumber\\
 &&=(2\pi)^6\exp\left(-\left(1+\frac{1}{2}\delta^{3}(0)\right)\int d^3{\bf k}~\ln\left(2\sinh \frac{\beta E_{\bf k}(\tau_1)}{2}\right)\right)\nonumber\eea
 \bea
 &&~~~~~~~~~~~~\left[ {\cal M}_4({\bf k}_1,{\bf k}_2,{\bf k}_3,{\bf k}_4;\tau_1,\tau_2)~\left\{\delta^3\left({\bf k}_1+{\bf k}_4\right)\delta^3\left({\bf k}_2+{\bf k}_3\right)+\delta^3\left({\bf k}_1+{\bf k}_3\right)\delta^3\left({\bf k}_2+{\bf k}_4\right)\right\}\right.\nonumber\\
  && \left.~~~~~~~~~~~~+ {\cal M}_6({\bf k}_1,{\bf k}_2,{\bf k}_3,{\bf k}_4;\tau_1,\tau_2)~\left\{\delta^3\left({\bf k}_1+{\bf k}_2\right)\delta^3\left({\bf k}_3+{\bf k}_4\right)+\delta^3\left({\bf k}_1+{\bf k}_4\right)\delta^3\left({\bf k}_2+{\bf k}_3\right)\right\}\right.\nonumber\\
  && \left.~~~~~~~~~~~~+{\cal M}_7({\bf k}_1,{\bf k}_2,{\bf k}_3,{\bf k}_4;\tau_1,\tau_2)~\left\{\delta^3\left({\bf k}_1+{\bf k}_2\right)\delta^3\left({\bf k}_3+{\bf k}_4\right)+\delta^3\left({\bf k}_1+{\bf k}_3\right)\delta^3\left({\bf k}_2+{\bf k}_4\right)\right\}\right.\nonumber\\
  && \left.~~~~~~~~~~~~+ {\cal M}_{10}({\bf k}_1,{\bf k}_2,{\bf k}_3,{\bf k}_4;\tau_1,\tau_2)~\left\{\delta^3\left({\bf k}_1+{\bf k}_2\right)\delta^3\left({\bf k}_3+{\bf k}_4\right)+\delta^3\left({\bf k}_1+{\bf k}_3\right)\delta^3\left({\bf k}_2+{\bf k}_4\right)\right\}\right.\nonumber\\
  && \left.~~~~~~~~~~~~+{\cal M}_{11}({\bf k}_1,{\bf k}_2,{\bf k}_3,{\bf k}_4;\tau_1,\tau_2)~\left\{\delta^3\left({\bf k}_1+{\bf k}_2\right)\delta^3\left({\bf k}_3+{\bf k}_4\right)+\delta^3\left({\bf k}_1+{\bf k}_4\right)\delta^3\left({\bf k}_2+{\bf k}_3\right)\right\}\right.\nonumber\\
  && \left.~~~~~~~~~~~~+{\cal M}_{13}({\bf k}_1,{\bf k}_2,{\bf k}_3,{\bf k}_4;\tau_1,\tau_2)~\left\{\delta^3\left({\bf k}_1+{\bf k}_3\right)\delta^3\left({\bf k}_2+{\bf k}_4\right)+\delta^3\left({\bf k}_1+{\bf k}_4\right)\delta^3\left({\bf k}_2+{\bf k}_3\right)\right\}\right],~~~~~~~~~~~\eea\bea
  &&\int d\Psi_{\bf BD}~\langle\Psi_{\bf BD}|\widehat{\cal V}_2({\bf k}_1,{\bf k}_2,{\bf k}_3,{\bf k}_4;\tau_1,\tau_2;\beta)|\Psi_{\bf BD}\rangle\nonumber\\
 &&=(2\pi)^6\exp\left(-\left(1+\frac{1}{2}\delta^{3}(0)\right)\int d^3{\bf k}~\ln\left(2\sinh \frac{\beta E_{\bf k}(\tau_1)}{2}\right)\right)\nonumber\\
 &&~~~~~~~~~~~~\left[ {\cal J}_4({\bf k}_1,{\bf k}_2,{\bf k}_3,{\bf k}_4;\tau_1,\tau_2)~\left\{\delta^3\left({\bf k}_1+{\bf k}_4\right)\delta^3\left({\bf k}_2+{\bf k}_3\right)+\delta^3\left({\bf k}_1+{\bf k}_3\right)\delta^3\left({\bf k}_2+{\bf k}_4\right)\right\}\right.\nonumber\\
  && \left.~~~~~~~~~~~~+ {\cal J}_6({\bf k}_1,{\bf k}_2,{\bf k}_3,{\bf k}_4;\tau_1,\tau_2)~\left\{\delta^3\left({\bf k}_1+{\bf k}_2\right)\delta^3\left({\bf k}_3+{\bf k}_4\right)+\delta^3\left({\bf k}_1+{\bf k}_4\right)\delta^3\left({\bf k}_2+{\bf k}_3\right)\right\}\right.\nonumber\\
  && \left.~~~~~~~~~~~~+{\cal J}_7({\bf k}_1,{\bf k}_2,{\bf k}_3,{\bf k}_4;\tau_1,\tau_2)~\left\{\delta^3\left({\bf k}_1+{\bf k}_2\right)\delta^3\left({\bf k}_3+{\bf k}_4\right)+\delta^3\left({\bf k}_1+{\bf k}_3\right)\delta^3\left({\bf k}_2+{\bf k}_4\right)\right\}\right.\nonumber\\
  && \left.~~~~~~~~~~~~+ {\cal J}_{10}({\bf k}_1,{\bf k}_2,{\bf k}_3,{\bf k}_4;\tau_1,\tau_2)~\left\{\delta^3\left({\bf k}_1+{\bf k}_2\right)\delta^3\left({\bf k}_3+{\bf k}_4\right)+\delta^3\left({\bf k}_1+{\bf k}_3\right)\delta^3\left({\bf k}_2+{\bf k}_4\right)\right\}\right.\nonumber\\
  && \left.~~~~~~~~~~~~+{\cal J}_{11}({\bf k}_1,{\bf k}_2,{\bf k}_3,{\bf k}_4;\tau_1,\tau_2)~\left\{\delta^3\left({\bf k}_1+{\bf k}_2\right)\delta^3\left({\bf k}_3+{\bf k}_4\right)+\delta^3\left({\bf k}_1+{\bf k}_4\right)\delta^3\left({\bf k}_2+{\bf k}_3\right)\right\}\right.\nonumber\\
  && \left.~~~~~~~~~~~~+{\cal J}_{13}({\bf k}_1,{\bf k}_2,{\bf k}_3,{\bf k}_4;\tau_1,\tau_2)~\left\{\delta^3\left({\bf k}_1+{\bf k}_3\right)\delta^3\left({\bf k}_2+{\bf k}_4\right)+\delta^3\left({\bf k}_1+{\bf k}_4\right)\delta^3\left({\bf k}_2+{\bf k}_3\right)\right\}\right],~~~~~~~~~~~\\
  &&\int d\Psi_{\bf BD}~\langle\Psi_{\bf BD}|\widehat{\cal V}_3({\bf k}_1,{\bf k}_2,{\bf k}_3,{\bf k}_4;\tau_1,\tau_2;\beta)|\Psi_{\bf BD}\rangle\nonumber\\
 &&=(2\pi)^6\exp\left(-\left(1+\frac{1}{2}\delta^{3}(0)\right)\int d^3{\bf k}~\ln\left(2\sinh \frac{\beta E_{\bf k}(\tau_1)}{2}\right)\right)\nonumber\\
 &&~~~~~~~~~~~~\left[ {\cal N}_4({\bf k}_1,{\bf k}_2,{\bf k}_3,{\bf k}_4;\tau_1,\tau_2)~\left\{\delta^3\left({\bf k}_1+{\bf k}_4\right)\delta^3\left({\bf k}_2+{\bf k}_3\right)+\delta^3\left({\bf k}_1+{\bf k}_3\right)\delta^3\left({\bf k}_2+{\bf k}_4\right)\right\}\right.\nonumber\\
  && \left.~~~~~~~~~~~~+ {\cal N}_6({\bf k}_1,{\bf k}_2,{\bf k}_3,{\bf k}_4;\tau_1,\tau_2)~\left\{\delta^3\left({\bf k}_1+{\bf k}_2\right)\delta^3\left({\bf k}_3+{\bf k}_4\right)+\delta^3\left({\bf k}_1+{\bf k}_4\right)\delta^3\left({\bf k}_2+{\bf k}_3\right)\right\}\right.\nonumber\\
  && \left.~~~~~~~~~~~~+{\cal N}_7({\bf k}_1,{\bf k}_2,{\bf k}_3,{\bf k}_4;\tau_1,\tau_2)~\left\{\delta^3\left({\bf k}_1+{\bf k}_2\right)\delta^3\left({\bf k}_3+{\bf k}_4\right)+\delta^3\left({\bf k}_1+{\bf k}_3\right)\delta^3\left({\bf k}_2+{\bf k}_4\right)\right\}\right.\nonumber\\
  && \left.~~~~~~~~~~~~+ {\cal N}_{10}({\bf k}_1,{\bf k}_2,{\bf k}_3,{\bf k}_4;\tau_1,\tau_2)~\left\{\delta^3\left({\bf k}_1+{\bf k}_2\right)\delta^3\left({\bf k}_3+{\bf k}_4\right)+\delta^3\left({\bf k}_1+{\bf k}_3\right)\delta^3\left({\bf k}_2+{\bf k}_4\right)\right\}\right.\nonumber\\
  && \left.~~~~~~~~~~~~+{\cal N}_{11}({\bf k}_1,{\bf k}_2,{\bf k}_3,{\bf k}_4;\tau_1,\tau_2)~\left\{\delta^3\left({\bf k}_1+{\bf k}_2\right)\delta^3\left({\bf k}_3+{\bf k}_4\right)+\delta^3\left({\bf k}_1+{\bf k}_4\right)\delta^3\left({\bf k}_2+{\bf k}_3\right)\right\}\right.\nonumber\\
  && \left.~~~~~~~~~~~~+{\cal N}_{13}({\bf k}_1,{\bf k}_2,{\bf k}_3,{\bf k}_4;\tau_1,\tau_2)~\left\{\delta^3\left({\bf k}_1+{\bf k}_3\right)\delta^3\left({\bf k}_2+{\bf k}_4\right)+\delta^3\left({\bf k}_1+{\bf k}_4\right)\delta^3\left({\bf k}_2+{\bf k}_3\right)\right\}\right],~~~~~~~~~~~\\
  &&\int d\Psi_{\bf BD}~\langle\Psi_{\bf BD}|\widehat{\cal V}_4({\bf k}_1,{\bf k}_2,{\bf k}_3,{\bf k}_4;\tau_1,\tau_2;\beta)|\Psi_{\bf BD}\rangle\nonumber\\
 &&=(2\pi)^6\exp\left(-\left(1+\frac{1}{2}\delta^{3}(0)\right)\int d^3{\bf k}~\ln\left(2\sinh \frac{\beta E_{\bf k}(\tau_1)}{2}\right)\right)\nonumber\\
 &&~~~~~~~~~~~~\left[ {\cal Q}_4({\bf k}_1,{\bf k}_2,{\bf k}_3,{\bf k}_4;\tau_1,\tau_2)~\left\{\delta^3\left({\bf k}_1+{\bf k}_4\right)\delta^3\left({\bf k}_2+{\bf k}_3\right)+\delta^3\left({\bf k}_1+{\bf k}_3\right)\delta^3\left({\bf k}_2+{\bf k}_4\right)\right\}\right.\nonumber\\
  && \left.~~~~~~~~~~~~+ {\cal Q}_6({\bf k}_1,{\bf k}_2,{\bf k}_3,{\bf k}_4;\tau_1,\tau_2)~\left\{\delta^3\left({\bf k}_1+{\bf k}_2\right)\delta^3\left({\bf k}_3+{\bf k}_4\right)+\delta^3\left({\bf k}_1+{\bf k}_4\right)\delta^3\left({\bf k}_2+{\bf k}_3\right)\right\}\right.\nonumber\\
  && \left.~~~~~~~~~~~~+{\cal Q}_7({\bf k}_1,{\bf k}_2,{\bf k}_3,{\bf k}_4;\tau_1,\tau_2)~\left\{\delta^3\left({\bf k}_1+{\bf k}_2\right)\delta^3\left({\bf k}_3+{\bf k}_4\right)+\delta^3\left({\bf k}_1+{\bf k}_3\right)\delta^3\left({\bf k}_2+{\bf k}_4\right)\right\}\right.\nonumber\\
  && \left.~~~~~~~~~~~~+ {\cal Q}_{10}({\bf k}_1,{\bf k}_2,{\bf k}_3,{\bf k}_4;\tau_1,\tau_2)~\left\{\delta^3\left({\bf k}_1+{\bf k}_2\right)\delta^3\left({\bf k}_3+{\bf k}_4\right)+\delta^3\left({\bf k}_1+{\bf k}_3\right)\delta^3\left({\bf k}_2+{\bf k}_4\right)\right\}\right.\nonumber\\
  && \left.~~~~~~~~~~~~+{\cal Q}_{11}({\bf k}_1,{\bf k}_2,{\bf k}_3,{\bf k}_4;\tau_1,\tau_2)~\left\{\delta^3\left({\bf k}_1+{\bf k}_2\right)\delta^3\left({\bf k}_3+{\bf k}_4\right)+\delta^3\left({\bf k}_1+{\bf k}_4\right)\delta^3\left({\bf k}_2+{\bf k}_3\right)\right\}\right.\nonumber\\
  && \left.~~~~~~~~~~~~+{\cal Q}_{13}({\bf k}_1,{\bf k}_2,{\bf k}_3,{\bf k}_4;\tau_1,\tau_2)~\left\{\delta^3\left({\bf k}_1+{\bf k}_3\right)\delta^3\left({\bf k}_2+{\bf k}_4\right)+\delta^3\left({\bf k}_1+{\bf k}_4\right)\delta^3\left({\bf k}_2+{\bf k}_3\right)\right\}\right],~~~~~~~~~~~
  \eea
  Consequently, the individual contributions can be computed in the normal ordered form as:
    \bea &&\int d\Psi_{\bf BD}~\langle\Psi_{\bf BD}|:\widehat{\cal V}_1({\bf k}_1,{\bf k}_2,{\bf k}_3,{\bf k}_4;\tau_1,\tau_2;\beta):|\Psi_{\bf BD}\rangle\nonumber\\
 &&=(2\pi)^6\exp\left(-\int d^3{\bf k}~\ln\left(2\sinh \frac{\beta E_{\bf k}(\tau_1)}{2}\right)\right)\nonumber\eea\bea
 &&~~~~~~~~~~~~\left[ {\cal M}_4({\bf k}_1,{\bf k}_2,{\bf k}_3,{\bf k}_4;\tau_1,\tau_2)~\left\{\delta^3\left({\bf k}_1+{\bf k}_4\right)\delta^3\left({\bf k}_2+{\bf k}_3\right)+\delta^3\left({\bf k}_1+{\bf k}_3\right)\delta^3\left({\bf k}_2+{\bf k}_4\right)\right\}\right.\nonumber\\
  && \left.~~~~~~~~~~~~+ {\cal M}_6({\bf k}_1,{\bf k}_2,{\bf k}_3,{\bf k}_4;\tau_1,\tau_2)~\left\{\delta^3\left({\bf k}_1+{\bf k}_2\right)\delta^3\left({\bf k}_3+{\bf k}_4\right)+\delta^3\left({\bf k}_1+{\bf k}_4\right)\delta^3\left({\bf k}_2+{\bf k}_3\right)\right\}\right.\nonumber\\
  && \left.~~~~~~~~~~~~+{\cal M}_7({\bf k}_1,{\bf k}_2,{\bf k}_3,{\bf k}_4;\tau_1,\tau_2)~\left\{\delta^3\left({\bf k}_1+{\bf k}_2\right)\delta^3\left({\bf k}_3+{\bf k}_4\right)+\delta^3\left({\bf k}_1+{\bf k}_3\right)\delta^3\left({\bf k}_2+{\bf k}_4\right)\right\}\right.\nonumber\\
  && \left.~~~~~~~~~~~~+ {\cal M}_{10}({\bf k}_1,{\bf k}_2,{\bf k}_3,{\bf k}_4;\tau_1,\tau_2)~\left\{\delta^3\left({\bf k}_1+{\bf k}_2\right)\delta^3\left({\bf k}_3+{\bf k}_4\right)+\delta^3\left({\bf k}_1+{\bf k}_3\right)\delta^3\left({\bf k}_2+{\bf k}_4\right)\right\}\right.\nonumber\\
  && \left.~~~~~~~~~~~~+{\cal M}_{11}({\bf k}_1,{\bf k}_2,{\bf k}_3,{\bf k}_4;\tau_1,\tau_2)~\left\{\delta^3\left({\bf k}_1+{\bf k}_2\right)\delta^3\left({\bf k}_3+{\bf k}_4\right)+\delta^3\left({\bf k}_1+{\bf k}_4\right)\delta^3\left({\bf k}_2+{\bf k}_3\right)\right\}\right.\nonumber\\
  && \left.~~~~~~~~~~~~+{\cal M}_{13}({\bf k}_1,{\bf k}_2,{\bf k}_3,{\bf k}_4;\tau_1,\tau_2)~\left\{\delta^3\left({\bf k}_1+{\bf k}_3\right)\delta^3\left({\bf k}_2+{\bf k}_4\right)+\delta^3\left({\bf k}_1+{\bf k}_4\right)\delta^3\left({\bf k}_2+{\bf k}_3\right)\right\}\right],~~~~~~~~~~~\\
  &&\int d\Psi_{\bf BD}~\langle\Psi_{\bf BD}|:\widehat{\cal V}_2({\bf k}_1,{\bf k}_2,{\bf k}_3,{\bf k}_4;\tau_1,\tau_2;\beta):|\Psi_{\bf BD}\rangle\nonumber\\
 &&=(2\pi)^6\exp\left(-\int d^3{\bf k}~\ln\left(2\sinh \frac{\beta E_{\bf k}(\tau_1)}{2}\right)\right)\nonumber\\
 &&~~~~~~~~~~~~\left[ {\cal J}_4({\bf k}_1,{\bf k}_2,{\bf k}_3,{\bf k}_4;\tau_1,\tau_2)~\left\{\delta^3\left({\bf k}_1+{\bf k}_4\right)\delta^3\left({\bf k}_2+{\bf k}_3\right)+\delta^3\left({\bf k}_1+{\bf k}_3\right)\delta^3\left({\bf k}_2+{\bf k}_4\right)\right\}\right.\nonumber\\
  && \left.~~~~~~~~~~~~+ {\cal J}_6({\bf k}_1,{\bf k}_2,{\bf k}_3,{\bf k}_4;\tau_1,\tau_2)~\left\{\delta^3\left({\bf k}_1+{\bf k}_2\right)\delta^3\left({\bf k}_3+{\bf k}_4\right)+\delta^3\left({\bf k}_1+{\bf k}_4\right)\delta^3\left({\bf k}_2+{\bf k}_3\right)\right\}\right.\nonumber\\
  && \left.~~~~~~~~~~~~+{\cal J}_7({\bf k}_1,{\bf k}_2,{\bf k}_3,{\bf k}_4;\tau_1,\tau_2)~\left\{\delta^3\left({\bf k}_1+{\bf k}_2\right)\delta^3\left({\bf k}_3+{\bf k}_4\right)+\delta^3\left({\bf k}_1+{\bf k}_3\right)\delta^3\left({\bf k}_2+{\bf k}_4\right)\right\}\right.\nonumber\\
  && \left.~~~~~~~~~~~~+ {\cal J}_{10}({\bf k}_1,{\bf k}_2,{\bf k}_3,{\bf k}_4;\tau_1,\tau_2)~\left\{\delta^3\left({\bf k}_1+{\bf k}_2\right)\delta^3\left({\bf k}_3+{\bf k}_4\right)+\delta^3\left({\bf k}_1+{\bf k}_3\right)\delta^3\left({\bf k}_2+{\bf k}_4\right)\right\}\right.\nonumber\\
  && \left.~~~~~~~~~~~~+{\cal J}_{11}({\bf k}_1,{\bf k}_2,{\bf k}_3,{\bf k}_4;\tau_1,\tau_2)~\left\{\delta^3\left({\bf k}_1+{\bf k}_2\right)\delta^3\left({\bf k}_3+{\bf k}_4\right)+\delta^3\left({\bf k}_1+{\bf k}_4\right)\delta^3\left({\bf k}_2+{\bf k}_3\right)\right\}\right.\nonumber\\
  && \left.~~~~~~~~~~~~+{\cal J}_{13}({\bf k}_1,{\bf k}_2,{\bf k}_3,{\bf k}_4;\tau_1,\tau_2)~\left\{\delta^3\left({\bf k}_1+{\bf k}_3\right)\delta^3\left({\bf k}_2+{\bf k}_4\right)+\delta^3\left({\bf k}_1+{\bf k}_4\right)\delta^3\left({\bf k}_2+{\bf k}_3\right)\right\}\right],~~~~~~~~~~~\\
  &&\int d\Psi_{\bf BD}~\langle\Psi_{\bf BD}|:\widehat{\cal V}_3({\bf k}_1,{\bf k}_2,{\bf k}_3,{\bf k}_4;\tau_1,\tau_2;\beta):|\Psi_{\bf BD}\rangle\nonumber\\
 &&=(2\pi)^6\exp\left(-\int d^3{\bf k}~\ln\left(2\sinh \frac{\beta E_{\bf k}(\tau_1)}{2}\right)\right)\nonumber\\
 &&~~~~~~~~~~~~\left[ {\cal N}_4({\bf k}_1,{\bf k}_2,{\bf k}_3,{\bf k}_4;\tau_1,\tau_2)~\left\{\delta^3\left({\bf k}_1+{\bf k}_4\right)\delta^3\left({\bf k}_2+{\bf k}_3\right)+\delta^3\left({\bf k}_1+{\bf k}_3\right)\delta^3\left({\bf k}_2+{\bf k}_4\right)\right\}\right.\nonumber\\
  && \left.~~~~~~~~~~~~+ {\cal N}_6({\bf k}_1,{\bf k}_2,{\bf k}_3,{\bf k}_4;\tau_1,\tau_2)~\left\{\delta^3\left({\bf k}_1+{\bf k}_2\right)\delta^3\left({\bf k}_3+{\bf k}_4\right)+\delta^3\left({\bf k}_1+{\bf k}_4\right)\delta^3\left({\bf k}_2+{\bf k}_3\right)\right\}\right.\nonumber\\
  && \left.~~~~~~~~~~~~+{\cal N}_7({\bf k}_1,{\bf k}_2,{\bf k}_3,{\bf k}_4;\tau_1,\tau_2)~\left\{\delta^3\left({\bf k}_1+{\bf k}_2\right)\delta^3\left({\bf k}_3+{\bf k}_4\right)+\delta^3\left({\bf k}_1+{\bf k}_3\right)\delta^3\left({\bf k}_2+{\bf k}_4\right)\right\}\right.\nonumber\\
  && \left.~~~~~~~~~~~~+ {\cal N}_{10}({\bf k}_1,{\bf k}_2,{\bf k}_3,{\bf k}_4;\tau_1,\tau_2)~\left\{\delta^3\left({\bf k}_1+{\bf k}_2\right)\delta^3\left({\bf k}_3+{\bf k}_4\right)+\delta^3\left({\bf k}_1+{\bf k}_3\right)\delta^3\left({\bf k}_2+{\bf k}_4\right)\right\}\right.\nonumber\\
  && \left.~~~~~~~~~~~~+{\cal N}_{11}({\bf k}_1,{\bf k}_2,{\bf k}_3,{\bf k}_4;\tau_1,\tau_2)~\left\{\delta^3\left({\bf k}_1+{\bf k}_2\right)\delta^3\left({\bf k}_3+{\bf k}_4\right)+\delta^3\left({\bf k}_1+{\bf k}_4\right)\delta^3\left({\bf k}_2+{\bf k}_3\right)\right\}\right.\nonumber\\
  && \left.~~~~~~~~~~~~+{\cal N}_{13}({\bf k}_1,{\bf k}_2,{\bf k}_3,{\bf k}_4;\tau_1,\tau_2)~\left\{\delta^3\left({\bf k}_1+{\bf k}_3\right)\delta^3\left({\bf k}_2+{\bf k}_4\right)+\delta^3\left({\bf k}_1+{\bf k}_4\right)\delta^3\left({\bf k}_2+{\bf k}_3\right)\right\}\right],~~~~~~~~~~~\\
  &&\int d\Psi_{\bf BD}~\langle\Psi_{\bf BD}|:\widehat{\cal V}_4({\bf k}_1,{\bf k}_2,{\bf k}_3,{\bf k}_4;\tau_1,\tau_2;\beta):|\Psi_{\bf BD}\rangle\nonumber\\
 &&=(2\pi)^6\exp\left(-\int d^3{\bf k}~\ln\left(2\sinh \frac{\beta E_{\bf k}(\tau_1)}{2}\right)\right)\nonumber\\
 &&~~~~~~~~~~~~\left[ {\cal Q}_4({\bf k}_1,{\bf k}_2,{\bf k}_3,{\bf k}_4;\tau_1,\tau_2)~\left\{\delta^3\left({\bf k}_1+{\bf k}_4\right)\delta^3\left({\bf k}_2+{\bf k}_3\right)+\delta^3\left({\bf k}_1+{\bf k}_3\right)\delta^3\left({\bf k}_2+{\bf k}_4\right)\right\}\right.\nonumber\\
  && \left.~~~~~~~~~~~~+ {\cal Q}_6({\bf k}_1,{\bf k}_2,{\bf k}_3,{\bf k}_4;\tau_1,\tau_2)~\left\{\delta^3\left({\bf k}_1+{\bf k}_2\right)\delta^3\left({\bf k}_3+{\bf k}_4\right)+\delta^3\left({\bf k}_1+{\bf k}_4\right)\delta^3\left({\bf k}_2+{\bf k}_3\right)\right\}\right.\nonumber\\
  && \left.~~~~~~~~~~~~+{\cal Q}_7({\bf k}_1,{\bf k}_2,{\bf k}_3,{\bf k}_4;\tau_1,\tau_2)~\left\{\delta^3\left({\bf k}_1+{\bf k}_2\right)\delta^3\left({\bf k}_3+{\bf k}_4\right)+\delta^3\left({\bf k}_1+{\bf k}_3\right)\delta^3\left({\bf k}_2+{\bf k}_4\right)\right\}\right.\nonumber\\
  && \left.~~~~~~~~~~~~+ {\cal Q}_{10}({\bf k}_1,{\bf k}_2,{\bf k}_3,{\bf k}_4;\tau_1,\tau_2)~\left\{\delta^3\left({\bf k}_1+{\bf k}_2\right)\delta^3\left({\bf k}_3+{\bf k}_4\right)+\delta^3\left({\bf k}_1+{\bf k}_3\right)\delta^3\left({\bf k}_2+{\bf k}_4\right)\right\}\right.\nonumber\\
  && \left.~~~~~~~~~~~~+{\cal Q}_{11}({\bf k}_1,{\bf k}_2,{\bf k}_3,{\bf k}_4;\tau_1,\tau_2)~\left\{\delta^3\left({\bf k}_1+{\bf k}_2\right)\delta^3\left({\bf k}_3+{\bf k}_4\right)+\delta^3\left({\bf k}_1+{\bf k}_4\right)\delta^3\left({\bf k}_2+{\bf k}_3\right)\right\}\right.\nonumber\\
  && \left.~~~~~~~~~~~~+{\cal Q}_{13}({\bf k}_1,{\bf k}_2,{\bf k}_3,{\bf k}_4;\tau_1,\tau_2)~\left\{\delta^3\left({\bf k}_1+{\bf k}_3\right)\delta^3\left({\bf k}_2+{\bf k}_4\right)+\delta^3\left({\bf k}_1+{\bf k}_4\right)\delta^3\left({\bf k}_2+{\bf k}_3\right)\right\}\right],~~~~~~~~~~~
  \eea
\section{Time dependent two-point amplitude in micro-canonical  OTOC}
We define the following momentum integrated time dependent amplitude:
\bea {\cal B}(T,\tau):&=&\int^{L}_{k_1=0}~k^2_1~dk_1~{\cal P}({\bf k}_1,-{\bf k}_1;T,\tau)\nonumber\\
&=&\int^{L}_{k_1=0}~k^2_1~dk_1~\left[ f^{*}_{{\bf -k}_1}(T)\Pi_{-{\bf k}_1}(\tau)+f_{{\bf k}_1}(T)\Pi^{*}_{{\bf k}_1}(\tau)\right.\nonumber\\
  && \left.~~~~~~~~~~~~~~~~~~~~~~~~~~~~~~~~~~~~~~-\Pi^{*}_{{\bf -k}_1}(\tau)f_{-{\bf k}_1}(T)- \Pi_{{\bf k}_1}(\tau)f^{*}_{{\bf k}_1}(T)\right] ~~~~~~~~~\nonumber\\
  &=&(-T)^{\frac{1}{2}-\nu}(-\tau)^{\frac{3}{2}-\nu}\left[Z_{(1)}(\tau_1,\tau_2)+Z_{(2)}(\tau_1,\tau_2)-Z_{(3)}(\tau_1,\tau_2)-Z_{(4)}(\tau_1,\tau_2)\right],~~~~~~~~\eea
  where we have introduced the time dependent four individual amplitudes, $Z_{(i)}(T,\tau)~\forall~i=1,2,3,4$:
  \bea && Z_{(1)}(T,\tau):=\int^{L}_{k_1=0}~k^2_1~dk_1~f_{{\bf k}_1}(T)\Pi^{*}_{{\bf k}_1}(\tau),\\
   && Z_{(2)}(T,\tau)=\int^{L}_{k_1=0}~k^2_1~dk_1~ f^{*}_{{\bf -k}_1}(T)\Pi_{-{\bf k}_1}(\tau),\\
   && Z_{(3)}(T,\tau):=\int^{L}_{k_1=0}~k^2_1~dk_1~ \Pi_{{\bf k}_1}(\tau)f^{*}_{{\bf k}_1}(T),\\
   && Z_{(4)}(T,\tau):=\int^{L}_{k_1=0}~k^2_1~dk_1~ \Pi^{*}_{{\bf -k}_1}(\tau)f_{-{\bf k}_1}(T),\eea
  which we are going to explicitly evaluate in this Appendix.
  
  Now before going to evaluate the individual contributions from the symmetry properties of the momentum dependent amplitudes we have derived the following results:
\bea && Z_{(2)}(T,\tau)=(-1)^{-(2\nu+1)}Z_{(1)}(T,\tau),\\
&&Z_{(4)}(T,\tau)=(-1)^{-(2\nu+1)}Z_{(3)}(T,\tau),\eea
using which the simplified form of the momentum integrated time dependent amplitude can be written as:
\bea {\cal B}(T,\tau):&=&(-T)^{\frac{1}{2}-\nu}(-\tau)^{\frac{3}{2}-\nu}\left[1+(-1)^{-(2\nu+1)}\right]\left(Z_{(1)}(T,\tau)-Z_{(3)}(T,\tau)\right).\eea
Consequently, the two-point OTOC can be computed as:
\bea Y^{f}(T,\tau)=-\frac{1}{2\pi^2} {\cal B}(T,\tau)=(-T)^{\frac{1}{2}-\nu}(-\tau)^{\frac{3}{2}-\nu}\left[1+(-1)^{-(2\nu+1)}\right]\left(Z_{(3)}(T,\tau)-Z_{(1)}(T,\tau)\right).~~~~~~~\eea
The expressions for $Z_{(1)}(T,\tau)+Z_{(3)}(T,\tau)$ is given by the following expressions:
\bea &&Z_{(1)}(T,\tau)+Z_{(3)}(T,\tau)=\frac{\left(A^2-B^2\right) L^{-2 \nu } }{2 \tau ^2 (\tau -T)^3}(T^3 \Gamma (-2 \nu ) (-i L (T-\tau ))^{2 \nu }\nonumber \\ && +2 \nu  \tau ^3 \Gamma (-2 \nu ) (-i L (T-\tau ))^{2 \nu }-\tau ^3 \Gamma (-2 \nu ) (-i L (T-\tau ))^{2 \nu }+3 T \tau ^2 \Gamma (-2 \nu ) (-i L (T-\tau ))^{2 \nu }\nonumber\\
&& -6 T \nu  \tau ^2 \Gamma (-2 \nu ) (-i L (T-\tau ))^{2 \nu }-2 T^3 \nu  \Gamma (-2 \nu ) (-i L (T-\tau ))^{2 \nu }-3 T^2 \tau  \Gamma (-2 \nu ) (-i L (T-\tau ))^{2 \nu }\nonumber\\
&&+6 T^2 \nu  \tau  \Gamma (-2 \nu ) (-i L (T-\tau ))^{2 \nu }-2 T \tau ^2 \Gamma (3-2 \nu ) (-i L (T-\tau ))^{2 \nu }\nonumber\\
&&-T^3 \Gamma (-2 \nu ,-i L (T-\tau )) (-i L (T-\tau ))^{2 \nu }-2 \nu  \tau ^3 \Gamma (-2 \nu ,-i L (T-\tau )) (-i L (T-\tau ))^{2 \nu }\nonumber\\
&&+\tau ^3 \Gamma (-2 \nu ,-i L (T-\tau )) (-i L (T-\tau ))^{2 \nu }-3 T \tau ^2 \Gamma (-2 \nu ,-i L (T-\tau )) (-i L (T-\tau ))^{2 \nu }\nonumber\\
&&+6 T \nu  \tau ^2 \Gamma (-2 \nu ,-i L (T-\tau )) (-i L (T-\tau ))^{2 \nu }+2 T^3 \nu  \Gamma (-2 \nu ,-i L (T-\tau )) (-i L (T-\tau ))^{2 \nu }\nonumber\\
&&+3 T^2 \tau  \Gamma (-2 \nu ,-i L (T-\tau )) (-i L (T-\tau ))^{2 \nu }-6 T^2 \nu  \tau  \Gamma (-2 \nu ,-i L (T-\tau )) (-i L (T-\tau ))^{2 \nu }\nonumber\\
&&-T^3 \Gamma (1-2 \nu ,-i L (T-\tau )) (-i L (T-\tau ))^{2 \nu }-2 \nu  \tau ^3 \Gamma (1-2 \nu ,-i L (T-\tau )) (-i L (T-\tau ))^{2 \nu }\nonumber\\
&&+\tau ^3 \Gamma (1-2 \nu ,-i L (T-\tau )) (-i L (T-\tau ))^{2 \nu }-3 T \tau ^2 \Gamma (1-2 \nu ,-i L (T-\tau )) (-i L (T-\tau ))^{2 \nu }\nonumber\\
&&+6 T \nu  \tau ^2 \Gamma (1-2 \nu ,-i L (T-\tau )) (-i L (T-\tau ))^{2 \nu }+2 T^3 \nu  \Gamma (1-2 \nu ,-i L (T-\tau )) (-i L (T-\tau ))^{2 \nu }\nonumber\\
&&+3 T^2 \tau  \Gamma (1-2 \nu ,-i L (T-\tau )) (-i L (T-\tau ))^{2 \nu }-6 T^2 \nu  \tau  \Gamma (1-2 \nu ,-i L (T-\tau )) (-i L (T-\tau ))^{2 \nu }\nonumber\\
&&-2 \tau ^3 \Gamma (2-2 \nu ,-i L (T-\tau )) (-i L (T-\tau ))^{2 \nu }+T \tau ^2 \Gamma (2-2 \nu ,-i L (T-\tau )) (-i L (T-\tau ))^{2 \nu }\nonumber\\
&&+2 T \nu  \tau ^2 \Gamma (2-2 \nu ,-i L (T-\tau )) (-i L (T-\tau ))^{2 \nu }+T^2 \tau  \Gamma (2-2 \nu ,-i L (T-\tau )) (-i L (T-\tau ))^{2 \nu }\nonumber\\
&&-2 T^2 \nu  \tau  \Gamma (2-2 \nu ,-i L (T-\tau )) (-i L (T-\tau ))^{2 \nu }+2 T \tau ^2 \Gamma (3-2 \nu ,-i L (T-\tau )) (-i L (T-\tau ))^{2 \nu }\nonumber\\
&&-T^3 (i L (T-\tau ))^{2 \nu } \Gamma (-2 \nu )+2 T^3 \nu  (i L (T-\tau ))^{2 \nu } \Gamma (-2 \nu )\nonumber\\
&&-2 \nu  (i L (T-\tau ))^{2 \nu } \tau ^3 \Gamma (-2 \nu )+(i L (T-\tau ))^{2 \nu } \tau ^3 \Gamma (-2 \nu )-3 T (i L (T-\tau ))^{2 \nu } \tau ^2 \Gamma (-2 \nu )\nonumber\\
&&+6 T \nu  (i L (T-\tau ))^{2 \nu } \tau ^2 \Gamma (-2 \nu )+3 T^2 (i L (T-\tau ))^{2 \nu } \tau  \Gamma (-2 \nu )\nonumber\\
&&-6 T^2 \nu  (i L (T-\tau ))^{2 \nu } \tau  \Gamma (-2 \nu )+(2 \nu -1) \left((i L (T-\tau ))^{2 \nu }-(-i L (T-\tau ))^{2 \nu }\right) (T-\tau )^3 \Gamma (1-2 \nu )\nonumber\\
&&+\left((-i L (T-\tau ))^{2 \nu }-(i L (T-\tau ))^{2 \nu }\right) (T-\tau ) \tau  (2 \nu  T-T-2 \tau ) \Gamma (2-2 \nu )\nonumber\\
&&+2 T (i L (T-\tau ))^{2 \nu } \tau ^2 \Gamma (3-2 \nu )+T^3 (i L (T-\tau ))^{2 \nu } \Gamma (-2 \nu ,i L (T-\tau ))\nonumber\\
&&-2 T^3 \nu  (i L (T-\tau ))^{2 \nu } \Gamma (-2 \nu ,i L (T-\tau ))+2 \nu  (i L (T-\tau ))^{2 \nu } \tau ^3 \Gamma (-2 \nu ,i L (T-\tau ))\nonumber\\
&&-(i L (T-\tau ))^{2 \nu } \tau ^3 \Gamma (-2 \nu ,i L (T-\tau ))+3 T (i L (T-\tau ))^{2 \nu } \tau ^2 \Gamma (-2 \nu ,i L (T-\tau ))\nonumber\\
&&-6 T \nu  (i L (T-\tau ))^{2 \nu } \tau ^2 \Gamma (-2 \nu ,i L (T-\tau ))-3 T^2 (i L (T-\tau ))^{2 \nu } \tau  \Gamma (-2 \nu ,i L (T-\tau ))\nonumber\\
&&+6 T^2 \nu  (i L (T-\tau ))^{2 \nu } \tau  \Gamma (-2 \nu ,i L (T-\tau ))+T^3 (i L (T-\tau ))^{2 \nu } \Gamma (1-2 \nu ,i L (T-\tau ))\nonumber\\
&&-2 T^3 \nu  (i L (T-\tau ))^{2 \nu } \Gamma (1-2 \nu ,i L (T-\tau ))+2 \nu  (i L (T-\tau ))^{2 \nu } \tau ^3 \Gamma (1-2 \nu ,i L (T-\tau ))\nonumber\\
&&-(i L (T-\tau ))^{2 \nu } \tau ^3 \Gamma (1-2 \nu ,i L (T-\tau ))+3 T (i L (T-\tau ))^{2 \nu } \tau ^2 \Gamma (1-2 \nu ,i L (T-\tau ))\nonumber\\
&&-6 T \nu  (i L (T-\tau ))^{2 \nu } \tau ^2 \Gamma (1-2 \nu ,i L (T-\tau ))-3 T^2 (i L (T-\tau ))^{2 \nu } \tau  \Gamma (1-2 \nu ,i L (T-\tau ))\nonumber\\
&&+6 T^2 \nu  (i L (T-\tau ))^{2 \nu } \tau  \Gamma (1-2 \nu ,i L (T-\tau ))+2 (i L (T-\tau ))^{2 \nu } \tau ^3 \Gamma (2-2 \nu ,i L (T-\tau ))\nonumber\\
&&-T (i L (T-\tau ))^{2 \nu } \tau ^2 \Gamma (2-2 \nu ,i L (T-\tau ))-2 T \nu  (i L (T-\tau ))^{2 \nu } \tau ^2 \Gamma (2-2 \nu ,i L (T-\tau ))\nonumber\\
&&-T^2 (i L (T-\tau ))^{2 \nu } \tau  \Gamma (2-2 \nu ,i L (T-\tau ))+2 T^2 \nu  (i L (T-\tau ))^{2 \nu } \tau  \Gamma (2-2 \nu ,i L (T-\tau ))\nonumber\\
&&-2 T (i L (T-\tau ))^{2 \nu } \tau ^2 \Gamma (3-2 \nu ,i L (T-\tau ))).\eea
  \newpage	 
\section{Time dependent four-point amplitudes in micro-canonical OTOC }
We define the following momenta integrated time dependent amplitudes:
  \bea && {\cal I}_1(\tau_1,\tau_2):=\int^{L}_{k_1=0} k^2_1~dk_1\int^{L}_{k_2=0} k^2_2~dk_2~2{\cal E}_4({\bf k}_1,{\bf k}_2,-{\bf k}_2,-{\bf k}_1;\tau_1,\tau_2),\\
  && {\cal I}_2(\tau_1,\tau_2):=\int^{L}_{k_1=0} k^2_1~dk_1\int^{L}_{k_2=0} k^2_2~dk_2~2{\cal E}_{13}({\bf k}_1,{\bf k}_2,-{\bf k}_2,-{\bf k}_1;\tau_1,\tau_2),\\
  && {\cal I}_3(\tau_1,\tau_2):=\int^{L}_{k_1=0} k^2_1~dk_1\int^{L}_{k_2=0} k^2_2~dk_2~{\cal E}_6({\bf k}_1,{\bf k}_2,-{\bf k}_2,-{\bf k}_1;\tau_1,\tau_2),\\
  && {\cal I}_4(\tau_1,\tau_2):=\int^{L}_{k_1=0} k^2_1~dk_1\int^{L}_{k_2=0} k^2_2~dk_2~{\cal E}_7({\bf k}_1,{\bf k}_2,-{\bf k}_1,-{\bf k}_2;\tau_1,\tau_2),\\
 &&  {\cal I}_5(\tau_1,\tau_2):=\int^{L}_{k_1=0} k^2_1~dk_1\int^{L}_{k_2=0} k^2_2~dk_2~{\cal E}_{10}({\bf k}_1,{\bf k}_2,-{\bf k}_1,-{\bf k}_2;\tau_1,\tau_2),\\
 && {\cal I}_6(\tau_1,\tau_2):=\int^{L}_{k_1=0} k^2_1~dk_1\int^{L}_{k_2=0} k^2_2~dk_2~{\cal E}_{11}({\bf k}_1,{\bf k}_2,-{\bf k}_2,-{\bf k}_1;\tau_1,\tau_2),\\
 && {\cal I}_7(\tau_1,\tau_2):=\int^{L}_{k_1=0} k^2_1~dk_1\int^{L}_{k_2=0} k^2_2~dk_2~{\cal E}_7({\bf k}_1,-{\bf k}_1,{\bf k}_2,-{\bf k}_2;\tau_1,\tau_2),\\
 && {\cal I}_8(\tau_1,\tau_2):=\int^{L}_{k_1=0} k^2_1~dk_1\int^{L}_{k_2=0} k^2_2~dk_2~{\cal E}_{10}({\bf k}_1,-{\bf k}_1,{\bf k}_2,-{\bf k}_2;\tau_1,\tau_2),\\
 && {\cal I}_9(\tau_1,\tau_2):=\int^{L}_{k_1=0} k^2_1~dk_1\int^{L}_{k_2=0} k^2_2~dk_2~{\cal E}_{10}({\bf k}_1,-{\bf k}_1,{\bf k}_2,-{\bf k}_2;\tau_1,\tau_2).
  \eea 
 From the symmetry properties of the momentum dependent amplitudes we have derived the following results:
  \bea  
  {\cal I}_2(\tau_1,\tau_2)&&=(-1)^{4\nu}{\cal I}_1(\tau_1,\tau_2)~~~~{\rm with~ weight}~~w_2=2,~~~\\
   {\cal I}_3(\tau_1,\tau_2)&&=(-1)^{2\nu}{\cal I}_1(\tau_1,\tau_2)~~~~{\rm with~ weight}~~w_3=1,~~~\\
    {\cal I}_4(\tau_1,\tau_2)&&=(-1)^{2\nu}{\cal I}_1(\tau_1,\tau_2)~~~~{\rm with~ weight}~~w_4=1,~~~\\
     {\cal I}_5(\tau_1,\tau_2)&&=(-1)^{2\nu}{\cal I}_1(\tau_1,\tau_2)~~~~{\rm with~ weight}~~w_5=1,~~~\\
      {\cal I}_6(\tau_1,\tau_2)&&=(-1)^{2\nu}{\cal I}_1(\tau_1,\tau_2)~~~~{\rm with~ weight}~~w_6=1,~~~\\
       {\cal I}_7(\tau_1,\tau_2)&&=(-1)^{2\nu}{\cal I}_1(\tau_1,\tau_2)~~~~{\rm with~ weight}~~w_7=1,~~~\\
        {\cal I}_8(\tau_1,\tau_2)&&=(-1)^{2\nu}{\cal I}_1(\tau_1,\tau_2)~~~~{\rm with~ weight}~~w_8=1,~~~\\
         {\cal I}_9(\tau_1,\tau_2)&&=(-1)^{2\nu}{\cal I}_1(\tau_1,\tau_2)~~~~{\rm with~ weight}~~w_9=1.~~~\eea
         The details of the all of these regularised four-point integral computations we have given in the following subsections. These computations are useful to construct the final expression for the cosmological OTOC. Also the detailed structure of these integrals in general, as well as super horizon and sub horizon limiting regions are useful to physically understand the role of these integrals to fix the out of equilibrium behaviour of quantum mechanical correlations in the context of Cosmology.
\subsection{Computation of ${\cal I}_{1}(\tau_1,\tau_2)$}
\bea  \hll{{\cal I}_1(T,\tau)=\int^{L}_{k_1=0} k^2_1dk_1\int^{L}_{k_2=0} k^2_2dk_2{\cal E}_4({\bf k}_1,{\bf k}_2,-{\bf k}_2,-{\bf k}_1;T,\tau)=\frac{(-T)^{1-2\nu}(-\tau)^{3-2\nu}}{(-1)^{4\nu}}\sum^{4}_{i=1}X^{(i)}_{1}(T,\tau)},~~~~~~~~\eea 
where we define four time dependent functions, $X^{(i)}_{1}(T,\tau)~\forall~~i=1,2,3,4$, which are given by the following expressions:
\bea X^{(1)}_1&=&\frac{L^{-4 \nu } }{4 \tau ^4}(B^2 (\Gamma (-2 \nu )-\Gamma (-2 \nu ,-i L (T-\tau ))) (-i L (T-\tau ))^{2 \nu }\nonumber\\
&&-2 B^2 \nu  (\Gamma (-2 \nu )-\Gamma (-2 \nu ,-i L (T-\tau ))) (-i L (T-\tau ))^{2 \nu }\nonumber\\
&&+\frac{B^2 \tau  (\Gamma (1-2 \nu )-\Gamma (1-2 \nu ,-i L (T-\tau ))) (-i L (T-\tau ))^{2 \nu }}{\tau -T}\nonumber\\
&&+\frac{2 B^2 T \nu  \tau  (\Gamma (2-2 \nu )-\Gamma (2-2 \nu ,-i L (T-\tau ))) (-i L (T-\tau ))^{2 \nu }}{(T-\tau )^2}\nonumber\\
&&-\frac{2 B^2 \tau ^2 (\Gamma (2-2 \nu )-\Gamma (2-2 \nu ,-i L (T-\tau ))) (-i L (T-\tau ))^{2 \nu }}{(T-\tau )^2}\nonumber\\
&&-\frac{B^2 T \tau  (\Gamma (2-2 \nu )-\Gamma (2-2 \nu ,-i L (T-\tau ))) (-i L (T-\tau ))^{2 \nu }}{(T-\tau )^2}\nonumber\\
&&-\frac{2 B^2 T \tau ^2 (\Gamma (3-2 \nu )-\Gamma (3-2 \nu ,-i L (T-\tau ))) (-i L (T-\tau ))^{2 \nu }}{(T-\tau )^3}\nonumber\\
&&-\frac{2 B^2 \nu  \tau  (\Gamma (1-2 \nu )-\Gamma (1-2 \nu ,-i L (T-\tau ))) (-i L (T-\tau ))^{2 \nu }}{\tau -T}\nonumber\\
&&+B^2 (-i) L T (\Gamma (1-2 \nu )-\Gamma (1-2 \nu ,-i L (T-\tau ))) (-i L (T-\tau ))^{2 \nu -1}\nonumber\\
&&+2 B^2 i L T \nu  (\Gamma (1-2 \nu )-\Gamma (1-2 \nu ,-i L (T-\tau ))) (-i L (T-\tau ))^{2 \nu -1}\nonumber\\
&&+A^2 (i L (T-\tau ))^{2 \nu } (\Gamma (-2 \nu )-\Gamma (-2 \nu ,i L (T-\tau )))\nonumber\\
&&-2 A^2 \nu  (i L (T-\tau ))^{2 \nu } (\Gamma (-2 \nu )-\Gamma (-2 \nu ,i L (T-\tau )))\nonumber\\
&&-A B (-i L (T+\tau ))^{2 \nu } (\Gamma (-2 \nu )-\Gamma (-2 \nu ,-i L (T+\tau )))\nonumber\\
&&+2 A B \nu  (-i L (T+\tau ))^{2 \nu } (\Gamma (-2 \nu )-\Gamma (-2 \nu ,-i L (T+\tau )))\nonumber\\
&&-A B (i L (T+\tau ))^{2 \nu } (\Gamma (-2 \nu )-\Gamma (-2 \nu ,i L (T+\tau )))\nonumber\\
&&+2 A B \nu  (i L (T+\tau ))^{2 \nu } (\Gamma (-2 \nu )-\Gamma (-2 \nu ,i L (T+\tau )))\nonumber\\
&&+A^2 i L T (i L (T-\tau ))^{2 \nu -1} (\Gamma (1-2 \nu )-\Gamma (1-2 \nu ,i L (T-\tau )))\nonumber\\
&&-2 i A^2 L T \nu  (i L (T-\tau ))^{2 \nu -1} (\Gamma (1-2 \nu )-\Gamma (1-2 \nu ,i L (T-\tau )))\nonumber\\
&&+\frac{A^2 (i L (T-\tau ))^{2 \nu } \tau  (\Gamma (1-2 \nu )-\Gamma (1-2 \nu ,i L (T-\tau )))}{\tau -T}\nonumber\eea

\bea &&+\frac{2 A B T \nu  (-i L (T+\tau ))^{2 \nu } (\Gamma (1-2 \nu )-\Gamma (1-2 \nu ,-i L (T+\tau )))}{T+\tau }\nonumber\\
&&+\frac{2 A B \nu  \tau  (-i L (T+\tau ))^{2 \nu } (\Gamma (1-2 \nu )-\Gamma (1-2 \nu ,-i L (T+\tau )))}{T+\tau }\nonumber\\
&&+A B i L T (-i L (T+\tau ))^{2 \nu -1} (\Gamma (1-2 \nu )-\Gamma (1-2 \nu ,-i L (T+\tau )))\nonumber\\
&&+A B i L \tau  (-i L (T+\tau ))^{2 \nu -1} (\Gamma (1-2 \nu )-\Gamma (1-2 \nu ,-i L (T+\tau )))\nonumber\\
&&+\frac{2 A B T \nu  (i L (T+\tau ))^{2 \nu } (\Gamma (1-2 \nu )-\Gamma (1-2 \nu ,i L (T+\tau )))}{T+\tau }\nonumber\\
&&+\frac{2 A B \nu  \tau  (i L (T+\tau ))^{2 \nu } (\Gamma (1-2 \nu )-\Gamma (1-2 \nu ,i L (T+\tau )))}{T+\tau }\nonumber\\
&&+\frac{2 A^2 T \nu  (i L (T-\tau ))^{2 \nu } \tau  (\Gamma (2-2 \nu )-\Gamma (2-2 \nu ,i L (T-\tau )))}{(T-\tau )^2}\nonumber\\
&&+\frac{2 A B \tau ^2 (-i L (T+\tau ))^{2 \nu } (\Gamma (2-2 \nu )-\Gamma (2-2 \nu ,-i L (T+\tau )))}{(T+\tau )^2}\nonumber\\
&&+\frac{2 A B T \nu  \tau  (-i L (T+\tau ))^{2 \nu } (\Gamma (2-2 \nu )-\Gamma (2-2 \nu ,-i L (T+\tau )))}{(T+\tau )^2}\nonumber\\
&&+\frac{2 A B \tau ^2 (i L (T+\tau ))^{2 \nu } (\Gamma (2-2 \nu )-\Gamma (2-2 \nu ,i L (T+\tau )))}{(T+\tau )^2}\nonumber\\
&&+\frac{2 A B T \nu  \tau  (i L (T+\tau ))^{2 \nu } (\Gamma (2-2 \nu )-\Gamma (2-2 \nu ,i L (T+\tau )))}{(T+\tau )^2}\nonumber\\
&&+\frac{2 A B T \tau ^2 (-i L (T+\tau ))^{2 \nu } (\Gamma (3-2 \nu )-\Gamma (3-2 \nu ,-i L (T+\tau )))}{(T+\tau )^3}\nonumber\\
&&+\frac{2 A B T \tau ^2 (i L (T+\tau ))^{2 \nu } (\Gamma (3-2 \nu )-\Gamma (3-2 \nu ,i L (T+\tau )))}{(T+\tau )^3}\nonumber\\
&&-\frac{2 A^2 (i L (T-\tau ))^{2 \nu } \tau ^2 (\Gamma (2-2 \nu )-\Gamma (2-2 \nu ,i L (T-\tau )))}{(T-\tau )^2}\nonumber\\
&&-\frac{A^2 T (i L (T-\tau ))^{2 \nu } \tau  (\Gamma (2-2 \nu )-\Gamma (2-2 \nu ,i L (T-\tau )))}{(T-\tau )^2}\nonumber\\
&&-\frac{2 A^2 T (i L (T-\tau ))^{2 \nu } \tau ^2 (\Gamma (3-2 \nu )-\Gamma (3-2 \nu ,i L (T-\tau )))}{(T-\tau )^3}\nonumber\\
&&-\frac{2 A^2 \nu  (i L (T-\tau ))^{2 \nu } \tau  (\Gamma (1-2 \nu )-\Gamma (1-2 \nu ,i L (T-\tau )))}{\tau -T}\nonumber\\
&&-\frac{A B T (i L (T+\tau ))^{2 \nu } (\Gamma (1-2 \nu )-\Gamma (1-2 \nu ,i L (T+\tau )))}{T+\tau }\nonumber\\
&&-\frac{A B \tau  (i L (T+\tau ))^{2 \nu } (\Gamma (1-2 \nu )-\Gamma (1-2 \nu ,i L (T+\tau )))}{T+\tau }\nonumber\\
&&-\frac{A B T \tau  (-i L (T+\tau ))^{2 \nu } (\Gamma (2-2 \nu )-\Gamma (2-2 \nu ,-i L (T+\tau )))}{(T+\tau )^2}\nonumber\\
&&-\frac{A B T \tau  (i L (T+\tau ))^{2 \nu } (\Gamma (2-2 \nu )-\Gamma (2-2 \nu ,i L (T+\tau )))}{(T+\tau )^2})\nonumber\eea

\bea && (A^2 (\Gamma (-2 \nu )-\Gamma (-2 \nu ,-i L (T-\tau ))) (-i L (T-\tau ))^{2 \nu }\nonumber\\
&&-2 A^2 \nu  (\Gamma (-2 \nu )-\Gamma (-2 \nu ,-i L (T-\tau ))) (-i L (T-\tau ))^{2 \nu }\nonumber\\
&&+\frac{A^2 \tau  (\Gamma (1-2 \nu )-\Gamma (1-2 \nu ,-i L (T-\tau ))) (-i L (T-\tau ))^{2 \nu }}{\tau -T}\nonumber\\
&&+\frac{2 A^2 T \nu  \tau  (\Gamma (2-2 \nu )-\Gamma (2-2 \nu ,-i L (T-\tau ))) (-i L (T-\tau ))^{2 \nu }}{(T-\tau )^2}\nonumber\\
&&-\frac{2 A^2 \tau ^2 (\Gamma (2-2 \nu )-\Gamma (2-2 \nu ,-i L (T-\tau ))) (-i L (T-\tau ))^{2 \nu }}{(T-\tau )^2}\nonumber\\
&&-\frac{A^2 T \tau  (\Gamma (2-2 \nu )-\Gamma (2-2 \nu ,-i L (T-\tau ))) (-i L (T-\tau ))^{2 \nu }}{(T-\tau )^2}\nonumber\\
&&-\frac{2 A^2 T \tau ^2 (\Gamma (3-2 \nu )-\Gamma (3-2 \nu ,-i L (T-\tau ))) (-i L (T-\tau ))^{2 \nu }}{(T-\tau )^3}\nonumber\\
&&-\frac{2 A^2 \nu  \tau  (\Gamma (1-2 \nu )-\Gamma (1-2 \nu ,-i L (T-\tau ))) (-i L (T-\tau ))^{2 \nu }}{\tau -T}\nonumber\\
&&+A^2 (-i) L T (\Gamma (1-2 \nu )-\Gamma (1-2 \nu ,-i L (T-\tau ))) (-i L (T-\tau ))^{2 \nu -1}\nonumber\\
&&+2 A^2 i L T \nu  (\Gamma (1-2 \nu )-\Gamma (1-2 \nu ,-i L (T-\tau ))) (-i L (T-\tau ))^{2 \nu -1}\nonumber\\
&&+B^2 (i L (T-\tau ))^{2 \nu } (\Gamma (-2 \nu )-\Gamma (-2 \nu ,i L (T-\tau )))\nonumber\\
&&-2 B^2 \nu  (i L (T-\tau ))^{2 \nu } (\Gamma (-2 \nu )-\Gamma (-2 \nu ,i L (T-\tau )))\nonumber\\
&&-A B (-i L (T+\tau ))^{2 \nu } (\Gamma (-2 \nu )-\Gamma (-2 \nu ,-i L (T+\tau )))\nonumber\\
&&+2 A B \nu  (-i L (T+\tau ))^{2 \nu } (\Gamma (-2 \nu )-\Gamma (-2 \nu ,-i L (T+\tau )))\nonumber\\
&&-A B (i L (T+\tau ))^{2 \nu } (\Gamma (-2 \nu )-\Gamma (-2 \nu ,i L (T+\tau )))\nonumber\\
&&+2 A B \nu  (i L (T+\tau ))^{2 \nu } (\Gamma (-2 \nu )-\Gamma (-2 \nu ,i L (T+\tau )))\nonumber\\
&&+B^2 i L T (i L (T-\tau ))^{2 \nu -1} (\Gamma (1-2 \nu )-\Gamma (1-2 \nu ,i L (T-\tau )))\nonumber\\
&&-2 i B^2 L T \nu  (i L (T-\tau ))^{2 \nu -1} (\Gamma (1-2 \nu )-\Gamma (1-2 \nu ,i L (T-\tau )))\nonumber\\
&&+\frac{B^2 (i L (T-\tau ))^{2 \nu } \tau  (\Gamma (1-2 \nu )-\Gamma (1-2 \nu ,i L (T-\tau )))}{\tau -T}\nonumber\\
&&+\frac{2 A B T \nu  (-i L (T+\tau ))^{2 \nu } (\Gamma (1-2 \nu )-\Gamma (1-2 \nu ,-i L (T+\tau )))}{T+\tau }\nonumber\\
&&+\frac{2 A B \nu  \tau  (-i L (T+\tau ))^{2 \nu } (\Gamma (1-2 \nu )-\Gamma (1-2 \nu ,-i L (T+\tau )))}{T+\tau }\nonumber\\
&&+A B i L T (-i L (T+\tau ))^{2 \nu -1} (\Gamma (1-2 \nu )-\Gamma (1-2 \nu ,-i L (T+\tau )))\nonumber\\
&&+A B i L \tau  (-i L (T+\tau ))^{2 \nu -1} (\Gamma (1-2 \nu )-\Gamma (1-2 \nu ,-i L (T+\tau )))\nonumber\\
&&+\frac{2 A B T \nu  (i L (T+\tau ))^{2 \nu } (\Gamma (1-2 \nu )-\Gamma (1-2 \nu ,i L (T+\tau )))}{T+\tau }\nonumber\\
&&+\frac{2 A B \nu  \tau  (i L (T+\tau ))^{2 \nu } (\Gamma (1-2 \nu )-\Gamma (1-2 \nu ,i L (T+\tau )))}{T+\tau }\nonumber\eea

\bea &&+\frac{2 B^2 T \nu  (i L (T-\tau ))^{2 \nu } \tau  (\Gamma (2-2 \nu )-\Gamma (2-2 \nu ,i L (T-\tau )))}{(T-\tau )^2}\nonumber\\
&&+\frac{2 A B \tau ^2 (-i L (T+\tau ))^{2 \nu } (\Gamma (2-2 \nu )-\Gamma (2-2 \nu ,-i L (T+\tau )))}{(T+\tau )^2}\nonumber\\
&&+\frac{2 A B T \nu  \tau  (-i L (T+\tau ))^{2 \nu } (\Gamma (2-2 \nu )-\Gamma (2-2 \nu ,-i L (T+\tau )))}{(T+\tau )^2}\nonumber\\
&&+\frac{2 A B \tau ^2 (i L (T+\tau ))^{2 \nu } (\Gamma (2-2 \nu )-\Gamma (2-2 \nu ,i L (T+\tau )))}{(T+\tau )^2}\nonumber\\
&&+\frac{2 A B T \nu  \tau  (i L (T+\tau ))^{2 \nu } (\Gamma (2-2 \nu )-\Gamma (2-2 \nu ,i L (T+\tau )))}{(T+\tau )^2}\nonumber\\
&&+\frac{2 A B T \tau ^2 (-i L (T+\tau ))^{2 \nu } (\Gamma (3-2 \nu )-\Gamma (3-2 \nu ,-i L (T+\tau )))}{(T+\tau )^3}\nonumber\\
&&+\frac{2 A B T \tau ^2 (i L (T+\tau ))^{2 \nu } (\Gamma (3-2 \nu )-\Gamma (3-2 \nu ,i L (T+\tau )))}{(T+\tau )^3}\nonumber\\
&&-\frac{2 B^2 (i L (T-\tau ))^{2 \nu } \tau ^2 (\Gamma (2-2 \nu )-\Gamma (2-2 \nu ,i L (T-\tau )))}{(T-\tau )^2}\nonumber\\
&&-\frac{B^2 T (i L (T-\tau ))^{2 \nu } \tau  (\Gamma (2-2 \nu )-\Gamma (2-2 \nu ,i L (T-\tau )))}{(T-\tau )^2}\nonumber\\
&&-\frac{2 B^2 T (i L (T-\tau ))^{2 \nu } \tau ^2 (\Gamma (3-2 \nu )-\Gamma (3-2 \nu ,i L (T-\tau )))}{(T-\tau )^3}\nonumber\\
&&-\frac{2 B^2 \nu  (i L (T-\tau ))^{2 \nu } \tau  (\Gamma (1-2 \nu )-\Gamma (1-2 \nu ,i L (T-\tau )))}{\tau -T}\nonumber\\
&&-\frac{A B T (i L (T+\tau ))^{2 \nu } (\Gamma (1-2 \nu )-\Gamma (1-2 \nu ,i L (T+\tau )))}{T+\tau }\nonumber\\
&&-\frac{A B \tau  (i L (T+\tau ))^{2 \nu } (\Gamma (1-2 \nu )-\Gamma (1-2 \nu ,i L (T+\tau )))}{T+\tau }\nonumber\\
&&-\frac{A B T \tau  (-i L (T+\tau ))^{2 \nu } (\Gamma (2-2 \nu )-\Gamma (2-2 \nu ,-i L (T+\tau )))}{(T+\tau )^2}\nonumber\\
&&-\frac{A B T \tau  (i L (T+\tau ))^{2 \nu } (\Gamma (2-2 \nu )-\Gamma (2-2 \nu ,i L (T+\tau )))}{(T+\tau )^2})\eea

\bea X^{(1)}_2&=&\frac{L^{-2 \nu } }{128 \tau ^4}(-\frac{8 A^2 L^4 T^2}{\nu -2}-\frac{8 A^2 L^2}{\nu -1}+\frac{A B 4^{\nu } (\nu -3) \Gamma (4-2 \nu ) \left((-i L T)^{2 \nu }+(i L T)^{2 \nu }\right)}{(\nu -1) T^2}\nonumber\\
&&-\frac{A B 4^{\nu +1} (-i L T)^{2 \nu } \Gamma (2-2 \nu ,-2 i L T)}{T^2}-\frac{A B 4^{\nu +1} (-i L T)^{2 \nu } \Gamma (3-2 \nu ,-2 i L T)}{T^2}\nonumber\\
&&-\frac{A B 4^{\nu } (-i L T)^{2 \nu } \Gamma (4-2 \nu ,-2 i L T)}{T^2}-\frac{A B 4^{\nu +1} (i L T)^{2 \nu } \Gamma (2-2 \nu ,2 i L T)}{T^2}\nonumber\eea
\bea
&&-\frac{A B 4^{\nu +1} (i L T)^{2 \nu } \Gamma (3-2 \nu ,2 i L T)}{T^2}-\frac{A B 4^{\nu } (i L T)^{2 \nu } \Gamma (4-2 \nu ,2 i L T)}{T^2}-\frac{8 B^2 L^4 T^2}{\nu -2}-\frac{8 B^2 L^2}{\nu -1})\nonumber\\
 && (12 A^2 \tau ^2 L^{-2 \nu }+12 B^2 \tau ^2 L^{-2 \nu }-4 A^2 \nu  \tau ^2 L^{-2 \nu }-4 B^2 \nu  \tau ^2 L^{-2 \nu }-\frac{5 A^2 \tau ^2 L^{-2 \nu }}{\nu }-\frac{5 B^2 \tau ^2 L^{-2 \nu }}{\nu }\nonumber\\
 &&-\frac{A^2 L^{-2 (\nu +1)}}{\nu +1}-\frac{B^2 L^{-2 (\nu +1)}}{\nu +1}-\frac{4 A^2 \nu ^2 L^{-2 (\nu +1)}}{\nu +1}-\frac{4 B^2 \nu ^2 L^{-2 (\nu +1)}}{\nu +1}+\frac{4 A^2 \nu  L^{-2 (\nu +1)}}{\nu +1}+\frac{4 B^2 \nu  L^{-2 (\nu +1)}}{\nu +1}\nonumber\\
 &&-\frac{4 A^2 \tau ^4 L^{2-2 \nu }}{\nu -1}-\frac{4 B^2 \tau ^4 L^{2-2 \nu }}{\nu -1}+\frac{2^{2 \nu +1} A B (11 \nu -5) \tau ^2 \left((-i \tau )^{2 \nu }+(i \tau )^{2 \nu }\right) \Gamma (-2 \nu )}{\nu +1}\nonumber\\
 &&+\frac{2^{2 \nu +1} A B \nu  \tau ^2 \left((-i \tau )^{2 \nu }+(i \tau )^{2 \nu }\right) \Gamma (1-2 \nu )}{\nu +1}+3\ 2^{2 \nu +1} A B \tau ^2 (-i \tau )^{2 \nu } \Gamma (-2 \nu ,-2 i L \tau )\nonumber\\
 &&-2^{2 \nu +3} A B \nu ^2 (-i \tau )^{2 \nu } \tau ^2 \Gamma (-2 \nu ,-2 i L \tau )-2^{2 \nu +3} A B \nu  (-i \tau )^{2 \nu } \tau ^2 \Gamma (-2 \nu ,-2 i L \tau )\nonumber\\
 &&+3\ 2^{2 \nu +1} A B \tau ^2 (i \tau )^{2 \nu } \Gamma (-2 \nu ,2 i L \tau )-2^{2 \nu +3} A B \nu ^2 (i \tau )^{2 \nu } \tau ^2 \Gamma (-2 \nu ,2 i L \tau )\nonumber\\
 &&-2^{2 \nu +3} A B \nu  (i \tau )^{2 \nu } \tau ^2 \Gamma (-2 \nu ,2 i L \tau )+2^{2 \nu +5} A B \nu  \tau ^2 (-i \tau )^{2 \nu } \Gamma (-2 (\nu +1),-2 i L \tau )\nonumber\\
 &&-2^{2 \nu +5} A B \nu ^2 (-i \tau )^{2 \nu } \tau ^2 \Gamma (-2 (\nu +1),-2 i L \tau )-2^{2 \nu +3} A B (-i \tau )^{2 \nu } \tau ^2 \Gamma (-2 (\nu +1),-2 i L \tau )\nonumber\\
 &&+2^{2 \nu +5} A B \nu  \tau ^2 (i \tau )^{2 \nu } \Gamma (-2 (\nu +1),2 i L \tau )-2^{2 \nu +5} A B \nu ^2 (i \tau )^{2 \nu } \tau ^2 \Gamma (-2 (\nu +1),2 i L \tau )\nonumber\\
 &&-2^{2 \nu +3} A B (i \tau )^{2 \nu } \tau ^2 \Gamma (-2 (\nu +1),2 i L \tau )+2^{2 \nu +5} A B \nu  \tau ^2 (-i \tau )^{2 \nu } \Gamma (-2 \nu -1,-2 i L \tau )\nonumber\\
 &&-2^{2 \nu +5} A B \nu ^2 (-i \tau )^{2 \nu } \tau ^2 \Gamma (-2 \nu -1,-2 i L \tau )-2^{2 \nu +3} A B (-i \tau )^{2 \nu } \tau ^2 \Gamma (-2 \nu -1,-2 i L \tau )\nonumber\\
 &&+2^{2 \nu +5} A B \nu  \tau ^2 (i \tau )^{2 \nu } \Gamma (-2 \nu -1,2 i L \tau )-2^{2 \nu +5} A B \nu ^2 (i \tau )^{2 \nu } \tau ^2 \Gamma (-2 \nu -1,2 i L \tau )\nonumber\\
 &&-2^{2 \nu +3} A B (i \tau )^{2 \nu } \tau ^2 \Gamma (-2 \nu -1,2 i L \tau )+4^{\nu +1} A B \tau ^2 (-i \tau )^{2 \nu } \Gamma (1-2 \nu ,-2 i L \tau )\nonumber\\
 &&-2^{2 \nu +3} A B \nu  (-i \tau )^{2 \nu } \tau ^2 \Gamma (1-2 \nu ,-2 i L \tau )+4^{\nu +1} A B \tau ^2 (i \tau )^{2 \nu } \Gamma (1-2 \nu ,2 i L \tau )\nonumber\\
 &&-2^{2 \nu +3} A B \nu  (i \tau )^{2 \nu } \tau ^2 \Gamma (1-2 \nu ,2 i L \tau )-2^{2 \nu +1} A B (-i \tau )^{2 \nu } \tau ^2 \Gamma (2-2 \nu ,-2 i L \tau )\nonumber\\
 &&-2^{2 \nu +1} A B (i \tau )^{2 \nu } \tau ^2 \Gamma (2-2 \nu ,2 i L \tau ))\eea
 
 \bea X^{(1)}_3&=&-\frac{L^{-2 \nu }}{128 \tau ^4} (-\frac{8 A^2 L^{2-2 \nu }}{\nu -1}-\frac{8 A^2 T^2 L^{4-2 \nu }}{\nu -2}-\frac{A B 4^{\nu +1} (-i T)^{2 \nu } \Gamma (2-2 \nu ,-2 i L T)}{T^2}\nonumber\\
 &&-\frac{A B 4^{\nu +1} (-i T)^{2 \nu } \Gamma (3-2 \nu ,-2 i L T)}{T^2}-\frac{A B 4^{\nu } (-i T)^{2 \nu } \Gamma (4-2 \nu ,-2 i L T)}{T^2}\nonumber\\
 &&-\frac{A B 4^{\nu +1} (i T)^{2 \nu } \Gamma (2-2 \nu ,2 i L T)}{T^2}-\frac{A B 4^{\nu +1} (i T)^{2 \nu } \Gamma (3-2 \nu ,2 i L T)}{T^2}\nonumber\\
 &&-\frac{A B 4^{\nu } (i T)^{2 \nu } \Gamma (4-2 \nu ,2 i L T)}{T^2}+\frac{A B 4^{\nu } (\nu -3) \Gamma (4-2 \nu ) \left((-i T)^{2 \nu }+(i T)^{2 \nu }\right)}{(\nu -1) T^2}\nonumber\\
 &&-\frac{8 B^2 L^{2-2 \nu }}{\nu -1}-\frac{8 B^2 T^2 L^{4-2 \nu }}{\nu -2})
 (-2^{2 \nu +3} A B \nu ^2 \tau ^2 \Gamma (-2 \nu ,-2 i L \tau ) (-i L \tau )^{2 \nu }\nonumber\\
 &&+3\ 2^{2 \nu +1} A B \tau ^2 \Gamma (-2 \nu ,-2 i L \tau ) (-i L \tau )^{2 \nu }-2^{2 \nu +3} A B \nu  \tau ^2 \Gamma (-2 \nu ,-2 i L \tau ) (-i L \tau )^{2 \nu }\nonumber\\
 &&-2^{2 \nu +5} A B \nu ^2 \tau ^2 \Gamma (-2 (\nu +1),-2 i L \tau ) (-i L \tau )^{2 \nu }-2^{2 \nu +3} A B \tau ^2 \Gamma (-2 (\nu +1),-2 i L \tau ) (-i L \tau )^{2 \nu }\nonumber\\
 &&+2^{2 \nu +5} A B \nu  \tau ^2 \Gamma (-2 (\nu +1),-2 i L \tau ) (-i L \tau )^{2 \nu }-2^{2 \nu +5} A B \nu ^2 \tau ^2 \Gamma (-2 \nu -1,-2 i L \tau ) (-i L \tau )^{2 \nu }\nonumber\eea
 
\bea && -2^{2 \nu +3} A B \tau ^2 \Gamma (-2 \nu -1,-2 i L \tau ) (-i L \tau )^{2 \nu }+2^{2 \nu +5} A B \nu  \tau ^2 \Gamma (-2 \nu -1,-2 i L \tau ) (-i L \tau )^{2 \nu }\nonumber\\
 &&+4^{\nu +1} A B \tau ^2 \Gamma (1-2 \nu ,-2 i L \tau ) (-i L \tau )^{2 \nu }-2^{2 \nu +3} A B \nu  \tau ^2 \Gamma (1-2 \nu ,-2 i L \tau ) (-i L \tau )^{2 \nu }\nonumber\\
 &&-2^{2 \nu +1} A B \tau ^2 \Gamma (2-2 \nu ,-2 i L \tau ) (-i L \tau )^{2 \nu }+12 A^2 \tau ^2+12 B^2 \tau ^2-4 A^2 \nu  \tau ^2-4 B^2 \nu  \tau ^2\nonumber\\
 &&+\frac{2^{2 \nu +1} A B (11 \nu -5) \tau ^2 \left((-i L \tau )^{2 \nu }+(i L \tau )^{2 \nu }\right) \Gamma (-2 \nu )}{\nu +1}+\frac{2^{2 \nu +1} A B \nu  \tau ^2 \left((-i L \tau )^{2 \nu }+(i L \tau )^{2 \nu }\right) \Gamma (1-2 \nu )}{\nu +1}\nonumber\\
 &&-2^{2 \nu +3} A B \nu ^2 \tau ^2 (i L \tau )^{2 \nu } \Gamma (-2 \nu ,2 i L \tau )+3\ 2^{2 \nu +1} A B \tau ^2 (i L \tau )^{2 \nu } \Gamma (-2 \nu ,2 i L \tau )\nonumber\\
 &&-2^{2 \nu +3} A B \nu  \tau ^2 (i L \tau )^{2 \nu } \Gamma (-2 \nu ,2 i L \tau )-2^{2 \nu +5} A B \nu ^2 \tau ^2 (i L \tau )^{2 \nu } \Gamma (-2 (\nu +1),2 i L \tau )\nonumber\\
 &&-2^{2 \nu +3} A B \tau ^2 (i L \tau )^{2 \nu } \Gamma (-2 (\nu +1),2 i L \tau )+2^{2 \nu +5} A B \nu  \tau ^2 (i L \tau )^{2 \nu } \Gamma (-2 (\nu +1),2 i L \tau )\nonumber\\
 &&-2^{2 \nu +5} A B \nu ^2 \tau ^2 (i L \tau )^{2 \nu } \Gamma (-2 \nu -1,2 i L \tau )-2^{2 \nu +3} A B \tau ^2 (i L \tau )^{2 \nu } \Gamma (-2 \nu -1,2 i L \tau )\nonumber\\
 &&+2^{2 \nu +5} A B \nu  \tau ^2 (i L \tau )^{2 \nu } \Gamma (-2 \nu -1,2 i L \tau )+4^{\nu +1} A B \tau ^2 (i L \tau )^{2 \nu } \Gamma (1-2 \nu ,2 i L \tau )\nonumber\\
 &&-2^{2 \nu +3} A B \nu  \tau ^2 (i L \tau )^{2 \nu } \Gamma (1-2 \nu ,2 i L \tau )-2^{2 \nu +1} A B \tau ^2 (i L \tau )^{2 \nu } \Gamma (2-2 \nu ,2 i L \tau )\nonumber\\
 &&-\frac{4 A^2 L^2 \tau ^4}{\nu -1}-\frac{4 B^2 L^2 \tau ^4}{\nu -1}-\frac{5 A^2 \tau ^2}{\nu }-\frac{5 B^2 \tau ^2}{\nu }-\frac{A^2}{L^2 (\nu +1)}-\frac{B^2}{L^2 (\nu +1)}\nonumber\\
 &&-\frac{4 A^2 \nu ^2}{L^2 (\nu +1)}-\frac{4 B^2 \nu ^2}{L^2 (\nu +1)}+\frac{4 A^2 \nu }{L^2 (\nu +1)}+\frac{4 B^2 \nu }{L^2 (\nu +1)})\eea
 
 \bea X^{(1)}_4&=&-\frac{L^{-4 \nu } }{4 \tau ^4}(B^2 (\Gamma (-2 \nu )-\Gamma (-2 \nu ,-i L (T-\tau ))) (-i L (T-\tau ))^{2 \nu }\nonumber\\
 &&-2 B^2 \nu  (\Gamma (-2 \nu )-\Gamma (-2 \nu ,-i L (T-\tau ))) (-i L (T-\tau ))^{2 \nu }\nonumber\\
 &&+\frac{B^2 \tau  (\Gamma (1-2 \nu )-\Gamma (1-2 \nu ,-i L (T-\tau ))) (-i L (T-\tau ))^{2 \nu }}{\tau -T}\nonumber\\
 &&+\frac{2 B^2 T \nu  \tau  (\Gamma (2-2 \nu )-\Gamma (2-2 \nu ,-i L (T-\tau ))) (-i L (T-\tau ))^{2 \nu }}{(T-\tau )^2}\nonumber\\
 &&-\frac{2 B^2 \tau ^2 (\Gamma (2-2 \nu )-\Gamma (2-2 \nu ,-i L (T-\tau ))) (-i L (T-\tau ))^{2 \nu }}{(T-\tau )^2}\nonumber\\
 &&-\frac{B^2 T \tau  (\Gamma (2-2 \nu )-\Gamma (2-2 \nu ,-i L (T-\tau ))) (-i L (T-\tau ))^{2 \nu }}{(T-\tau )^2}\nonumber\\
 &&-\frac{2 B^2 T \tau ^2 (\Gamma (3-2 \nu )-\Gamma (3-2 \nu ,-i L (T-\tau ))) (-i L (T-\tau ))^{2 \nu }}{(T-\tau )^3}\nonumber\\
 &&-\frac{2 B^2 \nu  \tau  (\Gamma (1-2 \nu )-\Gamma (1-2 \nu ,-i L (T-\tau ))) (-i L (T-\tau ))^{2 \nu }}{\tau -T}\nonumber\\
 &&+B^2 (-i) L T (\Gamma (1-2 \nu )-\Gamma (1-2 \nu ,-i L (T-\tau ))) (-i L (T-\tau ))^{2 \nu -1}\nonumber\\
 &&+2 B^2 i L T \nu  (\Gamma (1-2 \nu )-\Gamma (1-2 \nu ,-i L (T-\tau ))) (-i L (T-\tau ))^{2 \nu -1}\nonumber\\
 &&+A^2 (i L (T-\tau ))^{2 \nu } (\Gamma (-2 \nu )-\Gamma (-2 \nu ,i L (T-\tau )))\nonumber\eea
 
 \bea && -2 A^2 \nu  (i L (T-\tau ))^{2 \nu } (\Gamma (-2 \nu )-\Gamma (-2 \nu ,i L (T-\tau )))\nonumber\\
 &&-A B (-i L (T+\tau ))^{2 \nu } (\Gamma (-2 \nu )-\Gamma (-2 \nu ,-i L (T+\tau )))\nonumber\\
 &&+2 A B \nu  (-i L (T+\tau ))^{2 \nu } (\Gamma (-2 \nu )-\Gamma (-2 \nu ,-i L (T+\tau )))\nonumber\\
 &&-A B (i L (T+\tau ))^{2 \nu } (\Gamma (-2 \nu )-\Gamma (-2 \nu ,i L (T+\tau )))\nonumber\\
 &&+2 A B \nu  (i L (T+\tau ))^{2 \nu } (\Gamma (-2 \nu )-\Gamma (-2 \nu ,i L (T+\tau )))\nonumber\\
 &&+A^2 i L T (i L (T-\tau ))^{2 \nu -1} (\Gamma (1-2 \nu )-\Gamma (1-2 \nu ,i L (T-\tau )))\nonumber\\
 &&-2 i A^2 L T \nu  (i L (T-\tau ))^{2 \nu -1} (\Gamma (1-2 \nu )-\Gamma (1-2 \nu ,i L (T-\tau )))\nonumber\\
 &&+\frac{A^2 (i L (T-\tau ))^{2 \nu } \tau  (\Gamma (1-2 \nu )-\Gamma (1-2 \nu ,i L (T-\tau )))}{\tau -T}\nonumber\\
 &&+\frac{2 A B T \nu  (-i L (T+\tau ))^{2 \nu } (\Gamma (1-2 \nu )-\Gamma (1-2 \nu ,-i L (T+\tau )))}{T+\tau }\nonumber\\
 &&+\frac{2 A B \nu  \tau  (-i L (T+\tau ))^{2 \nu } (\Gamma (1-2 \nu )-\Gamma (1-2 \nu ,-i L (T+\tau )))}{T+\tau }\nonumber\\
 &&+A B i L T (-i L (T+\tau ))^{2 \nu -1} (\Gamma (1-2 \nu )-\Gamma (1-2 \nu ,-i L (T+\tau )))\nonumber\\
 &&+A B i L \tau  (-i L (T+\tau ))^{2 \nu -1} (\Gamma (1-2 \nu )-\Gamma (1-2 \nu ,-i L (T+\tau )))\nonumber\\
 &&+\frac{2 A B T \nu  (i L (T+\tau ))^{2 \nu } (\Gamma (1-2 \nu )-\Gamma (1-2 \nu ,i L (T+\tau )))}{T+\tau }\nonumber\\
 &&+\frac{2 A B \nu  \tau  (i L (T+\tau ))^{2 \nu } (\Gamma (1-2 \nu )-\Gamma (1-2 \nu ,i L (T+\tau )))}{T+\tau }\nonumber\\
 &&+\frac{2 A^2 T \nu  (i L (T-\tau ))^{2 \nu } \tau  (\Gamma (2-2 \nu )-\Gamma (2-2 \nu ,i L (T-\tau )))}{(T-\tau )^2}\nonumber\\
 &&+\frac{2 A B \tau ^2 (-i L (T+\tau ))^{2 \nu } (\Gamma (2-2 \nu )-\Gamma (2-2 \nu ,-i L (T+\tau )))}{(T+\tau )^2}\nonumber\\
 &&+\frac{2 A B T \nu  \tau  (-i L (T+\tau ))^{2 \nu } (\Gamma (2-2 \nu )-\Gamma (2-2 \nu ,-i L (T+\tau )))}{(T+\tau )^2}\nonumber\\
 &&+\frac{2 A B \tau ^2 (i L (T+\tau ))^{2 \nu } (\Gamma (2-2 \nu )-\Gamma (2-2 \nu ,i L (T+\tau )))}{(T+\tau )^2}\nonumber\\
 &&+\frac{2 A B T \nu  \tau  (i L (T+\tau ))^{2 \nu } (\Gamma (2-2 \nu )-\Gamma (2-2 \nu ,i L (T+\tau )))}{(T+\tau )^2}\nonumber\\
 &&+\frac{2 A B T \tau ^2 (-i L (T+\tau ))^{2 \nu } (\Gamma (3-2 \nu )-\Gamma (3-2 \nu ,-i L (T+\tau )))}{(T+\tau )^3}\nonumber\\
 &&+\frac{2 A B T \tau ^2 (i L (T+\tau ))^{2 \nu } (\Gamma (3-2 \nu )-\Gamma (3-2 \nu ,i L (T+\tau )))}{(T+\tau )^3}\nonumber\\
 &&-\frac{2 A^2 (i L (T-\tau ))^{2 \nu } \tau ^2 (\Gamma (2-2 \nu )-\Gamma (2-2 \nu ,i L (T-\tau )))}{(T-\tau )^2}\nonumber\\
 &&-\frac{A^2 T (i L (T-\tau ))^{2 \nu } \tau  (\Gamma (2-2 \nu )-\Gamma (2-2 \nu ,i L (T-\tau )))}{(T-\tau )^2}\nonumber\eea
 
 \bea &&-\frac{2 A^2 T (i L (T-\tau ))^{2 \nu } \tau ^2 (\Gamma (3-2 \nu )-\Gamma (3-2 \nu ,i L (T-\tau )))}{(T-\tau )^3}\nonumber\\
 &&-\frac{2 A^2 \nu  (i L (T-\tau ))^{2 \nu } \tau  (\Gamma (1-2 \nu )-\Gamma (1-2 \nu ,i L (T-\tau )))}{\tau -T}\nonumber\\
 &&-\frac{A B T (i L (T+\tau ))^{2 \nu } (\Gamma (1-2 \nu )-\Gamma (1-2 \nu ,i L (T+\tau )))}{T+\tau }\nonumber\\
 &&-\frac{A B \tau  (i L (T+\tau ))^{2 \nu } (\Gamma (1-2 \nu )-\Gamma (1-2 \nu ,i L (T+\tau )))}{T+\tau }\nonumber\\
 &&-\frac{A B T \tau  (-i L (T+\tau ))^{2 \nu } (\Gamma (2-2 \nu )-\Gamma (2-2 \nu ,-i L (T+\tau )))}{(T+\tau )^2}\nonumber\\
 &&-\frac{A B T \tau  (i L (T+\tau ))^{2 \nu } (\Gamma (2-2 \nu )-\Gamma (2-2 \nu ,i L (T+\tau )))}{(T+\tau )^2})\nonumber\\
 && (A^2 (\Gamma (-2 \nu )-\Gamma (-2 \nu ,-i L (T-\tau ))) (-i L (T-\tau ))^{2 \nu }\nonumber\\
 &&-2 A^2 \nu  (\Gamma (-2 \nu )-\Gamma (-2 \nu ,-i L (T-\tau ))) (-i L (T-\tau ))^{2 \nu }\nonumber\\
 &&+\frac{A^2 \tau  (\Gamma (1-2 \nu )-\Gamma (1-2 \nu ,-i L (T-\tau ))) (-i L (T-\tau ))^{2 \nu }}{\tau -T}\nonumber\\
 &&+\frac{2 A^2 T \nu  \tau  (\Gamma (2-2 \nu )-\Gamma (2-2 \nu ,-i L (T-\tau ))) (-i L (T-\tau ))^{2 \nu }}{(T-\tau )^2}\nonumber\\
 &&-\frac{2 A^2 \tau ^2 (\Gamma (2-2 \nu )-\Gamma (2-2 \nu ,-i L (T-\tau ))) (-i L (T-\tau ))^{2 \nu }}{(T-\tau )^2}\nonumber\\
 &&-\frac{A^2 T \tau  (\Gamma (2-2 \nu )-\Gamma (2-2 \nu ,-i L (T-\tau ))) (-i L (T-\tau ))^{2 \nu }}{(T-\tau )^2}\nonumber\\
 &&-\frac{2 A^2 T \tau ^2 (\Gamma (3-2 \nu )-\Gamma (3-2 \nu ,-i L (T-\tau ))) (-i L (T-\tau ))^{2 \nu }}{(T-\tau )^3}\nonumber\\
 &&-\frac{2 A^2 \nu  \tau  (\Gamma (1-2 \nu )-\Gamma (1-2 \nu ,-i L (T-\tau ))) (-i L (T-\tau ))^{2 \nu }}{\tau -T}\nonumber\\
 &&+A^2 (-i) L T (\Gamma (1-2 \nu )-\Gamma (1-2 \nu ,-i L (T-\tau ))) (-i L (T-\tau ))^{2 \nu -1}\nonumber\\
 &&+2 A^2 i L T \nu  (\Gamma (1-2 \nu )-\Gamma (1-2 \nu ,-i L (T-\tau ))) (-i L (T-\tau ))^{2 \nu -1}\nonumber\\
 &&+B^2 (i L (T-\tau ))^{2 \nu } (\Gamma (-2 \nu )-\Gamma (-2 \nu ,i L (T-\tau )))\nonumber\\
 &&-2 B^2 \nu  (i L (T-\tau ))^{2 \nu } (\Gamma (-2 \nu )-\Gamma (-2 \nu ,i L (T-\tau )))\nonumber\\
 &&-A B (-i L (T+\tau ))^{2 \nu } (\Gamma (-2 \nu )-\Gamma (-2 \nu ,-i L (T+\tau )))\nonumber\\
 &&+2 A B \nu  (-i L (T+\tau ))^{2 \nu } (\Gamma (-2 \nu )-\Gamma (-2 \nu ,-i L (T+\tau )))\nonumber\\
 &&-A B (i L (T+\tau ))^{2 \nu } (\Gamma (-2 \nu )-\Gamma (-2 \nu ,i L (T+\tau )))\nonumber\\
 &&+2 A B \nu  (i L (T+\tau ))^{2 \nu } (\Gamma (-2 \nu )-\Gamma (-2 \nu ,i L (T+\tau )))\nonumber\\
 &&+B^2 i L T (i L (T-\tau ))^{2 \nu -1} (\Gamma (1-2 \nu )-\Gamma (1-2 \nu ,i L (T-\tau )))\nonumber\\
 &&-2 i B^2 L T \nu  (i L (T-\tau ))^{2 \nu -1} (\Gamma (1-2 \nu )-\Gamma (1-2 \nu ,i L (T-\tau )))\nonumber\eea
 
 \bea &&+\frac{B^2 (i L (T-\tau ))^{2 \nu } \tau  (\Gamma (1-2 \nu )-\Gamma (1-2 \nu ,i L (T-\tau )))}{\tau -T}\nonumber\\
 &&+\frac{2 A B T \nu  (-i L (T+\tau ))^{2 \nu } (\Gamma (1-2 \nu )-\Gamma (1-2 \nu ,-i L (T+\tau )))}{T+\tau }\nonumber\\
 &&+\frac{2 A B \nu  \tau  (-i L (T+\tau ))^{2 \nu } (\Gamma (1-2 \nu )-\Gamma (1-2 \nu ,-i L (T+\tau )))}{T+\tau }\nonumber\\
 &&+A B i L T (-i L (T+\tau ))^{2 \nu -1} (\Gamma (1-2 \nu )-\Gamma (1-2 \nu ,-i L (T+\tau )))\nonumber\\
 &&+A B i L \tau  (-i L (T+\tau ))^{2 \nu -1} (\Gamma (1-2 \nu )-\Gamma (1-2 \nu ,-i L (T+\tau )))\nonumber\\
 &&+\frac{2 A B T \nu  (i L (T+\tau ))^{2 \nu } (\Gamma (1-2 \nu )-\Gamma (1-2 \nu ,i L (T+\tau )))}{T+\tau }\nonumber\\
 &&+\frac{2 A B \nu  \tau  (i L (T+\tau ))^{2 \nu } (\Gamma (1-2 \nu )-\Gamma (1-2 \nu ,i L (T+\tau )))}{T+\tau }\nonumber\\
 &&+\frac{2 B^2 T \nu  (i L (T-\tau ))^{2 \nu } \tau  (\Gamma (2-2 \nu )-\Gamma (2-2 \nu ,i L (T-\tau )))}{(T-\tau )^2}\nonumber\\
 &&+\frac{2 A B \tau ^2 (-i L (T+\tau ))^{2 \nu } (\Gamma (2-2 \nu )-\Gamma (2-2 \nu ,-i L (T+\tau )))}{(T+\tau )^2}\nonumber\\
 &&+\frac{2 A B T \nu  \tau  (-i L (T+\tau ))^{2 \nu } (\Gamma (2-2 \nu )-\Gamma (2-2 \nu ,-i L (T+\tau )))}{(T+\tau )^2}\nonumber\\
 &&+\frac{2 A B \tau ^2 (i L (T+\tau ))^{2 \nu } (\Gamma (2-2 \nu )-\Gamma (2-2 \nu ,i L (T+\tau )))}{(T+\tau )^2}\nonumber\\
 &&+\frac{2 A B T \nu  \tau  (i L (T+\tau ))^{2 \nu } (\Gamma (2-2 \nu )-\Gamma (2-2 \nu ,i L (T+\tau )))}{(T+\tau )^2}\nonumber\\
 &&+\frac{2 A B T \tau ^2 (-i L (T+\tau ))^{2 \nu } (\Gamma (3-2 \nu )-\Gamma (3-2 \nu ,-i L (T+\tau )))}{(T+\tau )^3}\nonumber\\
 &&+\frac{2 A B T \tau ^2 (i L (T+\tau ))^{2 \nu } (\Gamma (3-2 \nu )-\Gamma (3-2 \nu ,i L (T+\tau )))}{(T+\tau )^3}\nonumber\\
 &&-\frac{2 B^2 (i L (T-\tau ))^{2 \nu } \tau ^2 (\Gamma (2-2 \nu )-\Gamma (2-2 \nu ,i L (T-\tau )))}{(T-\tau )^2}\nonumber\\
 &&-\frac{B^2 T (i L (T-\tau ))^{2 \nu } \tau  (\Gamma (2-2 \nu )-\Gamma (2-2 \nu ,i L (T-\tau )))}{(T-\tau )^2}\nonumber\\
 &&-\frac{2 B^2 T (i L (T-\tau ))^{2 \nu } \tau ^2 (\Gamma (3-2 \nu )-\Gamma (3-2 \nu ,i L (T-\tau )))}{(T-\tau )^3}\nonumber\\
 &&-\frac{2 B^2 \nu  (i L (T-\tau ))^{2 \nu } \tau  (\Gamma (1-2 \nu )-\Gamma (1-2 \nu ,i L (T-\tau )))}{\tau -T}\nonumber\\
 &&-\frac{A B T (i L (T+\tau ))^{2 \nu } (\Gamma (1-2 \nu )-\Gamma (1-2 \nu ,i L (T+\tau )))}{T+\tau }\nonumber\\
 &&-\frac{A B \tau  (i L (T+\tau ))^{2 \nu } (\Gamma (1-2 \nu )-\Gamma (1-2 \nu ,i L (T+\tau )))}{T+\tau }\nonumber\\
 &&-\frac{A B T \tau  (-i L (T+\tau ))^{2 \nu } (\Gamma (2-2 \nu )-\Gamma (2-2 \nu ,-i L (T+\tau )))}{(T+\tau )^2}\nonumber\eea
 
 \bea &&-\frac{A B T \tau  (i L (T+\tau ))^{2 \nu } (\Gamma (2-2 \nu )-\Gamma (2-2 \nu ,i L (T+\tau )))}{(T+\tau )^2})\eea

In the super-horizon limit we get the following simplified results:
\bea X^{(1)}_1&=&\frac{L^{-4 \nu } }{4 \tau ^4}\left(B^2 (\Gamma (-2 \nu )-\Gamma (-2 \nu ,-i L (T-\tau ))\right.\nonumber\\
&&\left.-2\nu(\Gamma (-2 \nu )-\Gamma (-2 \nu ,-i L (T-\tau ))) (-i L (T-\tau ))^{2 \nu }\right.\nonumber\\
&&\left.-\frac{2 B^2 \tau ^2 (\Gamma (2-2 \nu )-\Gamma (2-2 \nu ,-i L (T-\tau ))) (-i L (T-\tau ))^{2 \nu }}{(T-\tau )^2}\right.\nonumber\\
&&\left.+A^2 (i L (T-\tau ))^{2 \nu } (\Gamma (-2 \nu )-\Gamma (-2 \nu ,i L (T-\tau ))-2\nu (\Gamma (-2 \nu )-\Gamma (-2 \nu ,i L (T-\tau ))))\right.\nonumber\\
&&\left.-A B (-i L (T+\tau ))^{2 \nu } (\Gamma (-2 \nu )-\Gamma (-2 \nu ,-i L (T+\tau ))\right.\nonumber\\
&&\left.-2 \nu  (\Gamma (-2 \nu )-\Gamma (-2 \nu ,-i L (T+\tau ))))\right.\nonumber\\
&&\left.-A B (i L (T+\tau ))^{2 \nu } (\Gamma (-2 \nu )-\Gamma (-2 \nu ,i L (T+\tau ))\right.\nonumber\\
&&\left.-2\nu  (\Gamma (-2 \nu )-\Gamma (-2 \nu ,i L (T+\tau ))))\right.\nonumber\\
&&\left.+\frac{2 A B \tau ^2 (-i L (T+\tau ))^{2 \nu } (\Gamma (2-2 \nu )-\Gamma (2-2 \nu ,-i L (T+\tau )))}{(T+\tau )^2}\right.\nonumber\\
&&\left.+\frac{2 A B \tau ^2 (i L (T+\tau ))^{2 \nu } (\Gamma (2-2 \nu )-\Gamma (2-2 \nu ,i L (T+\tau )))}{(T+\tau )^2}\right.\nonumber\\
&&\left.-\frac{2 A^2 (i L (T-\tau ))^{2 \nu } \tau ^2 (\Gamma (2-2 \nu )-\Gamma (2-2 \nu ,i L (T-\tau )))}{(T-\tau )^2}\right)\nonumber\\
&& \left(A^2 (\Gamma (-2 \nu )-\Gamma (-2 \nu ,-i L (T-\tau ))\right.\nonumber\\
&&\left.-2\nu  (\Gamma (-2 \nu )-\Gamma (-2 \nu ,-i L (T-\tau )))) (-i L (T-\tau ))^{2 \nu }\right.\nonumber\\
&&\left.-\frac{2 A^2 \tau ^2 (\Gamma (2-2 \nu )-\Gamma (2-2 \nu ,-i L (T-\tau ))) (-i L (T-\tau ))^{2 \nu }}{(T-\tau )^2}\right.\nonumber\\
&&\left.+B^2 (i L (T-\tau ))^{2 \nu } (\Gamma (-2 \nu )-\Gamma (-2 \nu ,i L (T-\tau ))\right.\nonumber\\
&&\left.-2\nu  (\Gamma (-2 \nu )-\Gamma (-2 \nu ,i L (T-\tau ))))\right.\nonumber\\
&&\left.-A B (-i L (T+\tau ))^{2 \nu } (\Gamma (-2 \nu )-\Gamma (-2 \nu ,-i L (T+\tau )))\right.\nonumber\\
&&\left.+2 A B \nu  (-i L (T+\tau ))^{2 \nu } (\Gamma (-2 \nu )-\Gamma (-2 \nu ,-i L (T+\tau )))\right.\nonumber\\
&&\left.-A B (i L (T+\tau ))^{2 \nu } (\Gamma (-2 \nu )-\Gamma (-2 \nu ,i L (T+\tau )))\right.\nonumber\\
&&\left.+2 A B \nu  (i L (T+\tau ))^{2 \nu } (\Gamma (-2 \nu )-\Gamma (-2 \nu ,i L (T+\tau )))\right.\nonumber\\
&&\left.+\frac{2 A B \tau ^2 (-i L (T+\tau ))^{2 \nu } (\Gamma (2-2 \nu )-\Gamma (2-2 \nu ,-i L (T+\tau )))}{(T+\tau )^2}\right.\nonumber\\
&&\left.+\frac{2 A B \tau ^2 (i L (T+\tau ))^{2 \nu } (\Gamma (2-2 \nu )-\Gamma (2-2 \nu ,i L (T+\tau )))}{(T+\tau )^2}\right.\nonumber\\
&&\left.-\frac{2 B^2 (i L (T-\tau ))^{2 \nu } \tau ^2 (\Gamma (2-2 \nu )-\Gamma (2-2 \nu ,i L (T-\tau )))}{(T-\tau )^2}\right)\eea

\bea X^{(2)}_1&=&\frac{L^{-2 \nu } }{16 \tau ^4}\left(-\frac{4^{\nu } A B (\Gamma (2-2 \nu )-\Gamma (2-2 \nu ,-2 i L T)) (-i L T)^{2 \nu }}{T^2}\right.\nonumber\\
&&\left.-\frac{4^{\nu } A B (i L T)^{2 \nu } (\Gamma (2-2 \nu )-\Gamma (2-2 \nu ,2 i L T))}{T^2}+\frac{2 A^2 L^2}{\nu -1}+\frac{2 B^2 L^2}{\nu -1}\right) \nonumber\\&&
\left(-4 (A^2+B^2) \tau ^2 L^{-2 \nu }+\frac{2 (A^2+B^2) \tau ^2 L^{-2 \nu }}{\nu }+\frac{2 (A^2+B^2) \nu ^2 L^{-2 (\nu +1)}}{\nu +1}\right.\nonumber\\
&&\left.-\frac{2 (A^2+B^2) \nu  L^{-2 (\nu +1)}}{\nu +1}+\frac{(A^2+B^2) L^{-2 (\nu +1)}}{2 (\nu +1)}+\frac{2 (A^2+B^2) \tau ^4 L^{2-2 \nu }}{\nu -1}\right.\nonumber\\
&&\left.+4^{\nu +1} A B \tau ^2 (-i \tau )^{2 \nu } (\Gamma (-2 \nu )-\Gamma (-2 \nu ,-2 i L \tau ))\right.\nonumber\\
&&\left.-2^{2 \nu +3} A B \nu  (-i \tau )^{2 \nu } \tau ^2 (\Gamma (-2 \nu )-\Gamma (-2 \nu ,-2 i L \tau ))\right.\nonumber\\
&&\left.+4^{\nu +1} A B \tau ^2 (i \tau )^{2 \nu } (\Gamma (-2 \nu )-\Gamma (-2 \nu ,2 i L \tau ))\right.\nonumber\\
&&\left.-2^{2 \nu +3} A B \nu  (i \tau )^{2 \nu } \tau ^2 (\Gamma (-2 \nu )-\Gamma (-2 \nu ,2 i L \tau ))\right.\nonumber\\
&&\left.+4^{\nu +2} A B \nu  \tau ^2 (-i \tau )^{2 \nu } (\Gamma (-2 (\nu +1))-\Gamma (-2 (\nu +1),-2 i L \tau ))\right.\nonumber\\
&&\left.+4^{\nu +2} A B \nu ^2 (-i \tau )^{2 (\nu +1)} (\Gamma (-2 (\nu +1))-\Gamma (-2 (\nu +1),-2 i L \tau ))\right.\nonumber\\
&&\left.+4^{\nu +1} A B (-i \tau )^{2 (\nu +1)} (\Gamma (-2 (\nu +1))-\Gamma (-2 (\nu +1),-2 i L \tau ))\right.\nonumber\\
&&\left.+4^{\nu +2} A B \nu  \tau ^2 (i \tau )^{2 \nu } (\Gamma (-2 (\nu +1))-\Gamma (-2 (\nu +1),2 i L \tau ))\right.\nonumber\\
&&\left.+4^{\nu +2} A B \nu ^2 (i \tau )^{2 (\nu +1)} (\Gamma (-2 (\nu +1))-\Gamma (-2 (\nu +1),2 i L \tau ))\right.\nonumber\\
&&\left.+4^{\nu +1} A B (i \tau )^{2 (\nu +1)} (\Gamma (-2 (\nu +1))-\Gamma (-2 (\nu +1),2 i L \tau ))\right.\nonumber\\
&&\left.-4^{\nu } A B (-i \tau )^{2 \nu } \tau ^2 (\Gamma (2-2 \nu )-\Gamma (2-2 \nu ,-2 i L \tau ))\right.\nonumber\\
&&\left.-4^{\nu } A B (i \tau )^{2 \nu } \tau ^2 (\Gamma (2-2 \nu )-\Gamma (2-2 \nu ,2 i L \tau ))\right)\\
 X^{(3)}_1&=&-\frac{L^{-2 \nu } }{8 \tau ^4}\left(\frac{4 (A^2+B^2) \nu ^2}{L^2 (\nu +1)}+\frac{4 (A^2+B^2) L^2 \tau ^4}{\nu -1}+\frac{(A^2+B^2)}{L^2 (\nu +1)}-\frac{4 (A^2+B^2) \nu }{L^2 (\nu +1)}+\frac{4 (A^2+B^2) \tau ^2}{\nu }\right.\nonumber\\
&&\left.-8 (A^2+B^2) \tau ^2+A B 2^{2 \nu +5} \nu ^2 \tau ^2 (-i L \tau )^{2 \nu } \Gamma (-2 (\nu +1),-2 i L \tau )\right.\nonumber\\
&&\left.+A B 2^{2 \nu +5} \nu ^2 \tau ^2 (i L \tau )^{2 \nu } \Gamma (-2 (\nu +1),2 i L \tau )\right.\nonumber\\
&&\left.-\frac{A B 2^{2 \nu +1} \left(2 \nu ^3+7 \nu ^2+9 \nu +1\right) \tau ^2 \Gamma (2-2 \nu ) \left((-i L \tau )^{2 \nu }+(i L \tau )^{2 \nu }\right)}{\nu  (\nu +1) (2 \nu +1)}\right.\nonumber\\
&&\left.+A B \left(-2^{2 \nu +3}\right) \tau ^2 (-i L \tau )^{2 \nu } \Gamma (-2 \nu ,-2 i L \tau )+A B 4^{\nu +2} \nu  \tau ^2 (-i L \tau )^{2 \nu } \Gamma (-2 \nu ,-2 i L \tau )\right.\nonumber\\
&&\left.+A B 2^{2 \nu +3} \tau ^2 (-i L \tau )^{2 \nu } \Gamma (-2 (\nu +1),-2 i L \tau )-A B 2^{2 \nu +5} \nu  \tau ^2 (-i L \tau )^{2 \nu } \Gamma (-2 (\nu +1),-2 i L \tau )\right.\nonumber\\
&&\left.+A B 2^{2 \nu +1} \tau ^2 (-i L \tau )^{2 \nu } \Gamma (2-2 \nu ,-2 i L \tau )-A B 2^{2 \nu +3} \tau ^2 (i L \tau )^{2 \nu } \Gamma (-2 \nu ,2 i L \tau )\right.\nonumber\\
&&\left.+A B 4^{\nu +2} \nu  \tau ^2 (i L \tau )^{2 \nu } \Gamma (-2 \nu ,2 i L \tau )+A B 2^{2 \nu +3} \tau ^2 (i L \tau )^{2 \nu } \Gamma (-2 (\nu +1),2 i L \tau )\right.\nonumber\\
&&\left.-A B 2^{2 \nu +5} \nu  \tau ^2 (i L \tau )^{2 \nu } \Gamma (-2 (\nu +1),2 i L \tau )+A B 2^{2 \nu +1} \tau ^2 (i L \tau )^{2 \nu } \Gamma (2-2 \nu ,2 i L \tau )\right) \nonumber\\
&&\left(\frac{(A^2+B^2) L^{2-2 \nu }}{2 (\nu -1)}+A B 4^{\nu -1} (-i T)^{2 (\nu -1)} (\Gamma (2-2 \nu )-\Gamma (2-2 \nu ,-2 i L T))\right.\nonumber\\
&&\left.+A B 4^{\nu -1} (i T)^{2 (\nu -1)} (\Gamma (2-2 \nu )-\Gamma (2-2 \nu ,2 i L T))\right)\eea

\bea X^{(4)}_1&=&-\frac{L^{-4 \nu }}{4 \tau ^4} \left(B^2 (\Gamma (-2 \nu )-\Gamma (-2 \nu ,-i L (T-\tau ))) (-i L (T-\tau ))^{2 \nu }\right.\nonumber\\
&&\left.-2 B^2 \nu  (\Gamma (-2 \nu )-\Gamma (-2 \nu ,-i L (T-\tau ))) (-i L (T-\tau ))^{2 \nu }\right.\nonumber\\
&&\left.-\frac{2 B^2 \tau ^2 (\Gamma (2-2 \nu )-\Gamma (2-2 \nu ,-i L (T-\tau ))) (-i L (T-\tau ))^{2 \nu }}{(T-\tau )^2}\right.\nonumber\\
&&\left.+A^2 (i L (T-\tau ))^{2 \nu } (\Gamma (-2 \nu )-\Gamma (-2 \nu ,i L (T-\tau )))\right.\nonumber\\
&&\left.-2 A^2 \nu  (i L (T-\tau ))^{2 \nu } (\Gamma (-2 \nu )-\Gamma (-2 \nu ,i L (T-\tau )))\right.\nonumber\\
&&\left.-A B (-i L (T+\tau ))^{2 \nu } (\Gamma (-2 \nu )-\Gamma (-2 \nu ,-i L (T+\tau )))\right.\nonumber\\
&&\left.+2 A B \nu  (-i L (T+\tau ))^{2 \nu } (\Gamma (-2 \nu )-\Gamma (-2 \nu ,-i L (T+\tau )))\right.\nonumber\\
&&\left.-A B (i L (T+\tau ))^{2 \nu } (\Gamma (-2 \nu )-\Gamma (-2 \nu ,i L (T+\tau )))\right.\nonumber\\
&&\left.+2 A B \nu  (i L (T+\tau ))^{2 \nu } (\Gamma (-2 \nu )-\Gamma (-2 \nu ,i L (T+\tau )))\right.\nonumber\\
&&\left.+\frac{2 A B \tau ^2 (-i L (T+\tau ))^{2 \nu } (\Gamma (2-2 \nu )-\Gamma (2-2 \nu ,-i L (T+\tau )))}{(T+\tau )^2}\right.\nonumber\\
&&\left.+\frac{2 A B \tau ^2 (i L (T+\tau ))^{2 \nu } (\Gamma (2-2 \nu )-\Gamma (2-2 \nu ,i L (T+\tau )))}{(T+\tau )^2}\right.\nonumber\\
&&\left.-\frac{2 A^2 (i L (T-\tau ))^{2 \nu } \tau ^2 (\Gamma (2-2 \nu )-\Gamma (2-2 \nu ,i L (T-\tau )))}{(T-\tau )^2}\right)\nonumber\\
&& \left(A^2 (\Gamma (-2 \nu )-\Gamma (-2 \nu ,-i L (T-\tau ))) (-i L (T-\tau ))^{2 \nu }\right.\nonumber\\
&&\left.-2 A^2 \nu  (\Gamma (-2 \nu )-\Gamma (-2 \nu ,-i L (T-\tau ))) (-i L (T-\tau ))^{2 \nu }\right.\nonumber\\
&&\left.-\frac{2 A^2 \tau ^2 (\Gamma (2-2 \nu )-\Gamma (2-2 \nu ,-i L (T-\tau ))) (-i L (T-\tau ))^{2 \nu }}{(T-\tau )^2}\right.\nonumber\\
&&\left.+B^2 (i L (T-\tau ))^{2 \nu } (\Gamma (-2 \nu )-\Gamma (-2 \nu ,i L (T-\tau )))\right.\nonumber\\
&&\left.-2 B^2 \nu  (i L (T-\tau ))^{2 \nu } (\Gamma (-2 \nu )-\Gamma (-2 \nu ,i L (T-\tau )))\right.\nonumber\\
&&\left.-A B (-i L (T+\tau ))^{2 \nu } (\Gamma (-2 \nu )-\Gamma (-2 \nu ,-i L (T+\tau )))\right.\nonumber\\
&&\left.+2 A B \nu  (-i L (T+\tau ))^{2 \nu } (\Gamma (-2 \nu )-\Gamma (-2 \nu ,-i L (T+\tau )))\right.\nonumber\\
&&\left.-A B (i L (T+\tau ))^{2 \nu } (\Gamma (-2 \nu )-\Gamma (-2 \nu ,i L (T+\tau )))\right.\nonumber\\
&&\left.+2 A B \nu  (i L (T+\tau ))^{2 \nu } (\Gamma (-2 \nu )-\Gamma (-2 \nu ,i L (T+\tau )))\right.\nonumber\\
&&\left.+\frac{2 A B \tau ^2 (-i L (T+\tau ))^{2 \nu } (\Gamma (2-2 \nu )-\Gamma (2-2 \nu ,-i L (T+\tau )))}{(T+\tau )^2}\right.\nonumber\\
&&\left.+\frac{2 A B \tau ^2 (i L (T+\tau ))^{2 \nu } (\Gamma (2-2 \nu )-\Gamma (2-2 \nu ,i L (T+\tau )))}{(T+\tau )^2}\right.\nonumber\\
&&\left.-\frac{2 B^2 (i L (T-\tau ))^{2 \nu } \tau ^2 (\Gamma (2-2 \nu )-\Gamma (2-2 \nu ,i L (T-\tau )))}{(T-\tau )^2}\right)\eea

In the sub-horizon limit we get the following simplified results:

\bea X^{(1)}_1&=&\frac{L^{-4 \nu } T^2 }{4 \tau ^2}\left(\frac{A^2 (i L (T-\tau ))^{2 \nu } (\Gamma (2-2 \nu )-\Gamma (2-2 \nu ,i L (T-\tau )))}{(T-\tau )^2}\right.\nonumber\\
&&\left.+\frac{2 A^2 \tau  (i L (T-\tau ))^{2 \nu } (\Gamma (3-2 \nu )-\Gamma (3-2 \nu ,i L (T-\tau )))}{(T-\tau )^3}\right.\nonumber\\
&&\left.-\frac{2 A^2 \nu  (i L (T-\tau ))^{2 \nu } (\Gamma (2-2 \nu )-\Gamma (2-2 \nu ,i L (T-\tau )))}{(T-\tau )^2}\right.\nonumber\\
&&\left.+\frac{A B (-i L (\tau +T))^{2 \nu } (\Gamma (2-2 \nu )-\Gamma (2-2 \nu ,-i L (T+\tau )))}{(\tau +T)^2}\right.\nonumber\\
&&\left.+\frac{A B (i L (\tau +T))^{2 \nu } (\Gamma (2-2 \nu )-\Gamma (2-2 \nu ,i L (T+\tau )))}{(\tau +T)^2}\right.\nonumber\\
&&\left.-\frac{2 A B \nu  (-i L (\tau +T))^{2 \nu } (\Gamma (2-2 \nu )-\Gamma (2-2 \nu ,-i L (T+\tau )))}{(\tau +T)^2}\right.\nonumber\\
&&\left.-\frac{2 A B \nu  (i L (\tau +T))^{2 \nu } (\Gamma (2-2 \nu )-\Gamma (2-2 \nu ,i L (T+\tau )))}{(\tau +T)^2}\right.\nonumber\\
&&\left.-\frac{2 A B \tau  (-i L (\tau +T))^{2 \nu } (\Gamma (3-2 \nu )-\Gamma (3-2 \nu ,-i L (T+\tau )))}{(\tau +T)^3}\right.\nonumber\\
&&\left.-\frac{2 A B \tau  (i L (\tau +T))^{2 \nu } (\Gamma (3-2 \nu )-\Gamma (3-2 \nu ,i L (T+\tau )))}{(\tau +T)^3}\right.\nonumber\\
&&\left.+\frac{B^2 (-i L (T-\tau ))^{2 \nu } (\Gamma (2-2 \nu )-\Gamma (2-2 \nu ,-i L (T-\tau )))}{(T-\tau )^2}\right.\nonumber\\
&&\left.+\frac{2 B^2 \tau  (-i L (T-\tau ))^{2 \nu } (\Gamma (3-2 \nu )-\Gamma (3-2 \nu ,-i L (T-\tau )))}{(T-\tau )^3}\right.\nonumber\\
&&\left.-\frac{2 B^2 \nu  (-i L (T-\tau ))^{2 \nu } (\Gamma (2-2 \nu )-\Gamma (2-2 \nu ,-i L (T-\tau )))}{(T-\tau )^2}\right)\nonumber\\
&&(\frac{A^2 (-i L (T-\tau ))^{2 \nu } (\Gamma (2-2 \nu )-\Gamma (2-2 \nu ,-i L (T-\tau )))}{(T-\tau )^2}\nonumber\\
&&+\frac{2 A^2 \tau  (-i L (T-\tau ))^{2 \nu } (\Gamma (3-2 \nu )-\Gamma (3-2 \nu ,-i L (T-\tau )))}{(T-\tau )^3}\nonumber\\
&&-\frac{2 A^2 \nu  (-i L (T-\tau ))^{2 \nu } (\Gamma (2-2 \nu )-\Gamma (2-2 \nu ,-i L (T-\tau )))}{(T-\tau )^2}\nonumber\\
&&+\frac{A B (-i L (\tau +T))^{2 \nu } (\Gamma (2-2 \nu )-\Gamma (2-2 \nu ,-i L (T+\tau )))}{(\tau +T)^2}\nonumber\\
&&+\frac{A B (i L (\tau +T))^{2 \nu } (\Gamma (2-2 \nu )-\Gamma (2-2 \nu ,i L (T+\tau )))}{(\tau +T)^2}\nonumber\\
&&-\frac{2 A B \nu  (-i L (\tau +T))^{2 \nu } (\Gamma (2-2 \nu )-\Gamma (2-2 \nu ,-i L (T+\tau )))}{(\tau +T)^2}\nonumber\\
&&-\frac{2 A B \nu  (i L (\tau +T))^{2 \nu } (\Gamma (2-2 \nu )-\Gamma (2-2 \nu ,i L (T+\tau )))}{(\tau +T)^2}\nonumber\eea

\bea &&-\frac{2 A B \tau  (-i L (\tau +T))^{2 \nu } (\Gamma (3-2 \nu )-\Gamma (3-2 \nu ,-i L (T+\tau )))}{(\tau +T)^3}\nonumber\\
&&-\frac{2 A B \tau  (i L (\tau +T))^{2 \nu } (\Gamma (3-2 \nu )-\Gamma (3-2 \nu ,i L (T+\tau )))}{(\tau +T)^3}\nonumber\\
&&+\frac{B^2 (i L (T-\tau ))^{2 \nu } (\Gamma (2-2 \nu )-\Gamma (2-2 \nu ,i L (T-\tau )))}{(T-\tau )^2}\nonumber\\
&&+\frac{2 B^2 \tau  (i L (T-\tau ))^{2 \nu } (\Gamma (3-2 \nu )-\Gamma (3-2 \nu ,i L (T-\tau )))}{(T-\tau )^3}\nonumber\\
&&-\frac{2 B^2 \nu  (i L (T-\tau ))^{2 \nu } (\Gamma (2-2 \nu )-\Gamma (2-2 \nu ,i L (T-\tau )))}{(T-\tau )^2}) \eea

\bea X^{(2)}_1&=&-\frac{L^{-4 \nu } T^2 }{128 (\nu -1) \nu  \tau ^2}(-\frac{8 A^2 L^4}{\nu -2}+\frac{A B 4^{\nu } (-i L T)^{2 \nu } (\Gamma (4-2 \nu )-\Gamma (4-2 \nu ,-2 i L T))}{T^4}\nonumber\\
&&+\frac{A B 4^{\nu } (i L T)^{2 \nu } (\Gamma (4-2 \nu )-\Gamma (4-2 \nu ,2 i L T))}{T^4}-\frac{8 B^2 L^4}{\nu -2})\nonumber\\
&&(4 A^2 L^2 \nu  \tau ^2+4 A^2 \nu ^3-8 A^2 \nu ^2+5 A^2 \nu -A^2+A B 2^{2 \nu +3} \nu ^4 L^{2 \nu } (-i \tau )^{2 \nu } \Gamma (-2 \nu ,-2 i L \tau )\nonumber\\
&&+A B 2^{2 \nu +3} \nu ^4 L^{2 \nu } (i \tau )^{2 \nu } \Gamma (-2 \nu ,2 i L \tau )-A B 4^{\nu +2} \nu ^3 L^{2 \nu } (-i \tau )^{2 \nu } \Gamma (-2 \nu ,-2 i L \tau )\nonumber\\
&&-A B 4^{\nu +2} \nu ^3 L^{2 \nu } (i \tau )^{2 \nu } \Gamma (-2 \nu ,2 i L \tau )+A B 2^{2 \nu +3} \nu ^3 L^{2 \nu } (i \tau )^{2 \nu } \Gamma (1-2 \nu ,2 i L \tau )\nonumber\\
&&+5 A B 2^{2 \nu +1} \nu ^2 L^{2 \nu } (-i \tau )^{2 \nu } \Gamma (-2 \nu ,-2 i L \tau )+5 A B 2^{2 \nu +1} \nu ^2 L^{2 \nu } (i \tau )^{2 \nu } \Gamma (-2 \nu ,2 i L \tau )\nonumber\\
&&+A B 4^{\nu +1} \nu  \left(2 \nu ^2-3 \nu +1\right) L^{2 \nu } (-i \tau )^{2 \nu } \Gamma (1-2 \nu ,-2 i L \tau )\nonumber\\
&&-3 A B 4^{\nu +1} \nu ^2 L^{2 \nu } (i \tau )^{2 \nu } \Gamma (1-2 \nu ,2 i L \tau )+A B 2^{2 \nu +1} \nu ^2 L^{2 \nu } (-i \tau )^{2 \nu } \Gamma (2-2 \nu ,-2 i L \tau )\nonumber\\
&&+A B 2^{2 \nu +1} \nu ^2 L^{2 \nu } (i \tau )^{2 \nu } \Gamma (2-2 \nu ,2 i L \tau )+A B 4^{\nu } (\nu -1) L^{2 \nu } \left((-i \tau )^{2 \nu }+(i \tau )^{2 \nu }\right) \Gamma (2-2 \nu )\nonumber\\
&&-A B 2^{2 \nu +1} \nu  L^{2 \nu } (-i \tau )^{2 \nu } \Gamma (-2 \nu ,-2 i L \tau )-A B 2^{2 \nu +1} \nu  L^{2 \nu } (i \tau )^{2 \nu } \Gamma (-2 \nu ,2 i L \tau )\nonumber\\
&&+A B 4^{\nu +1} \nu  L^{2 \nu } (i \tau )^{2 \nu } \Gamma (1-2 \nu ,2 i L \tau )-A B 2^{2 \nu +1} \nu  L^{2 \nu } (-i \tau )^{2 \nu } \Gamma (2-2 \nu ,-2 i L \tau )\nonumber\\
&&-A B 2^{2 \nu +1} \nu  L^{2 \nu } (i \tau )^{2 \nu } \Gamma (2-2 \nu ,2 i L \tau )+4 B^2 L^2 \nu  \tau ^2+4 B^2 \nu ^3-8 B^2 \nu ^2+5 B^2 \nu -B^2)~~~~~~~~~~~\eea

\bea X^{(3)}_1&=&-\frac{T^2 L^{-2 \nu }}{8 \tau ^2} \left(\frac{A^2 L^{4-2 \nu }}{4-2 \nu }+\frac{A B 4^{\nu -2} (-i T)^{2 \nu } (\Gamma (4-2 \nu )-\Gamma (4-2 \nu ,-2 i L T))}{T^4}\right.\nonumber\\
&& \left.+\frac{A B 4^{\nu -2} (i T)^{2 \nu } (\Gamma (4-2 \nu )-\Gamma (4-2 \nu ,2 i L T))}{T^4}+\frac{B^2 L^{4-2 \nu }}{4-2 \nu }\right)\nonumber\\
&&(-\frac{4 A^2 L^2 \tau ^2}{\nu -1}-4 A^2 \nu -\frac{A^2}{\nu }+4 A^2+A B \left(-2^{2 \nu +3}\right) \nu ^2 (-i L \tau )^{2 \nu } \Gamma (-2 \nu ,-2 i L \tau )\nonumber\\
&&-A B 2^{2 \nu +3} \nu ^2 (i L \tau )^{2 \nu } \Gamma (-2 \nu ,2 i L \tau )-\frac{A B 4^{\nu } \Gamma (2-2 \nu ) \left((-i L \tau )^{2 \nu }+(i L \tau )^{2 \nu }\right)}{\nu }\nonumber\eea

\bea &&-A B 2^{2 \nu +1} (-i L \tau )^{2 \nu } \Gamma (-2 \nu ,-2 i L \tau )+A B 2^{2 \nu +3} \nu  (-i L \tau )^{2 \nu } \Gamma (-2 \nu ,-2 i L \tau )\nonumber\\
&&-A B 4^{\nu +1} (2 \nu -1) (-i L \tau )^{2 \nu } \Gamma (1-2 \nu ,-2 i L \tau )-A B 2^{2 \nu +1} (-i L \tau )^{2 \nu } \Gamma (2-2 \nu ,-2 i L \tau )\nonumber\\
&&-A B 2^{2 \nu +1} (i L \tau )^{2 \nu } \Gamma (-2 \nu ,2 i L \tau )+A B 2^{2 \nu +3} \nu  (i L \tau )^{2 \nu } \Gamma (-2 \nu ,2 i L \tau )\nonumber\\
&&+A B 4^{\nu +1} (i L \tau )^{2 \nu } \Gamma (1-2 \nu ,2 i L \tau )-A B 2^{2 \nu +3} \nu  (i L \tau )^{2 \nu } \Gamma (1-2 \nu ,2 i L \tau )\nonumber\\
&&-A B 2^{2 \nu +1} (i L \tau )^{2 \nu } \Gamma (2-2 \nu ,2 i L \tau )-\frac{4 B^2 L^2 \tau ^2}{\nu -1}-4 B^2 \nu -\frac{B^2}{\nu }+4 B^2)\eea

\bea X^{(4)}_1&=&-\frac{T^2 L^{-4 \nu }}{4 \tau ^2} \left(\frac{A^2 (i L (T-\tau ))^{2 \nu } (\Gamma (2-2 \nu )-\Gamma (2-2 \nu ,i L (T-\tau )))}{(T-\tau )^2}\right.\nonumber\\
&&\left.+\frac{2 A^2 \tau  (i L (T-\tau ))^{2 \nu } (\Gamma (3-2 \nu )-\Gamma (3-2 \nu ,i L (T-\tau )))}{(T-\tau )^3}\right.\nonumber\\
&&\left.-\frac{2 A^2 \nu  (i L (T-\tau ))^{2 \nu } (\Gamma (2-2 \nu )-\Gamma (2-2 \nu ,i L (T-\tau )))}{(T-\tau )^2}\right.\nonumber\\
&&\left.+\frac{A B (-i L (\tau +T))^{2 \nu } (\Gamma (2-2 \nu )-\Gamma (2-2 \nu ,-i L (T+\tau )))}{(\tau +T)^2}\right.\nonumber\\
&&\left.+\frac{A B (i L (\tau +T))^{2 \nu } (\Gamma (2-2 \nu )-\Gamma (2-2 \nu ,i L (T+\tau )))}{(\tau +T)^2}\right.\nonumber\\
&&\left.-\frac{2 A B \nu  (-i L (\tau +T))^{2 \nu } (\Gamma (2-2 \nu )-\Gamma (2-2 \nu ,-i L (T+\tau )))}{(\tau +T)^2}\right.\nonumber\\
&&\left.-\frac{2 A B \nu  (i L (\tau +T))^{2 \nu } (\Gamma (2-2 \nu )-\Gamma (2-2 \nu ,i L (T+\tau )))}{(\tau +T)^2}\right.\nonumber\\
&&\left.-\frac{2 A B \tau  (-i L (\tau +T))^{2 \nu } (\Gamma (3-2 \nu )-\Gamma (3-2 \nu ,-i L (T+\tau )))}{(\tau +T)^3}\right.\nonumber\\
&&\left.-\frac{2 A B \tau  (i L (\tau +T))^{2 \nu } (\Gamma (3-2 \nu )-\Gamma (3-2 \nu ,i L (T+\tau )))}{(\tau +T)^3}\right.\nonumber\\
&&\left.+\frac{B^2 (-i L (T-\tau ))^{2 \nu } (\Gamma (2-2 \nu )-\Gamma (2-2 \nu ,-i L (T-\tau )))}{(T-\tau )^2}\right.\nonumber\\
&&\left.+\frac{2 B^2 \tau  (-i L (T-\tau ))^{2 \nu } (\Gamma (3-2 \nu )-\Gamma (3-2 \nu ,-i L (T-\tau )))}{(T-\tau )^3}\right.\nonumber\\
&&\left.-\frac{2 B^2 \nu  (-i L (T-\tau ))^{2 \nu } (\Gamma (2-2 \nu )-\Gamma (2-2 \nu ,-i L (T-\tau )))}{(T-\tau )^2}\right)\nonumber\\
&&(\frac{A^2 (-i L (T-\tau ))^{2 \nu } (\Gamma (2-2 \nu )-\Gamma (2-2 \nu ,-i L (T-\tau )))}{(T-\tau )^2}\nonumber\\
&&+\frac{2 A^2 \tau  (-i L (T-\tau ))^{2 \nu } (\Gamma (3-2 \nu )-\Gamma (3-2 \nu ,-i L (T-\tau )))}{(T-\tau )^3}\nonumber\eea

\bea &&-\frac{2 A^2 \nu  (-i L (T-\tau ))^{2 \nu } (\Gamma (2-2 \nu )-\Gamma (2-2 \nu ,-i L (T-\tau )))}{(T-\tau )^2}\nonumber\\
&&+\frac{A B (-i L (\tau +T))^{2 \nu } (\Gamma (2-2 \nu )-\Gamma (2-2 \nu ,-i L (T+\tau )))}{(\tau +T)^2}\nonumber\\
&&+\frac{A B (i L (\tau +T))^{2 \nu } (\Gamma (2-2 \nu )-\Gamma (2-2 \nu ,i L (T+\tau )))}{(\tau +T)^2}\nonumber\\
&&-\frac{2 A B \nu  (-i L (\tau +T))^{2 \nu } (\Gamma (2-2 \nu )-\Gamma (2-2 \nu ,-i L (T+\tau )))}{(\tau +T)^2}\nonumber\\
&&-\frac{2 A B \nu  (i L (\tau +T))^{2 \nu } (\Gamma (2-2 \nu )-\Gamma (2-2 \nu ,i L (T+\tau )))}{(\tau +T)^2}\nonumber\\
&&-\frac{2 A B \tau  (-i L (\tau +T))^{2 \nu } (\Gamma (3-2 \nu )-\Gamma (3-2 \nu ,-i L (T+\tau )))}{(\tau +T)^3}\nonumber\\
&&-\frac{2 A B \tau  (i L (\tau +T))^{2 \nu } (\Gamma (3-2 \nu )-\Gamma (3-2 \nu ,i L (T+\tau )))}{(\tau +T)^3}\nonumber\\
&&+\frac{B^2 (i L (T-\tau ))^{2 \nu } (\Gamma (2-2 \nu )-\Gamma (2-2 \nu ,i L (T-\tau )))}{(T-\tau )^2}\nonumber\\
&&+\frac{2 B^2 \tau  (i L (T-\tau ))^{2 \nu } (\Gamma (3-2 \nu )-\Gamma (3-2 \nu ,i L (T-\tau )))}{(T-\tau )^3}\nonumber\\
&&-\frac{2 B^2 \nu  (i L (T-\tau ))^{2 \nu } (\Gamma (2-2 \nu )-\Gamma (2-2 \nu ,i L (T-\tau )))}{(T-\tau )^2}) \eea
 Here we have introduced two factors $A$ and $B$ which are defined as:
 \bea A&=& 2^{\nu-\frac{3}{2}}~\frac{1}{\sqrt{2}}\left|\frac{\Gamma(\nu)}{\Gamma\left(\frac{3}{2}\right)}\right|~\exp\left(-i\left\{\frac{\pi}{2}\left(\nu+\frac{1}{2}\right)\right\}\right)C_1,~~~\\
 B&=&2^{\nu-\frac{3}{2}}~\frac{1}{\sqrt{2}}\left|\frac{\Gamma(\nu)}{\Gamma\left(\frac{3}{2}\right)}\right|~\exp\left(i\left\{\frac{\pi}{2}\left(\nu+\frac{1}{2}\right)\right\}\right)C_2.\eea
 Here for general $\alpha$ vacua we choose $C_1=\cosh\alpha$ and $C_2=\sinh\alpha$ and for $\alpha=0$ we get $C_1=1$ and $C_2=0$, which is the result for Bunch Davies vacuum state.
 
\subsection{Computation of ${\cal I}_{2}(\tau_1,\tau_2)$}
\bea  \hll{{\cal I}_2(T,\tau)=\int^{L}_{k_1=0} k^2_1dk_1\int^{L}_{k_2=0} k^2_2dk_2{\cal E}_{13}({\bf k}_1,{\bf k}_2,-{\bf k}_1,-{\bf k}_2;T,\tau)=(-T)^{1-2\nu}(-\tau)^{3-2\nu}\sum^{4}_{i=1}X^{(i)}_{2}(T,\tau)},~~~~~~~~\eea 
where we define four time dependent functions, $X^{(i)}_{2}(T,\tau)~\forall~~i=1,2,3,4$, which are given by the following expressions:
\bea X^{(1)}_2&=&X^{(1)}_1,\\
X^{(2)}_2&=&X^{(2)}_1,\\
 X^{(3)}_2&=&X^{(2)}_1,\\
 X^{(4)}_2&=&X^{(4)}_1.\eea
Consequently, one can write:
\bea  \hll{{\cal I}_2(T,\tau)=(-T)^{1-2\nu}(-\tau)^{3-2\nu}\sum^{4}_{i=1}X^{(i)}_{1}(T,\tau)=(-1)^{4\nu}{\cal I}_1(T,\tau)}.~~~\eea

\subsection{Computation of ${\cal I}_{3}(\tau_1,\tau_2)$}
\bea  \hll{{\cal I}_3(T,\tau)=\int^{L}_{k_1=0} k^2_1dk_1\int^{L}_{k_2=0} k^2_2dk_2{\cal E}_{6}({\bf k}_1,{\bf k}_2,-{\bf k}_2,-{\bf k}_1;T,\tau)=\frac{(-T)^{1-2\nu}(-\tau)^{3-2\nu}}{(-1)^{2\nu}}\sum^{4}_{i=1}X^{(i)}_{3}(T,\tau)},~~~~~~~~~~\eea 
where we define four time dependent functions, $X^{(i)}_{3}(T,\tau)~\forall~~i=1,2,3,4$, which are given by the following expressions:
\bea X^{(1)}_3&=&X^{(1)}_1,\\
X^{(2)}_3&=&X^{(2)}_1,\\
 X^{(3)}_3&=&X^{(2)}_1,\eea\bea
 X^{(4)}_3&=&X^{(4)}_1.\eea
Consequently, one can write:
\bea  \hll{{\cal I}_3(T,\tau)=\frac{(-T)^{1-2\nu}(-\tau)^{3-2\nu}}{(-1)^{2\nu}}\sum^{4}_{i=1}X^{(i)}_{1}(T,\tau)=(-1)^{2\nu}{\cal I}_1(T,\tau)}.~~~\eea

\subsection{Computation of ${\cal I}_{4}(\tau_1,\tau_2)$}
\bea  \hll{{\cal I}_4(T,\tau)=\int^{L}_{k_1=0} k^2_1dk_1\int^{L}_{k_2=0} k^2_2dk_2{\cal E}_{7}({\bf k}_1,{\bf k}_2,-{\bf k}_1,-{\bf k}_2;T,\tau)=\frac{(-T)^{1-2\nu}(-\tau)^{3-2\nu}}{(-1)^{2\nu}}\sum^{4}_{i=1}X^{(i)}_{4}(T,\tau)},~~~~~~~~~~\eea 
where we define four time dependent functions, $X^{(i)}_{4}(T,\tau)~\forall~~i=1,2,3,4$, which are given by the following expressions:
\bea X^{(1)}_4&=&X^{(1)}_1,\\
X^{(2)}_4&=&X^{(2)}_1,\\
 X^{(3)}_4&=&X^{(2)}_1,\\
 X^{(4)}_4&=&X^{(4)}_1.\eea
Consequently, one can write:
\bea  \hll{{\cal I}_4(T,\tau)=\frac{(-T)^{1-2\nu}(-\tau)^{3-2\nu}}{(-1)^{2\nu}}\sum^{4}_{i=1}X^{(i)}_{1}(T,\tau)=(-1)^{2\nu}{\cal I}_1(T,\tau)}.~~~\eea

\subsection{Computation of ${\cal I}_{5}(\tau_1,\tau_2)$}
\bea  \hll{{\cal I}_5(T,\tau)=\int^{L}_{k_1=0} k^2_1dk_1\int^{L}_{k_2=0} k^2_2dk_2{\cal E}_{10}({\bf k}_1,{\bf k}_2,-{\bf k}_1,-{\bf k}_2;T,\tau)=\frac{(-T)^{1-2\nu}(-\tau)^{3-2\nu}}{(-1)^{2\nu}}\sum^{4}_{i=1}X^{(i)}_{5}(T,\tau)},~~~~~~~~~~\eea 
where we define four time dependent functions, $X^{(i)}_{5}(T,\tau)~\forall~~i=1,2,3,4$, which are given by the following expressions:
\bea X^{(1)}_5&=&X^{(1)}_1,\\
X^{(2)}_5&=&X^{(2)}_1,\\
 X^{(3)}_5&=&X^{(2)}_1,\\
 X^{(4)}_5&=&X^{(4)}_1.\eea
Consequently, one can write:
\bea  \hll{{\cal I}_5(T,\tau)=\frac{(-T)^{1-2\nu}(-\tau)^{3-2\nu}}{(-1)^{2\nu}}\sum^{4}_{i=1}X^{(i)}_{1}(T,\tau)=(-1)^{2\nu}{\cal I}_1(T,\tau)}.~~~\eea

\subsection{Computation of ${\cal I}_{6}(\tau_1,\tau_2)$}
\bea  \hll{{\cal I}_6(T,\tau)=\int^{L}_{k_1=0} k^2_1dk_1\int^{L}_{k_2=0} k^2_2dk_2{\cal E}_{11}({\bf k}_1,{\bf k}_2,-{\bf k}_2,-{\bf k}_1;T,\tau)=\frac{(-T)^{1-2\nu}(-\tau)^{3-2\nu}}{(-1)^{2\nu}}\sum^{4}_{i=1}X^{(i)}_{6}(T,\tau)},~~~~~~~~~~\eea 
where we define four time dependent functions, $X^{(i)}_{6}(T,\tau)~\forall~~i=1,2,3,4$, which are given by the following expressions:
\bea X^{(1)}_6&=&X^{(1)}_1,\\
X^{(2)}_6&=&X^{(2)}_1,\\
 X^{(3)}_6&=&X^{(2)}_1,\\
 X^{(4)}_6&=&X^{(4)}_1.\eea
Consequently, one can write:
\bea  \hll{{\cal I}_6(T,\tau)=\frac{(-T)^{1-2\nu}(-\tau)^{3-2\nu}}{(-1)^{2\nu}}\sum^{4}_{i=1}X^{(i)}_{1}(T,\tau)=(-1)^{2\nu}{\cal I}_1(T,\tau)}.~~~\eea

\subsection{Computation of ${\cal I}_{7}(\tau_1,\tau_2)$}
\bea  \hll{{\cal I}_7(T,\tau)=\int^{L}_{k_1=0} k^2_1dk_1\int^{L}_{k_2=0} k^2_2dk_2{\cal E}_{7}({\bf k}_1,-{\bf k}_1,{\bf k}_2,-{\bf k}_2;T,\tau)=\frac{(-T)^{1-2\nu}(-\tau)^{3-2\nu}}{(-1)^{2\nu}}\sum^{4}_{i=1}X^{(i)}_{7}(T,\tau)},~~~~~~~~~~\eea 
where we define four time dependent functions, $X^{(i)}_{7}(T,\tau)~\forall~~i=1,2,3,4$, which are given by the following expressions:
\bea X^{(1)}_7&=&X^{(1)}_1,\\
X^{(2)}_7&=&X^{(2)}_1,\\
 X^{(3)}_7&=&X^{(2)}_1,\\
 X^{(4)}_7&=&X^{(4)}_1.\eea
Consequently, one can write:
\bea  \hll{{\cal I}_7(T,\tau)=\frac{(-T)^{1-2\nu}(-\tau)^{3-2\nu}}{(-1)^{2\nu}}\sum^{4}_{i=1}X^{(i)}_{1}(T,\tau)=(-1)^{2\nu}{\cal I}_1(T,\tau)}.~~~\eea
 
\subsection{Computation of ${\cal I}_{8}(\tau_1,\tau_2)$}
\bea  \hll{{\cal I}_8(T,\tau)=\int^{L}_{k_1=0} k^2_1dk_1\int^{L}_{k_2=0} k^2_2dk_2{\cal E}_{10}({\bf k}_1,-{\bf k}_1,{\bf k}_2,-{\bf k}_2;T,\tau)=\frac{(-T)^{1-2\nu}(-\tau)^{3-2\nu}}{(-1)^{2\nu}}\sum^{4}_{i=1}X^{(i)}_{8}(T,\tau)},~~~~~~~~~~\eea 
where we define four time dependent functions, $X^{(i)}_{8}(T,\tau)~\forall~~i=1,2,3,4$, which are given by the following expressions:
\bea X^{(1)}_8&=&X^{(1)}_1,\\
X^{(2)}_8&=&X^{(2)}_1,\\
 X^{(3)}_8&=&X^{(2)}_1,\\
 X^{(4)}_8&=&X^{(4)}_1.\eea
Consequently, one can write:
\bea  \hll{{\cal I}_8(T,\tau)=\frac{(-T)^{1-2\nu}(-\tau)^{3-2\nu}}{(-1)^{2\nu}}\sum^{4}_{i=1}X^{(i)}_{1}(T,\tau)=(-1)^{2\nu}{\cal I}_1(T,\tau)}.~~~\eea

\subsection{Computation of ${\cal I}_{9}(\tau_1,\tau_2)$}
\bea  \hll{{\cal I}_9(T,\tau)=\int^{L}_{k_1=0} k^2_1dk_1\int^{L}_{k_2=0} k^2_2dk_2{\cal E}_{11}({\bf k}_1,-{\bf k}_1,{\bf k}_2,-{\bf k}_2;T,\tau)=\frac{(-T)^{1-2\nu}(-\tau)^{3-2\nu}}{(-1)^{2\nu}}\sum^{4}_{i=1}X^{(i)}_{9}(T,\tau)},~~~~~~~~~~\eea 
where we define four time dependent functions, $X^{(i)}_{9}(T,\tau)~\forall~~i=1,2,3,4$, which are given by the following expressions:
\bea X^{(1)}_9&=&X^{(1)}_1,\\
X^{(2)}_9&=&X^{(2)}_1,\\
 X^{(3)}_9&=&X^{(2)}_1,\\
 X^{(4)}_9&=&X^{(4)}_1.\eea
Consequently, one can write:
\bea  \hll{{\cal I}_9(T,\tau)=\frac{(-T)^{1-2\nu}(-\tau)^{3-2\nu}}{(-1)^{2\nu}}\sum^{4}_{i=1}X^{(i)}_{1}(T,\tau)=(-1)^{2\nu}{\cal I}_1(T,\tau)}.~~~\eea 

\newpage
\section{Computation of the normalization factor in four-point micro-canonical OTOC}

\subsection{Normalization factor of four-point micro-canonical  OTOC computed from rescaled field variable}
Further, our aim is to compute the normalisation factor of OTOC computed from the rescaled field variable $f$, which is given by the following expression:
\bea \hll{{\cal N}^{f}(\tau_1,\tau_2):=\frac{1}{\langle \hat{f}(\tau_1)\hat{f}(\tau_1)\rangle_{\beta} \langle \hat{\Pi}(\tau_2)\hat{\Pi}(\tau_2)\rangle_{\beta}}},\eea
 for this we need to explicitly evaluate the denominator of the above mentioned expression.
 
 Now, the product of the two thermal two point function is evaluated as:
 \bea \displaystyle \hll{\langle \hat{f}(\tau_1)\hat{f}(\tau_1)\rangle_{\beta} =\displaystyle \frac{1}{Z_{\alpha}(\beta;\tau_1)}{\rm Tr}\left[e^{-\beta \hat{H}(\tau_1)}\hat{f}({\bf x},\tau_1)\hat{f}({\bf x},\tau_1)\right]_{(\alpha)}},\eea
 \bea \hll{\displaystyle \langle \hat{\Pi}(\tau_2)\hat{\Pi}(\tau_2)\rangle_{\beta} =\displaystyle \frac{1}{Z_{\alpha}(\beta;\tau_2)}{\rm Tr}\left[e^{-\beta \hat{H}(\tau_2)}\hat{\Pi}({\bf x},\tau_2)\hat{\Pi}({\bf x},\tau_2)\right]_{(\alpha)}},\eea
 where the thermal partition function for cosmology computed for $\alpha$ vacua can be expressed as:
 \bea \hll{Z_{\alpha}(\beta;\tau_i)=\frac{Z_{\bf BD}(\beta;\tau_i)}{|\cosh\alpha|}=\frac{1}{|\cosh\alpha|}\exp\left(-\left(1+\frac{1}{2}\delta^{3}(0)\right)\int d^3{\bf k}~\ln\left(2\sinh\frac{\beta E_{\bf k}(\tau_i)}{2}\right)\right)~\forall~~i=1,2}.~~~~~\eea
 Next, we compute the expression for the numerator with respect to the $\alpha$ vacua, which is given by:
 \bea &&{\rm Tr}\left[e^{-\beta \hat{H}(\tau_1)}\hat{f}({\bf x},\tau_1)\hat{f}({\bf x},\tau_1)\right]_{(\alpha)}\nonumber\\ 
&&=\int d\Psi_{\alpha}~\langle \Psi_{\alpha}|e^{-\beta \hat{H}(\tau_1)}\hat{f}(\tau_1)\hat{f}(\tau_1)|\Psi_{\alpha}\rangle\nonumber\\ 
&&=\frac{1}{|\cosh\alpha|}\int d\Psi_{\bf BD}~\langle \Psi_{\bf BD}|\left\{\exp\left(\frac{i}{2}\tanh \alpha~\int \frac{d^3{\bf k}_1}{(2\pi)^3}~a_{{\bf k}_1}a^{\dagger}_{{\bf k}_1}\right)\right.\nonumber\\
&&\left.~~~~~~~~~~~~\exp\left(-\beta\int d^3{\bf k}~\left(a^{\dagger}_{\bf k}a_{\bf k}+\frac{1}{2}\delta^{3}(0)\right)E_{\bf k}(\tau_1)\right)\right.\nonumber\\
&&\left.~~~~~~~~~~~~\int\frac{d^3{\bf k}_3}{(2\pi)^3}\int\frac{d^3{\bf k}_4}{(2\pi)^3}\hat{f}_{{\bf k}_3}(\tau_1)\hat{f}_{{\bf k}_4}(\tau_1)~\exp\left(-\frac{i}{2}\tanh \alpha~\int \frac{d^3{\bf k}_2}{(2\pi)^3}~a^{\dagger}_{{\bf k}_2}a_{{\bf k}_2}\right)\right\}|\Psi_{\bf BD}\rangle\nonumber\eea\bea
&&=\frac{1}{|\cosh\alpha|}\int d\Psi_{\bf BD}~\langle \Psi_{\bf BD}|\left\{\exp\left(\frac{i}{2}\tanh \alpha~\int \frac{d^3{\bf k}_1}{(2\pi)^3}~a_{{\bf k}_1}a^{\dagger}_{{\bf k}_1}\right)\right.\nonumber\\
&&\left.~~~~~~~~~~~~~~~~~~~~~~~~~~~~~\exp\left(-\beta\int d^3{\bf k}~\left(a^{\dagger}_{\bf k}a_{\bf k}+\frac{1}{2}\delta^{3}(0)\right)E_{\bf k}(\tau_1)\right)\right.\nonumber\\
&&\left.~\int\frac{d^3{\bf k}_3}{(2\pi)^3}\int\frac{d^3{\bf k}_4}{(2\pi)^3}\exp\left(\left({\bf k}_3+{\bf k}_4\right).{\bf x}\right)\left[f_{{\bf k}_3}(\tau_1)~a_{{\bf k}_3}+f^{*}_{{\bf -k}_3}(\tau_1)~a^{\dagger}_{-{\bf k}_3}\right]\left[f_{{\bf k}_4}(\tau_1)~a_{{\bf k}_4}+f^{*}_{{\bf -k}_4}(\tau_1)~a^{\dagger}_{-{\bf k}_4}\right]  \right.\nonumber\\
&&\left.~~~~~~~~~~~~~~~~~~~~~~~~~~~~~~~~~~~~~~~~~~~~~~~~~~~~~\exp\left(-\frac{i}{2}\tanh \alpha~\int \frac{d^3{\bf k}_2}{(2\pi)^3}~a^{\dagger}_{{\bf k}_2}a_{{\bf k}_2}\right)\right\}|\Psi_{\bf BD}\rangle\nonumber\\
&&=\frac{1}{|\cosh\alpha|}\int d\Psi_{\bf BD}~\langle \Psi_{\bf BD}|\left\{\exp\left(\frac{i}{2}\tanh \alpha~\int \frac{d^3{\bf k}_1}{(2\pi)^3}~a_{{\bf k}_1}a^{\dagger}_{{\bf k}_1}\right)\right.\nonumber\\
&&\left.~~~~~~~~~~~~~~~~~~~~~~~~~~\exp\left(-\beta\int d^3{\bf k}~\left(a^{\dagger}_{\bf k}a_{\bf k}+\frac{1}{2}\delta^{3}(0)\right)E_{\bf k}(\tau_1)\right)\right.\nonumber\\
&&\left.~~~~~~~~~~\int\frac{d^3{\bf k}_3}{(2\pi)^3}\int\frac{d^3{\bf k}_4}{(2\pi)^3}\exp\left(\left({\bf k}_3+{\bf k}_4\right).{\bf x}\right) \left[f_{{\bf k}_3}(\tau_1)f_{{\bf k}_4}(\tau_1)~a_{{\bf k}_3}a_{{\bf k}_4}+f^{*}_{{\bf -k}_3}(\tau_1)f_{{\bf k}_4}(\tau_1)~a^{\dagger}_{-{\bf k}_3}a_{{\bf k}_4}\right.\right.\nonumber\\
&&\left.\left.~~~~~~~~~~~~~~~~~~~~~~~~~~~~~~~~~~~~~~~~~~~~~~+f_{{\bf k}_3}(\tau_1)f^{*}_{-{\bf k}_4}(\tau_1)~a_{{\bf k}_3}a^{\dagger}_{-{\bf k}_4}+f^{*}_{-{\bf k}_3}(\tau_1)f^{*}_{-{\bf k}_4}(\tau_1)~a^{\dagger}_{-{\bf k}_3}a^{\dagger}_{-{\bf k}_4}\right]\right.\nonumber\\
&&\left.~~~~~~~~~~~~~~~~~~~~~~~~~~~~~~~~~~~~~~~~~~~~~~~~~~~~~\exp\left(-\frac{i}{2}\tanh \alpha~\int \frac{d^3{\bf k}_2}{(2\pi)^3}~a^{\dagger}_{{\bf k}_2}a_{{\bf k}_2}\right)\right\}|\Psi_{\bf BD}\rangle.~~~~~~~~~~~~\eea
Now we will explicitly compute the individual contributions, which are given by: 
\bea \exp\left(-\frac{i}{2}\tanh \alpha~\int \frac{d^3{\bf k}_2}{(2\pi)^3}~a^{\dagger}_{{\bf k}_2}a_{{\bf k}_2}\right)|\Psi_{\bf BD}\rangle &=&\sum^{\infty}_{n=0}\frac{(-1)^n}{n!}\left(\frac{i}{2}\tanh \alpha~\int \frac{d^3{\bf k}_2}{(2\pi)^3}~a^{\dagger}_{{\bf k}_2}a_{{\bf k}_2}\right)^n|\Psi_{\bf BD}\rangle\nonumber\\
&=&\sum^{\infty}_{n=0}\frac{(-1)^n}{n!}\left(\frac{i}{2}\tanh \alpha~\int \frac{d^3{\bf k}_2}{(2\pi)^3}~\right)^n|\Psi_{\bf BD}\rangle\nonumber\\
&=&\exp\left(-\frac{i}{2}\tanh \alpha~\int \frac{d^3{\bf k}_2}{(2\pi)^3}~\right)|\Psi_{\bf BD}\rangle.,\\
 \langle \Psi_{\bf BD}|\exp\left(\frac{i}{2}\tanh \alpha~\int \frac{d^3{\bf k}_2}{(2\pi)^3}~a_{{\bf k}_2}a^{\dagger}_{{\bf k}_2}\right)&=&\left[\exp\left(-\frac{i}{2}\tanh \alpha~\int \frac{d^3{\bf k}_2}{(2\pi)^3}~a^{\dagger}_{{\bf k}_2}a_{{\bf k}_2}\right)|\Psi_{\bf BD}\rangle\right]^{\dagger}\nonumber\\
&=&\left[\exp\left(-\frac{i}{2}\tanh \alpha~\int \frac{d^3{\bf k}_2}{(2\pi)^3}~\right)|\Psi_{\bf BD}\rangle\right]^{\dagger}\nonumber\\
&=& \langle \Psi_{\bf BD}|\exp\left(\frac{i}{2}\tanh \alpha~\int \frac{d^3{\bf k}_2}{(2\pi)^3}~\right),\eea
and we have the following sets of results:
\bea
&&\int d\Psi_{\bf BD}~\langle \Psi_{\bf BD}|\exp\left(-\beta\int d^3{\bf k}~\left(a^{\dagger}_{\bf k}a_{\bf k}+\frac{1}{2}\delta^{3}(0)\right)E_{\bf k}(\tau_1)\right)a_{{\bf k}_3}a_{{\bf k}_4}|\Psi_{\bf BD}\rangle=0,\eea
\bea
&&\int d\Psi_{\bf BD}~\langle \Psi_{\bf BD}|\exp\left(-\beta\int d^3{\bf k}~\left(a^{\dagger}_{\bf k}a_{\bf k}+\frac{1}{2}\delta^{3}(0)\right)E_{\bf k}(\tau_1)\right)a^{\dagger}_{-{\bf k}_3}a_{{\bf k}_4}|\Psi_{\bf BD}\rangle\nonumber\\
&&~~~~~~~~~=(2\pi)^3\exp\left(-\left(1+\frac{1}{2}\delta^{3}(0)\right)\int d^3{\bf k}~\ln\left(2\sinh\frac{\beta E_{\bf k}(\tau_1)}{2}\right)\right)~\delta^{3}({\bf k}_3+{\bf k}_4),~~~~~~\\
&&\int d\Psi_{\bf BD}~\langle \Psi_{\bf BD}|\exp\left(-\beta\int d^3{\bf k}~\left(a^{\dagger}_{\bf k}a_{\bf k}+\frac{1}{2}\delta^{3}(0)\right)E_{\bf k}(\tau_1)\right)~a_{{\bf k}_3}a^{\dagger}_{-{\bf k}_4}|\Psi_{\bf BD}\rangle\nonumber\\
&&~~~~~~~~~=(2\pi)^3\exp\left(-\left(1+\frac{1}{2}\delta^{3}(0)\right)\int d^3{\bf k}~\ln\left(2\sinh\frac{\beta E_{\bf k}(\tau_1)}{2}\right)\right)~\delta^{3}({\bf k}_3+{\bf k}_4),~~~~~~\\
&&\int d\Psi_{\bf BD}~\langle \Psi_{\bf BD}|\exp\left(-\beta\int d^3{\bf k}~\left(a^{\dagger}_{\bf k}a_{\bf k}+\frac{1}{2}\delta^{3}(0)\right)E_{\bf k}(\tau_1)\right)~a^{\dagger}_{-{\bf k}_3}a^{\dagger}_{-{\bf k}_4}|\Psi_{\bf BD}\rangle=0.~~~~.\eea
Consequently, we can simplify the final result of the previously mentioned trace as given by the following expression:
 \bea &&{\rm Tr}\left[e^{-\beta \hat{H}(\tau_1)}\hat{f}({\bf x},\tau_1)\hat{f}({\bf x},\tau_1)\right]_{(\alpha)}\nonumber\\ 
&&=Z_{\alpha}(\beta;\tau_1)\int\frac{d^3{\bf k}_3}{(2\pi)^3}\int\frac{d^3{\bf k}_4}{(2\pi)^3}(2\pi)^3\delta^{3}({\bf k}_3+{\bf k}_4)\left[f_{{\bf k}_3}(\tau_1)f^{*}_{-{\bf k}_4}(\tau_1)+f^{*}_{-{\bf k}_3}(\tau_1)f_{{\bf k}_4}(\tau_1)\right]\nonumber\\
&&=\frac{Z_{\alpha}(\beta;\tau_1)}{\pi^2}~{\cal F}^{(\alpha)}_1(\tau_1),\eea 
where we define a regularised time dependent function ${\cal F}^{(\alpha)}_1(\tau_1)$ as:
\bea &&{\cal F}^{(\alpha)}_1(\tau_1):=\int^{L}_{0}dk_3~k^2_3~|f_{{\bf k}_3}(\tau_1)|^2,\nonumber\\
&&~=\frac{iAB}{32\tau^3_1} \left(\frac{2}{L}\right)^{2 \nu } \left[\frac{32 (A^2+B^2)}{iAB} \tau^3_1L^3\left(\frac{L^2 \tau^2_1}{5-2 \nu }+\frac{1}{3-2 \nu }\right)+\frac{(2 \nu -7) \Gamma (5-2 \nu ) \left((-i L \tau_1)^{2 \nu }-(i L \tau_1)^{2 \nu }\right)}{(2 \nu -3)}\right.\nonumber\\&& \left.~~~~~~~~~~~~~~~~~~+(i L \tau_1)^{2 \nu } \left\{ 4\Gamma (3-2 \nu ,2 i L \tau_1)+\Gamma (5-2 \nu ,2 i L \tau_1)+4 \Gamma (4-2 \nu ,2 i L \tau_1)\right\}\right.\nonumber\\&& \left.~~~~~~~~~~~~~~~-(-i L \tau_1)^{2 \nu } \left\{  4\Gamma (3-2 \nu ,2 i L \tau_1)+\Gamma (5-2 \nu ,-2 i L \tau_1)+4\Gamma (4-2 \nu ,-2 i L \tau_1)\right\}\right].~~~~~~~~~~~~\eea
Similarly, following the same steps one can show that:
\bea &&{\rm Tr}\left[e^{-\beta \hat{H}(\tau_2)}\hat{\Pi}({\bf x},\tau_2)\hat{\Pi}({\bf x},\tau_2)\right]_{(\alpha)}\nonumber\\ 
&&=Z_{\alpha}(\beta;\tau_2)\int\frac{d^3{\bf k}_3}{(2\pi)^3}\int\frac{d^3{\bf k}_4}{(2\pi)^3}(2\pi)^3\delta^{3}({\bf k}_3+{\bf k}_4)\left[\Pi_{{\bf k}_3}(\tau_2)\Pi^{*}_{-{\bf k}_4}(\tau_2)+\Pi^{*}_{-{\bf k}_3}(\tau_2)\Pi_{{\bf k}_4}(\tau_2)\right]\nonumber\\
&&=\frac{Z_{\alpha}(\beta;\tau_2)}{\pi^2}~{\cal F}^{(\alpha)}_2(\tau_2),\eea 
where we define a regularised time dependent function ${\cal F}^{(\alpha)}_2(\tau_2)$ as:
\bea &&{\cal F}^{(\alpha)}_2(\tau_2):=\int^{L}_{0}dk_3~k^2_3~|\Pi_{{\bf k}_3}(\tau_2)|^2,\nonumber\\
&&=\frac{1}{4 \tau ^4_2}L^{-2 \nu } \left[-i 4^{\nu +1} A B \tau ^3_2 \Gamma (-2 \nu ,-2 i L \tau_2 ) (-i L \tau_2 )^{2 \nu }+2^{2 \nu +3} A B i \nu  \tau ^3_2 \Gamma (-2 \nu ,-2 i L \tau_2 ) (-i L \tau_2 )^{2 \nu }\right.\nonumber\\ &&\left.+2^{2 \nu +5} A B i \nu ^2 \tau ^3_2 \Gamma (-2 (\nu +1),-2 i L \tau_2 ) (-i L \tau_2 )^{2 \nu }+2^{2 \nu +3} A B i \tau ^3_2 \Gamma (-2 (\nu +1),-2 i L \tau_2 ) (-i L \tau_2 )^{2 \nu }\right.\nonumber\\ &&\left.-i 2^{2 \nu +5} A B \nu  \tau ^3_2 \Gamma (-2 (\nu +1),-2 i L \tau_2 ) (-i L \tau_2 )^{2 \nu }+2^{2 \nu +3} A B i (1-2 \nu )^2 \tau ^3_2 \Gamma (-2 \nu -3,-2 i L \tau_2 ) (-i L \tau_2 )^{2 \nu }\right.\nonumber\\ &&\left.+2^{2 \nu +3} A B i \nu ^2 \tau ^3_2 \Gamma (-2 \nu -1,-2 i L \tau_2 ) (-i L \tau_2 )^{2 \nu }-3 i 2^{2 \nu +1} A B \tau ^3_2 \Gamma (-2 \nu -1,-2 i L \tau_2 ) (-i L \tau_2 )^{2 \nu }\right.\nonumber\\ &&\left.+2^{2 \nu +3} A B i \nu  \tau ^3_2 \Gamma (-2 \nu -1,-2 i L \tau_2 ) (-i L \tau_2 )^{2 \nu }+2^{2 \nu +1} A B i \tau ^3_2 \Gamma (1-2 \nu ,-2 i L \tau_2 ) (-i L \tau_2 )^{2 \nu }\right.\nonumber\\ &&\left.+\frac{4 (A^2+B^2) L \tau ^4_2}{1-2 \nu }-4 i B^2 \tau ^3_2+\frac{2 B^2 i \tau ^3_2}{\nu }\right.\nonumber\\ &&\left.+\frac{\tau ^2_2[4\nu^2(2A^2+B^2)+B^2(4\nu-3)-5A^2]}{L(2 \nu+1)}+\frac{B^2 i \tau_2[4\nu(\nu-1)+1]}{L^2 (\nu +1)}\right.\nonumber\\ &&\left.+4^{\nu +1} A B i \tau ^3_2 \left(7 (-i L \tau_2 )^{2 \nu }-13 (i L \tau_2 )^{2 \nu }+4 \nu ^3 \left((-i L \tau_2 )^{2 \nu }+7 (i L \tau_2 )^{2 \nu }\right)\right.\right.\nonumber\\ &&\left.\left.+\nu  \left(9 (i L \tau_2 )^{2 \nu }-25 (-i L \tau_2 )^{2 \nu }\right)+\nu ^2 \left(68 (i L \tau_2 )^{2 \nu }-28 (-i L \tau_2 )^{2 \nu }\right)\right) \Gamma (-2 \nu -3)\right.\nonumber\\ &&\left.-i 2^{2 \nu +5} A B \nu ^2 \tau ^3_2 (i L \tau_2 )^{2 \nu } \Gamma (-2 \nu -3,2 i L \tau_2 )-i 2^{2 \nu +3} A B \tau ^3_2 (i L \tau_2 )^{2 \nu } \Gamma (-2 \nu -3,2 i L \tau_2 )\right.\nonumber\\ &&\left.+2^{2 \nu +5} A B i \nu  \tau ^3_2 (i L \tau_2 )^{2 \nu } \Gamma (-2 \nu -3,2 i L \tau )+2^{2 \nu +3} A B i \nu ^2 \tau ^3 (i L \tau )^{2 \nu } \Gamma (-2 \nu -1,2 i L \tau_2 )\right.\nonumber\\ &&\left.+5\ 2^{2 \nu +1} A B i \tau ^3_2 (i L \tau_2 )^{2 \nu } \Gamma (-2 \nu -1,2 i L \tau_2 )-3 i 2^{2 \nu +3} A B \nu  \tau ^3_2 (i L \tau_2 )^{2 \nu } \Gamma (-2 \nu -1,2 i L \tau_2 )\right.\nonumber\\ &&\left.-i 2^{2 \nu +1} A B \tau ^3_2 (i L \tau_2 )^{2 \nu } \Gamma (1-2 \nu ,2 i L \tau_2 )+\frac{4 (A^2+B^2) [\nu(1-\nu)-1] }{L^3 (2 \nu +3)}\right],\eea
Then we have found the following expression:
  \bea &&\hll{\displaystyle \langle \hat{f}(\tau_1)\hat{f}(\tau_1)\rangle_{\beta} =\displaystyle \frac{1}{Z_{\alpha}(\beta;\tau_1)}{\rm Tr}\left[e^{-\beta \hat{H}(\tau_1)}\hat{f}({\bf x},\tau_1)\hat{f}({\bf x},\tau_1)\right]_{(\alpha)}=\frac{1}{\pi^2}{\cal F}^{(\alpha)}_1(\tau_1)},\\
 &&\hll{\displaystyle \langle \hat{\Pi}(\tau_2)\hat{\Pi}(\tau_2)\rangle_{\beta} =\displaystyle \frac{1}{Z_{\alpha}(\beta;\tau_2)}{\rm Tr}\left[e^{-\beta \hat{H}(\tau_2)}\hat{\Pi}({\bf x},\tau_2)\hat{\Pi}({\bf x},\tau_2)\right]_{(\alpha)}=\frac{1}{\pi^2}{\cal F}^{(\alpha)}_2(\tau_2)} .\eea
Consequently, the normalisation factor of OTOC for the rescaled field variable can be computed as:
\bea \hll{{\cal N}^{f}(\tau_1,\tau_2)=\frac{1}{\langle \hat{f}(\tau_1)\hat{f}(\tau_1)\rangle_{\beta} \langle \hat{\Pi}(\tau_2)\hat{\Pi}(\tau_2)\rangle_{\beta}}=\frac{\pi^4}{{\cal F}^{(\alpha)}_1(\tau_1){\cal F}^{(\alpha)}_2(\tau_2)}}.\eea
 \subsection{Normalization factor of four-point micro-canonical OTOC computed from curvature perturbation field variable}
 Now, we are going to perform the similar computation when we express the normalisation factor of OTOC written  in terms of the scalar curvature perturbation field variable:
  \bea \hll{{\cal N}^{\zeta}(\tau_1,\tau_2):=\frac{1}{\langle \hat{\zeta}(\tau_1)\hat{\zeta}(\tau_1)\rangle_{\beta} \langle \hat{\Pi}_{\zeta}(\tau_2)\hat{\Pi}_{\zeta}(\tau_2)\rangle_{\beta}}},\eea
 for this we need to explicitly evaluate the denominator of the above mentioned expression.
 
 Now, the product of the two thermal two point function written in terms of curvature perturbation and its canonically conjugate momenta are evaluated as:
 \bea \hll{\displaystyle \langle \hat{\zeta}(\tau_1)\hat{\zeta}(\tau_1)\rangle_{\beta} =\displaystyle \frac{1}{Z_{\alpha}(\beta;\tau_1)}{\rm Tr}\left[e^{-\beta \hat{H}(\tau_1)}\hat{\zeta}({\bf x},\tau_1)\hat{\zeta}({\bf x},\tau_1)\right]_{(\alpha)}},\\
\hll{\displaystyle \langle \hat{\Pi}(\tau_2)\hat{\Pi}(\tau_1)\rangle_{\beta} =\displaystyle \frac{1}{Z_{\alpha}(\beta;\tau_2)}{\rm Tr}\left[e^{-\beta \hat{H}(\tau_2)}\hat{\Pi}({\bf x},\tau_2)\hat{\Pi}({\bf x},\tau_2)\right]_{(\alpha)}},\eea
  where the thermal partition function for cosmology in terms of curvature perturbation computed for $\alpha$ vacua can be expressed as:
 \bea \hll{Z^{\zeta}_{\alpha}(\beta;\tau_i)=\frac{Z^{\zeta}_{\bf BD}(\beta;\tau_i)}{|\cosh\alpha|}~~~~~~~\forall~~i=1,2}.~~~~~\eea
 Here the thermal partition function for cosmology in terms of curvature perturbation computed for Bunch Davies vacuum as:
 \bea \hll{Z^{\zeta}_{\bf BD}(\beta;\tau_i)\approx\exp\left(-\left(1+\frac{1}{2}\delta^{3}(0)\right)\int d^3{\bf k}~\ln\left(2\sinh\frac{\beta z^2(\tau_i)E_{{\bf k},\zeta}(\tau_i)}{2}\right)\right)~\forall~~i=1,2,}~~~~~~~~~~\eea
 where we define the time dependent energy dispersion relation in terms of the curvature perturbation variable as:
\bea E_{{\bf k},\zeta}(\tau_i):&=&\left|\Pi^{\zeta}_{\bf k}(\tau_i)\right|^2+\left(\omega^2_{\bf k}(\tau_i)+\left(\frac{1}{z(\tau_1)}\frac{dz(\tau_i)}{d\tau_i}\right)^2\right)|\zeta_{\bf k}(\tau_i)|^2,\nonumber\\
&=&\left|\Pi^{\zeta}_{\bf k}(\tau_i)\right|^2+\left(k^2-\frac{1}{z(\tau_i)}\frac{d^2z(\tau_i)}{d\tau^2_i}+\left(\frac{1}{z(\tau_i)}\frac{dz(\tau_i)}{d\tau_i}\right)^2\right)|\zeta_{\bf k}(\tau_i)|^2~\forall~~i=1,2~~~~~~~~~~\eea
Consequently, we can simplify the final result of the previously mentioned trace in terms of the curvature perturbation and its canonically conjugate momenta as given by the following expression:
\bea {\rm Tr}\left[e^{-\beta \hat{H}(\tau_1)}\hat{\zeta}({\bf x},\tau_1)\hat{\zeta}({\bf x},\tau_1)\right]_{(\alpha)}&=&\frac{Z^{\zeta}_{\alpha}(\beta;\tau_1)}{\pi^2z^2(\tau_1)}~{\cal F}^{(\alpha)}_1(\tau_1),\\
{\rm Tr}\left[e^{-\beta \hat{H}(\tau_2)}\hat{\Pi}_{\zeta}({\bf x},\tau_2)\hat{\Pi}_{\zeta}({\bf x},\tau_2)\right]_{(\alpha)}&=&\frac{Z^{\zeta}_{\alpha}(\beta;\tau_2)}{\pi^2z^2(\tau_2)}~{\cal F}^{(\alpha)}_2(\tau_2),\eea 
Then we have found the following expressions for the thermal two point functions:
  \bea &&\hll{\displaystyle \langle \hat{\zeta}(\tau_1)\hat{\zeta}(\tau_1)\rangle_{\beta} =\displaystyle \frac{1}{Z^{\zeta}_{\alpha}(\beta;\tau_1)}{\rm Tr}\left[e^{-\beta \hat{H}(\tau_1)}\hat{\zeta}({\bf x},\tau_1)\hat{\zeta}({\bf x},\tau_1)\right]_{(\alpha)}=\frac{1}{\pi^2z^2(\tau_1)}{\cal F}^{(\alpha)}_1(\tau_1)},\\
 &&\hll{\displaystyle \langle \hat{\Pi}_{\zeta}(\tau_2)\hat{\Pi}_{\zeta}(\tau_2)\rangle_{\beta} =\displaystyle \frac{1}{Z^{\zeta}_{\alpha}(\beta;\tau_2)}{\rm Tr}\left[e^{-\beta \hat{H}(\tau_2)}\hat{\Pi}_{\zeta}({\bf x},\tau_2)\hat{\Pi}_{\zeta}({\bf x},\tau_2)\right]_{(\alpha)}=\frac{1}{\pi^2z^2(\tau_2)}{\cal F}^{(\alpha)}_2(\tau_2)}~.~~~~~~~~~~\eea
 This further implies that the connection between the two point thermal correlation functions computed from the rescaled variable and curvature perturbation variable and their conjugate momenta are given by:
   \bea &&\hll{\displaystyle \langle \hat{f}(\tau_1)\hat{f}(\tau_1)\rangle_{\beta}=z^2(\tau_1)\langle \hat{\zeta}(\tau_1)\hat{\zeta}(\tau_1)\rangle_{\beta}},\eea\bea
&& \hll{\displaystyle \langle \hat{\Pi}(\tau_2)\hat{\Pi}(\tau_2)\rangle_{\beta}=z^2(\tau_2)\langle \hat{\Pi}_{\zeta}(\tau_2)\hat{\Pi}_{\zeta}(\tau_2)\rangle_{\beta}}.~~~~~~~\eea
   Consequently, the normalisation factor of OTOC for the curvature perturbation variable can be computed as:
  \bea\hll{ {\cal N}^{\zeta}(\tau_1,\tau_2)=\frac{1}{\langle \hat{\zeta}(\tau_1)\hat{\zeta}(\tau_1)\rangle_{\beta} \langle \hat{\Pi}_{\zeta}(\tau_2)\hat{\Pi}_{\zeta}(\tau_2)\rangle_{\beta}}=\frac{\pi^4z^2(\tau_1)z^2(\tau_2)}{{\cal F}^{(\alpha)}_1(\tau_1){\cal F}^{(\alpha)}_2(\tau_2)}=z^2(\tau_1)z^2(\tau_2){\cal N}^{f}(\tau_1,\tau_2)}.~~~~~~~~~~~\eea
  \section{Computation of the normalization factor in classical limit of four-point micro-canonical OTOC}

\subsection{Normalization factor of the classical version of four-point micro-canonical OTOC computed from rescaled field variable} 
Further, our aim is to compute the normalisation factor of the classical version of OTOC computed from the rescaled field variable $f$, which is given by the following expression: 
\bea \hll{{\cal N}^{f}_{\bf Classical}(\tau_1,\tau_2):=\frac{1}{\langle f(\tau_1)f(\tau_1)\rangle_{\beta} \langle{\Pi}(\tau_2{\Pi}(\tau_2)\rangle_{\beta}}},\eea
 for this we need to explicitly evaluate the denominator of the above mentioned expression.
 
 Now, the two thermal two point functions in the classical limit are evaluated as:
 \bea \displaystyle \hll{\langle{f}(\tau_1){f}(\tau_1)\rangle_{\beta} =\displaystyle \frac{1}{Z_{\bf Classical}(\beta;\tau_1)}\int\int\frac{{\cal D}f{\cal D}\Pi}{2\pi}e^{-\beta {H}(\tau_1)}\left\{{f}({\bf x},\tau_1),{f}({\bf x},\tau_1)\right\}_{\bf PB}},\eea
 \bea \hll{\displaystyle \langle {\Pi}(\tau_2){\Pi}(\tau_2)\rangle_{\beta} =\displaystyle \frac{1}{Z_{\bf Classical}(\beta;\tau_2)}\int\int\frac{{\cal D}f{\cal D}\Pi}{2\pi}e^{-\beta {H}(\tau_2)}\left\{{\Pi}({\bf x},\tau_2),{\Pi}({\bf x},\tau_2)\right\}_{\bf PB}},\eea
 where the thermal partition function for cosmology in the classical limit is computed as:
 \bea \hll{Z_{\bf Classical}(\beta;\tau_i)=\exp\left(-\int d^3{\bf k}~\ln\left(2\sinh\frac{\beta E_{\bf k}(\tau_i)}{2}\right)\right)~\forall~~i=1,2}.~~~~~\eea
 Now we compute the Poission brackets as:
\bea \left\{{f}({\bf x},\tau_1),{f}({\bf x},\tau_1)\right\}_{\bf PB}&=&\int\frac{d^3{\bf k}_1}{(2\pi)^3}\int\frac{d^3{\bf k}_2}{(2\pi)^3}~\exp(i({\bf k}_1+{\bf k}_2).{\bf x})\left\{f_{{\bf k}_1}(\tau_1),f_{{\bf k}_2}(\tau_1)\right\}_{\bf PB}\nonumber\\
&=&{\bf W}(0)\exp\left(-\lambda_{f}|\tau_1|\right)\int\frac{d^3{\bf k}_1}{(2\pi)^3}\int\frac{d^3{\bf k}_2}{(2\pi)^3}~\exp(i({\bf k}_1+{\bf k}_2).{\bf x})~(2\pi)^3~\delta^{3}({\bf k}_1+{\bf k}_2)\nonumber\\
&=&{\bf W}(0)\exp\left(-\lambda_{f}|\tau_1|\right)\int\frac{d^3{\bf k}_1}{(2\pi)^3}=\frac{L^3}{6\pi^2}{\bf W}(0)\exp\left(-\lambda_{f}|\tau_1|\right)\eea
\bea \left\{{\Pi}({\bf x},\tau_2),{\Pi}({\bf x},\tau_2)\right\}_{\bf PB}&=&\int\frac{d^3{\bf k}_1}{(2\pi)^3}\int\frac{d^3{\bf k}_2}{(2\pi)^3}~\exp(i({\bf k}_1+{\bf k}_2).{\bf x})\left\{\Pi_{{\bf k}_1}(\tau_2),\Pi_{{\bf k}_2}(\tau_2)\right\}_{\bf PB}\nonumber\\
&=&{\bf W}(0)\exp\left(-\lambda_{f}|\tau_2|\right)\int\frac{d^3{\bf k}_1}{(2\pi)^3}\int\frac{d^3{\bf k}_2}{(2\pi)^3}~\exp(i({\bf k}_1+{\bf k}_2).{\bf x})~(2\pi)^3~\delta^{3}({\bf k}_1+{\bf k}_2)\nonumber\\
&=&{\bf W}(0)\exp\left(-\lambda_{f}|\tau_2|\right)\int\frac{d^3{\bf k}_1}{(2\pi)^3}=\frac{L^3}{6\pi^2}{\bf W}(0)\exp\left(-\lambda_{f}|\tau_2|\right)\eea
Then we have found the following expression:
  \bea &&\hll{\displaystyle \langle {f}(\tau_1){f}(\tau_1)\rangle_{\beta} =\displaystyle \frac{L^3}{6\pi^2}{\bf W}(0)\exp\left(-\lambda_{f}|\tau_1|\right)},\\
 &&\hll{\displaystyle \langle {\Pi}(\tau_2){\Pi}(\tau_2)\rangle_{\beta} =\displaystyle \frac{L^3}{6\pi^2}{\bf W}(0)\exp\left(-\lambda_{f}|\tau_2|\right)} .\eea
Consequently, the normalisation factor of classical limit of OTOC for the rescaled field variable can be computed as:
\bea \hll{{\cal N}^{f}_{\bf Classical}(\tau_1,\tau_2)=\frac{36\pi^4}{L^6{\bf W}^2(0)\exp\left(-\lambda_{f}[|\tau_1|+|\tau_2|]\right)}=\frac{36\pi^4}{L^6{\bf G}_{\bf Kernel}(0)\exp\left(-\lambda_{f}[|\tau_1|+|\tau_2|]\right)}}~.~~~~~~\eea
Now, considering the examples of non-Gaussian coloured noise and Gaussian white noise we get the following answer for the normalization factor:
\begin{eqnarray}
&& \hll{{\cal N}^{f}_{\bf Classical}(\tau_1,\tau_2)= \large \left\{
     \begin{array}{lr}
   \displaystyle\frac{36\gamma \pi^4}{L^6{\bf A}\exp\left(-\lambda_{f}[|\tau_1|+|\tau_2|]\right)}~, &~ \text{\textcolor{red}{\bf Coloured~Noise}}\\  \\
   \displaystyle   0 & \text{\textcolor{red}{\bf White~Noise}}  \end{array}
   \right.}~~~~~~~~~~
\end{eqnarray}
 \subsection{Normalization factor of the classical version of four-point micro-canonical OTOC computed from curvature perturbation field variable} 
 Now, we are going to perform the similar computation when we express the normalisation factor of OTOC in the classical limit written  in terms of the scalar curvature perturbation field variable:
  \bea \hll{{\cal N}^{\zeta}_{\bf Classical}(\tau_1,\tau_2):=\frac{1}{\langle {\zeta}(\tau_1){\zeta}(\tau_1)\rangle_{\beta} \langle {\Pi}_{\zeta}(\tau_2){\Pi}_{\zeta}(\tau_2)\rangle_{\beta}}},\eea
 for this we need to explicitly evaluate the denominator of the above mentioned expression.
 
 Now, the product of the two thermal two point function written in terms of curvature perturbation and its canonically conjugate momenta are evaluated as:
 \bea \hll{\displaystyle \langle {\zeta}(\tau_1){\zeta}(\tau_1)\rangle_{\beta} =\displaystyle \frac{1}{Z_{\bf Classical}(\beta;\tau_1)}\int\int\frac{{\cal D}f{\cal D}\Pi}{2\pi}e^{-\beta {H}(\tau_1)}\left\{{\zeta}({\bf x},\tau_1),{\zeta}({\bf x},\tau_1)\right\}_{\bf PB}},\\
\hll{\displaystyle \langle \hat{\Pi}(\tau_2)\hat{\Pi}(\tau_2)\rangle_{\beta} =\displaystyle \frac{1}{Z_{\bf Classical}(\beta;\tau_2)}\int\int\frac{{\cal D}f{\cal D}\Pi}{2\pi}e^{-\beta {H}(\tau_2)}\left\{{\Pi}({\bf x},\tau_2),{\Pi}({\bf x},\tau_2)\right\}_{\bf PB}},\eea
  where the thermal partition function for cosmology in the classical limit in terms of curvature perturbation computed as:
 \bea \hll{Z^{\zeta}_{\bf Classical}(\beta;\tau_i)=\exp\left(-\int d^3{\bf k}~\ln\left(2\sinh\frac{\beta z^2(\tau_i)E_{{\bf k},\zeta}(\tau_i)}{2}\right)\right)~\forall~~i=1,2,}~~~~~~~~~~\eea
 where we define the time dependent energy dispersion relation in terms of the curvature perturbation variable as:
\bea E_{{\bf k},\zeta}(\tau_i):&=&\left|\Pi^{\zeta}_{\bf k}(\tau_i)\right|^2+\left(\omega^2_{\bf k}(\tau_i)+\left(\frac{1}{z(\tau_1)}\frac{dz(\tau_i)}{d\tau_i}\right)^2\right)|\zeta_{\bf k}(\tau_i)|^2,\nonumber\\
&=&\left|\Pi^{\zeta}_{\bf k}(\tau_i)\right|^2+\left(k^2-\frac{1}{z(\tau_i)}\frac{d^2z(\tau_i)}{d\tau^2_i}+\left(\frac{1}{z(\tau_i)}\frac{dz(\tau_i)}{d\tau_i}\right)^2\right)|\zeta_{\bf k}(\tau_i)|^2~\forall~~i=1,2~~~~~~~~~~\eea
 Now we compute the Poission brackets as:
\bea \left\{{\zeta}({\bf x},\tau_1),{\zeta}({\bf x},\tau_1)\right\}_{\bf PB}&=&\int\frac{d^3{\bf k}_1}{(2\pi)^3}\int\frac{d^3{\bf k}_2}{(2\pi)^3}~\exp(i({\bf k}_1+{\bf k}_2).{\bf x})\left\{\zeta_{{\bf k}_1}(\tau_1),\zeta_{{\bf k}_2}(\tau_1)\right\}_{\bf PB}\nonumber\\
&=&\frac{L^3}{6\pi^2 z^2(\tau_1)}{\bf W}(0)\exp\left(-\lambda_{f}|\tau_1|\right)\\
 \left\{{\Pi}_{\zeta}({\bf x},\tau_2),{\Pi}_{\zeta}({\bf x},\tau_2)\right\}_{\bf PB}&=&\int\frac{d^3{\bf k}_1}{(2\pi)^3}\int\frac{d^3{\bf k}_2}{(2\pi)^3}~\exp(i({\bf k}_1+{\bf k}_2).{\bf x})\left\{\Pi_{\zeta;{\bf k}_1}(\tau_2),\Pi_{\zeta;{\bf k}_2}(\tau_2)\right\}_{\bf PB}\nonumber\\
&=&\frac{L^3}{6\pi^2 z^2(\tau_2)}{\bf W}(0)\exp\left(-\lambda_{f}|\tau_2|\right)\eea 
Then we have found the following expressions for the thermal two point functions:
  \bea &&\hll{\displaystyle \langle{\zeta}(\tau_1){\zeta}(\tau_1)\rangle_{\beta} =\displaystyle \frac{L^3}{6\pi^2 z^2(\tau_1)}{\bf W}(0)\exp\left(-\lambda_{f}|\tau_1|\right)},\\
 &&\hll{\displaystyle \langle {\Pi}_{\zeta}(\tau_2){\Pi}_{\zeta}(\tau_2)\rangle_{\beta} =\displaystyle \frac{L^3}{6\pi^2 z^2(\tau_2)}{\bf W}(0)\exp\left(-\lambda_{f}|\tau_2|\right)}~.~~~~~~~~~~\eea
 This further implies that the connection between the two point thermal correlation functions computed from the rescaled variable and curvature perturbation variable and their conjugate momenta in the classical limit are given by:
   \bea &&\hll{\displaystyle \langle {f}(\tau_1){f}(\tau_1)\rangle_{\beta}=z^2(\tau_1)\langle{\zeta}(\tau_1){\zeta}(\tau_1)\rangle_{\beta}},\\
&& \hll{\displaystyle \langle {\Pi}(\tau_2){\Pi}(\tau_2)\rangle_{\beta}=z^2(\tau_2)\langle {\Pi}_{\zeta}(\tau_2){\Pi}_{\zeta}(\tau_2)\rangle_{\beta}}.~~~~~~~\eea
   Consequently, the normalisation factor of OTOC in the classical limit for the curvature perturbation variable can be computed as:
  \bea\hll{ {\cal N}^{\zeta}_{\bf Classical}(\tau_1,\tau_2)=\frac{1}{\langle {\zeta}(\tau_1){\zeta}(\tau_1)\rangle_{\beta} \langle {\Pi}_{\zeta}(\tau_2){\Pi}_{\zeta}(\tau_2)\rangle_{\beta}}=z^2(\tau_1)z^2(\tau_2){\cal N}^{f}_{\bf Classical}(\tau_1,\tau_2)}.~~~~~~~~~~~\eea

  \section{Thermal trace operation in terms of wave function of the universe in Cosmology}
  In this appendix, we explicitly mention about the trace operation which is appearing almost everywhere in the definition of two-point, four-point OTOC and during the computation of quantum partition function.
To illustrate this operation in detail let us first consider two operators $\hat{\bf O}_1(\tau_1)$ and $\hat{\bf O}_2(\tau_2)$, which are separated in time scale. Here the trace operation is defined as:
\bea \hll{{\rm Tr}\left[\exp(-\beta\hat{H})~\left[\hat{\bf O}_1(\tau_1),\hat{\bf O}_2(\tau_2)\right]\right]_{(\alpha)}=\int d\Psi_{\alpha}~\langle \Psi_{\alpha}|\exp(-\beta\hat{H})~\left[\hat{\bf O}_1(\tau_1),\hat{\bf O}_2(\tau_2)\right]|\Psi_{\alpha}\rangle}~.~~~~~~~~\eea
Here $\Psi_{\alpha}$ is the wave function of our universe in Cosmology which is define with respect $\alpha$-quantum vacuum state. Since the Hamiltonian under consideration do not have any eigenstate discrete representation, to define the trace operation we actually have to transform this in the continuous representation by incorporating the integral over the wave function of the universe. We have explicitly mention this definition to avoid any further confusion regarding the computation of the thermal trace over the micro-canonical statistical ensemble of the universe.

\newpage
\phantomsection
\addcontentsline{toc}{section}{References}
\bibliographystyle{utphys}

\end{document}